\documentclass[rmp,aps,amsfonts,amssymb,nofootinbib,twocolumn]{revtex4}
\usepackage{graphics}
\usepackage{hyperref}
\usepackage{epsfig}

\begin{document}
\title{Progress and perspectives on the electron-doped cuprates\footnote{\vspace{-1.75cm}{\bf Accepted Reviews of Modern Physics 2010}}}
\author{N.P. Armitage}
\affiliation{The Institute of Quantum Matter, Department of Physics and Astronomy, The Johns
Hopkins University, Baltimore, MD 21218, USA.}

\author{P. Fournier}
\affiliation{Regroupement qu\'eb\'ecois sur les mat\'eriaux de
pointe} \affiliation{D\'epartement de Physique, Universit\'e de
Sherbrooke, Sherbrooke, Qu\'ebec, CANADA, J1K 2R1.}

\author{R.L. Greene}
\affiliation{Center for Nanophysics and Advanced Materials, Department of
Physics, University of Maryland, College Park, MD 20742, USA.}

\date{\today}

\begin{abstract}

Although the vast majority of high-$T_c$ cuprate superconductors are hole-doped, a small family of electron-doped compounds exists. Under investigated until recently, there has been tremendous recent progress in their characterization.  A consistent view is being reached on a number of formerly contentious issues, such as their order parameter symmetry, phase diagram, and normal state electronic structure. Many other aspects have been revealed exhibiting both their similarities and differences with the hole-doped compounds. This review summarizes the current experimental status of these materials.  We attempt to synthesize this information into a consistent view on a number of topics important to both this material class as well as the overall cuprate phenomenology including the phase diagram, the superconducting order parameter symmetry,  electron-phonon coupling, phase separation,  the nature of the normal state, the role of competing orders, the spin-density wave mean-field description of the normal state, and pseudogap effects.

\end{abstract}

\maketitle

\tableofcontents

\section{Introduction}

It has now been over 20 years since the discovery of high
temperature superconductivity in the layered copper-oxide
perovskites by \textcite{Bednorz86}. Despite an almost
unprecedented material science effort, the origin of the
superconductivity or indeed even much consensus on their dominate
physics remains elusive\cite{Fischer07a,Scalapino95a, Alloul08a,Kastner98a,Timusk99a,Damascelli03a,Campuzano04a,Lee06a,Orenstein00a}.

%
\begin{figure}[htbp]
\begin{center}
\includegraphics[scale=0.3,angle=0]{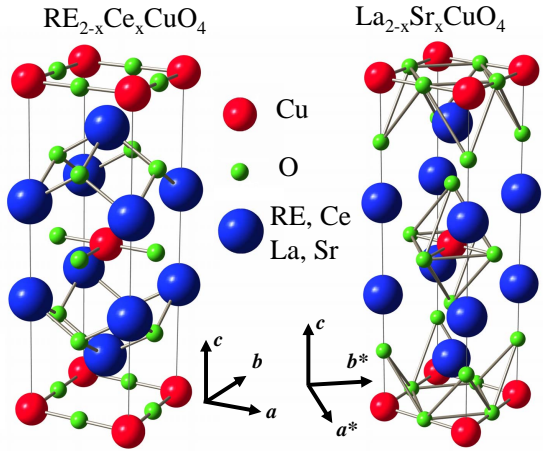}
\caption{A comparison of the crystal structures of the
electron-doped cuprate
RE$_{2-x}$Ce$_{x}$CuO$_4$ and of its closest
hole-doped counterpart La$_{2-x}$Sr$_{x}$CuO$_4$. Here RE is one of a number of rare earth ions, including Nd, Pr, Sm, or Eu.  One should note 
the different directions for the in-plane lattice parameters with 
respect to the Cu-O bonds.}
\label{unit.cell}
\end{center}
\end{figure}
%
%

The undoped parent compounds of high-temperature cuprate
superconductors are known to be antiferromagnetic (AFM) Mott
insulators. As the CuO$_2$ planes are doped with charge carriers,
the antiferromagnetic phase subsides and superconductivity
emerges. The symmetry, or the lack thereof, between doping with
electrons ($n$-type) or holes ($p$-type) has important theoretical
implications as most models implicitly assume symmetry.  One
possible route towards understanding the cuprate superconductors
may come through a detailed comparison of these two sides of the
phase diagram. However, most of what we know about these
superconductors comes from experiments performed on $p$-type
materials.  The much fewer measurements from $n$-type
compounds suggest that there may be both commonalities and
differences between these compounds. This issue of
electron/hole symmetry has not been seriously discussed, perhaps,
because until recently, the experimental database of $n$-type
results was very limited.  The case of electron doping provides an
important additional example of the result of introducing charge
into the CuO$_2$ planes.  The hope is that a detailed study
will give insight into what aspects of these compounds are
universal, what aspects are important for the existence of
superconductivity and the anomalous and perhaps non-Fermi liquid
normal state, what aspects are not universal, and how various
phenomena depend on the microscopics of the states involved.

\begin{figure}[t!]
\includegraphics[width=8.5cm,angle=0]{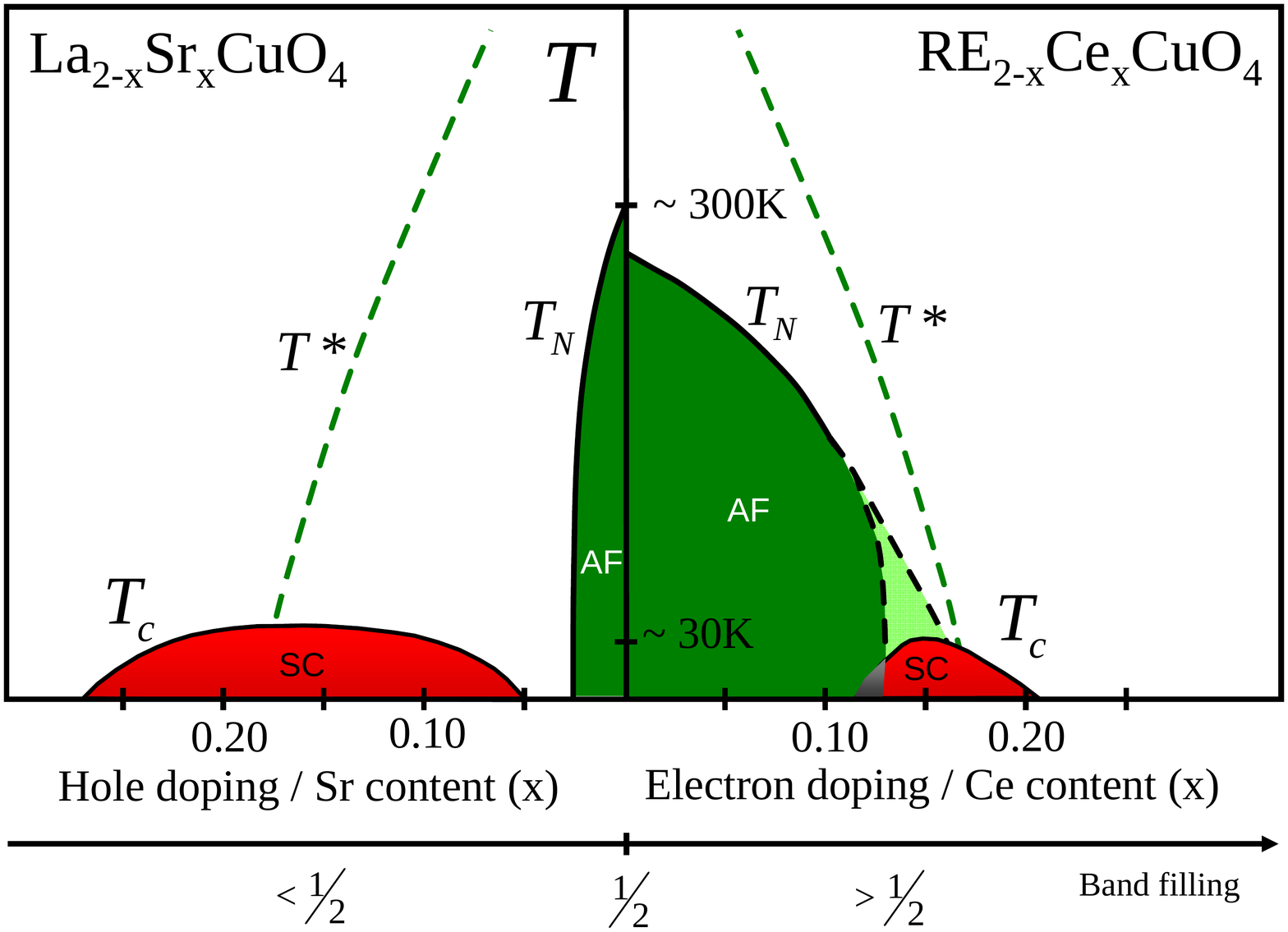}
\caption{Joint phase diagram of the LSCO/NCCO material systems.   The uncertainty regarding the extent of AF on the electron-doped side and its coexistence with superconductivity is shown by the white/green dotted area.  Maximum N\'{e}el temperatures have been reported as 270 K on the electron-doped side in NCO \cite{Mang04a}, 284 K in PCO  \cite{Sumarlin95a} and 320 K on the hole-doped side in LCO\cite{Keimer92a}. T$^*$ indicates the approximate extent of the pseudogap (PG) phase.  It is not clear if PG phenomena have the same origin on both sides of the phase diagram.   At low dopings on the hole-doped side, a spin-glass phase exists (not-shown).  There is as of yet no evidence for a spin-glass phase in the electron-doped compounds.} \label{Phdiagram}\end{figure}

The high-temperature cuprate superconductors are all based on a
certain class of ceramic perovskites.  They share the common feature of square planar
copper-oxygen layers separated by charge reservoir layers.
Fig.~\ref{unit.cell} presents the crystal structures for the canonical
single layer parent materials La$_{2}$CuO$_{4}$ (LCO). The undoped materials are
antiferromagnetic insulators.  With the substitution of Sr for La
in La$_{2}$CuO$_{4}$, holes are introduced into the CuO$_{2}$
planes.  The N\'{e}el temperature precipitously drops and the
material at some higher hole doping level becomes a superconductor
(Fig.~\ref{Phdiagram}). 

Although the majority of high-T$_c$ superconductors are hole-doped compounds there are a small number that can be doped with electrons \cite{Tokura89a,Takagi89a}.  Along with the mostly commonly
investigated compound Nd$_{2-x}$Ce$_{x}$CuO$_{4}$ (NCCO), most members of
this material class have the chemical formula
RE$_{2-x}$M$_{x}$CuO$_{4}$ where the lanthanide rare earth
(RE) substitution is Pr, Nd, Sm or Eu and M is Ce or Th
\cite{Maple90a,Dalichaouch93a}. These are single-layer
compounds which, unlike their other brethren 214 hole-doped
systems (for instance the $T$ crystal structured
La$_{2-x}$Sr$_{x}$CuO$_{4\pm \delta}$ discussed above), have a $T'$ crystal
structure that is characterized by a lack of oxygen in the apical
position (see Fig.~\ref{unit.cell} left).

The most dramatic and immediate difference between electron- and
hole-doped materials is in their phase diagrams.  Only an
approximate symmetry exists about zero doping between
$p$- and $n$-type, as the antiferromagnetic phase is much more
robust in the electron-doped material and persists to much higher
doping levels (Fig.~\ref{Phdiagram}).  Superconductivity occurs in a doping range that is
almost five times narrower.  In addition, these two ground states
occur in much closer proximity to each other and may even coincide
unlike in the hole-doped materials.  Additionally, in contrast to many $p-$type cuprates,  it is found that in doped compounds spin fluctuations remain commensurate\cite{Thurston90a,Yamada99a}.  Various other differences are found including a resistivity that goes like $\approx T^2$ near optimal doping, lower superconducting $T_c$'s and much smaller critical magnetic fields.   One of the other remarkable aspects of the $n$-type cuprates, is that a mean-field spin-density wave treatment of the normal metallic state near optimal doping describes many material properties quite well.  Such a description is not possible in the hole-doped compounds.  Whether this is a consequence of the close proximity of antiferromagnetism and superconductivity in the phase diagram,  smaller correlation effects than the $p$-type,  or the absence of other competing effects (stripes etc.) is unknown.   This issue will be addressed more completely in Sec.~\ref{sec:SDW}.

In the last few years incredible progress has been made both in regards to material quality as well as in the experimental understanding of these compounds.  Unfortunately even in the comparatively under-investigated and well-defined scope of the $n$-type cuprates, the experimental literature is vast and we cannot hope to cover all the excellent work. Important omissions are regrettable, but inevitable.

\section{Overview}

\subsection{General aspects of the phase diagram}

Like many great discoveries in material science the discovery of superconductivity in the
Nd$_{2-x}$RE$_{x}$CuO$_4$ material class came from a blend of careful systematic investigation and serendipity \cite{Khurana89a}.  Along with the intense activity on the hole-doped
compounds in the late 1980's, a number of groups had investigated
$n$-type substitutions.  The work on the
Nd$_{2-x}$RE$_{x}$CuO$_4$ system was motivated after the
discovery of 28K superconductivity in Nd$_{2-x-y}$Sr$_{x}$Ce$_{y}$CuO$_4$, by \textcite{Akimitsu88a}, where it was found that higher cerium concentrations eventually killed the superconducting $T_{c}$ \cite{Tokura89b}.

The
original work on NCCO at the University of Tokyo was done with the
likely result in mind that the material when doped with electrons
may become an $n$-type metal, but it would not become a
$superconductor$.  This would indicate the special role played by
superconducting holes.  Initial work seemed to back up this
prejudice.  The group found that indeed the conductivity seemed to
rise when increasing cerium concentration and that for well-doped
samples the behavior was metallic ($d\rho/dT>0$) for much of the
temperature range.  Hall effect measurements confirmed the
presence of mobile electrons, which underscored the suspicion that
cerium was substituting tetravalently ($+4$) for trivalent
neodymium ($+3$) and was donating electrons for conduction.

At the lowest temperatures the materials were not good metals and
showed residual semiconducting tendencies with $d\rho/dT<0$. In an
attempt to create a true metallic state at low temperature,
various growth conditions and sample compositions were tried.  A
breakthrough occurred when a student, H. Matsubara, accidently
quenched a sample in air from 900$^{\circ}$C to room temperature.
This sample, presumed to be destroyed by such a violent process,
actually showed superconductivity at 10K.   Later it was found that by
optimizing the conditions $T_{c}$ could be pushed as high as 24K
\cite{Tokura89a,Takagi89a}.

\par
\begin{figure}[htbp]
\begin{center}
\includegraphics[scale=0.3,angle=0]{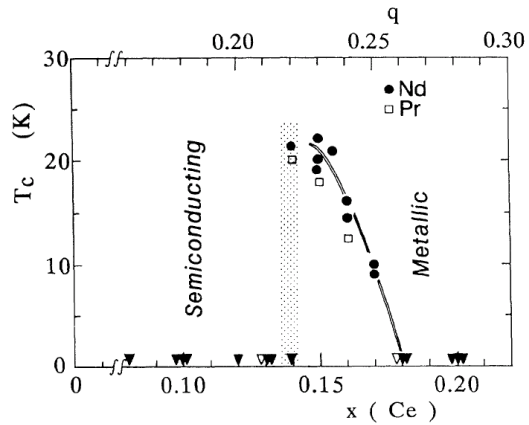}
\caption{Transition temperature T$_c$ as a function of the Ce
concentration in reduced NCCO (circles) and PCCO (squares). The
closed and open triangles indicate that bulk superconductivity was
not observed above 5 K for the Nd or Pr systems respectively.  From the original paper of 
\textcite{Takagi89a}.} \label{Takagi.Tcvsx}
\end{center}
\end{figure}

The first reports of superconductivity in
Nd$_{2-x}$Ce$_{x}$CuO$_4$ and Pr$_{2-x}$Ce$_{x}$CuO$_4$ by
\textcite{Takagi89a} also presented the first phase diagram of
this family (see Fig.~\ref{Takagi.Tcvsx}). In comparison to
La$_{2-x}$Sr$_{x}$CuO$_4$, the doping dependence of the critical
temperature (T$_c$) of these materials was sharply peaked around
$x_{opt} = 0.15$ (optimal doping) corresponding to the maximum
value of T$_{c,opt}$ $\sim 24K$. In fact, at first glance, the $T_c(x)$
relation showed no underdoped regime ($x < x_{opt}$) with $T_c$
rising from zero to its maximum value within a $\Delta x$ of $\sim
0.01$ (from $x = 0.13$ to 0.14). Even the overdoped regime ($x >
x_{opt}$) presents a sharp variation of $T_c$
($\frac{dT_c}{dx}\sim 600 K/$ Ce atom). As discussed below, such steep
dependence of $T_c(x)$ makes the exploration of the phase diagram
very difficult.
%
%
%
%
%
\par
An important difference in the phase diagram of
electron-doped cuprates with respect to their hole-doped cousins
is the close proximity of the antiferromagnetic phase to the
superconducting phase. Using muon spin resonance and rotation on
polycrystalline samples, \textcite{Luke90a} first
found that the Mott insulating parent compound Nd$_{2}$CuO$_4$
has a N\'eel temperature ($T_N$) of approximately 250K\footnote{$T_N$ is a very sensitive function of oxygen concentration.   Subsequent work has shown that the N\'eel temperature of ideally reduced NCO is probably closer to 270 K \cite{Mang04a}}. Upon
substitution of Nd by Ce, $T_N$ of Nd$_{2-x}$Ce$_{x}$CuO$_4$
decreases gradually to reach zero close to optimal doping
(x $\sim$ 0.15)\footnote{The precise extent of the AF state and its coexistence or not with SC is a matter of much debate.   See Sec.~\ref{QCPoint} for further details.}. As noted previously, this should be contrasted with the case of La$_{2-x}$Sr$_{x}$CuO$_4$ in which
antiferromagnetism collapses completely with dopings as small as
$x=0.02$ \cite{Luke90a,Kastner98a}.  In the case of NCCO, the
antiferromagnetic phase extends over a much wider range of cerium
doping. This difference gives us a first hint that electron doping
and hole doping are not affecting the electronic properties in the
exact same manner in the cuprates.  The proximity of
the AF phase to the superconducting one is
reminiscent of the situation in other strongly
correlated electronic systems like some organic
superconductors\cite{Lefebvre00a,McKenzie97a} and heavy fermion
compounds \cite{Joynt02a,Coleman07a,Pfleiderer09a}. However, it still remains unclear up to
now whether or not AF actually coexists with superconductivity. This will
also be discussed in detail below.

\subsection{Specific considerations of the cuprate electronic structure upon electron doping} \label{sec:edoping}

\begin{figure}[htbp]
\begin{center}
\includegraphics[width=2.4in,angle=0]{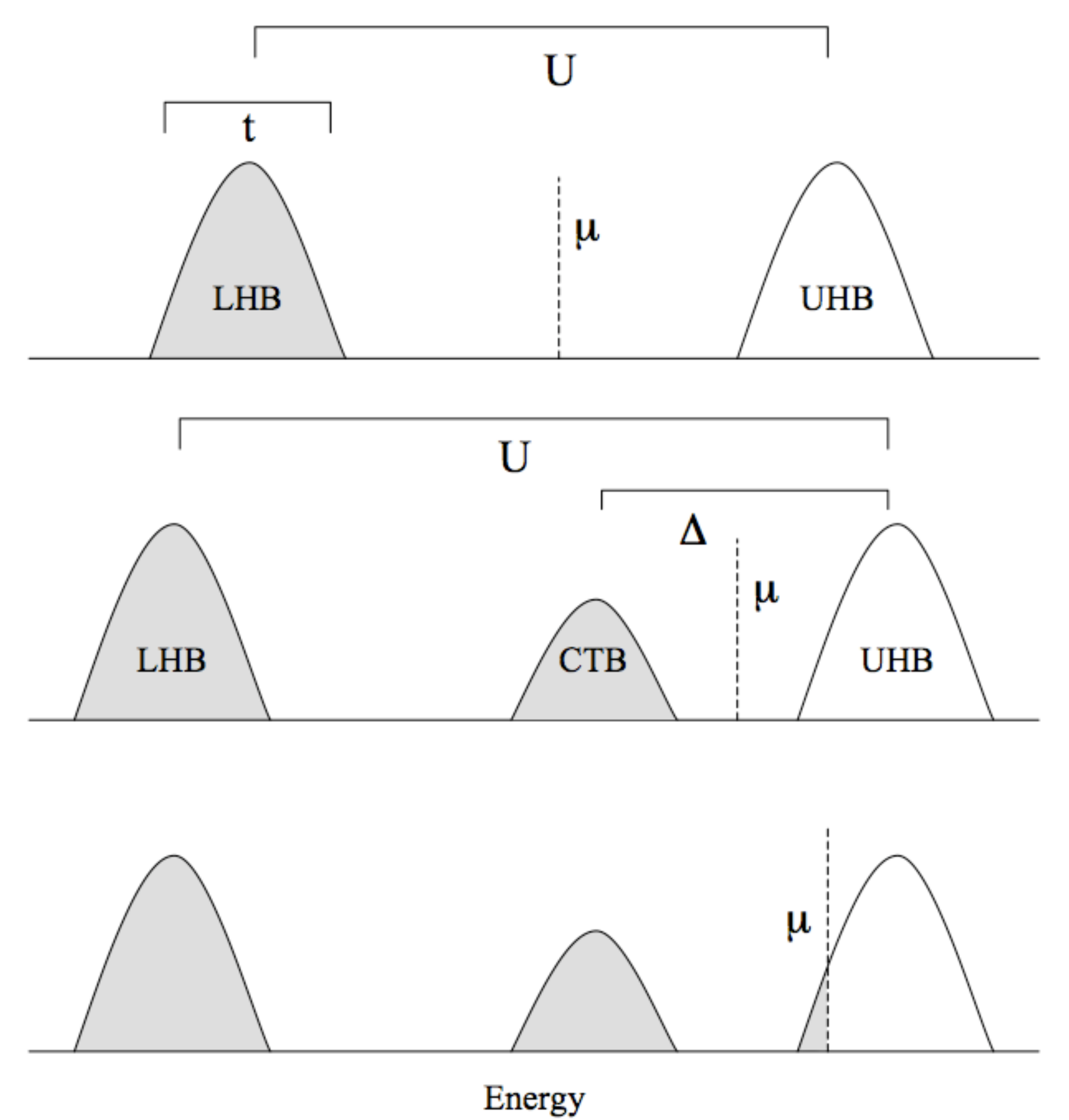}
\caption{(top) Schematic of the one band Hubbard model with $U \gg t$.  At half-filling the chemical potential $\mu$ lies in the middle of the Mott gap.  (middle) Schematic for a charge-transfer band insulator.   $\Delta_{pd}$ may play the role of an effective Hubbard $U$ with the charge transfer band (CTB) standing in for the lower Hubbard band.  (bottom) Upon doping the CTB insulator with electrons, the chemical potential $\mu$ presumably moves into the upper Hubbard band.} \label{Hubbard}
\end{center}
\end{figure}

As emphasized early on \cite{Anderson87a}, the central defining feature of all the cuprates is their ubiquitous CuO$_2$ layers and the resulting strong hybridization between Cu and O orbitals, which is the primary contributor to their magnetic and electronic properties.  It is believed that the states relevant for superconductivity are formed out of primarily in-plane Cu $d_{x^2 - y^2}$ and O $p_{x,y}$ orbitals.  Small admixtures of other orbitals like Cu $d_{z^2 - r^2}$ are also typically present, but these make usually less than a 10 $\% $ contribution \cite{Nucker89a,Pellegrin93a}.  The formal valences in the CuO$_2$ planes of the undoped parent compounds are Cu$^{+2}$ and O$^{-2}$.  With one hole per unit cell, band theory predicts the undoped parent compounds (for instance La$_2$CuO$_4$ and Nd$_2$CuO$_4$) of these materials to be metallic.  In fact they are insulators, which is believed to be driven by a strong local Coulomb interaction that suppresses charge fluctuations.  Mott insulators, where insulation is caused by a strong on-site correlation energy that discourages double occupation, are frequently described by the single-band Hubbard Hamiltonian $H = \sum_{ij} t_{ij} c_j^\dagger c_i   + \sum_i  U n_{i \uparrow } n_{i \downarrow }$.  If $U \gg t_{ij}$ the single band is split into two, the so-called upper and lower Hubbard bands (see Fig.~\ref{Hubbard} (top)) that are respectively empty and completely full at half-filling.  The Hubbard Hamiltonian is the minimal model that includes the strong local interactions, which are believed to be so central to these compounds.  Although frequently referred to as Mott insulators, the cuprates are more properly referred to as charge-transfer band insulators within the Zaanen-Sawatsky-Allen scheme \cite{Zaanen85a}.  Here the energy to overcome for charge motion is not the strong on-site Coulomb interaction on the Cu site, but instead the energy associated with the potential difference beween Cu $d_{x^2 - y^2}$ and O $p_{x,y}$ orbitals $\Delta_{pd}$ (Fig.~\ref{Hubbard} (middle)).  The optical gap in the undoped La$_2$CuO$_4$ is found to be $1.5 - 2$ eV \cite{Basov05a}, which is close to the expected value of $\Delta_{pd}$.  This means that doped holes preferentially reside in the so-called `charge transfer band' composed primarily of oxygen orbitals (with a local configuration primarily $3d^9\underline{L}$ where $\underline{L}$ is the oxygen `ligand' hole), whereas doped electron preferentially reside on the Cu sites (Fig.~\ref{Hubbard} (bottom)) (with a local configuration mostly $3d^{10}$).  In the half-filled cuprates, with a single electron on the Cu $d_{x^2 - y^2}$ orbital and filled O $p_{x,y}$ orbitals these compounds can be then described by a three band Hubbard model, which generally takes into account hopping, the onsite Coulomb interactions $U_d$, the energy difference between oxygen and copper orbitals  $\Delta_{pd}$, and intersite interactions $V_{pd}$ \cite{Varma87a}.

A number of simplifications of the three band model may be possible. \textcite{Zhang88a} argued that the maximum gain in hybridization energy is gained by placing doped holes in a linear combination of the O $p_{x,y}$  orbitals with the same symmetry as the existing hole in the Cu $d_{x^2 - y^2}$ orbital that they surround.  This requires an antisymmetry of the wave function in their spin coordinates, so that the two holes must form a singlet.  They argued that this split-off state retained its integrity even when intercell hopping is taken into account.  With this simplification, the separate degrees of freedom of the Cu and O orbitals are removed and the CuO$_2$ plaquette is real space renormalized to an effective site centered on Cu.  In this case, it may be possible to reduce the three band model to an effective single band one, where the role of the lower Hubbard band is played by the primarily oxygen based charge transfer band (of possibly singlet character) and an effective Hubbard gap given primarily by the charge transfer energy $\Delta_{pd}$.  Even though the local structure of the states is different upon hole or electron doping (Fig.~\ref{Hubbard} (bottom)), both would be of singlet character ($3d^9\underline{L}$ and $3d^{10}$ respectively).

In general, it may be possible to reduce the one band Hubbard model even further by taking the limit of large effective $U$ (the charge transfer energy $\Delta_{pd}$ in the cuprates).  One can find an effective Hamiltonian in the subspace of only singly occupied sites.  Localized electrons with oppositely aligned spins on adjacent sites can still reduce their kinetic energy by undergoing virtual hops to neighboring sites.  As such, hops are only allowed with neighboring electrons being anti-aligned, this gives an effective spin exchange interaction which favors antiferromagnetism.   The effect of the upper Hubbard band comes in only through these virtual hops.   Second order perturbation theory gives an energy lowering for oppositely directed spins of $4 t^2/U$.  By neglecting correlated hopping terms, the one band Hubbard model can then be replaced by the so-called $`t-J'$ model which is a possible minimal model for the cuprates.  The $t-J$ model can be refined by the inclusion of next-nearest $t'$ and next-next-nearest $t''$ neighbor hopping terms.

We should note that the reduction of the three band model to models of the $t-U$ or $t-J$ variety is still very controversial.  \textcite{Emery88a} argued that in fact the quasiparticles of the three band model have both charge and a spin of $1/2$, in contrast to the singlets of Zhang and Rice and that the $t-J$ model is incomplete.  Similarly, Varma has proposed that one must consider the full three band Hubbard model \cite{Varma97a,Varma99a} and that non-trivial phase factors between the bands become possible at low energies, which leads to a state with orbital currents on the O-Cu-O triangular plaquettes.  The order associated with these currents has been proposed to be a candidate for the pseudogap phase.   Moreover, questions regarding even  the validity of the parameter $J$ remain.  It is derived for the insulating case of localized electrons.  Is it still a valid parameter when many holes or electrons have been introduced?  Although throughout this review we will frequently appeal to the insight given by these simpler models (single band Hubbard or $t-J$), we caution that it is far from clear whether or not these models are missing some very important physics.  

As mentioned, doped electrons are believed to reside primarily on Cu sites.  This nominal $3d^{10}$ atomic configuration of an added electron has been confirmed $via$ a number of resonant photoemission studies which show a dominant Cu $3d$ character at the Fermi level \cite{Allen90a,Sakisaka90a} in electron-doped compounds.   Within the context of the single band Hubbard or $t-J$ models, the effective orbital of a hole-doped into the CuO$_2$ plane $3d^9 \underline{L}$  may be approximated as a singlet formed between the local Cu$^{+2}$ spin and the hole on the oxygen atoms.  This is viewed as symmetry equivalent to the state of a spinless hole $3d^8$ on the copper atom (albeit with different effective parameters). Although their actual local character is different, such an approximation makes the effective model describing holes and electrons doped into $3d^9$ virtually identical between the $p-$ and $n-$type cuprates leading to the prediction of an electron-hole symmetry.  The noted $asymmetry$ between the two sides of the phase diagram means that there are specific extra considerations that must go into the two different cases.

There are a number of such considerations regarding the cuprate electronic structure that apply specifically to the case of electron doping.   One approach
to understanding the relative robustness of antiferromagnetism in the electron-doped compounds has been to consider {\it
spin-dilution models}.  It was shown $via$ neutron scattering
that Zn doping into La$_{2}$CuO$_{4}$ reduces the N\'{e}el
temperature at a similar rate as Ce doping in
Pr$_{2-x}$Ce$_{x}$CuO$_{4\pm \delta}$ \cite{Keimer92a}.  Since Zn
substitutes in a configuration that is nominally a localized $d^{10}$ filled
shell, it can be regarded as a spinless impurity.  In this regard
Zn substitution can be seen as simple dilution of the spin system.
The similarity with the case of Ce doping into
Pr$_{2-x}$Ce$_{x}$CuO$_{4\pm \delta}$ implies that one of the effects of electron doping is to dilute the spin system by neutralizing the spin on a
$d^{9}$ Cu site.  It subsequently was shown that the reduction of
the N\'{e}el temperature in Nd$_{2-x}$Ce$_{x}$CuO$_{4\pm \delta}$
comes through a continuous reduction of the spin stiffness
$\rho_{s}$ which is consistent with this model \cite{Matsuda92a}.
In contrast, in the hole-doped case \textcite{Aharony88a} have
proposed that only a small number of holes is required to
suppress antiferromagnetism because they primarily exist on the
in-plane oxygen atoms and result in not spin-$dilution$ but
instead spin-$frustration$. The oxygen-hole/copper-hole
interaction, whether ferromagnetic or antiferromagnetic, induces an effective ferromagnetic Cu-Cu interaction.  This interaction
competes with the antiferromagnetic superexchange and frustrates
the N\'{e}el order so a small density of doped holes has a
catastrophic effect on the long-range order.  This
additional frustration does not occur with electron doping as
electrons are primarily introduced onto Cu sites.  This comparison of Ce with Zn doping is compelling, but cannot be exact as Zn does not add itinerant charge carriers like Ce does,
as its $d^{10}$ electrons are tightly bound and can more
efficiently frustrate the spin order.   Models or analysis which takes into account electron itinerancy must be used to describe the phase diagrams.   But this simple model gives some indication how the principle interactions can be very different between electron and hole-doping if one considers physics beyond the $t-J$ or single band Hubbard models.

Alternatively,  it has been argued that the observed asymmetry between hole and electron doping (say for instance in their phase diagrams) can be understood within single band models by considering the fact that  the hopping terms, which have values $t>0$, $t'<0$, and $t''>0$ for hole doping assume values $t<0$, $t'>0$, and $t''<0$ for electron doping.  These sign reversals  arise in the particle-hole transformation $c^\dagger_{i \sigma}  \rightarrow (-1)^i c_{i \sigma}$  where $c$ and $c^\dagger$ are particle creation and annihilation operators and $i$ is a site index\footnote{Note that for the case of long-range AF order, the final results in either doping case are invariant with respect to the sign of $t$, as a change in the sign of $t$ is equivalent to a shift of the momentum by the AF reciprocal lattice vector ($\pi$,$\pi$). }.

Since the next nearest $t'$ and next-next nearest neighbor $t''$ terms facilitate hopping on the same sub-lattice of the N\'eel state, the energy and stability of antiferromagnetic order is very sensitive to their values.  For instance, it has been argued that the greater stability of antiferromagnetism in the electron-doped compounds is primarily a consequence of $t'>0$.   This scenario is supported by a number of numerical calculations and analytical treatments  \cite{Tohyama94a,Tohyama01a,Singh02a,Gooding94a,Pathak09a}.  

Models that account for the effective sign change of the hopping parameter, successfully account for the fact that the lowest energy hole addition (electron removal) states are near ($\pi/2$, $\pi/2$) \cite{Wells95a}, while the lowest energy electron addition states (hole removal) are near  ($\pi$, 0) \cite{Armitage02a}.   This manifests as a small hole-like Fermi arc (or perhaps pocket \cite{Doiron07a, LeBoeuf07a}) for low hole dopings near the ($\pi/2$, $\pi/2$) point and a small electron-pocket near the  ($\pi$, 0) point for low electron dopings \cite{Armitage02a}.  Such considerations also mean that the insulating gap of the parent compounds is indirect.  Aspects of $t-J$ type models applied to both sign of charge carriers have been reviewed by \textcite{Tohyama04a}.

Although it may be that such a mapping can be applied so that the same model (for instance $t-J$) captures aspects of the physics for both hole and electron doping, the object that is undergoing hopping in each case has very different spatial structure and local character ($3d^9\underline{L}$ ZR singlet vs. $3d^{10}$ respectively).  It is reasonable to expect that the values for their effective hopping parameters could be very different\footnote{It is also reasonable to expect that their interaction with degrees of freedom not explicitly considered in these electronic models, such as the strength of their lattice coupling, could also be very different.  Aspects related to lattice coupling are discussed in Sec.~\ref{sec:ephonon} below.}.  In this regard, \textcite{Hozoi08a} found $via$ $ab$ $initio$ quantum chemical calculations very different values for the $bare$ hopping parameters of the $3d^{10}$ and $3d^9\underline{L}$ states.  They found $|t|=0.29$, $|t'| = 0.13$, and $|t''| = 0.045$ for  $3d^{10}$ and $|t|=0.54$, $|t'| = 0.305$, and $|t''| = 0.115$ for $3d^9\underline{L}$.  Interestingly, they find that after including interactions the $renormalized$ values are much closer to each other, but still with some significant differences. ($|t|=0.115$, $|t'| = 0.13$, and $|t''| = 0.015$  for $3d^{10}$ and $|t|=0.135$, $|t'| = 0.1$, and $|t''| = 0.075$ for $3d^9\underline{L}$.   All values in eV.).  Similar magnitudes of $t$ and $t'$  for electron and hole doping have also been found in Cu-O cluster calculations.  \textcite{Hybertsen90a} used $ab$ $initio$ local-density-functional theory to generate input parameters for the three-band Hubbard model and computed spectral functions exactly on finite clusters using the three band Hubbard model and compared the results with the spectra of the one band Hubbard and the $t-t'-J$ models.  The extracted effective nearest neighbor and next-nearest neighbor hopping parameters were found to be almost identical at $t=0.41$ and $|t'|=0.07$ eV for electron doping and $t=0.44$ and $|t'|=0.06$ eV for hole doping.   $J$ was found in this study to be 128 $\pm$ 5 meV which is in reasonable agreement with neutron \cite{Mang04a} and two-magnon Raman scattering \cite{Lyons88a,Singh89a,Sulewski90a,Blumberg96a}.  Somewhat similar results for the hole-doped case were obtained by \textcite{Bacci91a}.  However these results conflicted with those of \textcite{Eskes89a} who found slightly different values between hole ($t=- 0.44$, $t' = 0.18$ eV) and electron doping ($t=0.40$, $t' = - 0.10$ eV) in their numerical diagonalization study of Cu$_2$O$_7$ and Cu$_2$O$_8$ clusters.  In a similar calculation but with slightly different parameters and also taking into account the apical oxygen for the hole-doped case \textcite{Tohyama90a} found the even more different $t= -0.224$ and $t'=0.124$ eV for the $p$-type and $t= 0.3$ and $t'= - 0.06$ eV for the $n$-type cases.  This shows the strong sensitivity that these effective parameters likely have on the local energies and the presence of apical oxygen.  Despite differences in the estimates for these parameters it is still remarkable that in all these studies the values of the hopping parameters for holes and electrons are so close to each other considering the large differences in these states' local character.  This shows the principal importance that correlations have in both cases in renormalizing their dispersions.

It is interesting to note that although very different behavior of the electronic structure is expected and indeed found at low dopings (FS pockets around ($\pi$,0) vs.  ($\pi$/2,$\pi$/2)),  it appears that at higher dopings in both systems the set of small Fermi pockets go away and a large Fermi surface centered around the ($\pi$, $\pi$) point emerges \cite{King93a,Anderson93a,Armitage02a}.  In the electron-doped materials, aside from the `hot-spot' effect discussed in detail below, the Fermi surface resembles the one calculated $via$ LDA band structure calculations.

A number of workers \cite{Kyung04a,Kyung03a,Kusko02a,Tremblay06a}  have pointed out that due to the different size of the effective onsite repulsion $U$ and the electronic bandwidth $W$ in the $n$-type systems, the expansion parameter $U/W$ is less than unity, which puts the electron-doped cuprates in a weaker correlated regime than the $p$-type compounds.   Among other things, this makes Hubbard model-like calculations more amenable\footnote{In a reanalysis of optical and ARPES data \textcite{Xiang08a} in fact have argued that the charge transfer gap $\Delta$, which is the effective onsite Hubbard repulsion is even smaller than usually assumed in the electron-doped compounds.   They claim $\Delta \approx$ 0.5 eV.}.  Smaller values of $U/W$ physically derive from better screening and the Madelung potential differences noted above, as well as a larger occupied bandwidth.   This weaker coupling may allow for more realistic comparisons between theoretical models and experiment and even serve as a check on what models are most appropriate for the more correlated hole-doped materials.   The fact that we may be able to regard the $n$-type systems as somewhat weaker correlated is manifest in a number of ways, including the remarkable fact that a mean-field spin-density wave (SDW) like treatment of the normal state near optimal doping can capture many of the gross features of transport, optics, and photoemission quite well (Sec~\ref{sec:SDW}).  It also makes the issue of AF fluctuations easier to incorporate.  For instance, the Two-Particle Self-Consistent \cite{Kyung04a} approach to the Hubbard model allows one to predict the momentum dependence of the PG in the ARPES spectra of the $n$-type cuprates, the onset temperature of the pseudogap $T^*$, and the temperature and doping dependence of the AF correlation length.   A similar treatment fails in hole-doped compounds with their corresponding larger values of $U/W$.  The $n$-type compounds appear to be the first cuprate superconductors whose normal state lends itself to such a detailed theoretical treatment.  Recent work by \textcite{Weber09a}, even claims that $U/W$ in the electron-doped cuprates 
is low enough to be below the critical value for the Mott transition and hence that the $x \rightarrow 0$ insulating behavior must derive from antiferromagnetism.   To distinguish from the $p$-type Mott systems, they call such a system a Slater insulator.  These issues are dealt with in more detail below.

\subsection{Crystal structure and solid-state chemistry}
\label{sec:phasediag}
\par
RE$_2$CuO$_4$ with RE = Nd, Pr, Sm, Eu, Gd crystallizes in the
so-called $T'$ crystal structure and are typically doped with Ce\footnote{There is at least one more class of superconducting electron-doped cuprates, the so-called infinite layer compounds.  Sr$_{0.9}$La$_{0.1}$CuO$_2$ (SLCO) has been known for almost as long as the (RE)CCO material class \cite{Siegrist88a}. It has the highest T$_c$ ($\approx$ 42K) of any $n-$doped cuprate. However, there has been comparatively little research performed on it due to
difficulties in sample preparation.  This system will be touched only briefly here.} \footnote{Another intriguing path to doping in the (RE)CCO electron-doped family is Nd$_2$CuO$_{4-y}$F$_y$, which uses an under investigated fluorine substitution for oxygen \cite{James89a} and no Ce doping}.  These compounds are tetragonal with typical lattice parameters of a = b $\sim$
3.95 {\rm \AA} and c $\sim$ 12.15 {\rm \AA} . Their structure is a close cousin of $T$ structure
La$_2$CuO$_4$ (LCO) compound.  $T'$ is represented by the $D^{17}_{4h}$ point
group (I4/mmm).  It has a body-centered unit cell where the copper
ions of adjacent copper-oxygen CuO$_2$ layers are displaced by
(a/2,a/2) with respect to each other \cite{Kastner98a}. In Fig.
~\ref{unit.cell}, we compare the crystal structure of these
parent compounds.

Although aspects of the LCO and NCO crystal structures are
similar, closer inspection \cite{Kwei89a,Marin93a} reveals notable
differences. First, the coordination number of the in-plane copper
is different.  The $T'$ structure has no apical oxygen above the in-plane Cu and hence only four oxygen ions O(1)  surround each copper.  The $T$ structure has 6 surrounding O atoms, two of which are in the apical position.

The different relative positions of the reservoir oxygens O(2) with respect to the $T$-structure results in an expanded in-plane unit cell with respect to La$_2$CuO$_4$ that allows a further 
decrease of the unit cell volume with decreasing rare earth ionic 
radius (see Table~\ref{ionic.size}). While La$_{2}$CuO$_4$ with 
the $T$-structure has typical in-plane lattice parameters on the order of a
= b $\sim$ 3.81 {\rm \AA} and c $\sim$ 13.2 {\rm \AA}
\cite{Kastner98a}\footnote{Note that the real crystallographic in-plane lattice parameters of LCO 
are actually a$^\ast$ $\sim$ b$^\ast = \sqrt{2}$a as shown in 
Fig.~\ref{unit.cell}. In this case, a$^\ast$ and b$^\ast$ are 45$^{\circ}$ with respect to the Cu-O bonds.} and a unit cell volume of 191.6 {\rm \AA}$^3$, the largest undoped $T'$ phase cuprate,
Pr$_{2}$CuO$_4$, has a = b $\sim$ 3.96 {\rm \AA} and c $\sim$ 12.20
{\rm \AA} and similar unit cell volume of 191.3 {\rm \AA}$^3$.  The smallest, Eu$_{2}$CuO$_4$, has a = b $\sim$ 3.90
{\rm \AA} and c $\sim$ 11.9  {\rm \AA} \cite{Nedilko82a,Uzumaki91a,Fontcuberta96a,Vigoureux95a}\footnote{The other (RE)CCO compound in this series Gd$_{2}$CuO$_4$ is not a superconductor upon Ce doping}.   The second notable difference arising from the expanded in-plane lattice parameters is that the rare-earth and oxygen ions in the
reservoirs are not positioned in the same plane.
\par

Until recently it was believed that only
$T'$ crystal structures without apical oxygen can be
electron doped.  This was understood within a Madelung
potential analysis, where the local ionic potential on the Cu site
is influenced strongly by the presence of an O$^{-2}$ ion in the
apical site immediately above it \cite{Ohta91a,Torrance89a}.  As doped
electrons are expected to primarily occupy the Cu site, while
doped holes primarily occupy in-plane O sites the local ionic
potentials play a strong role in determining which sites mobile
charges can occupy. Recent developments may
not be entirely consistent with this scenario as there has been a report of superconductivity in 
$T$ phase La$_{2-x}$Ce$_x$CuO$_4$  \cite{Oka03a}.   However this report contrasts with various thin film studies, which claim that although $T$ phase La$_{2-x}$Ce$_x$CuO$_4$ can be electron-doped (i.e with Ce in valence state $+4$) it does not become a superconductor \cite{Tsukada05a,Tsukada07a}.   There has also been the recent remarkable work by \textcite{Segawa06a} who reported ambipolar doping of the (Y$_{1-z}$La$_z$)(Ba$_{1-y}$La$_y$)$_2$Cu$_3$O$_y$ (YLBLCO) system.  They found that La$^{+3}$ substitutes for Ba$^{+2}$ at the 13 $\%$ level.   By varying oxygen content $y$ between 6.21 and 6.95 by controlled annealing, they could tune the in-plane resistivity through a maximum at 6.32.   This was interpreted as an ability to tune the material through the Mott insulating state from hole- to electron-doping.   Electron-doping was confirmed by negative Hall and Seebeck coefficients for $y < 6.32$.   Subsequent photoemission work has shown that the chemical potential crosses a CT gap of $\sim$ 0.8 eV upon cross the $n/p$ threshold \cite{Ikeda09c} .  This work represents the first demonstration of ambipolarity in a single material system, which deserves further investigation.

\par
\begin{figure}[htbp]
\begin{center}
\includegraphics[width=8.5cm,angle=0]{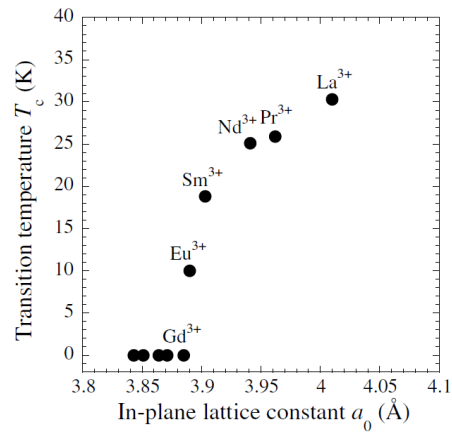}
\caption{The highest superconducting onset temperature versus in  
plane lattice constant in RE$_{2-x}$Ce$_{x}$CuO$_{4+\delta}$.  Note  
that $T'$ phase  La$_{2-x}$Ce$_{x}$CuO$_{4+\delta}$ can only  be 
stabilized in thin film form as discussed in the text.  From  
\textcite{Naito02a}.}
\label{RETcprogression}
\end{center}
\end{figure}
\begin{figure}[htbp]
\begin{center}
\includegraphics[scale=0.30,angle=0]{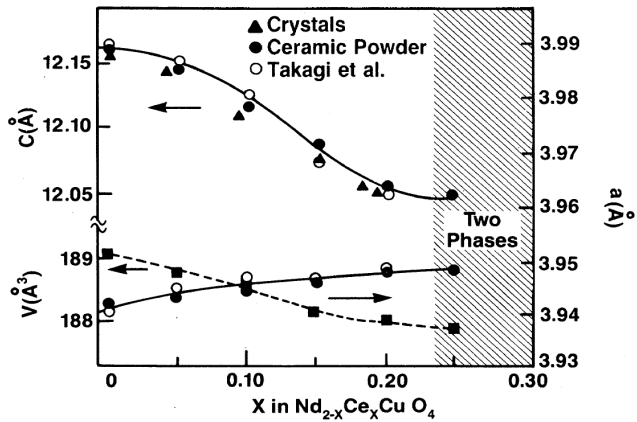}
\caption{The lattice parameters of NCCO single crystals and ceramic  
powders as a function
of cerium content $x$ showing the decreasing unit cell volume with  
increasing $x$. Solid circles
and triangles refer to powder samples and single crystals  
respectively. Open circles are the
results from \cite{Takagi89a}. From  \textcite{Tarascon89a}.}
\label{abc.vs.x}
\end{center}
\end{figure}

As observed originally by \textcite{Takagi89a} and expanded upon by
\textcite{Fontcuberta96a}, the actual phase diagram of the (RE)CCO electron-doped family is sensitive to the rare-earth ion size. The smaller the ionic radius of the rare earth the smaller the optimal $T_{c,max}$ (see Table~\ref{ionic.size}, Fig.~\ref{RETcprogression}, and \textcite
{Fontcuberta96a,Vigoureux95a} and references therein).  The most  
obvious effect of the decreasing ionic size (Table ~\ref
{ionic.size})  is a decrease of roughly 2.6$\%$ of the c-axis and 1.5$
\%$ of the in-plane lattice constant across the series. One should note that the  
lattice also contracts with Ce substitution in Pr$_{2-x}$Ce$_x$CuO$_4$ (PCCO) and NCCO  
\cite{Tarascon89a,Fontcuberta96a,Vigoureux95a} as shown in Fig. \ref{abc.vs.x}.
%
%
\begin{table}[htbp]
     \begin{center}
         \begin{tabular}{|c||c|c|c|c|c|c|c|}
         \hline
                     & $La^{3+}$ & $Pr^{3+}$ & $Nd^{3+}$ & $Sm^{3+}$  
& $Eu^{3+}$ & $Gd^{3+}$ & \textbf{$Ce^{4+}$}    \\
         \hline
         \hline
               Ionic     &   &  &  &  &  &  &    \\
                     radius ({\rm \AA})  & 1.30 & 1.266 & 1.249 &  
1.219 & 1.206 & 1.193 & \textbf{1.11}  \\
         \hline
                     a ({\rm \AA}) & $-$ & 3.9615    & 3.942 & 3.915  
& 3.901 & 3.894 & - \\
         \hline
                     c ({\rm \AA})   & $-$   & 12.214    & 12.16 &  
11.97 & 11.90 & 11.88 & - \\
         \hline
                     t   & $-$   & 0.856 & 0.851 & 0.841 & 0.837 &  
0.832 & - \\
         \hline
                     T$_{c,max} (K)$ & $-$*  & 22 & 24   & 20    &  
13    & 0 & - \\
         \hline

         \end{tabular}
     \caption{Dependence on ionic radius of unit cell parameters of  
the parent compound, the tolerance factor $t$ \cite
{Cox89a,Nedilko82a,Uzumaki91a,Muller75a}, and the maximum transition  
temperature $T_{c,max}$ obtained by cerium doping for $x \sim 0.15$  
\cite{Maple90a,Fontcuberta96a}. The  $T'$ (La,Ce)$_2$CuO$_4$ can only  
be easily stabilized in thin films, giving $T_{c,max} \sim$ 25K \cite
{Naito02a}. Ionic radii are given for a coordination of 8 according to \textcite{Shannon76a} 
and references therein. 
See Sec.~\ref{sec.materials}.}
     \label{ionic.size}
     \end{center}
\end{table}

%
%
%
%
%
%
%
%
\par
As the RE-O distance gradually decreases with decreasing ionic
radius, the crystal structure is subjected to increasing internal
stress indicated by a decreasing tolerance factor $t \equiv
\frac{r_{RE}+r_{O}}{\sqrt{2}(r_{Cu}+r_{O})}$ where $r_{RE}$,
$r_{O}$ and $r_{Cu}$ are respectively the ionic sizes of the rare
earth, the oxygen and the copper ions (see Table
\ref{ionic.size}).  Several neutron and high-resolution x-ray
scattering studies report structural distortions when the Cu-O
bond length becomes too large with respect to the shrinking RE-O
ionic distance which promotes Cu-O-Cu bond angles that deviate from
180$^{\circ}$. The most striking result is the distorted structure
of (non-superconducting) Gd$_2$CuO$_4$ with its commensurate  
distortion corresponding
to the rigid rotation of the four planar oxygen atoms around each
copper sites\cite{Braden94a,Vigoureux97a}. This distortion leads
to antisymmetric exchange term of Dzyaloshinski-Moriya
type that may account for the weak ferromagnetism of Gd$_2$CuO$_4$
\cite{Oseroff90a,Stepanov93a}.
\par
\begin{figure}[htbp]
\begin{center}
\hspace{5mm}
\includegraphics[scale=0.35,angle=0]{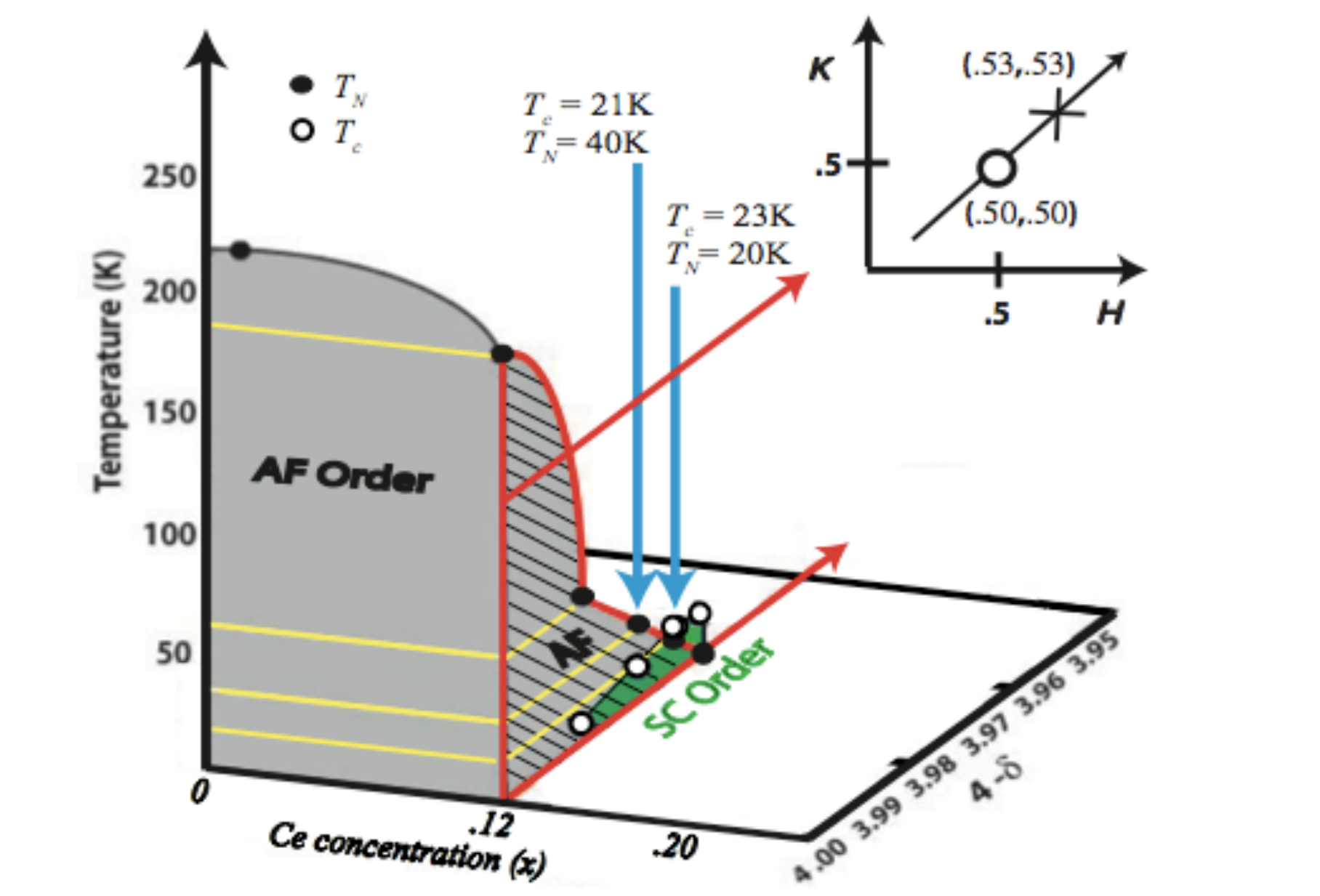} \\
\includegraphics[scale=0.35,angle=0]{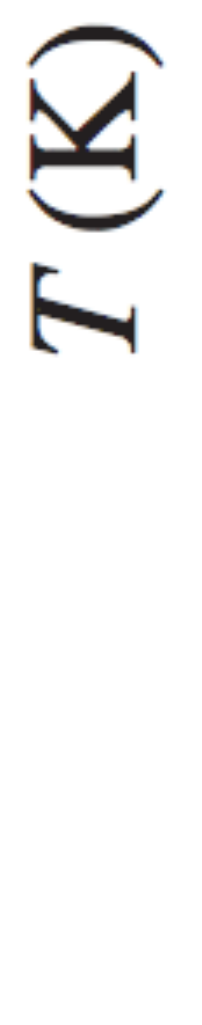}
\includegraphics[scale=0.35,angle=0]{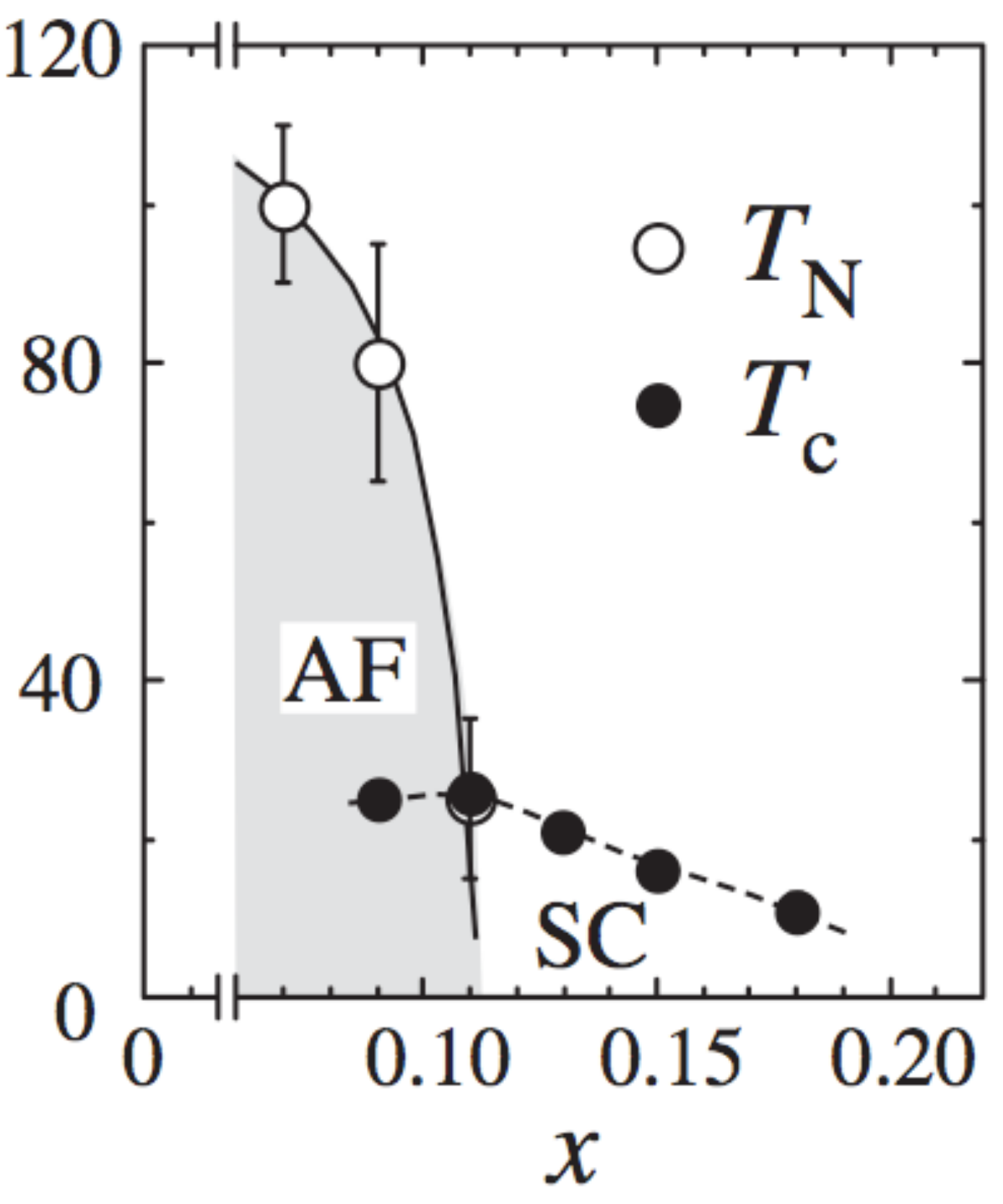}
\caption{(top) $\delta- x$ phase diagram of PLCCO  \cite{Wilson06c}. (bottom) Phase diagram for PLCCO as a function of Ce content $x$ \cite{Fujita08a} as determined by neutron scattering and SQUID measurements.  } 
\label{PLCCOpd}
\end{center}
\end{figure}

At the other extreme, for large ionic radius, the crystal
structure approaches the $T'$ to $T$ structural transition, which is  
expected for
an ionic radius between those of Pr and La \cite{Fontcuberta96a}.
PCO is at the limit of the bulk $T'$ phase: the next compound in the RE
series with a larger atomic radius is LCO which crystallizes not
in the $T'$ phase, but instead in the more compressed $T$-phase form
where the out-of plane oxygens are in apical positions as mentioned  
previously.  It does seem to be possible to stabilize a doped $T'$  
phase of La$_{2-x}$Ce$_{x}$CuO$_{4+\delta}$ by substitution of La by  
the smaller Ce ion, although bulk crystals are not of that high  
quality due to the low growth temperatures required \cite
{Yamada94a}.  However it has been shown by \textcite
{Naito00a,Naito02a,Wu09a} that the $T'$ phase of La$_{2-x}$Ce$_{x}$CuO
$_{4+\delta}$ (LCCO) can be strain stabilized in thin film form  
leading to high quality superconducting materials with $T_c$ as high  
as 27K.  One can also drive the $T'$ structure even closer to the  
structural instability by only partial substitution of Pr by La \cite
{Koike92a,Fontcuberta96a} as Pr$_{1-y-x}$La$_y$Ce$_{x}$CuO$_{4 \pm  
\delta}$. This substitution provokes a significant modification to  
the phase diagram,
with an optimal T$_{c,opt} \sim$ 25K for Pr$_{1-x}$LaCe$_{x}$CuO$_{4  
\pm \delta}$ at $x \sim 0.11$ and superconductivity extending as low  
as $x=0.09$ and as high as $x=0.20$ \cite{Fujita03a}  (See Fig. \ref
{PLCCOpd}).  The mechanism leading to a different phase diagram  in PLCCO remains a mystery,
but it has been suggested that it corresponds to the ability to  
remove a larger amount of oxygen during the necessary reduction process compared
to PCCO and NCCO, which leads to larger electron
concentrations \cite{Kuroshima03a}.

\par

There have been many fewer studies of 
the infinite layer class of electron-doped cuprate superconductors 
Sr$_{1-x}$Nd$_{x}$CuO$_2$ (SNCO)\cite{Smith91a} and Sr$_{1-x}$La$_{x}$CuO$_2$ (SLCO)\cite{Kikkawa92a}.  To date no single crystals have been produced and the data up to now relies on 
ceramic samples\cite{Jorgensen93a,Ikeda93a,Kim02a,Khasanov08a} and 
thin films\cite{Naito02a,Nie03a,Karimoto04a,Leca06a,ZZLi08a}. 
Its simple structure is based 
on alternating of CuO$_2$ planes with Sr (La) layers with lattice parameters 
a = b $\sim$ 3.94 {\rm \AA} and c $\sim$ 3.40 {\rm \AA}. Electron doping is 
suggested because the La (Nd) nominal valence is $+3$ as compared to Sr's $+2$ valence. 
Electron-type doping is supported by a negative thermopower\cite{Kikkawa92a} 
and XANES results \cite{Liu01a} confirming the presence of Cu$^{1+}$ ions. 
However, a systematic study of the Hall effect with doping is still lacking, which makes difficult
any comparison with the well-established behavior of transport for $T'$ electron-doped 
(RE)CCO (see Section~\ref{Transport}).

\begin{figure}[htbp]
\begin{center}
\includegraphics[width=7.5cm,angle=0]{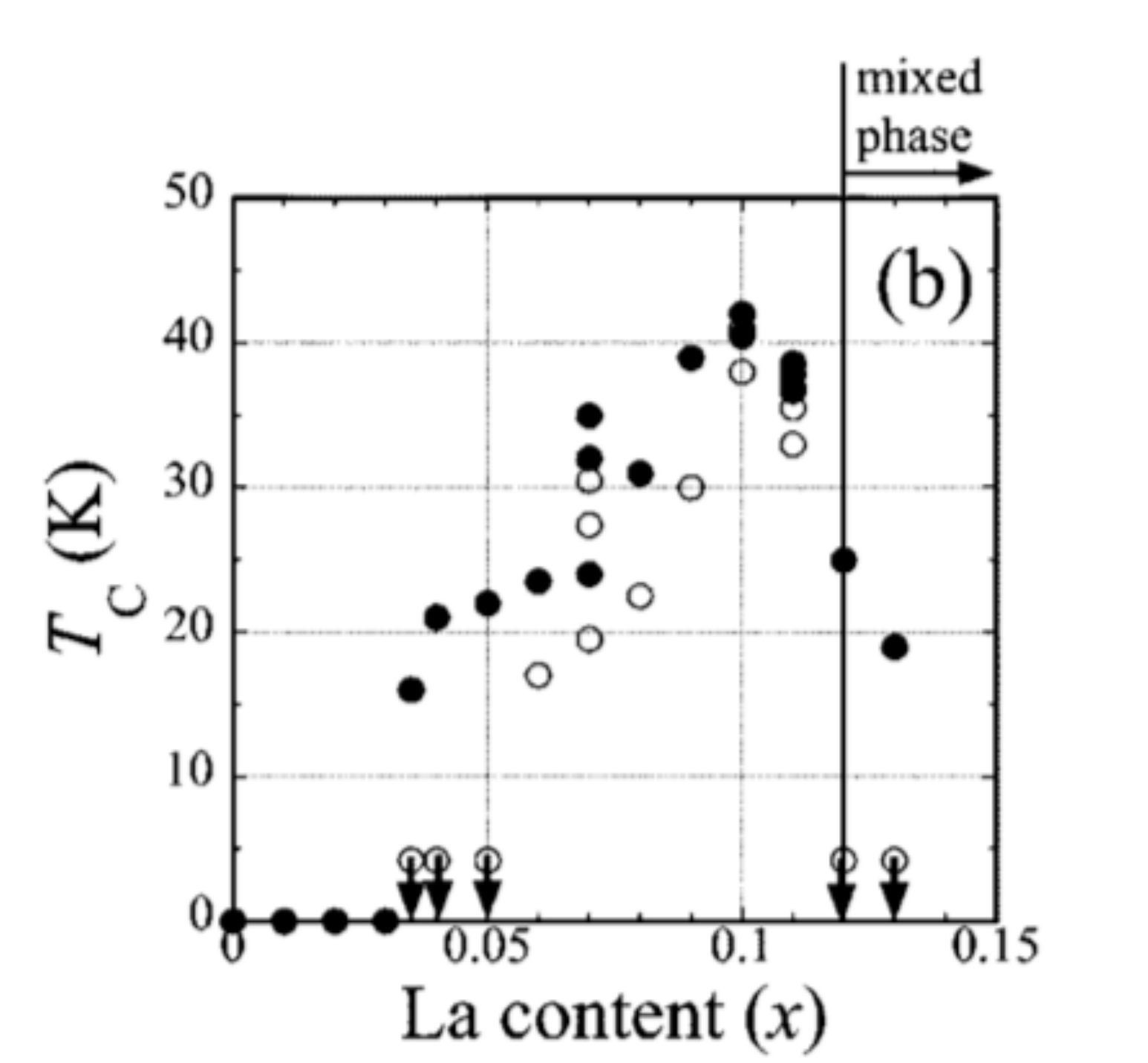}
\caption{Extracted values of $T_c$ for SLCO as a function of doping. Solid and open circles are for the onset and zero resistivity respectively. From  \textcite{Karimoto04a}.} 
\label{SLCO.PhD}
\end{center}
\end{figure}

The most complete phase diagram for the 
Sr$_{1-x}$La$_x$CuO$_2$ system has been established from the MBE 
film studies \cite{Karimoto04a}. In this exploration the ab-plane resistivity shows 
that superconductivity exists in the doping range $0.08<x<0.15$ with the maximum
T$_c \sim$ 40K for x$\sim$0.1 as shown in Fig.~\ref{SLCO.PhD}.  Because of the limited sample size, very little is known about the possibility of an antiferromagnetic phase in the lightly doped materials and a complete phase diagram showing antiferromagnetic and superconducting phase boundaries has not been produced.  However, muon spin rotation (uSR) measurements \cite{shengelaya05a} on ceramic samples have claimed that magnetism and SC do not coexist at the La=0.1 doping and that the superfluid density is four times larger than in $p$-type cuprates with comparable T$_c$ (i.e. off the `Uemura line' \cite{Uemura89a,Uemura91a}).  A similar doping dependence of $T_c$ with substitution of Pr\cite{Smith91a}, Sm and Gd\cite{Ikeda93a} rules out the possibility that superconductivity  in SLCO arises due to the intercalation of the (La,Sr)$_2$CuO$_4$ phase. 
Early neutron scattering studies on bulk materials have shown that
superconducting SLCO is perfectly stoichiometric and presents no 
excess (interstitial) oxygen in the Sr(La) layers \cite{Jorgensen93a}. 
Thus, neither oxygen vacancies nor interstitial oxygen seem to play 
a role in the doping of this compound, although a recent report on thin films may be indicating a required reduction process \cite{ZZLi08a}.

%
%
%
%
%


\subsection{Materials growth} \label{sec.materials}

The growth of electron-doped cuprate materials has been a challenge since their 
discovery. Because there are in principle two doping degrees of freedom (cerium and 
oxygen), the optimisation of their growth and annealing parameters is tedious and has been the source of great variability in their physical properties.   For instance, it took almost 10 years after the discovery of the $n$-type compounds until superconducting crystals of sufficient size and quality could be prepared to perform inelastic neutron scattering \cite{Yamada99a}.  It is obvious that many properties, for example the temperature dependence of the resistivity, are strongly affected by grain boundaries.  Due to these difficulties we focus our attention here on the growth of single crystals and epitaxial thin films.

\subsubsection{Single crystals}

\begin{figure}[htbp]
\begin{center}
\includegraphics[width=8.5cm,angle=0]{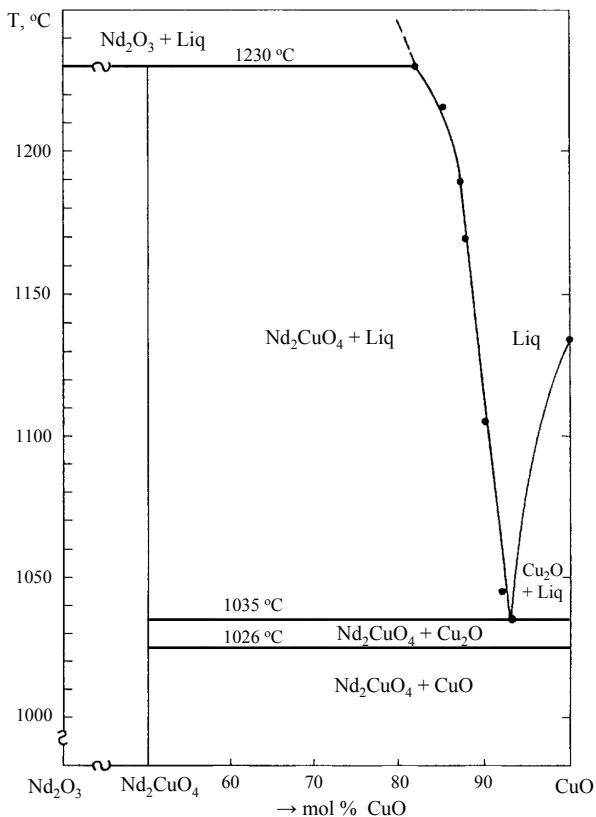}
\caption{Phase diagram of the Nd$_2$O$_3$-CuO binary system. Without Ce, the eutectic point corresponds to 
approximatively 90$\%$ CuO content in the flux. From \textcite{Maljuk96a}.} 
\label{eutectic.growth}
\end{center}
\end{figure}

Two main techniques have been used to grow single crystals of the $n$-type family: in-flux solidification and traveling-solvent floating zone (TSFZ).   The first single crystals of Nd$_{2-x}$Ce$_x$CuO$_4$ were grown using the directional solidification flux technique taking advantage of the stability of the NCCO $T'$ phase in a flux of CuO close to an eutectic point\cite{Tarascon89a}.  As shown in Fig.~\ref{eutectic.growth}, the $T-x$ phase diagram of the NdCeO-CuO mixture presents a large region between 1030 and 1250$^{\circ}$C for which the growth of NCCO crystals is possible within a liquid phase\cite {Pinol90b,Oka90a,Maljuk96a}. Typical crucibles used for the flux growth of the electron-doped 
cuprates are high purity alumina\cite{Sadowski90a,Peng91a,Dalichaouch93a,Brinkmann96b}, magnesia, zirconia\cite{Kaneko99a} and platinum\cite{Tarascon89a,Matsuda91a,Kaneko99a}. 
After reaching temperatures high enough for melting the whole content of a crucible (above 1250$^{\circ}$C following the phase diagram in Fig.~\ref{eutectic.growth}), the temperature is slowly ramped down with typical rates of 1 to 6$^{\circ}$C/h) while imposing a temperature gradient at the crucible position promoting the growth of the CuO$_2$ planes along its direction. As the crucible is further cooled down, the flux solidifies leaving the NCCO single crystals usually embedded in a solid matrix. Platelet crystals can reach sizes on the order of several millimeters in the $a-b$ direction, with the c axis limited to a few tens to several hundred microns.

%
%
%
%

When their growth and annealing processes are under control, flux grown single crystals present very high crystalline quality with few defects. They also have well-defined faces which necessitate little 
cutting and polishing to prepare for most experiments.  Flux grown crystals can have however a cerium content that can vary substantially even within the same batch\cite{Dalichaouch93a} and moreover, the thickest crystals have been shown to have an inhomogeneous cerium distribution along their thickness\cite{Skelton94a}. Finally, since the flux properties change considerably with composition, it remains quite difficult to vary the cerium content substantially around optimal 
doping and preserve narrow transitions. A variant of this directional flux technique, top seeded solution, has also been developed\cite{Cassanho89a,Maljuk00a} which leads to large single crystals with apparently more uniform cerium content. 

These millimeter-size crystals are large enough for many experiments, however, their 
limited volume is a drawback for others like neutron scattering.  As is also the case for the $p$-type cuprates, larger single crystals can be grown by the TSFZ technique using image furnaces\cite{Tanaka91a,Gamayunov94b,Kurahashi02a}.  Large boules of electron-doped cuprates several centimeters in length and half a centimeter in diameter\cite{Tanaka91a} can be produced with close to stoichiometric flux in various atmospheres and pressures. Using such conditions, NCCO crystals with $x$ as large as 0.18, the solubility limit, could be grown and studied by neutron scattering\cite{Mang04a,Motoyama07a}.  Large TSFZ single crystals of Pr$_{1-y-x}$La$_y$Ce$_{x}$CuO$_{4 \pm \delta}$ have also been grown successfully in recent years  (see Ref. \cite{Kuroshima03a,Wilson06a} and references therein). Interestingly, it appears that the presence of La stabilizes their growth\cite{Fujita03a,Lavrov04a}.

\subsubsection{Role of the reduction process and effects of oxygen stochiometry}
\label{sec:reduction}

Superconductivity in the electron-doped cuprates can $only$ be achieved after reducing the as-grown materials \cite{Tokura89a,Takagi89a}. Unannealed crystals are never superconducting.  This reduction process removes only a small fraction of the oxygen atoms as measured by many techniques\cite{Tarascon89a,Moran89a,Radaelli94a,Schultz96a,Klamut97a,Navarro01a}, but has dramatic consequences for its conducting and magnetic properties.  The  oxygen removed in general ranges between 0.1 and 2$\%$ and generally decreases with increasing cerium 
content \cite{Takayama89a,Suzuki90a,Kim93a,Schultz96a}.   The exact effect of oxygen reduction is still unknown.  Although reduction in principle should contribute electrons, it clearly has additional effects as it is not possible to compensate for a lack of reduction by the addition of extra Ce.  

\begin{figure}[htbp]
\begin{center}
\includegraphics[scale=0.26,angle=0]{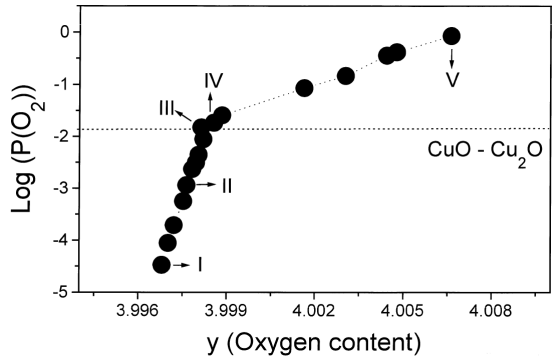}
\caption{Equilibrium oxygen partial pressure p(O$_2$) as a function of
oxygen content \textit{y} for Nd$_{1.85}$Ce$_{0.15}$CuO$_{y}$ at 900$^{\circ}$C. 
The dashed line indicates the Cu$^{2+}$/Cu$^{1+}$ transition. Samples 
below this line are superconducting, while those above are not. From  \textcite{Navarro01a}.} 
\label{CuOCu2O.crossover}
\end{center}
\end{figure}

\begin{figure}[htbp]
\begin{center}
\includegraphics[width= 8cm,angle=0]{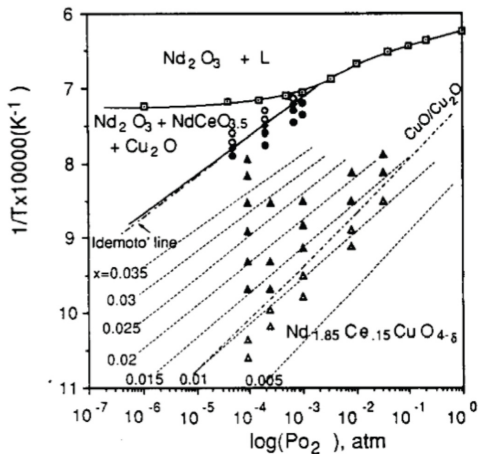}
\caption{ The phase stability diagram for x=0.15 NCCO.  The filled diamond symbols are obtained by thermogravimetric analysis and the filled and open circles are compounds which lie within and outside the field of stability respectively.  The filled triangles represent superconducting samples and the open triangles at the bottom represent nonsuperconducting oxides.  The dash-dot-dash line is for the Cu$_2$O - CuO transition and the dotted lines are isocompositions. From \textcite{Kim93a}.} \label{KimGaskell}
\end{center}
\end{figure}

There are many different procedures mentioned in the literature for the reduction process. In general, the single crystals are annealed at high temperature (850 to 1080$^{\circ}$C in flowing inert gas or vacuum) for tens of hours to several days. In some of these annealing procedures, the single crystals are also covered by polycrystalline materials, powder and pellets, in order to protect them against decomposition\cite{Brinkmann96b}.   As revealed by a thermogravimetric study\cite{Navarro01a} of polycrystalline Nd$_{1.85}$Ce$_{0.15}$CuO$_{4+\delta}$,  the annealing process in small oxygen partial pressures at a fixed temperature (900$^{\circ}$) consists of two distinct regimes as shown in Fig.~\ref{CuOCu2O.crossover}: a first one at high pressure leading to non-superconducting materials and a second one at low pressure inducing superconductivity. Interestingly, the separation of these two regimes coincides with the phase stability line between CuO and Cu$_2$O with their respective Cu$^{2+}$ and Cu$^{1+}$ oxidation states\cite{Navarro01a}. A similar conclusion was reported by  \textcite{Kim93a} in the phase stability diagram shown in Fig.~\ref{KimGaskell}.  The coincidence of the Cu$^{2+}$/Cu$^{1+}$ (CuO/Cu$_2$O) transition and the onset of superconductivity may be interpreted as a sign that oxygen reduction removes oxygen atoms in the CuO$_2$ planes leaving behind localized electrons on the Cu sites in proximity to the oxygen vacancies (these Cu ions then have oxidation state +1).  Fig.~\ref{KimGaskell} also shows that annealing electron-doped cuprates in lower pressures and/or higher temperatures leads eventually to the decomposition of the materials into a mixture of Nd$_2$O$_3$, NdCeO$_{3.5}$ and Cu$_2$O.  As emphasized by \textcite{Kim93a} and more recently by \textcite{Mang04b} it is interesting that the highest T$_c$ samples are found when the reduction conditions push the crystal almost to the limit of decomposition (Fig.~\ref{KimGaskell}).  This underlines the difficulty of achieving high quality reduction when it requires exploring annealing conditions on the verge of decomposition.

%
%
%
%

%
%
%
%

This annealing and small changes in oxygen content have a dramatic impact on the physical properties. As-grown materials are typically antiferromagnetic with a N\'eel temperature T$_N$ above 100K for x = 0.15\cite{Mang04a,Uefuji01a}. It shows fairly large resistivity with a low temperature upturn (see section~\ref{Transport}). After reduction, antiferromagnetism is suppressed and superconductivity emerges. As mentioned, there is still no consensus on the exact mechanism for this striking sensitivity to oxygen stoichiometry and the annealing process.  There are three main (not necessarily exclusive) proposals to explain how as small a change as $0.1\%$ change in oxygen content can have such an important effect, which is similar to the impact of changing the cerium doping by $\Delta x \sim$ 0.05 - 0.10. 

The first proposal and historically the mostly widely assumed, proposes that apical oxygen atoms, an interstitial defect observed in the $T'$ structure by neutron scattering in Nd$_2$CuO$_4$\cite{Radaelli94a}, acts as a strong scattering center (increasing resistivity) and as a source of pair breaking\cite{Xu96a}.   By Madelung potential consideration, one expects that apical oxygen may strongly perturb the local ionic potential on the Cu site immediately below it \cite{Torrance89a,Ohta91a}.  \textcite{Radaelli94a} showed that reduction leads to a decrease in apical occupancy to approximately 0.04 from 0.1 for the \textit{undoped compounds} \cite{Radaelli94a}.  In doped compounds, the oxygen loss is less and almost at the detection limit of the diffraction experiments, however \textcite{Schultz96a} claimed that their results in Nd$_{1.85}$Ce$_{0.15}$CuO$_{4+\delta}$ were consistent  with a loss of a small amount of oxygen at the apical position.

However, there are several recent reports that favor a second scenario in which only oxygen ions on the intrinsic sites [O(1) in-plane and O(2) out-of-plane in Fig.~\ref{unit.cell}] are removed.  It was found that a local Raman mode which is associated with the presence of apical oxygen is not affected at all by reduction in cerium-doped crystals  \cite{Riou01a,Riou04a,Richard04a}. This appears to indicate that reduction does not change the apical site's oxygen occupation as originally believed.  In the same reports, crystal-field spectroscopy of the Nd or Pr ions on their low symmetry site show also that the excitations associated with the interstitial oxygen ions are not changed by reduction while
new sets of excitations appear\cite{Riou01a,Riou04a,Richard04a}.
These new excitations were naturally related to the creation of
O(1) and O(2) vacancies; in-plane O(1) vacancies appear to be
favored at large cerium doping. Such a surprising conclusion was
first formulated by \textcite{Brinkmann96a} from the results of
a wide exploration of the cerium and oxygen doping dependence of
transport properties in single crystals. In order to explain the appearance
of a minimum in resistivity as a function of oxygen content (for a
fixed cerium content), these authors proposed that the increasing
scattering rate (increasing $\rho_{xx}$) with decreasing oxygen
content for extreme annealing conditions could only be due to an
increasing density of defects (vacancies) into or in close
proximity to the CuO$_2$ planes. They
targeted the reservoir O(2) as the likely site for
vacancies.

\begin{figure}[htbp]
\begin{center}
\includegraphics[scale=0.24,angle=0]{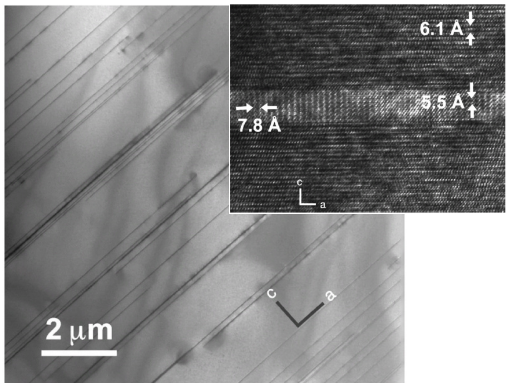}
\caption{HRTEM images of a reduced
Nd$_{1.84}$Ce$_{0.16}$CuO$_{4+\delta}$ single crystal showing the
intercalated layers.  The (Nd,Ce)$_2$O$_3$ layers are found to be
parallel to the CuO$_2$ planes. From  \textcite{Mang04b}.}
\label{HRTEM.images}
\end{center}
\end{figure}

Finally, a third scenario has been suggested by recent detailed studies of the microstructure of
Nd$_{1.85}$Ce$_{0.15}$CuO$_{4+\delta}$.  \textcite{Kurahashi02a} reported the appearance and disappearance of an unknown impurity phase associated with an annealing/re-oxygenation process.  \textcite{Mang04b} showed that this phase was  (Nd,Ce)$_2$O$_3$, which grew in epitaxial register with material under reduction.  This observation has an important repercussions on the interpretation of neutron scattering experiments (Sec.~\ref{NeutronsScattering}),  but it also suggests a scenario for the role of reduction in this family. In Fig.~\ref{HRTEM.images}, high
resolution transmission electron microscopy (HRTEM) images reveal
the presence of narrow bands of this parasitic phase about 60{\rm \AA}
thick on average extending well over 1$\mu m$ along the CuO$_2$
planes. This phase represents approximately 1$\%$ of the entire
volume. Since this phase is claimed to appear with reduction and
to disappear surprisingly with oxygenation, it was
proposed that these zones act as copper
reservoirs to cure intrinsic Cu vacancies in the as-grown CuO$_2$ planes \cite{Kurahashi02a,Kang07a}.
Within this scenario,  during the reduction process Cu atoms migrate from these layers
to the NCCO structure to ``repair" defects present in the as-grown
materials resulting in Cu deficient regions with the epitaxial
(Nd,Ce)$_2$O$_3$ intercalation. Thus, the decreasing density of Cu
vacancies in the CuO$_2$ planes removes pair-breaking sites
favoring superconductivity.

%
%
%
%
%

\par

\begin{figure}[htbp]
\begin{center}
\includegraphics[width=8cm,angle=0]{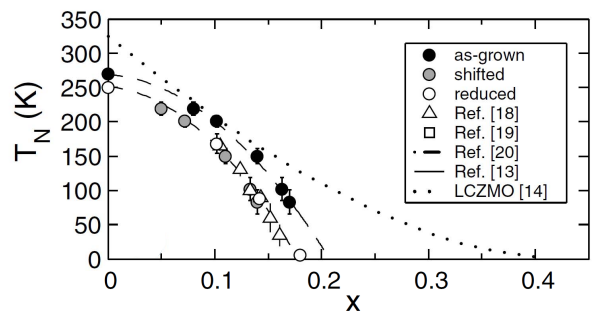}\\
\vspace{-0.7cm}
\caption{(a) Phase diagram for Nd$_{2-x}$Ce$_{x}$CuO$_{4+\delta}$
single crystals as determined by neutron scattering. It
shows the N\'eel temperature as a function of cerium content (x).
Full circles:  as-grown (oxygenated) samples. Open symbols :
reduced samples. Grey circles are the data for the as-grown
crystals shifted to simulate the carrier density change with
reduction.  From \textcite{Mang04a}. } \label{Mang.PhDiag}
\end{center}
\end{figure}

In addition to whatever role it plays in enabling superconductivity, the oxygen reduction process probably also adds charge carriers.  Using neutron scattering on TSFZ single crystals with various values of $x$ and annealing conditions to tune the presence of
superconductivity, \textcite{Mang04a} confirmed \textcite{Luke90a}'s results for reduced samples, with the $T_N(x)$ line plunging to zero at $x \sim 0.17$ as shown in
Fig.~\ref{Mang.PhDiag}(a).  They found that for unreduced samples 
$T_N(x)$ extrapolated somewhere around $x=0.21$.  For a fixed $x$ value below 0.17,
reduction lowers $T_N$ and the corresponding staggered in-plane
magnetization, while promoting
superconductivity.  The change in $T_N(x)$  with oxygen content was interpreted as a direct consequence of carrier doping i.e. that
removal of oxygen acts exactly like cerium substitution \cite{Mang04a}, because
one could simply rigidly shift the as-grown T$_N(x)$ line by $\Delta x \approx$ 0.03 to overlay
the reduced one.  However, the conclusions of  \textcite{Mang04a}  may be called into question by later work of \textcite{Motoyama07a}, who claim that the AF state terminates at $x$ approximately 0.134 for reduced samples.   It may be then that this picture of shifting the T$_N(x)$ line by an amount corresponding to the added electron contribution from reduction is only valid at low dopings.   Different physics may come into play near superconducting compositions.  \textcite{Arima93a} had found that reduced and unreduced infrared spectra which differed by $\Delta x \approx$ 0.05 could be overlayed on top of each other.   If one considers the reduction of oxygen content corresponds to an addition of electron carriers to the CuO$_2$ plane this implies an oxygen reduction of 0.02-0.03, which is consistent with thermogravimetric studies  \cite{Arima93a}.

%
%
%
%

\subsubsection{Thin films}

Thin film growth offers additional control on the stoichiometry of the electron-doped system, for both cerium and oxygen content.  Thin films have been grown using most of the usual techniques for the deposition of other cuprates and oxides, including pulsed-laser deposition (PLD) \cite{Gupta89a,Mao92a,Maiser98a,Gauthier07a} and molecular beam epitaxy (MBE) \cite{Naito97a,Naito02a}. Both techniques lead to a single crystalline phase with the cerium content accuracies better than 3$\%$.  Since film thicknesses are in the range of 10 to 500 nm and the flux and proportion of each constituent can be accurately controlled during deposition, their cerium content is more homogeneous in contrast to single crystals. Moreover, since oxygen diffusion along the c-axis is easier they can be reduced much more uniformly and efficiently with post-annealing periods on the order of one to several tens of minutes\cite{Mao92a,Maiser98a}.  As a consequence of the greater stoichiometry control, superconducting transition widths as small as $\Delta T_c \sim 0.3$K (from AC susceptibility) have been regularly reported\cite{Maiser98a}.  For PLD films, growth in a nitrous oxide (N$_2$O) atmosphere \cite{Mao92a,Maiser98a} instead of molecular oxygen\cite{Gupta89a} has also been used in an effort to decrease the time needed for reduction. Unlike single crystals, it is possible to finely control the oxygen content using in-situ post-annealing in low pressure of O$_2$. Within a narrow range of increasing pressure, the resulting films show a gradual decrease of T$_c$ and related changes in transport properties\cite{Gauthier07a} (see section~\ref{disordertransport} and Fig.~\ref{Gauthier07}).

Since they are grown on single crystalline substrates with closely matching lattice parameters (LaAlO$_3$, SrTiO$_3$, etc.), films are generally epitaxial with a highly ordered (001) structure with their $c$ axis oriented normal to the substrate and providing the needed template for the exploration of in-plane transport and optical properties.  Unlike other high-T$_c$ cuprates like YBa$_2$Cu$_3$O$_7$ \cite{Covington96a}, there have been very few reports on films with other orientations. There is evidence that films with (110) and (103) orientations can be grown on selected substrates as confirmed by x-ray diffraction and anisotropic resistivity\cite{Ponomarev04a,HWu06a},  but the width of their superconducting transition ($\Delta T_c \sim$1K) shows that there is room for further optimisation as compared to c-axis films. These particular film orientations could be of interest for directional tunneling experiments \cite{Covington96a}.

\begin{figure}[htbp]
\begin{center}
\includegraphics[scale=0.3,angle=0]{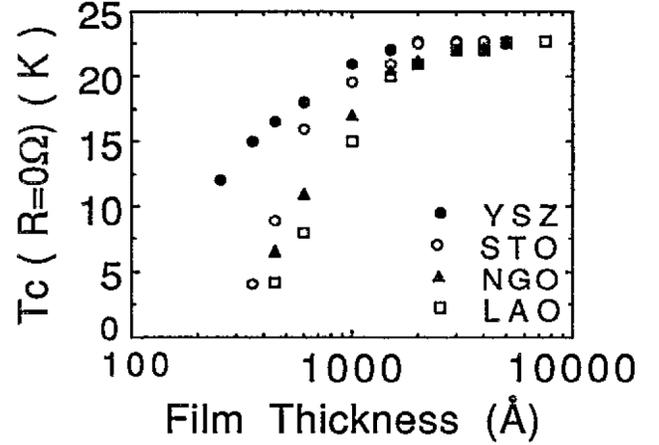}
\caption{Variation of the transition temperature with the thickness of the PLD films on various substrates. From  \textcite{Mao94b}.} 
\label{Mao.thickness}
\end{center}
\end{figure}

A number of drawbacks to thin films do exist. There has been reports of parasitic phases detected by x-ray diffraction in PLD films\cite{Gupta89a,Mao92a,Maiser98a,Lanfredi06a} and HRTEM\cite{Beesabathina93a,Roberge09a}. These phases have been indexed to other crystalline orientations\cite{Maiser98a,Prijamboedi06a} or Cu-poor intercalated phases\cite{Mao92a,Beesabathina93a,Lanfredi06a,Roberge09a}, which have also been observed in single crystals\cite{Mang04b}. These parasitic phases are mostly absent in MBE films\cite{Naito02a} except for extreme cases when the substrate/film lattice mismatch becomes important. This microstructural difference between MBE and PLD films may be at the origin of the difference in the magnitude of their in-plane resistivity\cite{Naito02a} as was confirmed recently by \textcite{Roberge09a} in a new set of films grown with off-stochiometric targets to remove the parasitic phase. Moreover, a significant effect of a strain-induced shift of T$_c$ shown in Fig.~\ref{Mao.thickness} has been observed as it decreases with decreasing thickness\cite{Mao94b}.

%
%
%
%

\begin{figure}[htbp]
\begin{center}
\includegraphics[width=7.5cm,angle=0]{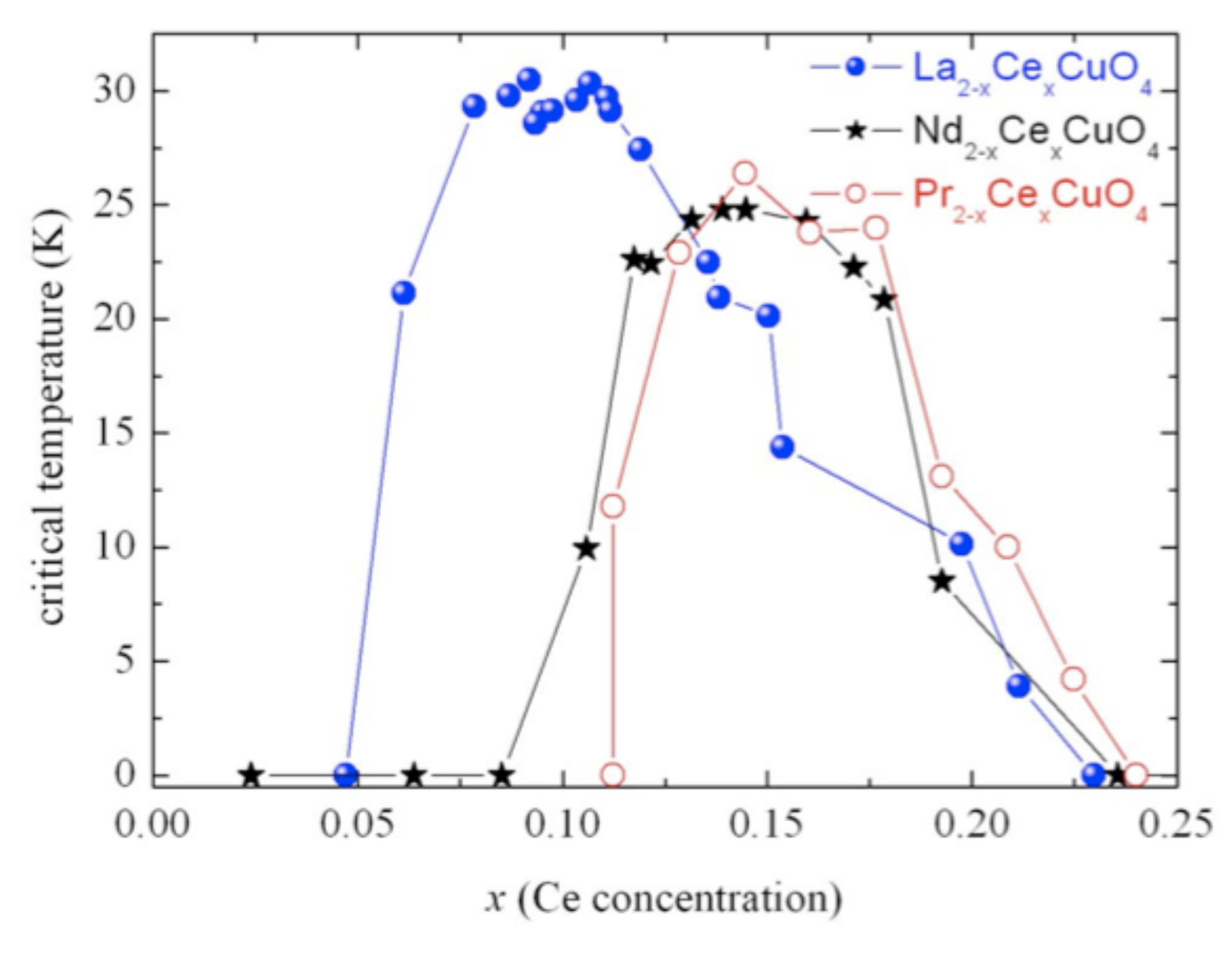}
\caption{The transition temperature as a function of cerium doping for $T'$ La$_{2-x}$Ce$_{x}$CuO$_{4+\delta}$, NCCO, and PCCO thin films grown by MBE.  From  \textcite{Krockenberger08a}.} 
\label{MBE.TcLCCO}
\end{center}
\end{figure}

Strain from the substrate however can also play a crucial role to help stabilizing the $T'$ structure.  As mentioned previously, usual bulk LCO grows in the $T$ phase. It was shown by Naito \textit{et al.} that La$_{2-x}$Ce$_{x}$CuO$_{4+\delta}$ (LCCO) can actually be grown successfully in $T'$ by MBE leading to superconducting materials with $T_c$ as high as 27K\cite{Naito00a,Naito02a,Krockenberger08a}.    These electron-doped  films also exhibit a modified phase diagram with superconductivity extending to $x$ values below 0.10 as shown in Fig.~\ref{MBE.TcLCCO}, which is fairly similar to that of the (Pr,La)$_{2-x}$Ce$_{x}$CuO$_{4+\delta}$ compounds\cite{Fujita03a,Fujita08a,Fontcuberta96a}. The LCCO $T'$ phase has also been successfully grown by DC magnetron sputtering\cite{Zhao04a} and PLD \cite{Sawa02a}.

%
%
%
%

Recently, \textcite{Matsumoto09a} have shown that it is possible to grow
$superconducting$ thin films of the undoped $T'$ structure RE$_2$CuO$_4$ with RE = Pr, Nd, Sm,
Eu, Gd with T$_c$'s as high as 30K using metal-organic deposition (MOD). This is
obviously quite a different behavior compared to the abovementioned trends, in
particular the observation of superconductivity in Gd$_2$CuO$_4$.  The authors
claim that a complete removal of all the apical oxygen acting as a scatterer
and a source of pair breaking during the reduction process explains the
observation of superconductivity in these undoped compound. It could also be that such films are the  $T'$ electron-doped analog of superconducting  La$_{2}$CuO$_{4+\delta}$ where $\delta$ is excess interstitial `staged' oxygen.  Obviously, such behavior is intriguing and may raise important questions on the actual mechanism of superconductivity and definitely deserves further investigation.


\subsection{Unique aspects of the copper and rare earth magnetism} \label{sec:magnetism}
\par

Irrespective of the actual superconductivity mechanism, it is clear that the magnetism of the high-T$_c$ superconductors dominates their phenomenology.  The magnetic properties of the electron-doped cuprates are unusually complex and intriguing and demand special consideration. On top of the usual AF order of the in-plane Cu spins observed for example in Nd$_2$CuO$_4$ at T$_{N,Cu} \sim$ 270K \cite{Mang04a}, additional magnetism arises
from the response of the rare-earth ions to the local crystal
field.  In the $T'$ structure, rare-earth ions sit on a low-symmetry site (point group
C$_{4v}$) where they experience a local electric field leading to a
splitting of their $4f$ atomic levels\cite{Nekvasil01a,Sachidanandam97a}.  In
Table~\ref{Re.moments}, we present the estimated magnetic moment
of the most common RE ions used in the $n$-type family.  Some of these magnetic moments are large and interactions between the RE ions and localized Cu spins give rise to a rich set of properties and signatures of magnetic order.  As an emblematic example, Nd magnetic moments are known to grow as the temperature is decreased since it is a Kramers
doublet \cite{Kramers30a}, implying a complex
temperature-dependent interaction with the Cu sub-lattice and other Nd ions.  
Among other things, these growing Nd moments at low temperature have an impact on several
low temperature properties that are used to characterize the
pairing symmetry (see Section~\ref{sec:pendepth}).  
Here we summarize the different magnetic states observed in the
electron-doped cuprates. We first focus on the N\'eel order of the
Cu spins, and then follow with a quick overview of its interaction
with the RE moments.

%
%
\begin{table}[htbp]
    \begin{center}
        \begin{tabular}{|c||c|c|c|c|c|c|}
        \hline
                      & PCO & NCO   & SCO   & ECO & GCO  & PLCO   \\
        \hline
        \hline
                    J   & 4 & 9/2 & 5/2 & 0 & 7/2 &  \\
        \hline
        						effective &     &     &     &    &    &  \\
                    moment & 3.65$\mu_B$    & 3.56$\mu_B$   & 0.5$\mu_B$ & 0$\mu_B$ & 7.8$\mu_B$  &  \\
                    Curie-Weiss &    &   &  &    & &  \\
        \hline
                    ordered &     &     &     &    &    &  \\
                    moment & 0.08$\mu_B$    & 1.23$\mu_B$   & 0.37$\mu_B$   & 0$\mu_B$  & 6.5$\mu_B$  & 0.08$\mu_B$ \\
                    measured &  &   &   &   &  & \\
        \hline
                    RE N\'{e}el & $-$   & 1.7   & 5.95  & $-$   & 6.7  & $-$ \\
                    Temp. (K)   &   &   &   &   & &  \\
        \hline
        \end{tabular}
    \caption{Table summarizing the magnetic properties arising from RE moments.  The RE effective moment is from a fit of the high-temperature susceptibility to the Curie-Weiss law while the ordered moment is estimated at low temperature from 0.4 to 10K mostly from neutron scattering experiments. The N\'eel temperature corresponding to the magnetic ordering of the RE moments was determined using specific heat. From \textcite{Vigoureux95a,Matsuda90a,Ghamaty89a,Lynn01a} and references therein.}
    \label{Re.moments}
    \end{center}
\end{table}
%
%

\subsubsection{Cu spin order} \label{sec:CuSpins}
\begin{figure}[htbp]
\begin{center}
\includegraphics[scale=0.3,angle=0]{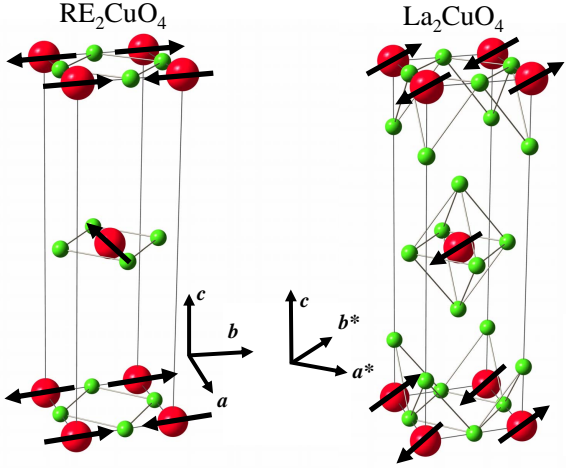}
\caption{Cu spin structures for the non-collinear 
phase of NCO in Phase II  at 30 K $<$ T $<$ 75 K or SCO.  Nd Phases I and III are equivalent to a structure with the central Cu spin of the figure flipped 180$^\circ$.  (left) The collinear structure of
La$_2$CuO$_4$ (right). The moments (arrows) are aligned along the 
nearest neighbor Cu along the Cu-O bonds [(100) and (010)] for RE$_2$CuO$_4$ while 
they point toward the next-nearest neighbor Cu at 45$^\circ$ with respect to the 
Cu-O bonds [along (110)] for La$_2$CuO$_4$.} 
\label{Cumag.structure}
\end{center}
\end{figure}

The commensurate antiferromagnetic order of the Cu spins observed for the
parent compounds of the electron-doped family is quite different
from that of La$_2$CuO$_4$, despite close values of $T_{N,Cu}
\sim$ 300K, similar crystal structures and Cu-O bond lengths\footnote{Note that the maximum N\'{e}el temperature of NCO is reported differently in various studies, which is presumably due to a strong sensitivity to oxygen content.  For instance, \textcite{Matsuda90a} report 255 K,  \textcite{Bourges97a} report 243 K, whereas \textcite{Mang04a} report $\approx$ 270 K.  The maximum reported T$_N$ for PCO appears to be 284 K \cite{Sumarlin95a}.  In contrast, the maximum reported  $T_{N}$ for LCO is 320 K \cite{Keimer92a}.}.  In Fig.~\ref{Cumag.structure}, we compare the magnetic orders 
deduced from elastic neutron scattering for both families. Although the 
magnetic moments lie in the CuO$_2$ planes for both systems with fairly 
strong intraplane AF exchange interaction, the in-plane alignment differs as the spins lie 
along the Cu-O bonds in the case of electron-doped cuprates\cite{Skanthakumar93b,Skanthakumar95a} 
while they point at roughly 45$^{\circ}$ to the Cu-O bond directions for LCO\cite{Kastner98a}.  Since the resulting isotropic exchange between planes cancels out due to this in-plane alignment and crystal symmetry, the 3D magnetic order in the case of the electron-doped cuprates is governed by a delicate balance of RE-Cu coupling, superexchange, spin-orbit, and Coulomb exchange interactions \cite{Sachidanandam97a,Petitgrand99a,Lynn01a,Yildirim94a,Yildirim96a}.  They lead to a spin configuration where the in-plane magnetization alternates in directions between adjacent layers \cite{Sumarlin95a,Sachidanandam97a,Lynn01a} in a non-collinear structure, which is compatible with the tetragonal crystal structure as shown in Fig.~\ref{Cumag.structure}.   For orthorhombic LCO, the spin structure is collinear along the c-axis\cite{Kastner98a}.  A non-collinear structure of the Cu spins has been confirmed for NCO,  Sm$_{2}$CuO$_4$ (SCO),  Eu$_{2}$CuO$_4$ (ECO), PCO and PLCCO using elastic neutron scattering\cite{Skanthakumar95a,Skanthakumar93b,Skanthakumar93a,Sumarlin95a,Lavrov04a,Chattopadhyay94a}.  Various different non-collinear Cu spin patterns are stabilized depending on the nature of the RE-Cu interaction.  The Cu spin wave spectrum is gapped due to anisotropy by about 5 meV in PCO \cite{Sumarlin95a} and \cite{Bourges92a}, which can be compared with the anisotropy gap of 2.5 meV in LCO \cite{Peters88a}.  Magnetic exchange constants are of the same order as the hole-doped compound.   See for instance the two magnon Raman data of \textcite{Sulewski90a}, who find exchange constants of 128, 108 and 110 meV for LCO, NCO and SCO respectively.  These values are similar to those found by fits to spin-wave theory \cite{Sumarlin95a}.  One may expect these numbers to be refined as new time-of-flight neutron spectrometers come online.

%
%

\par
\begin{figure}[htbp]
\begin{center}
\includegraphics[scale=0.35,angle=0]{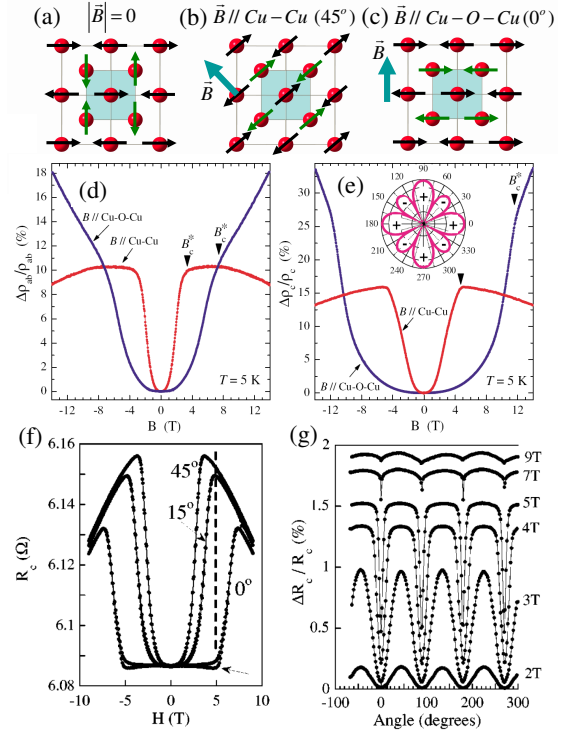}
\caption{In-plane Cu spin structures for (a) zero applied magnetic field and
a large magnetic field beyond the spin-flop field applied at (b) 45$^{\circ}$ and (c) 0$^{\circ}$
with respect to the Cu-O-Cu bonds. (d) in-plane and (e) c-axis magnetoresistance of lightly 
doped Pr$_{1.3}$La$_{0.7}$Ce$_x$CuO$_4$ ($x = 0.01$) single crystals at 5K for a magnetic 
field applied along the (100) or (010) Cu-O bonds and along the (110) Cu-Cu 
direction (from  \textcite{Lavrov04a}).  (f) c-axis magnetoresistance at 5K for 
non-superconducting as-grown Pr$_{1.85}$Ce$_{0.15}$CuO$_4$ as a function of field for three selected in-plane orientations (0, 15 and 45$^{\circ}$) and (g) as a function of angle at selected magnetic fields below and above the spin-flop field of 5T (from Ref.  \textcite{Fournier04a}).} \label{SFlop.MR}
\end{center}
\end{figure}

For at least small cerium doping (x $\sim$ 0.01 - 0.03), non-collinear commensurate magnetic
structure persists and has a detectable impact on the electronic
properties, in particular electrical transport in large magnetic fields, 
indicate the coupling of the free charge carriers to the underlying
antiferromagnetism \cite{Lavrov04a}. The carriers
couple strongly to the AF structure leading to large angular magnetoresistance (MR)
oscillations for both in-plane and out-of-plane resistivity when a large 
magnetic field is rotated in the CuO$_2$ plane\cite{Lavrov04a,Li05a,TWu08a,XHChen05a,WYu07b}.  
Although originally thought to be related to magnetic domains \cite{Fournier04a}, these oscillations are now believed to be related to the first-order spin-flop transition at a magnetic field of order 5T observed in magnetization and elastic neutron scattering measurements\cite{Cherny92a,Plakhty03a}.  At that field applied along the Cu-O bonds, the in-plane and c-axis MRs change dramatically
as the magnetic structure changes from the non-collinear order to a collinear one
\cite{Cherny92a,Lavrov04a} as shown in Fig.~\ref{SFlop.MR}.  Similar signatures but with smaller 
amplitudes were also observed at higher doping\cite{Fournier04a,Yu07a} for as-grown non-superconducting $x = 0.15$ PCCO crystals indicating that AF correlations are preserved over a wide range of doping in these as-grown materials\footnote{Recently, \textcite{Jin09a} found no four-fold effect in the in-plane angular magnetoresistance of thin films of $T'$ LCCO.   This is interesting in view of the fact that La has no RE moment and the role that RE-Cu coupling plays in determining the stability of the particular forms of non-collinear order as discussed below.}.  The effect of doping on the Cu spin structure is dealt with in more detail below.

%
%
%
%
%
%
%

\subsubsection{Effects of rare earth ions on magnetism} \label{sec:ReSpins}
\par
Additional magnetism arises from the large magnetic moments that
can exist at the RE sites. Because of their different spin magnitudes different RE ions lead to very different magnetic structures, some of which with well-defined order.  For more details on the magnetism of the rare earths in these compounds, see the excellent review by \textcite{Lynn01a}.  We will discuss the various cases separately, but briefly.
\par
\begin{figure}[htbp]
\begin{center}
\includegraphics[width=6cm,angle=0]{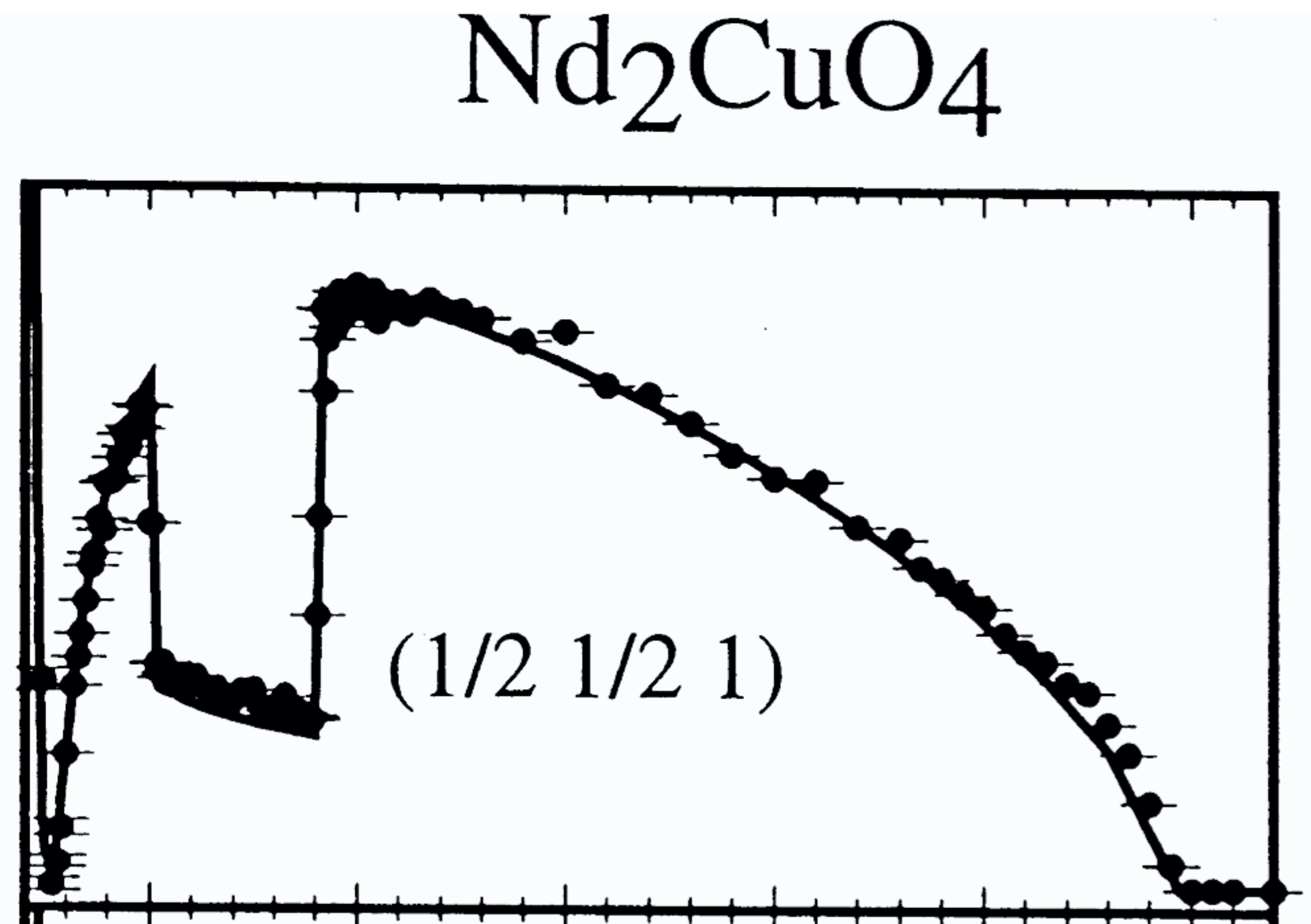}
\includegraphics[width=6cm,angle=0]{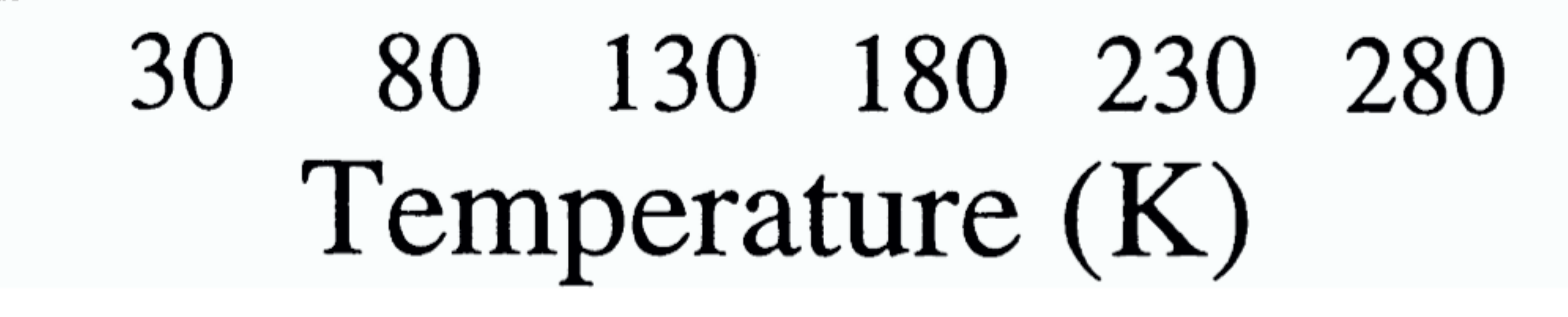}
\caption{Elastic neutron scattering intensity as a function of temperature at the (1/2 1/2 0)
reciprocal space position for as-grown Nd$_2$CuO$_4$. The sudden changes in intensity occur at the transition from Type-I to Type-II, then Type-II to Type-III Nd-Cu moment configurations with decreasing temperature. 
From  \textcite{Lynn01a} and references therein.} \label{NdCu.phases}
\end{center}
\end{figure}

In the case of RE = Nd, the fairly strong magnetic moment of the
Nd ion was found early on to couple to the Cu spins
sub-lattice\cite{Lynn90a,Cherny92a}.   A number of successive Cu spin transitions can be observed in Nd$_{2}$CuO$_{4}$ with decreasing temperature using neutron scattering\cite{Endoh89a,Matsuda90a,Skanthakumar93b,Matsuura03a,Skanthakumar95a}. 
These transitions are seen as sharp changes of intensity for specific magnetic Bragg reflections (see Fig.
~\ref{NdCu.phases}) and reveal a growing interaction between the Cu and Nd 
spins as the temperature decreases.  First, Cu spins order below T$_{N1} \approx$ 276 K in a non-coliinear structure defined as phase I.  At still lower temperatures there are two successive spin reorientations transitions at T$_{N2}$ = 75 K (shown in Fig.~\ref{Cumag.structure}) and again at T$_{N3}$ = 30 K.  At T$_{N2}$ the Cu spins rotate by 90 degree about the c axis (phase II).  The rotation direction is opposite for two successive Cu planes.  At T$_{N3}$ they realign back to their initial direction (phase III).  Phases I and III are identical with the exception that the Nd magnetic moment is larger at low temperature since it is a Kramers doublet.  Finally (inset of Fig.~\ref{NdCu.phases}) additional Bragg intensity is detected below 1K arising from the AF ordering 
of the Nd moments in the same structure as the Cu. This feature is a clear indication that substantial Nd-Nd interaction is present on top of the Nd-Cu ones that lead to the transitions at T$_{N2}$ and T$_{N3}$.  These reorientations are the result of the competition between three energy scales : 1) the Cu-Cu ; 2) the Nd-Nd ; and 3) the Nd-Cu interactions. Since the Nd moment grows
with decreasing temperature, the contributions from 2) and 3) grow
accordingly.

%
%
%
%
%

\par
The reordering and the low temperature interaction of Nd with Cu
can be observed in various other ways. The reorientations were first
observed by muon spin resonance and rotation
experiments \cite{Luke90a} and were confirmed later on by
crystal-field spectroscopy\cite{Jandl99a}, and more recently by
ultrasound propagation experiments\cite{Richard05a}.  The growing competition between the
three energy scales leads also to a wide variety of anomalies at
low temperature\cite{Cherny92a,Richard05a,Richard05b,Li05a,TWu08a,Li05b}.   

The larger moments at the Sm sites in SCCO order quite
differently than in NCCO. Sm$_2$CuO$_4$ shows a
well-defined antiferromagnetic order below $T_{N,Sm}$ = 6K with 
a transition easily observed by specific heat \cite{Hundley89a,Dalichaouch93a,Cho01a}, magnetization\cite{Dalichaouch93a} and elastic neutron scattering\cite{Sumarlin92a}.  Here ferromagnetically aligned in-plane RE moments orient themselves antiferromagnetically along the c-axis \cite{Sumarlin92a}. This special arrangement should lead to no significant coupling between the Sm and the Cu moments.   A lack of coupling is supported by the absence of spin transitions in the Cu moments like in NCO.   The non-collinear Cu spin order is the same as NCO in Phase II.

Pr$_{2-x}$Ce$_x$CuO$_4$ and Pr$_{1-y-x}$La$_y$Ce$_{x}$CuO$_{4 \pm \delta}$ exhibit the same non-collinear c-axis spin order as Phase I NCO.  At low doping the magnetic moments at the Pr site have been shown to be small, but non-zero due to exchange mixing with a value of roughly 0.08$\mu_B$/Pr \cite{Sumarlin95a,Lavrov04a}.  Due to the small moment the magnetic transitions associated with RE-Cu and RE-RE interaction in NCO do not appear to take place in PCO \cite{Matsuda90a}.   Nevertheless, there is evidence for Pr-Pr interactions in both the in-plane and out-of-plane directions\cite{Sumarlin95a} mediated by Cu spins.    This is supported by the onset  of a weak polarization of the Pr moments at the N\'eel temperature for Cu spin ordering (T$_N \sim$ 270K for Pr$_2$CuO$_4$ and T$_N \sim$ 236K for Pr$_{1.29}$La$_{0.7}$Ce$_{0.01}$CuO$_{4 \pm \delta}$).  Despite this induced magnetic moments at the Pr sites, PCO and PLCCO have a very small uniform magnetic susceptibility on the order of 1 $\%$ that of NCCO\cite{Fujita03a}. This is a great advantage in the study of their magnetic and superconducting properties without the influence from the RE moments. For instance, the small DC susceptibility of PCCO as compared to NCCO allows precision measurements of the superconducting penetration depth and symmetry of the order parameter (see Section~\ref{sec:SpecHeat}).   

\par
Finally, there has been less effort directed towards Eu$_{2}$CuO$_4$ (ECO) and Gd$_{2}$CuO$_4$ (GCO), but both systems present evidence of weak ferromagnetism, albeit with different sources.  There are indications that the small size of the rare earth ions and subsequent lattice distortions are playing a crucial role in both cases\cite{Thompson89a,Mira95a,Alvarenga96a}. For GCO, specific heat and magnetization anomalies at 9K demonstrate antiferromagnetic order on the Gd sublattice with signatures very similar to SCO.  It is believed that like SCO,  the Gd RE spins orient ferromagnetically in-plane and anti-ferromagnetically out-of-plane.  A notable difference however is that the Gd spin direction is in-plane.  A large anisotropy of the DC susceptibility with its onset at the Cu spin order temperature ($\sim$ 260K) indicates the contribution of a Dzyaloshinskii-Moriya (DM) interaction between the Cu spins leading to the weak ferromagnetism.  For ECO, there have been reports of weak ferromagnetism Cu correlations\cite{Alvarenga96a}, but clearly no RE ordering as the Eu ion is non-mangnetic.  Eu exhibits the same non-collinear Cu spin order as SCO due also to an absence of a RE-Cu coupling.

\section{Experimental survey}

\subsection{Transport}
\label{Transport}

\subsubsection{Resistivity and Hall effect}\label{TransportResHall}

The ab-plane electrical resistivity ($\rho_{ab}$) and Hall effect
for the $n$-type cuprates have been studied by many groups. The
earliest work found $\rho_{ab} = \rho_{o}+AT^2$ at optimal doping over the temperature range from $T_c$ to approximately 250K \cite{Tsuei89a} and a temperature dependent
Hall number in the same temperature range \cite{Wang91a}. The
T$^2$ behavior is in contrast to the linear in T behavior found
for the optimal hole-doped cuprates. Although $\rho \sim T^2$ is a
behavior consistent with electron-electron scattering in a normal
(i.e., Fermi liquid) metal, it is quite unusual to find such
behavior at temperatures above 20K. This suggested that there is
some anomalous scattering in the $n$-type cuprates, but that phonons
do not make a major contribution to the resistivity up to 250K.

\begin{figure}[htb]
\hspace{0cm}
\includegraphics[width=9.5cm,angle=0]{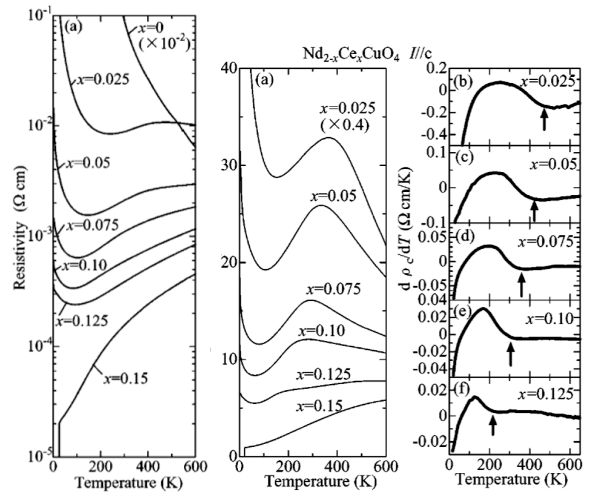}
\caption{(left) Temperature dependence of the in-plane resistivity of
NCCO crystals at various doping levels x. (center)  The temperature dependence of the out-of-plane resistivity of NCCO at various doping. (right) The temperature derivative of the
out-of-plane resistivity ($d \rho_c /dT$).  The nominal T$^*$ is indicated by the arrow.  From  \textcite{Onose04a}.} \label{OnoseRho}\end{figure}

The general doping and temperature evolution of the ab-plane
resistivity is illustrated in $\rho_{ab}$ data on NCCO
crystals as shown in Fig.~\ref{OnoseRho} (left) \cite{Onose04a}. These
data shows that even at rather low doping (i.e., in the AFM state)
a ``metallic''-like resistivity is observed at higher temperatures
which becomes ``insulator-like" at lower temperatures. The
temperature of the minimum resistivity decreases as the doping
increases and it extrapolates to less than $T_c$  near optimal doping. The
development of ``metallic" resistivity at low doping is consistent
with the ARPES data, which shows electron states near the Fermi
level around ($\pi$, 0) for $x > 0.04$ \cite{Armitage02a} and an
increased Fermi energy density of states in other regions of the
Brillouin zone (BZ) as doping increases (see ARPES discussion below). Recent work \cite{Dagan07a} showed a scaling of the T$^2$ resistivity above 100K for dopings x=0.11 to 0.19.  \textcite{Sun04a} emphasizes that despite the upturns in the ab-plane resistivity, the mobility over much of the temperature range is still quite high in even lightly doped AFM samples (5 cm$^2$/V $\cdot$ sec).   They interpreted this as consistent with the formation of metallic stripe domains.

The dependence of the high field ``insulator to metal" crossover with Ce doping
at low temperature (T $\ll$ T$_c$, H $\gg$ H$_{c2}$) has been studied in detail by
\textcite{fournier98a} and \textcite{Dagan04a}. Important
aspects of their data to note are: 1) the linear in T resistivity from 35mK to
10K at one particular doping ($x= 0.17$ in \cite{fournier98a}), 2)
the crossover from insulator to metal occurs at a $k_Fl$ value of
order 20, 3) the resistivity follows a T$^2$  dependence for all
Ce doping at temperatures above the minimum or above 40K, and 4)
the resistivity follows T$^{\beta}$ with $\beta < 2$ in the
temperature range less than 40K for samples in which there is no
resistivity minimum.

The doping dependent ``insulator to metal" crossover in the
resistivity data appears very similar to behavior found in the
hole-doped cuprates \cite{Boebinger96a}. However, electron-doped
cuprates are much more convenient to investigate such physics as
much larger magnetic fields are needed to suppress the
superconductivity in $p$-type compounds.  In the few cases that
sufficient fields have been used in the hole-doped compounds the
low temperature upturn in resistivity occurs in samples near
optimal doping with similar $k_ Fl$ values of order 20. The
behavior of the resistivity at low T below is very similar in hole
and electron-doped materials ($\rho \sim \log 1/T$) but the exact
cause of the upturn is not known at present. Work by \textcite{Dagan05a} suggests that it is related to the onset of AFM in the $n-$doped cuprates. Disorder may also play a role in the appearance of the resistivity upturn (and metal-insulator crossover) as recently suggested for hole-doped cuprates \cite{Rullier08a}.

An insulator-metal crossover can also be obtained at a fixed Ce
concentration by varying the oxygen reduction
conditions \cite{Gauthier07a,Jiang94a,Fournier97a,Gollnik98a,Tanda92a,Gantmakher03a}. Under these conditions the crossover occurs at a $k_Fl$  value of order unity and near a 2D sheet resistance (treating a single copper-oxide plane as the 2D conductor) appropriate for a
superconductor to insulator transition (SIT) \cite{Goldman98a}.
Some authors have interpreted their data as giving convincing
evidence for a SIT \cite{Tanda92a} while others have argued against this view \cite{Gantmakher03a}.  More detailed study will be needed to resolve this issue.

\begin{figure}[htb]
\hspace{0cm}
\includegraphics[width=7cm,angle=0]{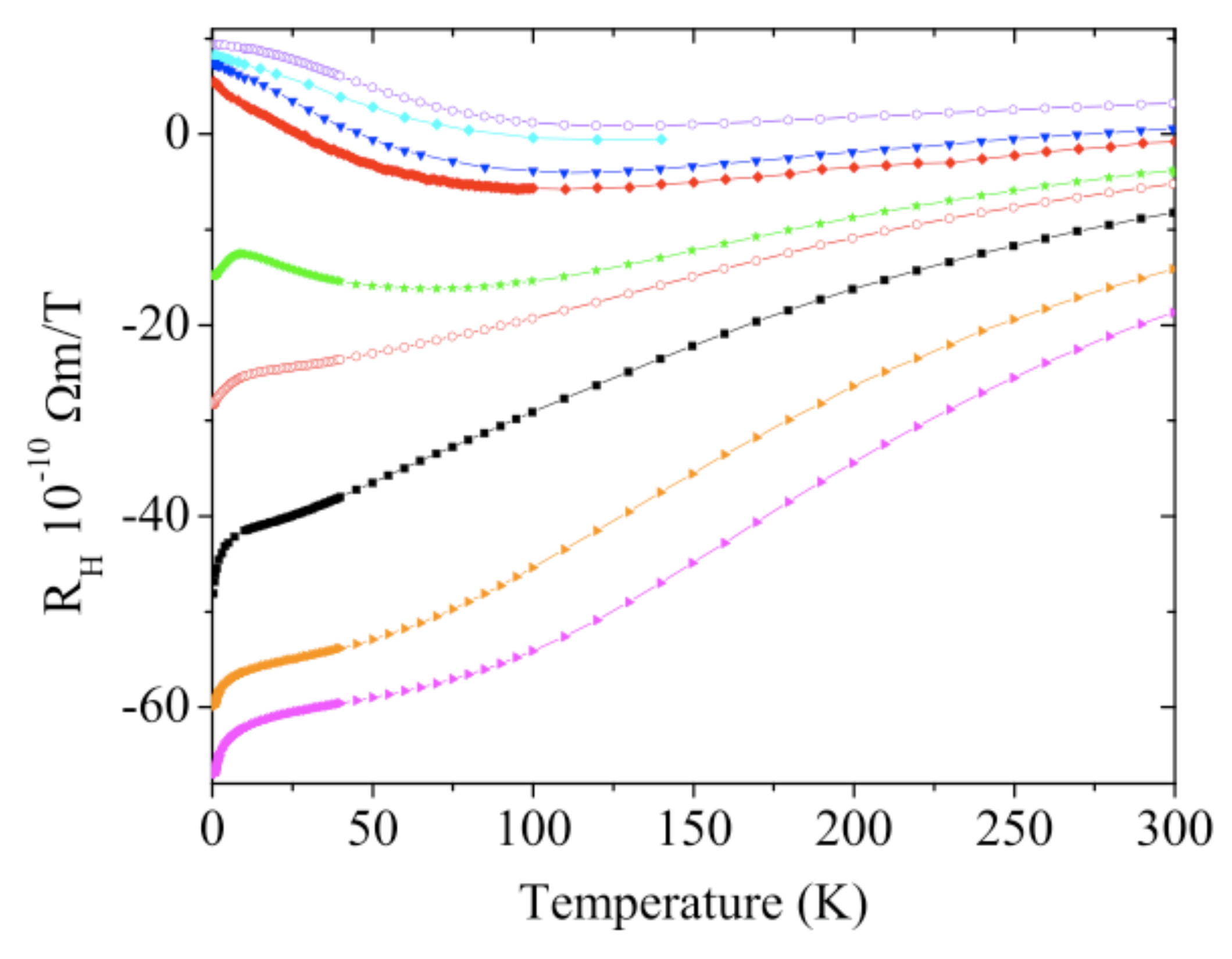}
\caption{The Hall coefficient R$_H$ in  Pr$_{2-x}$Ce$_x$CuO$_4$ films as function of  temperature for the various doping levels (top to bottom):  x = 0.19, x = 0.18, x = 0.17, x = 0.16, x = 0.15, x = 0.14, x = 0.13, x = 0.12, and x = 0.11 \cite{Dagan04b}.}
\label{Dagan04HallA}\end{figure}

\begin{figure}[htb]
\hspace{0cm}
\includegraphics[width=5cm,angle=270]{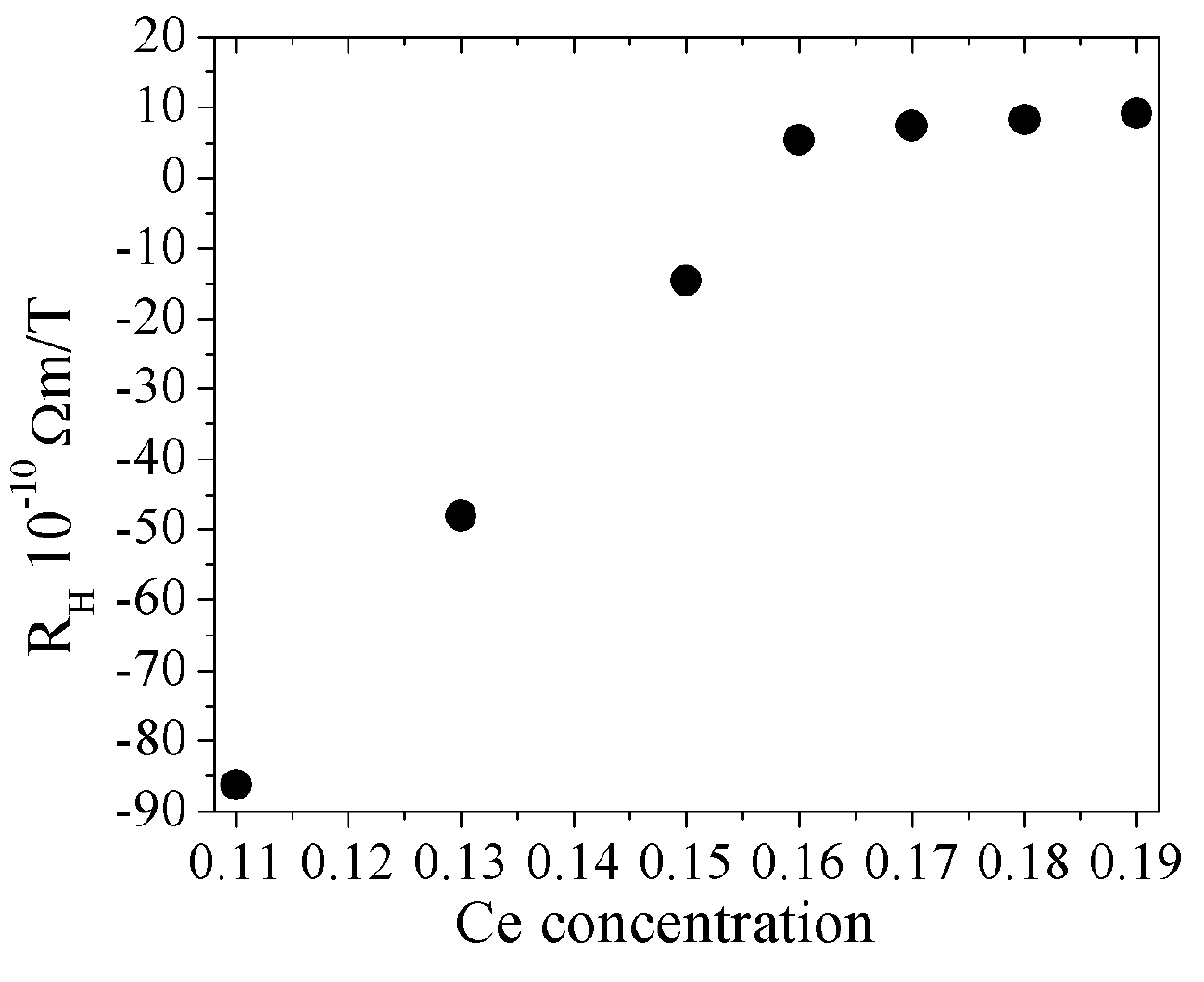}
\caption{The Hall coefficient at 0.35K (using the data from Fig.
\ref{Dagan04HallA}). A distinct kink in the Hall coefficient is
seen between $x=0.16$ and $x=0.17$. The error on the concentration
is approximately 0.003. The error in R$_H$ comes primarily from
the error in the film thickness; it is approximately the size
of the data points\cite{Dagan04a}.
\label{Dagan04HallB}}\end{figure}

The doping and temperature dependence of the normal state ($H>H_{c2}$) ab-plane Hall
coefficient ($R_H$) are shown in Fig~\ref{Dagan04HallA}
\cite{Dagan04a} for PCCO films. These recent results agree with previous work
\cite{Gollnik98a,Wang91a,Fournier97a} but cover a wider temperature and doping
range. Notable features of these data are the significant
temperature dependence for all but the most overdoped samples and
the change in sign from negative to positive near optimal doping
at low temperature. This latter behavior is most dramatically seen
by plotting $R_H$ versus Ce doping at 350mK (the lowest
temperature measured) as shown in Fig~\ref{Dagan04HallB}
\cite{Dagan04a}. At this low temperature one expects that only
elastic scattering will contribute to $\rho_{xy}$ and
$R_H$ and thus the behavior seen in Fig.~\ref{Dagan04HallB}
suggests some significant change in the Fermi surface near optimal
doping. Qualitatively, the behavior of $R_H$  is consistent with
the Fermi surface evolution shown $via$ ARPES in Fig.
\ref{PeterARPES}, which suggests that a SDW-like band structure rearrangement occurs, which breaks up the Fermi surface into electron and hole regions \cite{Armitage01b,Armitage02b,Zimmers05a,Matsui07a}. A mean field
calculation of the $T \rightarrow 0$ limit of the Hall conductance
showed that the data are qualitatively consistent with the
reconstruction of the Fermi surface expected upon density wave
ordering \cite{Lin05a}.  A convincing demonstration of such a FS reconstruction is the recent observation of Shubinkov-de Haas oscillations in NCCO by \textcite{Helm09a}, who found quantum oscillations consistent with a small FS pocket for $x=0.15$ and a large FS for $x=0.17$.  We will discuss these results and two-band transport in more detail below (Secs.~\ref{sec:SDW} and~\ref{QCPoint}).

The Hall angle ($\theta_H$) follows a behavior different than the
well-known $T^2$ dependence found in the $p$-doped cuprates. Several
groups \cite{Woods02a,Fournier97a,Dagan07a,CHWang05a} have found an
approximately $T^4$ behavior for $\cot \theta_H$ in optimal $n$-type cuprates. \textcite{Dagan07a} (but not \textcite{CHWang05a}) find the power law dependence on temperature of $\cot\theta_H$ becomes less than 4 for underdoped materials but cannot be fit to any power law for overdoped.  This change may be related to the purported QCP which occurs near x=0.16, but more detailed studies will be needed to verify this. The unusual power law dependence for the Hall
angle agrees with the theoretical model of \textcite{abrahams03a} at optimal doping. These authors showed that the Hall angle is proportional to the square of the scattering rate if this rate is measured by the T dependence of the ab-plane resistivity. Since a resistivity proportional to $T^2$ is found at all dopings for T above 100K \cite{Dagan07a}, but the Hall angle does not vary as $T^4$ for all dopings in this range this theoretical model can only be valid at optimal doping.  The origin of the temperature dependence for other dopings is not understood.

\subsubsection{Nernst effect, thermopower and magnetoresistance}
\label{Nernst}

The Nernst effect has given important information about the normal
and superconducting states in the cuprates. It is the thermal
analog of the Hall Effect, whereby a thermal gradient in the
$\hat{x}$ direction and a magnetic field in the $\hat{z}$
direction, induces an electric field in the $\hat{y}$ direction.
The induced field comes from the thermal drift of carriers and
their deflection by the magnetic field or by the Josephson
mechanism if moving vortices exist (for details see \textcite{YayuWang06a} and references therein). In conventional superconductors one finds a large Nernst signal in the superconducting state from
vortex motion and a very small signal in the normal state from the
carriers. Boltzmann theory predicts a zero Nernst signal
from a single band of carriers with energy independent scattering
\cite{Sondheimer48a,Wang01a}.  Surprisingly, a large Nernst signal
is found in the normal state of both electron and hole-doped
cuprates. However, the origin of this signal appears to be quite
different in the two cases. For $p$-type cuprates the large normal state Nernst effect has been attributed to superconducting fluctuations in a
large temperature region above $T_c$, especially in the range of
doping where the pseudogap exists \cite{Wang01a}. For $n$-type
cuprates the large Nernst signal was attributed to two types of
carriers in the normal
state\cite{Jiang94a,Fournier97a,Balci03a,Gollnik98a,Li07b}. The evidence
for a difference in behavior between $p$- and $n$-type cuprates is
persuasive.

\begin{figure}[t!]
\includegraphics[width=4.2cm,angle=0]{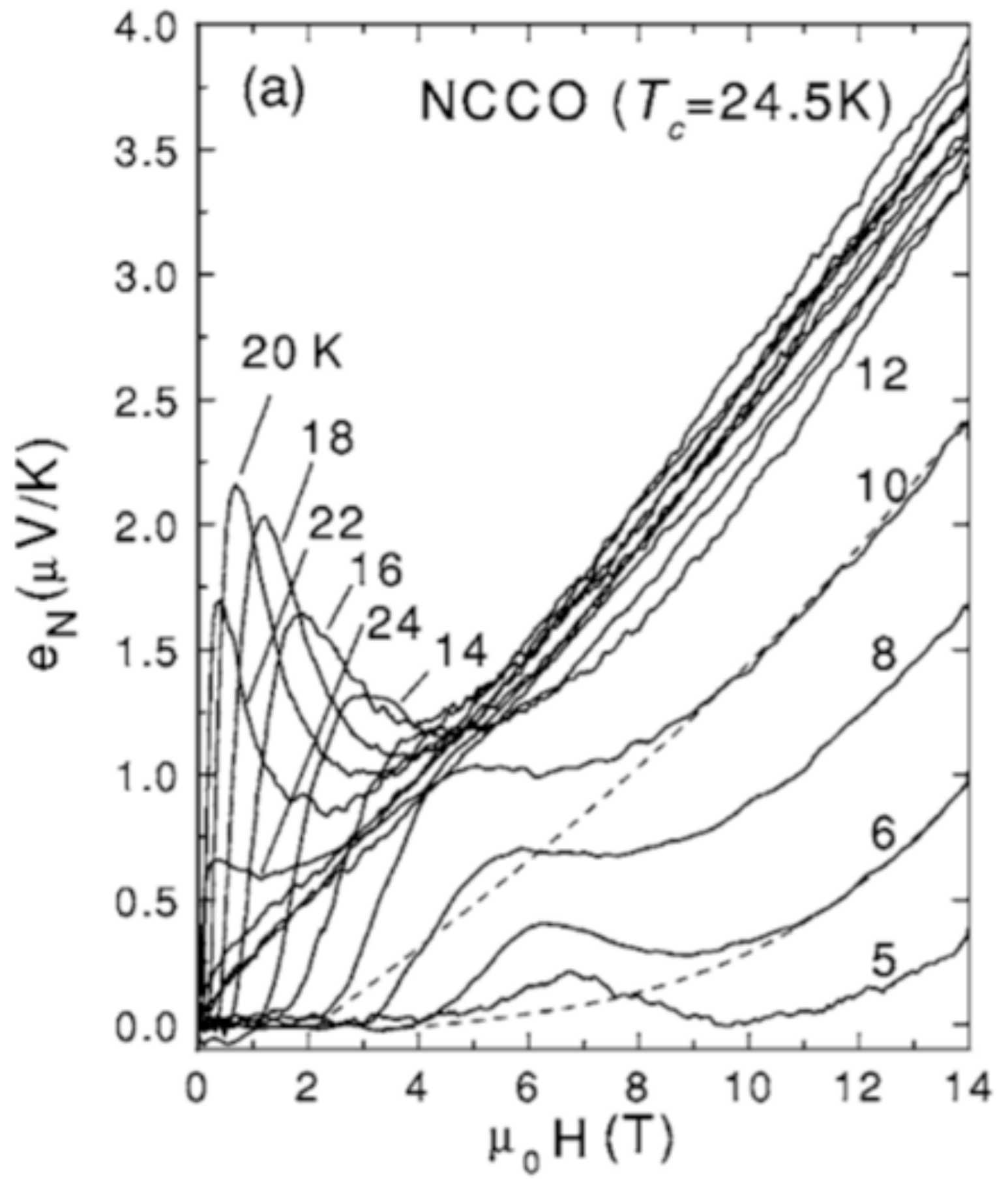}
\includegraphics[width=4.2cm,angle=0]{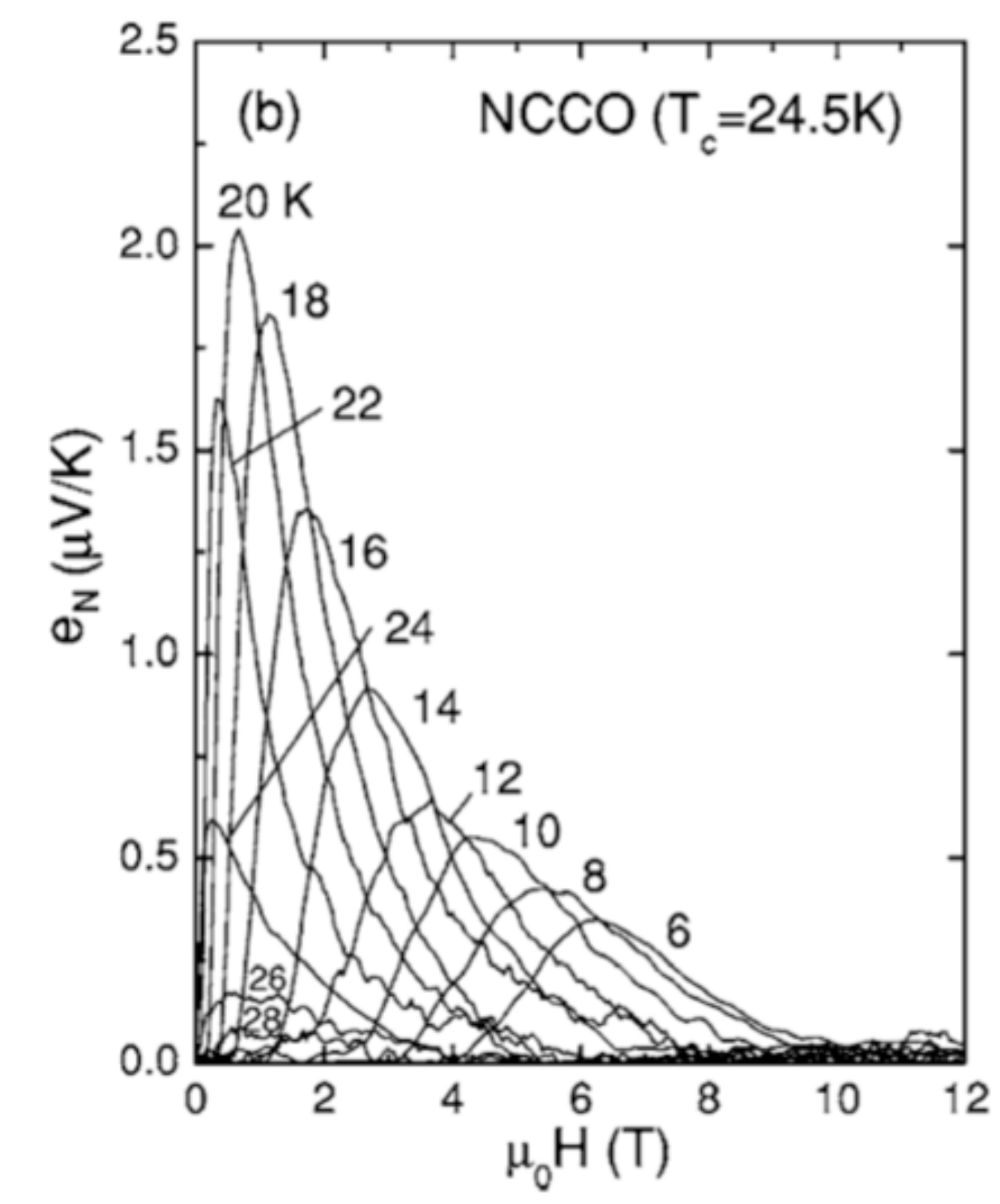}
\caption{(a) The experimentally measured Nernst signal e$_N$ vs H in optimally doped $x = 0.15$ NCCO and T$_c$ = 24.5 K from temperatures of 5 K to 30 K. The dashed lines are fits of the high-field segments to a quasiparticle term of the form $e_N^n(T,H) = c_1 H + c_3 H^3$ as detailed in Ref. \cite{YayuWang06a}.  (b)  The vortex contribution to the Nernst effect $e_N^s$ as extracted from the data of panel (a) as also detailed in Ref. \cite{YayuWang06a}.}
\label{OngNCCONernst}\end{figure}

The Nernst signal as a function of magnetic field at various
temperatures for optimal-doped NCCO is shown in Fig.~\ref{OngNCCONernst} \cite{YayuWang06a}. A vortex signal non-linear in field is seen for $H < H_{c2}$ for $T < T_c$ whereas for $T > T_c$ a linear in H normal state dependence is found. This
is behavior typical of low-$T_c$ superconductors, i.e., the
non-linear superconducting vortex Nernst signal disappears for $T
> T_c$ and $H > H_{c2}$. There is evidence for a modest temperature range of
superconducting (SC) fluctuations just above $T_c$ in the
underdoped compositions \cite{Balci03a,Li07a}. However, these data
contrasts dramatically from the data found in most hole-doped
cuprates. In those cuprates there is a very wide parameter range (in both T
and H) of Nernst signal due to SC fluctuations, interpreted as
primarily vortex-like phase fluctuations\cite{YayuWang06a}. This
interpretation of the large Nernst signal above $T_c$ in the
hole-doped cuprates is supported by recent theory (see \textcite{Podolsky07a} and references therein)\footnote{This interpretation has been recently challenged by a detailed Nernst effect measurement in Eu-LSCO as a function of Sr doping for which the Nernst signature assigned to the onset of the PG disappears at a doping where superconductivity persists \cite{Cyr09a}.}.  The inference that phase fluctuations are larger in hole-doped cuprates than the electron-doped is consistent with ``phase
fluctuation" models and estimates from various material parameters
\cite{Emery95a}.

\begin{figure}[t!]
\includegraphics[width=6.5cm,angle=0]{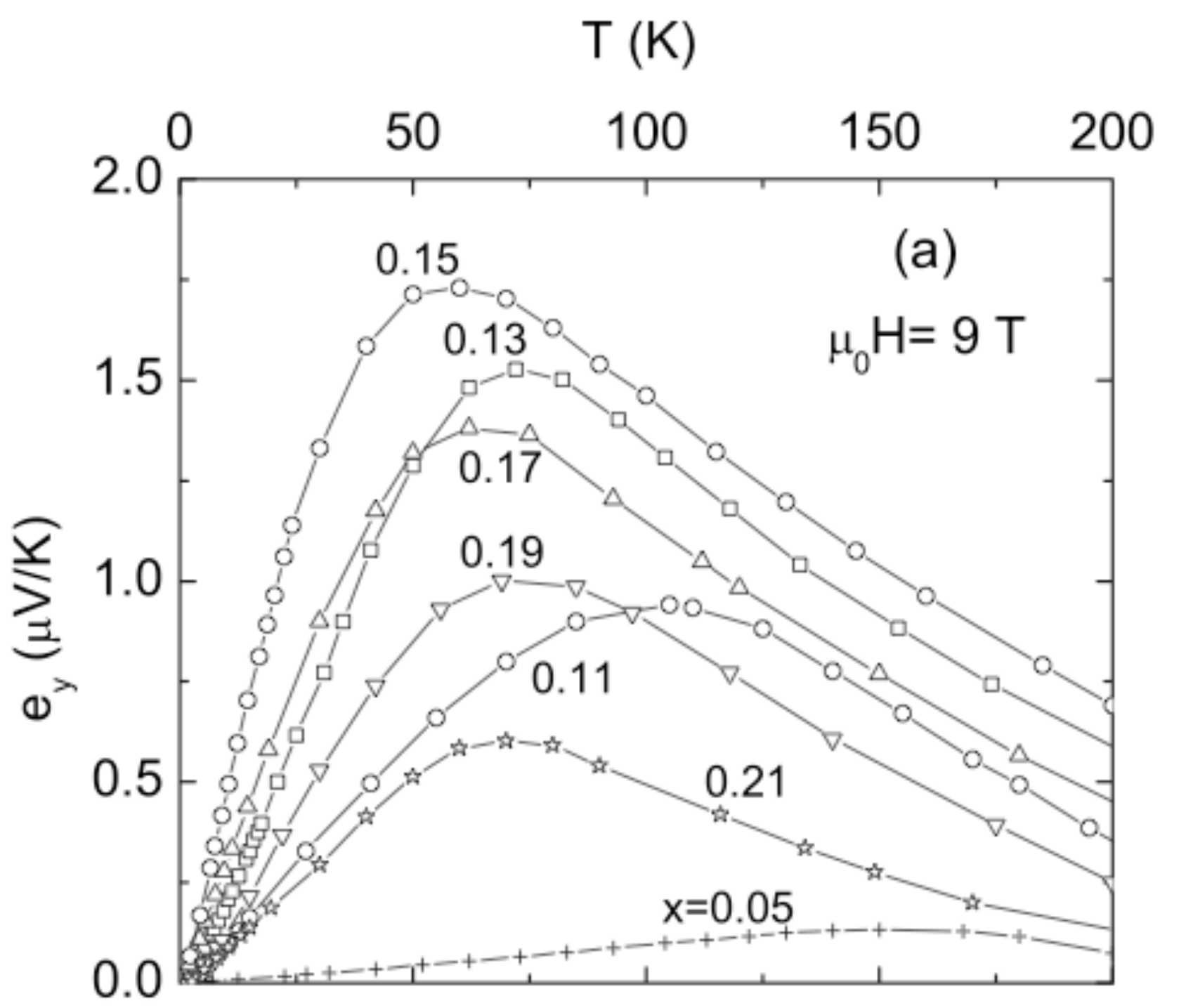}
\caption{Temperature dependence of normal-state Nernst signal at $\mu_0$H = 9 T for all the doped PCCO films from \textcite{Li07a}.}
\label{LiFig6PRB07}\end{figure}

However, the magnitude of the Nernst signal is large for
$T > T_c$ for both $n-$ and $p$-type cuprates for most doping levels
above and below optimal doping. The temperature dependence of the
Nernst signal at 9T ($ H \|_{c} > H_{c2}$) for several PCCO
dopings is shown in Fig.~\ref{LiFig6PRB07} \cite{Li07a}. It was found that at fixed temperature the signal is linear in field,  so that it can be assigned to normal state
carriers. In contrast, in the $p$-doped materials the field
dependence is non-linear for a wide T range above $T_c$, which
suggests a SC origin for the large Nernst signal. As mentioned
earlier, the large signal in the normal state of the $n-$doped
materials has been interpreted as arising from two carrier types.  This is consistent with the ARPES and optics data, which shows that both electron and hole regions of the Fermi surface (FS)
exist for dopings near optimal \cite{Armitage01b,Armitage02b,Zimmers05a}.  See \textcite{Li07a} for a thorough discussion on these points.  Recent theoretical work of \textcite{Hackl09a}, shows that the effects of FS reconstruction due to SDW order can give a magnitude and doping dependence of the low temperature Nernst signal that agrees with the measurements of Li and Greene.  However, further work will be needed for a quantitative explanation of the complete temperature dependence of the Nernset signal shown in Fig.~\ref{LiFig6PRB07}.

The ab-plane thermoelectric power (TEP) of the $n-$doped cuprates
has been measured by many authors \cite{Gollnik98a,Fournier97a,Xu96a,Budhani02a,CHWang05a,Li07a,PLi07a} (for work prior to 1995 see
\textcite{Fontcuberta96a}). To date, there has been few quantitative interpretation of
the temperature, doping and field dependence of the TEP in the cuprates \cite{Gasumyants95a, Sun08a}. However, a number of qualitative conclusions have been reached for the $n$-type compounds. The doping dependence of the low temperature magnitude and sign of the TEP is consistent with the evolution of the FS from electron-like at low doping to two-carrier-like near optimal doping to hole-like at the highest doping.  For example, at low doping ($x=0.03$) the low-T TEP is ``metallic-like" and negative \cite{Hagen91a,CHWang05a} even though the resistivity has an ``insulator-like" temperature
dependence. This reveals the presence of significant density of states at Fermi level and is consistent with the small pocket of electrons seen in ARPES and a possible 2D localization of these electrons at low temperature.  A recent detailed study of the doping dependence of
the low-temperature normal state TEP has given additional
evidence for a quantum phase transition (QPT) that occurs near x=0.16 doping \cite{PLi07a}.    It is remarkable that the appearance of superconductivity in the compounds is almost coincident with the appearance of a hole-like contribution to their transport.    Note that the existence of holes in this otherwise electron-doped metal is essential to the existence of superconductivity within some theories \cite{Hirsch89a}. 

The unusual and large magnetoresistance found in the $n-$doped cuprates has been
studied by a number of authors. The most striking behavior is the
large negative MR found for optimal and underdoped compositions at
low temperature ($T < T_{min}$). This has been interpreted as
arising from 2D weak localization\cite{Hagen91a,Fournier00a}, 3D
Kondo scattering from Cu$^{+2}$ spins in the CuO$_2$ plane
\cite{Sekitani03a}, or scattering from unknown magnetic entities
associated with the AFM state\cite{Dagan05a}. At low
doping ($x \leq 0.05$) the MR is dominated by an anisotopic
effect, largest for $H \| c$, and can reasonably be interpreted as
an orbital 2D weak localization effect (especially since the
ab-plane resistivity lead to $k_Fl < 1$ and follows a log T
temperature dependence). At dopings between x=0.1 and 0.17 the
negative MR is dominated by an isotropic effect and the orbital
contribution becomes weaker as the doping increases.

\textcite{Dagan05a} have isolated the isotropic MR and shown that
it disappears for $x \sim 0.16$. This suggests that this MR is
associated with a QCP occurring at this doping and is caused by
some heretofore unknown isotropic magnetic scattering related to the AFM state. \textcite{Dagan05a} also showed that the isotropic negative MR disappears above a $T_{min}$ and this suggests that
the upturn in the ab-plane resistivity is associated with the AFM state.   Recent high-field transverse magnetoresistance measurements (i.e., H applied along the c-axis)\cite{PLi07c} and angular magnetoresistance measurements (H rotated in the ab plane)\cite{Yu07a} support the picture of a AFM to PM quantum phase transition near $x=0.165$ doping \footnote{The existence of a QPT associated with the termination of the AF state at this doping is at odds with the work of  \textcite{Motoyama07a} who concluded $via$ inelastic neutron scattering that the spin stiffness $\rho_s$ fell to zero at a doping level of $x \approx 0.134$ (Fig.~\ref{GrevenStiffness}a).   This issue is discussed in more detail in Sec.~\ref{QCPoint}.}.

\subsubsection{c-axis transport}

The temperature and field dependence of the DC c-axis resistivity
has been studied in single crystals by many authors [for earlier
work see the review of \textcite{Fontcuberta96a}].  The
behavior of the $n-$doped c-axis resistivity is quite different than
that found in $p$-type cuprates.  Some representative
data as a function of doping and temperature is shown in Fig
\ref{OnoseRho} (center) and (right) \cite{Onose04a}.  This may reflect the
different gapped parts of the FS in $n-$ and $p$-type, since c-axis
transport is dominated by specific FS dependent matrix elements
\cite{Chakravarty93a,andersen95a} which peak near ($\pi$,0) and the SDW-like
state in the $n$-type as opposed to the unknown nature of the
pseudogap in the $p$-type. As shown by \textcite{Onose04a} (Fig
\ref{OnoseRho}) the c-axis resistivity has a distinct
change from ``insulating-like" to ``metallic-like" below a temperature T$^*$, near
the temperature at which the SDW gap is observed in optical
experiments \cite{Onose04a,Zimmers05a}, before going insulating at
the lowest temperature for the most underdoped samples. Below T$^*$,
the T dependence of the c-axis and ab-plane resistivity are
similar although with an anisotropy ratio of 1000-10000. This
behavior is strikingly different than that found in $p$-type
cuprates. In $p$-type compounds the c-axis resistivity becomes
``insulator-like" below the pseudogap temperature while the
ab-plane resistivity remains ``metallic" (down to $T_{min}$ in
underdoped compositions \cite{Ando01a}.
The interpretation of the c-axis resistivity upturn as a signature
of the pseudogap formation in $p$-type cuprates has been reinforced
by magnetic field studies, where the T$^*$ is suppressed by field in
a Zeeman-splitting-like manner \cite{Shibauchi01a}. A recent field-dependent study of $n$-type SCCO near optimal doping has been interpreted as for the
$p$-type cuprates \cite{Kawakami05a,Kawakami06a}. However,
the T$^*$ found in this work is much lower than that found in the
optical studies, casting some doubt on this interpretation. In
contrast, \textcite{Yu06a} have interpreted their field dependent
c-axis resistivity results in terms of superconducting
fluctuations. The origin of the c-axis resistivity upturn and its
relation to the ab-plane upturn requires more investigation.

\subsubsection{Effects of disorder on transport}
\label{disordertransport}

\begin{figure}[htb]
\includegraphics[width=0.5cm,angle=0]{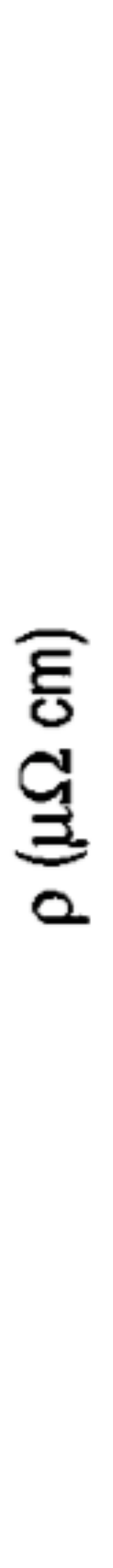}
\includegraphics[width=7.5cm,angle=0]{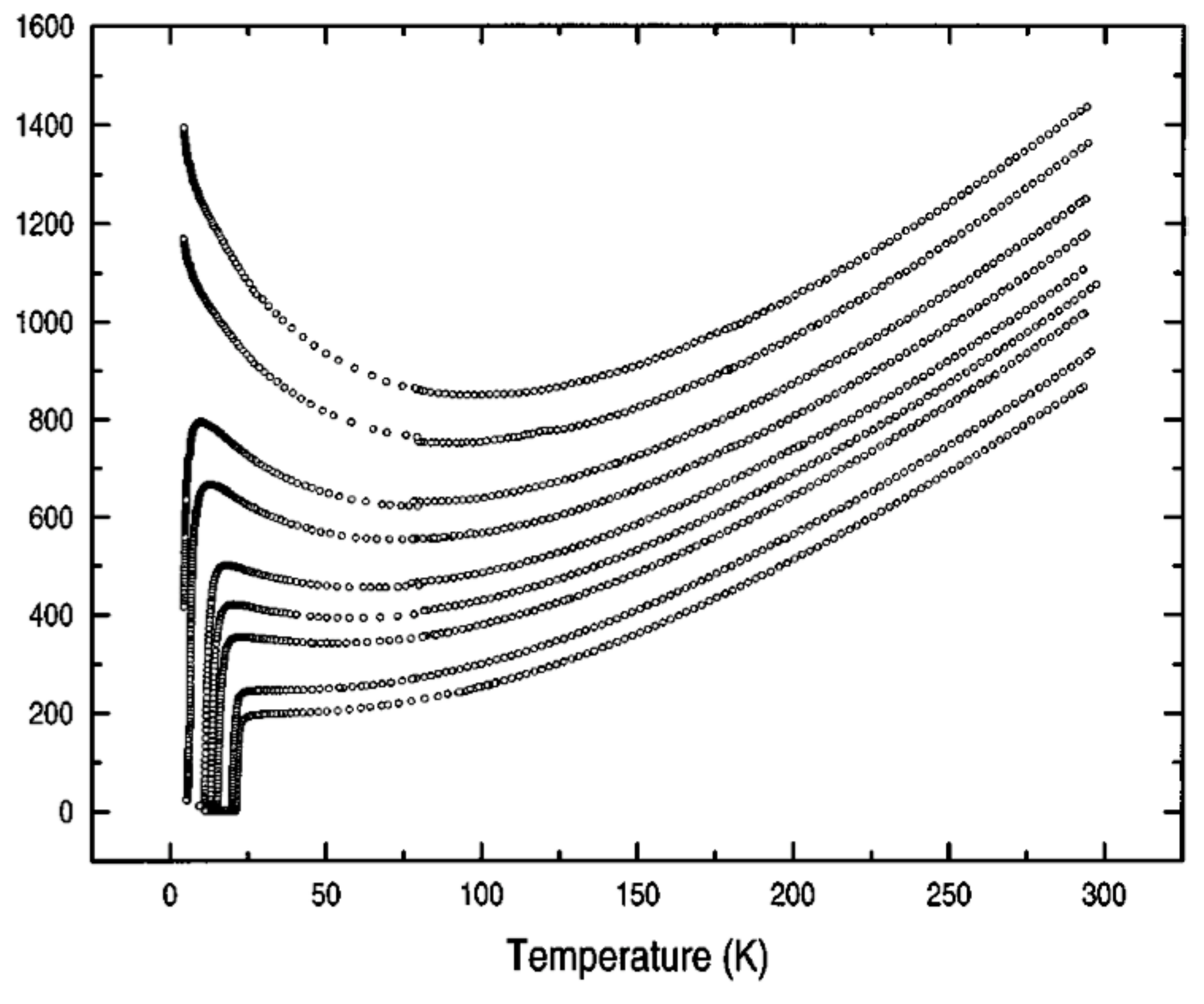}
\caption{Temperature dependent resistivity for NCCO $x=0.14$
films damaged with He$^+$ ions. From bottom to top, ion fluences are
0, 0.5, 1, 1.5, 2, 2.5, 3, 4, and 4.5 $* 10^{14}$ ions/cm$^2$.   From \textcite{Woods98a}.} \label{WoodsPRB}\end{figure}

\begin{figure}[htb]
\includegraphics[width=8cm,angle=0]{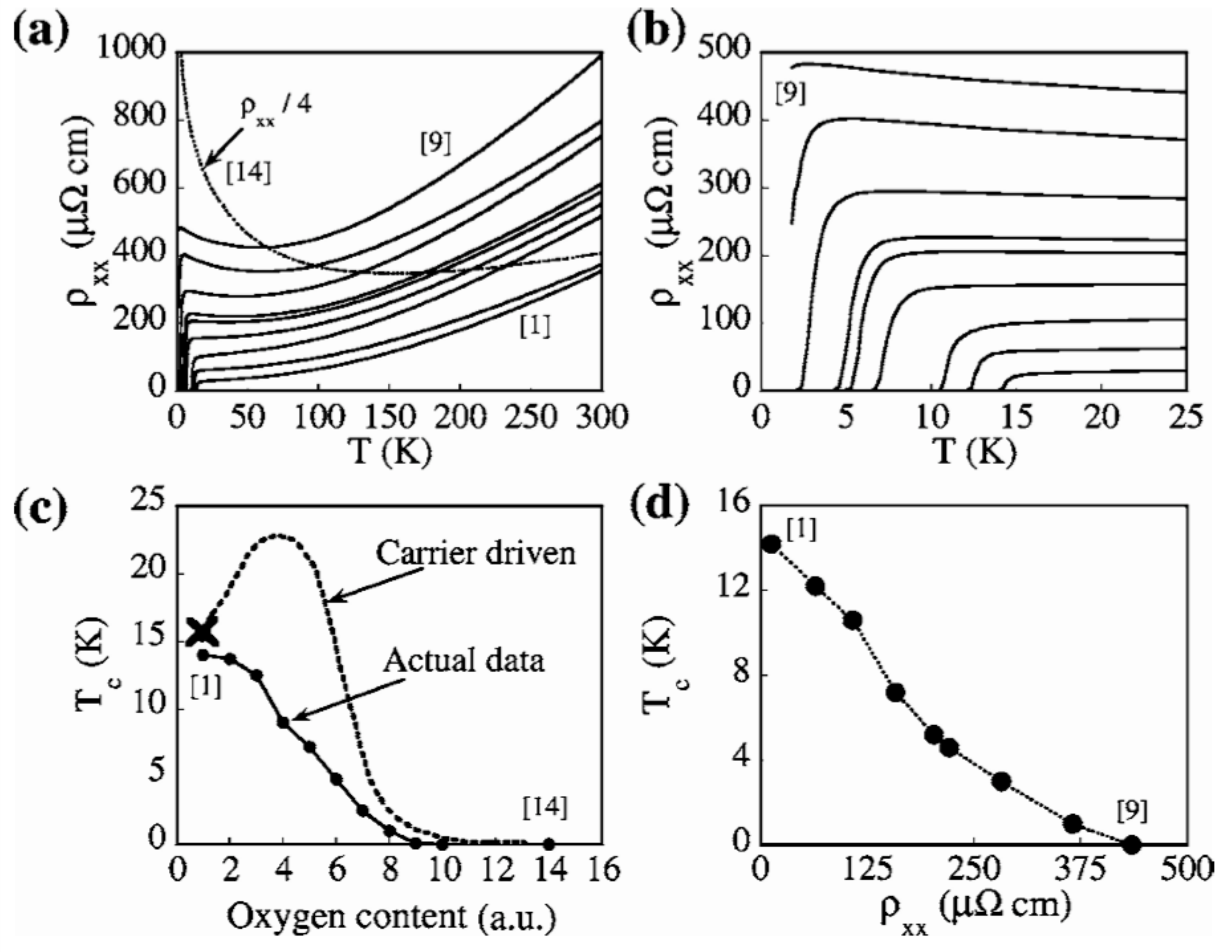}
\caption{(a) Resistivity as a function of temperature for $x = 0.17$ thin films with various oxygen contents. (b) Low-temperature region of the same data. (c) Critical temperature T$_c$ as a function of oxygen content for x = 0.17 for films grown in oxygen full circles, solid line is a guide to the eye. Cross: highest T$_c$ under N$_2$O.  Dashed line: schematic of the expected behavior for a carrier driven T$_c$ (see \cite{Gauthier07a}). (d) T$_c$ as a function of the in-plane resistivity at 30 K.   From \textcite{Gauthier07a}.} \label{Gauthier07}\end{figure}

Disorder has a significant impact on the $ab$-plane transport
properties of the cuprates. This has been studied most extensively
in the $p$-type materials [\textcite{Rullier08a,Alloul08a} and references
therein]. The results obtained to date on the $n$-type cuprates seem
to agree qualitatively with those found in $p$-type.
Disorder in these compounds is caused by the cerium doping itself, the annealing
process (where oxygen may be removed from some sites), doping of Zn or
Ni for Cu, and by ion or electron irradiation. The general
behavior of $\rho_{ab}(T)$ as defects are introduced by
irradiation is shown in Fig.~\ref{WoodsPRB} \cite{Woods98a} for
optimally doped NCCO.  The general trends are: $T_c$ is decreased,
the residual resistivity increases while the metallic T dependence
at higher T remains roughly the same, and an ``insulator-like"
upturn appears at low temperature. As the irradiation level
increases the superconductivity is eventually completely suppressed and the
upturn dominates the low temperature resistivity. The decrease of
$T_c$ is linearly proportional to the residual resistivity and
extrapolates to zero at $R_\Box$ per unit layer of ~5-10 k$\Omega$,
which is near the quantum of resistance for Cooper pairs
\cite{Woods98a}, similar to behavior seen in YBCO
\cite{Rullier03a}. Defects introduced by irradiation do not appear
to change the carrier concentration since the Hall coefficient is
basically unchanged. Oxygen defects (vacancies or impurity site
occupancy) can cause both changes in carrier concentration and in
impurity scattering.  In two recent papers \cite{Gauthier07a,higgins06a} have studied the effects of oxygen on the Hall effect and $\rho_{ab}$ of slightly overdoped PCCO.  As oxygen is added to an
optimally prepared x=0.17 film the $\rho(T)$ behavior (Fig.~\ref{Gauthier07}) becomes
quite similar to the $\rho(T)$ data under increased irradiation as
shown in Fig~\ref{WoodsPRB}.   \textcite{Gauthier07a}  attribute the role of oxygen not to changing the carrier concentration significantly but to having a
dramatic impact on the quasiparticle scattering rate.
\textcite{higgins06a} compare resistivity and Hall effect for
films with oxygen variation and with irradiation. They conclude
that oxygen changes both the carrier concentration and the
scattering rate.  The exact origin of all these disorder effects
on $T_c$ and the transport properties has not yet been determined.  However, various recent proposals \cite{Rullier08a,Alloul08a} for how defects influence the properties of hole-doped cuprates are probably valid for $n-$doped cuprates as well.

\subsubsection{Normal state thermal conductivity}
\label{NSThermalConductivity}

In general the ab plane thermal conductivity $\kappa$ of the $n-$type materials resembles that of the hole-doped compounds.  In the best crystals an increase in $\kappa_{ab}$ is found at $T_c$ and can be attributed to a change in electron-phonon scattering as in the hole-doped cuprates \cite{Yu92a}.  The most significant $\kappa$ data has been taken below $T_c$ at temperatures down to 50mK for $H>H_{c2}$. A striking result was the report of a violation of the  Wiedemann-Franz law below 1K
in slightly underdoped PCCO \cite{Hill01a} samples, which was
interpreted as a possible signature of a non-Fermi liquid normal state.  This will be discussed in more detail in section~\ref{nonFL} below.

\textcite{Sun04a} measured the $ab$-plane and c-axis thermal
conductivity for underdoped crystals of
Pr$_{1.3-x}$La$_{0.7}$Ce$_x$CuO$_4$. They found that the low T phonon
conductivity $\kappa$ has a very anisotropic evolution upon
electron doping; namely, the low-T peak of $\kappa _c$ was much
more rapidly suppressed with doping than the peak in $\kappa
_{ab}$. Over the same doping range the ab-plane resistivity
develops a ``high mobility" metallic transport in the AFM state.
They interpret these two peculiar transport features as evidence
for stripe formation in the underdoped $n$-type cuprates.
Essentially the same features are seen in underdoped $p$-type
cuprates \cite{Ando01a} where the evidence for stripe formation is stronger.

In the underdoped $n$-type compounds, phonons, magnons and electronic carriers (quasiparticles) all contribute to the thermal conductivity.  Only at very low temperature it is possible to separate out the various contributions. However since phonons and magnons both have a $T^3$ variation, it has been necessary in undoped and AFM Nd$_2$CuO$_4$ to use the magnetic field induced spin-flop transition to switch on and off the acoustic Nd magnons and hence separate the magnon and phonon contributions to the heat transport \cite{Li05b}.

\subsection{Tunneling}
\label{TransportTunneling}

Tunnelling experiments on $n-$doped cuprates have been difficult and
controversial. This is likely due to the problems associated with
preparing adequate tunnel barriers and the sensitivity of the
electron-doped material to preparation conditions. Some of these
difficulties have been discussed by \textcite{Yamamoto97a}.
Improvements have been made in recent years and we will focus on
the most recent results.  It is important to keep in mind that the
surface layer being probed by tunnelling is very thin (of order
the coherence length) and the surface may have properties
different than the bulk because the oxygen reduction conditions at
the barrier may not be the same as the interior. Experiments
that show a bulk $T_c$ or $H_{c2}$ in their tunnel spectra are
most likely to represent properties of the bulk. We only discuss
what appear to be measurements representative of the bulk. Tunnelling
experiments have been performed on films and single crystals using
four methods; natural barriers with metals such as Pb, Sn, Al,
In, and Au, point contact spectroscopy with Au or Pt alloy
tips, bi-crystal grain boundary Josephson junctions (GBJ) on
STO substrates, and Scanning Tunnelling Measurements (STM).  Thus, these experiments are either in superconductor-insulator-superconductor (SIS),
superconductor-insulator-normal metal (SIN), or SN configurations. 

\begin{figure}[htb]
\includegraphics[width=8cm,angle=0]{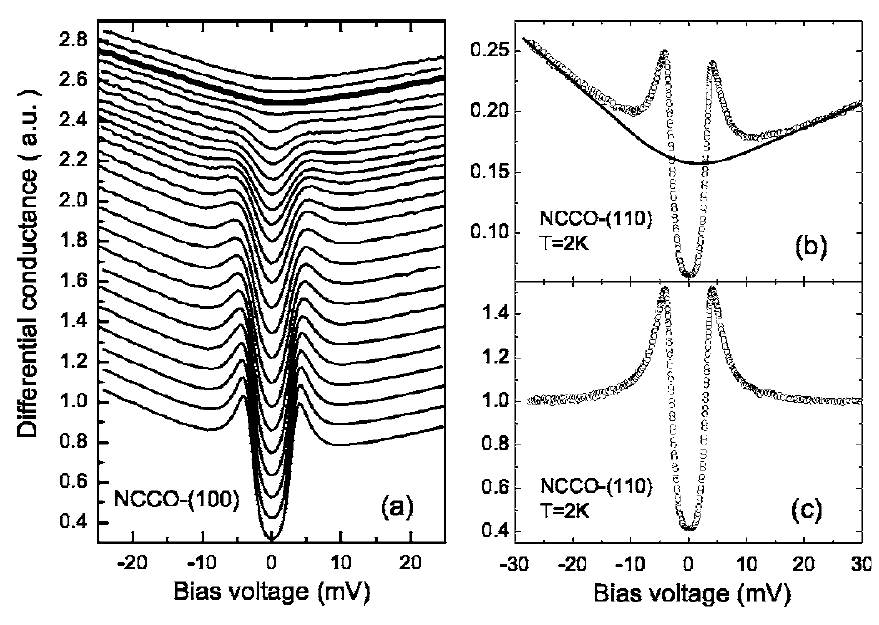}
\caption{Raw data of the directional tunneling measurements for optimally doped NCCO. (a)
Temperature dependence of the tunnelling spectra measured along
(100) direction. The curves have been shifted for clarity. The
temperature increases from the bottom upwards in steps of 1 K
(from 2 to 22 K) and then 2 K (from 22 to 30 K). The thick solid
line denotes the data at 26 K which is approximately $T_c$. (b) An
illustration of the constructed normal conductance background
above $T_c$. (c) The normalized 2 K spectrum in the (110)
direction.  From \textcite{Shan05a}.} \label{ShanPRB}\end{figure}

The aim of the tunnelling experiments has been to determine the SC
energy gap, find evidence for bosonic coupling, the SC pairing
symmetry, and evidence for a normal state gap (pseudogap). Typical quasi-particle
conductance G(V)=dI/dV spectra on optimal-doped NCCO using point contact spectroscopy are shown in Fig~\ref{ShanPRB}.  Similar spectra are found for Pb/PCCO natural
barrier junctions \cite{Dagan05a} and GB junctions
\cite{Alff98a,Chesca05a}. The main features of the $n-$doped tunnel
spectra are: prominent coherence peaks (which give an energy gap
of order 4 meV at 1.8K), an asymmetric linear background G(V) for
voltage well above the energy gap, a characteristic 'V' shape,
coherence peaks which disappear completely by $T \approx T_c$ at
H=0 (and by $H \approx H_{c2}$ for T=1.8K), and typically the absence of a
zero bias conductance peak (ZBCP) at V=0.  Issues related to the determination of the order parameter are discussed in more detail in Sec.~\ref{sec:tunnel} below.

\begin{figure}[htb]
\includegraphics[width=6cm,angle=0]{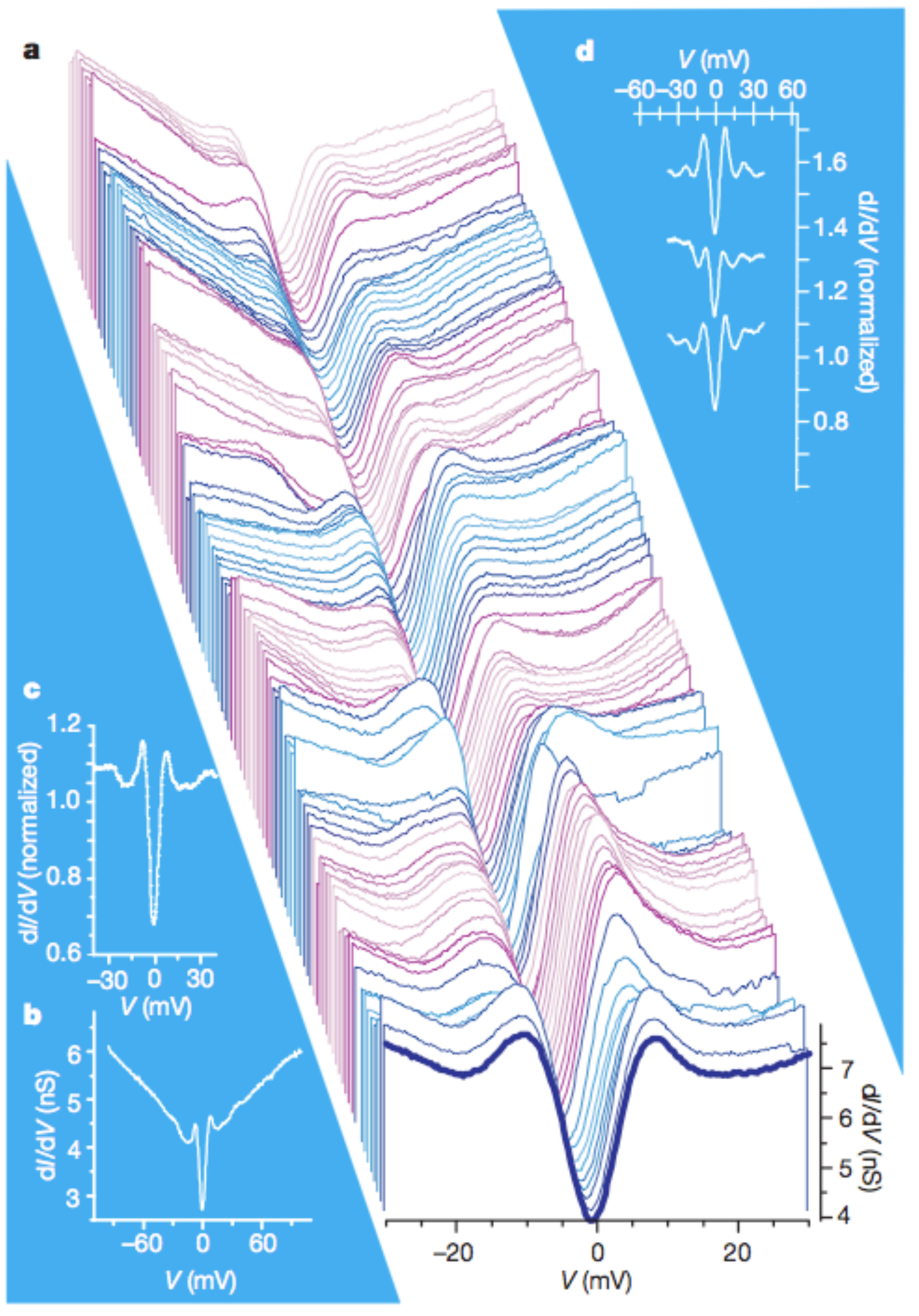}
\caption{(Color)  (a) A 200 {\rm \AA} linecut that shows the variations in coherence peak heights and gap magnitude ($\Delta$) The spectra have been offset for clarity. The gap magnitude, which is
defined as half the energy separation between the coherence peaks varies from 5 meV to 8 meV  in this linecut. (b) A representative $\pm$100-mV range (dI/dV) spectrum that illustrates the dominate `V'-shaped background. (c) The spectrum in (b) after division by a linear V-shaped function. (d) Additional examples of dI/dV spectra that demonstrate the clearly resolved coherence peaks and modes resulting from a V-shaped division.  From \textcite{Niestemski07a}.}\label{PLCCOSTM}\end{figure}

Tunneling experiments have also found evidence for a normal state
energy gap with energy $\sim$5 meV at 2K, which is of the same
order as the superconducting gap energy
\cite{Kleefisch01a,Biswas01a}. This normal state gap (NSG) is
found in SIS experiments and  point contact spectroscopy experiments which probe the
ab-plane by applying a c-axis magnetic field greater than
$H_{c2}$. The low energy NSG is distinctly different than the high
energy ($\sim$ 100 meV) ``pseudogap" seen in ARPES and optical
experiments \cite{Armitage01b,Zimmers05a} and most recently in a local tunneling spectroscopy experiment \cite{Zimmers07a}. The high energy gap is suggested to be
associated with SDW-like gapping of the FS. The origin of the low
energy NSG is not conclusively determined at this time. Proposed
explanations include: Coulomb gap from electron-electron
interactions \cite{Biswas01a}; hidden and competing order
parameter under the SC dome which vanishes near optimal doping \cite{Alff03a}; and preformed SC singlet pairs \cite{Dagan05b}.  \textcite{Dagan05b} claim to rule out the Coulomb
gap and competing order scenarios. They find that the NSG is
present at all dopings from 0.11 to 0.19 and the temperature at
which it disappears correlates with $T_c$, at least on the
overdoped side of the SC dome.  However, the NSG also survives to
surprisingly high magnetic fields and this is not obviously
explained by the preformed pair (SC fluctuation) picture either
\cite{Kleefisch01a,Biswas01a,Yu06a}.  In contrast, \textcite{Shan08b} reported that the NSG and the SC gap are distinct entities at all dopings, which is consistent with the `two-gap' scenario in the underdoped $p$-type cuprates.

Very few STM studies have
been performed on the $n$-type compounds as compared to the
extensive measurements on the hole-doped materials
\cite{Fischer07a}.  \textcite{Niestemski07a} reported reproducible high resolution STM measurements of $x=0.12$ PLCCO (T$_c=24$ K) (Fig.~\ref{PLCCOSTM}).  The extremely inhomogeneous nature of doped transition metal oxides makes spatially resolved STM an essential tool for probing local energy scales. Statistics of the superconducting gap spatial variation were obtained through thousands of mappings in various regions of the sample.  Previous STM measurements on NCCO gave gaps of 3.5 to 5 meV, but no obvious coherence peaks \cite{Kashiwaya98a}.  The linecut  (Fig.~\ref{PLCCOSTM}a) shows spectra that vary from ones with sharp coherence peaks to a few with more pseudogap-like features and no coherence peaks.  Although most measured samples at this doping (9 out of 13 mappings) gave gaps in the range of 6.5 - 7.0 meV, the average gap over all measured maps was $7.2 \pm 1.2$ meV, which gives a $2 \Delta/k_B T_c$ ratio of 7.5, which is consistent with a strong coupling scenario.  However, this ratio strongly differs with point contact \cite{Shan08a} and SIS planar tunneling results \textcite{Dagan07b} as well as Raman scattering \cite{Qazilbash05a}, which have given a  $2 \Delta/k_B T_c$ ratio of approximately 3.5 for PCCO at $x = 0.15$. This may be pointing to a $2\Delta / k_B T_c$ ratio that varies significantly with $x$ as seen in the $p$-type compounds \cite{Deutscher99a}.

The STM spectra have a very notable `$V$' shaped higher energy background.  When this background is divided out a number of other features become visible.  Similar to the hole-doped compounds \cite{Fischer07a},  the claim is that features in the tunneling spectra can be related to an electron-bosonic mode coupling at energies of $10.5 \pm 2.5$ meV. This energy is consistent with an inferred magnetic resonance mode energy in PLCCO \cite{Wilson06b} as measured by inelastic neutron scattering as well as low-energy acoustic phonon modes, but differs substantially from the oxygen vibrational mode identified $via$ STM as coupling to charge in BSCCO\cite{JLee06a}.  The analysis of \textcite{Niestemski07a} of the variation of the local mode energy and intensity with the local gap energy scale was interpreted as being consistent with an electronic origin of the mode consistent with spin-excitations rather than phonons.

\subsection{ARPES} \label{ARPES}

The first angle resolved photoemission (ARPES) studies of
the electron-doped cuprates  appeared in adjoining 1993 Physical Review Letters \cite{Anderson93a,King93a}.  Both reported the existence of a large Fermi surface centered around the
($\pi,\pi$) position in Nd$_{1.85}$Ce$_{0.15}$CuO$_4$.  It had a
volume that scaled approximately with the number of charge
carriers thereby satisfying Luttinger's theorem and a shape
similar to existing band structure calculations \cite{Massidda89a}.
It was pointed out by \textcite{King93a} that the extended Van Hove
states at the $(\pi,0)$ point located at approximately 350
meV binding energy contrasted with the hole-doped case,
where these states were located within tens of meV of E$_F$.  It was
speculated at that time that the lack of a large near-E$_F$ density of states
may be responsible for some of the very different hole and
electron-doped compound properties.   Subsequent calculations also emphasize this difference as a route to explaining the differences between electron- and hole-doped compounds \cite{Manske01b}.

Recently there have been a number of electron-doped ARPES studies which
take advantage of dramatic advances in photoemission technology,
including the vastly improved energy ($< 10$ meV) and momentum ($<
1\%$ of $\pi/a$ for a typical cuprate) resolution as well as
the utility provided by parallel angle scanning in $Scienta$-style
detectors \cite{Armitage01a,Armitage01b,Armitage02a,Armitage03a,Sato01a,Matsui05a,Matsui05b}.   The contribution of ARPES to the study of the superconducting order parameter is detailed below in Sec.~\ref{sec:ARPES-symm}.

\begin{figure}[t!]
\includegraphics[width=9cm,angle=0]{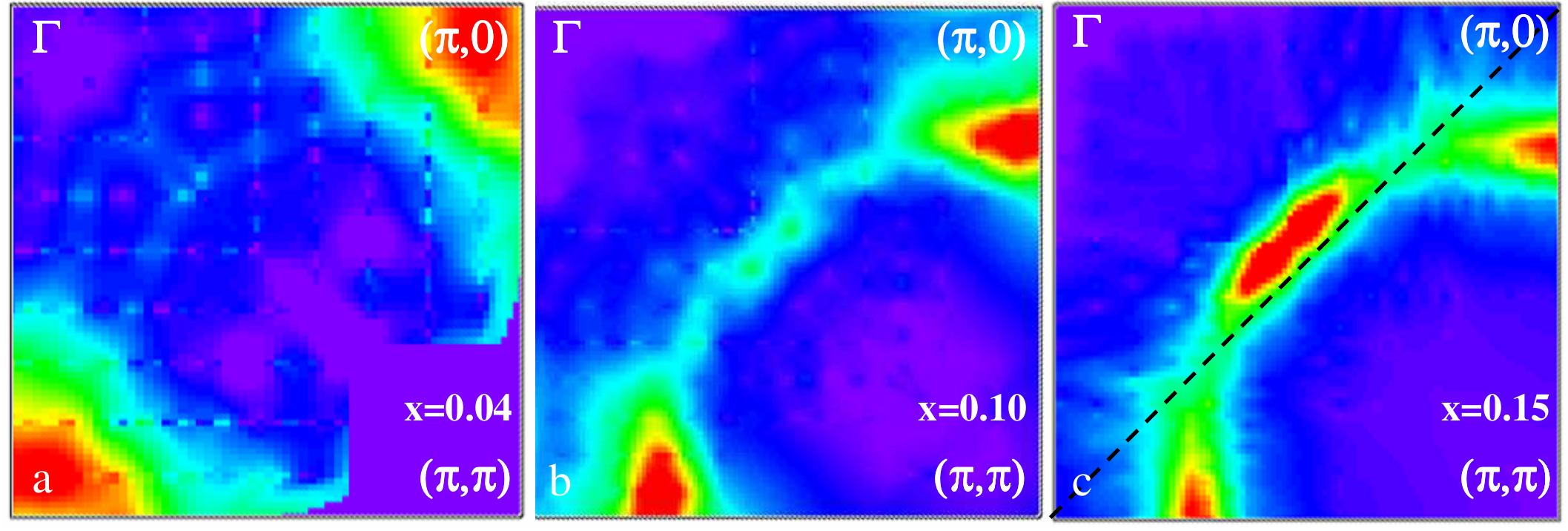}
\caption{(Color) Fermi surface plot: (a) $x=0.04$,(b) $x=0.10$,
and (c) $x=0.15$.  EDCs integrated in a 60meV window
(-40meV,+20meV) plotted as a function of $\vec{k}$. Data were
typically taken in the displayed upper octant and symmetrized
across the zone diagonal.  Adapted from \textcite{Armitage03a}. }
\label{PeterARPES}\end{figure}

In studies concerning the overall electronic structure, the large
Fermi surface around the ($\pi,\pi$) position was confirmed in the
later high resolution studies by \textcite{Armitage01b}, but it was
also found that there are anomalous regions on the Fermi surface
where the near $E_F$ intensity is suppressed (Fig.~\ref{PeterARPES}(c)). A detailed look at
the Energy Distribution Curves (EDCs) through the suppressed
region of the Fermi surface reveals that the electronic peak
initially approaches E$_F$, but then monotonically loses weight
despite the fact that its maximum never comes closer than ~100 meV
to E$_F$. Such behavior with broad features and suppression of
low-energy spectral weight is similar to the high-energy pseudogap
seen in the extreme underdoped $p$-type materials \cite{marshall96a}, although in the present case it is observed to be maximum near ($0.65\pi,0.3\pi$) and not at ($\pi,0$), the maximum of the $d$-wave
functional form.

As noted by \textcite{Armitage01b,Armitage02b} these regions of
momentum space with the unusual low-energy behavior fall close to
the intersection of the underlying FS with the antiferromagnetic Brillouin zone (AFBZ) boundary, as
shown by the dashed lines in Fig.~\ref{PeterARPES}(c). This
suppression of low-energy spectral weight and the large scattering
rate in certain regions on the FS is reminiscent of various theoretical
predictions that emphasize a coupling of charge carriers to a
low-energy collective mode or order parameter with characteristic momentum
($\pi,\pi$). A simple phase space argument shows that it is those
charge carriers which lie at the intersection of the FS with the
AFBZ boundary that suffer the largest effect of anomalous
($\pi,\pi$) scattering as these are the only FS locations that can
have low-energy coupling with $q \approx (\pi,\pi$).  These
regions were those later inferred by
\textcite{Blumberg02a,Matsui05b} to have the largest
superconducting gap as well.  Although it is the natural choice, due to the close proximity of
antiferromagnetism and superconductivity, this low-energy
scattering channel need not be antiferromagnetic for the role
played by the AFBZ boundary to hold; other possibilities such as
$d$-density wave exist \cite{Chakravarty01a}.  It is only necessary
that its characteristic wave vector be ($\pi,\pi$).
 These heavily scattered regions of the FS have been referred to in
the literature as ``hot spots" \cite{hlubina95}. It has been
suggested that the large backscattering felt by charge carriers in
the hot spots is the origin of the pseudogap in the underdoped
hole-type materials\footnote{We should mention that the observation of ``hot-spots" has been
disputed \cite{claesson04a} in an ARPES study that used
higher energy photons, thereby gaining marginally more bulk
sensitivity over other measurements.  It is unclear however, whether this study's
relatively poor energy resolution (140 meV as compared to
$\approx$ 10 meV in other studies) coupled with a large near-E$_F$
integration window (136 meV) can realistically give any insight
into this matter regarding low energy spectral suppression when
the near $E_F$ suppression is observed primarily at energies below
70 meV.}.

The gross features of the ARPES spectra in the optimally doped
$n$-type compounds can be approximately described by a two band
model exhibiting long range SDW order
\cite{Armitage02b,Matsui05a,Park07a}. Such a model reflects the
folding of the underlying band structure across the AFBZ boundary and hybridization between bands $via$ a potential $V_{\pi,\pi}$ (see Sec.~\ref{sec:SDW} below).  It gives the two components (peak-dip-hump (PDH) structure) in the spectra near the ($\pi,0$) position
\cite{Armitage01a,Sato01a,Armitage03a,Matsui05a}, the location of
the hot spots, and perhaps more subtle features showing back folded
sections of the FS.  As will be discussed below in
more detail, such a scenario does shed light on a number of other
aspects of $n$-type compounds, including one long outstanding issue in
transport where both hole and electron contributions to the Hall
coefficient have been resolved \cite{Dagan04a,Gollnik98a,Wang91a,Fournier97a}.  Additionally, such a scenario appears to be consistent with aspects of the optical data \cite{Zimmers05a}.  

\begin{figure}[t!]
\includegraphics[width=8cm,angle=0]{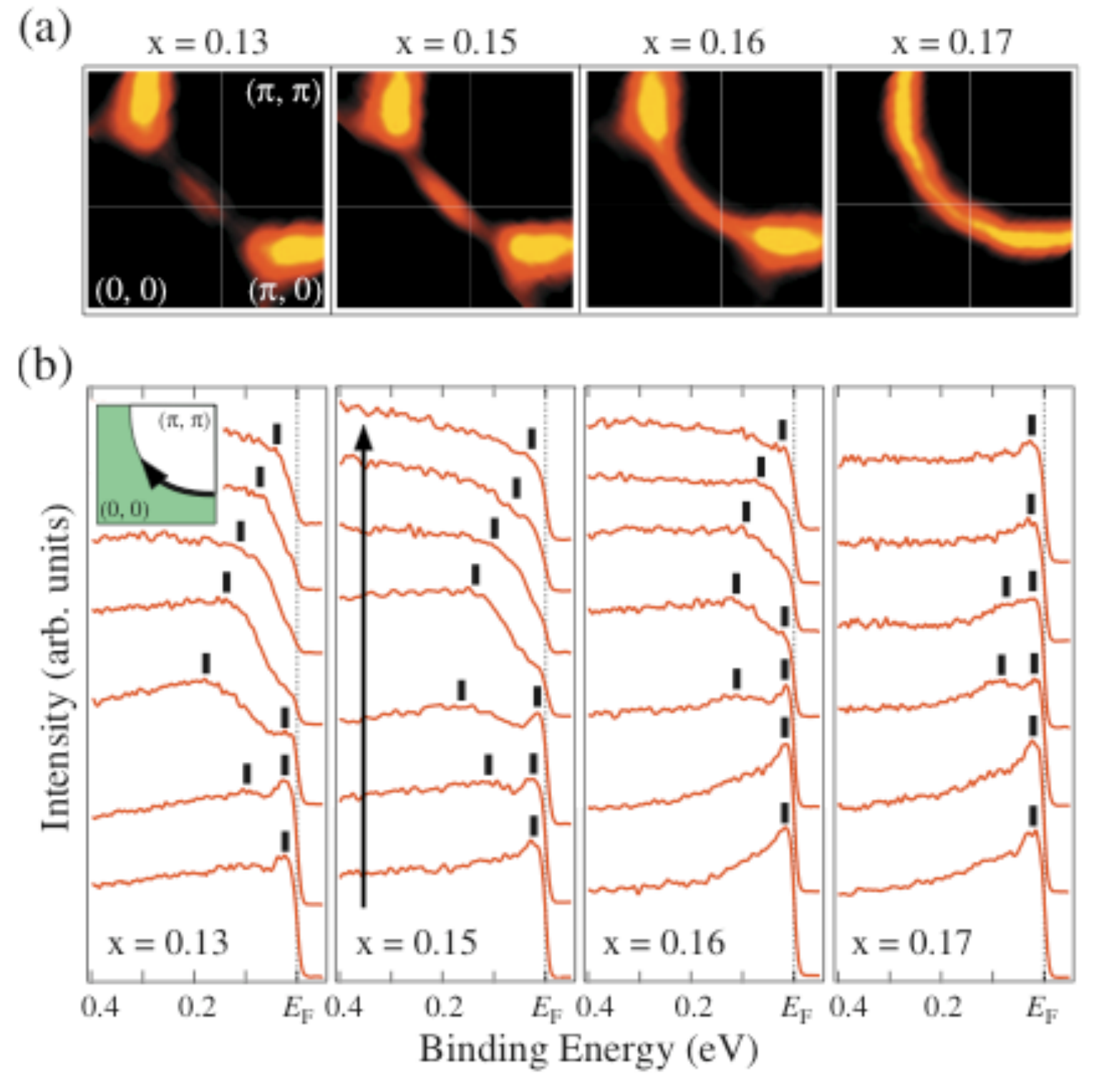}
\caption{(Color) (a) Doping dependence of the FS in NCCO, 
obtained by plotting the ARPES intensity integrated over $\pm$20 meV 
with respect to E$_F$ as a function of momentum. 
The intensity is normalized to that at 400 meV binding energy and symmetrized with respect to the ($0,0$) -  ($\pi,\pi$) direction.  (b) Doping dependence of a set of ARPES spectra measured at 
several k points around the FS at several dopings. From \textcite{Matsui07a}. } \label{MatsuiFS}\end{figure}

\begin{figure}[t!]
\includegraphics[width=5cm,angle=0]{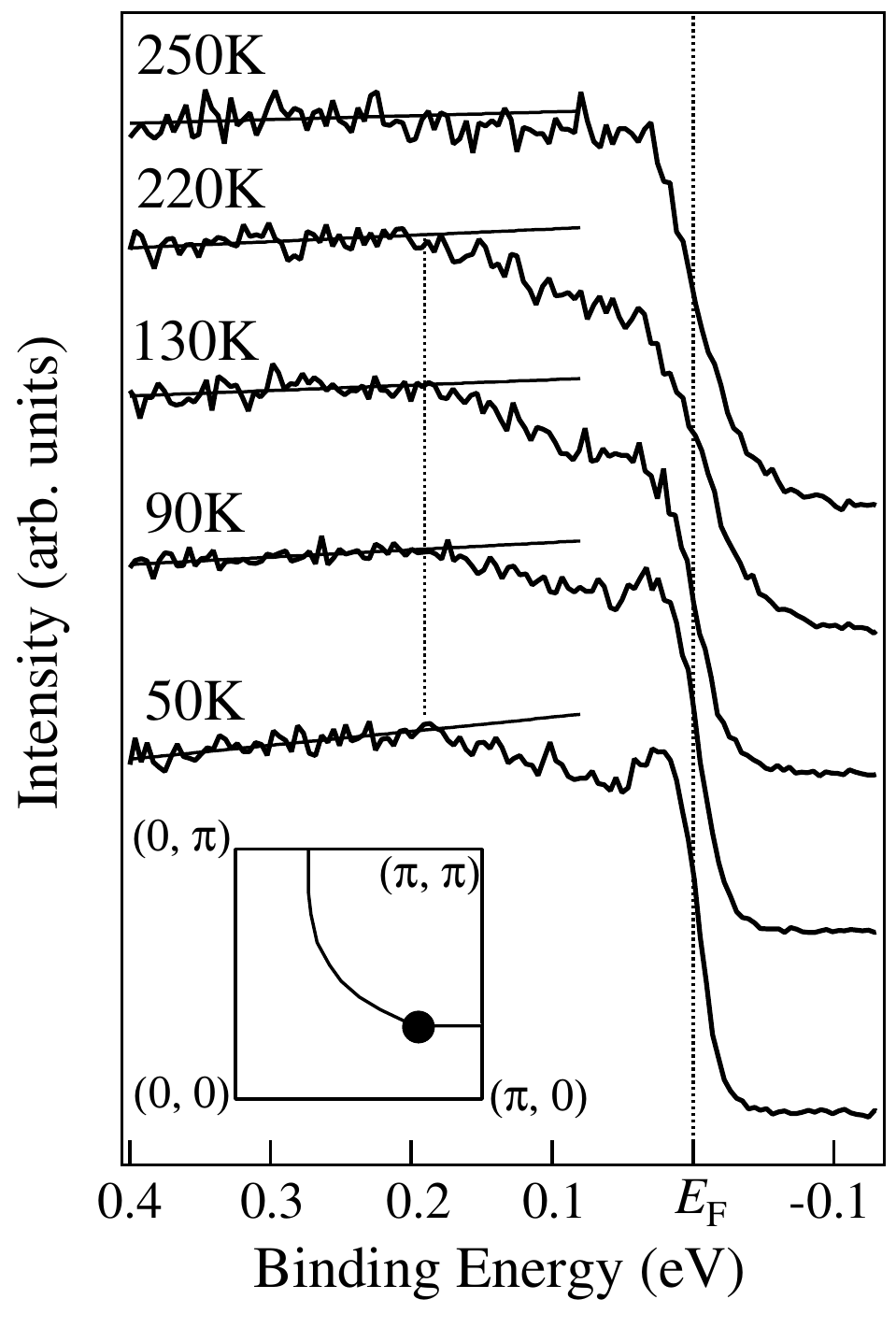}
\caption{Temperature dependence of the ARPES spectrum of NCCO (x =
0.13) measured in the ``hot spot" (at the position on the Fermi
surface shown by a circle in the inset) where the two component
structure is observed clearly. The solid straight lines on the
spectra show the linear fits to the high-energy region (0.2–0.5
eV) showing that it doesn't change with
temperature. From \textcite{Matsui05a}. } \label{MatsuiTemp}\end{figure}

\textcite{Matsui05a} found that the
lineshape in these ``hot-spots" in NCCO x=0.13 have a strong
temperature dependence, giving more credence to the idea that this
suppression is due to spin density wave formation
\cite{Matsui05a}.  As shown in Fig.~\ref{MatsuiFS}, \textcite{Matsui07a} also demonstrate that the hot-spot effect largely goes away by x=0.17 doping in NCCO, with the high-energy pseudogap filling in at the antiferromagnet-superconductor phase boundary.  The magnitude ($\Delta_{PG}$) and the temperature (T$^*$) at which the pseudogap fills in shows a close relation to the effective magnetic coupling energy (J$_{eff}$) and the spin-correlation length ($\xi_{AF}$), respectively again suggesting the magnetic origin of the pseudogap and hot-spot effect.  As seen in Fig.~\ref{MatsuiTemp}, it was shown that the lowest energy sharp peak had largely disappeared by the N\'{e}el temperature $T_N = 110K$ for the $x=0.13$ sample while the near-E$_F$ spectral weight suppression persisted until a higher temperature
scale.  These authors also claimed that the overall k-space
dependence of their data was best understood within a spin density
wave model with a non-uniform SDW gap in k-space.

In contrast, \textcite{Park07a} in a comprehensive BZ wide study
on SCCO claim that it was not that the gap was non-uniform, but
that there appeared to be remnant bands reflective of the bare
band structure that dispersed uninterrupted through the AFBZ.
Through a simple model they showed how this might be reflective of short-range magnetic ordering.  Moreover, these authors showed that the hot-spot effect in SCCO is so strong that the zone diagonal
states were actually pushed below $E_F$ raising the possibility of
nodeless $d$-wave superconductivity in this compound.  This observation may shed light on reports of nodeless superconductivity found in PCCO films grown on buffered substrates \cite{Kim03a}.  Such effects had been theoretically anticipated \cite{Yuan06a,Das07a}.

\textcite{Richard07a} have made a detailed comparison of the ARPES spectra of as-grown and oxygen-reduced PCCO and PLCCO materials.  They claim that to within their error bars (estimated by us to be approximately 1 $\%$)  neither the band filling nor the tight binding parameters are significantly affected by the reduction process in which a small amount of oxygen was removed ($\approx 1 \%$).  They demonstrated that the main observable effect of reduction was to remove an anisotropic leading edge gap around the Fermi surface.

Much recent discussion regarding ARPES spectra of the hole-doped
cuprates has concerned a ``kink" or mass renormalization in the electronic dispersion which has been found ubiquitously in the $p$-type materials (at $\approx$ 70 meV) \cite{Bogdanov00a,Lanzara01a}.  Its origin is a matter of much current debate \cite{Damascelli03a,Campuzano04a}, with various phononic or magnetic scenarios being argued for or against.  Its existence on the electron-doped side of the phase diagram has been controversial.    \textcite{Armitage03a} claimed that there was no kink feature along the zone diagonal and that the zone diagonal was best characterized by a smooth concave downward dispersion.   Although apparent mass renormalizations were found along the zone face \cite{Sato01a,Armitage03a,Matsui05a}, it was claimed these were related to the ``hot spot" effect and therefore of different origin than the $p$-type kink\footnote{\textcite{Armitage03a} gave effective Fermi velocities that did not include the effects of any kinks, by using a $\omega \rightarrow 0$ extrapolation of a function fitted to the dispersion at higher energy ($\omega > 100 $ meV).  Analyzing the data in this fashion gives velocities of ${\vec v_{F}^{0 \medskip eff}(\omega \rightarrow 0) } = 4.3 \times 10 ^{5}$ m/sec (2.3 eV$\cdot a/ \hbar \pi$) for the $\Gamma$-($\pi, \pi$) FS crossing and ${\vec v_{F}^{0 \medskip eff}(\omega \rightarrow 0) } = 3.4 \times 10 ^{5}$ m/sec (1.8 eV$\cdot a/ \hbar \pi$) for the ($\pi, 0$)-($\pi, \pi$) FS crossing.}.  More recently, it has been claimed that a weak kink around 60 - 70 meV in fact exists along both relevant symmetry directions in NCCO with even a stronger kink found in SCCO \cite{Liu08b,Park08a,Schmitt08a}.  Coupling constants are reported in the range of 0.2-0.8 which is a similar magnitude as in the hole doped compounds \cite{Lanzara01a}.  All these groups make the point that unlike the hole-doped compounds where magnetic modes and phonons exist at similar energies, in the electron-doped cuprates the magnetic resonance mode appears to be found at much lower energies \cite{Wilson06b,Zhao07a,Yu08a}.  A kink has also been found in recent soft x-ray angle resolved photoemission \cite{Tsunekawa08a}.  As phonon anomalies associated with the oxygen half-breathing mode are found in the 60 meV energy range, the mass renormalization is reasonably associated with electron-phonon interaction.  As discussed below (Sec.~\ref{sec:ephonon}), this work gives additional evidence that the electron-phonon interaction is not so different on the two sides of the phase diagram.

Another item of recent interest in the photoemission spectra of cuprates is that of an almost universal `high-energy kink' in the dispersion of the hole-doped cuprates that manifests as an almost vertical drop in the dispersion curve around 300 meV \cite{Ronning05a,Meevasana07a,Graf07a}.  \textcite{Pan06a} found a similar anomaly in PLCCO at energies around 600 meV that they termed a quasiparticle coherence-incoherence crossover.  \textcite{Moritz08a} showed a drop in the dispersion of $x=0.17$ NCCO around 600 meV  that confirms a high energy kink in the electron-doped cuprates found at an energy approximately twice that of the hole-doped compounds.  \textcite{Pan06a} claimed that this result ruled out the super exchange interaction $J$ as the driving interaction as the energy scales of the high energy kink were so different, yet the scale of $J$ so similar between the two sides of the phase diagram.  Through their quantum Monte Carlo calculations within the one band Hubbard model  \textcite{Moritz08a} assign the anomaly to a cross-over when following the dispersion from a quasi-particle-like band at low binding energy near-E$_F$ to an incoherent Hubbard band-like features.  These features are at higher energies in the electron-doped cuprates due to the presence of the charge transfer gap on the occupied side of the spectrum.   

\begin{figure}[t!]
\includegraphics[width=8.5cm,angle=0]{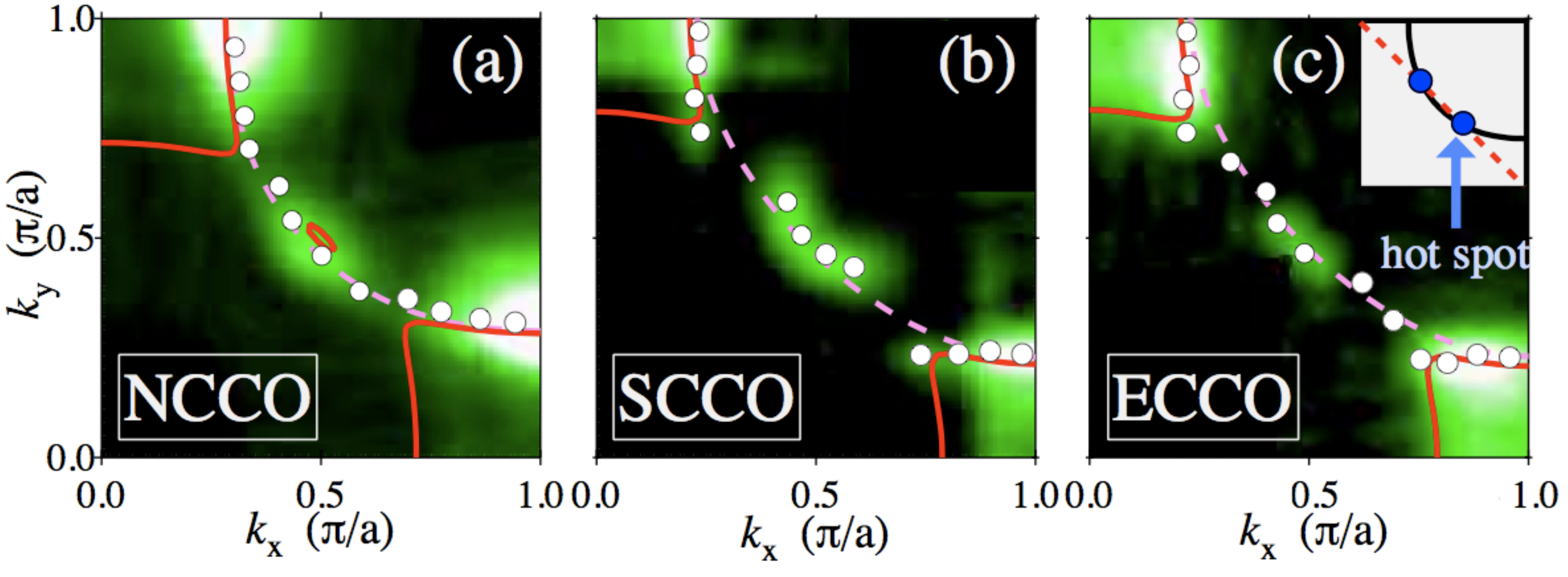}
\caption{(Color) ARPES intensity within $\pm$30 meV of E$_F$ plotted in the BZ quandrant space for nominally $x=0.15$ NCCO, SCCO, and ECCO.  White circles show the peak positions of momentum distribution curves (MDCs) at E$_F$, indicating the underlying Fermi surface. Solid red curves and dashed pink curves show the Fermi surface obtained by tight-binding fit to the ARPES data assuming the AFM and paramagnetic band structures, respectively.  The FS exhibit significantly less curvature in ECCO as compared to NCCO.  Inset: Schematic diagram of the hot spot. Black curve and red dashed line represent the Fermi surface and the antiferromagnetic Brillouin zone boundary, respectively.  From \textcite{Ikeda09a}. } \label{ReARPESSeries}\end{figure}

We should point out that although the general ``hot-spot" phenomena are exhibited in all studied electron-doped cuprates close to optimal doping, the details can be considerably different.  In NCCO \cite{Matsui05a} and PLCCO  \cite{Matsui05b}  there is an actual peak at E$_F$ with greatly reduced spectral weight in the hot spot.  In  contrast in underdoped SCCO  \cite{Park07a} and ECCO \cite{Ikeda07a,Ikeda09a} there is a clear gap at the hot spot and no sign of near-E$_F$ quasi-particle.   These differences may be directly related to changes in chemical pressure caused by different rare earth ion radii and its effect on band structure parameters like the $t'/t$ ratio or indirectly by chemical pressure by causing the extent of antiferromagnetism (and for instance the strength of $V_{\pi,\pi}$) to be different.  Note that these differences may also be due to the differences in the optimal reduction conditions for different compounds, which are known to exist as one goes from PCCO to SCCO and ECCO.  \textcite{Ikeda09a} performed a systematic ARPES study of the Nd, Sm, Eu series of rare earth substitutions, which due to decreasing ion size corresponds to increasing chemical pressure.  In- and out-of-plane lattice constants as well as T$_c$ decreases across this series \cite{Markert90a,Uzumaki91a}.  Ikeda $et$ $al.$ found that the underlying Fermi surface shape changes considerably (Fig.~\ref{ReARPESSeries}) and exhibits significantly less curvature when going from Nd to Eu, which is consistent with a decreasing $|t'/t|$ ratio.  Fitting to a tight bonding band structure with nearest and next-nearest neighbors they found $|-t'/t|$ = 0.40, 0.23, and 0.21 for NCCO, SCCO, and ECCO, respectively\footnote{Note that, fitting not just to the FS positions, but also $v_F$  \textcite{Armitage08a} find slightly different values for $x=0.15$ NCCO.  Using a dispersion relation $E_k= \mu + 2t(\cos{k_x}+\cos{k_y}) +  4t'(\cos{k_x}\cos{k_y})$ he gets parameters $\mu = 0.081\pm 0.02 $, $t  = -0.319 \pm 0.015   $, and  $t' = 0.099 \pm0.01$ (all in eV) when one includes the mass renormalization from the kink and  $\mu = 0.04\pm 0.04 $, $t  = -0.31 \pm 0.01   $, and  $t' = 0.068 \pm0.01$  when one does not.}.   The decreasing ratio was associated with a strong dependence on the in-plane lattice constant.  The hot spot effects also change considerably within this series as seen by the increasing suppression of the near-E$_F$ intensity in Fig.~\ref{ReARPESSeries}.  Ikeda $et$ $al.$  attributed this to an increasing $V_{\pi,\pi}$, which was associated with the decreasing out-of-plane lattice constant and a strengthening of 3D antiferromagnetism.   The $V_{\pi,\pi}$ undoubtedly increases across this series, however at least part of the differences in the hot-spot phenomena may be due to whether or not different $x=0.15$ samples near the AF phase boundary exhibit long range SDW order or just strong fluctuations of it.   One expects that a true gap forms at the hot spot only in the case of true long range order.  These issues are discussed in more detail in Sec.~\ref{sec:SDW} below.

Finally, with regards to the doping dependence,
\textcite{Armitage02a} showed dramatic changes of the ARPES
spectra as the undoped AF parent compound NCO is doped with
electrons away from half filling towards the optimally doped metal
as shown in Fig.~\ref{PeterARPES}.  It was found that the spectral
weight was lost from the charge transfer band (CTB) or lower
Hubbard band feature observed by \textcite{Wells95a,Ronning98a} and
transferred to low energies as expected for a doped Mott insulator
\cite{Meinders93a}. One interesting feature about performing a
photoemission study on an electron-doped material is that - in
principle - the doping evolution of the Mott gap is observable due
to it being below the chemical potential (See Fig.~\ref{Hubbard}).  In hole-doped compounds, such information is only available $via$ inverse photoemission.  At the lowest doping
levels, $x=0.04$, it was observed that the electrons reside in small
`Fermi' patches near the ($\pi$,0) position, at an energy position near the
bottom of the upper Hubbard band (as inferred from optics \cite{Tokura90a}).   This is consistent with many models in which the lowest electron addition states to the insulator are found near  ($\pi$,0)  \cite{Tohyama04a}.  Importantly mid-gap spectral weight also develops.  At higher
dopings the band near ($\pi$,0) becomes deeper and the midgap
spectral weight becomes sharper and moves toward the chemical
potential, eventually contacting the Fermi energy and forming the
large Fermi surface observed in the highest-$T_c$ compounds.  

These observations showed for the very first time, at least
phenomenologically, how the metallic state can develop out of the
Mott insulator. Note that there was some evidence that the CT gap was renormalized to smaller energies upon electron doping as the energy from the CTB onset to the chemical potential (0.8 eV) is smaller than the energy onset of the optical gap in the undoped compound.  However, these data clearly showed that the CT gap does $not$ collapse or close with electron addition \cite{Kusko02a} and instead `fills in.'  A gap that mostly fills in and does not collapse with doping is also consistent with optical experiments \cite{Arima93a,Onose04a}.  Such behavior is reproduced within slave-boson approaches \cite{Yuan05a} as well as numerical calculations within the Hubbard model  \cite{Senechal04a,Aichhorn06a,Macridin06a,Tohyama04a,Kancharla08a} that show most features of the FS development can be reproduced with a doping independent CT gap.

\subsection{Optics}
\label{Optics}

As in the hole-doped cuprates, optical and infrared spectroscopy
has contributed greatly to our knowledge of electronic dynamics in
the $n$-type materials.  The first detailed comparison between
electron- and hole-doped insulating parent compounds was reported
by \textcite{Tokura90a}.  Interestingly, they found an onset in the optical conductivity around 1 eV and a peak around 1.5 eV, which is about 0.5 eV smaller than that found in the analogous $T$ phase La$_2$CuO$_4$.   This optical gap was associated with a charge transfer (CT) gap of 1 - 1.5 eV in the $T'$ structure compound Nd$_2$CuO$_4$.  The smaller charge transfer gap energy was correlated with the lack of oxygen in the apical oxygen-free $T'$
structure compound and its effect on the local Madelung potential.

\begin{figure}[t!]
\includegraphics[width=6cm,angle=0]{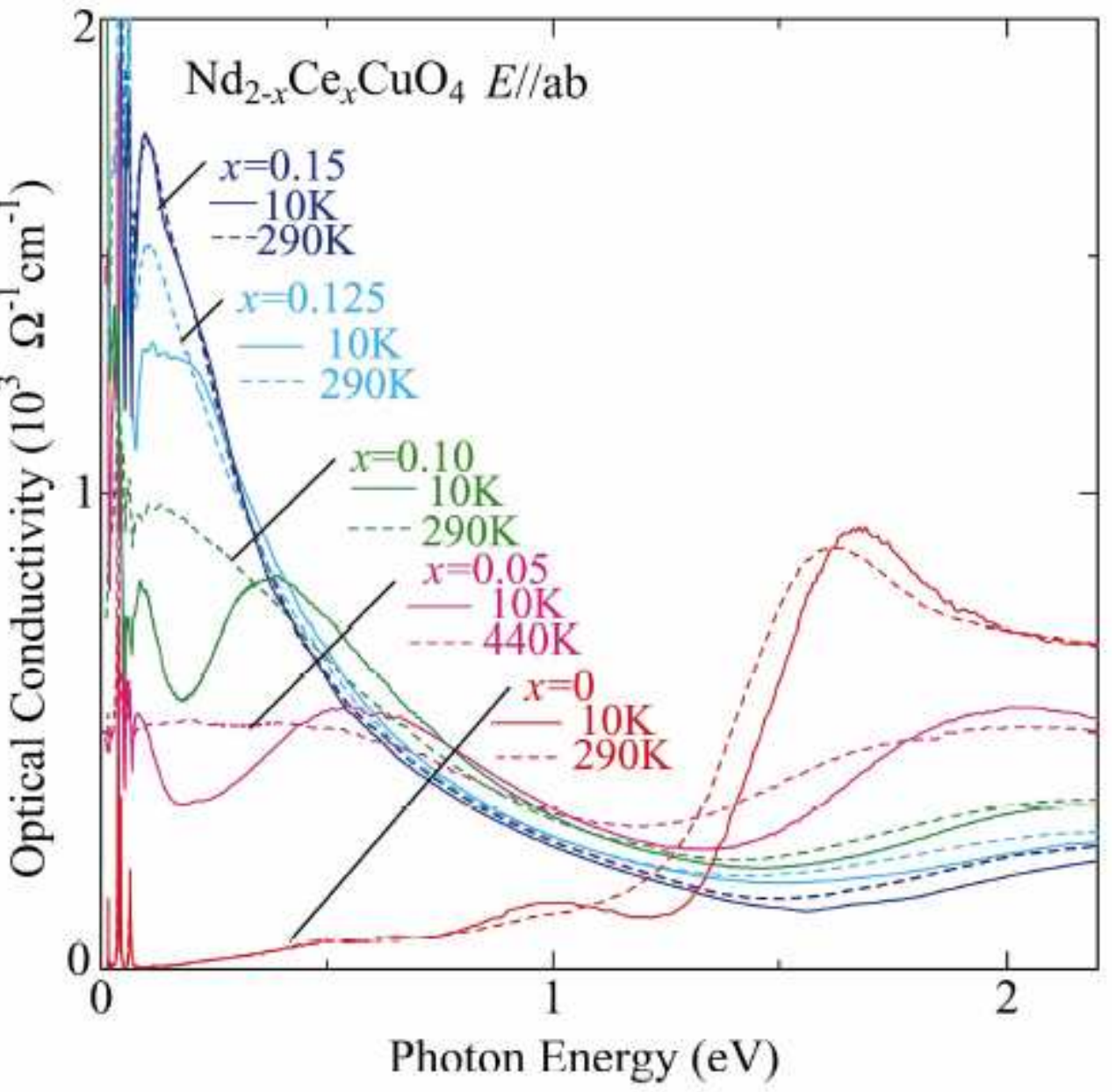}
\caption{(Color) Doping dependence of optical conductivity spectra
for Nd$_{2-x}$Ce$_{x}$CuO$_4$ crystals with x=0 - 0.15 at 10 K and a
sufficient high temperature (440 K) for the x=0.05 crystal and 290
K for the others. From \textcite{Onose04a} }
\label{OnoseOpticsPRB}\end{figure}

In one of the first detailed studies of the optical spectra's
doping dependence \textcite{Arima93a} found in
Pr$_{2-x}$Ce$_{x}$CuO$_4$ mid-gap states that grew
in intensity with doping similar to, but slightly slower than the hole
doped compounds.  They also found a remnant of the CT band at
doping levels almost as high as optimal doping. More
recently the infrared and optical conductivity has been
investigated by a number of groups
\cite{Onose99a,Onose01a,Homes97a,Onose04a,Singley01a,Lupi99a,Zimmers05a,Wang06a}.
It is found generally, that upon rare earth substitution, a
transfer of spectral weight from the CT band to lower frequencies
takes place. A broad peak in the mid-infrared (4000-5000
cm$^{-1}$ or approximately 0.6 eV) is first formed at low doping
levels, with a Drude component emerging at higher dopings.  Fig.
\ref{OnoseOpticsPRB} shows typical behavior.  It bears
a passing resemblance to the hole-doped compounds except that
despite softening with Ce doping the mid-IR band can still be
resolved as a distinct feature in the highest $T_c$ samples
(x=0.15).

Other important differences exist.  For instance, \textcite{Onose01a,Onose04a} found that this notable
`pseudogapped' mid-infrared feature ($\Delta_{pg} = 0.2 - 0.4$ eV)
appeared directly in the optical conductivity spectrum for
metallic but non-superconducting crystals of
Nd$_{2-x}$Ce$_{x}$CuO$_4$ below a characteristic temperature
T$^*$.   They found that $\Delta_{pg} = 10k_B$T$^*$ and that both
decrease with increasing doping. Moreover, the low temperature
$\Delta_{pg}$ was comparable to the magnitude of the pseudogap
measured by \textcite{Armitage02a} $via$ photoemission spectroscopy,
which indicates that the pseudogap appearing in the optical
spectra is the same as that in photoemission.  Such a distinct
pseudogap (PG) in the optical spectrum is not found in underdoped
$p$-type superconductors where instead only an onset in the
frequency dependent scattering-rate $1/ \tau( \omega)$ derived by an extended Drude
model analysis is assigned to a PG
\cite{Puchkov96a}.  \textcite{Singley01a} found that the
frequency dependent scattering rate in the electron-doped compounds is depressed below 650 cm$^{-1}$, which is similar to the behavior which has been ascribed to the pseudogap
state in the hole-doped materials \cite{Puchkov96a}. However,
whereas in the underdoped $p$-type compounds the energy scales
associated with the pseudogap and superconducting states can be
quite similar, these authors showed that in
Nd$_{1.85}$Ce$_{0.15}$CuO$_4$ the two scales differ by more than
an order of magnitude.  In this case, the origin of pseudogap
formation was ascribed to the strong T-dependent evolution of
antiferromagnetic correlations in the electron-doped cuprates.  It has been claimed that it is actually the maximum in the scattering rate and not the visible gap in the optical conductivity that correlates with the ARPES gap best \cite{Wang06a}.

\textcite{Zimmers05a} found that the magnitude of the PG
$\Delta_{pg}$ extrapolates to zero at concentration of x = 0.17 in
Pr$_{2-x}$Ce$_{x}$CuO$_4$ films, implying the coexistence of
magnetism and superconductivity in the highest $T_c$ samples and
the existence of a quantum critical point around this doping.   Moreover, they
performed a detailed analysis of their optical spectra over an
extended doping range and found that a simple spin density wave
model similar to the one dicussed in the context of photoemission
above with ($\pi,\pi$) commensurate order with frequency- and
temperature-dependent self energies could describe many of the
principal features of the data.  Note however, that \textcite{Motoyama07a} gives convincing evidence the AF state terminates around $x=0.134$ doping in NCCO.  In this regard it is possible that the PG observed by \textcite{Zimmers05a} and others corresponds to the buildup of appreciable AF correlations and not the occurrence of long-range anti-ferromagnetic behavior.   For instance,  \textcite{Wang06a}'s optical data clearly show the existence of a large pseudogap in underdoped samples at temperatures well above the N\'eel temperature.

\textcite{Onose99a} found that although the temperature
dependence for reduced superconducting crystals was weak, for unreduced
Nd$_{1.85}$Ce$_{0.15}$CuO$_4$ the large pseudogap structure
evolves around 0.3 eV, but also that activated infrared and Raman
Cu-O phonon modes grew in intensity with decreasing temperature.
This was interpreted as being due to a charge ordering instability
promoted by a small amount of apical oxygen.
\textcite{Singley01a} also found a low energy peak in the in-plane
charge response at $50 - 110$ cm$^{-1}$ of even superconducting Nd$_{1.85}$Ce$_{0.15}$CuO$_4$ crystals, possibly indicative of residual charge localizing tendencies 

\textcite{Singley01a} also showed that in contrast to the $ab$-plane
optical conductivity, the c-axis showed very little difference
between reduced superconducting x=0.15 and as-grown
samples.  This is in contrast to the expectation for hole-doped cuprates where large changes in the c-axis response are observed below the pseudogap temperature (see \textcite{Basov05a} and references therein).  Since the matrix element for interlayer transport is
believed to be largest near the $(\pi,0)$ position and zero along
the zone diagonal, interlayer transport ends up being a sensitive
probe in changes of FS topology.  The polarized c-axis results
indicate that the biggest effects of oxygen reduction should be
found along the zone diagonal.  Using low frequency THz
\textcite{Pimenov00} have shown that the out of plane low
frequency conductivity closely follows the dependence of the
in-plane.  This is, again, likely the result of an interplane
tunnelling matrix elements and the lack of a PG near $(\pi,0)$.
Using these techniques they also found that there was no apparent
anomaly in the quasi-particle scattering rate at $T_c$, unlike in
some hole-doped cuprates \cite{Bonn96a}.

In the superconducting state of Nd$_{1.85}$Ce$_{0.15}$CuO$_4$ \textcite{Singley01a} found that the c-axis spectral weight which collapses into the condensate peak, was drawn from an
anomalously large energy range ($E> 8 \Delta$) similar to that of
the hole-doped cuprates.  In contrast, \textcite{Zimmers04a} claimed
that the $in$-$plane$ Ferrell-Glover-Tinkham spectral weight sum rule
was satisfied in their Pr$_{2-x}$Ce$_{x}$CuO$_4$ thin films at a
conventional energy scale $4\Delta_{max}$ much less than that of
the hole-doped cuprates\cite{Zimmers04a}.  If true, the discrepancy between
out-of- and in-plane sum rule `violation' is unlike the $p$-type
cuprates and is unexplained.  It would be worthwhile to repeat these measurements on the same sample, perhaps with the benefit of higher accuracy far infrared ellipsometry.

Finally, \textcite{Homes06a} has made the observation of
a kink in the frequency dependent reflectivity of
Pr$_{1.85}$Ce$_{0.15}$CuO$_4$  at $T_c$.  This is
interpreted as a signature of the superconducting gap whose
presence in the optical spectra is consistent with their
observation that scattering rate $1/\tau$ is bigger than $2 \Delta
$ and hence that these materials are in the dirty limit.  It was
argued that the ability to see the gap is enhanced as
consequence of its non-monotonic $d$-wave nature (see Sec.~\ref{sec:symmetry} below).  The extracted gap frequency $ \Delta_0 \approx 35$ cm$^{-1}$ (4.3 meV) gives a $2
\Delta /k_B T_c$ ratio of approximately 5, which is in good agreement
with other techniques such as tunnelling \cite{Shan05a}.  \textcite{Schachinger08a} recently reanalyzed the data of  \textcite{Homes06a} as well as \textcite{Zimmers04a,Zimmers05a} to generate a boson-electron coupling function $I^2 \chi(\omega)$.   They find that the optical conductivity can be modeled with a coupling function with peaks at 10 meV and 44 meV.  They identified this lower peak with the magnetic resonance mode found by \textcite{Wilson06b} in PLCCO at 11 meV and draw attention to the correspondence of this energy scale with the 10.5 meV feature in STM \cite{Niestemski07a}.

\subsection{Raman spectroscopy}
\label{RamanNS}

Raman spectroscopy has been extensively used for the investigation of both normal state and superconducting properties of the cuprate superconductors \cite{Devereaux07a}.  It is a sensitive probe of quasiparticle properties, phonon structure, superconducting order parameter symmetry, and charge order.  In the electron-doped compounds, both phonons \cite{Heyen91a} and crystal-field excitations \cite{Jandl93a,Jandl96a} were studied early.

\begin{figure}[t!]
\includegraphics[width=8cm,angle=0]{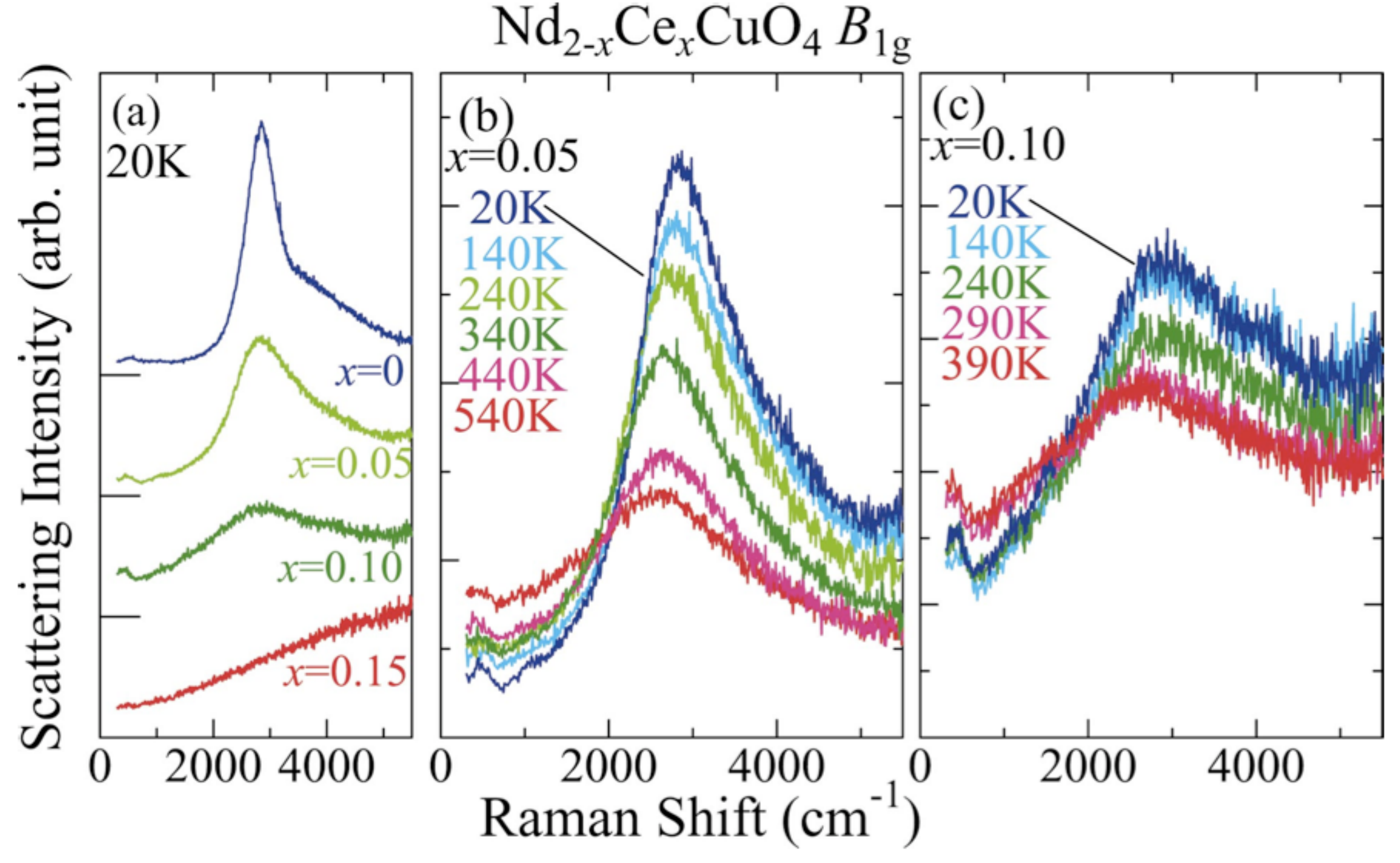}
\caption{(Color)  (a) Doping dependence of the B$_{1g}$ two-magnon peak Raman spectra at 20 K 
for crystals of  Nd$_{2-x}$Ce$_{x}$CuO$_4$.  (b),(c) Temperature variation of the B$_{1g}$ Raman spectra for (b) $x= 0.05$ and (c) $x= 0.10$  crystals.  From \textcite{Onose04a}.}
\label{OnoseRaman}\end{figure}

 \textcite{Onose99a} found that activated infrared and Raman Cu-O phonon modes grew in intensity with decreasing temperature in unreduced crystals.  This was interpreted in terms of a charge ordering instability induced by a minute amount of interstitial apical oxygen.  \textcite{Onose04a} found some of the most definitive evidence that antiferromagnetic correlations manifest themselves in transport anomalies and signatures in the charge spectra (ARPES, optics etc.).  As shown in Fig.~\ref{OnoseRaman}, the B$_{1g}$ two-magnon peak, which is found at 2800 cm$^{-1}$ in the x=0 compound \cite{Sugai89a}, broadens and loses intensity with Ce doping.  The peak energy itself shows little doping dependence.  They found that the peak's integrated intensity shows a sudden onset below T$^*$  - the same temperature where the optical and ARPES pseudogaps develop and there is a crossover in the out-of-plane resistivity.

\textcite{Koitzsch03a} specifically studied the pseudogap state of Nd$_{1.85}$Ce$_{0.15}$CuO$_4$.  They observed the suppression of spectral weight below 850 cm$^{-1}$ for the B$_{2g}$ Raman response and identify it as an anisotropic PG in the vicinity of  ($\pi/2,\pi/2$) points of the BZ.  This was consistent with a model of the pseudogap which originated in enhanced AF interactions in the hot spot region which are closer to the ($\pi/2,\pi/2$) points in these materials than in the hole-doped compounds.  They also observed a narrow Drude-like coherent peak in the B$_{2g}$ channel in the pseudogap phase below T$^*$, which reveals the emergence of long-lived excitations in the vicinity of the ($\pi/2,\pi/2$) points.  Interestingly these excitations do not seem to contribute to the optical conductivity, as it is the B$_{1g}$ response (sensitive to the ($\pi,0$) region) which closely tracks the optical response.

Although the original Raman measurements of the superconducting gap \cite{Stadlober95a} found evidence for an s-wave order parameter, more recent measurements have been interpreted in terms of an $d$-wave order parameter, which is non-monotonic with angle around the FS\cite{Blumberg02a,Qazilbash05a}.  This will be discussed in more detail in Sec.~\ref{sec:Raman-symm}.  In PCCO and NCCO, \textcite{Qazilbash05a} have also determined both an effective upper critical field H$_{c2}^*(T,x)$ at which the superfluid stiffness vanishes and an H$_{c2}^{2 \Delta}(T,x)$ at which the SC gap amplitude is suppressed.  H$_{c2}^{2 \Delta}(T,x)$ is larger than H$_{c2}^*(T,x)$ for all doping concentrations. The difference between the two quantities suggests the presence of phase fluctuations that are larger for $x <  0.15$.  The ability of a magnetic field to suppress the Raman gap linearly at even small fields is unlike the hole-doped compounds \cite{Blumberg97a} or even conventional s-wave NbSe$_2$ \cite{Sooryakumar80a,Sooryakumar81a} and may be related to the non-monotonic $d$-wave gap where points of maximum gap amplitude are close to each other in reciprocal space.  From the doping dependence of H$_{c2}^{2 \Delta}(T,x)$  \textcite{Qazilbash05a} extracted the Ginzburg-Landau coherence length $\xi_{GL} = \sqrt{\Phi_0/ 2 \pi H_{c2}^{2 \Delta}(T,x)}$.  $\xi_{GL} $  is almost an order of magnitude larger than the $p$-doped compounds, giving $k_F \xi_{GL}$ values between 40 and 150 (or E$_F / \Delta \approx  6-24$).  This larger Cooper pair size requires higher order pair interactions to be taken into account and supports the existence of the nonmonotonic $d$-wave functional form.

\subsection{Neutron scattering}
\label{NeutronsScattering}
\subsubsection{Commensurate magnetism and doping dependence}

Neutron scattering has been the central tool for investigating magnetic and
lattice degrees of freedom in the cuprates
\cite{Kastner98a,Bourges99a}. In this section we concentrate on
their contribution towards our understanding of the magnetism of the electron-doped cuprates.
Their important contribution to the understanding of electron-phonon
coupling in the $n$-type compounds (See for instance \textcite{Braden05a}) will be
discussed in Sec.~\ref{sec:ephonon}.

It was found early on \cite{Thurston90a,Matsuda92a} in
doped, but not superconducting materials that the spin
response of the $n$-type systems remained commensurate at
($\pi,\pi$) unlike the hole-doped compounds, which develop a large
incommensurability.  This commensurability is shared by all doped compounds in this material class.  Doping also appears to preserve the unusual non-collinear c-axis spin arrangement \cite{Sumarlin95a}.  Due primarily to the lack of large superconducting single phase
crystals, it wasn't until 1999 that \textcite{Yamada99a} showed the
existence of well-defined commensurate spin fluctuations in a reduced superconducting
sample. The magnetic scattering intensity was peaked at
($\pi,\pi$) as in the as-grown antiferromagnetic materials, but
with a broader $q$-width.  It was suggested by \textcite{Yamada03a}
 that the commensurate dynamic ($>$ 4 meV) short range spin correlations in the SC
phase of the $n$-type cuprate reflect an inhomogeneous distribution
of doped electrons in the form of droplets/bubbles in the CuO$_2$
planes, rather than organizing into one-dimensional stripes as the
doped holes may in many $p$-type cuprates. They estimated the 
low temperature (8 K) in-plane and out-of-plane dynamic magnetic correlation
lengths to be $\xi_{ab}$ = 150{\rm \AA} and $\xi_{c}$ =
80 {\rm \AA} respectively for a {\it T}$_{c}$ = 25 K sample.

It has been emphasized by \textcite{Kruger07a} that within a fermiology approach the commensurate magnetic response of the doped compounds is even more at odds with their experimentally determined FS than a commensurate response would be for hole-doped compounds (which are actually incommensurate).   They demonstrated that with a momentum independent Coulomb repulsion (which derives from the dominate hard core, local repulsion inherited from the microscopic Hubbard U) the magnetic spectrum will be strongly incommensurate\footnote{In contrast, \textcite{Ismer07a} have claimed that the magnetic spectrum can be fit well within a fermiology RPA approach.  However they use a Coulomb repulsion U($\vec{k}$) which is peaked strongly at ($\pi,\pi$), which essentially ensures the excellent fit.  \textcite{Li03a} used a slave-boson mean-field approach to the $t-J$ model and include the antiferromagnetic spin fluctuations $via$ the random-phase approximation.   They claim that one does expect strong commensurate spin fluctuations in NCCO $via$ nesting between FS sections near ($\pi/2$,$\pi/2$) and symmetry related points}.   Indeed based on the nesting wavevectors between ($\pi$,0) regions of the Fermi surfaces \cite{Armitage01b}, one might expect that their magnetic response to be even $more$ $incommensurate$ than the hole-doped.   The commensurability shows the central role that strong coupling and local interactions play in these compounds.

As mentioned in Sec.~\ref{sec:edoping}, one approach to understanding the relatively robust extent of the antiferromagnetic phase in the $n$-type compounds has  been to consider {\it spin-dilution models}.  \textcite{Keimer92a} showed that Zn doping into La$_{2}$CuO$_{4}$ reduces the N\'{e}el temperature at roughly the same rate as Ce doping in Pr$_{2-x}$Ce$_{x}$CuO$_{4\pm \delta}$.   As Zn substitutes as a spinless impurity in $d^{10}$ configuration and serves to dilute the spin system, this implies that Ce does a similar thing.  Although this comparison of Ce with Zn doping is compelling it cannot be exact as the charge carriers added by Ce doping are itinerant and cannot  decrease the spin stiffness as efficiently as localized Zn.  \textcite{Mang04a} found in as grown non-superconducting
Nd$_{2-x}$Ce$_{x}$CuO$_{4\pm \delta}$ that by looking
at the instantaneous correlation length [obtained by integrating
the dynamic structure factor S(q$_{2D}$, $\omega$)] the effects of
itinerancy could apparently be mitigated.  An almost
quantitative agreement was found with quantum Monte Carlo
calculations of the randomly site-diluted
nearest-neighbor spin 1/2 square-lattice Heisenberg
antiferromagnet.\footnote{However other observables showed worse agreement
(for instance the ordered moment), pointing to the strong role
that dynamics play and that fluctuations manifest themselves
differently for different observables.}

\begin{figure}[t!]
\includegraphics[width=0.4cm,angle=0]{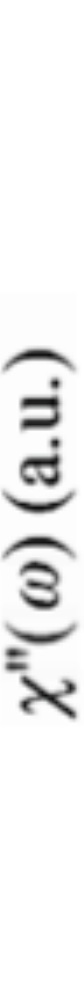}
\includegraphics[width=7cm,angle=0]{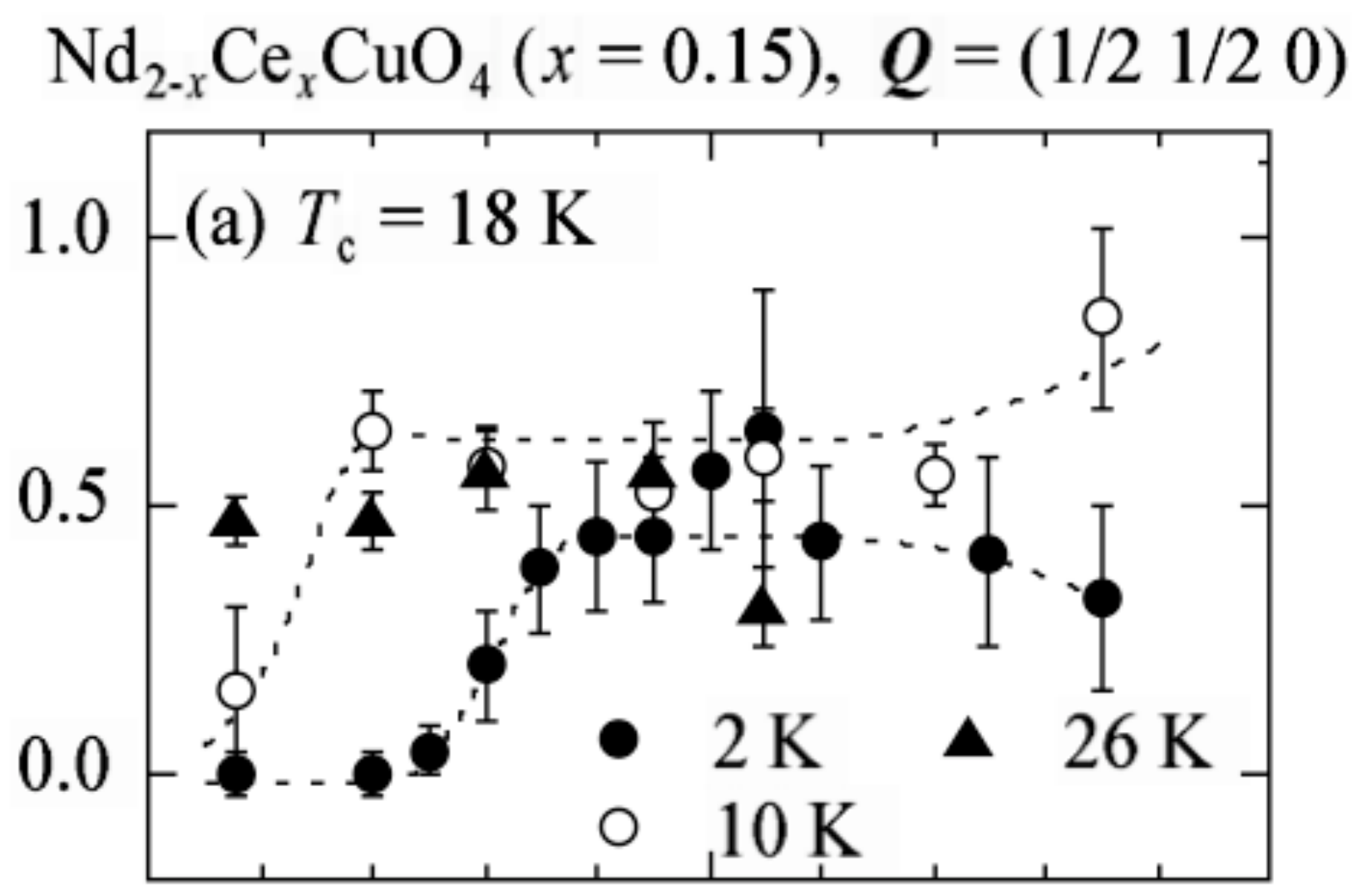}
\includegraphics[width=7.5cm,angle=0]{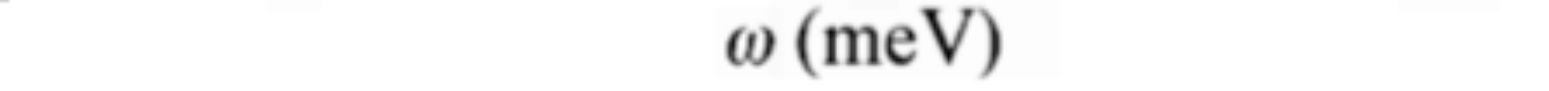}
\caption{Energy spectra of $\chi''(\omega)$ of NCCO obtained from the normal and SC phases (a) x=0.15, T$_c$ = 18K.  From  \cite{Yamada03a}.}
\label{Yamadaneutron}\end{figure}

In NCCO's superconducting state, \textcite{Yamada03a} showed that in addition to the
commensurate elastic response, a gap-like feature opens up in the
inelastic signal  (Fig.~\ref{Yamadaneutron}).  A similar spin gap with a magnitude of $6-7$ meV has also been reported in the $p$-type LSCO system near optimal doping. The maximum gap $2\Delta$ behaves linearly with the SC temperature scale C$k_BT_c$ with $C
\approx 2$ irrespective of carrier type. However, \textcite{Yamada03a} claim that whereas the spin
pseudogap behavior in the SC state of the $p$-type cuprates has a
temperature independent gap energy and slowly ``fills in'' upon
warming, in $x=0.15$ NCCO the gap slowly closes from 4 meV as the
temperature decreases from $T_c$ to 2 K.  `Filling-in' behavior has been associated with phase separation and its absence argues against such phenomena in the $n$-type cuprates.  Interestingly, \textcite{Motoyama06a} found that the superconducting magnetic gap's magnetic field dependence shows an analogous trend as the temperature dependence when comparing hole- and electron-doping.   Magnetic field causes a rigid shift towards lower energies of the $n$-type compound's gap.  Such behavior contrasts with the case of optimally-doped and over-doped LSCO, in which an applied field induces in-gap states and the gap slowly fills in \cite{Lake01a,Tranquada04b,Gilardi04a}\footnote{As discussed below (Sec.~\ref{resonancesection})  \textcite{Yu08a} have disputed the claim of an approximately 4 meV spin gap and claim that the spectra is better understood as an $\approx$ 6.4 meV spin gap and an $\approx$ 4.5 meV resonance.   If true, this would necessitate a reinterpretation of some of the results presented above.}.

\begin{figure}[t!]
\includegraphics[width=7.5cm,angle=0]{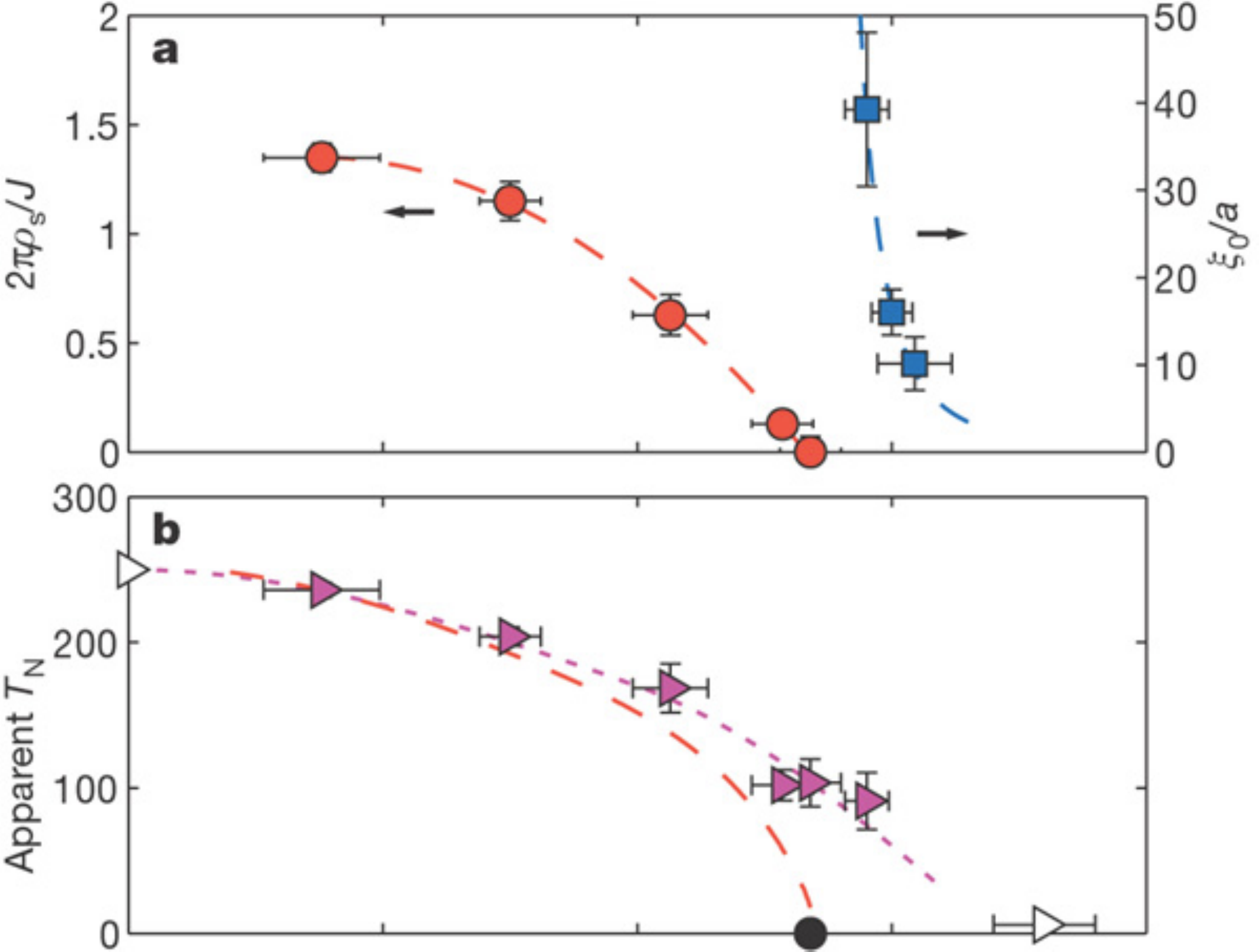} \\ \hspace{1mm}
\includegraphics[width=6.3cm,angle=0]{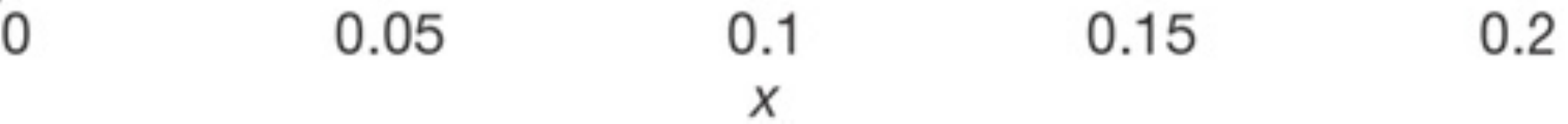}
\caption{(Color)(a)  Doping dependence of the spin stiffness $\rho_s$ normalized to the AF superexchange ($J = 125$ meV for the undoped Mott insulator Nd$_2$CuO$_4$) as $2\pi \rho_s/J$ as well as the low-temperature spin correlation length $\xi_0$.  The spin stiffness decreases smoothly with doping and reaches zero in an approximately linear fashion around $x_{AF} \approx 0.134$. The ground state for $x < x_{AF}$ has long-range AF order as indicated by the diverging $\xi_0$.  (b)  The apparent N\'eel temperature T$_N$, as determined from elastic scattering, as a function of doping given by the dotted curve.  The dashed curve is the extrapolated contour of $\xi/a = 400$.  Adapted from \textcite{Motoyama07a}.}
\label{GrevenStiffness}\end{figure}

With regards to a coexistence of antiferromagnetism and superconductivity, \textcite{Motoyama07a} concluded $via$ inelastic scattering that the spin stiffness $\rho_s$ fell to zero at a doping level of  approximately $x=0.13$ (Fig.~\ref{GrevenStiffness}a) in NCCO which is close to the onset of superconductivity.    They concluded that the actual antiferromagnetic phase boundary terminates at $x \approx 0.134$, and that the magnetic Bragg peaks observed at higher Ce concentrations
originate from rare portions of the sample which were insufficiently oxygen reduced (Fig.~\ref{GrevenStiffness}b).  This issue of the precise extent of antiferromagnetism, the presence of a quantum phase transition, and coexistence regimes will be dealt with in more detail below.

\textcite{Wilson06c} reported inelastic neutron scattering measurements on Pr$_{1.88}$LaCe$_{0.12}$CuO$_{4 - \delta}$ in which they tracked the response from the long-range ordered antiferromagnet into the superconducting sample $via$ oxygen annealing.    This is along the $\delta$ axis in Fig.~\ref{PLCCOpd} (top).   As discussed elsewhere (Sec.~\ref{sec:reduction}), in general oxygen annealing creates an RE$_2$O$_3$ impurity phase in these systems.  An advantage of PLCCO is that its impurity phase has a much weaker magnetic signal due to the small RE magnetic moment.  They find that the spin gap of the antiferromagnet (finite in the insulator due to anisotropy) decreases rapidly with decreasing oxygen concentration, eventually resulting in a gapless low energy spectrum in this material.    Note that superconducting PLCCO compounds do not exhibit a spin gap found in NCCO  below T$_c$ \cite{Yamada03a}\footnote{We note that a spin gap was not observed in hole-doped LSCO crystals until sample quality improved sufficiently \cite{Yamada95a}.  Whether the lack of spin gap in PLCCO is due to the current sample quality of single crystals or is an intrinsic effect is unknown.}.  The linewidths of the excitations broaden dramatically with doping, and thus the spin-stiffness effectively weakens as the system is tuned toward optimal doping. The low energy response of PLCCO is characterized by two regimes.   At higher temperatures and frequencies, the dynamic spin susceptibility $\chi"(\omega,T)$ can be scaled as a function of $\frac{ \omega}{T}$ at AF ordering wavevectors. The low energy cut-off of the scaling regime is connected to the onset of AF order.   The fact that this energy scale comes down as the antiferromagnetic phase is suppressed leads to an association of this behavior with a QCP near optimal doping.

\textcite{Fujita08a} performed an inelastic study on PLCCO over a wide Ce doping range that spanned the antiferromagnetic ($x=0.07$) to superconducting regimes ($x=0.18$).  For all concentrations measured, the low energy spectra were commensurate and centered at ($\pi$, $\pi$).  Although they found a small coexistence regime between superconductivity and antiferromagnetism around $x=0.11$, some characteristics, such as the relaxation rate and spin-stiffness decreases rapidly when one enters the superconducting phase.   The static AF response is absent at  $x > 0.13$.  The spin stiffness appears to extrapolate to zero around $x=0.21$ when superconductivity disappears\footnote{Here the spin stiffness is defined as $\omega / \Delta q $ where $\Delta q$ is the momentum width of a peak at a frequency $\omega$, and is given by the slope of the $\Delta q$ vs. $\omega$ relation.   This is a different definition than that given by \textcite{Motoyama07a}.}.  This indicates a close relation between spin fluctuations and the superconductivity in the electron-doped system.  Interestingly other quantities, like the spectral weight ($\omega$ integration of $\chi''(\omega)$) do not show much doping dependence.  This is unlike the $p$-type systems and was associated by these authors with a lack of phase separation in the $n$-type compounds.

In contrast to the Òhour-glassÓ-type dispersion observed in hole-doped cuprates \cite{arai99a,tranquada04a}, the dispersion at higher energies in optimally doped PLCCO $T_c = 21 - 25.5 K$  looks like a more conventional spin wave response centered around the commensurate position, which disperses outward in a ring-like pattern at higher energy transfers  \cite{Wilson06a,Wilson06c,Fujita06a}.  It can be described in terms of three basic energy regimes \cite{Wilson06b} .  At the lowest energies $\omega < 20 $ meV the system shows the essentially over damped spin wave behavior discussed above with a small nearest neighbor spin coupling J$_1$ of approximately 29 $\pm$ 2.5 meV. At intermediate energies 50 meV $<
\omega <$ 80 meV the excitations are broad and only weakly
dispersing.  At energies above 100 meV, the fluctuations are again
spin wave like with a J$_1$ of 162 $\pm$ 13 meV. This is
substantially larger than the undoped compounds (121 meV for PCO
\cite{Bourges97a} and 104 meV for LCO \cite{Coldea01a}).   A similar situation with a high energy response centered around the commensurate position has also been observed in overdoped PLCCO (T$_c = 16 K$) \cite{Fujita08b}.

\begin{figure}[htb]
\includegraphics[width=8cm,angle=0]{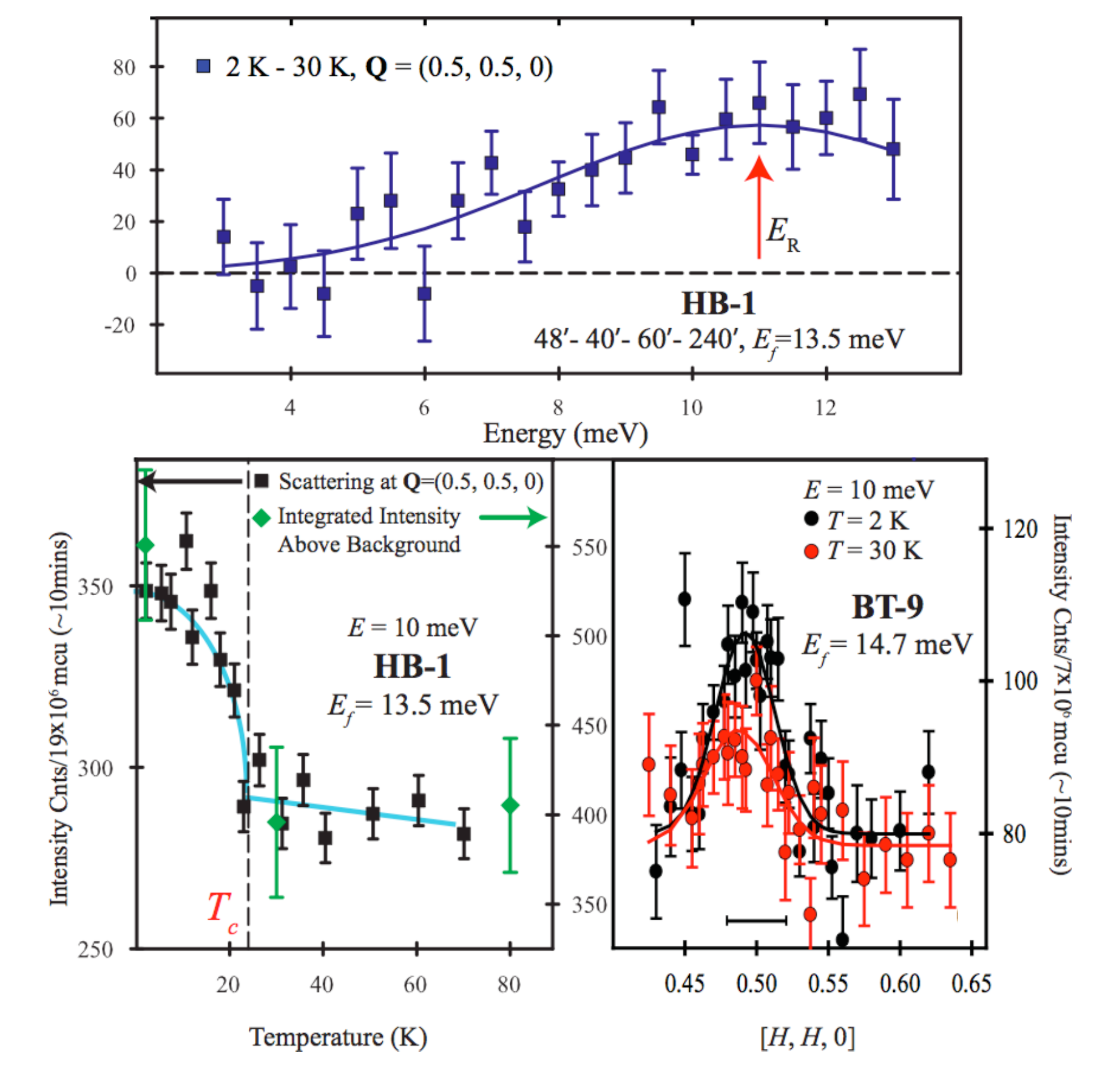}
\caption{(Color) (top) Temperature difference spectrum between 2 K and 30 K suggests a resonance-like enhancement at $\sim$11 meV.   (bottom left)  Temperature dependence of the neutron intensity ($\sim$ 1 hour/point) at (1/2, 1/2, 0) and 10 meV in black squares. Green diamonds are 
integrated intensity of the localized signal centered around Q= (1/2, 1/2, 0) 
above backgrounds.  (bottom right)  Q-scans at $\omega = 
10$ meV above and below the superconducting transitions.  From \textcite{Wilson06b}.} \label{PLCCOresonance}\end{figure}

\subsubsection{The magnetic `resonance'}
\label{resonancesection}

In the superconducting state, \textcite{Wilson06b} found an enhancement of 
peak in the inelastic neutron scattering response of PLCCO (Fig.~\ref{PLCCOresonance}) at approximately 11 meV at (1/2,1/2,0) (equivalent to ($\pi,\pi$)) in the superconducting state.  This was interpreted to be the much heralded `resonance' peak \cite{rossat91a} found in many of the hole-doped cuprates,
perhaps indicating that it is an essential part of superconductivity
in all these compounds.  They find that it has the same $E_r = 5.8 k_B
T_c$ relationship as other cuprates, but that it does not
derive from incommensurate `hour-glass' peaks that merge together as in YBCO
and LSCO \cite{arai99a,tranquada04a}.   Instead it appears to rise
out of the commensurate (1/2,1/2,0) features found in the electron-doped
systems \cite{Yamada99a}.  The inferred resonance energy also scales with the different T$_c$'s for different annealing conditions \cite{Li08a}.   It is important to note that as mentioned above superconducting PLCCO spectra are essentially gapless below T$_c$ \cite{Yamada03a} and in fact, except for the resonance, show very little temperature dependence at all below 30 K.   Supporting evidence for this feature being `the resonance' comes also from \textcite{Niestemski07a} who have - as mentioned above - found signatures of a bosonic mode coupling to charge in their STM spectra at 10.5 meV $\pm$ 2.5 meV (Fig.~\ref{PLCCOSTM}) and \textcite{Schachinger08a} who find a feature in the electron-boson coupling function $I^2\chi(\omega)$ extracted from the optical conductivity at 10 meV.   Additionally \textcite{Wilson07a} have shown that a magnetic field suppresses the superconducting condensation energy and this resonance feature in PLCCO in a remarkably similar way.

\begin{figure}[htb]
\includegraphics[width=7cm,angle=0]{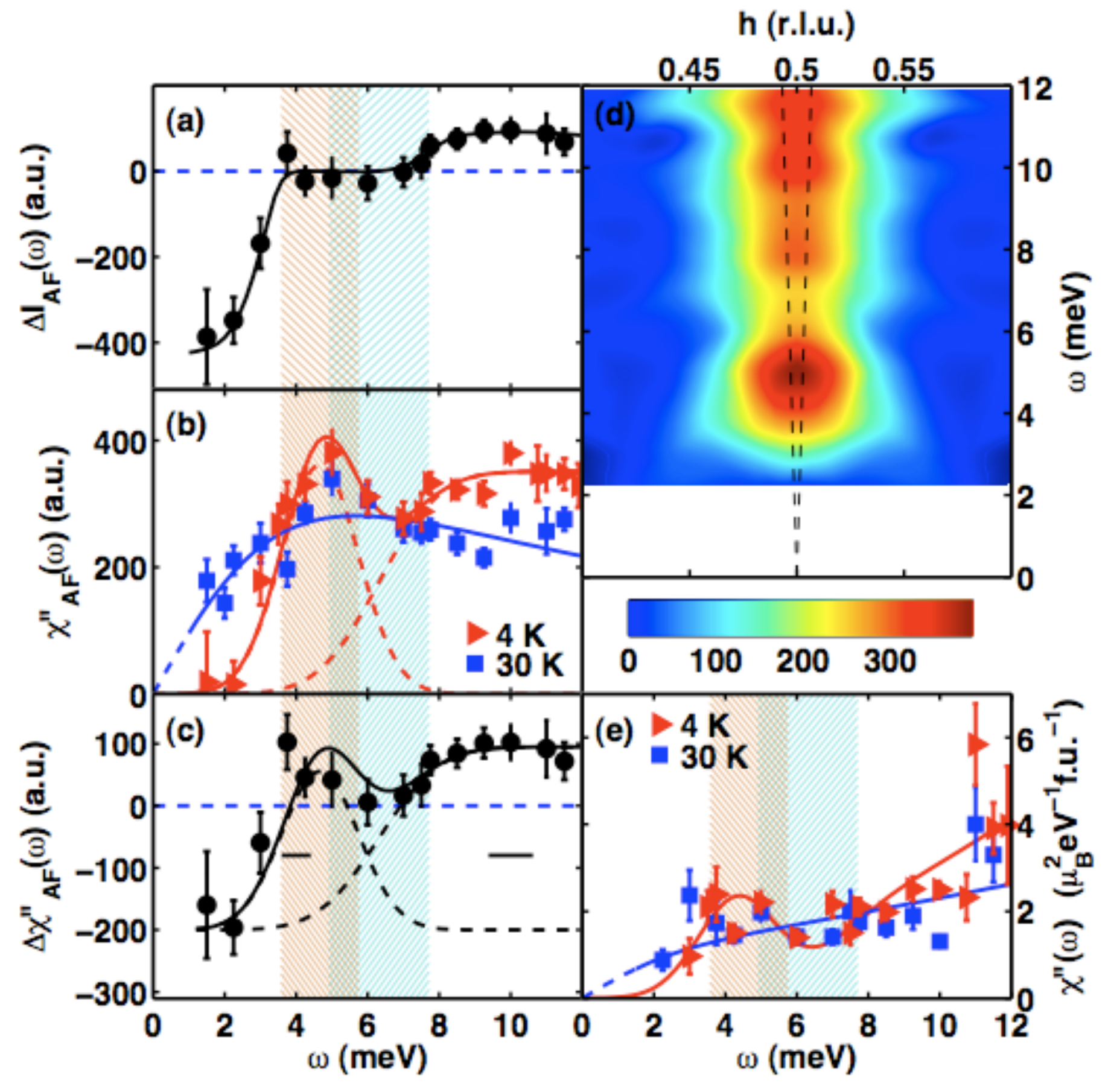}
\caption{(a) Change in scattering intensity between 4 K and 30 K at  the antiferromagnetic wavevector (1/2,1/2,0).   (b) Dynamic susceptibility $\chi''(Q,\omega)$ which shows two peaks after correcting the measured intensity for the thermal factor. (c) Relative change from 30 K to 4 K in susceptibility at the AF wavevector.  (d) Contour plot of $\chi''(Q,\omega)$ at 4K, made by interpolation of symmetrized momentum scans through the AF zone center with a constant background removed.  (e) Local susceptibility in absolute units from the momentum-integral of the dynamic susceptibility by comparing with the measured intensity of acoustic phonons.  The shaded vertical bands in (a) - (c) indicates the range of values of $2\Delta_\mathrm{el}$ from Raman scattering \cite{Qazilbash05a} corresponding to the author's estimation of the distribution of gap sizes from chemical inhomogeneity.   From \textcite{Yu08a}.} \label{GrevenNCCOplot}\end{figure}

In continuing work \textcite{Zhao07a} have claimed that optimally doped NCCO has a resonance at 9.5 meV, which also obeys the $E_r = 5.8 k_B T_c$ relation.  However, their assignment of this intensity enhancement has been disputed by \textcite{Yu08a}, who claim that their full $\omega$ scans show the spectra are better described by an inhomogeneity broadened spin gap at $\approx$ 6.4 meV and a sub gap resonance at the much smaller energy of  $\approx$ 4.5 meV as shown in Fig.~\ref{GrevenNCCOplot}. This scenario has a number of appealing features.    Both energy positions show sudden increases in intensity below T$_c$.  Moreover, the spin gap they assign is to within error bars equal to the full electronic gap maximum $2 \Delta$ (as measured by techniques like Raman scattering \cite{Qazilbash05a}), suggesting that - $unlike$ the hole-doped cuprates - the commensurate response allows electronic features to be directly imaged in the magnetic scattering as  ($\pi,\pi$) bridges these portions of the FS.  $Like$ the hole-doped cuprates the resonance they find is at energies less than the full superconducting gap, which is a reasonable condition for the stability of spin-exciton-like excitations.  This interpretation is at odds with the original observation of the spin gap in NCCO by \textcite{Yamada03a} and as the authors point out necessitates a reinterpretation of that data as well as some of their own previous work.  These authors  are careful to state that their result does not necessarily invalidate the claim of a resonance peak at the larger energy of 11 meV in PLCCO as the superconducting gap may be much larger in PLCCO  \cite{Niestemski07a} and may allow a stable coherent excitation at this energy.

\subsubsection{Magnetic field dependence}
\label{MagFieldDep}

The dependence of the ordered spin structure  on magnetic field of
superconducting samples and the possibility of field
induced antiferromagnetism has become of intense interest.  These
studies parallel those on underdoped LSCO, where neutron
scattering has shown that a c-axis-aligned magnetic field not only
can suppress superconductivity but also creates a static
incommensurate spin density wave order, thus implying that such an
order directly competes with the superconducting state
\cite{Katano00a,Lake01a,Lake02a,Khaykovich02a}.  The effect of field on $n$-type 
superconducting and reduced samples is a matter of some controversy.  While
experiments by \textcite{Matsuda02a} found that a 10-T
c-axis-aligned field has no effect on the AF signal in their
superconducting NCCO x=0.14 samples, \textcite{Kang03a}
demonstrated in similar x=0.15 samples antiferromagnetic related Bragg reflections such as (1/2,1/2,0) grew in intensity until a field close to the critical field $B_{c2}$ and then decreased. The experiments were interpreted as demonstrating that a quantum phase transition from the superconducting state to an
antiferromagnetic state is induced at $B_{c2}$.

\begin{figure}[t!]
\includegraphics[width=6.5cm,angle=0]{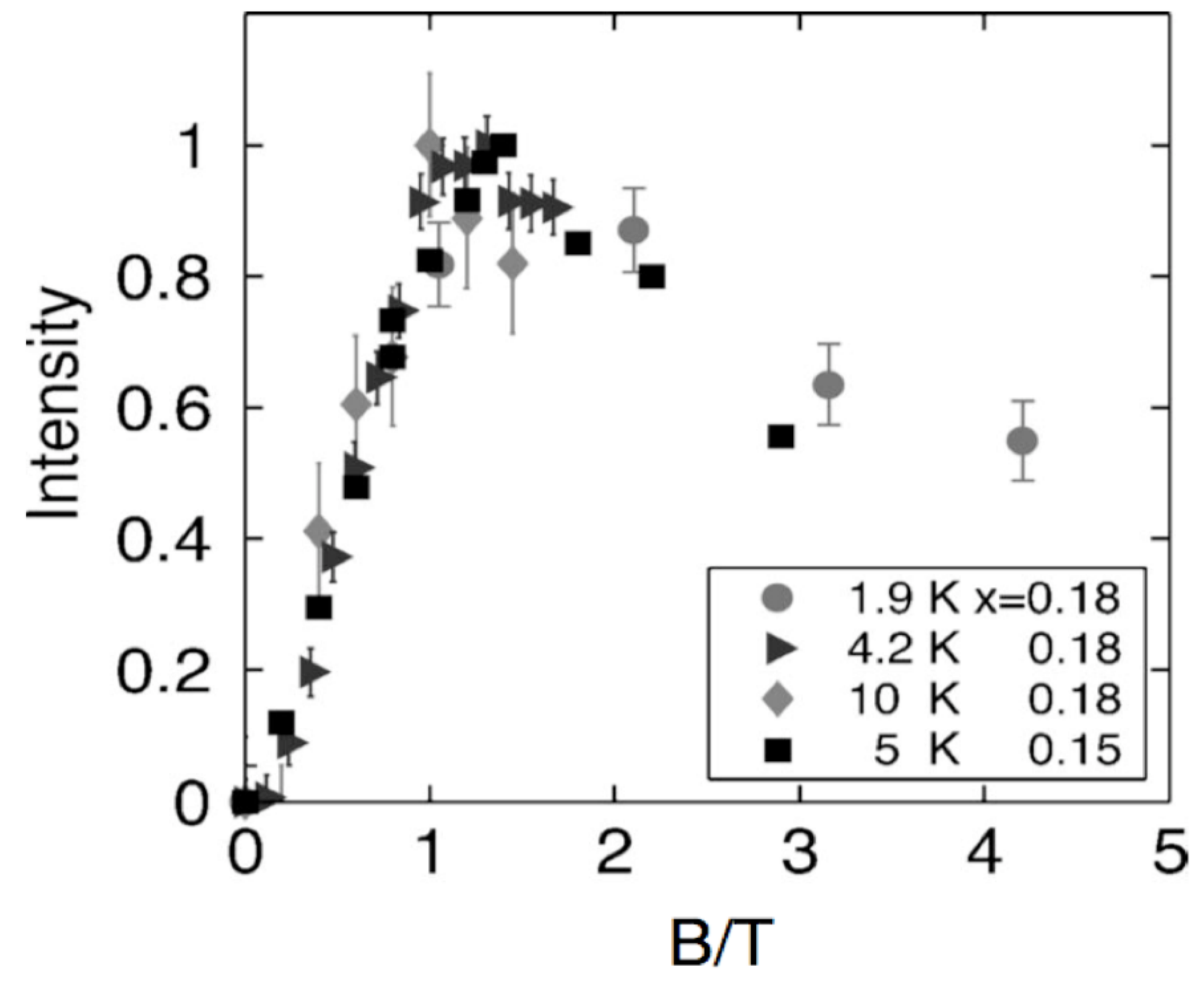}
\caption{Scaled scattering intensity at (1/2,1/2,0) for a
superconducting sample of NCCO ($x=0.18$; $T_c$ = 20 K), plotted
as a function of B/T.  Field direction is [0,0,1].  Data is compared with the results at $T = 5$ K of
\textcite{Kang03a} ($x=0.15$; $T_c$=25 K).  Adapted from \textcite{Mang04b}.} \label{MangComment}\end{figure}

Although their raw data are similar to \textcite{Kang03a}, this interpretation was
disputed by \textcite{mang03a} who found that additional
magnetic intensity comes from a secondary phase of (Nd,Ce)$_2$O$_3$.  As noted above, a severe oxygen reduction procedure always has to be applied to as-grown crystals to induce
superconductivity.  \textcite{mang03a} discovered
that the reduction process decomposes a small amount
of NCCO (0.1 - 1.0 $\%$ by volume fraction).  The resultant
(Nd,Ce)$_2$O$_3$ secondary phase has a complex cubic bixbyite
structure, with a lattice constant approximately $2\sqrt{2}$ times
the planar lattice constant of tetragonal NCCO. The
(Nd,Ce)$_2$O$_3$ impurity phase grows in epitaxial register with
the host lattice in sheets on average five unit cells thick.  Because of the simple $2\sqrt{2}$ relationship
between the lattice constants of NCCO and (Nd,Ce)$_2$O$_3$ the
structural reflections of the impurity phase - for instance the
cubic $(2,0,0)_c$ - can be observed at the commensurate NCCO
positions (1/2,1/2,0).  However the c axis is different and
there is approximately a 10$\%$ mismatch between
(Nd,Ce)$_2$O$_3$'s lattice constant and $a_c$ of NCCO, and
therefore the impurity phase's $(0,0,2)_c$ can be indexed as
(0,0,2.2).  Moreover,  \textcite{Mang04b}
found that the field effects reported by \textcite{Kang03a} are
observable in non-superconducting, but still oxygen-reduced,
x=0.10 samples, both at the previously reported lattice positions
and at positions unrelated to NCCO but equivalent in the cubic
lattice of (Nd,Ce)$_2$O$_3$.   \textcite{Mang04b}
interpreted the non-monotonic field dependence of the scattering
amplitude as a consequence of the two inequivalent crystalline
sites of the Nd atoms in Nd$_2$O$_3$ and in accordance with such a model, showed
that the intensity scales as a function of B/T as shown in Fig.
\ref{MangComment}.

Dai and coworkers subsequently confirmed the presence
of a cubic impurity phase, but feel additional results support
their original scenario. \textcite{Kang03b,Matsuura03a} pointed
out, that while one would expect the field induced intensity of the
impurity phase to be the same along all axis directions due to its
cubic symmetry, the effect at (1/2,1/2,0) is much larger when B is
parallel to the c-axis.  This is consistent with the much smaller
upper critical field along the c-axis. Moreover the (1/2,1/2,3)
peak has a $z$ index which cannot be contaminated by the impurity
phase and yet shows an induced antiferromagnetic component when
the field is along the c-axis and hence superconductivity is
strongly suppressed, but not when in-plane and superconductivity
is only weakly affected. \cite{Matsuura03a}.

It is difficult to draw generalized conclusions about the field dependence of the neutron scattering response in the electron-doped cuprates as important differences
exist between the measurements of NCCO and PLCCO.  Optimally doped PLCCO \cite{Fujita04a,Kang05a} has no residual AF order (like LSCO \cite{Kastner98a}) while a 3D AF order has been inferred to coexist with superconductivity in NCCO even for optimally doped samples\footnote{As noted above and discussed in more detail below, \textcite{Motoyama07a} have concluded that true long-range order in NCCO terminates at $x=0.13$ and that the Bragg peaks seen near optimal doping are due to insufficiently reduced portions of the sample.}.  Additionally, the magnetic field of the maximum induced intensity in PLCCO is independent of temperature, in contrast to the peak position scaled by H/T in NCCO and in opposition to the impurity model proposed by \textcite{mang03a}.  A c-axis magnetic field enhances not only the scattering signal in optimally doped NCCO at (1/2,1/2,0) but also at the 3D AF Bragg positions such as (1/2,3/2,0) and (1/2,1/2,3), whereas
for underdoped PLCCO, there is no observable effect on (1/2,3/2,0)
(and related) 3D peaks up to 14 T \cite{Kang05a}.   At this point the effect of magnetic field on the SC state of the electron-doped compounds still has to be regarded as an open question.

\subsection{Local magnetic probes: $\mu$SR and NMR}
\label{muSR.NMR}

Nuclear Magnetic Resonance (NMR) \cite{Asayama96a} and muon spin resonance and rotation ($\mu$SR)  \cite{Sonier00a,Luke90a} measurements are sensitive probes of $local$ magnetic structure and have been used widely in the cuprate superconductors.  Using $\mu$SR  on
polycrystalline samples, \textcite{Luke90a} first showed that the Mott insulating parent compound Nd$_{2}$CuO$_4$ has a N\'eel temperature ($T_N$) of approximately 250K, which decreases gradually upon substitution of Nd by Ce to reach a zero value close to optimal doping ($x \sim 0.15$).  \textcite{Fujita03a}  performed a comprehensive $\mu$SR study, which established the phase diagram of PLCCO.  They found bulk superconductivity from $x=0.09$ to 0.2 and only a weak dependence of T$_c$ on $x$ for much of that range.  The antiferromagnetic state was found to terminate right at the edge of the superconducting region, which was interpreted as a competitive relationship between the two phases.  Only a very narrow coexistence regime was observed ($\approx$ 0.01 wide).  Although changes in the form of the muon relaxation were observed below a temperature T$_{N1}$ where elastic neutron Bragg peaks have been observed \cite{Fujita03a}, there was no evidence for a static internal field until a lower temperature T$_{N2}$.  At the lowest temperatures, it was found that the magnitude of the internal field decreased upon electron doping, showing a continuous and apparently spatially uniform degradation of magnetism.  This is in contrast to the hole-doped system where in the N\'{e}el state ($x < 0.02$) the internal field was constant \cite{Harshman88a,Borsa95a}, which has been taken as evidence for phase separation \cite{Chou93a,Matsuda02a}.  Extensive NMR measurements have also been done by \textcite{Williams05a,Zamborszky04a,Bakharev04a} that have given important information about inhomogeneity in these systems.   These measurements are discussed in more detail in Sec.~\ref{Inhomo}.

\textcite{Zheng03a} showed that when the superconducting state was suppressed in $x=0.11$ PLCCO  with a large out-of-plane magnetic field the NMR spin relaxation rate obeyed the Fermi-liquid Korringa law $1/T_1 \propto T$ over 2 decades in temperature.  We discuss this result in more detail below (Sec.~\ref{nonFL}). \textcite{Zheng03a} also found no sign of a spin pseudogap opening up at temperatures much larger than T$_c$, which is a hallmark of NMR in the underdoped $p$-type cuprates.  Here they found that above the superconducting T$_c$  1/$T_1 T$ showed only a weak increase, consistent with   antiferromagnetic correlation.

Related to the neutron scattering studies in field detailed above, under a weak perpendicular field \textcite{Sonier03a} observed $via$ $\mu$SR the onset of a substantial magnetic order signal (Knight shift) which was static on the $\mu$SR time scales in the superconducting state of optimally doped PCCO single crystals.  The data was consistent with moments as large as 0.4$\mu$ being induced by fields as small as 90 Oe.  There was evidence that the antiferromagnetism was not confined to the vortex cores, since nearly all the muons saw an increase in the internal field and the vortex density was so low and so again the magnetism looked uniform.  It has been argued however that this study overestimated the induced Cu moments by not explicitly taking into account the superexchange coupling between Pr and Cu ions as well as an unconventional hyperfine interaction between the Pr ions and the muons \cite{Kadono04a,Kadono05a}.  \textcite{Kadono04a,Kadono05a} have interpreted their measurements as then consistent with only a weak field induced Cu magnetism in $x=0.11$ PLCCO (near the AF boundary of $x \approx$0.10) which becomes even smaller at $x=0.15$.

Overall $\mu$SR results in the field applied state of the electron-doped cuprates appear to show substantial differences from the $p$-type compounds.  At the onset of superconductivity, there is a well-defined Knight shift whereas in the hole-doped materials, superconductivity under applied field only evinces from an enhancement in the spin relaxation rate \cite{Kakuyanagi02a,Savici05a,Mitrovic01a} or changes in the field profile of the vortex cores \cite{Kadono04b,Miller02a}.  This again indicates that the induced polarization of Cu ions in the electron-doped compounds appears to be relatively uniform over the sample volume, whereas it appears to be more localized to the vortex cores in the hole-doped materials.

\section{Discussion}


\subsection{Symmetry of the superconducting order parameter} \label{sec:symmetry}
\par

There is a consensus picture emerging for the order parameter
symmetry for the $n$-type cuprates.  The original generation of measurements on 
polycrystals, single crystals and thin films seemed to favor s-wave symmetry, but experiments on improved samples including tricrystal measurements\cite{Tsuei00a}, penetration depth\cite{Kokales00a,Prozorov00b,Cote08a}, ARPES\cite{Armitage01a,Sato01a,Matsui05a}, and others favor a $d_{x^2 - y^2}$  symmetry over most of the phase diagram, albeit with an interesting non-monotonic functional form.  Below, we give an overview of the main results of their order parameter and discuss similarities and differences with respect to the hole-doped cuprates.  This is a subject that deserves a comprehensive review that sorts through the multitude of experiments.  We give only a comparatively brief overview here.

\subsubsection{Penetration depth} \label{sec:pendepth}
\par
In the mid 90's, penetration depth $\lambda$ measurements on high quality
YBa$_2$Cu$_3$O$_7$ crystals gave some of the first clear
signatures for an anomalous order parameter in the cuprates.  The
linear temperature dependence of $\Delta \lambda (T)$ (related to
the superfluid density) was a clear demonstration that the density
of states of this material was linear for sub-gap energies ($E <
20$ meV), in agreement with the behavior expected from a $d$-wave
symmetry of the order parameter with nodes \cite{Hardy93a}.

\par
Early $\Delta \lambda (T)$ data obtained on single crystals and thin films of
optimally doped NCCO showed no such temperature
dependence\cite{Wu93a,Andreone94a,Anlage94b,Schneider94a}, not
even the expected dirty $d$-wave behavior characterized by a $\Delta
\lambda \propto T^2$ dependence at low
temperature\cite{Hirschfeld93a} seen for example in thin films of
YBa$_2$Cu$_3$O$_7$\cite{Ma93a}. The NCCO data was best fit to a BCS
s-wave-like temperature dependence down to $T / T_c \sim 0.1$ with
unusually small values of $2 \Delta_o / k_BT_c \sim 1.5 - 2.5$. Later, Cooper
proposed that the temperature dependence of the
superfluid density measured with these techniques had
been masked by the strong Nd magnetic response (see
Section~\ref{sec:magnetism}) at low T \cite{Cooper96a}. Using the data
of \textcite{Wu93a} and correcting for
the contribution of the low temperature magnetic permeability $\mu_{DC}(T)$ in 
NCCO \cite{Dalichaouch93a}, he reached the conclusion
that the real temperature dependence of $\Delta \lambda (T)$ could
be close to $T^2$ at low temperature.
%
%
%
%
\par
\begin{figure}[htbp]
\begin{center}
\includegraphics[width=7cm,angle=0]{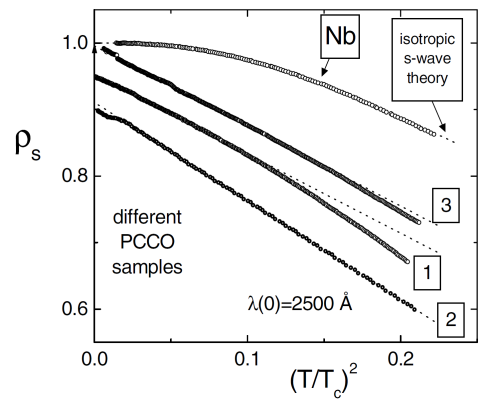}
\caption{Tunnel-diode driven
LC resonator data for three different PCCO single crystals showing power law
behavior of the superfluid density. From \textcite{Prozorov00b}.} 
\label{delta.lambda.d}
\end{center}
\end{figure}

To circumvent the inherent magnetism of Nd ions in NCCO, 
slightly different experimental probes were used by \textcite{Kokales00a} and
\textcite{Prozorov00b} to evaluate $\Delta \lambda (T)$ and the superfluid density (Fig.~\ref{delta.lambda.d}) in 
Pr$_{1.85}$Ce$_{0.15}$CuO$_4$ single crystals which has much
weaker RE magnetism. Both
experiments showed for the first time that $\Delta \lambda (T)$
follows a $\sim T^2$ behavior at low temperatures in PCCO, in agreement
with the dirty $d$-wave scenario.  Moreover, by extending the
temperature range of the measurements for NCCO, they showed the presence of
an upturn in the magnetic response due to Nd residual magnetism,
confirming Cooper's interpretation.

%
%
%
%

\par
More recent reports targeting the doping dependence
of the superfluid density on certain specifically prepared thin films give a still controversial
picture however.  \textcite{Skinta02a} observed that the temperature
dependence of $\Delta \lambda (T)$ evolves with increasing cerium
doping. Using PCCO and LCCO thin films grown by Molecular
Beam Epitaxy (MBE) \cite{Naito02a}, the low temperature data present the gradual development of a
gapped-like behavior for increasing doping\cite{Skinta02a} observed as 
a flattening of $\Delta \lambda (T)$ at low temperature. The growth of 
this T-independent s-wave-like behavior was interpreted
as a possible signature of a transition from a pure $d$-wave symmetry on the underdoped 
regime to a $d-$ and $s-$wave admixture on the overdoped regime. A similar trend was also 
deduced by \textcite{Pronin03a} from a quasioptical transmission measurement of $\Delta \lambda (T)$ 
at millimeter wavelengths (far-infrared). Another report \cite{Kim03a} on MBE-grown buffered PCCO
thin films from under- to overdoping range claimed that $\lambda (T)$ can only 
be explained with a fully gapped density of states with a $d+is$-wave 
admixture for all doping.  In contrast, \textcite{Snezhko04a}
showed that the $T^2$ behavior of thin films grown by pulsed-laser
ablation deposition (PLD) is preserved even in the overdoped regime. These very conflicting
results have yet to be explained, but the answers may lie partly
in the different growth techniques, the quality of films, the presence of parasitic 
phases (Section~\ref{sec.materials}) and the differences in the
experimental probes.  It has been proposed that the presence of 
electron and hole Fermi surface pockets, as observed by ARPES (Section~\ref{ARPES}) 
and confirmed by electrical transport (Section~\ref{TransportResHall}), 
could result in an s-wave-like contribution despite that 
the dominant pairing channel has a $d_{x^2-y^2}$ symmetry\cite{Luo05a}. The 
variability between different kinds of samples may reflect the influence of
different oxygen content on the presence and the contribution of these 
pockets (arcs) as shown by ARPES\cite{Richard07a}.

\par
As a possible demonstration of other material-related issues, \textcite{Cote08a} recently compared the 
penetration depth measurements by the microwave perturbation technique of optimally doped PCCO thin films grown by PLD with very similar $T_c$'s but with different quality as characterized by their different normal-state resistivity close to $T_c$.   They found that lower quality films show a flat 
$\lambda_1 (T)$ at low temperature, showing that oxygen reduction and the presence of defects may be of crucial importance in determining the actual symmetry using penetration depth measurements.

\par
Another avenue for the estimation of the temperature dependence of the penetration depth 
relies on the properties of grain boundary junctions (GBJ) made on SrTiO$_3$ bicrystal 
substrates\cite{Hilgenkamp02a}. Using the maximum critical current density $J_c$ of small 
Josephson junctions, \textcite{Alff99b} estimated $\Delta \lambda_{ab} / \lambda_{ab}$ 
as a function of temperature for both NCCO and PCCO GBJ's using thin films made by MBE. This 
scheme assumes that $J_c \propto n_s$, thus $\lambda_{ab}\propto 1/ \sqrt{n_s} \propto 1/\sqrt{J_c}$.
The striking aspect of this data is the upturn of the estimated effective 
$\lambda_{ab}$ for NCCO due to Nd magnetism. Using the same correction scheme as that proposed 
by \textcite{Cooper96a}, the NCCO data could be superimposed on top of the PCCO GBJ data\cite{Alff99b}. However, it was concluded that the penetration depth followed an s-wave-like exponential temperature dependence with $2 \Delta_o / k_B T_c \sim 3$, in agreement with the initial penetration depth measurements and indicating a nodeless gap. This result together with the unresolved doping dependence controversy mentioned above may arise from the different sample preparations leading to many superimposed extrinsic contributions. 

\subsubsection{Tunnelling spectroscopy} \label{sec:tunnel}
\par
There are two main signatures in tunnelling spectroscopy that can reveal the presence of 
$d$-wave symmetry. The first is related to their `V-shaped' density of states. Unlike the conductance characteristic observed for tunnelling between a metal and a conventional s-wave superconductor at T = 0, which shows zero conductance until a threshold voltage $V = \Delta_o /e$ is reached\cite{Tinkham96a}, tunnelling into $d$-wave superconductors reveals substantial conductance at subgap energies even at T$\rightarrow$0. The second signature, a zero-bias conductance 
peak (ZBCP), reveals the presence of an Andreev quasiparticle bound state (ABS) 
at the interface of a $d$-wave superconductor arising from the phase change of the order 
parameter as a function of angle in $\vec{k}$-space\cite{Hu94a,Kashiwaya95a,Lofwander01a,Deutscher05a}.  This bound state occurs for all interface orientations with projection on the (110) direction. 
The ZBCP can also split under an increasing 
magnetic field\cite{Beck04a,Deutscher05a} and, in some instances, it
is reported to even show splitting at zero magnetic field in the holed doped cuprates.\cite{Covington97a,Fogelstrom97a,Deutscher05a}. 
\par

As discussed above in Sec.~\ref{TransportTunneling}, tunnelling experiments on $n-$doped cuprates have been particularly difficult, which is presumably related to difficulties in preparing high quality tunnel junctions.  Typical quasi-particle
conductance G(V)=dI/dV spectra on optimal-doped NCCO \cite{Shan05a} are shown in
Figure~\ref{ShanPRB}.  Similar spectra are found for Pb/I/PCCO (where I is a natural
barrier) \cite{Dagan05a}, and GB junctions
\cite{Alff98a,Chesca05a}. The main features of the $n-$doped tunnel
spectra are: prominent coherence peaks which reveal an energy gap
of order 4 meV at 1.8K for optimal doping, an asymmetric linear background G(V) for
voltage well above the energy gap, a characteristic `V' shape,
coherence peaks which disappear completely by $T \approx T_c$ at
H=0 (and by $H \approx H_{c2}$ for T=1.8K), and typically the absence of a
zero bias conductance peak (ZBCP) at V=0.

\par
Tunneling has given conflicting views of the pairing
symmetry in $n-$doped cuprates. The characteristic '$V$' shape of
G(V) cannot be fit by an isotropic s-wave BCS behavior and closely resembles
that of $d$-wave hole-doped cuprates \cite{Fischer07a}.  On
the other hand the ZBCP has been observed only sporadically \cite{Biswas02a,Qazilbash03a,Chesca05a,Wagenknecht08a}. 
Its absence in most spectra of tunnel junctions with large barriers may be the consequence of the coherence length ($\sim 50 {\rm \AA}$) being comparable to the mean free path \cite{Biswas02a} similar to the effect observed in YBCO \cite{Aprili98a}.  Its absence has also been attributed to the coexistence of AFM and SC orders \cite{Liu07a}.

\par
Point contact spectroscopy data have shown a ZBCP in underdoped (x = 0.13) PCCO films,
while it is absent for optimal and overdoped compositions
\cite{Biswas02a,Qazilbash03a}. Combined with an analysis of the G(V) data based 
on Blonder-Tinkham-Klapwijk theory \cite{Blonder82a,Tanaka95a}, this
result has been interpreted as a signature of a $d$- to $s$-wave symmetry transition
with increasing doping. However, there has been a more recent claim that all such tunneling spectra are better fit with a non-monotonic $d$-wave functional form \cite{Dagan07b}  over the entire 
doping range of superconductivity.  This may explain in part the many reports claiming that 
the tunnelling spectra from several experimental configurations can not be fit with either 
pure $d$-wave or $s$-wave gaps [see for example \textcite{Kashiwaya98a,Alff98b,Shan05a}]. The SIS planar tunnelling work of \textcite{Dagan07b} and a detailed point contact tunnelling study as a 
function of doping of \textcite{Shan08a} also provide strong evidence that the $n-$doped 
cuprates are weak coupling, $d$-wave BCS superconductors over 
the whole phase diagram.  This is in agreement with other techniques including Raman scattering \cite{Qazilbash05a}.
\par
As discussed above (Sec.~\ref{TransportTunneling}) \textcite{Niestemski07a} reported the first reproducible high resolution STM measurements of PLCCO (T$_c=24$ K) (Fig.~\ref{PLCCOSTM}).  The linecut  (Fig.~\ref{PLCCOSTM}a) shows spectra that vary from ones with sharp coherence peaks to a few with more pseudogap-like features and no coherence peaks.  However almost all spectra show the very notable `V' shaped higher energy background, which is consistent with $d$-wave symmetry.

\textcite{Chesca05a} used a bicrystal GBJ
with optimal doped LCCO films (a SIS junction) and measured both
Josephson tunnelling and quasiparticle tunnelling below T$_c \sim$ 29K. A
ZBCP was clearly seen in their quasiparticle tunnelling spectrum and it has
the magnetic field and temperature dependence expected for a
$d$-wave symmetry, ABS-induced, zero energy peak. These authors
argue that it requires extremely high-quality GB junctions-- to
reduce disorder at the barrier and to have a large enough critical
current-- in order to observe the ZBCP. Given the sporadic
observations of a ZBCP in $n-$doped cuprates
[\textcite{Shan05a} and references therein], the authors suggest that perhaps the
observation of a ZBCP rather than its absence should be regarded
as a true test of the pairing symmetry.

\par
Finally, similar GB junctions of LCCO have also revealed an intriguing behavior with the observation 
of a ZBCP for magnetic field much larger than the usual upper critical field measured on the same 
film using in-plane resistivity \cite{Wagenknecht08a}.  With increasing temperature T, they find that the ZBCP vanishes at the critical temperature T$_c$ = 29K if B = 0, and at T = 12K for B = 16T.  These observations may suggest that the real upper critical field is larger than the one inferred from transport.   They estimate H$_{c2} \approx$ 25 T at T=0.  However, this is in complete disagreement with the bulk upper critical field that has been estimated to remain below 10T at 2K for all dopings using specific heat\cite{Balci04c} and the Nernst effect\cite{Balci03a,Li07a}.  These differences are not yet explained.


\subsubsection{Low-energy spectroscopy using Raman scattering}
\label{sec:Raman-symm}
\par

Raman scattering is sensitive to the anisotropy of the
superconducting gap, as particular polarization configurations probe specific
regions of momentum space.  It is possible to isolate signatures
related to the superconducting gap, and in
particular demonstrate anisotropy and zeros in the gap
function \cite{Devereaux94a,Devereaux07a}. As observed in hole-doped cuprates
\cite{Stadlober95a}, peaks related to the magnitude of the gap are
extracted in two specific polarization configurations, B$_{1g}$
and B$_{2g}$.  With a $d$-wave gap anisotropy, these peaks
are expected to be found at different frequencies in different polarizations.
Moreover, the presence of low-energy excitations below the maximum
gap value (down to zero energy in the case
of lines of nodes for $d$-wave symmetry) implies that the Raman
response follows very specific power law frequency dependencies
for these various polarizations \cite{Devereaux94a, Devereaux07a}. In the
original Raman work on NCCO's order parameter, \textcite{Stadlober95a} showed that
the peaks in the B$_{1g}$ and B$_{2g}$ channels were positioned at close to the same energy, much like older works on s-wave classical superconductors like Nb$_3$Sn
\cite{Dierker83a}.

\par
\begin{figure}[htbp]
\begin{center}
\includegraphics[width=6.5cm,angle=0]{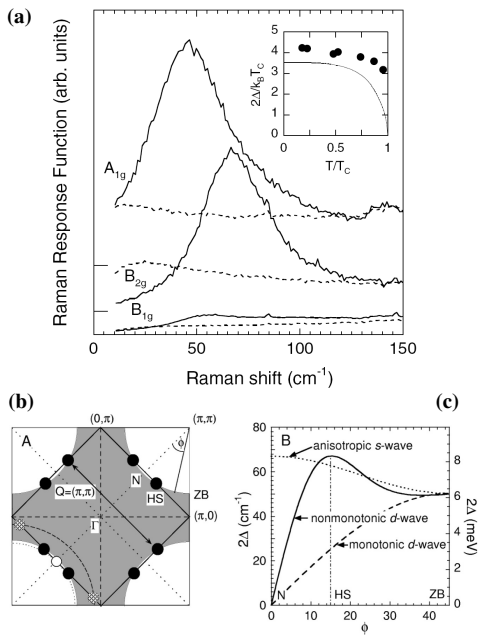}
\caption{(a) Electronic Raman scattering results comparing the
response above (35K - dashed line) and below (11K - solid line)
the critical temperature in the B$_{1g}$, the B$_{2g}$ and the
A$_{1g}$ configurations; (b) a sketch of the position of the hot
spots (HS) on the Fermi surface where the gap maximum also occurs;
(c) a comparison of the angular dependence of the non-monotonic $d$-wave
gap (solid line) with monotonic $d$-wave (dashed line) and anisotropic 
s-wave gap (dotted line). From \textcite{Blumberg02a}.} 
\label{raman.nmd}
\end{center}
\end{figure}

However, more recent experiments on single crystals and thin films
reveal a more complicated picture.  The low-frequency behavior of the
B$_{1g}$ and B$_{2g}$ channels approach power laws consistent with
the presence of lines of nodes in the gap function\cite{Kendziora01a}. These
power laws, although not perfect, indicate the presence of
low energy excitations. Moreover, in some instances, the peak
energy values in the B$_{1g}$ and B$_{2g}$ channels can be
different\cite{Kendziora01a}, and in some others, they are
virtually identical \cite{Blumberg02a,Qazilbash05a}. In all these
recent data however, the low energy spectrum continues to follow
 the expected power laws for lines of nodes. To reconcile
the fact that these power laws are always observed and that some
samples present peaks at identical energies in both channels,
\textcite{Blumberg02a} first proposed that a non-monotonic $d$-wave
gap function could explain this anomalous response \cite{Blumberg02a,Qazilbash05a}.  In Fig.~\ref{raman.nmd}, we show a representative dataset in the B$_{1g}$, 
B$_{2g}$ and A$_{1g}$ channels, together with the non-monotonic gap
function proposed by \textcite{Blumberg02a}. In this picture, the
maximum value of the gap function ($\Delta_{max} \sim$ 4 meV) coincides 
with the `hot spots' on the Fermi surface (\textbf{HS}
in Fig.~\ref{raman.nmd}), namely the position in $\vec{k}$-space
where the Fermi surface crosses the AFBZ as found by \textcite{Armitage01b}.  At the zone boundary
(ZB in Fig.~\ref{raman.nmd}), the gap value drops to $\sim$3 meV\footnote{It has been argued recently that the non-monotonic gap proposed by \textcite{Blumberg02a} and others is not purely the superconducting one, but in fact reflects a coexistence of antiferromagnetic and superconducting orders \cite{Yuan06b}}.   The non-monotonic gap has been found to be consistent with recent ARPES and tunneling results \cite{Matsui05b,Dagan07b}.
%
%
%
%
\par
\textcite{Venturini03a} countered that the basis for the conclusion of \textcite{Blumberg02a} was insufficient and so an s-wave form can still not be ruled out. They argued that since the Raman scattering amplitudes are finite at the maximum of the proposed gap function for all symmetries, the spectra in all these symmetries should exhibit multiple structures at the same energies in the limit of low damping as opposed to simply different size gaps in the different geometries (i.e. peaks should appear for energies corresponding to $\partial \Delta / \partial \phi = 0$).  \textcite{Blumberg03a} stand by their original interpretation and replied that no sharp threshold gap structures had ever been observed in any electron-doped cuprates even at the lowest temperature and frequencies and therefore irrespective of any other arguments an $s$ wave symmetry can be definitively ruled out.
\par
A very detailed study on single crystals and thin films has been reported by \textcite{Qazilbash05a} who followed the doping dependence of PCCO and NCCO's Raman response.  The authors extracted the magnitude of the gap as a function of doping and concluded that the smooth continuous decrease of the Raman response below the gap signatures (coherence peaks) is a sign that the superconducting gap preserves its lines of nodes throughout the whole doping range from under- to overdoping.  Obviously, this non-monotonic $d$-wave gap function should have a definite impact on properties sensitive to the low energy spectrum.  


\subsubsection{ARPES} \label{sec:ARPES-symm}
\par

ARPES provided some of the first dramatic evidence for an 
anisotropic superconducting gap in the hole-doped cuprates  \cite{Shen93a}. Comparing the photoemission response close to the Fermi energy on the same
sample for temperatures above and below T$_c$, one can clearly
distinguish a shift of the intensity in the spectral function for momentum
regions near ($\pi,0$). This ``leading-edge" shift gets its origin
from the opening of the superconducting gap and one can then map
it as a function of $\vec{k}$ on the Fermi surface in the
Brillouin zone (BZ). In the case of hole-doped cuprates, the first
$\Delta (\vec{k})$ mapping was obtained with
Bi$_2$Sr$_2$CaCu$_2$O$_{8+\delta}$ \cite{Shen93a}, which is easily
cleaved due to its weakly coupled Bi-O planes.  Gap values consistent with zero were observed along the diagonal directions in the BZ, i.e. along the (0,0) to ($\pi$,$\pi$) line \cite{Ding96a}. Away from the zone diagonal, the magnitude of the gap tracks the $\vec{k}$-dependence of the monotonic 
$d$-wave functional form.  

\par
\begin{figure}[htbp]
\begin{center}
\includegraphics[width=5cm,angle=0]{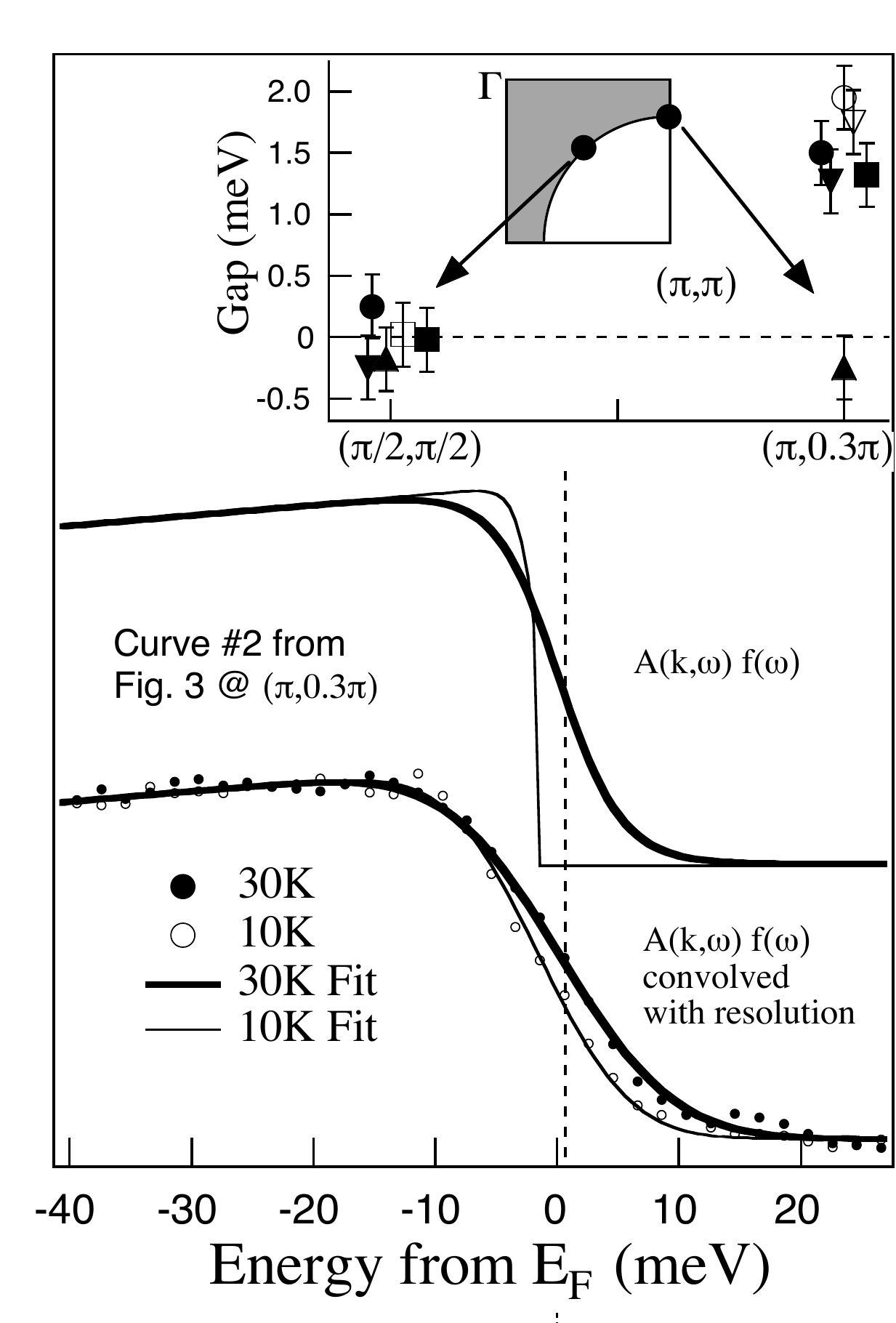}
\caption{Bottom curves are near E$_F$ ARPES EDCs of optimally doped NCCO from $\vec{k}_F$ close to ($\pi$,0.3$\pi$).  Open and solid circles are the experimental data at 
10 and 30K respectively, while solid lines are fits.  Upper curves are the experimental fits without resolution convolution.  "Curve $\#$2 from Fig. 3" refers to figures in \textcite{Armitage01a}.  Upper panel:  Gap values extracted from fits at the two $\vec{k}$-space positions using the difference between the 10 and 30K data.  Different symbols are for different samples.   From \textcite{Armitage01a}.} 
\label{ARPES.aniso-d}
\end{center}
\end{figure}

Until modern advances in the technology, the smaller energy gap of the electron-doped
cuprates, on the order of 5 meV for optimal doping, was at the
limit of ARPES resolution.   The first reports of
a measured superconducting gap in NCCO were presented by
\textcite{Armitage01a} and reported independently by \textcite{Sato01a} and are shown in Fig.
\ref{ARPES.aniso-d}. They found an
gap anisotropy with a negligible gap value along the zone diagonal
directions and a leading-edge shift of $\sim $ 2-3 meV along the Cu-O bond directions \cite{Sato01a,Armitage01a}. Such behavior was consistent with an order parameter of $d$-wave symmetry.  Using a model taking into account thermal broadening and the finite energy resolution,
Sato \textit{et al.} estimated the maximum gap value to be on the order of
4 to 5 meV, in close agreement with the values observed by
tunnelling (see Section~\ref{sec:tunnel}) and Raman Sec.~\ref{sec:Raman-symm}.
%
%
%
%
%
\par
\begin{figure}[htbp]
\begin{center}
\includegraphics[scale=0.3,angle=0]{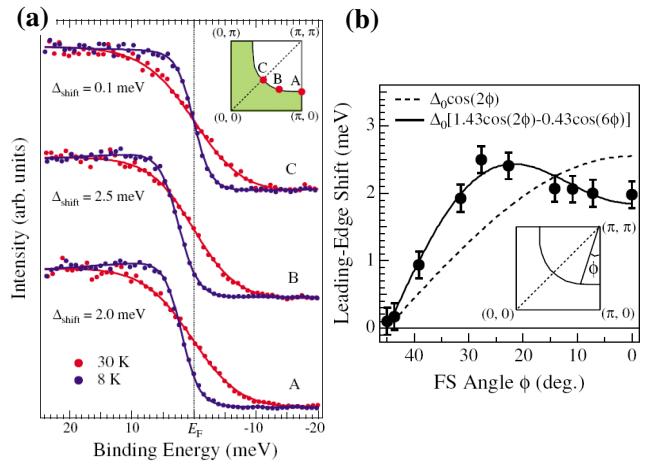}
\caption{(a) EDCs from ARPES measurements for
temperatures above (30K : red) and below (8K : blue) the
transition temperature for Pr$_{0.89}$LaCe$_{0.11}$CuO$_4$ single
crystals at three distinct points in k-space on the Fermi surface;
(b) Leading-edge shift determined as a function of position
(angle) on the Fermi surface showing that it fits a non-monotonic
$d$-wave symmetry. From \textcite{Matsui05b}.}
\label{ARPES.aniso-nmd}
\end{center}
\end{figure}

In these early studies, the limited number of momentum space positions measured could not
give the explicit shape of the gap function.  Matsui \textit{et
al.} followed a few years later with more comprehensive results on
Pr$_{0.89}$LaCe$_{0.11}$CuO$_4$ (PLCCO) that mapped out the
explicit momentum dependence of the superconducting gap. Their
data, shown in Fig.~\ref{ARPES.aniso-nmd}, confirm the presence of
a very anisotropic gap function with zeros along the diagonal
directions \cite{Matsui05b} as in BSCCO.  They also
concluded that the gap function is non-monotonic as found by
\textcite{Blumberg02a} $via$ Raman, with the maximum gap value
coinciding with the position of the hot spots in the BZ.  Matsui
\textit{et al.} fit their data with the function: $\Delta =
\Delta_o [1.43 \cos{2\phi} - 0.43 \cos{6\phi}]$, with $\Delta_o = $
1.9 meV . Intriguingly, the maximum value of the gap extracted
from the ARPES data ($\Delta_{max} \sim$ 2.5 meV)  seems to fall
short from the values obtained from other experiments, in
particular in comparison to the data of Blumberg \textit{et al.}
in Fig.~\ref{raman.nmd}, but also tunnelling data giving a
maximum gap value of 4 meV for optimal doping (see Section
~\ref{sec:tunnel}). This could be related to the worse resolution of ARPES, but perhaps also the different materials used for the separate experiments
(NCCO vs PLCCO) present slightly different properties.  The 
possibility also exists that the non-monotonic gap reflects a superposition   
of superconducting and antiferromagnetic order parameter gaps \cite{Yuan06b}. 
%
%
%
%
%


\subsubsection{Specific heat} \label{sec:SpecHeat}
\par
Specific heat measurements probe the low energy excitations of the bulk and are 
not sensitive to surface quality.  Its temperature and field dependencies away from $T_c$ and
$H_{c2}$ in hole-doped cuprates are sensitive
to the energy dependence of the density of states below the gap
energy. The specific heat for a pure $d$-wave superconductor with
line nodes and a linear density of states
should have an electronic contribution given by $c_{el}(T) =
\gamma_n T^2 /T_c$ where $\gamma_n T$ is the expected normal state
electronic contribution to the specific heat\cite{Volovik93a,Scalapino95a}.  In the presence of a
magnetic field at fixed temperature, this electronic
contribution should grow as $c_{el}(H) \propto \sqrt{H}$. This is the so-called `Volovik
effect' \cite{Volovik93a} for a clean $d$-wave superconductor.  Moler \textit{et al.} showed that the electronic specific heat of YBCO, has the expected square root dependence on magnetic field.  The temperature dependence exhibited a non-zero \textit{linear} (not T$^2$) term down to zero temperature \cite{Moler94a,Moler97a}, which is consistent with various `dirty $d$-wave' scenarios.

\par
For the electron-doped cuprates, extracting similar information about the 
electronic contribution to the specific heat is challenging because of its relatively small magnitude with respect to the phonon contribution\cite{Marcenat93a,Marcenat94a}, 
the magnitude of $T_c$ and the relatively small value for $H_{c2} \sim 10T$ [see Refs.
\cite{Balci03a,Qazilbash05a,Fournier03a} and references therein].
Moreover, rare-earth magnetism gives rise to additional anomalies
at low temperature that makes it difficult to extract the
electronic contribution.  For this reason, most recent studies of
the electronic specific heat to unravel the symmetry of the gap
have been performed with PCCO single crystals\cite{Balci02a,Balci04c,Yu05a} 
with its weaker RE magnetism (Sect.~\ref{sec:magnetism}). 
The most recent results demonstrate that the field dependence
follows very closely the expected $\gamma(H) \propto \sqrt{H}$ for
all superconducting dopant concentrations\cite{Balci02a,Balci04c,Yu05a}.
\par
\begin{figure}[htbp]
\begin{center}
\includegraphics[width=4.1cm,angle=0]{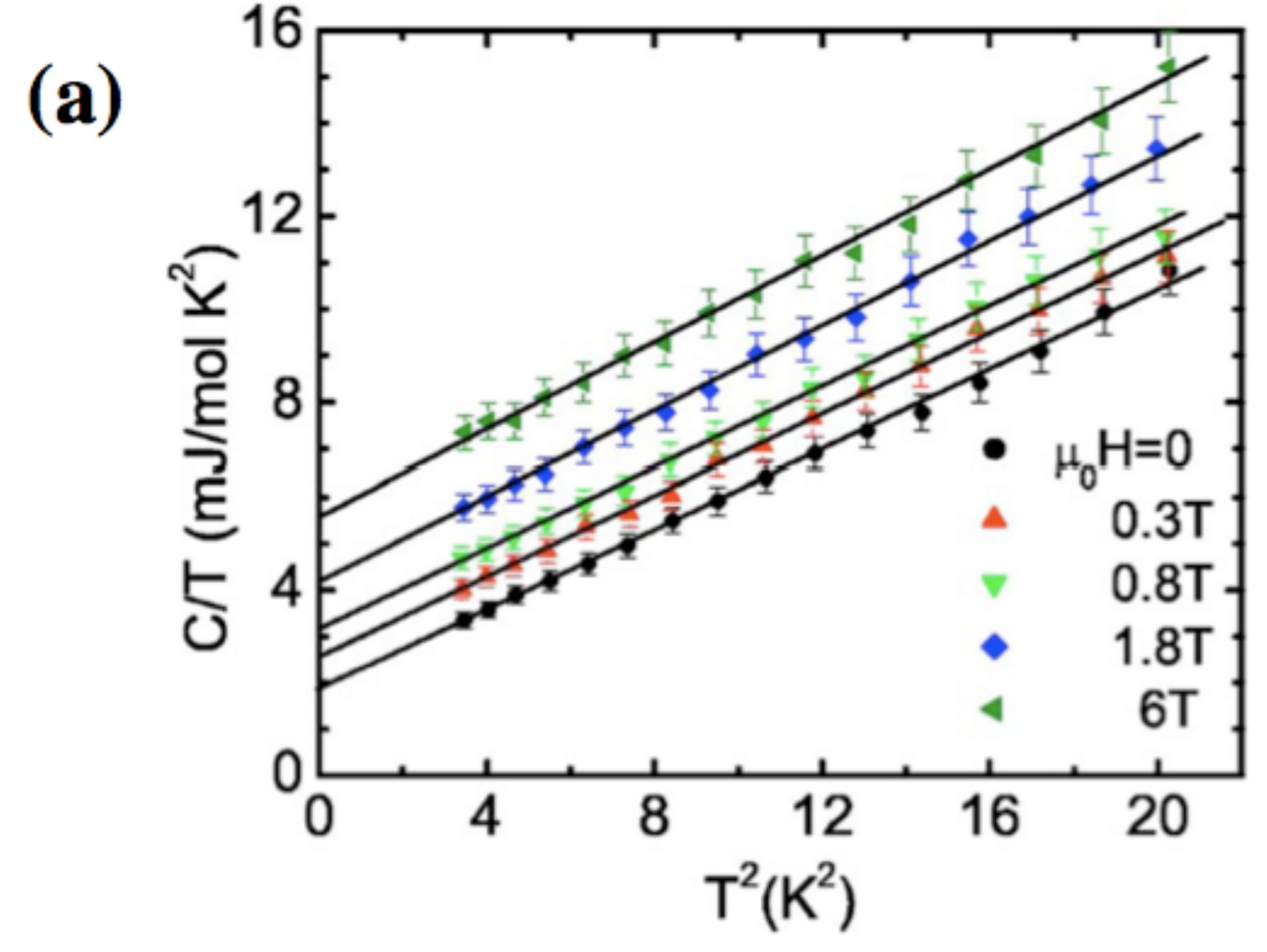}
\includegraphics[width=4.1cm,angle=0]{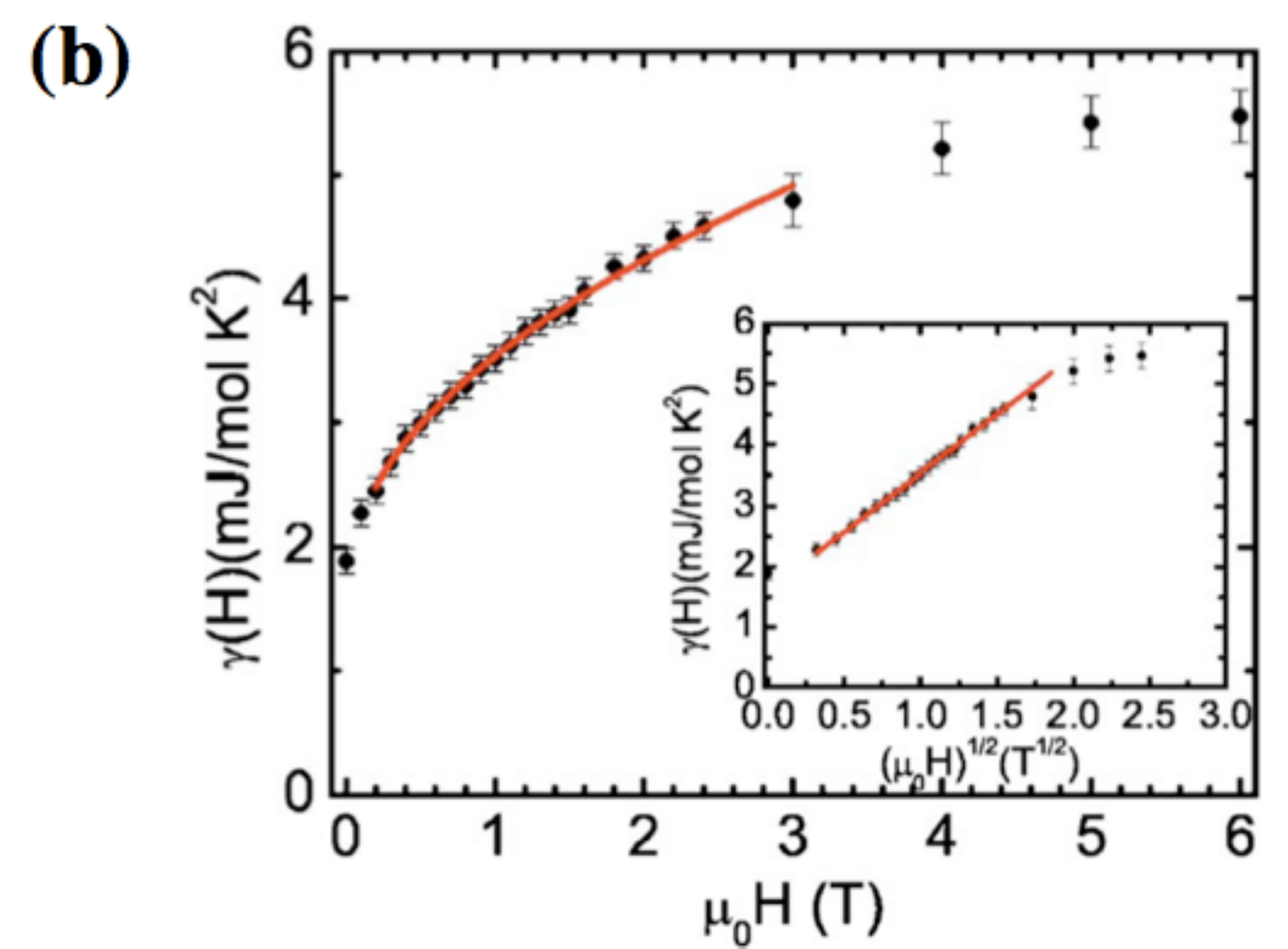}
\caption{Specific heat data from \textcite{Yu05a} on a single crystal of 
Pr$_{1.85}$Ce$_{0.15}$CuO$_4$. In (a), the temperature 
dependence at various magnetic fields is used to extract the linear-T electronic 
contribution. (b) Field dependence of the linear in T coefficient that shows 
close to $\sqrt{H}$ dependence in a magnetic field range considerably below $H_{c2}$. The red line is a fit to $\gamma (H) = \gamma_o + A \sqrt{H}$.  These data show also the saturation of the electronic specific heat at roughly 6T interpreted as the bulk upper critical field.} \label{Spec.heat.vsH}
\end{center}
\end{figure}

The initial measurements on optimally doped PCCO (x = 0.15) showed
a large non-zero linear in temperature electronic contribution
down to the lowest temperature ($T / T_c \sim $ 0.1) very similar to 
YBCO\cite{Moler94a,Moler97a}. Furthermore, it presented a
magnetic field dependence approaching $\sqrt{H}$ over a 2 - 7K
temperature range as long as the field was well below
$H_{c2}$\cite{Balci02a}. Similar to hole-doped cuprates, these
features were interpreted as evidence for line nodes in the
gap function. However, a subsequent study from the same group
seemed to reveal that the temperature range over which $c_{el}(H)
\propto \sqrt{H}$ is limited to high temperatures, and that a
possible transition (from $d$- to $s$-wave) is observed as the
temperature is lowered\cite{Balci04c}.  However, a different measurement scheme that removes the vortex pinning contribution  through field cooling reveals  (Fig.~\ref{Spec.heat.vsH}) that the anomalies 
interpreted as a possible $d$- to $s$-wave transition are actually resulting from 
the thermomagnetic history of the samples \cite{Yu05a}. Thus, the $c_{el}(H) \propto \sqrt{H}$ 
behavior is preserved down to the lowest temperatures for all dopant concentrations. It extends over a limited field region followed by a saturation 
at approximately $\mu_o H \sim$ 6T interpreted as a value close to the bulk upper 
critical field. From a quantitative point of view, the analysis of the 
field dependence using a clean $d$-wave scenario according to 
$c_{el}/T \equiv \gamma (H) = \gamma_o + A \sqrt{H}$ yields 
$A \sim $1.92 mJ/mol K$^2$ T$^{1/2}$. In the clean limit, this $A$ parameter 
can be related to the normal state electronic specific heat measured at high magnetic fields leading to
$A = \gamma_n \left( 8 a^2 / \pi H_{c2} \right)^{1/2}$\cite{Wang01a} where $a$ is a constant approaching 0.7.  With $H_{c2} \sim$ 6T, one gets $\gamma_n \sim$ 4.1 mJ/mol K$^2$ and $\gamma_o + \gamma_n \sim$ 5.7 mJ/mol K$^2$  approaching the normal state Sommerfeld constant measured at 6T.  These results confirm that the bulk of the electron-doped cuprates presents specific heat behaviors in full agreement with a dominant $d$-wave symmetry over the whole range of doping at all temperatures explored.


%
%
%
%
%
\subsubsection{Thermal conductivity} \label{sec:ThCond}
\par
Thermal conductivity at very low temperatures is a sensitive probe
of the very lowest energy excitations of a system \cite{Durst00a}.
The electronic contribution to the thermal conductivity
given usually by $\kappa_{el} = \frac{1}{3} c_{el} v_F l$ (where $c_{el}$
is the electronic specific heat, $v_F$ is the Fermi velocity and
$l$ is the mean-free path of the carriers) becomes a very
sensitive test of the presence of zeroes in the gap function. In the case of
conventional BCS s-wave superconductor, the fully gapped Fermi
surface leads to an exponentially suppressed number of electronic thermal excitations as $T \rightarrow
0$. On the contrary, a non-zero electronic contribution is expected down to the lowest
temperatures in a $d$-wave superconductor with line nodes. To
extract this part from the total thermal 
conductivity that includes also a phonon contribution, a plot of $\kappa /T$ as a function of $T^2$ yields
a non-zero intercept at $T = 0$ \cite{Taillefer97a}. Assuming that 
$\kappa = \kappa_{el}+\kappa_{ph} = AT + CT^3$, one can compare the measured value 
of $A$ to the theoretical predictions that relates it to the slope of the
gap function at the nodes [$S \equiv
\left(\frac{d\Delta}{d\phi}\right)_{node}$], i.e. its angular
dependence along the Fermi surface. \textcite{Durst00a} showed
that the electronic part is given by :
\begin{equation}
  \frac{\kappa_{el}}{T} = \frac{k_B^2}{3 \hbar} \frac{n}{d}\left(\frac{v_F}{v_2}+\frac{v_2}{v_F} \right)\\
    \label{eq.kappa}
\end{equation}
where $\frac{d}{n}$ is the average distance between CuO$_2$ planes. 
The first term of Eq.~\ref{eq.kappa} is expected to give the primary contribution (for
example, $\frac{v_F}{v_2} \sim 14$ in YBCO \cite{Chiao00a}), such
that $\kappa_{el} / T \approx \frac{k_B^2}{3 \hbar}
\frac{n}{d}\left(\frac{v_F}{v_2}\right)$ where $v_2 = S /\hbar
k_F$. For a monotonic $d$-wave gap function, $\Delta = \Delta_o
\cos(2\phi)$ such that $\kappa_{el} / T \propto 1 / S \propto 1 /
\Delta_o$. This linear temperature dependence and its link to $v_F
/ v_2$ (which is sample-dependent) were confirmed in hole-doped cuprates by
\textcite{Chiao00a} for example. Similar to the specific heat, the
Volovik effect should give rise also to $\kappa_{el}(H) \propto
\sqrt{H}$ as was observed in YBCO \cite{Chiao99a}.
\par

\begin{figure}[htbp]
\begin{center}
\includegraphics[width=8cm,angle=0]{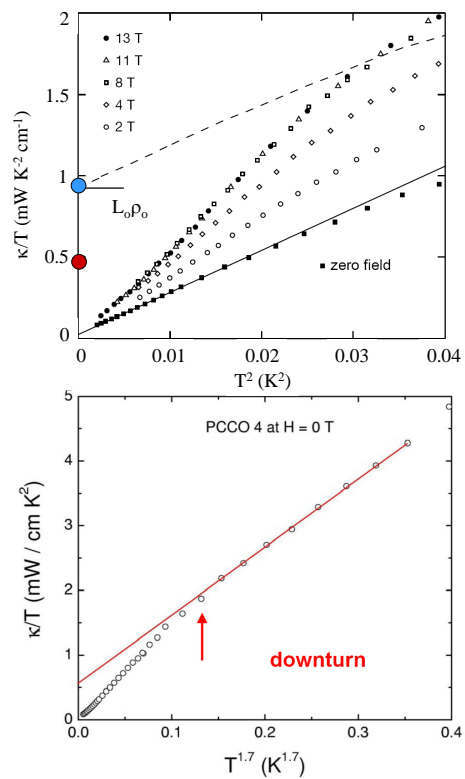}
\caption{(top) Thermal conductivity of PCCO for a heat current in the basal plane, plotted as
$\kappa /T$ versus $T^2$, at different values of the magnetic field applied normal to the plane. The
solid line is a linear fit to the zero-field data below 130 mK. The dashed line shows the
behaviour of a Fermi liquid consisting of the expected electronic part extracted from the 
Wiedemann-Franz law using the residual resistivity $\rho_o$ for this sample obtained at high magnetic 
and a phonon contribution given by the solid line (zero-field data). From \textcite{Hill01a}.   The blue (near (0,1)) and red (near 0,0.5) dots are estimates for the coefficient of the electronic contribution for monotonic and non-monotonic gap functions as given in the text.  The experimental data show a zero $y$-intercept because of electron-phonon decoupling at low temperature.  (bottom) Thermal conductivity of an optimal PCCO single crystal for a heat current in the basal plane, plotted as $\kappa /T$ versus $T^{1.7}$. These data show the downturn to the decoupling of the electron and phonons.   Note that the data above the downturn goes as a power of 1.7 and not 2.  Courtesy of L. Taillefer.} 
\label{kappa.vsT}
\end{center}
\end{figure}

In the case of the electron-doped cuprates, the low temperature
data obtained by \textcite{Hill01a} show a significant
phonon contribution at low temperature as observed in a plot of
$\kappa / T$ as a function of $T^2$ as evinced by the straight lines in
Figure~\ref{kappa.vsT}. Moreover, a substantial increase
of thermal conductivity with the magnetic field confirms the presence
of a large electronic contribution growing towards saturation
at large fields (roughly 8T), in agreement with the above mentioned specific
heat data.  However, as demonstrated by the lack of a $y$-intercept the observed electronic contribution does not extend down to the lowest temperatures as in YBCO\cite{Chiao99a}.  Instead  a clear downturn is observed below 200mK that has recently been
attributed to thermal decoupling of the charge carriers and the
phonons\cite{Smith05a}.  The electrons and the phonons
that carry heat are not reaching thermal equilibrium at low temperature
because of a poor electron-phonon coupling. This decoupling is obviously a major
drawback for a direct extraction of the electronic contribution without the use of a
theory\cite{Smith05a} and makes it difficult to confirm
the presence of a non-zero value at zero field in the electron-doped
cuprates.

%
%

One can make a crude estimate of the expected linear coefficient of the specific heat $A$ parameter (discussed in Sec.~\ref{sec:SpecHeat}) using $v_F \sim$ 270 km/s \cite{Schmitt08a,Park08a} 
for nodal quasipartil/k0cle excitations and $v_2 = 2 \Delta_o /
\hbar k_F$ assuming a monotonic $d$-wave gap with the tunnelling
maximum value of $\Delta_o \sim $ 4 meV for optimal
doping\cite{Biswas02a}.  This gives $v_F / v_2 \sim $ 96 and a
$\kappa_{el} / T \approx $ 0.96 mW/K$^2$-cm, which is shown as a blue circle 
in Fig.~\ref{kappa.vsT}. Assuming instead a non-monotonic $d$-wave gap 
with $\Delta = \Delta_o [1.43 \cos{2\phi} - 0.43 \cos{6\phi}$\cite{Matsui05a}, with $\Delta_o = $
3 meV \cite{Blumberg02a,Qazilbash05a} leads to $v_F / v_2 \sim $ 47, and $\kappa_{el} / T \approx $ 0.47 mW/K$^2$-cm, which is shown as the red circle in Fig.~\ref{kappa.vsT}\footnote{Note that we have used the non-monotonic gap function from \textcite{Matsui05a} but the maximum gap values obtained 
by \textcite{Blumberg02a,Qazilbash05a} to evaluate $v_F / v_2$. This takes into account the inconsistency between the absolute values of the gap maximum measured by various probes as discussed in section~\ref{sec:Raman-symm}.}.  In Figure~\ref{kappa.vsT}, an unpublished analysis of the thermal conductivity (courtesy of L. Taillefer) of an optimally doped PCCO sample allows one to isolate the linear term at low temperature as $\kappa_{el} / T \approx $ 0.60 mW/K$^2$-cm by taking into account the thermal decoupling of the charge carriers and the phonons mentioned above\cite{Smith05a}.  This value is intermediate to the estimates given above for monotonic and non-monotonic $d$-wave superconductors (red and blue dots of Fig.~\ref{kappa.vsT}).   In Fig.~\ref{kappa.vsT} one can see the clear downturn from electron-phonon decoupling around 300 mK, which prevents the explicit measurement of $\kappa /T$.

\subsubsection{Nuclear Magnetic Resonance}

\begin{figure}[htbp]
\begin{center}
\includegraphics[width=6.5cm,angle=0]{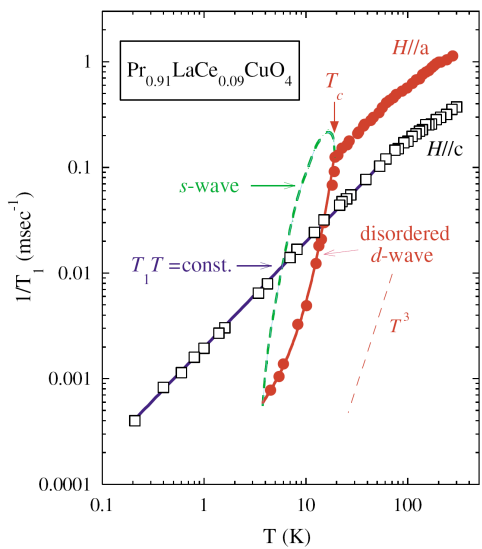}
\caption{$^{63}$Cu spin relaxation rate $1 /T_1$ as a function of temperature in the superconducting and the normal states of a  Pr$_{0.91}$LaCe$_{0.09}$CuO$_{4-y}$ single crystal. Red solid circles: data in the superconducting state measured with a magnetic field of 6.2T parallel to the CuO$_2$ planes. Open black circles : data in the normal state measured with an out-of-plane magnetic field of 15.3T. The red solid line is a fit using a d$_{x^2 - y^2}$ order parameter leading with $2 \Delta_0$ = 3.8 $k_B  T_c$. The solid line is a fit to the Korringa law, which is consistent with Fermi liquid behavior. From \textcite{Zheng03a}.} 
\label{NMR.dwave}
\end{center}
\end{figure}

Measuring the nuclear magnetic resonance (NMR) response of electron-doped cuprates is also a difficult task because of the large magnetic contribution of the rare earth ions.  It leads to dipolar and quadripolar local field that makes interpretation difficult. For this reason, only measurements with Pr and La (and eventually Eu) as the rare earth atoms have been of real interest to extract the symmetry of the order parameter. \textcite{Zheng03a} have shown explicitly that the spin relaxation rate $1/T_1$ of $^{63}$Cu in x=0.11 PLCCO falls dramatically in the superconducting state over some temperature range following a power law close to $T^3$ as shown in Figure~\ref{NMR.dwave}. This temperature dependence is consistent with the existence of line nodes and a $d$-wave superconducting order parameter as was observed in hole-doped cuprates\cite{Asayama96a}.  At the lowest temperatures the relaxation rate deviates from $T^3$ behavior which was interpreted by \textcite{Zheng03a} as a consequence of disorder scattering.  Also consistent with $d$-wave, there was no sign of a Hebel-Schlicter peak just below T$_c$ (see also Fig.~\ref{NMR.dwave}) which is a signature of class II coherence factors and s-wave superconductivity.  A comparison of the data with calculation using a d$_{x^2 - y^2}$ order parameter reveals a superconducting gap $2 \Delta_0$ = 3.8 $k_B  T_c$, which is consistent with many other probes. 
%
%
%
%


\subsubsection{Neutron scattering}

As discussed above in~\ref{resonancesection},   in the inelastic neutron scattering response of PLCCO \textcite{Wilson06b} found an enhancement of the intensity (Fig.~\ref{PLCCOresonance}) at approximately 11 meV at (1/2,1/2,0) (equivalent to ($\pi,\pi$) in the superconducting state.  This was interpreted as being the analog of the much heralded `resonance' peak \cite{rossat91a} found in many of the hole-doped compounds.  \textcite{Zhao07a} has claimed that optimally doped NCCO has a similar resonance at 9.5 meV although this has been disputed by \textcite{Yu08a}, who claim that a sub gap resonance is found at the much smaller energy of  $\approx$ 4.5 meV as shown in Fig.~\ref{GrevenNCCOplot}.  Irrespective of where exactly this peak is found, the observation is strong evidence for $d$-wave superconductivity, as the superconducting coherence factors impose that a ($\pi,\pi$)  excitation can only exist in the superconducting state if the OP changes sign under this momentum translation \cite{Manske01a,Bulut96a}.   The resonance feature in the $n$-type compounds appears to turn on at T$_c$ as seen in Fig.~\ref{PLCCOresonance}.

\subsubsection{Phase sensitive measurements} \label{sec:PhaseSensors}
\par
Some of the most convincing and definitive experiments to
demonstrate the $d$-wave pairing symmetry in the hole-doped cuprates
measure the phase of the order parameter directly instead of its
magnitude. Such techniques are sensitive to changes in the sign of the pair wave-function in
momentum space. Most are based on the fact that the current
flowing through a Josephson junction is sensitive to the phase
difference between superconducting electrodes\cite{Tinkham96a}. By
designing very special geometries of junctions and SQUIDs
(Superconducting Quantum Interference Devices) that incorporate
high-T$_c$ and possibly conventional superconductors, one
can demonstrate the presence of the sign change in the order
parameter\cite{vanHarlingen95a,Tsuei00b}. Quasiparticle tunnelling
can also be sensitive to the sign change. The presence of the
so-called Andreev bound states at the interface of
normal-insulator-superconductor (N-I-S) tunnel junctions is a
direct consequence of the particular symmetry of the high-T$_c$
cuprates.

\begin{figure}[htbp]
\begin{center}
\includegraphics[width=8cm,angle=0]{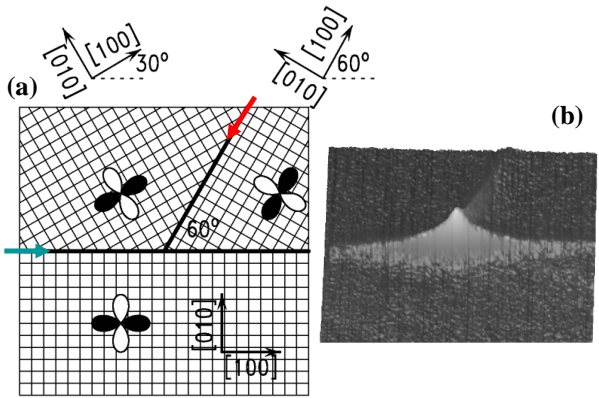}
\caption{(a) Tri-crystal geometry used to force phase frustration and 
spontaneous generation of a half-flux quantum at the tri-crystal junction point. The red and the blue arrows indicate the diagonal and horizontal grain boundaries respectively; (b) 3D 
image of the flux threading the film.  Adapted from Ref. \cite{Tsuei00a}.} 
\label{fig.halfquflux}
\end{center}
\end{figure}

The most convincing phase sensitive measurement for the
electron-doped cuprates has been reported by \textcite{Tsuei00a}
who observed a spontaneous half flux quantum ($\phi_o / 2$)
trapped at the intersection of a tri-crystal thin film. This
epitaxial thin-film-based experiment has been used extensively by
the same authors to demonstrate the universality of the $d$-wave
order parameter for hole-doped cuprates\cite{Tsuei00b}. It relies
on the measurement of the magnetic flux threading a thin film
using a scanning SQUID microscope. When the epitaxial thin film is
deposited on a tri-crystal substrate with carefully chosen
geometry as in Figure~\ref{fig.halfquflux}(a), Josephson junctions are formed in the films at the grain
boundaries of the substrates\cite{Hilgenkamp02a}.  The presence of
spontaneous currents induced by phase frustration at the
tri-crystal junction point is a definitive test of a sign change
in the order parameter.
\par
In the case of the electron-doped cuprates, \textcite{Tsuei00a}
showed using a fit of the magnetic field \cite{Kirtley96a} as a function of position that the magnetic flux at the tri-crystal junction in Fig.~\ref{fig.halfquflux}(b) corresponds to half a
flux quantum \cite{Tsuei00a}.  This observation was made despite very small critical current densities for the junctions along the grain boundaries, implying very weak coupling and very long
penetration depth of the field along the grain boundary junctions.
Similar to hole-doped cuprates \cite{vanHarlingen95a,Tsuei00b}, this observation is
consistent with pure $d$-wave pairing symmetry.

%
%
%
%
%
\par

\begin{figure}[htbp]
\begin{center}
\includegraphics[width=7.5cm,angle=0]{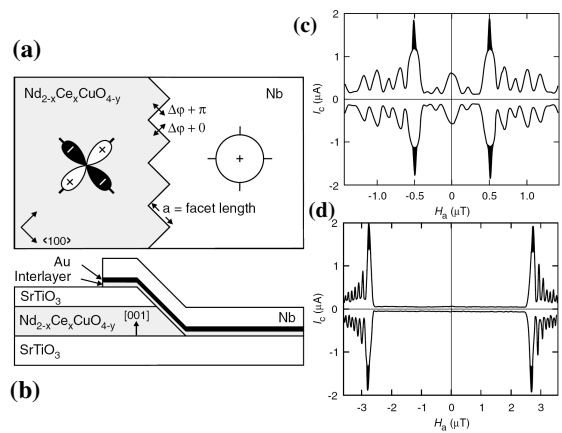}
\caption{Zigzag ramp-edge Josephson junctions made of Nd$_{2-x}$Ce$_x$CuO$_4$ and conventional s-wave Nb. (a) Top view of the zigzag design; (b) Cross-section view of a ramp-edge junction including a thin interlayer of NCCO. Examples of anomalous field modulations of the critical current arising from the $d$-wave symmetry for devices made of (c) 8 facets of 25$\mu$m width and (d) 80 facets of 5$\mu$m width. From \textcite{Ariando05a}.} 
\label{fig.zigzag}
\end{center}
\end{figure}

\textcite{Ariando05a}  fabricated ramp-edge
junctions between NCCO (x = 0.15 and 0.165) and Nb in a special
zigzag geometry as shown in Figure~\ref{fig.zigzag}.   Since the critical current density of the
NCCO/Au/Nb ramp-edge junctions is very small ($J_c \sim$ 30
A/cm$^2$), the zigzag geometry presented by Ariando \textit{et
al.} is in the small junction limit and one expects an anomalous magnetic field dependence in the $d$-wave case.  For instance, one can note that the critical current
density of this zigzag junction is suppressed at zero field. As
one applies a small magnetic field to this junction, the critical
current grows and then oscillates as the first quanta of flux
penetrate the zigzag junction.   Ariando \textit{et al.} demonstrate an order parameter consistent with $d$-wave symmetry.   In a similar experiment \textcite{Chesca03a} patterned a 500$\mu m$ thick LCCO film made by MBE on a tetracrystal substrate to create a $\pi$-SQUID at the junction point.  The minimum in critical current at zero field for the  $\pi$-design is consistent with $d$-wave pairing symmetry\cite{Chesca03a}.

%
%
%
%
%

\subsubsection{Order parameter of the infinite layer compounds}

Although measurements of the normal state
properties of the infinite layer compound SLCO are rare, there have been
a few experiments on their pairing symmetry. In general, measurements on the infinite layer compounds have been hampered by a lack of single crystals or thin films and give conflicting conclusions.  The spatial independence of the STM spectra observed when
performing a line scan across many randomly oriented grains on a
polycrystalline sample has, along with the lack of ZBCP, been
interpreted as being consistent with an s-wave
symmetry\cite{Chen02a}.  Low temperature specific heat
\cite{Liu05a} also suggests a conventional s-wave pairing symmetry
while NMR suggests an unconventional, non s-wave, symmetry
\cite{imai95a}. Stronger suppression of T$_c$ by the magnetic impurity
Ni than the non-magnetic impurity Zn is indirect supporting evidence
for s-wave \cite{Jung02a}.  \textcite{Chen02a} also concluded in
their STM study that the suppression of their tunneling coherence peaks with
Ni doping, in contrast to the much smaller effect with Zn doping,
was consistent with an s-wave symmetry.  This is a system that certainly needs
further investigation, both on the materials side as well as high
quality experimentation.  Measurements like $\mu$SR have not been able to measure the T dependence of the penetration depth to determine the pairing symmetry as they require single crystal \cite{shengelaya05a}.

\subsection{Position of the chemical potential and midgap states}   \label{ChemPot}

One long outstanding issue in the cuprates is the position of the chemical potential $\mu$ upon doping.  In the case that the cuprates are in fact described by some Mott-Hubbard-like model, the simplest scenario is that the chemical potential shifts into the lower Hubbard band (or CTB) upon hole doping and into the upper Hubbard band upon electron doping (see Fig.~\ref{Hubbard}).   This is the exact result for the one dimensional Hubbard model \cite{Woynarovich82a}.  In contrast, dynamic mean field theory (DMFT) calculations show that, at least for infinitely coordinated Mott-Hubbard system, for doped systems $\mu$ lies in coherent mid-gap states  \cite{Fisher95a}.  In a similar fashion, systems which have a tendency towards phase separation and inhomogeneity will generically generate mid-gap states in which the chemical potential will reside \cite{Emery92a}.  The position of the chemical potential and its movement upon doping is an absolutely central issue, as its resolution will shed light on the local character of the states involved in superconductivity, the issue of whether or not the physics of these materials can in fact be captured by Mott-Hubbard like models, and the fundamental problem of how the electronic structure evolves from that of a Mott insulator to a metal with a large Luttinger theorem \cite{Luttinger60a} respecting Fermi surface.

In the first detailed photoemission measurements of the $n$-type cuprates, \textcite{Allen90a} claimed  that $\mu$  did not cross the insulator's gap upon going from hole to electron doping and lies in states that fill the gap.  This inference was based on a comparison of the angle integrated valence band resonant photoemission spectra of Nd$_{2-x}$Ce$_x$CuO$_{4}$ at x=0 and 0.15 with La$_{2-x}$Sr$_x$CuO$_{4}$ which showed that the Fermi level lies at nearly the same energy in both cases as compared to the valence band maximum.  Similar conclusions based on x-ray photoemission have been reached by other authors \cite{Namatame90a,Matsuyama89a}.  However, these results have been called into question by \textcite{Steeneken03a} who showed that due to the large $4f$ electron occupation of Nd$_{2-x}$Ce$_x$CuO$_{4-y}$ (not to mention the crystal structure differences) the valence band maximum in NCCO is dominated by $4f$ electrons, making it a poor energy reference for the chemical potential.  They proposed instead that the appropriate reference for the internal energies of the copper oxygen plane across material classes was the peak at the $3d^8$ final states of the photoemission process, which represents configurations where the hole left from electron removal has its majority weight on the copper site instead of an oxygen (a  $3d^9 \underline{L}$ configuration).

\begin{figure}[htbp]
\begin{center}
\includegraphics[width= 7.3cm,angle=0]{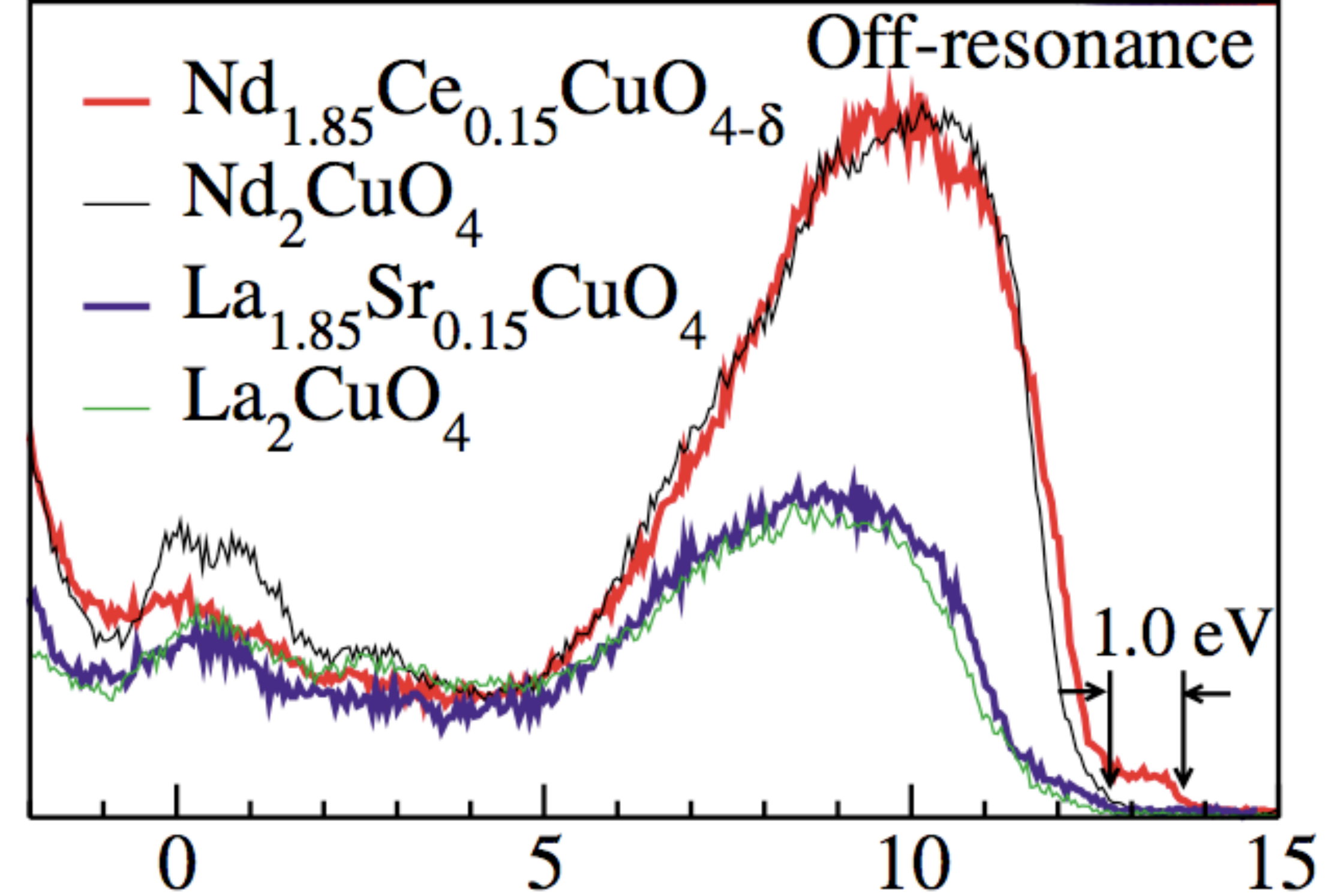}
\includegraphics[width= 7.3cm,angle=0]{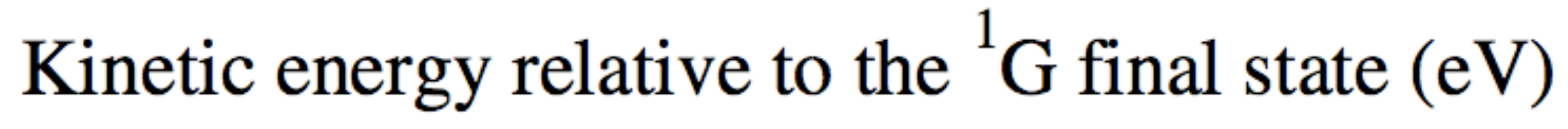}
\caption{(color) Photoemission valence band spectra of Nd$_{1.85}$Ce$_{0.15}$CuO$_{4-\delta}$, Nd$_{2}$CuO$_{4}$, La$_{2}$CuO$_{4}$,  and La$_{1.85}$Sr$_{0.15}$CuO$_{4}$  taken 5eV below the Cu $L_3$ edge.  Energies of the spectra are aligned with respect to the Cu $3d^8$ $^1G$ final states. From \textcite{Steeneken03a}.} \label{Steeneken}
\end{center}
\end{figure}

These $3d^9 \rightarrow$ $3d^8$ electron removal states are expected to be found at an energy approximately $U$ (the microscopic onsite Hubbard interaction energy) below the $3d^{10} \rightarrow$ $3d^9$ states and so give a good energy reference that refers directly to the electronic structure of the CuO$_2$ plane.   $3d^{10}$ initial states are assumed to be the primary result of electron doping, as electrons are believed to mostly be doped to Cu orbitals.  Lining up valence band spectra to the $3d^8$ states  (see Fig.~\ref{Steeneken}), \textcite{Steeneken03a} found that both Nd$_{2}$CuO$_{4}$, La$_{2}$CuO$_{4}$,  and La$_{1.85}$Sr$_{0.15}$CuO$_{4}$ showed a spectral weight onset at the same energy (approximately 13 eV above the $3d^8$ reference), which was presumably the CTB.  However, Nd$_{1.85}$Ce$_{0.15}$CuO$_{4-\delta}$ showed an onset approximately 1eV higher and so it was concluded that the chemical potential shifts by approximately this amount (across the charge transfer gap) going from lightly hole- to electron-doped materials.  As the onset in the optical charge transfer gap is of this order (1 -1.5 eV \cite{Tokura90a}), it was concluded that  $\mu$  lies near the bottom of the conduction band (presumably the upper Hubbard band) of the Ce doped system and near the top of the valence band (presumably the primarily oxygen derived CTB) for the Sr doped system.    A similar conclusion was reached in hard x-ray photoemission \cite{Taguchi05a}, which is more bulk sensitive.  The study of  \textcite{Steeneken03a} also concluded that the local character of the $3d^{10}$ near E$_F$ states was singlet.

\begin{figure}[htbp]
\begin{center}
\includegraphics[width= 7.5cm,angle=0]{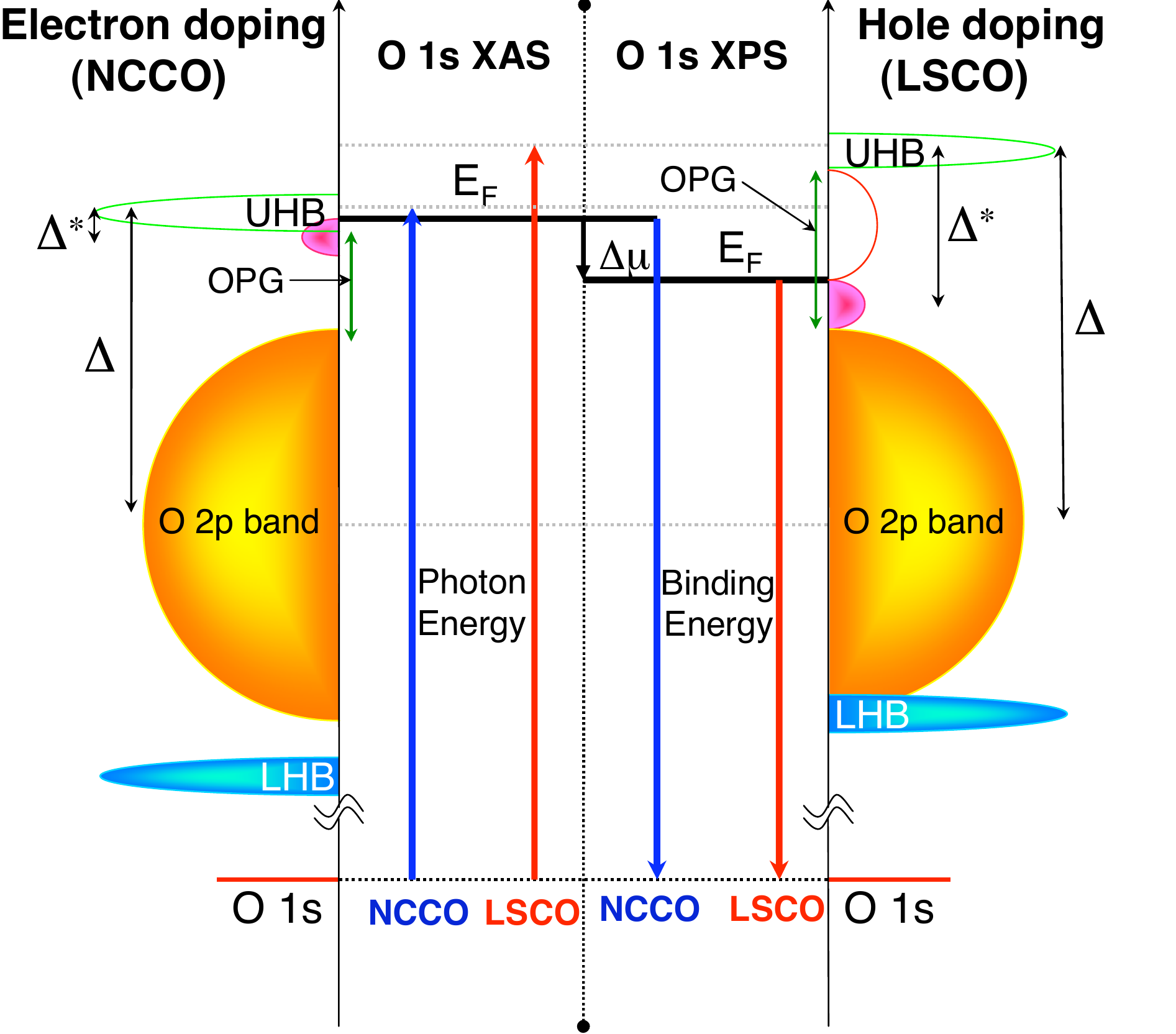}
\caption{(Color) Schematic illustration of the energy levels of LSCO and NCCO obtained from Anderson impurity model calculations and x-ray photoemission.   OPG is the optical gap from undoped materials. Shaded regions represent occupied density of states.  The manifold of O 2p bands are found to be displaced relative to the primarily Cu derived Hubbard-like bands between LSCO and NCCO.  This explains the very small shift in the O 1s core level data when comparing hole- and electron-doped data.  From \textcite{Taguchi05a}.} \label{Taguchixray}
\end{center}
\end{figure}

This conclusion with respect to hole doping is different than was inferred from ARPES studies on La$_{2-x}$Sr$_{x}$CuO$_{4}$, which posited that $\mu$ was found in mid-gap states derived from inhomogeneities \cite{Ino00a}.  It is however consistent with work on the Na$_{2-x}$Ca$_x$CuO$_2$Cl$_2$ system which finds that with light hole doping the chemical potential is always found near the top of the valence band \cite{Shen04a, Ronning03a}.  It is also consistent with the ARPES work of  \textcite{Armitage02a} who found that for lightly electron-doped Nd$_{1.86}$Ce$_{0.04}$CuO$_{4}$ the chemical potential sat an energy approximately 1 eV above the onset of the CTB (which could be imaged simultaneously).  As discussed above, there was evidence for in-gap states, but these filled in the gap at energies below $E_F$.  Additionally at this low doping, the near $E_F$ states formed a Fermi pocket at ($\pi$, 0), which is the expectation upon electron doping for many Hubbard-like models (see for instance \cite{Tohyama04a} and references therein).  Note that the studies of \textcite{Steeneken03a} and \textcite{Armitage02a} do not rule out scenarios where the chemical potential lies pinned in doping induced in-gap states.  They only show that this pinning does not take the chemical potential very far from the band edges.   The scenario proposed by \textcite{Taguchi05a} is shown in Fig.~\ref{Taguchixray}.

\subsection{How do we even know for sure it is $n$-type?}

Related to the issue of the position of the chemical potential is the even more basic issue of whether these materials can even be considered truly $n$-type.   It is usually assumed (and been assumed throughout this review) that these compounds are the electron-doped analog of the more commonly studied hole-doped compound.   However, there is no reason to believe $a$ $priori$ that the class of RE$_{2-x}$Ce$_{x}$CuO$_{4}$ compounds must be  understood in this fashion.   The effects of Ce doping could be of a completely different nature.  For instance, an analysis based on the aqueous chemical redox potentials shows that Ce$^{+3}$ may have difficulty reducing Cu$^{+2}$ to Cu$^{+1}$  \cite{Cummins93a}.   And even if electrons are actually introduced to the CuO$_{2}$ plane by Ce substitution, there is no guarantee that the correct way to think about its effects is by adding electrons into an upper Hubbard band in a fashion exactly analogous to adding holes to an effective lower Hubbard band.  There have been been suggestions for instance that due to structural considerations the effect of Ce doping is to liberate holes \cite{Hirsch95a,Billinge93a}, or that doped electrons instead go into a band formed of extended relatively uncorrelated Cu $4s$ states \cite{Okada90a} or an impurity band \cite{Tan90a}.   Meanwhile there is also the evidence discussed extensively above for simultaneous electron and hole contributions to transport \cite{Gollnik98a,Wang91a,Fournier97a} and claims that these compounds have a `negative charge transfer gap' \cite{Cummins93a}.  Are these compounds really $n$-type/electron-doped?  There are a few different ways to phrase and answer this question.

\textit{Do the CuO$_2$ planes of the doped compounds have local charge densities greater than the insulator? -}    This is perhaps the most basic definition of electron doping.  In principle high energy spectroscopies like x-ray core level photoemission (XPS), x-ray absorption (XAS) and electron energy loss spectroscopy (EELS) can probe the valence state of local orbitals (See \textcite{Yu98a} for a good overview).  As discussed above, naively one expects that electrons liberated from doped Ce will reside primarily in $3d$ orbitals nominally giving Cu $3d^{10}$.   This appears to be the case.

The first Cu K-edge x-ray absorption study from \textcite{Tranquada89a} concluded that $3d^{10}$ states formed upon Ce substitution, however a number of other early studies gave conflicting results  \cite{Nucker89a,Ishii89a,Fujimori90a}.  Note that in general, most of these kind of spectroscopies suffer from a strong sensitivity to surface quality.   In many of these measurements surfaces were prepared by scraping poly-crystalline ceramics (resulting in significant contamination signal as judged by the appearance of a shoulder on the high energy side of the main O 1s peak) or by high-temperature annealing which undoubtedly changes the surface composition \cite{Cummins93a}.   In contrast, in most of the measurements emphasized here, surfaces were prepared by breaking single crystals open in vacuum or the technique was inherently bulk sensitive (EELS or XAS in transmission or fluorescence yield mode for example).

\begin{figure}[htbp]
\begin{center}
\includegraphics[width= 6.5cm,angle=0]{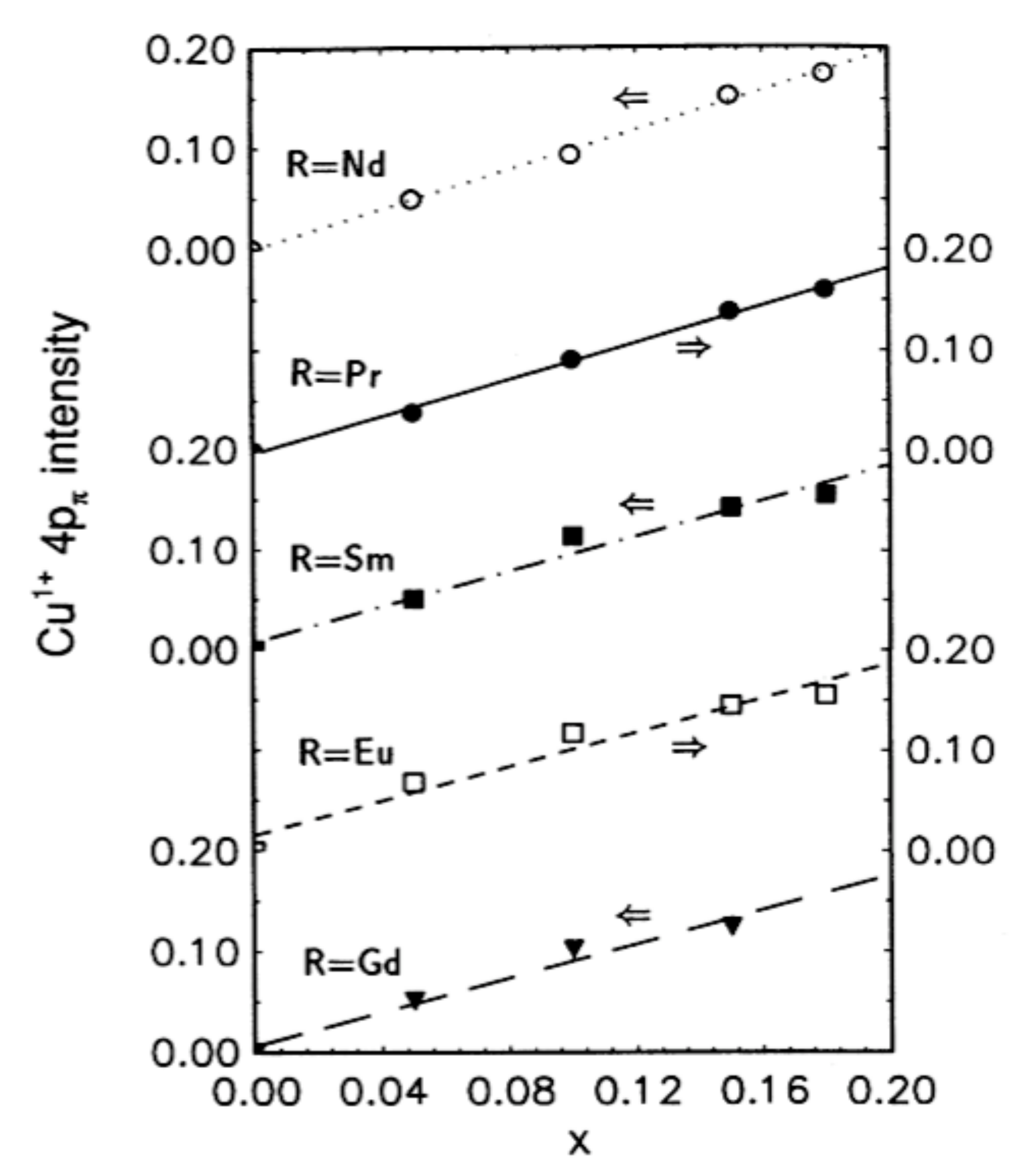}
\caption{Peak height of the Cu$^{+1}$ 4p$_\pi$ spectral feature in the Cu K-edge XAS spectra as a function of Ce concentration for various (RE)CCO compounds.   Its intensity is approximately proportional to the $3d^{10}$ occupation.  From \textcite{Liang94a}.} \label{LiangXAS}
\end{center}
\end{figure}

$Via$ Ce core level photoemission \textcite{Cummins93a}  demonstrated that Ce substitutes as Ce$^{+4}$  rather than  Ce$^{+3}$ across the full doping range showing that the effects of Ce subsitution is electron donation.  With EELS \textcite{Alexander91a} observed that Th doping into Nd$_{1.85}$Th$_{0.15}$CuO$_{4}$ caused a 14$ \% $ reduction in the relative intensity of the Cu $2p_{3/2}$ excitonic feature and only minor changes to the O $2p$ states, which is consistent with doping the Cu sites.   Similarly, \textcite{Liang94a} found that across the family of RE$_{2-x}$Th$_{x}$CuO$_{4}$  (RE = Pr, Nd, Sm, Edu, and Gd) that Cu 3$d^{10}$ features in the XAS Cu K edge spectra increases linearly with Ce doping as shown in Fig~\ref{LiangXAS}.  A similar picture was arrived at by \textcite{Pellegrin93a}.  In x-ray core level photoemission \textcite{Steeneken03a} observed that the $\underline{2p}3d^9$  ``satellites" decreased in intensity with increasing Ce content, while new structures like $\underline{2p}3d^{10}$ appear (where $\underline{2p}$ denotes a photoemission final state with a Cu core hole).  Additionally they found that the Cu L$_3$ x-ray absorption spectra intensity decreases with Ce doping.  As this absorption is dominated by the transition $3d^9 + h \nu \rightarrow \underline{2p} 3d^{10}$,   these results also imply that Ce doping results in a decrease of the Cu$^{+2}$ and increase in Cu$^{+1}$ content.  The nominal $3d^{10}$ configuration of an added electron has also been confirmed $via$ a number of resonant photoemission studies which show a Cu $3d$ character at the Fermi level \cite{Allen90a,Sakisaka90a}.  All these studies provide strong support for the expected increase in the mean $3d^{10}$ electron count with Ce doping\footnote{Similar studies on the hole-doped compounds in contrast give no evidence for $3d^{10}$ occupation and instead signatures of O 2p holes are found, which is consistent with the picture presented above in which doped holes reside primarily on the in-plane oxygen atoms.   See for instance, \textcite{Kuiper88a,Alp87a} and references therein.}.

As discussed above, in neutron scattering, \textcite{Mang04a} found that the $instantaneous$ AF correlation length of doped unreduced NCCO can be described by a quantum Monte Carlo calculations for the randomly site-diluted nearest-neighbor spin 1/2 square-lattice Heisenberg antiferromagnet.   Setting the number of non-magnetic sites to within $\Delta x \approx 0.02$ of the nominal Ce concentration gave quantitative agreement with their calculation.   This also shows that every Ce atom donates approximately one electron to the CuO$_2$ planes.

\textit{Does the enclosed volume of the FS reflect a metallic band that is greater than half-filled? -}   This is an equivalent question to the one immediately above if one agrees that there is a single metallic band that crosses E$_F$ which is formed out of Cu d$_{x^2- y^2}$ and O $2p_{x,y}$ states.   However, this specific issue can be addressed in a different, but very direct fashion from the area of the FS as measured by ARPES.  If one neglects the issue of 'hot-spots' and speaks only of the underlying FS the Luttinger sum rule appears to be approximately obeyed \cite{King93a}.  \textcite{Armitage01b} found that in NCCO the enclosed volume is 1.09 for a nominally $x=0.15$ NCCO sample.   Others have found FS volumes closer to that expected \cite{Park07a,Santander09a}, but in all cases the Luttinger sum rule is consistent with a band greater than half-filling (approximately $1+x$).   As hole-doped systems seem to have a Luttinger volume which closely reflects the number of doped holes $1-x$ \cite{Kordyuk02a}, in this sense also (RE)CCO systems can be regarded as electron-doped.

\textit{What is the nature of charge carriers from transport? -}   It was pointed out early on that there are both hole and electron contributions to transport \cite{Gollnik98a,Wang91a,Fournier97a}.  At low concentrations an electron contribution is naively expected within a model of electrons being doped into a semiconductor.  At higher dopings these observations were at odds with the shape of the experimentally determined FS, which \textcite{King93a} had found to be a large hole pocket centered at ($\pi$,$\pi$).   Later, it was found by \textcite{Armitage01b,Armitage02a} that at low dopings the FS was a small electron pocket around the ($\pi$,0) position.  At higher dopings there is a rearrangement of the electronic structure and a large Fermi surface is developed, which derives from electron- and hole-like sections of the FS and may retain remnant signatures of them.   Therefore the hole-like experimental signatures may result from electron doping itself.   These issues are discussed in more detail above in Sec.~\ref{TransportResHall} and~\ref{ARPES}.

\textit{Do doped electrons occupy electronic states analogous to those occupied by holes in the $p$-type compounds? -}    As discussed in a number of places in this review (see Section~\ref{sec:edoping}), although the local orbital character of doped electrons is undoubtedly different than doped holes within certain models, under certain circumstances one can map the hole and electron addition states to an effective Hubbard model with an approximate electron-hole symmetry.  Although mid-gap states are undoubtedly also created upon doping, it appears (Sec.~\ref{ChemPot}) that the chemical potential crosses the effective Hubbard gap (formally the CT gap) upon moving from hole to electron doping.  Additional evidence for the existence of an effective upper Hubbard band in NCO comes from \textcite{Alexander91a} who found the same prepeak in undoped (RE)CO and doped (RE)CCO O $1s \rightarrow 2p$ EELS absorption spectra as found in LCO.   To first approximation this absorption probes the local unoccupied O density of states.   Here however this prepeak is not interpreted as holes in a nominally filled $2p^6$ local configuration and instead has been interpreted as excitations into a Hubbard band of predominantly Cu $3d$ character with a small O $2p$ admixture as expected.  In this way Ce doping may be described as the addition of electrons to an effective upper Hubbard band, just as hole doping is the addition of holes to an effective lower Hubbard band.   In this sense also these systems may be considered as electron-doped.

\subsection{Electron-phonon interaction}\label{sec:ephonon}

There has been increasing discussion on the subject of strong electron-phonon coupling in the cuprate high temperature superconductors.  This has been inferred from both possible phonon signatures in the charge spectra \cite{Lanzara01a,JLee06a},  as well as directly in the doping induced softening of a number of high-frequency oxygen bond-stretching modes in many $p-$type cuprates as observed by neutron and x-ray scattering \cite{Pintschovius99a,McQueeney99a,McQueeney01a,Uchiyama04a,Fukuda05a,Pintschovius06a}.

In NCCO \textcite{Kang02a} did find changes with doping in the generalized phonon density of states around $\approx$ 70 meV by neutron scattering.  Although this is a similar energy scale as the softening is found on the hole=doped side, on general grounds one may expect a number of differences in the electron-phonon coupling between $p$- and $n$-type doping.  Since the purported soft phonon is the oxygen half-breathing mode, one might naively expect a weaker coupling for these modes with electron doping, as Madelung potential considerations\cite{Torrance89a,Ohta91a} indicate that doped electrons will preferentially sit on the Cu site, whereas doped holes have primarily oxygen character.  The biggest changes in the phonon density of states probed by \textcite{Kang02a} happen at similar doping levels in La$_{2-x}$Sr$_{x}$CuO$_4$ and Nd$_{2-x}$Ce$_{x}$CuO$_4$ ($x \approx 0.04$),  however the doping levels are at very different relative positions in the phase diagram, with $x = 0.04$ being still well into the antiferromagnetic and more insulating phase for the electron-doped compound.  As such modifications in the phonon spectrum may be associated generally with screening changes (and hence electron-phonon coupling) with the onset of metallicity, this demonstrates the possibility that the changes in the NCCO phonon spectrum, although superficially similar in the electron and hole-doped materials, may have some differences.

Irrespective of these expectations and differences, a number of similar signatures of phonon anomalies have been found in the electron-doped compounds.   As mentioned above, although initial measurements of the electron-phonon coupling in the ARPES spectra seemed to give little sign of the `kink' in the angle-resolved photoemission spectra \cite{Armitage03a}, which has been taken to be indicative of strong electron-phonon coupling on the hole-doped side of the phase diagram, more recent measurements show evidence for such a kink  \cite{Liu08b,Park08a,Schmitt08a}.  This work gives some evidence that electron-phonon interaction may not be so different on the two sides of the phase diagram.  

A number of features in the optical conductivity also have been assigned to polaron-like absorptions and Fano antiresonance features \cite{Calvani96a,Lupi98a,Lupi99a}.   For instance, pseudogap features and renormalizations in the infrared response have been interpreted \cite{Cappelluti08a} in terms of  lattice polaron formation within the Holstein$-t-J$ model in the context of DMFT.   \textcite{Cappelluti08a} point that the moderately large electron-phonon coupling of $\lambda \approx 0.7$ they extract is still not large enough to induce lattice polaron effects in the absence of exchange coupling.  This means that if lattice polaronic features exist in these compounds, they can be found only in the presence of (short-range) AF correlations.  The disappearance of pseudogap features near the termination of the AF phase is then consistent with this interpretation.

\begin{figure}[htbp]
\begin{center}
\includegraphics[width=8cm,angle=0]{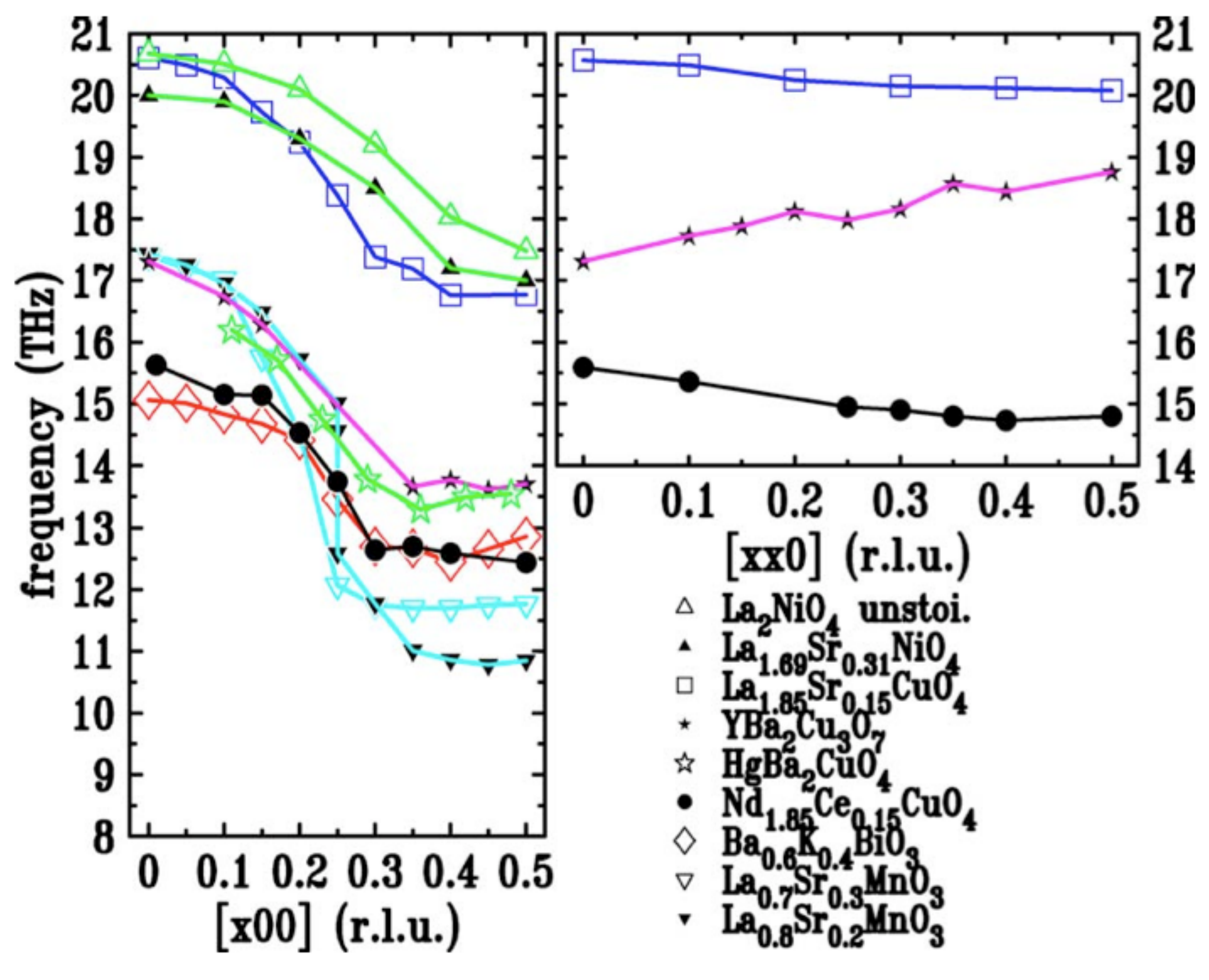}
\caption{(Color)  Comparison of the phonon anomaly in the bond-stretching branches observed $via$ neutron scattering in a number of metallic oxide perovskites.  (left) [100] direction. (right) [110] direction.  From \textcite{Braden05a}.}
\label{ephononcompare}
\end{center}
\end{figure}

 \textcite{dAstuto02a} measured NCCO's phonon dispersions $via$ inelastic x-ray scattering and assigned the softening in a similar 55 - 75 meV energy range to the same oxygen half-breathing mode in which softening is found in the $p$-type materials.  They claimed that the general softening of the phonon dispersion appears in a roughly similar way at 
a similar energy scale as in the $p-$types compounds, and that differences in the precise shape of
the anomaly in the phonon dispersion are due to an anti-crossing behavior of the bond-stretching modes with the lower energy O(2) mode in the [100] direction.    \textcite{Braden05a} later confirmed that with higher accuracy neutron scattering measurements (which present similar oxygen and heavy-ion dynamic structure factors) that all cuprates including NCCO are found to have roughly similar phonon anomalies along the [100] direction, showing a drop of $\approx 3$ THz (12.4 meV) as shown in Fig.~\ref{ephononcompare}.  The differences between compounds are larger along the [110] direction (Fig.~\ref{ephononcompare}), but still small overall. This indicates that all these systems (as well as many other perovskites (Fig.~\ref{ephononcompare}) have aspects of their electron-phonon coupling that have roughly similar character.

We should point out that very recent even higher resolution scattering experiments have shown that there are some specific phonon features in $p$-type compounds that are likely related to charge ordering instabilities (i.e. stripes) \cite{Reznik06a,dAstuto08a}.  These studies show that in the phonon dispersions one of the two normally degenerate components follows the smoother cosine-like dispersion while the other presents a much sharper dip, which has been interpreted as a Kohn-like anomaly at the stripe ordering wavevector.  Considering that there is little evidence for stripes in the electron-doped compounds, it would be very interesting to do such similar high resolution measurements on $n-$type materials.

\subsection{Inhomogeneous charge distributions} 
\label{Inhomo}

Doping a Mott insulator does not automatically result in a spatially homogenous state \cite{Lee03a,Emery93a}.  The magnetic energy loss that comes from breaking magnetic bonds by doping charge can be minimized by forming segregated charge rich regions.  This tendency towards phase separation can  lead to scenarios in which charges phase separate into various structures including charge puddles, stripes \cite{Machida89a,Zaanen89a,Emery93a,Carlson03a,Kivelson03a,Tranquada95a,Mook98a,Mook02a}, or checkerboard patterns \cite{Seo07a,Hanaguri04a}.  There is extensive evidence for such correlations in the hole-doped cuprates \cite{Kivelson03a}.

The situation is much less clear in the $n$-type compounds.  As some of the strongest evidence for `stripe' correlations in the hole-doped materials has come from the preponderance of incommensurate spin and charge correlations, the commensurate magnetic response has been taken as evidence for a lack of such forms of phase separation in these materials.   However,  \textcite{Yamada03a} has pointed out that the commensurate short range spin correlations detected by neutron scattering in the SC phase of the $n$-type cuprates can reflect an inhomogeneous distribution of doped electrons in the form of droplets/bubbles in the CuO$_2$ planes.  The commensurate magnetic signatures may also arise from `in-phase' stripe domains as contrasted to the `antiphase' domains of stripes in the $p$-type compounds \cite{Sun04a}.   This issue of stripe-like structures in the $n$-type materials has been investigated theoretically and numerically (See \textcite{Aichhorn05a,Sadori00a, Kusko02a} for instance.).   There is some evidence that numerical models which give stripes on the hole-doped side, give a homogeneous state on the electron-doped side \cite{Aichhorn06a}.  One can then consider the possibility of phase separation and inhomogeneity an open issue.

There has been a number of studies that have argued for an inhomogeneous state in the electron-doped curpates.  \textcite{Sun04a} found in Pr$_{1.3-x}$La$_{0.7}$Ce$_x$CuO$_4$ the same unusual transport features that have been argued to be evidence for stripe formation in LSCO \cite{Ando01a}.   They measured the $ab$-plane and c-axis thermal conductivity and found an anomalous damping of the c-axis phonons, which has been associated with scattering off of lattice distortions induced by stripes which are relatively well ordered in the plane, but disordered along the c axis.  In the AF state the $ab$-plane resistivity is consistent  with ``high mobility" metallic transport, consistent with motion along ``rivers of charge."  They interpret these peculiar transport features as evidence for stripe formation in the underdoped $n$-type cuprates.  In Pr$_{1.85}$Ce$_{0.15}$CuO$_{4}$  \textcite{Zamborszky04a} found signatures in the NMR spin-echo decay rate (1/T$_2$) for static inhomogeneous electronic structure.   Similarly, \textcite{Bakharev04a} found $via$ Cu NMR evidence for an inhomogeneous ``blob-phase" (bubble) in reoxygenated superconducting Nd$_{1.85}$Sr$_{0.15}$CuO$_4$.  Moreover,  \textcite{Granath04a} has shown that some unusual aspects of the doping evolution of the FS found by ARPES \cite{Armitage02a} could be explained by an inhomogeneous in-phase stripes or `bubble' phases.  `Bubble' phases, where the doped charge is confined to small zero-dimensional droplets instead of the one-dimensional  stripes, arise naturally instead of stripes in $t-J$ type models with long-range Coulomb repulsion in the limit of  $t<<J$, because of the lower magnetic energy \cite{Granath04a}.  They may be favored in the electron-doped materials, which have more robust antiferromagnetism than the hole-doped materials.   From their neutron scattering \textcite{Dai05a} argue that $x=0.12$ PLCCO is electronically phase separated and has a superconducting state, which coexists with both a 3D AF state and a 2D commensurate magnetic order that is consistent with in-phase stripes.  \textcite{Onose99a} found infrared and Raman Cu-O phonon modes that grew in intensity with decreasing temperature in unreduced crystals.  This was interpreted as being due to a charge ordering instability promoted by a small amount of apical oxygen.

In contrast to these measurements, \textcite{Williams05a} found no sign of the Cu NMR ``wipe out" effect  in x=0.15 PCCO which has been interpreted to be consistent with spatial inhomogeneity in La$_{2-x}$Sr$_x$CuO$_4$ \cite{Singer99a}.  Similarly, in the first spatially resolved STM measurements \textcite{Niestemski07a} showed that $T_c = 12 K$ PLCCO had a relatively narrow gap distribution of 6.5 - 7.0 meV (Fig.~\ref{PLCCOSTM}), with no signs of the gross inhomogeneity of some $p$-type compounds \cite{Lang02a,Howald01a}.  In neutron scattering \textcite{Motoyama06a} found that the field induced response at low temperature is momentum resolution limited, which implies that the dynamic magnetic correlations are long-range (>200 {\rm \AA}) with correlation lengths that span vortex-core and SC regions. This provides further evidence that NCCO forms a uniform state.   Circumstantial evidence for a homogeneous doped state also comes from other neutron measurements, where it has been found that the spin pseudogap closes with increasing temperature and field, in contrast to the hole-doped material where it is better described as ``filling in" \cite{Yamada03a,Motoyama07a}.  This `filling-in' behavior has been associated with phase separation and so its absence argues against such phenomena in the $n$-type cuprates.

\begin{figure}[htbp]
\begin{center}
\includegraphics[width= 7.5cm,angle=0]{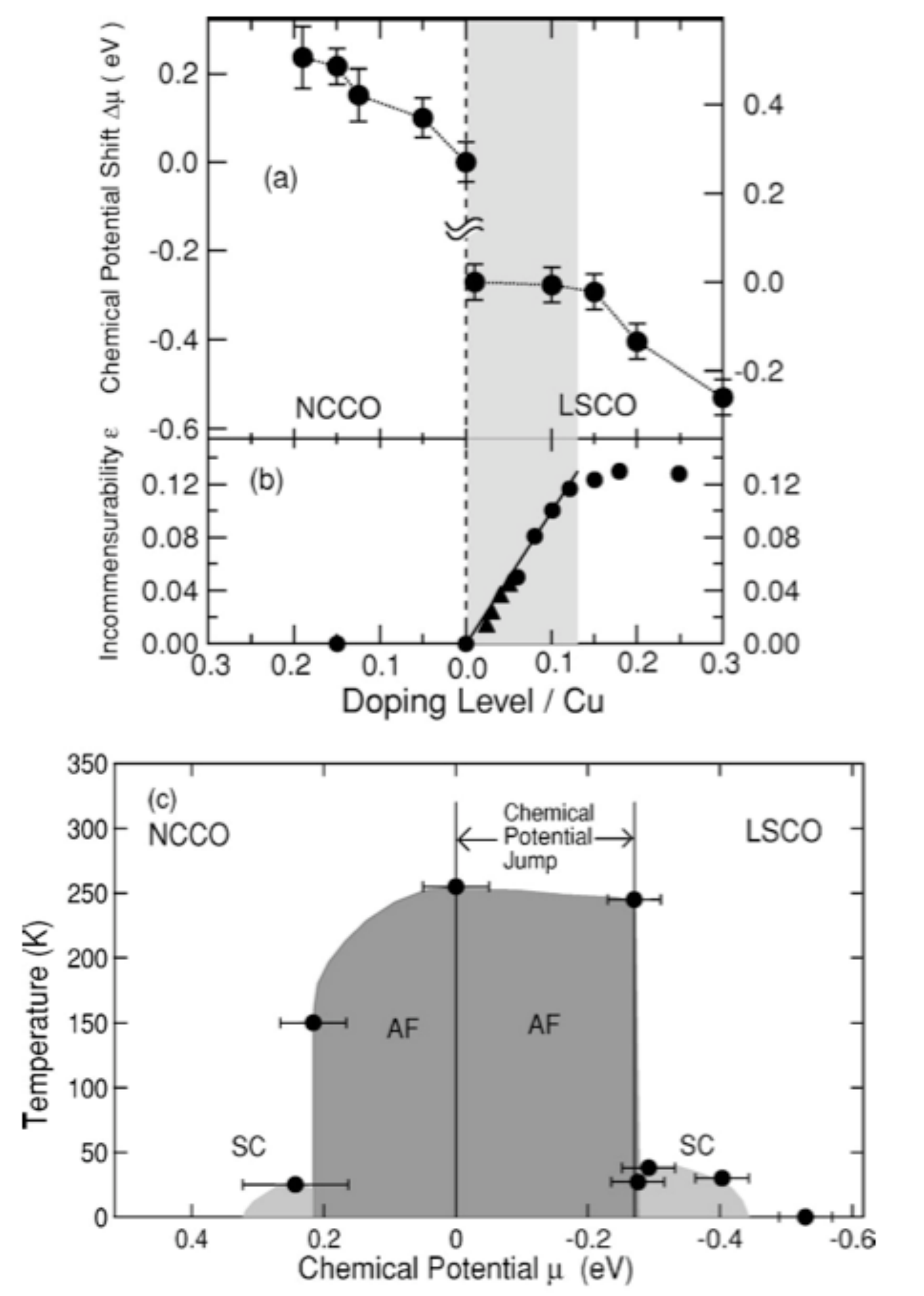}
\caption{(top) Chemical potential shift $\mu$ in NCCO and LSCO.   (middle) Incommensurability measured by inelastic neutron scattering experiments as given in Refs. \cite{Yamada98a} and \cite{Yamada03a}.  In the hatched region, the incommensurability varies linearly and $\Delta \mu$ is constant as functions of doping level.  (bottom) $\mu$-T phase diagram of NCCO and LSCO.  Note that there is an uncertainty in the absolute value of the chemical potential jump between NCO and LCO.   From \textcite{Harima01a}.} \label{HarimaChemPot}
\end{center}
\end{figure}

Finally there is the very interesting result of \textcite{Harima01a} (Fig.~\ref{HarimaChemPot})  who demonstrated that the chemical potential shifts very differently with hole and electron doping, which argues against phase separation in the $n$-type compounds.  \textcite{Harima01a} compared the chemical potential shifts in NCCO and LSCO $via$ measurements of core-level photoemission spectra.   Although the relative shift between LCO and NCO was uncertain in such measurements due to different crystal structures, they found that the chemical potential monotonously increases with electron doping, in contrast to the case of hole doping, where the shift is suppressed in the underdoped region (Fig.~\ref{HarimaChemPot} (top)).  The differences were ascribed to a tendency towards phase separation and mid-gap states in LSCO as compared to NCCO in this doping region  (Fig.~\ref{HarimaChemPot} (middle)).  We should note however that this suppression of the chemical potential shift in the hole-doped compounds does not seem to be universal as Bi2212 shows a much smaller suppression \cite{Harima03a}  and Na doped CCOC \cite{Yagi06a} apparently none at all.  Interestingly however, they found that the previously discussed electron-hole asymmetry of the NCCO/LSCO phase diagram with respect to the extent of antiferromagnetism and superconductivity is actually symmetric if plotted in terms of chemical potential  (Fig.~\ref{HarimaChemPot} (bottom)).  This is a fascinating result that deserves further investigation.

\subsection{Nature of normal state near optimal doping}  \label{nonFL}

A central subject of debate in the field of cuprate superconductivity is the nature of the `normal' state.  Is the metal above T$_c$ well described by Fermi liquid theory or are interactions such as to drive the system into a non-Fermi liquid state of some variety?   One of the problems with the resolution of this question experimentally is the ``unfortunate" intervention of superconductivity at relatively high temperatures and energy scales.  The matter of whether a material is or is not a Fermi liquid can only be resolved definitively at low energy scales as the criteria to have well-defined quasiparticles will always break down at sufficiently high temperatures or energies.  The advantage of trying to answer these question for the electron-doped cuprates as opposed to the $p$-type materials is that superconductivity can be suppressed by modest magnetic fields ($\approx 10$ T) allowing access to the low temperature behavior of the normal state.

\begin{figure}[htbp]
\begin{center}
\includegraphics[width= 7cm,angle=0]{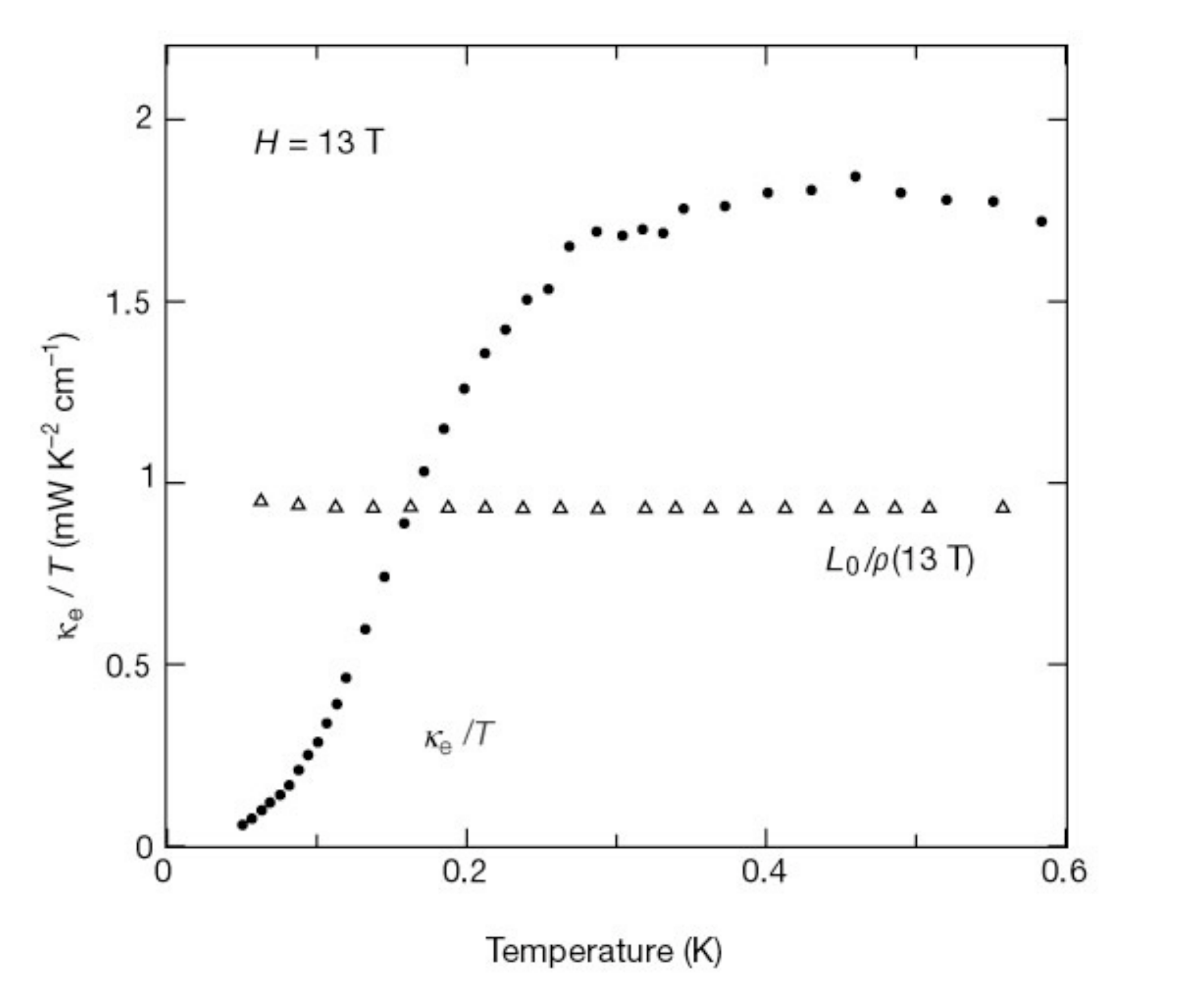}
\caption{A comparison of charge conductivity $\sigma(T) = 1/ \rho(T)$, plotted as $L_0/\rho(T)$ (triangles) (i.e. given by the Wiedemann-Franz expectation), and electronic contribution to the heat conductivity $\kappa_e$, plotted as $\kappa_e/T$ (circles), as a function of temperature in the normal state at H = 13 T.    A clear violation of the Wiedemann-Franz law is found \cite{Hill01a}.  The downturn below 300 mK is an artifact of thermal decoupling of the electronic and phononic degrees of freedom \cite{Smith05a}, but an approximately factor of two discrepancy remains in the magnitude of the thermal conductivity and the value inferred from the charge conductivity at low temperature.} \label{WFviolation}
\end{center}
\end{figure}

This issue has been discussed frequently in the context of the electron-doped cuprates due to the approximately quadratic dependence of the resistivity above T$_c$ \cite{Hidaka89a,Tsuei89a,fournier98a}.  For further discussion see Sec.~\ref{TransportResHall}.  The conventional wisdom is that this is evidence of a ``more Fermi liquid-like" normal state (T$^2$ being the nominal functional form for electron-electron scattering in a conventional metal at low temperature).  It is not.  While it is certainly true that the quadratic temperature dependence is very different than the remarkable linear dependence found in the hole-doped materials \cite{Gurvitch87a}, it is not likely evidence for a Fermi liquid state.  The temperature range over which T$^2$ is found (from T$_c$ to room temperature) is much larger than that ever expected for purely electron-electron scattering to manifest.  Within conventional transport theory, one will almost invariably have a phonon contribution that in certain limits will give a linear dependence to the resistivity and destroy the T$^2$ form except at the lowest temperatures.  Moreover, realistic treatments for electron-electron scattering give functional forms for various temperature ranges that depend on such factors as the Fermi surface geometry \cite{Hodges71a} and it is seldom that a pure T$^2$ functional form is observed even in conventional metals.  Whatever is causing the T$^2$ functional form almost certainly cannot be electron-electron scattering of a conventional variety and is therefore not evidence of a Fermi liquid ground state.  In a similar fashion the quadratic frequency dependence of the effective scattering rate that is found by \textcite{Wang06a} up to the high frequency scale of 6000 cm$^{-1}$ (0.74eV) in overdoped NCCO  is also unlikely to be evidence for a Fermi-liquid.

Recently however this issue of the Fermi liquid nature of the electron-doped cuprates has been put on more rigorous ground with sensitive measurements of the thermal conductivity of x=0.15 PCCO.  Taking advantange of the low critical magnetic fields of these compounds, \textcite{Hill01a} measured the thermal and electric conductivity of the normal state and discovered a clear violation of the ``Wiedemann-Franz law" (Fig.~\ref{WFviolation}).   The Wiedemann-Franz law is one of the defining experimental signatures of Fermi liquids and states that the ratio $\kappa / \sigma T$ where $\kappa$ is the thermal conductivity and $\sigma$ is the electrical conductivity should be universally close to Sommerfeld's value for the Lorenz ratio $L_0 = \pi^2  / 3 (k_B/e)^2 = 2.45 \times 10^{-8} W \Omega K^{-2} $.  This relation reflects the fact that at low temperature the particles that carry charge are the same as those that carry heat.  No known metal has been found to be in violation of it\footnote{Subsequent to the measurements described herein, violations of the Wiedemann-Franz law has been found near heavy-electron quantum critical points \cite{Tantar07a}.}.  \textcite{Hill01a} demonstrated that there was no correspondence between thermal and electrical conductivities in PCCO at low temperature.  For much of the temperature range, the heat conductivity was found to be greater than expected.  Because the Wiedemann-Franz law is a natural property of Fermi liquids, this violation had consequences for understanding the ground state and elementary exciations of these materials.  It implies that charge and heat are not carried by the same electronic excitations.  A similar violation of the WF law has now been reported in underdoped Bi$_{2+x}$Sr$_{2-x}$CuO$_{6-\delta}$ \cite{Proust05a}.  On the other hand, agreement with the WF law is found in some overdoped cuprates  \cite{Proust02a,Nakamae03a}.  In counter to these measurements \textcite{Jin03a} have pointed out that rare earth ordering and crystal field levels can affect heat as well as electrical transport.  They found large changes in the non-electronic portion of the thermal conductivity of NCO in a magnetic field, which may be responsible for the larger than expected heat current.

In NMR \textcite{Zheng03a} have measured a similar ratio that should also show universal behavior in a Fermi liquid.  They demonstrated that when the superconducting state was suppressed by magnetic field in x=0.11 PLCCO the spin relaxation rate obeyed the Fermi-liquid Korringa law $1/T_1 \propto T$ down to the lowest measured temperature (0.2K).  With the measured value for the Knight shift  $K_s$, it was found that the even stronger condition $T_1 T K_s^2$ = constant was obeyed below 55K albeit with a small  $T_1 T K_s^2$ value of $7.5 \times 10^{-8}$ sec K, which is 50 times less than the non-interacting value.  This points to the significance of strong correlations, but gives indication that the ground state revealed by application of a strong magnetic field is actually a Fermi liquid.

Clearly, this is a subject that deserves much more in-depth investigation.  It would be worthwhile to search for both Wiedemann-Franz and Korringa law violations over the larger phase diagram of electron-doped cuprates to see for what doping ranges - if any - violations exist.

\subsection{Spin-density wave description of the normal state}\label{sec:SDW}

As originally noticed by \textcite{Armitage01a}, electron-doped samples near  optimal doping present a FS, that while very close to that predicted by band structure calculations, have near-E$_F$ ARPES spectral weight that is strongly suppressed (pseudogapped) at the momentum space positions where the underlying Fermi surface (FS) contour crosses the AF Brillouin zone boundary.   This suggests the existence of a ($\pi$,$\pi$) scattering channel and a strong importance of this wavevector.

As discussed by \textcite{Armitage02b,Armitage01a}, one possible way to view the results  - at least qualitatively - for samples near optimal doping is as a manifestation of a $\sqrt{2}\times\sqrt{2}$ band reconstruction from a static (or slowly fluctuating) spin density wave (SDW) or similar order with characteristic wavevector ($\pi$,$\pi$).  This distortion or symmetry reduction is such that if the order is long-range and static the BZ decreases in volume by 1/2 and rotates by 45$^{\circ}$.   The AFBZ boundary becomes the new BZ boundary and gaps form at the BZ edge in the usual way.  Although an SDW is the natural choice based on the close proximity of the  antiferromagnetic phase, the data are consistent with any ordering of characteristic wave vector ($\pi$,$\pi$) such as DDW \cite{Chakravarty01a}.

\begin{figure}[htb]
\begin{center}
\includegraphics[width= 8cm,angle=0]{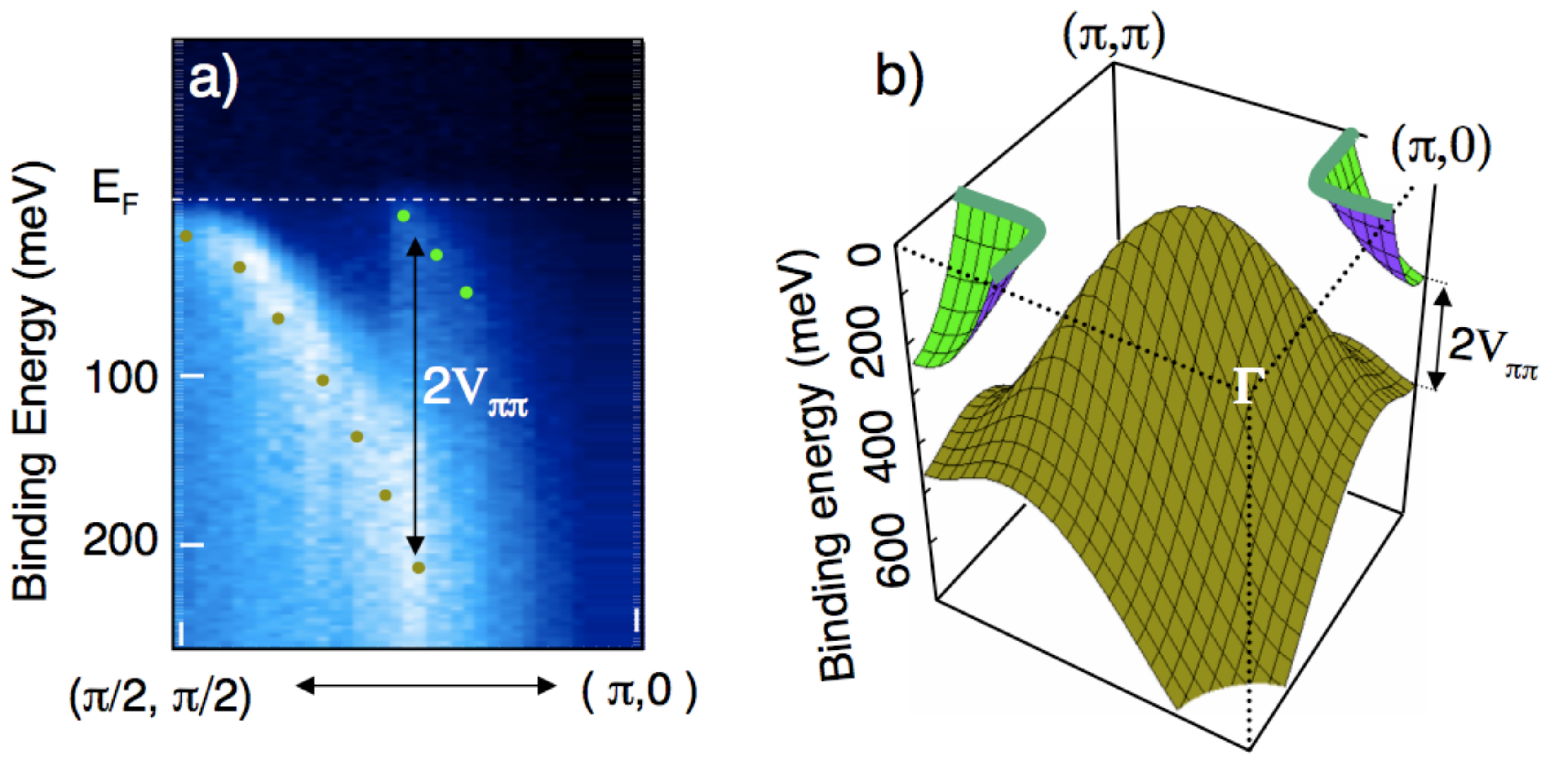}
\caption{(a) Measured ARPES spectral function along the AFBZ as given by the arrow in (b).   The SDW gap 2$V_{\pi,\pi}$  is readily visible in the raw spectra.   (b) Schematic of the band structure from a $\sqrt{2} \times \sqrt{2}$ reconstruction.   Adapted from \textcite{Park07a}.  } \label{FoldedDispersion}
\end{center}
\end{figure}

The $\sqrt{2}\times\sqrt{2}$ reconstructed band structure can be obtained $via$ simple degenerate perturbation theory \textcite{Armitage02b,Matsui05b,Park07a}.  This treatment gives

\begin{eqnarray}
E_k=E_0+4t'(\cos{k_x}\cos{k_y})+2t''(\cos{2k_x}+\cos{2k_y}) \nonumber \\
\pm\sqrt{4t^2(\cos{k_x}+\cos{k_y})^2+|V_{\pi\pi}|^2}
\label{FoldingEq}
\end{eqnarray}

where $V_{\pi\pi}$ is the strength of the effective ($\pi$,$\pi$) scattering, and $t$, $t'$ and $t''$ are nearest, next-nearest, and next-next-nearest hopping amplitudes.   The even and odd solutions correspond to new band sheets that appear due to the additional Bragg scattering potential. With realistic hopping parameters for the cuprates (as discussed in Sec.~\ref{sec:edoping}) a small hole pocket centered around ($\pi/2$,$\pi/2$)  and a small electron pocket around ($\pi$,0) appears at optimal doping as shown in Fig.~\ref{FoldedDispersion}b.   All measured $n$-type cuprates near optimal doping show a FS phenomenology roughly consistent with this band structure \cite{Armitage01b,Armitage02b,Zimmers05a,Matsui07a}\footnote{Small differences between material classes do exist (Fig.~\ref{ReARPESSeries}).   See the discussion in Sec.~\ref{ARPES}.}.  The 2$V_{\pi,\pi}$ splitting between the two band sheets in Fig~\ref{FoldedDispersion}b can be measured directly in a measurement of the ARPES spectral function along the AFBZ boundary as shown for SCCO in Fig.~\ref{FoldedDispersion}a.  Note that within this picture one expects differences with the $p$-type compounds as due to their smaller Luttinger volume the underlying `bare' FS comes closer to the ($\pi$,0) position.

This derivation is for a potential with long-range order, which according the \textcite{Motoyama07a} does not exist above $x \approx 0.134$.  Due to the ambiguity associated with the exact position of the phase boundary, possibly more relevant to the typical experimental case may be a situation where true long range order of the $\sqrt{2}\times\sqrt{2}$ phase does not exist, but where the material still has strong (but slow) fluctuations of this order. In this case more complicated treatments are necessary for quantitative treatments. An analysis in the spirit of the above is then much harder, but as long as the fluctuations are slow, then some aspects of the above zone folding picture should remain.  For instance depending on their particular time scales, some experiments may be sensitive to the formation of an electron pocket around ($\pi$,0).

\begin{figure}[htb]
\includegraphics[width=6.5cm,angle=0]{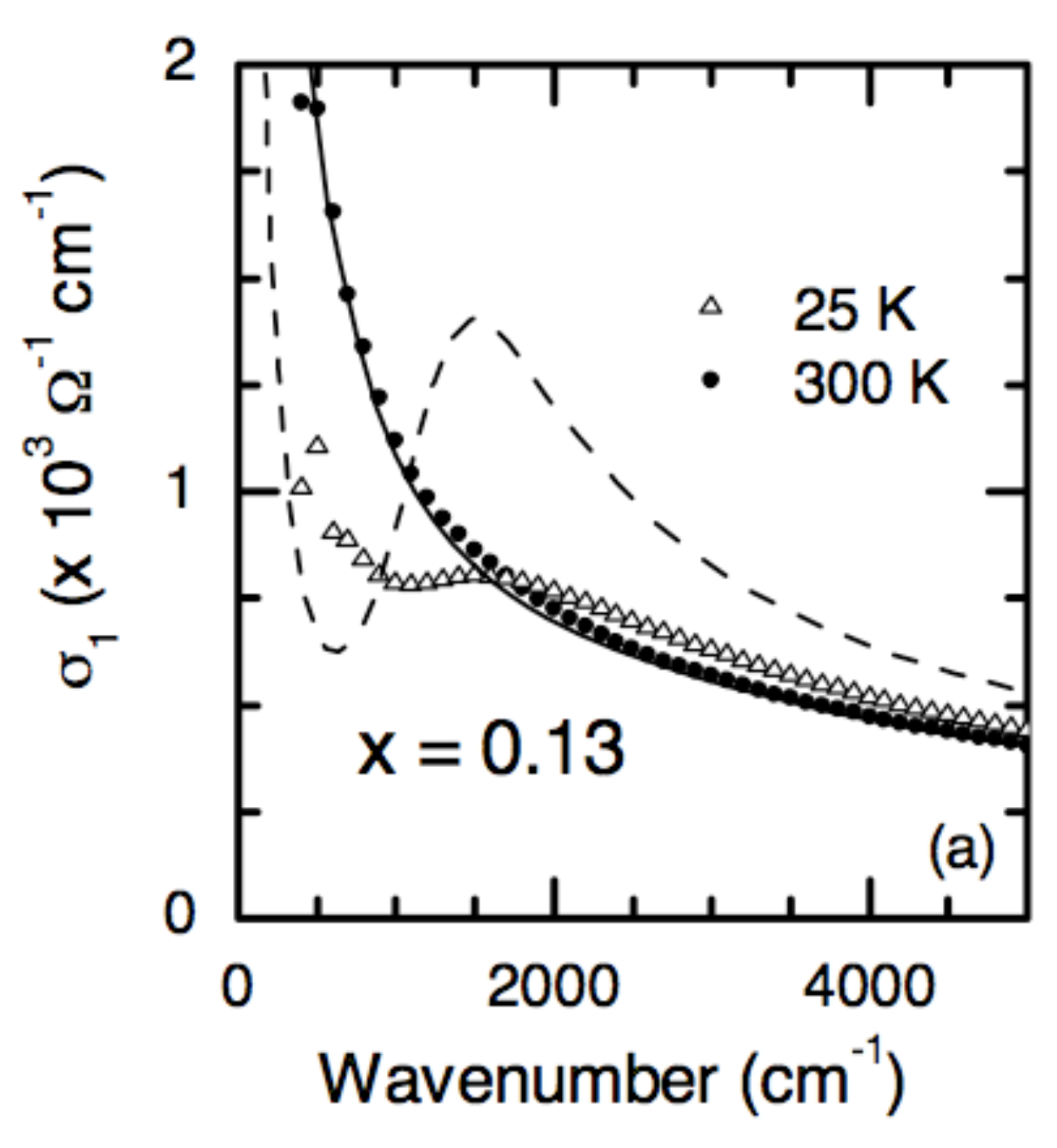}
\caption{Calculation of the optical conductivity based on the SDW band structure in Fig.~\ref{FoldedDispersion}.  Spectra were calculated for a x=0.13 doping with a value of 2$V_{\pi\pi}$ = 0.25 eV and a gap opening temperature of 170K.  The symbols are the measured optical conductivity for x=0.13 and the lines the spin density wave model calculation.  Compare also to the $x=0.10$ data in Fig.~\ref{OnoseOpticsPRB}. From \textcite{Zimmers05a}.}
\label{SDWOptics}\end{figure}

An interpretation based on such a zone folding scheme enables one to understand - at least qualitatively -  issues such as the sign change in the Hall coefficient \cite{Dagan04a,Dagan07a,Gollnik98a,Wang91a,Fournier97a}.   It had been a long standing mystery how a simply connected hole-like FS centered around ($\pi$,$\pi$) (originally thought to be the case from the first ARPES experiments of \textcite{King93a,Anderson93a}) could give both positive and negative contribution to the Hall coefficient and thermopower.   A mean field calculation of the Hall conductance based on the band structure in Eq.~\ref{FoldingEq} shows that the data are qualitatively consistent with the reconstruction of the Fermi surface expected upon density wave ordering \cite{Lin05a}, although the calculation has difficulty reproducing the $R_H$ values precisely\footnote{In these calculations long range order has been assumed up to $x=0.16$.   It is difficult to reconcile the reasonable agreement of the data at optimal doping and the mean-field model with the termination of the AF phase at $x \approx 0.134$ as inferred by  \textcite{Motoyama07a}.   More theoretical work and the explicit calculation of transport coefficients for systems with short range order and AF fluctuations are needed.  It is possible however the magnetic field used to suppress superconductivity stabilizes the magnetic state.} \footnote{More recent calculations by \textcite{Jenkins09b} using a band structure that takes into account the very anisotropic Fermi velocities observed experimentally in ARPES results has even worse quantitative agreement with the experimental R$_H$. However, they claim that one can describe the spectra with the inclusion of vertex corrections within the FLEX approximation \cite{Kotani08a}.}.

\textcite{Zimmers05a} showed that the notable pseudogaps in the optical conductivity as well as its overall shape can be reasonably modeled by a calculation based on the band structure in Eq.~\ref{FoldingEq} and Fig.~\ref{FoldedDispersion}. As seen in Fig.~\ref{SDWOptics} the overall temperature and frequency dependence matches well to the experimental data seen in the x=0.10 curves in Fig.~\ref{OnoseOpticsPRB} for instance.

Although it works best for samples near optimal doping, the SDW mean-field picture can also be used to understand the doping dependence of the FS for a limited range near optimal doping.  As materials are progressively underdoped and the antiferromagnetic phase is approached, antiferromagnetic correlations become stronger and the ``hot spot'' regions spread so that the zone-diagonal spectral weight is gapped by the approximate  ($\pi, \pi$)  nesting of the ($\pi/2, \pi/2$) section of FS with the ($-\pi/2,-\pi/2$) section of FS.  However a scheme based on nesting obviously breaks down as one approaches the Mott state, where the zone diagonal spectral weight is not only gapped, but also vanishes.  Experimentally, the near-E$_F$ spectral weight near ($\pi/2$,$\pi/2$) becomes progressively gapped  with underdoping and by $x=0.04$ in NCCO only an electron FS pocket exists around the ($\pi$,0) point (Fig.~\ref{PeterARPES}) \cite{Armitage02a}.  On the overdoped side, \textcite{Matsui07a} have extrapolated that that this hot spot effect would largely disappear in the ARPES spectra by $x=0.20$ as expected by the virtual disappearance of $V_{\pi,\pi}$ at that doping.  That the FS is no longer reconstructed for overdoped samples, was also inferred in the quantum oscillations experiments of \textcite{Helm09a}.

In an infrared Hall effect study \textcite{Zimmers07b} found
that at the lowest temperatures their data for $x=0.12, 0.15, 0.18$ was $qualitatively$
consistent with the simple SDW model.  However, \textcite{Jenkins09b} demonstrated strong $quantitative$ discrepancies  for underdoped materials of such a model with far infrared Hall measurements when using as input parameters the experimentally measured band structure from ARPES.  Additionally,  ARPES, IR and IR Hall measurements as high as 300K do not suggest a simple closing of the SDW gap (and hence formation of an unreconstructed Fermi surface around ($\pi$, $\pi$)) above T$_N$ \cite{Wang06a,Jenkins09b,Zimmers07b,Onose04a,Matsui05a}. For instance, there is a strong temperature dependence of the electron contribution to the Hall angle through the whole range up to and even above T$_N$.  \textcite{Jenkins09b} ascribed this to the role played by AF fluctuations.  It seems clear that experiments like optics, which measure at finite frequency, cannot necessarily distinguish between fluctuations and true long-range order in a model independent way.  For instance, despite the success of the extended Drude model in describing  $\sigma_{xx}$ of overdoped Pr$_{2-x}$Ce$_{x}$CuO$_4$, \textcite{Zimmers07b} found strong deviations in its description of $\sigma_{xy}$ in an x=0.18 sample showing that fluctuation effects are playing a role even at this doping.   Later work of this group using lower energy far infrared Hall data concludes that these deviations can in general be described by a model that incorporates vertex corrections due to antiferromagnetic fluctuations within the FLEX approximation \cite{Jenkins09a,Kotani08a}.   These observations show the obvious limits of a simple mean-field picture to understand all aspects of the data.  As discussed elsewhere, the observation of AF-like spectral gap in parts of the phase diagram, which don't exhibit long-range AF might be understandable within models that propose that a PG evinces in the charge spectra when the AF correlation length exceeds the thermal de Broglie wavelength \cite{Kyung04a}.

\subsection{Extent of antiferromagnetism and existence of a quantum critical point}
\label{QCPoint}

While it has long been known that antiferromagnetism extends to much higher doping level in the $n$-type as compared to the $p$-type compounds, reports differ on what doping level the AF phase actually terminates and whether it coexists or not with superconductivity  \cite{Kang05a,Fujita03a,Motoyama07a}.  There are at least two important questions here:   Do the intrinsic regimes of superconductivity and AF coincide?  And does the AF regime at higher dopings terminate in a second order transition and a T=0 QCP that manifests itself in the `scaling forms' of response functions and in physical observables like transport and susceptibility?  These are issues of utmost importance with regards to data interpretation in both $n-$ and $p$-type compounds.  Their resolution impinges on issues of the impact of quantum criticality \cite{Sachdev03a}, coupling of electrons to antiferromagnetism \cite{Carbotte99a,Abanov01a,Schachinger03a,Maier08a,Kyung09a}, and $SO(5)$ symmetry \cite{Zhang97a,Chen04a}  - yet a complete understanding requires weighing the competing claims of different neutron scattering groups, the information provided by $\mu$SR, as well as materials growth and oxygen reduction issues.

It has long been known that samples at superconducting stochiometries show a substantial AF magnetic response, as in for instance the existence of commensurate Bragg peaks \cite{Yamada99a,Yamada03a}.   Whether this is because phases truly coexist, or because samples are (chemically or intrinsically) spatially inhomogenous has been unclear\footnote{In this regard see also Sec.~\ref{Inhomo} that addresses the question of intrinsic charge inhomogeneity}.  Recently, \textcite{Motoyama07a} have concluded that they can distinguish these scenarios $via$ inelastic scattering by following the spin stiffness $\rho_s$ that sets the instantaneous correlation length.  They find it falls to zero at a doping level of $x \approx 0.134$ (Fig.~\ref{GrevenStiffness}a) in NCCO at the onset of superconductivity\footnote{Note that the definitions for the spin stiffness of  \textcite{Fujita08a}  and \textcite{Motoyama07a} differ, which may account for their differences of where  $\rho_s$  extrapolates to zero.  \textcite{Motoyama07a} derived it from the $T$ dependence of the linewidth of the \textit{instantaneous} spin correlations over a wide range of temperatures, while \textcite{Fujita08a} get it from the $\omega$ dependence of the peak width at a particular $T$.} and hence there is no intrinsic AF/SC coexistence regime. They found that the instantaneous spin-spin correlation length at low temperature remains at some small, but non-neglible value well into the superconducting regime showing the AF correlations are finite but not long range ordered in the superconductor (Fig.~\ref{GrevenStiffness}a).  As other inelastic neutron scattering experiments have clearly shown the presence of a superconducting magnetic gap \cite{Yamada03a}, (despite the presence of Bragg peaks in the elastic response) \textcite{Motoyama07a} concluded that the actual antiferromagnetic phase boundary terminates at $x \approx 0.134$, and that magnetic Bragg peaks observed at higher Ce concentrations
originate from rare portions of the sample which were insufficiently oxygen reduced (Fig.~\ref{GrevenStiffness}b)\footnote{In a related, but ultimately different interpretation, \textcite{Yamada03a} interpreted their narrow coexistence regime as evidence that the AF/SC phase boundary is first order and therefore these systems should lack a QCP and the associated critical fluctuations}.  This group had previously shown that the inner core of large TSFZ  annealed crystals have a different oxygen concentration than the outer shell \cite{Mang04a}.  They speculate that the antiferromagnetism of an ideally reduced NCCO sample would terminate in a 1st order transition [possibly rendered 2nd order by quenched randomness \cite{Hui89a,Imry79a,Aizenman90a}], which would give no intrinsic QCP.

A similar inference about the termination of AF state near the superconducting phase boundary can be reached from the neutron and $\mu$SR PLCCO data of \textcite{Fujita08a,Fujita03a}.  \textcite{Fujita08a} found only a narrow coexistence regime near the SC phase boundary ($\Delta x \approx 0.01$ near $x \approx 0.1$) which could also be a consequence of rare slightly less reduced regions.  They also find a dramatic decrease in AF signatures near this doping level.  However,  \textcite{Li08a} caution that since both the superconducting coherence length and spin-spin correlation length are both strongly affected by the oxygen annealing process, this issue of the true extent of AF and its coexistence with SC in the $n$-type cuprates may not be completely solved and there may be some oxygen reduction conditions where superconductivity and antiferromagnetism can genuinely coexist.  It is undoubtedly true that the annealing conditions depend on Ce concentration and in this regard it may be challenging to settle the question definitively about whether or not AF and superconductivity intrinsically coexist in any regions of phase space.  In support of a scenario of an AF QCP somewhere nearby in PLCCO, \textcite{Wilson06c} showed that at higher temperatures and frequencies, the dynamical spin susceptibility $\chi(\omega,T)$ of an $x=0.12$ sample can be scaled as a function of $\frac{ \omega}{T}$ at AF ordering wavevectors.  The low energy cut-off of the scaling regime is connected to the onset of AF order, which comes down as the antiferromagnetic phase is suppressed by oxygen reduction.  

In seeming contrast to these magnetic measurements that give evidence of no coexistence regime, based on their transport data \textcite{Dagan04a} claimed that an AF quantum phase transition (QPT) exists at dopings near optimal in PCCO. Their evidence for a quantum critical point (QCP) at $x \approx 0.165$ was: 1) the kink in $R_H$ at 350mK (see Fig~\ref{Dagan04HallB}), 2) the doping dependence of the resistivity's temperature dependent exponent $\beta$ in the temperature range 0.35 - 20K, 3)  the reduced temperature region near x=0.165  over which a T$^2$ dependence is observed, and 4) the disappearance of the low T resistivity upturn.  More recent very high-field (up to 60T) Hall effect and resistivity results support this scenario \cite{Li07b}.

\begin{figure}[t!]
\includegraphics[width=7cm,angle=0]{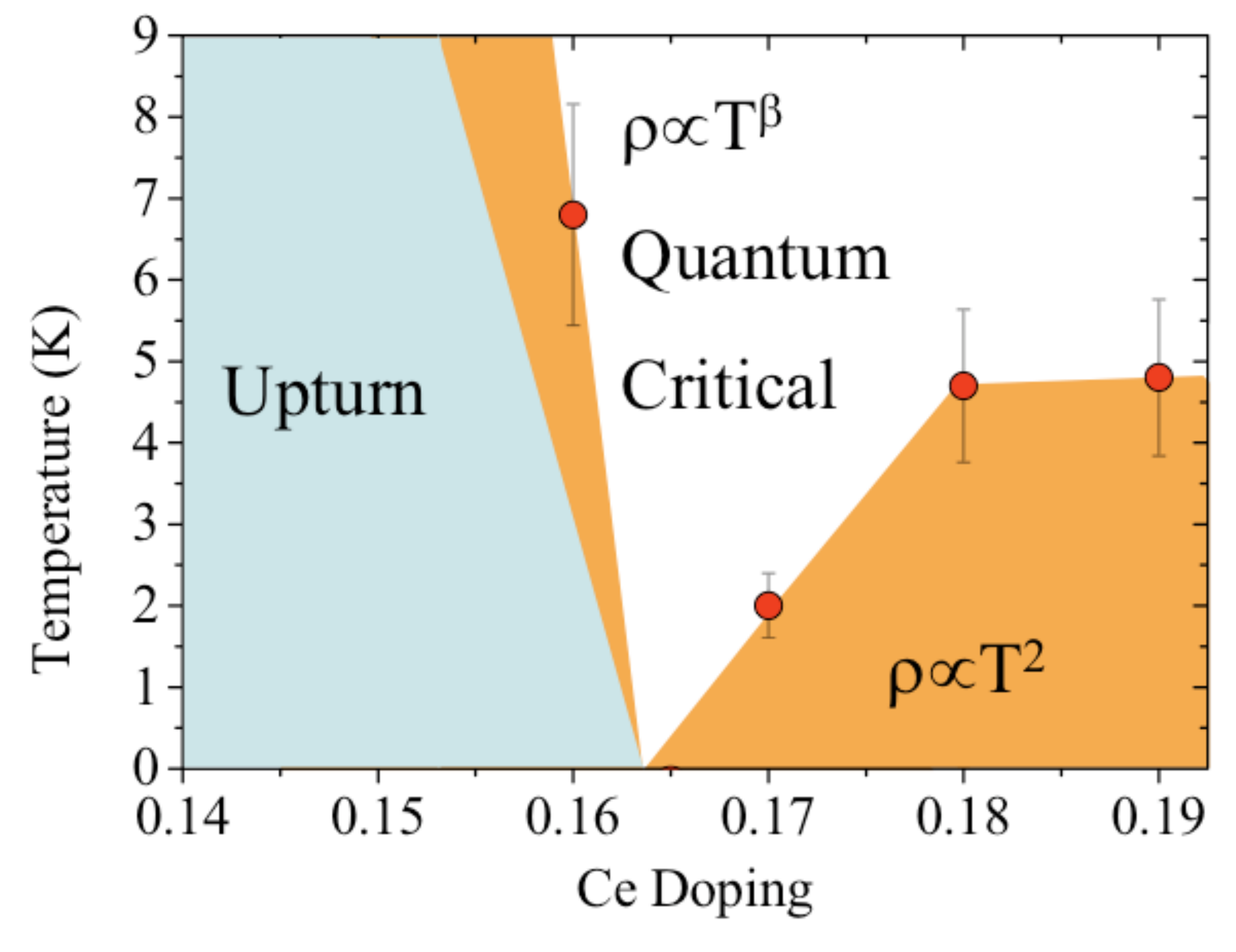}
\caption{(top)  Schematic illustration of the phase diagram of Pr$_{2-x}$Ce$_x$CuO$_4$ from resistivity measurements in magnetic field high enough to suppress superconductivity. Plotted as red dots is T$_0$, the temperature below which the T$^2$ behavior is observed (in orange).   At dopings lower than that of the nominal QCP the resistivity shows a low temperature upturn \cite{Dagan04b}.}
\label{Dagan04C}\end{figure}

The `funnel-like' dependence of the threshold  T$_0$ below which T$^2$ resistivity is observed shown in Fig~\ref{Dagan04C} (top), is precisely the behavior expected near a quantum phase transition \cite{Dagan04a}. This is particularly striking in the $n$-type cuprates where the resistivity has a $T^2$ dependence for all dopings at temperatures above $\approx$30K (or the resistivity minimum).  The phase diagram looks qualitatively similar to quantum phase transition diagrams found in the heavy fermions (see for instance \textcite{Custers03a}).  One may also take as evidence the striking linear in T resistivity found from 35mK to 10 K for x=0.17 PCCO \cite{Fournier98b} as evidence for a QCP near this doping.  A recent study of the doping dependence of the low-temperature normal state thermoelectric power has also been interpreted as evidence for a quantum phase transition (QPT) that occurs near x=0.16 doping \cite{PLi07a}.  And as discussed elsewhere, a number of other experiments such as optical conductivity \cite{Onose04a,Zimmers05a}, ARPES \cite{Matsui07a} and angular magnetoresistance \cite{Yu07a} experiments have also suggested that there is a phase transition at a higher doping.

Clearly, the inference of a QCP in PCCO and NCCO at $x \approx 0.165$ dopings is in disagreement with the conclusion of \textcite{Motoyama07a} who have found that the AF phase terminates at $x \approx 0.134$, before the occurrence of SC.  There are a number of possible different explanations for this.

It may be that the QCP of \textcite{Dagan04a} and others is due not to the disappearance of the magnetic phase $per$ $se$, but instead due to the occurrence of something like a Fermi surface topological transition.   For instance, it could be associated with the emergence of the full Fermi surface around the ($\pi,\pi$) position from the pockets around ($\pi,0$) in a manner unrelated to the loss of the AF phase.   Such behavior, is consistent with the kink-like behavior in the Hall coefficient (Fig~\ref{Dagan04HallB}).  It is also consistent with recent magnetic quantum oscillation experiments, which show a drastic change in FS topology between $x=0.16$ and $x=0.17$ doping \cite{Helm09a}\footnote{One interesting aspect of these quantum oscillation experiments is the lack of evidence for a small electron pocket on the low doping side, as one would expect that the band structure in Eq. \ref{FoldingEq} would give both electron and hole contributions.  \textcite{Eun09a} has shown that the contribution of the electron pocket is extremely sensitive to disorder.}.   Such a transition could occur just as a result of the natural evolution of the FS with doping, or it may be that this 2nd transition signifies the termination of an additional order parameter hidden within the superconducting dome\cite{Alff03a}, such as for instance DDW \cite{Chakravarty01a} or other orbital current states \cite{Varma99a}.  However a transition involving only charge degrees of freedom is superficially at odds with experiments that show a relationship of this transition to magnetism such as the sharp change in angular magnetoresistance at $x \approx 0.165$ \cite{Dagan05a,Yu07a}.  

\begin{figure}[t!]
\includegraphics[width=7cm,angle=0]{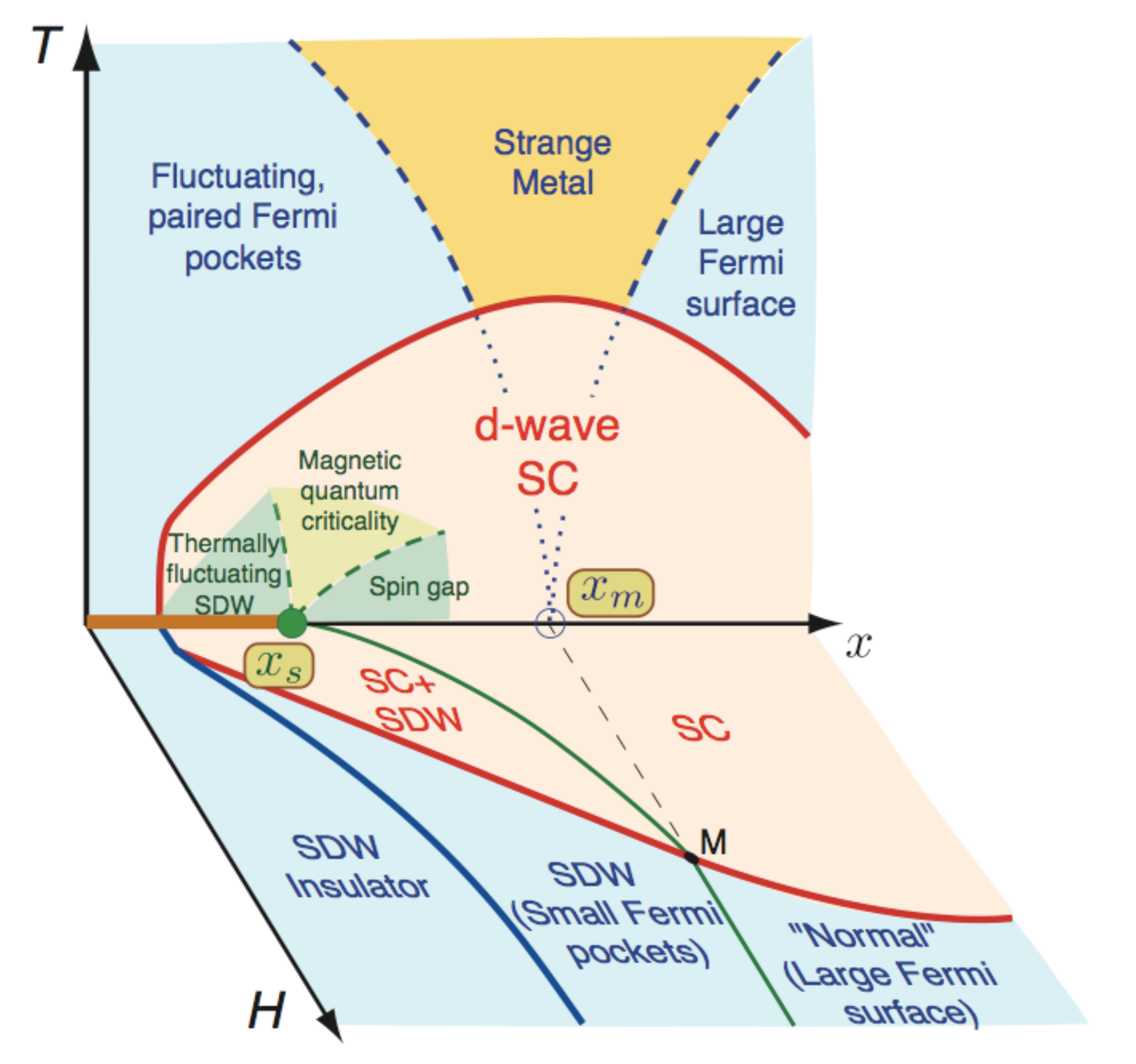}
\caption{Proposed phase diagram for field stabilized magnetism for the cuprates.  The phase diagram includes a transition from a large FS to a small FS at fields above H$_{c2}$.  The transition happens at a doping $x_m$ which is greater than the doping where the SDW is suppressed at zero field.   From \cite{Sachdev09a}.}
\label{Sachdev}\end{figure}

An alternative, but to our minds very natural scenario is that it is the magnetic field used to suppress superconductivity to reveal the low temperature normal state that stabilizes the SDW state.   Such a situation is believed to be the case in the hole-doped materials \cite{Moon09a,Demler01a,Lake01a,Lake02a,Khaykovich02a}.   The situation in the electron-doped materials is inconclusive (see the discussion in Sec.~\ref{MagFieldDep}), but it has been argued elsewhere that magnetic field enhances the magnetic ordered state in a somewhat similar fashion \cite{Kang03b,Matsuura03a,Kang05a}.   Recent calculations by \textcite{Moon09a,Sachdev09a} give a phase diagram (Fig.~\ref{Sachdev}) that is consistent with a zero-field SDW transition at $x \approx 0.134$ and a transition to a large FS at $x \approx 0.165$ at dopings above H$_{c2}$.  Such a scenario naturally explains the doping dependence of quantum oscillation \cite{Helm09a} and Hall effect measurements \cite{Dagan04a}.

However, a scenario of a field induced SDW does not simply explain measurements like ARPES and optics, which have inferred the existence of a QCP by the extrapolated doping level where a magnetic pseugogap closes.  However, as mentioned above it is likely that such measurements may be primarily sensitive to the development of short range order or fluctuations.  As pointed out by \textcite{Onose04a,Wang06a} optical data clearly show the existence of a large pseudogap in underdoped samples at temperatures well above the N\'eel temperature.  \cite{Jenkins09a,Zimmers07b} showed that gap-like features still appear in infrared Hall angle measurements even above those of the nominal QCP.  As this implies that only short-range order is necessary for the existence of  relatively well defined magnetic pseudogap in the charge spectra, it calls into question the utility of inferring the critical concentration of a magnetic QCP from such experiments.  Theoretically, in Hubbard model calculations  \textcite{Kyung04a} have shown that a pseudogap can develop in the photoemission spectra when the AF correlation length exceeds the thermal de Broglie wavelength (see Sec.~\ref{PGsection} below) i.e. long range order in the ground state is not necessary to develop a PG.   However, it seems difficult to imagine however that calculations which only incorporate short range order and fluctuations can reproduce the sharp anomalies in DC transport (Hall effect etc.) found near $x=0.165$.    For this, it seems likely that some sort of long-range order must be involved.  These effects are probably due to AF order stabilized by magnetic field as discussed above.  More work on this issue is clearly needed;  we are not aware of any measurements that show QPT-like anomalies in transport near $x = 0.13$ and so even more detailed studies should be done.

\subsection{Existence of a pseudogap in the electron-doped cuprates?}
\label{PGsection}

The pseudogap of the $p$-type cuprates is one of the most enigmatic aspects of the high-T$_c$ problem.  Below a temperature scale T$^*$,  underdoped cuprates are dominated by a suppression in various low-energy excitation spectra \cite{Timusk99a,Randeria97a,Norman05a}.  It has been a matter of much long standing debate whether this pseudogap is a manifestation of precursor superconductivity at temperatures well above T$_c$, or rather is indicative of some competing ordered phase.

An answer to the question of whether or not `The Pseudogap' in the $n$-type cuprates exists is difficult to address conclusively because of a large ambiguity in its definition in the $p$-type materials.   Moreover, even in the $p$-type compounds the precise boundary depends on the material system and the experimental probe.  Additionally, there has frequently been the distinction made between a `high-energy' PG, which is associated with physics on the scale of the magnetic exchange $J$ and a `low-energy' PG, which is of the same order of the superconducting gap.  What is clear is that there are undoubtedly a number of competing effects in underdoped cuprates.  These have all frequently been confusingly conflated under the rubric of pseudogap phenomena.  Here we concentrate on a number of manifestations of the phenomenology which can be directly compared to the $p$-type side.   A number of similarities and differences are found.  

At the outset of our discussion, it is interesting to point out that much of the pseudogap phenomena in the electron-doped cuprates seems to be related to antiferromagnetism and this phase's relative robustness in these materials.  The issue of whether the PG exists is of course then intimately related to the issues presented above in Secs.~\ref{QCPoint} and~\ref{sec:SDW} on the extent of antiferromagnetism and the SDW description of the normal state.

\begin{figure}[htb]
\includegraphics[width=6cm,angle=0]{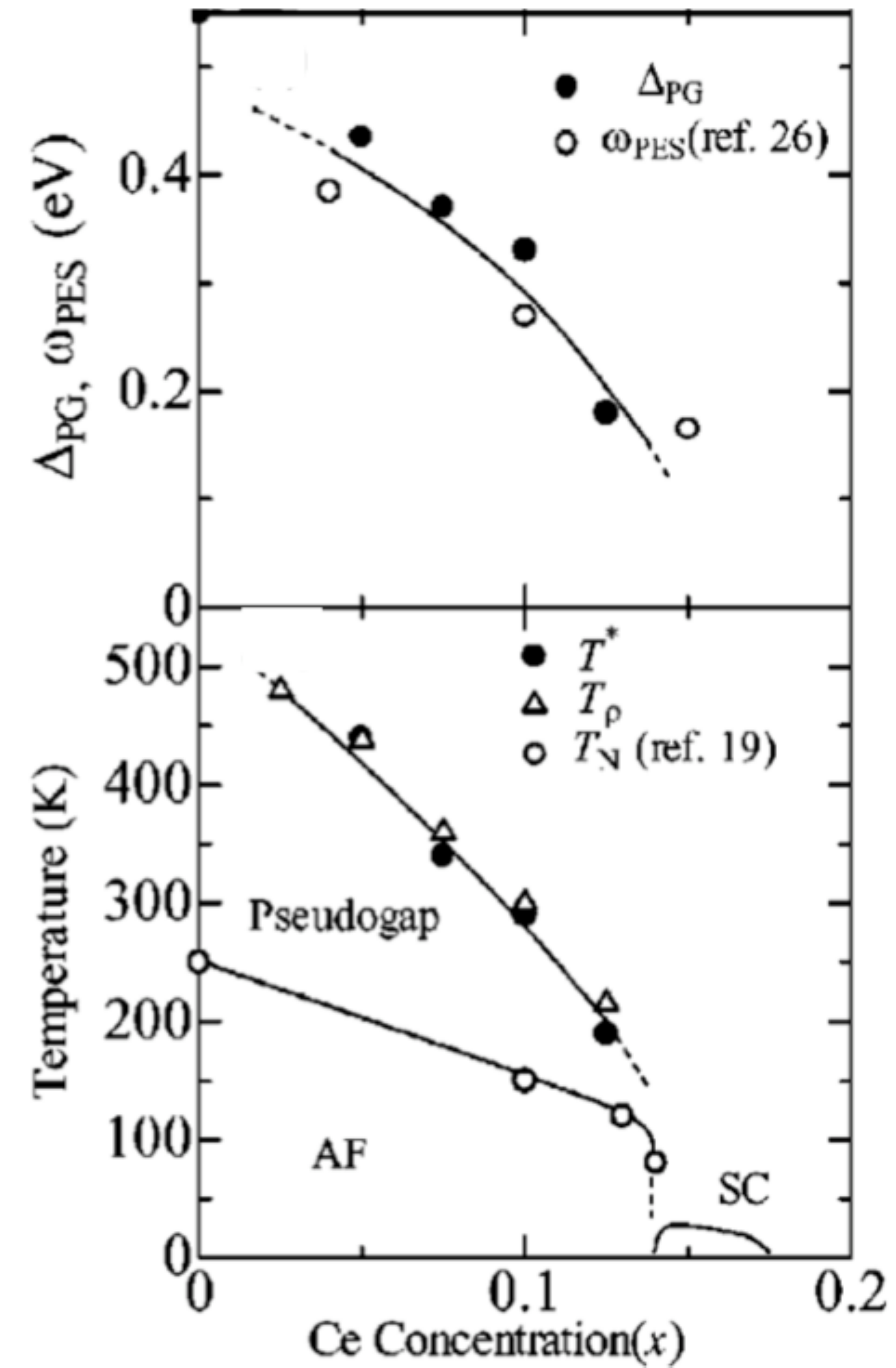}
\caption{(top)  The $x$ variation of the pseudogap magnitude  $\Delta_{PG}$ as
defined by the higher-lying isosbetic (equal-absorption) point in the
temperature-dependent conductivity spectra and the magnitude of
the pseudogap ($\omega_{PES}$ ) in the photoemission spectra \cite{Armitage02a} (Ref. 26 in the figure). The $\omega_{PES}$ is defined as the maximum energy of the quasiparticle peak on the putative large Fermi surface in the ARPES spectra shown in the Figs. 2(c) - (e) of \textcite{Armitage02a}.  (bottom) The obtained phase diagram.  The onset temperature of pseudogap formation
T$^*$ and the crossover temperature of out-of-plane resistivity T$_\rho$ (as given by the arrows in Fig.~\ref{OnoseRho} (right) are plotted against $x$ together with the N\'eel temperature T$_N$ reported
previously by \textcite{Luke90a} (Ref. 19 in the figure).}
\label{OnosePG}\end{figure}

As mentioned above, both optical conductivity and ARPES of underdoped single crystals ($x = 0$ to 0.125) shows the opening of a high energy gap-like structure at temperatures well above the N\'{e}el temperature \cite{Onose04a,Matsui05a}.  It can be viewed directly in the optical conductivity, which is in contrast to the hole-doped side, where gap-like features do not appear in the $ab$-plane optical conductivity itself and a `pseudogap' is only exhibited in the frequency dependent scattering rate \cite{Puchkov96a}.   The gap closes gradually with doping and vanishes by superconducting concentrations of approximately $x=0.15$ \cite{Onose04a} to $x=0.17$ \cite{Zimmers05a}.   \textcite{Onose04a} found that both its magnitude ($\Delta_{PG}$) and its onset temperature (T$^*$) obeys the approximate relation  $\Delta_{PG}) =$ 10$k_B$T$^*$ (Fig.~\ref{OnosePG}).   The magnitude of $\Delta_{PG}$ is comparable in magnitude to the pseudogap near ($\pi/2,\pi/2$) in the photoemission spectra reported by \cite{Armitage02a} (also Fig.~\ref{OnosePG}), which indicates that the pseudogap appearing in the optical spectra is the same as that found in photoemission.   Note that this is the same gap-like feature, of which a remarkable number of aspects can be modeled at low T by the SDW band structure as given in Sec.~\ref{sec:SDW}.   \textcite{Onose04a} identify the pseudogap with the buildup of antiferromagnetic correlations because:    (a)  In the underdoped region long range AF order develops at a temperature T$_N$ approximately half of T$^*$.  (b)   The intrinsic scale of the AF  exchange interaction $J$ is on the scale of the pseudogap magnitude  (c)  The gap anisotropy found in photoemission is consistent with that expected for 2D AF correlations with characteristic wavevector $(\pi,\pi)$ as pointed out by \textcite{Armitage02a}.

These PG phenomena may be analogous to the `high-energy' PG found in the hole-doped cuprates, although there are a number of differences as emphasized by  \textcite{Onose04a}.  (a) The large pseudogap of the hole-doped system is maximal near $(\pi,0)$  in contrast with one more centered around $(\pi/2,\pi/2)$ of the $n$-type cuprate.  (b) As mentioned, the pseudogap feature is not discernible in the $ab$-plane optical conductivity itself in the hole-doped cuprate, which may be because it is weaker than that in the electron-doped compound.  (c)  The ground state in the underdoped $n$-type system, where the pseudogap formation is observed strongest, is antiferromagnetic, while the superconducting phase is present even for underdoped samples in the hole-doped cuprate.  

As pointed out above,  it is interesting that for a PG related to AF, it forms at a temperature well above T$_N$.    This is presumably related to the fact that the spin correlation length $\xi$ is found to be quite large at temperatures even 100 K above T$_N$, which follows from the quasi-2D nature of the magnetism.  In this regard \textcite{Motoyama07a} found that at the PG temperature $T^*$ (as defined from the optical spectra) the spin correlation length $\xi$ becomes of order the estimated thermal de Broglie wavelegth $\xi_{th} = \hbar v_F/\pi k_B T$.   This is a condition for the onset of the PG  consistent with a number of theoretical calculations based on $t-t'-t"-U$ models  \cite{Kyung04a} that emphasize the build-up of AF correlations.  In these models, the weaker coupling regime (smaller $U/W$) of the electron-doped cuprates allows the identification of the pseudogap with long AF  correlation lengths.   These theories make quantitative predictions of the momentum dependence of the PG in the ARPES spectra, the  pseudogap onset temperature $T^*$, and the temperature and doping dependence of the AF correlation length that are in accord with experiment.    The hole-doped compounds appear to have stronger coupling and similar treatments give a pseudogap that is tied to the stronger local repulsive interaction and has different attributes \cite{Kyung04a,Kyung03a}.  Although aspects (such as the PG's momentum space location) are in qualitative agreement with experiment, quantities like the AF correlation length are in strong quantitative disagreement with neutron scattering results.   

At lower energy scales, there have been a number of tunneling experiments that have found evidence for a small normal state energy gap (NSG) ($\sim$5 meV) that seems somewhat analogous to the low-energy pseudogap found in the $p$-type materials.  This normal state gap (NSG) is probed in the ab-plane by applying a c-axis magnetic field greater than $H_{c2}$ \cite{Kleefisch01a,Biswas01a,Dagan05b,Alff03a}.  \textcite{Dagan05b} find that it is present at all dopings from 0.11 to 0.19 and the temperature at which it disappears correlates with $T_c$, at least on the overdoped side of the SC dome.  However, the NSG survives to very large magnetic fields and this is not obviously explained by the preformed Cooper pair scenario \cite{Kleefisch01a,Biswas01a}.  Recently, \textcite{Shan08b} conclude that the NSG and the SC gap are different across the phase diagram, which is consistent with various `two-gap' scenario in the underdoped $p$-type cuprates.  In PLCCO NMR \textcite{Zheng03a} found no sign of a $spin$ pseudogap opening up at temperatures much larger than T$_c$, which is a hallmark of underdoped $p$-type cuprates and has been interpreted as singlet formation at high temperatures.   Likewise the spin pseudogap observed in neutron scattering appears to close at T$_c$ \cite{Yamada03a} and not at some higher temperature.   This is consistent with a more mean field-like superconducting transition in these compounds, which may be tied to their apparently larger relative superfluid stiffness (4-15 times as compared to hole-doped compounds of similar T$_c$ \cite{shengelaya05a,Emery95a,Wu93a}).

In summary, although there are substantial signatures of PG effects in the electron-doped compounds, it is not clear if it is of the same character as in the hole-doped materials.   The bulk of PG phenomena in the electron-doped compounds appears to be related to AF correlations.  There is, as of yet, no evidence for many of the phenomena associated with pseudogap physics in the $p$-type materials, such as `spin pseudogaps' \cite{Alloul89a,Curro97a}, orbital currents \cite{Fauque06a,YLi08a}, stripes \cite{Tranquada95a,Mook98a,Mook02a},  or time-reverasal symmetry breaking \cite{Xia08a}.  It may be that such phenomena are obscurred by AF or it may just be that the $p$- and $n$-type compounds are just very different when it comes to these effects.

\section{Concluding remarks}

Our understanding of the electron-doped cuprates has advanced tremendously in recent years. Still, some important issues remain unresolved and more research will be needed to gain a full understanding.   For instance, the role of oxygen annealing is an important unresolved issue.  In the superconducting state, the evidence is now strong that the pairing symmetry in both $n$- and $p$-type cuprates is predominately $d$-wave, although of a non-monotonic form in the $n$-type.  In both $n$- and $p$-type cuprates, AFM gives way to SC upon doping and eventually the systems turn to a metallic, non-SC Fermi liquid-like state.  For both dopings, the normal state at the SC compositions is anomalous and is not yet well understood, although it is obvious that there is significant and important coupling to antiferromagnetism on at least the electron-doped side. Clearly an understanding of the metallic state on both sides is crucial to an understanding of the mechanism of the high-T$_c$ SC.  Similarly, there is convincing evidence for a pseudogap which derives from AFM in the $n-$type compounds.  This is in contrast to the pseudogap in the hole-doped compounds, which is as of yet of unknown origin.  The issue of whether an additional competing order parameter co-exists with SC and ends at a critical point just before or within the SC dome is still unresolved for both hole and electron doping.  Interaction effects play a central role in both classes of cuprates, although they may be weaker on the $n$-doped side.  For instance, numerical cluster calculations have been able to explain the gross features of the $n$-type phase diagram, pseudogap, and the evolution of the Fermi surface, in a manner not possible on the hole-doped side.

Detailed comparisons between the properties of $n$- and $p$-type cuprates will continue to be important areas of future investigation. Such studies should also prove themselves useful for understanding new classes of superconducting materials, such as the recently discovered iron-pnictides, which also show electron- and hole-doped varieties.  With regards to the high-T$_c$ problem, our hope is that systematic comparisons between the two sides of the cuprate phase diagram will give unique insight into what aspects of these compounds are universal, what aspects are not universal, and what aspects are crucial for the existence of high temperature superconductivity.

\section*{Acknowledgments}

We would like to thank our various close collaborators on this
subject including, S.M. Anlage, P. Bach, F. F. Balakirev, H. Balci, K. Behnia, J. Beauvais, A.  Biswas, G. Blumberg, N. Bontemps,  R. Budhani, S. Charlebois,  S. Charpentier,  G. C\^{o}t\'{e}, Y. Dagan, A.
Damascelli, H.D. Drew, D.L. Feng, J. Gauthier, R.G. Gianneta,
S. G.-Proulx, M.\`{E}. Gosselin, M. Greven, I. Hetel, J. Higgins, R. Hill,
C.C. Homes, S. Jandl, C. Kim, C.A. Kendziora,  X. B.-Lagani\`{e}re, P. Li, B. Liang, C. Lobb,
R.P.S.M. Lobo, D.H. Lu,  E. Maiser, P.K. Mang,  A. Millis,  Y. Onose,
M. Poirier, R. Prozorov, M. Qazilbash, P. Rauwel, J. Renaud, P. Richard,
G. Riou, G. Roberge, F. Ronning, A.F. Santander-Syro, K.M. Shen, Z.-X.
Shen,  J. Sonier, L. Taillefer, F. Tafuri, I. Takeuchi,  
Y. Tokura, K.D. Truong, W.Yu, T. Venkatesan, and A. Zimmers.

We would also like to thank M.F. Armitage, D. Basov, C. Bourbonnais,  S.
Brown, S. Chakravarty, A. Chubukov, Y. Dagan, P. Dai, M. d'Astuto, T. Deveraux, H.D. Drew, N. Drichko, M. Greven,  P.Goswami, J. Hirsch, L. Hozoi, G.S. Jenkins, M.-H. Julien, H. Jung, C. Kim, F. Kruger, A. Kuzmenko, H.-G. Luo, J. Li, J. Lynn, F. Marsiglio, E. Motoyama, M. Naito, J. Paglione, P. Richard, T. Sato, F. Schmitt, D. S\'{e}n\'{e}chal, K.M. Shen, M.P. Singh, T. Takahashi, Z. Tesanovic, A.-M. Tremblay, C. Varma, S. Wilson,  W. Yu, Q. Yuan, J. Zhao, and A. Zimmers for various helpful conversations on these topics and/or their careful reading of this manuscript.  Our work has been supported by the Sloan
Foundation, DOE DE-FG02-08ER46544,  NSF DMR-0847652 [NPA],
NSERC (Canada), CIfAR, CFI and FQRNT [PF], and NSF DMR-0653535 [RG].

\bibliographystyle{apsrmp}
\bibliography{RMP_ntypeBibliov6}

\begin{thebibliography}{552}
\expandafter\ifx\csname natexlab\endcsname\relax\def\natexlab#1{#1}\fi
\expandafter\ifx\csname bibnamefont\endcsname\relax
  \def\bibnamefont#1{#1}\fi
\expandafter\ifx\csname bibfnamefont\endcsname\relax
  \def\bibfnamefont#1{#1}\fi
\expandafter\ifx\csname citenamefont\endcsname\relax
  \def\citenamefont#1{#1}\fi
\expandafter\ifx\csname url\endcsname\relax
  \def\url#1{\texttt{#1}}\fi
\expandafter\ifx\csname urlprefix\endcsname\relax\def\urlprefix{URL }\fi
\providecommand{\bibinfo}[2]{#2}
\providecommand{\eprint}[2][]{\url{#2}}

\bibitem[{\citenamefont{Abanov} \emph{et~al.}(2001)\citenamefont{Abanov,
  Chubukov, and Schmalian}}]{Abanov01a}
\bibinfo{author}{\bibnamefont{Abanov}, \bibfnamefont{A.}},
  \bibinfo{author}{\bibfnamefont{A.~V.} \bibnamefont{Chubukov}}, and
  \bibinfo{author}{\bibfnamefont{J.}~\bibnamefont{Schmalian}},
  \bibinfo{year}{2001}, \bibinfo{journal}{Phys.\ Rev.\ B}
  \textbf{\bibinfo{volume}{63}}(\bibinfo{number}{18}), \bibinfo{pages}{180510}.

\bibitem[{\citenamefont{Abrahams and Varma}(2003)}]{abrahams03a}
\bibinfo{author}{\bibnamefont{Abrahams}, \bibfnamefont{E.}}, and
  \bibinfo{author}{\bibfnamefont{C.~M.} \bibnamefont{Varma}},
  \bibinfo{year}{2003}, \bibinfo{journal}{Phys.\ Rev.\ B}
  \textbf{\bibinfo{volume}{68}}(\bibinfo{number}{9}), \bibinfo{pages}{094502}.

\bibitem[{\citenamefont{Aharony} \emph{et~al.}(1988)\citenamefont{Aharony,
  Birgeneau, Coniglio, Kastner, and Stanley}}]{Aharony88a}
\bibinfo{author}{\bibnamefont{Aharony}, \bibfnamefont{A.}},
  \bibinfo{author}{\bibfnamefont{R.~J.} \bibnamefont{Birgeneau}},
  \bibinfo{author}{\bibfnamefont{A.}~\bibnamefont{Coniglio}},
  \bibinfo{author}{\bibfnamefont{M.~A.} \bibnamefont{Kastner}}, and
  \bibinfo{author}{\bibfnamefont{H.~E.} \bibnamefont{Stanley}},
  \bibinfo{year}{1988}, \bibinfo{journal}{Phys.\ Rev.\ Lett.}
  \textbf{\bibinfo{volume}{60}}(\bibinfo{number}{13}), \bibinfo{pages}{1330}.

\bibitem[{\citenamefont{Aichhorn and Arrigoni}(2005)}]{Aichhorn05a}
\bibinfo{author}{\bibnamefont{Aichhorn}, \bibfnamefont{M.}}, and
  \bibinfo{author}{\bibfnamefont{E.}~\bibnamefont{Arrigoni}},
  \bibinfo{year}{2005}, \bibinfo{journal}{Europhys.\ Lett.}
  \textbf{\bibinfo{volume}{72}}(\bibinfo{number}{1}), \bibinfo{pages}{117}.

\bibitem[{\citenamefont{Aichhorn} \emph{et~al.}(2006)\citenamefont{Aichhorn,
  Arrigoni, Potthoff, and Hanke}}]{Aichhorn06a}
\bibinfo{author}{\bibnamefont{Aichhorn}, \bibfnamefont{M.}},
  \bibinfo{author}{\bibfnamefont{E.}~\bibnamefont{Arrigoni}},
  \bibinfo{author}{\bibfnamefont{M.}~\bibnamefont{Potthoff}}, and
  \bibinfo{author}{\bibfnamefont{W.}~\bibnamefont{Hanke}},
  \bibinfo{year}{2006}, \bibinfo{journal}{Phys.\ Rev.\ B}
  \textbf{\bibinfo{volume}{74}}(\bibinfo{number}{23}), \bibinfo{eid}{235117}.

\bibitem[{\citenamefont{Aizenman and Wehr}(1990)}]{Aizenman90a}
\bibinfo{author}{\bibnamefont{Aizenman}, \bibfnamefont{M.}}, and
  \bibinfo{author}{\bibfnamefont{J.}~\bibnamefont{Wehr}}, \bibinfo{year}{1990},
  \bibinfo{journal}{Communications in Mathematical Physics}
  \textbf{\bibinfo{volume}{130}}(\bibinfo{number}{3}), \bibinfo{pages}{489}.

\bibitem[{\citenamefont{Akimitsu} \emph{et~al.}(1988)\citenamefont{Akimitsu,
  Suzuki, Watanabe, and Sawa}}]{Akimitsu88a}
\bibinfo{author}{\bibnamefont{Akimitsu}, \bibfnamefont{J.}},
  \bibinfo{author}{\bibfnamefont{S.}~\bibnamefont{Suzuki}},
  \bibinfo{author}{\bibfnamefont{M.}~\bibnamefont{Watanabe}}, and
  \bibinfo{author}{\bibfnamefont{H.}~\bibnamefont{Sawa}}, \bibinfo{year}{1988},
  \bibinfo{journal}{Jpn.\ J.\ Appl.\ Phys}
  \textbf{\bibinfo{volume}{27}}(\bibinfo{number}{19}), \bibinfo{pages}{L1859}.

\bibitem[{\citenamefont{Alexander} \emph{et~al.}(1991)\citenamefont{Alexander,
  Romberg, N\"ucker, Adelmann, Fink, Markert, Maple, Uchida, Takagi, Tokura,
  James, and Murphy}}]{Alexander91a}
\bibinfo{author}{\bibnamefont{Alexander}, \bibfnamefont{M.}},
  \bibinfo{author}{\bibfnamefont{H.}~\bibnamefont{Romberg}},
  \bibinfo{author}{\bibfnamefont{N.}~\bibnamefont{N\"ucker}},
  \bibinfo{author}{\bibfnamefont{P.}~\bibnamefont{Adelmann}},
  \bibinfo{author}{\bibfnamefont{J.}~\bibnamefont{Fink}},
  \bibinfo{author}{\bibfnamefont{J.~T.} \bibnamefont{Markert}},
  \bibinfo{author}{\bibfnamefont{M.~B.} \bibnamefont{Maple}},
  \bibinfo{author}{\bibfnamefont{S.}~\bibnamefont{Uchida}},
  \bibinfo{author}{\bibfnamefont{H.}~\bibnamefont{Takagi}},
  \bibinfo{author}{\bibfnamefont{Y.}~\bibnamefont{Tokura}},
  \bibinfo{author}{\bibfnamefont{A.~C. W.~P.} \bibnamefont{James}}, and
  \bibinfo{author}{\bibfnamefont{D.~W.} \bibnamefont{Murphy}},
  \bibinfo{year}{1991}, \bibinfo{journal}{Phys.\ Rev.\ B}
  \textbf{\bibinfo{volume}{43}}(\bibinfo{number}{1}), \bibinfo{pages}{333}.

\bibitem[{\citenamefont{Alff}
  \emph{et~al.}(1998{\natexlab{a}})\citenamefont{Alff, Beck, Gross, Marx,
  Kleefisch, Bauch, Sato, Naito, and Koren}}]{Alff98a}
\bibinfo{author}{\bibnamefont{Alff}, \bibfnamefont{L.}},
  \bibinfo{author}{\bibfnamefont{A.}~\bibnamefont{Beck}},
  \bibinfo{author}{\bibfnamefont{R.}~\bibnamefont{Gross}},
  \bibinfo{author}{\bibfnamefont{A.}~\bibnamefont{Marx}},
  \bibinfo{author}{\bibfnamefont{S.}~\bibnamefont{Kleefisch}},
  \bibinfo{author}{\bibfnamefont{T.}~\bibnamefont{Bauch}},
  \bibinfo{author}{\bibfnamefont{H.}~\bibnamefont{Sato}},
  \bibinfo{author}{\bibfnamefont{M.}~\bibnamefont{Naito}}, and
  \bibinfo{author}{\bibfnamefont{G.}~\bibnamefont{Koren}},
  \bibinfo{year}{1998}{\natexlab{a}}, \bibinfo{journal}{Phys.\ Rev.\ B}
  \textbf{\bibinfo{volume}{58}}(\bibinfo{number}{17}), \bibinfo{pages}{11197}.

\bibitem[{\citenamefont{Alff}
  \emph{et~al.}(1998{\natexlab{b}})\citenamefont{Alff, Kleefisch, Schoop,
  Zittartz, Kemen, Bauch, Marx, and Gross}}]{Alff98b}
\bibinfo{author}{\bibnamefont{Alff}, \bibfnamefont{L.}},
  \bibinfo{author}{\bibfnamefont{S.}~\bibnamefont{Kleefisch}},
  \bibinfo{author}{\bibfnamefont{U.}~\bibnamefont{Schoop}},
  \bibinfo{author}{\bibfnamefont{M.}~\bibnamefont{Zittartz}},
  \bibinfo{author}{\bibfnamefont{T.}~\bibnamefont{Kemen}},
  \bibinfo{author}{\bibfnamefont{T.}~\bibnamefont{Bauch}},
  \bibinfo{author}{\bibfnamefont{A.}~\bibnamefont{Marx}}, and
  \bibinfo{author}{\bibfnamefont{R.}~\bibnamefont{Gross}},
  \bibinfo{year}{1998}{\natexlab{b}}, \bibinfo{journal}{Euro.\ Phys.\ J.\ B -
  Cond.\ Matt.\ Comp.\ Syst.}
  \textbf{\bibinfo{volume}{5}}(\bibinfo{number}{3}), \bibinfo{pages}{423}.

\bibitem[{\citenamefont{Alff} \emph{et~al.}(2003)\citenamefont{Alff,
  Krockenberger, Welter, Schonecke, Gross, Manske, and Naito}}]{Alff03a}
\bibinfo{author}{\bibnamefont{Alff}, \bibfnamefont{L.}},
  \bibinfo{author}{\bibfnamefont{Y.}~\bibnamefont{Krockenberger}},
  \bibinfo{author}{\bibfnamefont{B.}~\bibnamefont{Welter}},
  \bibinfo{author}{\bibfnamefont{M.}~\bibnamefont{Schonecke}},
  \bibinfo{author}{\bibfnamefont{R.}~\bibnamefont{Gross}},
  \bibinfo{author}{\bibfnamefont{D.}~\bibnamefont{Manske}}, and
  \bibinfo{author}{\bibfnamefont{M.}~\bibnamefont{Naito}},
  \bibinfo{year}{2003}, \bibinfo{journal}{Nature}
  \textbf{\bibinfo{volume}{422}}, \bibinfo{pages}{698}.

\bibitem[{\citenamefont{Alff} \emph{et~al.}(1999)\citenamefont{Alff, Meyer,
  Kleefisch, Schoop, Marx, Sato, Naito, and Gross}}]{Alff99b}
\bibinfo{author}{\bibnamefont{Alff}, \bibfnamefont{L.}},
  \bibinfo{author}{\bibfnamefont{S.}~\bibnamefont{Meyer}},
  \bibinfo{author}{\bibfnamefont{S.}~\bibnamefont{Kleefisch}},
  \bibinfo{author}{\bibfnamefont{U.}~\bibnamefont{Schoop}},
  \bibinfo{author}{\bibfnamefont{A.}~\bibnamefont{Marx}},
  \bibinfo{author}{\bibfnamefont{H.}~\bibnamefont{Sato}},
  \bibinfo{author}{\bibfnamefont{M.}~\bibnamefont{Naito}}, and
  \bibinfo{author}{\bibfnamefont{R.}~\bibnamefont{Gross}},
  \bibinfo{year}{1999}, \bibinfo{journal}{Phys.\ Rev.\ Lett.}
  \textbf{\bibinfo{volume}{83}}(\bibinfo{number}{13}), \bibinfo{pages}{2644}.

\bibitem[{\citenamefont{Allen} \emph{et~al.}(1990)\citenamefont{Allen, Olson,
  Maple, Kang, Liu, Park, Anderson, Ellis, Markert, Dalichaouch, and
  Liu}}]{Allen90a}
\bibinfo{author}{\bibnamefont{Allen}, \bibfnamefont{J.~W.}},
  \bibinfo{author}{\bibfnamefont{C.~G.} \bibnamefont{Olson}},
  \bibinfo{author}{\bibfnamefont{M.~B.} \bibnamefont{Maple}},
  \bibinfo{author}{\bibfnamefont{J.-S.} \bibnamefont{Kang}},
  \bibinfo{author}{\bibfnamefont{L.~Z.} \bibnamefont{Liu}},
  \bibinfo{author}{\bibfnamefont{J.-H.} \bibnamefont{Park}},
  \bibinfo{author}{\bibfnamefont{R.~O.} \bibnamefont{Anderson}},
  \bibinfo{author}{\bibfnamefont{W.~P.} \bibnamefont{Ellis}},
  \bibinfo{author}{\bibfnamefont{J.~T.} \bibnamefont{Markert}},
  \bibinfo{author}{\bibfnamefont{Y.}~\bibnamefont{Dalichaouch}}, and
  \bibinfo{author}{\bibfnamefont{R.}~\bibnamefont{Liu}}, \bibinfo{year}{1990},
  \bibinfo{journal}{Phys.\ Rev.\ Lett.}
  \textbf{\bibinfo{volume}{64}}(\bibinfo{number}{5}), \bibinfo{pages}{595}.

\bibitem[{\citenamefont{Alloul} \emph{et~al.}(2009)\citenamefont{Alloul,
  Bobroff, Gabay, and Hirschfeld}}]{Alloul08a}
\bibinfo{author}{\bibnamefont{Alloul}, \bibfnamefont{H.}},
  \bibinfo{author}{\bibfnamefont{J.}~\bibnamefont{Bobroff}},
  \bibinfo{author}{\bibfnamefont{M.}~\bibnamefont{Gabay}}, and
  \bibinfo{author}{\bibfnamefont{P.~J.} \bibnamefont{Hirschfeld}},
  \bibinfo{year}{2009}, \bibinfo{journal}{Rev.\ Mod.\ Phys.}
  \textbf{\bibinfo{volume}{81}}(\bibinfo{number}{1}), \bibinfo{eid}{45}.

\bibitem[{\citenamefont{Alloul} \emph{et~al.}(1989)\citenamefont{Alloul, Ohno,
  and Mendels}}]{Alloul89a}
\bibinfo{author}{\bibnamefont{Alloul}, \bibfnamefont{H.}},
  \bibinfo{author}{\bibfnamefont{T.}~\bibnamefont{Ohno}}, and
  \bibinfo{author}{\bibfnamefont{P.}~\bibnamefont{Mendels}},
  \bibinfo{year}{1989}, \bibinfo{journal}{Phys.\ Rev.\ Lett.}
  \textbf{\bibinfo{volume}{63}}(\bibinfo{number}{16}), \bibinfo{pages}{1700}.

\bibitem[{\citenamefont{Alp} \emph{et~al.}(1987)\citenamefont{Alp, Shenoy,
  Hinks, II, Soderholm, Schuttler, Guo, Ellis, Montano, and
  Ramanathan}}]{Alp87a}
\bibinfo{author}{\bibnamefont{Alp}, \bibfnamefont{E.~E.}},
  \bibinfo{author}{\bibfnamefont{G.~K.} \bibnamefont{Shenoy}},
  \bibinfo{author}{\bibfnamefont{D.~G.} \bibnamefont{Hinks}},
  \bibinfo{author}{\bibfnamefont{D.~W.~C.} \bibnamefont{II}},
  \bibinfo{author}{\bibfnamefont{L.}~\bibnamefont{Soderholm}},
  \bibinfo{author}{\bibfnamefont{H.-B.} \bibnamefont{Schuttler}},
  \bibinfo{author}{\bibfnamefont{J.}~\bibnamefont{Guo}},
  \bibinfo{author}{\bibfnamefont{D.~E.} \bibnamefont{Ellis}},
  \bibinfo{author}{\bibfnamefont{P.~A.} \bibnamefont{Montano}}, and
  \bibinfo{author}{\bibfnamefont{M.}~\bibnamefont{Ramanathan}},
  \bibinfo{year}{1987}, \bibinfo{journal}{Phys.\ Rev.\ B}
  \textbf{\bibinfo{volume}{35}}(\bibinfo{number}{13}), \bibinfo{pages}{7199}.

\bibitem[{\citenamefont{Alvarenga} \emph{et~al.}(1996)\citenamefont{Alvarenga,
  Rao, Sanjurjo, Granado, Torriani, Rettori, Oseroff, Sarrao, and
  Fisk}}]{Alvarenga96a}
\bibinfo{author}{\bibnamefont{Alvarenga}, \bibfnamefont{A.~D.}},
  \bibinfo{author}{\bibfnamefont{D.}~\bibnamefont{Rao}},
  \bibinfo{author}{\bibfnamefont{J.~A.} \bibnamefont{Sanjurjo}},
  \bibinfo{author}{\bibfnamefont{E.}~\bibnamefont{Granado}},
  \bibinfo{author}{\bibfnamefont{I.}~\bibnamefont{Torriani}},
  \bibinfo{author}{\bibfnamefont{C.}~\bibnamefont{Rettori}},
  \bibinfo{author}{\bibfnamefont{S.}~\bibnamefont{Oseroff}},
  \bibinfo{author}{\bibfnamefont{J.}~\bibnamefont{Sarrao}}, and
  \bibinfo{author}{\bibfnamefont{Z.}~\bibnamefont{Fisk}}, \bibinfo{year}{1996},
  \bibinfo{journal}{Phys.\ Rev.\ B}
  \textbf{\bibinfo{volume}{53}}(\bibinfo{number}{2}), \bibinfo{pages}{837}.

\bibitem[{\citenamefont{Andersen} \emph{et~al.}(1995)\citenamefont{Andersen,
  Liechtenstein, Jepsen, and Paulsen}}]{andersen95a}
\bibinfo{author}{\bibnamefont{Andersen}, \bibfnamefont{O.~K.}},
  \bibinfo{author}{\bibfnamefont{A.~I.} \bibnamefont{Liechtenstein}},
  \bibinfo{author}{\bibfnamefont{O.}~\bibnamefont{Jepsen}}, and
  \bibinfo{author}{\bibfnamefont{F.}~\bibnamefont{Paulsen}},
  \bibinfo{year}{1995}, \bibinfo{journal}{J.\ Phys.\ Chem.\ Sol.}
  \textbf{\bibinfo{volume}{56}}(\bibinfo{number}{12}), \bibinfo{eid}{1573}.

\bibitem[{\citenamefont{Anderson}(1987)}]{Anderson87a}
\bibinfo{author}{\bibnamefont{Anderson}, \bibfnamefont{P.~W.}},
  \bibinfo{year}{1987}, \bibinfo{journal}{Science}
  \textbf{\bibinfo{volume}{235}}(\bibinfo{number}{4793}),
  \bibinfo{pages}{1196}.

\bibitem[{\citenamefont{Anderson} \emph{et~al.}(1993)\citenamefont{Anderson,
  Claessen, Allen, Olson, Janowitz, Liu, Park, Maple, Dalichaouch, de~Andrade,
  Jardim, Early} \emph{et~al.}}]{Anderson93a}
\bibinfo{author}{\bibnamefont{Anderson}, \bibfnamefont{R.~O.}},
  \bibinfo{author}{\bibfnamefont{R.}~\bibnamefont{Claessen}},
  \bibinfo{author}{\bibfnamefont{J.~W.} \bibnamefont{Allen}},
  \bibinfo{author}{\bibfnamefont{C.~G.} \bibnamefont{Olson}},
  \bibinfo{author}{\bibfnamefont{C.}~\bibnamefont{Janowitz}},
  \bibinfo{author}{\bibfnamefont{L.~Z.} \bibnamefont{Liu}},
  \bibinfo{author}{\bibfnamefont{J.-H.} \bibnamefont{Park}},
  \bibinfo{author}{\bibfnamefont{M.~B.} \bibnamefont{Maple}},
  \bibinfo{author}{\bibfnamefont{Y.}~\bibnamefont{Dalichaouch}},
  \bibinfo{author}{\bibfnamefont{M.~C.} \bibnamefont{de~Andrade}},
  \bibinfo{author}{\bibfnamefont{R.~F.} \bibnamefont{Jardim}},
  \bibinfo{author}{\bibfnamefont{E.~A.} \bibnamefont{Early}}, \emph{et~al.},
  \bibinfo{year}{1993}, \bibinfo{journal}{Phys.\ Rev.\ Lett.}
  \textbf{\bibinfo{volume}{70}}(\bibinfo{number}{20}), \bibinfo{pages}{3163}.

\bibitem[{\citenamefont{Ando} \emph{et~al.}(2001)\citenamefont{Ando, Lavrov,
  Komiya, Segawa, and Sun}}]{Ando01a}
\bibinfo{author}{\bibnamefont{Ando}, \bibfnamefont{Y.}},
  \bibinfo{author}{\bibfnamefont{A.~N.} \bibnamefont{Lavrov}},
  \bibinfo{author}{\bibfnamefont{S.}~\bibnamefont{Komiya}},
  \bibinfo{author}{\bibfnamefont{K.}~\bibnamefont{Segawa}}, and
  \bibinfo{author}{\bibfnamefont{X.~F.} \bibnamefont{Sun}},
  \bibinfo{year}{2001}, \bibinfo{journal}{Phys.\ Rev.\ Lett.}
  \textbf{\bibinfo{volume}{87}}(\bibinfo{number}{1}), \bibinfo{pages}{017001}.

\bibitem[{\citenamefont{Andreone} \emph{et~al.}(1994)\citenamefont{Andreone,
  Cassinese, Di~Chiara, Vaglio, Gupta, and Sarnelli}}]{Andreone94a}
\bibinfo{author}{\bibnamefont{Andreone}, \bibfnamefont{A.}},
  \bibinfo{author}{\bibfnamefont{A.}~\bibnamefont{Cassinese}},
  \bibinfo{author}{\bibfnamefont{A.}~\bibnamefont{Di~Chiara}},
  \bibinfo{author}{\bibfnamefont{R.}~\bibnamefont{Vaglio}},
  \bibinfo{author}{\bibfnamefont{A.}~\bibnamefont{Gupta}}, and
  \bibinfo{author}{\bibfnamefont{E.}~\bibnamefont{Sarnelli}},
  \bibinfo{year}{1994}, \bibinfo{journal}{Phys.\ Rev.\ B}
  \textbf{\bibinfo{volume}{49}}(\bibinfo{number}{9}), \bibinfo{pages}{6392}.

\bibitem[{\citenamefont{Anlage} \emph{et~al.}(1994)\citenamefont{Anlage, Wu,
  Mao, Mao, Xi, Venkatesan, Peng, and Greene}}]{Anlage94b}
\bibinfo{author}{\bibnamefont{Anlage}, \bibfnamefont{S.~M.}},
  \bibinfo{author}{\bibfnamefont{D.-H.} \bibnamefont{Wu}},
  \bibinfo{author}{\bibfnamefont{J.}~\bibnamefont{Mao}},
  \bibinfo{author}{\bibfnamefont{S.~N.} \bibnamefont{Mao}},
  \bibinfo{author}{\bibfnamefont{X.~X.} \bibnamefont{Xi}},
  \bibinfo{author}{\bibfnamefont{T.}~\bibnamefont{Venkatesan}},
  \bibinfo{author}{\bibfnamefont{J.~L.} \bibnamefont{Peng}}, and
  \bibinfo{author}{\bibfnamefont{R.~L.} \bibnamefont{Greene}},
  \bibinfo{year}{1994}, \bibinfo{journal}{Phys.\ Rev.\ B}
  \textbf{\bibinfo{volume}{50}}(\bibinfo{number}{1}), \bibinfo{pages}{523}.

\bibitem[{\citenamefont{Aprili} \emph{et~al.}(1998)\citenamefont{Aprili,
  Covington, Paraoanu, Niedermeier, and Greene}}]{Aprili98a}
\bibinfo{author}{\bibnamefont{Aprili}, \bibfnamefont{M.}},
  \bibinfo{author}{\bibfnamefont{M.}~\bibnamefont{Covington}},
  \bibinfo{author}{\bibfnamefont{E.}~\bibnamefont{Paraoanu}},
  \bibinfo{author}{\bibfnamefont{B.}~\bibnamefont{Niedermeier}}, and
  \bibinfo{author}{\bibfnamefont{L.~H.} \bibnamefont{Greene}},
  \bibinfo{year}{1998}, \bibinfo{journal}{Phys.\ Rev.\ B}
  \textbf{\bibinfo{volume}{57}}(\bibinfo{number}{14}), \bibinfo{pages}{R8139}.

\bibitem[{\citenamefont{Arai} \emph{et~al.}(1999)\citenamefont{Arai, Nishijima,
  Endoh, Egami, Tajima, Tomimoto, Shiohara, Takahashi, Garrett, and
  Bennington}}]{arai99a}
\bibinfo{author}{\bibnamefont{Arai}, \bibfnamefont{M.}},
  \bibinfo{author}{\bibfnamefont{T.}~\bibnamefont{Nishijima}},
  \bibinfo{author}{\bibfnamefont{Y.}~\bibnamefont{Endoh}},
  \bibinfo{author}{\bibfnamefont{T.}~\bibnamefont{Egami}},
  \bibinfo{author}{\bibfnamefont{S.}~\bibnamefont{Tajima}},
  \bibinfo{author}{\bibfnamefont{K.}~\bibnamefont{Tomimoto}},
  \bibinfo{author}{\bibfnamefont{Y.}~\bibnamefont{Shiohara}},
  \bibinfo{author}{\bibfnamefont{M.}~\bibnamefont{Takahashi}},
  \bibinfo{author}{\bibfnamefont{A.}~\bibnamefont{Garrett}}, and
  \bibinfo{author}{\bibfnamefont{S.~M.} \bibnamefont{Bennington}},
  \bibinfo{year}{1999}, \bibinfo{journal}{Phys.\ Rev.\ Lett.}
  \textbf{\bibinfo{volume}{83}}(\bibinfo{number}{3}), \bibinfo{pages}{608}.

\bibitem[{\citenamefont{Ariando} \emph{et~al.}(2005)\citenamefont{Ariando,
  Darminto, Smilde, Leca, Blank, Rogalla, and Hilgenkamp}}]{Ariando05a}
\bibinfo{author}{\bibnamefont{Ariando}},
  \bibinfo{author}{\bibfnamefont{D.}~\bibnamefont{Darminto}},
  \bibinfo{author}{\bibfnamefont{H.~J.~H.} \bibnamefont{Smilde}},
  \bibinfo{author}{\bibfnamefont{V.}~\bibnamefont{Leca}},
  \bibinfo{author}{\bibfnamefont{D.~H.~A.} \bibnamefont{Blank}},
  \bibinfo{author}{\bibfnamefont{H.}~\bibnamefont{Rogalla}}, and
  \bibinfo{author}{\bibfnamefont{H.}~\bibnamefont{Hilgenkamp}},
  \bibinfo{year}{2005}, \bibinfo{journal}{Phys.\ Rev.\ Lett.}
  \textbf{\bibinfo{volume}{94}}(\bibinfo{number}{16}), \bibinfo{eid}{167001}.

\bibitem[{\citenamefont{Arima} \emph{et~al.}(1993)\citenamefont{Arima, Tokura,
  and Uchida}}]{Arima93a}
\bibinfo{author}{\bibnamefont{Arima}, \bibfnamefont{T.}},
  \bibinfo{author}{\bibfnamefont{Y.}~\bibnamefont{Tokura}}, and
  \bibinfo{author}{\bibfnamefont{S.}~\bibnamefont{Uchida}},
  \bibinfo{year}{1993}, \bibinfo{journal}{Phys.\ Rev.\ B}
  \textbf{\bibinfo{volume}{48}}(\bibinfo{number}{9}), \bibinfo{pages}{6597}.

\bibitem[{\citenamefont{Armitage}(2001)}]{Armitage02b}
\bibinfo{author}{\bibnamefont{Armitage}, \bibfnamefont{N.~P.}},
  \bibinfo{year}{2001}, \bibinfo{journal}{Stanford University Phd. thesis} .

\bibitem[{\citenamefont{Armitage}(2008)}]{Armitage08a}
\bibinfo{author}{\bibnamefont{Armitage}, \bibfnamefont{N.~P.}},
  \bibinfo{year}{2008}, \bibinfo{journal}{unpublished} .

\bibitem[{\citenamefont{Armitage}
  \emph{et~al.}(2001{\natexlab{a}})\citenamefont{Armitage, Lu, Feng, Kim,
  Damascelli, Shen, Ronning, Shen, Onose, Taguchi, and Tokura}}]{Armitage01a}
\bibinfo{author}{\bibnamefont{Armitage}, \bibfnamefont{N.~P.}},
  \bibinfo{author}{\bibfnamefont{D.~H.} \bibnamefont{Lu}},
  \bibinfo{author}{\bibfnamefont{D.~L.} \bibnamefont{Feng}},
  \bibinfo{author}{\bibfnamefont{C.}~\bibnamefont{Kim}},
  \bibinfo{author}{\bibfnamefont{A.}~\bibnamefont{Damascelli}},
  \bibinfo{author}{\bibfnamefont{K.~M.} \bibnamefont{Shen}},
  \bibinfo{author}{\bibfnamefont{F.}~\bibnamefont{Ronning}},
  \bibinfo{author}{\bibfnamefont{Z.-X.} \bibnamefont{Shen}},
  \bibinfo{author}{\bibfnamefont{Y.}~\bibnamefont{Onose}},
  \bibinfo{author}{\bibfnamefont{Y.}~\bibnamefont{Taguchi}}, and
  \bibinfo{author}{\bibfnamefont{Y.}~\bibnamefont{Tokura}},
  \bibinfo{year}{2001}{\natexlab{a}}, \bibinfo{journal}{Phys.\ Rev.\ Lett.}
  \textbf{\bibinfo{volume}{86}}(\bibinfo{number}{6}), \bibinfo{pages}{1126}.

\bibitem[{\citenamefont{Armitage}
  \emph{et~al.}(2001{\natexlab{b}})\citenamefont{Armitage, Lu, Kim, Damascelli,
  Shen, Ronning, Feng, Bogdanov, Shen, Onose, Taguchi, Tokura}
  \emph{et~al.}}]{Armitage01b}
\bibinfo{author}{\bibnamefont{Armitage}, \bibfnamefont{N.~P.}},
  \bibinfo{author}{\bibfnamefont{D.~H.} \bibnamefont{Lu}},
  \bibinfo{author}{\bibfnamefont{C.}~\bibnamefont{Kim}},
  \bibinfo{author}{\bibfnamefont{A.}~\bibnamefont{Damascelli}},
  \bibinfo{author}{\bibfnamefont{K.~M.} \bibnamefont{Shen}},
  \bibinfo{author}{\bibfnamefont{F.}~\bibnamefont{Ronning}},
  \bibinfo{author}{\bibfnamefont{D.~L.} \bibnamefont{Feng}},
  \bibinfo{author}{\bibfnamefont{P.}~\bibnamefont{Bogdanov}},
  \bibinfo{author}{\bibfnamefont{Z.-X.} \bibnamefont{Shen}},
  \bibinfo{author}{\bibfnamefont{Y.}~\bibnamefont{Onose}},
  \bibinfo{author}{\bibfnamefont{Y.}~\bibnamefont{Taguchi}},
  \bibinfo{author}{\bibfnamefont{Y.}~\bibnamefont{Tokura}}, \emph{et~al.},
  \bibinfo{year}{2001}{\natexlab{b}}, \bibinfo{journal}{Phys.\ Rev.\ Lett.}
  \textbf{\bibinfo{volume}{87}}(\bibinfo{number}{14}), \bibinfo{eid}{147003}.

\bibitem[{\citenamefont{Armitage} \emph{et~al.}(2003)\citenamefont{Armitage,
  Lu, Kim, Damascelli, Shen, Ronning, Feng, Bogdanov, Zhou, Yang, Hussain,
  Mang} \emph{et~al.}}]{Armitage03a}
\bibinfo{author}{\bibnamefont{Armitage}, \bibfnamefont{N.~P.}},
  \bibinfo{author}{\bibfnamefont{D.~H.} \bibnamefont{Lu}},
  \bibinfo{author}{\bibfnamefont{C.}~\bibnamefont{Kim}},
  \bibinfo{author}{\bibfnamefont{A.}~\bibnamefont{Damascelli}},
  \bibinfo{author}{\bibfnamefont{K.~M.} \bibnamefont{Shen}},
  \bibinfo{author}{\bibfnamefont{F.}~\bibnamefont{Ronning}},
  \bibinfo{author}{\bibfnamefont{D.~L.} \bibnamefont{Feng}},
  \bibinfo{author}{\bibfnamefont{P.}~\bibnamefont{Bogdanov}},
  \bibinfo{author}{\bibfnamefont{X.~J.} \bibnamefont{Zhou}},
  \bibinfo{author}{\bibfnamefont{W.~L.} \bibnamefont{Yang}},
  \bibinfo{author}{\bibfnamefont{Z.}~\bibnamefont{Hussain}},
  \bibinfo{author}{\bibfnamefont{P.~K.} \bibnamefont{Mang}}, \emph{et~al.},
  \bibinfo{year}{2003}, \bibinfo{journal}{Phys.\ Rev.\ B}
  \textbf{\bibinfo{volume}{68}}(\bibinfo{number}{6}), \bibinfo{eid}{064517}.

\bibitem[{\citenamefont{Armitage} \emph{et~al.}(2002)\citenamefont{Armitage,
  Ronning, Lu, Kim, Damascelli, Shen, Feng, Eisaki, Shen, Mang, Kaneko, Greven}
  \emph{et~al.}}]{Armitage02a}
\bibinfo{author}{\bibnamefont{Armitage}, \bibfnamefont{N.~P.}},
  \bibinfo{author}{\bibfnamefont{F.}~\bibnamefont{Ronning}},
  \bibinfo{author}{\bibfnamefont{D.~H.} \bibnamefont{Lu}},
  \bibinfo{author}{\bibfnamefont{C.}~\bibnamefont{Kim}},
  \bibinfo{author}{\bibfnamefont{A.}~\bibnamefont{Damascelli}},
  \bibinfo{author}{\bibfnamefont{K.~M.} \bibnamefont{Shen}},
  \bibinfo{author}{\bibfnamefont{D.~L.} \bibnamefont{Feng}},
  \bibinfo{author}{\bibfnamefont{H.}~\bibnamefont{Eisaki}},
  \bibinfo{author}{\bibfnamefont{Z.-X.} \bibnamefont{Shen}},
  \bibinfo{author}{\bibfnamefont{P.~K.} \bibnamefont{Mang}},
  \bibinfo{author}{\bibfnamefont{N.}~\bibnamefont{Kaneko}},
  \bibinfo{author}{\bibfnamefont{M.}~\bibnamefont{Greven}}, \emph{et~al.},
  \bibinfo{year}{2002}, \bibinfo{journal}{Phys.\ Rev.\ Lett.}
  \textbf{\bibinfo{volume}{88}}(\bibinfo{number}{25}), \bibinfo{eid}{257001}.

\bibitem[{\citenamefont{Asayama} \emph{et~al.}(1996)\citenamefont{Asayama,
  Kitaoka, qing Zheng, and Ishida}}]{Asayama96a}
\bibinfo{author}{\bibnamefont{Asayama}, \bibfnamefont{K.}},
  \bibinfo{author}{\bibfnamefont{Y.}~\bibnamefont{Kitaoka}},
  \bibinfo{author}{\bibfnamefont{G.}~\bibnamefont{qing Zheng}}, and
  \bibinfo{author}{\bibfnamefont{K.}~\bibnamefont{Ishida}},
  \bibinfo{year}{1996}, \bibinfo{journal}{Progress in Nuclear Magnetic
  Resonance Spectroscopy} \textbf{\bibinfo{volume}{28}}(\bibinfo{number}{3-4}),
  \bibinfo{pages}{221 }.

\bibitem[{\citenamefont{B.~Kyung}(2009)}]{Kyung09a}
\bibinfo{author}{\bibnamefont{B.~Kyung}, \bibfnamefont{A.-M.~T.,
  D.~S\'en\'echal}}, \bibinfo{year}{2009}, \bibinfo{journal}{arXiv:0812.1228} .

\bibitem[{\citenamefont{Bacci} \emph{et~al.}(1991)\citenamefont{Bacci,
  Gagliano, Martin, and Annett}}]{Bacci91a}
\bibinfo{author}{\bibnamefont{Bacci}, \bibfnamefont{S.~B.}},
  \bibinfo{author}{\bibfnamefont{E.~R.} \bibnamefont{Gagliano}},
  \bibinfo{author}{\bibfnamefont{R.~M.} \bibnamefont{Martin}}, and
  \bibinfo{author}{\bibfnamefont{J.~F.} \bibnamefont{Annett}},
  \bibinfo{year}{1991}, \bibinfo{journal}{Phys.\ Rev.\ B}
  \textbf{\bibinfo{volume}{44}}(\bibinfo{number}{14}), \bibinfo{pages}{7504}.

\bibitem[{\citenamefont{Bakharev} \emph{et~al.}(2004)\citenamefont{Bakharev,
  Abu-Shiekah, Brom, Nugroho, McCulloch, and Zaanen}}]{Bakharev04a}
\bibinfo{author}{\bibnamefont{Bakharev}, \bibfnamefont{O.~N.}},
  \bibinfo{author}{\bibfnamefont{I.~M.} \bibnamefont{Abu-Shiekah}},
  \bibinfo{author}{\bibfnamefont{H.~B.} \bibnamefont{Brom}},
  \bibinfo{author}{\bibfnamefont{A.~A.} \bibnamefont{Nugroho}},
  \bibinfo{author}{\bibfnamefont{I.~P.} \bibnamefont{McCulloch}}, and
  \bibinfo{author}{\bibfnamefont{J.}~\bibnamefont{Zaanen}},
  \bibinfo{year}{2004}, \bibinfo{journal}{Phys.\ Rev.\ Lett.}
  \textbf{\bibinfo{volume}{93}}(\bibinfo{number}{3}), \bibinfo{eid}{037002}.

\bibitem[{\citenamefont{Balci and Greene}(2004)}]{Balci04c}
\bibinfo{author}{\bibnamefont{Balci}, \bibfnamefont{H.}}, and
  \bibinfo{author}{\bibfnamefont{R.~L.} \bibnamefont{Greene}},
  \bibinfo{year}{2004}, \bibinfo{journal}{Phys.\ Rev.\ Lett.}
  \textbf{\bibinfo{volume}{93}}(\bibinfo{number}{6}), \bibinfo{eid}{067001}.

\bibitem[{\citenamefont{Balci} \emph{et~al.}(2003)\citenamefont{Balci, Hill,
  Qazilbash, and Greene}}]{Balci03a}
\bibinfo{author}{\bibnamefont{Balci}, \bibfnamefont{H.}},
  \bibinfo{author}{\bibfnamefont{C.~P.} \bibnamefont{Hill}},
  \bibinfo{author}{\bibfnamefont{M.~M.} \bibnamefont{Qazilbash}}, and
  \bibinfo{author}{\bibfnamefont{R.~L.} \bibnamefont{Greene}},
  \bibinfo{year}{2003}, \bibinfo{journal}{Phys.\ Rev.\ B}
  \textbf{\bibinfo{volume}{68}}(\bibinfo{number}{5}), \bibinfo{eid}{054520}.

\bibitem[{\citenamefont{Balci} \emph{et~al.}(2002)\citenamefont{Balci,
  Smolyaninova, Fournier, Biswas, and Greene}}]{Balci02a}
\bibinfo{author}{\bibnamefont{Balci}, \bibfnamefont{H.}},
  \bibinfo{author}{\bibfnamefont{V.~N.} \bibnamefont{Smolyaninova}},
  \bibinfo{author}{\bibfnamefont{P.}~\bibnamefont{Fournier}},
  \bibinfo{author}{\bibfnamefont{A.}~\bibnamefont{Biswas}}, and
  \bibinfo{author}{\bibfnamefont{R.~L.} \bibnamefont{Greene}},
  \bibinfo{year}{2002}, \bibinfo{journal}{Phys.\ Rev.\ B}
  \textbf{\bibinfo{volume}{66}}, \bibinfo{pages}{174510}.

\bibitem[{\citenamefont{Basov and Timusk}(2005)}]{Basov05a}
\bibinfo{author}{\bibnamefont{Basov}, \bibfnamefont{D.~N.}}, and
  \bibinfo{author}{\bibfnamefont{T.}~\bibnamefont{Timusk}},
  \bibinfo{year}{2005}, \bibinfo{journal}{Rev.\ Mod.\ Phys.}
  \textbf{\bibinfo{volume}{77}}(\bibinfo{number}{2}), \bibinfo{eid}{721}.

\bibitem[{\citenamefont{Beck} \emph{et~al.}(2004)\citenamefont{Beck, Dagan,
  Milner, Gerber, and Deutscher}}]{Beck04a}
\bibinfo{author}{\bibnamefont{Beck}, \bibfnamefont{R.}},
  \bibinfo{author}{\bibfnamefont{Y.}~\bibnamefont{Dagan}},
  \bibinfo{author}{\bibfnamefont{A.}~\bibnamefont{Milner}},
  \bibinfo{author}{\bibfnamefont{A.}~\bibnamefont{Gerber}}, and
  \bibinfo{author}{\bibfnamefont{G.}~\bibnamefont{Deutscher}},
  \bibinfo{year}{2004}, \bibinfo{journal}{Phys.\ Rev.\ B}
  \textbf{\bibinfo{volume}{69}}(\bibinfo{number}{14}), \bibinfo{pages}{144506}.

\bibitem[{\citenamefont{Bednorz and M\"{u}ller}(1986)}]{Bednorz86}
\bibinfo{author}{\bibnamefont{Bednorz}, \bibfnamefont{J.}}, and
  \bibinfo{author}{\bibfnamefont{K.~A.} \bibnamefont{M\"{u}ller}},
  \bibinfo{year}{1986}, \bibinfo{journal}{Z. Phys. B}
  \textbf{\bibinfo{volume}{64}}, \bibinfo{eid}{189}.

\bibitem[{\citenamefont{Beesabathina}
  \emph{et~al.}(1993)\citenamefont{Beesabathina, Salamanca-Riba, Mao, Xi, and
  Venkatesan}}]{Beesabathina93a}
\bibinfo{author}{\bibnamefont{Beesabathina}, \bibfnamefont{D.~P.}},
  \bibinfo{author}{\bibfnamefont{L.}~\bibnamefont{Salamanca-Riba}},
  \bibinfo{author}{\bibfnamefont{S.~N.} \bibnamefont{Mao}},
  \bibinfo{author}{\bibfnamefont{X.~X.} \bibnamefont{Xi}}, and
  \bibinfo{author}{\bibfnamefont{T.}~\bibnamefont{Venkatesan}},
  \bibinfo{year}{1993}, \bibinfo{journal}{Appl.\ Phys.\ Lett.}
  \textbf{\bibinfo{volume}{62}}(\bibinfo{number}{23}), \bibinfo{pages}{3022}.

\bibitem[{\citenamefont{Billinge and Egami}(1993)}]{Billinge93a}
\bibinfo{author}{\bibnamefont{Billinge}, \bibfnamefont{S.~J.~L.}}, and
  \bibinfo{author}{\bibfnamefont{T.}~\bibnamefont{Egami}},
  \bibinfo{year}{1993}, \bibinfo{journal}{Phys.\ Rev.\ B}
  \textbf{\bibinfo{volume}{47}}(\bibinfo{number}{21}), \bibinfo{pages}{14386}.

\bibitem[{\citenamefont{Biswas} \emph{et~al.}(2002)\citenamefont{Biswas,
  Fournier, Qazilbash, Smolyaninova, Balci, and Greene}}]{Biswas02a}
\bibinfo{author}{\bibnamefont{Biswas}, \bibfnamefont{A.}},
  \bibinfo{author}{\bibfnamefont{P.}~\bibnamefont{Fournier}},
  \bibinfo{author}{\bibfnamefont{M.~M.} \bibnamefont{Qazilbash}},
  \bibinfo{author}{\bibfnamefont{V.~N.} \bibnamefont{Smolyaninova}},
  \bibinfo{author}{\bibfnamefont{H.}~\bibnamefont{Balci}}, and
  \bibinfo{author}{\bibfnamefont{R.~L.} \bibnamefont{Greene}},
  \bibinfo{year}{2002}, \bibinfo{journal}{Phys.\ Rev.\ Lett.}
  \textbf{\bibinfo{volume}{88}}(\bibinfo{number}{20}), \bibinfo{pages}{207004}.

\bibitem[{\citenamefont{Biswas} \emph{et~al.}(2001)\citenamefont{Biswas,
  Fournier, Smolyaninova, Budhani, Higgins, and Greene}}]{Biswas01a}
\bibinfo{author}{\bibnamefont{Biswas}, \bibfnamefont{A.}},
  \bibinfo{author}{\bibfnamefont{P.}~\bibnamefont{Fournier}},
  \bibinfo{author}{\bibfnamefont{V.~N.} \bibnamefont{Smolyaninova}},
  \bibinfo{author}{\bibfnamefont{R.~C.} \bibnamefont{Budhani}},
  \bibinfo{author}{\bibfnamefont{J.~S.} \bibnamefont{Higgins}}, and
  \bibinfo{author}{\bibfnamefont{R.~L.} \bibnamefont{Greene}},
  \bibinfo{year}{2001}, \bibinfo{journal}{Phys.\ Rev.\ B}
  \textbf{\bibinfo{volume}{64}}, \bibinfo{pages}{104519}.

\bibitem[{\citenamefont{Blonder} \emph{et~al.}(1982)\citenamefont{Blonder,
  Tinkham, and Klapwijk}}]{Blonder82a}
\bibinfo{author}{\bibnamefont{Blonder}, \bibfnamefont{G.~E.}},
  \bibinfo{author}{\bibfnamefont{M.}~\bibnamefont{Tinkham}}, and
  \bibinfo{author}{\bibfnamefont{T.~M.} \bibnamefont{Klapwijk}},
  \bibinfo{year}{1982}, \bibinfo{journal}{Phys.\ Rev.\ B}
  \textbf{\bibinfo{volume}{25}}(\bibinfo{number}{7}), \bibinfo{pages}{4515}.

\bibitem[{\citenamefont{Blumberg} \emph{et~al.}(1996)\citenamefont{Blumberg,
  Abbamonte, Klein, Lee, Ginsberg, Miller, and Zibold}}]{Blumberg96a}
\bibinfo{author}{\bibnamefont{Blumberg}, \bibfnamefont{G.}},
  \bibinfo{author}{\bibfnamefont{P.}~\bibnamefont{Abbamonte}},
  \bibinfo{author}{\bibfnamefont{M.~V.} \bibnamefont{Klein}},
  \bibinfo{author}{\bibfnamefont{W.~C.} \bibnamefont{Lee}},
  \bibinfo{author}{\bibfnamefont{D.~M.} \bibnamefont{Ginsberg}},
  \bibinfo{author}{\bibfnamefont{L.~L.} \bibnamefont{Miller}}, and
  \bibinfo{author}{\bibfnamefont{A.}~\bibnamefont{Zibold}},
  \bibinfo{year}{1996}, \bibinfo{journal}{Phys.\ Rev.\ B}
  \textbf{\bibinfo{volume}{53}}(\bibinfo{number}{18}), \bibinfo{pages}{R11930}.

\bibitem[{\citenamefont{Blumberg} \emph{et~al.}(1997)\citenamefont{Blumberg,
  Kang, and Klein}}]{Blumberg97a}
\bibinfo{author}{\bibnamefont{Blumberg}, \bibfnamefont{G.}},
  \bibinfo{author}{\bibfnamefont{M.}~\bibnamefont{Kang}}, and
  \bibinfo{author}{\bibfnamefont{M.~V.} \bibnamefont{Klein}},
  \bibinfo{year}{1997}, \bibinfo{journal}{Phys.\ Rev.\ Lett.}
  \textbf{\bibinfo{volume}{78}}(\bibinfo{number}{12}), \bibinfo{pages}{2461}.

\bibitem[{\citenamefont{Blumberg} \emph{et~al.}(2002)\citenamefont{Blumberg,
  Koitzsch, Gozar, Dennis, Kendziora, Fournier, and Greene}}]{Blumberg02a}
\bibinfo{author}{\bibnamefont{Blumberg}, \bibfnamefont{G.}},
  \bibinfo{author}{\bibfnamefont{A.}~\bibnamefont{Koitzsch}},
  \bibinfo{author}{\bibfnamefont{A.}~\bibnamefont{Gozar}},
  \bibinfo{author}{\bibfnamefont{B.~S.} \bibnamefont{Dennis}},
  \bibinfo{author}{\bibfnamefont{C.~A.} \bibnamefont{Kendziora}},
  \bibinfo{author}{\bibfnamefont{P.}~\bibnamefont{Fournier}}, and
  \bibinfo{author}{\bibfnamefont{R.~L.} \bibnamefont{Greene}},
  \bibinfo{year}{2002}, \bibinfo{journal}{Phys.\ Rev.\ Lett.}
  \textbf{\bibinfo{volume}{88}}, \bibinfo{pages}{107002}.

\bibitem[{\citenamefont{Blumberg} \emph{et~al.}(2003)\citenamefont{Blumberg,
  Koitzsch, Gozar, Dennis, Kendziora, Fournier, and Greene}}]{Blumberg03a}
\bibinfo{author}{\bibnamefont{Blumberg}, \bibfnamefont{G.}},
  \bibinfo{author}{\bibfnamefont{A.}~\bibnamefont{Koitzsch}},
  \bibinfo{author}{\bibfnamefont{A.}~\bibnamefont{Gozar}},
  \bibinfo{author}{\bibfnamefont{B.~S.} \bibnamefont{Dennis}},
  \bibinfo{author}{\bibfnamefont{C.~A.} \bibnamefont{Kendziora}},
  \bibinfo{author}{\bibfnamefont{P.}~\bibnamefont{Fournier}}, and
  \bibinfo{author}{\bibfnamefont{R.~L.} \bibnamefont{Greene}},
  \bibinfo{year}{2003}, \bibinfo{journal}{Phys.\ Rev.\ Lett.}
  \textbf{\bibinfo{volume}{90}}(\bibinfo{number}{14}), \bibinfo{pages}{149702}.

\bibitem[{\citenamefont{Boebinger} \emph{et~al.}(1996)\citenamefont{Boebinger,
  Ando, Passner, Kimura, Okuya, Shimoyama, Kishio, Tamasaku, Ichikawa, and
  Uchida}}]{Boebinger96a}
\bibinfo{author}{\bibnamefont{Boebinger}, \bibfnamefont{G.~S.}},
  \bibinfo{author}{\bibfnamefont{Y.}~\bibnamefont{Ando}},
  \bibinfo{author}{\bibfnamefont{A.}~\bibnamefont{Passner}},
  \bibinfo{author}{\bibfnamefont{T.}~\bibnamefont{Kimura}},
  \bibinfo{author}{\bibfnamefont{M.}~\bibnamefont{Okuya}},
  \bibinfo{author}{\bibfnamefont{J.}~\bibnamefont{Shimoyama}},
  \bibinfo{author}{\bibfnamefont{K.}~\bibnamefont{Kishio}},
  \bibinfo{author}{\bibfnamefont{K.}~\bibnamefont{Tamasaku}},
  \bibinfo{author}{\bibfnamefont{N.}~\bibnamefont{Ichikawa}}, and
  \bibinfo{author}{\bibfnamefont{S.}~\bibnamefont{Uchida}},
  \bibinfo{year}{1996}, \bibinfo{journal}{Phys.\ Rev.\ Lett.}
  \textbf{\bibinfo{volume}{77}}(\bibinfo{number}{27}), \bibinfo{pages}{5417}.

\bibitem[{\citenamefont{Bogdanov} \emph{et~al.}(2000)\citenamefont{Bogdanov,
  Lanzara, Kellar, Zhou, Lu, Zheng, Gu, Shimoyama, Kishio, Ikeda, Yoshizaki,
  Hussain} \emph{et~al.}}]{Bogdanov00a}
\bibinfo{author}{\bibnamefont{Bogdanov}, \bibfnamefont{P.~V.}},
  \bibinfo{author}{\bibfnamefont{A.}~\bibnamefont{Lanzara}},
  \bibinfo{author}{\bibfnamefont{S.~A.} \bibnamefont{Kellar}},
  \bibinfo{author}{\bibfnamefont{X.~J.} \bibnamefont{Zhou}},
  \bibinfo{author}{\bibfnamefont{E.~D.} \bibnamefont{Lu}},
  \bibinfo{author}{\bibfnamefont{W.~J.} \bibnamefont{Zheng}},
  \bibinfo{author}{\bibfnamefont{G.}~\bibnamefont{Gu}},
  \bibinfo{author}{\bibfnamefont{J.-I.} \bibnamefont{Shimoyama}},
  \bibinfo{author}{\bibfnamefont{K.}~\bibnamefont{Kishio}},
  \bibinfo{author}{\bibfnamefont{H.}~\bibnamefont{Ikeda}},
  \bibinfo{author}{\bibfnamefont{R.}~\bibnamefont{Yoshizaki}},
  \bibinfo{author}{\bibfnamefont{Z.}~\bibnamefont{Hussain}}, \emph{et~al.},
  \bibinfo{year}{2000}, \bibinfo{journal}{Phys.\ Rev.\ Lett.}
  \textbf{\bibinfo{volume}{85}}(\bibinfo{number}{12}), \bibinfo{pages}{2581}.

\bibitem[{\citenamefont{Bonn and Hardy}(1996)}]{Bonn96a}
\bibinfo{author}{\bibnamefont{Bonn}, \bibfnamefont{D.}}, and
  \bibinfo{author}{\bibfnamefont{W.}~\bibnamefont{Hardy}},
  \bibinfo{year}{1996}, in \emph{\bibinfo{booktitle}{Physical Properties of
  High Temperature Superconductors V}}, edited by
  \bibinfo{editor}{\bibfnamefont{D.}~\bibnamefont{Ginsberg}}
  (\bibinfo{publisher}{World Scientific, Singapore}), p.~\bibinfo{pages}{7}.

\bibitem[{\citenamefont{Borsa} \emph{et~al.}(1995)\citenamefont{Borsa, Carreta,
  Cho, Chou, Hu, Johnston, Lascialfari, Torgeson, Gooding, Salem, and
  Vos}}]{Borsa95a}
\bibinfo{author}{\bibnamefont{Borsa}, \bibfnamefont{F.}},
  \bibinfo{author}{\bibfnamefont{P.}~\bibnamefont{Carreta}},
  \bibinfo{author}{\bibfnamefont{J.~H.} \bibnamefont{Cho}},
  \bibinfo{author}{\bibfnamefont{F.~C.} \bibnamefont{Chou}},
  \bibinfo{author}{\bibfnamefont{Q.}~\bibnamefont{Hu}},
  \bibinfo{author}{\bibfnamefont{D.~C.} \bibnamefont{Johnston}},
  \bibinfo{author}{\bibfnamefont{A.}~\bibnamefont{Lascialfari}},
  \bibinfo{author}{\bibfnamefont{D.~R.} \bibnamefont{Torgeson}},
  \bibinfo{author}{\bibfnamefont{R.~J.} \bibnamefont{Gooding}},
  \bibinfo{author}{\bibfnamefont{N.~M.} \bibnamefont{Salem}}, and
  \bibinfo{author}{\bibfnamefont{K.~J.~E.} \bibnamefont{Vos}},
  \bibinfo{year}{1995}, \bibinfo{journal}{Phys.\ Rev.\ B}
  \textbf{\bibinfo{volume}{52}}(\bibinfo{number}{10}), \bibinfo{pages}{7334}.

\bibitem[{\citenamefont{Bourges}(1999)}]{Bourges99a}
\bibinfo{author}{\bibnamefont{Bourges}, \bibfnamefont{P.}},
  \bibinfo{year}{1999}, in \emph{\bibinfo{booktitle}{The gap Symmetry and
  Fluctuations in High Temperature Superconductors: Proceedings of NATO ASI
  summer school held September 1-13, 1997 in Carg`ese, France}}, edited by
  \bibinfo{editor}{\bibfnamefont{D.~P.} \bibnamefont{J.~Bok},
  \bibfnamefont{G.~Deutscher}} and
  \bibinfo{editor}{\bibfnamefont{S.}~\bibnamefont{Wolf}}
  (\bibinfo{publisher}{Plenum Press, 1998}), pp. \bibinfo{pages}{349--371}.

\bibitem[{\citenamefont{Bourges} \emph{et~al.}(1992)\citenamefont{Bourges,
  Boudarene, Petitgrand, and Galez}}]{Bourges92a}
\bibinfo{author}{\bibnamefont{Bourges}, \bibfnamefont{P.}},
  \bibinfo{author}{\bibfnamefont{L.}~\bibnamefont{Boudarene}},
  \bibinfo{author}{\bibfnamefont{D.}~\bibnamefont{Petitgrand}}, and
  \bibinfo{author}{\bibfnamefont{P.}~\bibnamefont{Galez}},
  \bibinfo{year}{1992}, \bibinfo{journal}{Physica B}
  \textbf{\bibinfo{volume}{180}}, \bibinfo{pages}{447}.

\bibitem[{\citenamefont{Bourges} \emph{et~al.}(1997)\citenamefont{Bourges,
  Casalta, Ivanov, and Petitgrand}}]{Bourges97a}
\bibinfo{author}{\bibnamefont{Bourges}, \bibfnamefont{P.}},
  \bibinfo{author}{\bibfnamefont{H.}~\bibnamefont{Casalta}},
  \bibinfo{author}{\bibfnamefont{A.~S.} \bibnamefont{Ivanov}}, and
  \bibinfo{author}{\bibfnamefont{D.}~\bibnamefont{Petitgrand}},
  \bibinfo{year}{1997}, \bibinfo{journal}{Phys.\ Rev.\ Lett.}
  \textbf{\bibinfo{volume}{79}}(\bibinfo{number}{24}), \bibinfo{pages}{4906}.

\bibitem[{\citenamefont{Braden} \emph{et~al.}(1994)\citenamefont{Braden,
  Paulius, Cousson, Vigoureux, Heger, Goukassov, Bourges, and
  Petitgrand}}]{Braden94a}
\bibinfo{author}{\bibnamefont{Braden}, \bibfnamefont{M.}},
  \bibinfo{author}{\bibfnamefont{W.}~\bibnamefont{Paulius}},
  \bibinfo{author}{\bibfnamefont{A.}~\bibnamefont{Cousson}},
  \bibinfo{author}{\bibfnamefont{P.}~\bibnamefont{Vigoureux}},
  \bibinfo{author}{\bibfnamefont{G.}~\bibnamefont{Heger}},
  \bibinfo{author}{\bibfnamefont{A.}~\bibnamefont{Goukassov}},
  \bibinfo{author}{\bibfnamefont{P.}~\bibnamefont{Bourges}}, and
  \bibinfo{author}{\bibfnamefont{D.}~\bibnamefont{Petitgrand}},
  \bibinfo{year}{1994}, \bibinfo{journal}{Europhys.\ Lett.}
  \textbf{\bibinfo{volume}{25}}, \bibinfo{pages}{625 }.

\bibitem[{\citenamefont{Braden} \emph{et~al.}(2005)\citenamefont{Braden,
  Pintschovius, Uefuji, and Yamada}}]{Braden05a}
\bibinfo{author}{\bibnamefont{Braden}, \bibfnamefont{M.}},
  \bibinfo{author}{\bibfnamefont{L.}~\bibnamefont{Pintschovius}},
  \bibinfo{author}{\bibfnamefont{T.}~\bibnamefont{Uefuji}}, and
  \bibinfo{author}{\bibfnamefont{K.}~\bibnamefont{Yamada}},
  \bibinfo{year}{2005}, \bibinfo{journal}{Phys.\ Rev.\ B}
  \textbf{\bibinfo{volume}{72}}(\bibinfo{number}{18}), \bibinfo{eid}{184517}.

\bibitem[{\citenamefont{Brinkmann}
  \emph{et~al.}(1996{\natexlab{a}})\citenamefont{Brinkmann, Rex, Bach, and
  Westerholt}}]{Brinkmann96b}
\bibinfo{author}{\bibnamefont{Brinkmann}, \bibfnamefont{M.}},
  \bibinfo{author}{\bibfnamefont{T.}~\bibnamefont{Rex}},
  \bibinfo{author}{\bibfnamefont{H.}~\bibnamefont{Bach}}, and
  \bibinfo{author}{\bibfnamefont{K.}~\bibnamefont{Westerholt}},
  \bibinfo{year}{1996}{\natexlab{a}}, \bibinfo{journal}{J.\ Crystal Growth}
  \textbf{\bibinfo{volume}{163}}(\bibinfo{number}{4}), \bibinfo{pages}{369 }.

\bibitem[{\citenamefont{Brinkmann}
  \emph{et~al.}(1996{\natexlab{b}})\citenamefont{Brinkmann, Rex, Stief, Bach,
  and Westerholt}}]{Brinkmann96a}
\bibinfo{author}{\bibnamefont{Brinkmann}, \bibfnamefont{M.}},
  \bibinfo{author}{\bibfnamefont{T.}~\bibnamefont{Rex}},
  \bibinfo{author}{\bibfnamefont{M.}~\bibnamefont{Stief}},
  \bibinfo{author}{\bibfnamefont{H.}~\bibnamefont{Bach}}, and
  \bibinfo{author}{\bibfnamefont{K.}~\bibnamefont{Westerholt}},
  \bibinfo{year}{1996}{\natexlab{b}}, \bibinfo{journal}{Physica C}
  \textbf{\bibinfo{volume}{269}}(\bibinfo{number}{1-2}), \bibinfo{pages}{76}.

\bibitem[{\citenamefont{Budhani} \emph{et~al.}(2002)\citenamefont{Budhani,
  Sullivan, Lobb, and Greene}}]{Budhani02a}
\bibinfo{author}{\bibnamefont{Budhani}, \bibfnamefont{R.~C.}},
  \bibinfo{author}{\bibfnamefont{M.~C.} \bibnamefont{Sullivan}},
  \bibinfo{author}{\bibfnamefont{C.~J.} \bibnamefont{Lobb}}, and
  \bibinfo{author}{\bibfnamefont{R.~L.} \bibnamefont{Greene}},
  \bibinfo{year}{2002}, \bibinfo{journal}{Phys.\ Rev.\ B}
  \textbf{\bibinfo{volume}{65}}(\bibinfo{number}{10}), \bibinfo{eid}{100517}.

\bibitem[{\citenamefont{Bulut and Scalapino}(1996)}]{Bulut96a}
\bibinfo{author}{\bibnamefont{Bulut}, \bibfnamefont{N.}}, and
  \bibinfo{author}{\bibfnamefont{D.~J.} \bibnamefont{Scalapino}},
  \bibinfo{year}{1996}, \bibinfo{journal}{Phys. Rev. B}
  \textbf{\bibinfo{volume}{53}}(\bibinfo{number}{9}), \bibinfo{pages}{5149}.

\bibitem[{\citenamefont{Calvani} \emph{et~al.}(1996)\citenamefont{Calvani,
  Capizzi, Lupi, Maselli, Paolone, and Roy}}]{Calvani96a}
\bibinfo{author}{\bibnamefont{Calvani}, \bibfnamefont{P.}},
  \bibinfo{author}{\bibfnamefont{M.}~\bibnamefont{Capizzi}},
  \bibinfo{author}{\bibfnamefont{S.}~\bibnamefont{Lupi}},
  \bibinfo{author}{\bibfnamefont{P.}~\bibnamefont{Maselli}},
  \bibinfo{author}{\bibfnamefont{A.}~\bibnamefont{Paolone}}, and
  \bibinfo{author}{\bibfnamefont{P.}~\bibnamefont{Roy}}, \bibinfo{year}{1996},
  \bibinfo{journal}{Phys.\ Rev.\ B}
  \textbf{\bibinfo{volume}{53}}(\bibinfo{number}{5}), \bibinfo{pages}{2756}.

\bibitem[{\citenamefont{Campuzano} \emph{et~al.}(2004)\citenamefont{Campuzano,
  Norman, and Randeria}}]{Campuzano04a}
\bibinfo{author}{\bibnamefont{Campuzano}, \bibfnamefont{J.~C.}},
  \bibinfo{author}{\bibfnamefont{M.~R.} \bibnamefont{Norman}}, and
  \bibinfo{author}{\bibfnamefont{M.}~\bibnamefont{Randeria}},
  \bibinfo{year}{2004}, in \emph{\bibinfo{booktitle}{Physics of Conventional
  and Unconventional Superconductors}}, edited by
  \bibinfo{editor}{\bibfnamefont{K.~H.} \bibnamefont{Bennemann}} and
  \bibinfo{editor}{\bibfnamefont{J.~B.} \bibnamefont{Ketterson}}
  (\bibinfo{publisher}{Springer-Verlag, Berlin}), volume~\bibinfo{volume}{II},
  pp. \bibinfo{pages}{167--273}.

\bibitem[{\citenamefont{Cappelluti}
  \emph{et~al.}(2009)\citenamefont{Cappelluti, Ciuchi, and
  Fratini}}]{Cappelluti08a}
\bibinfo{author}{\bibnamefont{Cappelluti}, \bibfnamefont{E.}},
  \bibinfo{author}{\bibfnamefont{S.}~\bibnamefont{Ciuchi}}, and
  \bibinfo{author}{\bibfnamefont{S.}~\bibnamefont{Fratini}},
  \bibinfo{year}{2009}, \bibinfo{journal}{Phys.\ Rev.\ B}
  \textbf{\bibinfo{volume}{79}}(\bibinfo{number}{1}), \bibinfo{eid}{012502}.

\bibitem[{\citenamefont{Carbotte} \emph{et~al.}(1999)\citenamefont{Carbotte,
  Basov, and Schachinger}}]{Carbotte99a}
\bibinfo{author}{\bibnamefont{Carbotte}, \bibfnamefont{J.}},
  \bibinfo{author}{\bibfnamefont{D.~N.} \bibnamefont{Basov}}, and
  \bibinfo{author}{\bibfnamefont{E.}~\bibnamefont{Schachinger}},
  \bibinfo{year}{1999}, \bibinfo{journal}{Nature}
  \textbf{\bibinfo{volume}{401}}, \bibinfo{pages}{354}.

\bibitem[{\citenamefont{Carlson} \emph{et~al.}(2003)\citenamefont{Carlson,
  Emery, Kivelson, and Orgad}}]{Carlson03a}
\bibinfo{author}{\bibnamefont{Carlson}, \bibfnamefont{E.~W.}},
  \bibinfo{author}{\bibfnamefont{V.~J.} \bibnamefont{Emery}},
  \bibinfo{author}{\bibfnamefont{S.~A.} \bibnamefont{Kivelson}}, and
  \bibinfo{author}{\bibfnamefont{D.}~\bibnamefont{Orgad}},
  \bibinfo{year}{2003}, \bibinfo{journal}{The Physics of Conventional and
  Unconventional Superconductors, edited by K. H. Bennemann, and J. B.
  Ketterson} .

\bibitem[{\citenamefont{Cassanho} \emph{et~al.}(1989)\citenamefont{Cassanho,
  Gabbe, and Jenssen}}]{Cassanho89a}
\bibinfo{author}{\bibnamefont{Cassanho}, \bibfnamefont{A.}},
  \bibinfo{author}{\bibfnamefont{D.~R.} \bibnamefont{Gabbe}}, and
  \bibinfo{author}{\bibfnamefont{H.~P.} \bibnamefont{Jenssen}},
  \bibinfo{year}{1989}, \bibinfo{journal}{J.\ Crystal Growth}
  \textbf{\bibinfo{volume}{96}}(\bibinfo{number}{4}), \bibinfo{pages}{999}.

\bibitem[{\citenamefont{Chakravarty}
  \emph{et~al.}(2001)\citenamefont{Chakravarty, Laughlin, Morr, and
  Nayak}}]{Chakravarty01a}
\bibinfo{author}{\bibnamefont{Chakravarty}, \bibfnamefont{S.}},
  \bibinfo{author}{\bibfnamefont{R.~B.} \bibnamefont{Laughlin}},
  \bibinfo{author}{\bibfnamefont{D.~K.} \bibnamefont{Morr}}, and
  \bibinfo{author}{\bibfnamefont{C.}~\bibnamefont{Nayak}},
  \bibinfo{year}{2001}, \bibinfo{journal}{Phys.\ Rev.\ B}
  \textbf{\bibinfo{volume}{63}}(\bibinfo{number}{9}), \bibinfo{pages}{094503}.

\bibitem[{\citenamefont{Chakravarty}
  \emph{et~al.}(1993)\citenamefont{Chakravarty, Sudbo, Anderson, and
  Strong}}]{Chakravarty93a}
\bibinfo{author}{\bibnamefont{Chakravarty}, \bibfnamefont{S.}},
  \bibinfo{author}{\bibfnamefont{A.}~\bibnamefont{Sudbo}},
  \bibinfo{author}{\bibfnamefont{P.~W.} \bibnamefont{Anderson}}, and
  \bibinfo{author}{\bibfnamefont{S.}~\bibnamefont{Strong}},
  \bibinfo{year}{1993}, \bibinfo{journal}{Science}
  \textbf{\bibinfo{volume}{261}}, \bibinfo{eid}{337}.

\bibitem[{\citenamefont{Chattopadhyay}
  \emph{et~al.}(1994)\citenamefont{Chattopadhyay, Lynn, Rosov, Grigereit,
  Barilo, and Zhigunov}}]{Chattopadhyay94a}
\bibinfo{author}{\bibnamefont{Chattopadhyay}, \bibfnamefont{T.}},
  \bibinfo{author}{\bibfnamefont{J.~W.} \bibnamefont{Lynn}},
  \bibinfo{author}{\bibfnamefont{N.}~\bibnamefont{Rosov}},
  \bibinfo{author}{\bibfnamefont{T.~E.} \bibnamefont{Grigereit}},
  \bibinfo{author}{\bibfnamefont{S.~N.} \bibnamefont{Barilo}}, and
  \bibinfo{author}{\bibfnamefont{D.~I.} \bibnamefont{Zhigunov}},
  \bibinfo{year}{1994}, \bibinfo{journal}{Phys.\ Rev.\ B}
  \textbf{\bibinfo{volume}{49}}(\bibinfo{number}{14}), \bibinfo{pages}{9944}.

\bibitem[{\citenamefont{Chen} \emph{et~al.}(2002)\citenamefont{Chen, Seneor,
  Yeh, Vasquez, Bell, Jung, Kim, Park, Kim, and Lee}}]{Chen02a}
\bibinfo{author}{\bibnamefont{Chen}, \bibfnamefont{C.-T.}},
  \bibinfo{author}{\bibfnamefont{P.}~\bibnamefont{Seneor}},
  \bibinfo{author}{\bibfnamefont{N.-C.} \bibnamefont{Yeh}},
  \bibinfo{author}{\bibfnamefont{R.~P.} \bibnamefont{Vasquez}},
  \bibinfo{author}{\bibfnamefont{L.~D.} \bibnamefont{Bell}},
  \bibinfo{author}{\bibfnamefont{C.~U.} \bibnamefont{Jung}},
  \bibinfo{author}{\bibfnamefont{J.~Y.} \bibnamefont{Kim}},
  \bibinfo{author}{\bibfnamefont{M.-S.} \bibnamefont{Park}},
  \bibinfo{author}{\bibfnamefont{H.-J.} \bibnamefont{Kim}}, and
  \bibinfo{author}{\bibfnamefont{S.-I.} \bibnamefont{Lee}},
  \bibinfo{year}{2002}, \bibinfo{journal}{Phys.\ Rev.\ Lett.}
  \textbf{\bibinfo{volume}{88}}(\bibinfo{number}{22}), \bibinfo{pages}{227002}.

\bibitem[{\citenamefont{Chen} \emph{et~al.}(2004)\citenamefont{Chen, Vafek,
  Yazdani, and Zhang}}]{Chen04a}
\bibinfo{author}{\bibnamefont{Chen}, \bibfnamefont{H.-D.}},
  \bibinfo{author}{\bibfnamefont{O.}~\bibnamefont{Vafek}},
  \bibinfo{author}{\bibfnamefont{A.}~\bibnamefont{Yazdani}}, and
  \bibinfo{author}{\bibfnamefont{S.-C.} \bibnamefont{Zhang}},
  \bibinfo{year}{2004}, \bibinfo{journal}{Phys.\ Rev.\ Lett.}
  \textbf{\bibinfo{volume}{93}}(\bibinfo{number}{18}), \bibinfo{eid}{187002}.

\bibitem[{\citenamefont{Chen} \emph{et~al.}(2005)\citenamefont{Chen, Wang,
  Wang, Luo, Luo, Liu, and Wang}}]{XHChen05a}
\bibinfo{author}{\bibnamefont{Chen}, \bibfnamefont{X.~H.}},
  \bibinfo{author}{\bibfnamefont{C.~H.} \bibnamefont{Wang}},
  \bibinfo{author}{\bibfnamefont{G.~Y.} \bibnamefont{Wang}},
  \bibinfo{author}{\bibfnamefont{X.~G.} \bibnamefont{Luo}},
  \bibinfo{author}{\bibfnamefont{J.~L.} \bibnamefont{Luo}},
  \bibinfo{author}{\bibfnamefont{G.~T.} \bibnamefont{Liu}}, and
  \bibinfo{author}{\bibfnamefont{N.~L.} \bibnamefont{Wang}},
  \bibinfo{year}{2005}, \bibinfo{journal}{Phys.\ Rev.\ B}
  \textbf{\bibinfo{volume}{72}}(\bibinfo{number}{6}), \bibinfo{eid}{064517}.

\bibitem[{\citenamefont{Cherny} \emph{et~al.}(1992)\citenamefont{Cherny,
  Khats\char39{}ko, Chouteau, Louis, Stepanov, Wyder, Barilo, and
  Zhigunov}}]{Cherny92a}
\bibinfo{author}{\bibnamefont{Cherny}, \bibfnamefont{A.~S.}},
  \bibinfo{author}{\bibfnamefont{E.~N.} \bibnamefont{Khats\char39{}ko}},
  \bibinfo{author}{\bibfnamefont{G.}~\bibnamefont{Chouteau}},
  \bibinfo{author}{\bibfnamefont{J.~M.} \bibnamefont{Louis}},
  \bibinfo{author}{\bibfnamefont{A.~A.} \bibnamefont{Stepanov}},
  \bibinfo{author}{\bibfnamefont{P.}~\bibnamefont{Wyder}},
  \bibinfo{author}{\bibfnamefont{S.~N.} \bibnamefont{Barilo}}, and
  \bibinfo{author}{\bibfnamefont{D.~I.} \bibnamefont{Zhigunov}},
  \bibinfo{year}{1992}, \bibinfo{journal}{Phys.\ Rev.\ B}
  \textbf{\bibinfo{volume}{45}}(\bibinfo{number}{21}), \bibinfo{pages}{12600}.

\bibitem[{\citenamefont{Chesca} \emph{et~al.}(2003)\citenamefont{Chesca,
  Ehrhardt, Mossle, Straub, Koelle, Kleiner, and Tsukada}}]{Chesca03a}
\bibinfo{author}{\bibnamefont{Chesca}, \bibfnamefont{B.}},
  \bibinfo{author}{\bibfnamefont{K.}~\bibnamefont{Ehrhardt}},
  \bibinfo{author}{\bibfnamefont{M.}~\bibnamefont{Mossle}},
  \bibinfo{author}{\bibfnamefont{R.}~\bibnamefont{Straub}},
  \bibinfo{author}{\bibfnamefont{D.}~\bibnamefont{Koelle}},
  \bibinfo{author}{\bibfnamefont{R.}~\bibnamefont{Kleiner}}, and
  \bibinfo{author}{\bibfnamefont{A.}~\bibnamefont{Tsukada}},
  \bibinfo{year}{2003}, \bibinfo{journal}{Phys.\ Rev.\ Lett.}
  \textbf{\bibinfo{volume}{90}}(\bibinfo{number}{5}), \bibinfo{eid}{057004}.

\bibitem[{\citenamefont{Chesca} \emph{et~al.}(2005)\citenamefont{Chesca,
  Seifried, Dahm, Schopohl, Koelle, Kleiner, and Tsukada}}]{Chesca05a}
\bibinfo{author}{\bibnamefont{Chesca}, \bibfnamefont{B.}},
  \bibinfo{author}{\bibfnamefont{M.}~\bibnamefont{Seifried}},
  \bibinfo{author}{\bibfnamefont{T.}~\bibnamefont{Dahm}},
  \bibinfo{author}{\bibfnamefont{N.}~\bibnamefont{Schopohl}},
  \bibinfo{author}{\bibfnamefont{D.}~\bibnamefont{Koelle}},
  \bibinfo{author}{\bibfnamefont{R.}~\bibnamefont{Kleiner}}, and
  \bibinfo{author}{\bibfnamefont{A.}~\bibnamefont{Tsukada}},
  \bibinfo{year}{2005}, \bibinfo{journal}{Phys.\ Rev.\ B}
  \textbf{\bibinfo{volume}{71}}(\bibinfo{number}{10}), \bibinfo{eid}{104504}.

\bibitem[{\citenamefont{Chiao} \emph{et~al.}(1999)\citenamefont{Chiao, Hill,
  Lupien, Popi\ifmmode~\acute{c}\else \'{c}\fi{}, Gagnon, and
  Taillefer}}]{Chiao99a}
\bibinfo{author}{\bibnamefont{Chiao}, \bibfnamefont{M.}},
  \bibinfo{author}{\bibfnamefont{R.~W.} \bibnamefont{Hill}},
  \bibinfo{author}{\bibfnamefont{C.}~\bibnamefont{Lupien}},
  \bibinfo{author}{\bibfnamefont{B.}~\bibnamefont{Popi\ifmmode~\acute{c}\else
  \'{c}\fi{}}}, \bibinfo{author}{\bibfnamefont{R.}~\bibnamefont{Gagnon}}, and
  \bibinfo{author}{\bibfnamefont{L.}~\bibnamefont{Taillefer}},
  \bibinfo{year}{1999}, \bibinfo{journal}{Phys.\ Rev.\ Lett.}
  \textbf{\bibinfo{volume}{82}}(\bibinfo{number}{14}), \bibinfo{pages}{2943}.

\bibitem[{\citenamefont{Chiao} \emph{et~al.}(2000)\citenamefont{Chiao, Hill,
  Lupien, Taillefer, Lambert, Gagnon, and Fournier}}]{Chiao00a}
\bibinfo{author}{\bibnamefont{Chiao}, \bibfnamefont{M.}},
  \bibinfo{author}{\bibfnamefont{R.~W.} \bibnamefont{Hill}},
  \bibinfo{author}{\bibfnamefont{C.}~\bibnamefont{Lupien}},
  \bibinfo{author}{\bibfnamefont{L.}~\bibnamefont{Taillefer}},
  \bibinfo{author}{\bibfnamefont{P.}~\bibnamefont{Lambert}},
  \bibinfo{author}{\bibfnamefont{R.}~\bibnamefont{Gagnon}}, and
  \bibinfo{author}{\bibfnamefont{P.}~\bibnamefont{Fournier}},
  \bibinfo{year}{2000}, \bibinfo{journal}{Phys.\ Rev.\ B}
  \textbf{\bibinfo{volume}{62}}(\bibinfo{number}{5}), \bibinfo{pages}{3554}.

\bibitem[{\citenamefont{Cho} \emph{et~al.}(2001)\citenamefont{Cho, Kim, Kim, O,
  Kim, and Stewart}}]{Cho01a}
\bibinfo{author}{\bibnamefont{Cho}, \bibfnamefont{B.~K.}},
  \bibinfo{author}{\bibfnamefont{J.~H.} \bibnamefont{Kim}},
  \bibinfo{author}{\bibfnamefont{Y.~J.} \bibnamefont{Kim}},
  \bibinfo{author}{\bibfnamefont{B.-h.} \bibnamefont{O}},
  \bibinfo{author}{\bibfnamefont{J.~S.} \bibnamefont{Kim}}, and
  \bibinfo{author}{\bibfnamefont{G.~R.} \bibnamefont{Stewart}},
  \bibinfo{year}{2001}, \bibinfo{journal}{Phys.\ Rev.\ B}
  \textbf{\bibinfo{volume}{63}}(\bibinfo{number}{21}), \bibinfo{pages}{214504}.

\bibitem[{\citenamefont{Chou} \emph{et~al.}(1993)\citenamefont{Chou, Borsa,
  Cho, Johnston, Lascialfari, Torgeson, and Ziolo}}]{Chou93a}
\bibinfo{author}{\bibnamefont{Chou}, \bibfnamefont{F.~C.}},
  \bibinfo{author}{\bibfnamefont{F.}~\bibnamefont{Borsa}},
  \bibinfo{author}{\bibfnamefont{J.~H.} \bibnamefont{Cho}},
  \bibinfo{author}{\bibfnamefont{D.~C.} \bibnamefont{Johnston}},
  \bibinfo{author}{\bibfnamefont{A.}~\bibnamefont{Lascialfari}},
  \bibinfo{author}{\bibfnamefont{D.~R.} \bibnamefont{Torgeson}}, and
  \bibinfo{author}{\bibfnamefont{J.}~\bibnamefont{Ziolo}},
  \bibinfo{year}{1993}, \bibinfo{journal}{Phys.\ Rev.\ Lett.}
  \textbf{\bibinfo{volume}{71}}(\bibinfo{number}{14}), \bibinfo{pages}{2323}.

\bibitem[{\citenamefont{Claesson} \emph{et~al.}(2004)\citenamefont{Claesson,
  Mnsson, Dallera, Venturini, Nadai, Brookes, and Tjernberg}}]{claesson04a}
\bibinfo{author}{\bibnamefont{Claesson}, \bibfnamefont{T.}},
  \bibinfo{author}{\bibfnamefont{M.}~\bibnamefont{Mnsson}},
  \bibinfo{author}{\bibfnamefont{C.}~\bibnamefont{Dallera}},
  \bibinfo{author}{\bibfnamefont{F.}~\bibnamefont{Venturini}},
  \bibinfo{author}{\bibfnamefont{C.~D.} \bibnamefont{Nadai}},
  \bibinfo{author}{\bibfnamefont{N.~B.} \bibnamefont{Brookes}}, and
  \bibinfo{author}{\bibfnamefont{O.}~\bibnamefont{Tjernberg}},
  \bibinfo{year}{2004}, \bibinfo{journal}{Phys.\ Rev.\ Lett.}
  \textbf{\bibinfo{volume}{93}}(\bibinfo{number}{13}), \bibinfo{eid}{136402}.

\bibitem[{\citenamefont{Coldea} \emph{et~al.}(2001)\citenamefont{Coldea,
  Hayden, Aeppli, Perring, Frost, Mason, Cheong, and Fisk}}]{Coldea01a}
\bibinfo{author}{\bibnamefont{Coldea}, \bibfnamefont{R.}},
  \bibinfo{author}{\bibfnamefont{S.~M.} \bibnamefont{Hayden}},
  \bibinfo{author}{\bibfnamefont{G.}~\bibnamefont{Aeppli}},
  \bibinfo{author}{\bibfnamefont{T.~G.} \bibnamefont{Perring}},
  \bibinfo{author}{\bibfnamefont{C.~D.} \bibnamefont{Frost}},
  \bibinfo{author}{\bibfnamefont{T.~E.} \bibnamefont{Mason}},
  \bibinfo{author}{\bibfnamefont{S.-W.} \bibnamefont{Cheong}}, and
  \bibinfo{author}{\bibfnamefont{Z.}~\bibnamefont{Fisk}}, \bibinfo{year}{2001},
  \bibinfo{journal}{Phys.\ Rev.\ Lett.}
  \textbf{\bibinfo{volume}{86}}(\bibinfo{number}{23}), \bibinfo{pages}{5377}.

\bibitem[{\citenamefont{Coleman}(2007)}]{Coleman07a}
\bibinfo{author}{\bibnamefont{Coleman}, \bibfnamefont{P.}},
  \bibinfo{year}{2007}, in \emph{\bibinfo{booktitle}{Handbook of Magnetism and
  Advanced Magnetic Materials}}, edited by
  \bibinfo{editor}{\bibfnamefont{H.}~\bibnamefont{Kronmuller}} and
  \bibinfo{editor}{\bibfnamefont{S.}~\bibnamefont{Parkin}}
  (\bibinfo{publisher}{J. Wiley and Sons}), volume~\bibinfo{volume}{1}, pp.
  \bibinfo{pages}{95--148}.

\bibitem[{\citenamefont{Cooper}(1996)}]{Cooper96a}
\bibinfo{author}{\bibnamefont{Cooper}, \bibfnamefont{J.~R.}},
  \bibinfo{year}{1996}, \bibinfo{journal}{Phys.\ Rev.\ B}
  \textbf{\bibinfo{volume}{54}}(\bibinfo{number}{6}), \bibinfo{pages}{R3753}.

\bibitem[{\citenamefont{C\^{o}t\'{e}}
  \emph{et~al.}(2008)\citenamefont{C\^{o}t\'{e}, Poirier, and
  Fournier}}]{Cote08a}
\bibinfo{author}{\bibnamefont{C\^{o}t\'{e}}, \bibfnamefont{G.}},
  \bibinfo{author}{\bibfnamefont{M.}~\bibnamefont{Poirier}}, and
  \bibinfo{author}{\bibfnamefont{P.}~\bibnamefont{Fournier}},
  \bibinfo{year}{2008}, \bibinfo{journal}{J.\ of Appl.\ Phys.}
  \textbf{\bibinfo{volume}{104}}(\bibinfo{number}{12}), \bibinfo{eid}{123914}.

\bibitem[{\citenamefont{Covington} \emph{et~al.}(1997)\citenamefont{Covington,
  Aprili, Paraoanu, Greene, Xu, Zhu, and Mirkin}}]{Covington97a}
\bibinfo{author}{\bibnamefont{Covington}, \bibfnamefont{M.}},
  \bibinfo{author}{\bibfnamefont{M.}~\bibnamefont{Aprili}},
  \bibinfo{author}{\bibfnamefont{E.}~\bibnamefont{Paraoanu}},
  \bibinfo{author}{\bibfnamefont{L.~H.} \bibnamefont{Greene}},
  \bibinfo{author}{\bibfnamefont{F.}~\bibnamefont{Xu}},
  \bibinfo{author}{\bibfnamefont{J.}~\bibnamefont{Zhu}}, and
  \bibinfo{author}{\bibfnamefont{C.~A.} \bibnamefont{Mirkin}},
  \bibinfo{year}{1997}, \bibinfo{journal}{Phys.\ Rev.\ Lett.}
  \textbf{\bibinfo{volume}{79}}(\bibinfo{number}{2}), \bibinfo{pages}{277}.

\bibitem[{\citenamefont{Covington} \emph{et~al.}(1996)\citenamefont{Covington,
  Scheuerer, Bloom, and Greene}}]{Covington96a}
\bibinfo{author}{\bibnamefont{Covington}, \bibfnamefont{M.}},
  \bibinfo{author}{\bibfnamefont{R.}~\bibnamefont{Scheuerer}},
  \bibinfo{author}{\bibfnamefont{K.}~\bibnamefont{Bloom}}, and
  \bibinfo{author}{\bibfnamefont{L.~H.} \bibnamefont{Greene}},
  \bibinfo{year}{1996}, \bibinfo{journal}{Appl.\ Phys.\ Lett.}
  \textbf{\bibinfo{volume}{68}}(\bibinfo{number}{12}), \bibinfo{pages}{1717}.

\bibitem[{\citenamefont{Cox} \emph{et~al.}(1989)\citenamefont{Cox, Goldman,
  Subramanian, Gopalakrishnan, and Sleight}}]{Cox89a}
\bibinfo{author}{\bibnamefont{Cox}, \bibfnamefont{D.~E.}},
  \bibinfo{author}{\bibfnamefont{A.~I.} \bibnamefont{Goldman}},
  \bibinfo{author}{\bibfnamefont{M.~A.} \bibnamefont{Subramanian}},
  \bibinfo{author}{\bibfnamefont{J.}~\bibnamefont{Gopalakrishnan}}, and
  \bibinfo{author}{\bibfnamefont{A.~W.} \bibnamefont{Sleight}},
  \bibinfo{year}{1989}, \bibinfo{journal}{Phys.\ Rev.\ B}
  \textbf{\bibinfo{volume}{40}}(\bibinfo{number}{10}), \bibinfo{pages}{6998}.

\bibitem[{\citenamefont{Cummins and Egdell}(1993)}]{Cummins93a}
\bibinfo{author}{\bibnamefont{Cummins}, \bibfnamefont{T.~R.}}, and
  \bibinfo{author}{\bibfnamefont{R.~G.} \bibnamefont{Egdell}},
  \bibinfo{year}{1993}, \bibinfo{journal}{Phys.\ Rev.\ B}
  \textbf{\bibinfo{volume}{48}}(\bibinfo{number}{9}), \bibinfo{pages}{6556}.

\bibitem[{\citenamefont{Curro} \emph{et~al.}(1997)\citenamefont{Curro, Imai,
  Slichter, and Dabrowski}}]{Curro97a}
\bibinfo{author}{\bibnamefont{Curro}, \bibfnamefont{N.~J.}},
  \bibinfo{author}{\bibfnamefont{T.}~\bibnamefont{Imai}},
  \bibinfo{author}{\bibfnamefont{C.~P.} \bibnamefont{Slichter}}, and
  \bibinfo{author}{\bibfnamefont{B.}~\bibnamefont{Dabrowski}},
  \bibinfo{year}{1997}, \bibinfo{journal}{Phys.\ Rev.\ B}
  \textbf{\bibinfo{volume}{56}}(\bibinfo{number}{2}), \bibinfo{pages}{877}.

\bibitem[{\citenamefont{Custers} \emph{et~al.}(2003)\citenamefont{Custers,
  Gegenwart, Wilhelm, Neumaier, Tokiwa, Trovarelli, Geibel, Steglich, P\'epin,
  and Coleman}}]{Custers03a}
\bibinfo{author}{\bibnamefont{Custers}, \bibfnamefont{J.}},
  \bibinfo{author}{\bibfnamefont{P.}~\bibnamefont{Gegenwart}},
  \bibinfo{author}{\bibfnamefont{H.}~\bibnamefont{Wilhelm}},
  \bibinfo{author}{\bibfnamefont{K.}~\bibnamefont{Neumaier}},
  \bibinfo{author}{\bibfnamefont{Y.}~\bibnamefont{Tokiwa}},
  \bibinfo{author}{\bibfnamefont{O.}~\bibnamefont{Trovarelli}},
  \bibinfo{author}{\bibfnamefont{C.}~\bibnamefont{Geibel}},
  \bibinfo{author}{\bibfnamefont{F.}~\bibnamefont{Steglich}},
  \bibinfo{author}{\bibfnamefont{C.}~\bibnamefont{P\'epin}}, and
  \bibinfo{author}{\bibfnamefont{P.}~\bibnamefont{Coleman}},
  \bibinfo{year}{2003}, \bibinfo{journal}{Nature}
  \textbf{\bibinfo{volume}{424}}, \bibinfo{pages}{524}.

\bibitem[{\citenamefont{Cyr-Choiniere}
  \emph{et~al.}(2009)\citenamefont{Cyr-Choiniere, Daou, Laliberte, LeBoeuf,
  Doiron-Leyraud, Chang, Yan, Cheng, Zhou, Goodenough, Pyon, Takayama}
  \emph{et~al.}}]{Cyr09a}
\bibinfo{author}{\bibnamefont{Cyr-Choiniere}, \bibfnamefont{O.}},
  \bibinfo{author}{\bibfnamefont{R.}~\bibnamefont{Daou}},
  \bibinfo{author}{\bibfnamefont{F.}~\bibnamefont{Laliberte}},
  \bibinfo{author}{\bibfnamefont{D.}~\bibnamefont{LeBoeuf}},
  \bibinfo{author}{\bibfnamefont{N.}~\bibnamefont{Doiron-Leyraud}},
  \bibinfo{author}{\bibfnamefont{J.}~\bibnamefont{Chang}},
  \bibinfo{author}{\bibfnamefont{J.-Q.} \bibnamefont{Yan}},
  \bibinfo{author}{\bibfnamefont{J.-G.} \bibnamefont{Cheng}},
  \bibinfo{author}{\bibfnamefont{J.-S.} \bibnamefont{Zhou}},
  \bibinfo{author}{\bibfnamefont{J.~B.} \bibnamefont{Goodenough}},
  \bibinfo{author}{\bibfnamefont{S.}~\bibnamefont{Pyon}},
  \bibinfo{author}{\bibfnamefont{T.}~\bibnamefont{Takayama}}, \emph{et~al.},
  \bibinfo{year}{2009}, \bibinfo{journal}{Nature}
  \textbf{\bibinfo{volume}{458}}(\bibinfo{number}{7239}), \bibinfo{pages}{743}.

\bibitem[{\citenamefont{Dagan}
  \emph{et~al.}(2005{\natexlab{a}})\citenamefont{Dagan, Barr, Fisher, Beck,
  Dhakal, Biswas, and Greene}}]{Dagan05b}
\bibinfo{author}{\bibnamefont{Dagan}, \bibfnamefont{Y.}},
  \bibinfo{author}{\bibfnamefont{M.~C.} \bibnamefont{Barr}},
  \bibinfo{author}{\bibfnamefont{W.~M.} \bibnamefont{Fisher}},
  \bibinfo{author}{\bibfnamefont{R.}~\bibnamefont{Beck}},
  \bibinfo{author}{\bibfnamefont{T.}~\bibnamefont{Dhakal}},
  \bibinfo{author}{\bibfnamefont{A.}~\bibnamefont{Biswas}}, and
  \bibinfo{author}{\bibfnamefont{R.~L.} \bibnamefont{Greene}},
  \bibinfo{year}{2005}{\natexlab{a}}, \bibinfo{journal}{Phys.\ Rev.\ Lett.}
  \textbf{\bibinfo{volume}{94}}(\bibinfo{number}{5}), \bibinfo{eid}{057005}.

\bibitem[{\citenamefont{Dagan} \emph{et~al.}(2007)\citenamefont{Dagan, Beck,
  and Greene}}]{Dagan07a}
\bibinfo{author}{\bibnamefont{Dagan}, \bibfnamefont{Y.}},
  \bibinfo{author}{\bibfnamefont{R.}~\bibnamefont{Beck}}, and
  \bibinfo{author}{\bibfnamefont{R.~L.} \bibnamefont{Greene}},
  \bibinfo{year}{2007}, \bibinfo{journal}{Phys.\ Rev.\ Lett.}
  \textbf{\bibinfo{volume}{99}}(\bibinfo{number}{14}), \bibinfo{eid}{147004}.

\bibitem[{\citenamefont{Dagan and Greene}(2004)}]{Dagan04b}
\bibinfo{author}{\bibnamefont{Dagan}, \bibfnamefont{Y.}}, and
  \bibinfo{author}{\bibfnamefont{R.~L.} \bibnamefont{Greene}},
  \bibinfo{year}{2004}, \bibinfo{journal}{unpublished} .

\bibitem[{\citenamefont{Dagan and Greene}(2007)}]{Dagan07b}
\bibinfo{author}{\bibnamefont{Dagan}, \bibfnamefont{Y.}}, and
  \bibinfo{author}{\bibfnamefont{R.~L.} \bibnamefont{Greene}},
  \bibinfo{year}{2007}, \bibinfo{journal}{Phys.\ Rev.\ B}
  \textbf{\bibinfo{volume}{76}}(\bibinfo{number}{2}), \bibinfo{eid}{024506}.

\bibitem[{\citenamefont{Dagan}
  \emph{et~al.}(2005{\natexlab{b}})\citenamefont{Dagan, Qazilbash, and
  Greene}}]{Dagan05a}
\bibinfo{author}{\bibnamefont{Dagan}, \bibfnamefont{Y.}},
  \bibinfo{author}{\bibfnamefont{M.~M.} \bibnamefont{Qazilbash}}, and
  \bibinfo{author}{\bibfnamefont{R.~L.} \bibnamefont{Greene}},
  \bibinfo{year}{2005}{\natexlab{b}}, \bibinfo{journal}{Phys.\ Rev.\ Lett.}
  \textbf{\bibinfo{volume}{94}}(\bibinfo{number}{18}), \bibinfo{pages}{187003}.

\bibitem[{\citenamefont{Dagan} \emph{et~al.}(2004)\citenamefont{Dagan,
  Qazilbash, Hill, Kulkarni, and Greene}}]{Dagan04a}
\bibinfo{author}{\bibnamefont{Dagan}, \bibfnamefont{Y.}},
  \bibinfo{author}{\bibfnamefont{M.~M.} \bibnamefont{Qazilbash}},
  \bibinfo{author}{\bibfnamefont{C.~P.} \bibnamefont{Hill}},
  \bibinfo{author}{\bibfnamefont{V.~N.} \bibnamefont{Kulkarni}}, and
  \bibinfo{author}{\bibfnamefont{R.~L.} \bibnamefont{Greene}},
  \bibinfo{year}{2004}, \bibinfo{journal}{Phys.\ Rev.\ Lett.}
  \textbf{\bibinfo{volume}{92}}(\bibinfo{number}{16}), \bibinfo{eid}{167001}.

\bibitem[{\citenamefont{Dai} \emph{et~al.}(2005)\citenamefont{Dai, Kang, Mook,
  Matsuura, Lynn, Kurita, Komiya, and Ando}}]{Dai05a}
\bibinfo{author}{\bibnamefont{Dai}, \bibfnamefont{P.}},
  \bibinfo{author}{\bibfnamefont{H.~J.} \bibnamefont{Kang}},
  \bibinfo{author}{\bibfnamefont{H.~A.} \bibnamefont{Mook}},
  \bibinfo{author}{\bibfnamefont{M.}~\bibnamefont{Matsuura}},
  \bibinfo{author}{\bibfnamefont{J.~W.} \bibnamefont{Lynn}},
  \bibinfo{author}{\bibfnamefont{Y.}~\bibnamefont{Kurita}},
  \bibinfo{author}{\bibfnamefont{S.}~\bibnamefont{Komiya}}, and
  \bibinfo{author}{\bibfnamefont{Y.}~\bibnamefont{Ando}}, \bibinfo{year}{2005},
  \bibinfo{journal}{Phys.\ Rev.\ B}
  \textbf{\bibinfo{volume}{71}}(\bibinfo{number}{10}), \bibinfo{eid}{100502}.

\bibitem[{\citenamefont{Dalichaouch}
  \emph{et~al.}(1993)\citenamefont{Dalichaouch, de~Andrade, and
  Maple}}]{Dalichaouch93a}
\bibinfo{author}{\bibnamefont{Dalichaouch}, \bibfnamefont{Y.}},
  \bibinfo{author}{\bibfnamefont{M.}~\bibnamefont{de~Andrade}}, and
  \bibinfo{author}{\bibfnamefont{M.}~\bibnamefont{Maple}},
  \bibinfo{year}{1993}, \bibinfo{journal}{Physica C}
  \textbf{\bibinfo{volume}{218}}(\bibinfo{number}{1-2}), \bibinfo{pages}{309 }.

\bibitem[{\citenamefont{Damascelli}
  \emph{et~al.}(2003)\citenamefont{Damascelli, Hussain, and
  Shen}}]{Damascelli03a}
\bibinfo{author}{\bibnamefont{Damascelli}, \bibfnamefont{A.}},
  \bibinfo{author}{\bibfnamefont{Z.}~\bibnamefont{Hussain}}, and
  \bibinfo{author}{\bibfnamefont{Z.-X.} \bibnamefont{Shen}},
  \bibinfo{year}{2003}, \bibinfo{journal}{Rev.\ Mod.\ Phys.}
  \textbf{\bibinfo{volume}{75}}(\bibinfo{number}{2}), \bibinfo{eid}{473}.

\bibitem[{\citenamefont{Das} \emph{et~al.}(2007)\citenamefont{Das, Markiewicz,
  and Bansil}}]{Das07a}
\bibinfo{author}{\bibnamefont{Das}, \bibfnamefont{T.}},
  \bibinfo{author}{\bibfnamefont{R.~S.} \bibnamefont{Markiewicz}}, and
  \bibinfo{author}{\bibfnamefont{A.}~\bibnamefont{Bansil}},
  \bibinfo{year}{2007}, \bibinfo{journal}{Phys.\ Rev.\ Lett.}
  \textbf{\bibinfo{volume}{98}}(\bibinfo{number}{19}), \bibinfo{eid}{197004}.

\bibitem[{\citenamefont{d'Astuto} \emph{et~al.}(2008)\citenamefont{d'Astuto,
  Dhalenne, Graf, Hoesch, Giura, Krisch, Berthet, Lanzara, and
  Shukla}}]{dAstuto08a}
\bibinfo{author}{\bibnamefont{d'Astuto}, \bibfnamefont{M.}},
  \bibinfo{author}{\bibfnamefont{G.}~\bibnamefont{Dhalenne}},
  \bibinfo{author}{\bibfnamefont{J.}~\bibnamefont{Graf}},
  \bibinfo{author}{\bibfnamefont{M.}~\bibnamefont{Hoesch}},
  \bibinfo{author}{\bibfnamefont{P.}~\bibnamefont{Giura}},
  \bibinfo{author}{\bibfnamefont{M.}~\bibnamefont{Krisch}},
  \bibinfo{author}{\bibfnamefont{P.}~\bibnamefont{Berthet}},
  \bibinfo{author}{\bibfnamefont{A.}~\bibnamefont{Lanzara}}, and
  \bibinfo{author}{\bibfnamefont{A.}~\bibnamefont{Shukla}},
  \bibinfo{year}{2008}, \bibinfo{journal}{Phys.\ Rev.\ B}
  \textbf{\bibinfo{volume}{78}}(\bibinfo{number}{14}), \bibinfo{eid}{140511}.

\bibitem[{\citenamefont{d'Astuto} \emph{et~al.}(2002)\citenamefont{d'Astuto,
  Mang, Giura, Shukla, Ghigna, Mirone, Braden, Greven, Krisch, and
  Sette}}]{dAstuto02a}
\bibinfo{author}{\bibnamefont{d'Astuto}, \bibfnamefont{M.}},
  \bibinfo{author}{\bibfnamefont{P.~K.} \bibnamefont{Mang}},
  \bibinfo{author}{\bibfnamefont{P.}~\bibnamefont{Giura}},
  \bibinfo{author}{\bibfnamefont{A.}~\bibnamefont{Shukla}},
  \bibinfo{author}{\bibfnamefont{P.}~\bibnamefont{Ghigna}},
  \bibinfo{author}{\bibfnamefont{A.}~\bibnamefont{Mirone}},
  \bibinfo{author}{\bibfnamefont{M.}~\bibnamefont{Braden}},
  \bibinfo{author}{\bibfnamefont{M.}~\bibnamefont{Greven}},
  \bibinfo{author}{\bibfnamefont{M.}~\bibnamefont{Krisch}}, and
  \bibinfo{author}{\bibfnamefont{F.}~\bibnamefont{Sette}},
  \bibinfo{year}{2002}, \bibinfo{journal}{Phys.\ Rev.\ Lett.}
  \textbf{\bibinfo{volume}{88}}(\bibinfo{number}{16}), \bibinfo{pages}{167002}.

\bibitem[{\citenamefont{Demler} \emph{et~al.}(2001)\citenamefont{Demler,
  Sachdev, and Zhang}}]{Demler01a}
\bibinfo{author}{\bibnamefont{Demler}, \bibfnamefont{E.}},
  \bibinfo{author}{\bibfnamefont{S.}~\bibnamefont{Sachdev}}, and
  \bibinfo{author}{\bibfnamefont{Y.}~\bibnamefont{Zhang}},
  \bibinfo{year}{2001}, \bibinfo{journal}{Phys.\ Rev.\ Lett.}
  \textbf{\bibinfo{volume}{87}}(\bibinfo{number}{6}), \bibinfo{pages}{067202}.

\bibitem[{\citenamefont{Deutscher}(1999)}]{Deutscher99a}
\bibinfo{author}{\bibnamefont{Deutscher}, \bibfnamefont{G.}},
  \bibinfo{year}{1999}, \bibinfo{journal}{Nature}
  \textbf{\bibinfo{volume}{397}}, \bibinfo{pages}{410}.

\bibitem[{\citenamefont{Deutscher}(2005)}]{Deutscher05a}
\bibinfo{author}{\bibnamefont{Deutscher}, \bibfnamefont{G.}},
  \bibinfo{year}{2005}, \bibinfo{journal}{Rev.\ Mod.\ Phys.}
  \textbf{\bibinfo{volume}{77}}(\bibinfo{number}{1}), \bibinfo{eid}{109}.

\bibitem[{\citenamefont{Devereaux} \emph{et~al.}(1994)\citenamefont{Devereaux,
  Einzel, Stadlober, Hackl, Leach, and Neumeier}}]{Devereaux94a}
\bibinfo{author}{\bibnamefont{Devereaux}, \bibfnamefont{T.~P.}},
  \bibinfo{author}{\bibfnamefont{D.}~\bibnamefont{Einzel}},
  \bibinfo{author}{\bibfnamefont{B.}~\bibnamefont{Stadlober}},
  \bibinfo{author}{\bibfnamefont{R.}~\bibnamefont{Hackl}},
  \bibinfo{author}{\bibfnamefont{D.~H.} \bibnamefont{Leach}}, and
  \bibinfo{author}{\bibfnamefont{J.~J.} \bibnamefont{Neumeier}},
  \bibinfo{year}{1994}, \bibinfo{journal}{Phys.\ Rev.\ Lett.}
  \textbf{\bibinfo{volume}{72}}(\bibinfo{number}{3}), \bibinfo{pages}{396}.

\bibitem[{\citenamefont{Devereaux and Hackl}(2007)}]{Devereaux07a}
\bibinfo{author}{\bibnamefont{Devereaux}, \bibfnamefont{T.~P.}}, and
  \bibinfo{author}{\bibfnamefont{R.}~\bibnamefont{Hackl}},
  \bibinfo{year}{2007}, \bibinfo{journal}{Rev.\ Mod.\ Phys.}
  \textbf{\bibinfo{volume}{79}}(\bibinfo{number}{1}), \bibinfo{eid}{175}.

\bibitem[{\citenamefont{Dierker} \emph{et~al.}(1983)\citenamefont{Dierker,
  Klein, Webb, and Fisk}}]{Dierker83a}
\bibinfo{author}{\bibnamefont{Dierker}, \bibfnamefont{S.~B.}},
  \bibinfo{author}{\bibfnamefont{M.~V.} \bibnamefont{Klein}},
  \bibinfo{author}{\bibfnamefont{G.~W.} \bibnamefont{Webb}}, and
  \bibinfo{author}{\bibfnamefont{Z.}~\bibnamefont{Fisk}}, \bibinfo{year}{1983},
  \bibinfo{journal}{Phys.\ Rev.\ Lett.}
  \textbf{\bibinfo{volume}{50}}(\bibinfo{number}{11}), \bibinfo{pages}{853}.

\bibitem[{\citenamefont{Ding} \emph{et~al.}(1996)\citenamefont{Ding, Norman,
  Campuzano, Randeria, Bellman, Yokoya, Takahashi, Mochiku, and
  Kadowaki}}]{Ding96a}
\bibinfo{author}{\bibnamefont{Ding}, \bibfnamefont{H.}},
  \bibinfo{author}{\bibfnamefont{M.~R.} \bibnamefont{Norman}},
  \bibinfo{author}{\bibfnamefont{J.~C.} \bibnamefont{Campuzano}},
  \bibinfo{author}{\bibfnamefont{M.}~\bibnamefont{Randeria}},
  \bibinfo{author}{\bibfnamefont{A.~F.} \bibnamefont{Bellman}},
  \bibinfo{author}{\bibfnamefont{T.}~\bibnamefont{Yokoya}},
  \bibinfo{author}{\bibfnamefont{T.}~\bibnamefont{Takahashi}},
  \bibinfo{author}{\bibfnamefont{T.}~\bibnamefont{Mochiku}}, and
  \bibinfo{author}{\bibfnamefont{K.}~\bibnamefont{Kadowaki}},
  \bibinfo{year}{1996}, \bibinfo{journal}{Phys.\ Rev.\ B}
  \textbf{\bibinfo{volume}{54}}(\bibinfo{number}{14}), \bibinfo{pages}{R9678}.

\bibitem[{\citenamefont{Doiron-Leyraud}
  \emph{et~al.}(2007)\citenamefont{Doiron-Leyraud, Proust, LeBoeuf, Levallois,
  Bonnemaison, Liang, Bonn, Hardy, and Taillefer}}]{Doiron07a}
\bibinfo{author}{\bibnamefont{Doiron-Leyraud}, \bibfnamefont{N.}},
  \bibinfo{author}{\bibfnamefont{C.}~\bibnamefont{Proust}},
  \bibinfo{author}{\bibfnamefont{D.}~\bibnamefont{LeBoeuf}},
  \bibinfo{author}{\bibfnamefont{J.}~\bibnamefont{Levallois}},
  \bibinfo{author}{\bibfnamefont{J.-B.} \bibnamefont{Bonnemaison}},
  \bibinfo{author}{\bibfnamefont{R.}~\bibnamefont{Liang}},
  \bibinfo{author}{\bibfnamefont{D.}~\bibnamefont{Bonn}},
  \bibinfo{author}{\bibfnamefont{W.}~\bibnamefont{Hardy}}, and
  \bibinfo{author}{\bibfnamefont{L.}~\bibnamefont{Taillefer}},
  \bibinfo{year}{2007}, \bibinfo{journal}{Nature}
  \textbf{\bibinfo{volume}{447}}, \bibinfo{pages}{565}.

\bibitem[{\citenamefont{Durst and Lee}(2000)}]{Durst00a}
\bibinfo{author}{\bibnamefont{Durst}, \bibfnamefont{A.~C.}}, and
  \bibinfo{author}{\bibfnamefont{P.~A.} \bibnamefont{Lee}},
  \bibinfo{year}{2000}, \bibinfo{journal}{Phys.\ Rev.\ B}
  \textbf{\bibinfo{volume}{62}}(\bibinfo{number}{2}), \bibinfo{pages}{1270}.

\bibitem[{\citenamefont{Emery and Kivelson}(1992)}]{Emery92a}
\bibinfo{author}{\bibnamefont{Emery}, \bibfnamefont{V.~J.}}, and
  \bibinfo{author}{\bibfnamefont{S.~A.} \bibnamefont{Kivelson}},
  \bibinfo{year}{1992}, \bibinfo{journal}{J.\ Phys.\ Chem.\ Sol.}
  \textbf{\bibinfo{volume}{53}}(\bibinfo{number}{12}), \bibinfo{pages}{1499}.

\bibitem[{\citenamefont{Emery and Kivelson}(1993)}]{Emery93a}
\bibinfo{author}{\bibnamefont{Emery}, \bibfnamefont{V.~J.}}, and
  \bibinfo{author}{\bibfnamefont{S.~A.} \bibnamefont{Kivelson}},
  \bibinfo{year}{1993}, \bibinfo{journal}{Physica C}
  \textbf{\bibinfo{volume}{209}}, \bibinfo{pages}{597}.

\bibitem[{\citenamefont{Emery and Kivelson}(1995)}]{Emery95a}
\bibinfo{author}{\bibnamefont{Emery}, \bibfnamefont{V.~J.}}, and
  \bibinfo{author}{\bibfnamefont{S.~A.} \bibnamefont{Kivelson}},
  \bibinfo{year}{1995}, \bibinfo{journal}{Nature}
  \textbf{\bibinfo{volume}{374}}, \bibinfo{eid}{434}.

\bibitem[{\citenamefont{Emery and Reiter}(1988)}]{Emery88a}
\bibinfo{author}{\bibnamefont{Emery}, \bibfnamefont{V.~J.}}, and
  \bibinfo{author}{\bibfnamefont{G.}~\bibnamefont{Reiter}},
  \bibinfo{year}{1988}, \bibinfo{journal}{Phys.\ Rev.\ B}
  \textbf{\bibinfo{volume}{38}}(\bibinfo{number}{16}), \bibinfo{pages}{11938}.

\bibitem[{\citenamefont{Endoh} \emph{et~al.}(1989)\citenamefont{Endoh, Matsuda,
  Yamada, Kakurai, Hidaka, Shirane, and Birgeneau}}]{Endoh89a}
\bibinfo{author}{\bibnamefont{Endoh}, \bibfnamefont{Y.}},
  \bibinfo{author}{\bibfnamefont{M.}~\bibnamefont{Matsuda}},
  \bibinfo{author}{\bibfnamefont{K.}~\bibnamefont{Yamada}},
  \bibinfo{author}{\bibfnamefont{K.}~\bibnamefont{Kakurai}},
  \bibinfo{author}{\bibfnamefont{Y.}~\bibnamefont{Hidaka}},
  \bibinfo{author}{\bibfnamefont{G.}~\bibnamefont{Shirane}}, and
  \bibinfo{author}{\bibfnamefont{R.~J.} \bibnamefont{Birgeneau}},
  \bibinfo{year}{1989}, \bibinfo{journal}{Phys.\ Rev.\ B}
  \textbf{\bibinfo{volume}{40}}(\bibinfo{number}{10}), \bibinfo{pages}{7023}.

\bibitem[{\citenamefont{Eskes} \emph{et~al.}(1989)\citenamefont{Eskes,
  Sawatzky, and Feiner}}]{Eskes89a}
\bibinfo{author}{\bibnamefont{Eskes}, \bibfnamefont{H.}},
  \bibinfo{author}{\bibfnamefont{G.}~\bibnamefont{Sawatzky}}, and
  \bibinfo{author}{\bibfnamefont{L.}~\bibnamefont{Feiner}},
  \bibinfo{year}{1989}, \bibinfo{journal}{Physica C}
  \textbf{\bibinfo{volume}{160}}, \bibinfo{pages}{424}.

\bibitem[{\citenamefont{Eun} \emph{et~al.}(2009)\citenamefont{Eun, Jia, and
  Chakravarty}}]{Eun09a}
\bibinfo{author}{\bibnamefont{Eun}, \bibfnamefont{J.}},
  \bibinfo{author}{\bibfnamefont{X.}~\bibnamefont{Jia}}, and
  \bibinfo{author}{\bibfnamefont{S.}~\bibnamefont{Chakravarty}},
  \bibinfo{year}{2009}, \bibinfo{journal}{arXiv:0912.0728v2} .

\bibitem[{\citenamefont{Fauqu\'{e}}
  \emph{et~al.}(2006)\citenamefont{Fauqu\'{e}, Sidis, Hinkov, Pailh\`{e}s, Lin,
  Chaud, and Bourges}}]{Fauque06a}
\bibinfo{author}{\bibnamefont{Fauqu\'{e}}, \bibfnamefont{B.}},
  \bibinfo{author}{\bibfnamefont{Y.}~\bibnamefont{Sidis}},
  \bibinfo{author}{\bibfnamefont{V.}~\bibnamefont{Hinkov}},
  \bibinfo{author}{\bibfnamefont{S.}~\bibnamefont{Pailh\`{e}s}},
  \bibinfo{author}{\bibfnamefont{C.~T.} \bibnamefont{Lin}},
  \bibinfo{author}{\bibfnamefont{X.}~\bibnamefont{Chaud}}, and
  \bibinfo{author}{\bibfnamefont{P.}~\bibnamefont{Bourges}},
  \bibinfo{year}{2006}, \bibinfo{journal}{Phys.\ Rev.\ Lett.}
  \textbf{\bibinfo{volume}{96}}(\bibinfo{number}{19}), \bibinfo{eid}{197001}.

\bibitem[{\citenamefont{Fischer} \emph{et~al.}(2007)\citenamefont{Fischer,
  Kugler, Maggio-Aprile, Berthod, and Renner}}]{Fischer07a}
\bibinfo{author}{\bibnamefont{Fischer}, \bibfnamefont{O.}},
  \bibinfo{author}{\bibfnamefont{M.}~\bibnamefont{Kugler}},
  \bibinfo{author}{\bibfnamefont{I.}~\bibnamefont{Maggio-Aprile}},
  \bibinfo{author}{\bibfnamefont{C.}~\bibnamefont{Berthod}}, and
  \bibinfo{author}{\bibfnamefont{C.}~\bibnamefont{Renner}},
  \bibinfo{year}{2007}, \bibinfo{journal}{Rev.\ Mod.\ Phys.}
  \textbf{\bibinfo{volume}{79}}(\bibinfo{number}{1}), \bibinfo{eid}{353}.

\bibitem[{\citenamefont{Fisher} \emph{et~al.}(1995)\citenamefont{Fisher,
  Kotliar, and Moeller}}]{Fisher95a}
\bibinfo{author}{\bibnamefont{Fisher}, \bibfnamefont{D.~S.}},
  \bibinfo{author}{\bibfnamefont{G.}~\bibnamefont{Kotliar}}, and
  \bibinfo{author}{\bibfnamefont{G.}~\bibnamefont{Moeller}},
  \bibinfo{year}{1995}, \bibinfo{journal}{Phys.\ Rev.\ B}
  \textbf{\bibinfo{volume}{52}}(\bibinfo{number}{24}), \bibinfo{pages}{17112}.

\bibitem[{\citenamefont{Fogelstr\"om}
  \emph{et~al.}(1997)\citenamefont{Fogelstr\"om, Rainer, and
  Sauls}}]{Fogelstrom97a}
\bibinfo{author}{\bibnamefont{Fogelstr\"om}, \bibfnamefont{M.}},
  \bibinfo{author}{\bibfnamefont{D.}~\bibnamefont{Rainer}}, and
  \bibinfo{author}{\bibfnamefont{J.~A.} \bibnamefont{Sauls}},
  \bibinfo{year}{1997}, \bibinfo{journal}{Phys.\ Rev.\ Lett.}
  \textbf{\bibinfo{volume}{79}}(\bibinfo{number}{2}), \bibinfo{pages}{281}.

\bibitem[{\citenamefont{Fontcuberta and Fabrega}(1996)}]{Fontcuberta96a}
\bibinfo{author}{\bibnamefont{Fontcuberta}, \bibfnamefont{J.~T.}}, and
  \bibinfo{author}{\bibfnamefont{L.}~\bibnamefont{Fabrega}},
  \bibinfo{year}{1996}, in \emph{\bibinfo{booktitle}{Studies of High
  Temperature Superconductors}}, edited by
  \bibinfo{editor}{\bibfnamefont{A.~V.} \bibnamefont{Narlikar}}
  (\bibinfo{publisher}{Nova Science Publishers, Commack, NY}),
  volume~\bibinfo{volume}{16}, p. \bibinfo{pages}{185}.

\bibitem[{\citenamefont{Fournier} \emph{et~al.}(2004)\citenamefont{Fournier,
  Gosselin, Savard, Renaud, Hetel, Richard, and Riou}}]{Fournier04a}
\bibinfo{author}{\bibnamefont{Fournier}, \bibfnamefont{P.}},
  \bibinfo{author}{\bibfnamefont{M.-E.} \bibnamefont{Gosselin}},
  \bibinfo{author}{\bibfnamefont{S.}~\bibnamefont{Savard}},
  \bibinfo{author}{\bibfnamefont{J.}~\bibnamefont{Renaud}},
  \bibinfo{author}{\bibfnamefont{I.}~\bibnamefont{Hetel}},
  \bibinfo{author}{\bibfnamefont{P.}~\bibnamefont{Richard}}, and
  \bibinfo{author}{\bibfnamefont{G.}~\bibnamefont{Riou}}, \bibinfo{year}{2004},
  \bibinfo{journal}{Phys.\ Rev.\ B} \textbf{\bibinfo{volume}{69}},
  \bibinfo{pages}{220501}.

\bibitem[{\citenamefont{Fournier and Greene}(2003)}]{Fournier03a}
\bibinfo{author}{\bibnamefont{Fournier}, \bibfnamefont{P.}}, and
  \bibinfo{author}{\bibfnamefont{R.~L.} \bibnamefont{Greene}},
  \bibinfo{year}{2003}, \bibinfo{journal}{Phys.\ Rev.\ B}
  \textbf{\bibinfo{volume}{68}}, \bibinfo{pages}{094507}.

\bibitem[{\citenamefont{Fournier} \emph{et~al.}(2000)\citenamefont{Fournier,
  Higgins, Balci, Maiser, Lobb, and Greene}}]{Fournier00a}
\bibinfo{author}{\bibnamefont{Fournier}, \bibfnamefont{P.}},
  \bibinfo{author}{\bibfnamefont{J.}~\bibnamefont{Higgins}},
  \bibinfo{author}{\bibfnamefont{H.}~\bibnamefont{Balci}},
  \bibinfo{author}{\bibfnamefont{E.}~\bibnamefont{Maiser}},
  \bibinfo{author}{\bibfnamefont{C.~J.} \bibnamefont{Lobb}}, and
  \bibinfo{author}{\bibfnamefont{R.~L.} \bibnamefont{Greene}},
  \bibinfo{year}{2000}, \bibinfo{journal}{Phys.\ Rev.\ B}
  \textbf{\bibinfo{volume}{62}}, \bibinfo{pages}{11 993}.

\bibitem[{\citenamefont{Fournier} \emph{et~al.}(1997)\citenamefont{Fournier,
  Jiang, Jiang, Mao, Venkatesan, Lobb, and Greene}}]{Fournier97a}
\bibinfo{author}{\bibnamefont{Fournier}, \bibfnamefont{P.}},
  \bibinfo{author}{\bibfnamefont{X.}~\bibnamefont{Jiang}},
  \bibinfo{author}{\bibfnamefont{W.}~\bibnamefont{Jiang}},
  \bibinfo{author}{\bibfnamefont{S.~N.} \bibnamefont{Mao}},
  \bibinfo{author}{\bibfnamefont{T.}~\bibnamefont{Venkatesan}},
  \bibinfo{author}{\bibfnamefont{C.~J.} \bibnamefont{Lobb}}, and
  \bibinfo{author}{\bibfnamefont{R.~L.} \bibnamefont{Greene}},
  \bibinfo{year}{1997}, \bibinfo{journal}{Phys.\ Rev.\ B}
  \textbf{\bibinfo{volume}{56}}, \bibinfo{pages}{14149}.

\bibitem[{\citenamefont{Fournier}
  \emph{et~al.}(1998{\natexlab{a}})\citenamefont{Fournier, Maiser, and
  Greene}}]{fournier98a}
\bibinfo{author}{\bibnamefont{Fournier}, \bibfnamefont{P.}},
  \bibinfo{author}{\bibfnamefont{E.}~\bibnamefont{Maiser}}, and
  \bibinfo{author}{\bibfnamefont{R.~L.} \bibnamefont{Greene}},
  \bibinfo{year}{1998}{\natexlab{a}}, in \emph{\bibinfo{booktitle}{The Gap
  Symmetry and Fluctuations in High-T$_c$ Superconductors}}, edited by
  \bibinfo{editor}{\bibfnamefont{J.}~\bibnamefont{Bok}},
  \bibinfo{editor}{\bibfnamefont{G.}~\bibnamefont{Deutscher}},
  \bibinfo{editor}{\bibfnamefont{D.}~\bibnamefont{Pavuna}}, and
  \bibinfo{editor}{\bibfnamefont{S.}~\bibnamefont{Wolf}}
  (\bibinfo{publisher}{NATO ASI Series B}), volume \bibinfo{volume}{371}, p.
  \bibinfo{pages}{145}.

\bibitem[{\citenamefont{Fournier}
  \emph{et~al.}(1998{\natexlab{b}})\citenamefont{Fournier, Mohanty, Maiser,
  Darzens, Venkatesan, Lobb, Czjzek, Webb, and Greene}}]{Fournier98b}
\bibinfo{author}{\bibnamefont{Fournier}, \bibfnamefont{P.}},
  \bibinfo{author}{\bibfnamefont{P.}~\bibnamefont{Mohanty}},
  \bibinfo{author}{\bibfnamefont{E.}~\bibnamefont{Maiser}},
  \bibinfo{author}{\bibfnamefont{S.}~\bibnamefont{Darzens}},
  \bibinfo{author}{\bibfnamefont{T.}~\bibnamefont{Venkatesan}},
  \bibinfo{author}{\bibfnamefont{C.~J.} \bibnamefont{Lobb}},
  \bibinfo{author}{\bibfnamefont{G.}~\bibnamefont{Czjzek}},
  \bibinfo{author}{\bibfnamefont{R.~A.} \bibnamefont{Webb}}, and
  \bibinfo{author}{\bibfnamefont{R.~L.} \bibnamefont{Greene}},
  \bibinfo{year}{1998}{\natexlab{b}}, \bibinfo{journal}{Phys.\ Rev.\ Lett.}
  \textbf{\bibinfo{volume}{81}}, \bibinfo{pages}{4720}.

\bibitem[{\citenamefont{Fujimori} \emph{et~al.}(1990)\citenamefont{Fujimori,
  Tokura, Eisaki, Takagi, Uchida, and Takayama-Muromachi}}]{Fujimori90a}
\bibinfo{author}{\bibnamefont{Fujimori}, \bibfnamefont{A.}},
  \bibinfo{author}{\bibfnamefont{Y.}~\bibnamefont{Tokura}},
  \bibinfo{author}{\bibfnamefont{H.}~\bibnamefont{Eisaki}},
  \bibinfo{author}{\bibfnamefont{H.}~\bibnamefont{Takagi}},
  \bibinfo{author}{\bibfnamefont{S.}~\bibnamefont{Uchida}}, and
  \bibinfo{author}{\bibfnamefont{E.}~\bibnamefont{Takayama-Muromachi}},
  \bibinfo{year}{1990}, \bibinfo{journal}{Phys.\ Rev.\ B}
  \textbf{\bibinfo{volume}{42}}(\bibinfo{number}{1}), \bibinfo{pages}{325}.

\bibitem[{\citenamefont{Fujita} \emph{et~al.}(2003)\citenamefont{Fujita, Kubo,
  Kuroshima, Uefuji, Kawashima, Yamada, Watanabe, and Nagamine}}]{Fujita03a}
\bibinfo{author}{\bibnamefont{Fujita}, \bibfnamefont{M.}},
  \bibinfo{author}{\bibfnamefont{T.}~\bibnamefont{Kubo}},
  \bibinfo{author}{\bibfnamefont{S.}~\bibnamefont{Kuroshima}},
  \bibinfo{author}{\bibfnamefont{T.}~\bibnamefont{Uefuji}},
  \bibinfo{author}{\bibfnamefont{K.}~\bibnamefont{Kawashima}},
  \bibinfo{author}{\bibfnamefont{K.}~\bibnamefont{Yamada}},
  \bibinfo{author}{\bibfnamefont{I.}~\bibnamefont{Watanabe}}, and
  \bibinfo{author}{\bibfnamefont{K.}~\bibnamefont{Nagamine}},
  \bibinfo{year}{2003}, \bibinfo{journal}{Phys.\ Rev.\ B}
  \textbf{\bibinfo{volume}{67}}(\bibinfo{number}{1}), \bibinfo{eid}{014514}.

\bibitem[{\citenamefont{Fujita} \emph{et~al.}(2006)\citenamefont{Fujita,
  Matsuda, Fak, Frost, and Yamada}}]{Fujita06a}
\bibinfo{author}{\bibnamefont{Fujita}, \bibfnamefont{M.}},
  \bibinfo{author}{\bibfnamefont{M.}~\bibnamefont{Matsuda}},
  \bibinfo{author}{\bibfnamefont{B.}~\bibnamefont{Fak}},
  \bibinfo{author}{\bibfnamefont{C.~D.} \bibnamefont{Frost}}, and
  \bibinfo{author}{\bibfnamefont{K.}~\bibnamefont{Yamada}},
  \bibinfo{year}{2006}, \bibinfo{journal}{J.\ Phys.\ Soc.\ Jap.}
  \textbf{\bibinfo{volume}{75}}, \bibinfo{pages}{093704}.

\bibitem[{\citenamefont{Fujita} \emph{et~al.}(2004)\citenamefont{Fujita,
  Matsuda, Katano, and Yamada}}]{Fujita04a}
\bibinfo{author}{\bibnamefont{Fujita}, \bibfnamefont{M.}},
  \bibinfo{author}{\bibfnamefont{M.}~\bibnamefont{Matsuda}},
  \bibinfo{author}{\bibfnamefont{S.}~\bibnamefont{Katano}}, and
  \bibinfo{author}{\bibfnamefont{K.}~\bibnamefont{Yamada}},
  \bibinfo{year}{2004}, \bibinfo{journal}{Phys.\ Rev.\ Lett.}
  \textbf{\bibinfo{volume}{93}}(\bibinfo{number}{14}), \bibinfo{eid}{147003}.

\bibitem[{\citenamefont{Fujita}
  \emph{et~al.}(2008{\natexlab{a}})\citenamefont{Fujita, Matsuda, Lee,
  Nakagawa, and Yamada}}]{Fujita08a}
\bibinfo{author}{\bibnamefont{Fujita}, \bibfnamefont{M.}},
  \bibinfo{author}{\bibfnamefont{M.}~\bibnamefont{Matsuda}},
  \bibinfo{author}{\bibfnamefont{S.-H.} \bibnamefont{Lee}},
  \bibinfo{author}{\bibfnamefont{M.}~\bibnamefont{Nakagawa}}, and
  \bibinfo{author}{\bibfnamefont{K.}~\bibnamefont{Yamada}},
  \bibinfo{year}{2008}{\natexlab{a}}, \bibinfo{journal}{Phys.\ Rev.\ Lett.}
  \textbf{\bibinfo{volume}{101}}(\bibinfo{number}{10}), \bibinfo{eid}{107003}.

\bibitem[{\citenamefont{Fujita}
  \emph{et~al.}(2008{\natexlab{b}})\citenamefont{Fujita, Nakagawa, Frost, and
  Yamada}}]{Fujita08b}
\bibinfo{author}{\bibnamefont{Fujita}, \bibfnamefont{M.}},
  \bibinfo{author}{\bibfnamefont{M.}~\bibnamefont{Nakagawa}},
  \bibinfo{author}{\bibfnamefont{C.~D.} \bibnamefont{Frost}}, and
  \bibinfo{author}{\bibfnamefont{K.}~\bibnamefont{Yamada}},
  \bibinfo{year}{2008}{\natexlab{b}}, \bibinfo{journal}{Journal of Physics:
  Conference Series} \textbf{\bibinfo{volume}{108}}, \bibinfo{pages}{012006
  (4pp)}.

\bibitem[{\citenamefont{Fukuda} \emph{et~al.}(2005)\citenamefont{Fukuda,
  Zalalutdinov, Kovacik, Minoguchi, Obata, Kubota, and Sonin}}]{Fukuda05a}
\bibinfo{author}{\bibnamefont{Fukuda}, \bibfnamefont{M.}},
  \bibinfo{author}{\bibfnamefont{M.~K.} \bibnamefont{Zalalutdinov}},
  \bibinfo{author}{\bibfnamefont{V.}~\bibnamefont{Kovacik}},
  \bibinfo{author}{\bibfnamefont{T.}~\bibnamefont{Minoguchi}},
  \bibinfo{author}{\bibfnamefont{T.}~\bibnamefont{Obata}},
  \bibinfo{author}{\bibfnamefont{M.}~\bibnamefont{Kubota}}, and
  \bibinfo{author}{\bibfnamefont{E.~B.} \bibnamefont{Sonin}},
  \bibinfo{year}{2005}, \bibinfo{journal}{Phys.\ Rev.\ B}
  \textbf{\bibinfo{volume}{71}}(\bibinfo{number}{21}), \bibinfo{eid}{212502}.

\bibitem[{\citenamefont{Gamayunov} \emph{et~al.}(1994)\citenamefont{Gamayunov,
  Tanaka, and Kojima}}]{Gamayunov94b}
\bibinfo{author}{\bibnamefont{Gamayunov}, \bibfnamefont{K.}},
  \bibinfo{author}{\bibfnamefont{I.}~\bibnamefont{Tanaka}}, and
  \bibinfo{author}{\bibfnamefont{H.}~\bibnamefont{Kojima}},
  \bibinfo{year}{1994}, \bibinfo{journal}{Physica C}
  \textbf{\bibinfo{volume}{228}}(\bibinfo{number}{1-2}), \bibinfo{pages}{58}.

\bibitem[{\citenamefont{Gantmakher}
  \emph{et~al.}(2003)\citenamefont{Gantmakher, Ermolov, Tsydynzhapov, Zhukov,
  and Baturina}}]{Gantmakher03a}
\bibinfo{author}{\bibnamefont{Gantmakher}, \bibfnamefont{V.~F.}},
  \bibinfo{author}{\bibfnamefont{S.~N.} \bibnamefont{Ermolov}},
  \bibinfo{author}{\bibfnamefont{G.~E.} \bibnamefont{Tsydynzhapov}},
  \bibinfo{author}{\bibfnamefont{A.~A.} \bibnamefont{Zhukov}}, and
  \bibinfo{author}{\bibfnamefont{T.~I.} \bibnamefont{Baturina}},
  \bibinfo{year}{2003}, \bibinfo{journal}{JETP Lett.}
  \textbf{\bibinfo{volume}{77}}(\bibinfo{number}{8}), \bibinfo{pages}{424}.

\bibitem[{\citenamefont{Gasumyants}
  \emph{et~al.}(1995)\citenamefont{Gasumyants, Kaidanov, and
  Vladimirskaya}}]{Gasumyants95a}
\bibinfo{author}{\bibnamefont{Gasumyants}, \bibfnamefont{V.}},
  \bibinfo{author}{\bibfnamefont{V.}~\bibnamefont{Kaidanov}}, and
  \bibinfo{author}{\bibfnamefont{E.}~\bibnamefont{Vladimirskaya}},
  \bibinfo{year}{1995}, \bibinfo{journal}{Physica C: Superconductivity and its
  applications} \textbf{\bibinfo{volume}{248}}(\bibinfo{number}{3-4}),
  \bibinfo{pages}{255}.

\bibitem[{\citenamefont{Gauthier} \emph{et~al.}(2007)\citenamefont{Gauthier,
  Gagn\'{e}, Renaud, Gosselin, Fournier, and Richard}}]{Gauthier07a}
\bibinfo{author}{\bibnamefont{Gauthier}, \bibfnamefont{J.}},
  \bibinfo{author}{\bibfnamefont{S.}~\bibnamefont{Gagn\'{e}}},
  \bibinfo{author}{\bibfnamefont{J.}~\bibnamefont{Renaud}},
  \bibinfo{author}{\bibfnamefont{M.-E.} \bibnamefont{Gosselin}},
  \bibinfo{author}{\bibfnamefont{P.}~\bibnamefont{Fournier}}, and
  \bibinfo{author}{\bibfnamefont{P.}~\bibnamefont{Richard}},
  \bibinfo{year}{2007}, \bibinfo{journal}{Phys.\ Rev.\ B}
  \textbf{\bibinfo{volume}{75}}(\bibinfo{number}{2}), \bibinfo{eid}{024424}.

\bibitem[{\citenamefont{Ghamaty} \emph{et~al.}(1989)\citenamefont{Ghamaty, Lee,
  Markert, Early, Bjørnholm, Seaman, and Maple}}]{Ghamaty89a}
\bibinfo{author}{\bibnamefont{Ghamaty}, \bibfnamefont{S.}},
  \bibinfo{author}{\bibfnamefont{B.}~\bibnamefont{Lee}},
  \bibinfo{author}{\bibfnamefont{J.}~\bibnamefont{Markert}},
  \bibinfo{author}{\bibfnamefont{E.}~\bibnamefont{Early}},
  \bibinfo{author}{\bibfnamefont{T.}~\bibnamefont{Bjørnholm}},
  \bibinfo{author}{\bibfnamefont{C.}~\bibnamefont{Seaman}}, and
  \bibinfo{author}{\bibfnamefont{M.}~\bibnamefont{Maple}},
  \bibinfo{year}{1989}, \bibinfo{journal}{Physica C}
  \textbf{\bibinfo{volume}{160}}(\bibinfo{number}{2}), \bibinfo{pages}{217 }.

\bibitem[{\citenamefont{Gilardi} \emph{et~al.}(2004)\citenamefont{Gilardi,
  Hiess, Momono, Oda, Ido, and Mesot}}]{Gilardi04a}
\bibinfo{author}{\bibnamefont{Gilardi}, \bibfnamefont{R.}},
  \bibinfo{author}{\bibfnamefont{A.}~\bibnamefont{Hiess}},
  \bibinfo{author}{\bibfnamefont{N.}~\bibnamefont{Momono}},
  \bibinfo{author}{\bibfnamefont{M.}~\bibnamefont{Oda}},
  \bibinfo{author}{\bibfnamefont{M.}~\bibnamefont{Ido}}, and
  \bibinfo{author}{\bibfnamefont{J.}~\bibnamefont{Mesot}},
  \bibinfo{year}{2004}, \bibinfo{journal}{Europhys.\ Lett.}
  \textbf{\bibinfo{volume}{66}}(\bibinfo{number}{6}), \bibinfo{eid}{840}.

\bibitem[{\citenamefont{Goldman and Markovic}(1998)}]{Goldman98a}
\bibinfo{author}{\bibnamefont{Goldman}, \bibfnamefont{A.~M.}}, and
  \bibinfo{author}{\bibfnamefont{N.}~\bibnamefont{Markovic}},
  \bibinfo{year}{1998}, \bibinfo{journal}{Physics Today}
  \textbf{\bibinfo{volume}{51}}(\bibinfo{number}{11}), \bibinfo{pages}{39}.

\bibitem[{\citenamefont{Gollnik and Naito}(1998)}]{Gollnik98a}
\bibinfo{author}{\bibnamefont{Gollnik}, \bibfnamefont{F.}}, and
  \bibinfo{author}{\bibfnamefont{M.}~\bibnamefont{Naito}},
  \bibinfo{year}{1998}, \bibinfo{journal}{Phys.\ Rev.\ B}
  \textbf{\bibinfo{volume}{58}}(\bibinfo{number}{17}), \bibinfo{pages}{11734}.

\bibitem[{\citenamefont{Gooding} \emph{et~al.}(1994)\citenamefont{Gooding, Vos,
  and Leung}}]{Gooding94a}
\bibinfo{author}{\bibnamefont{Gooding}, \bibfnamefont{R.~J.}},
  \bibinfo{author}{\bibfnamefont{K.~J.~E.} \bibnamefont{Vos}}, and
  \bibinfo{author}{\bibfnamefont{P.~W.} \bibnamefont{Leung}},
  \bibinfo{year}{1994}, \bibinfo{journal}{Phys.\ Rev.\ B}
  \textbf{\bibinfo{volume}{50}}(\bibinfo{number}{17}), \bibinfo{pages}{12866}.

\bibitem[{\citenamefont{Graf} \emph{et~al.}(2007)\citenamefont{Graf, Gweon,
  McElroy, Zhou, Jozwiak, Rotenberg, Bill, Sasagawa, Eisaki, Uchida, Takagi,
  Lee} \emph{et~al.}}]{Graf07a}
\bibinfo{author}{\bibnamefont{Graf}, \bibfnamefont{J.}},
  \bibinfo{author}{\bibfnamefont{G.-H.} \bibnamefont{Gweon}},
  \bibinfo{author}{\bibfnamefont{K.}~\bibnamefont{McElroy}},
  \bibinfo{author}{\bibfnamefont{S.~Y.} \bibnamefont{Zhou}},
  \bibinfo{author}{\bibfnamefont{C.}~\bibnamefont{Jozwiak}},
  \bibinfo{author}{\bibfnamefont{E.}~\bibnamefont{Rotenberg}},
  \bibinfo{author}{\bibfnamefont{A.}~\bibnamefont{Bill}},
  \bibinfo{author}{\bibfnamefont{T.}~\bibnamefont{Sasagawa}},
  \bibinfo{author}{\bibfnamefont{H.}~\bibnamefont{Eisaki}},
  \bibinfo{author}{\bibfnamefont{S.}~\bibnamefont{Uchida}},
  \bibinfo{author}{\bibfnamefont{H.}~\bibnamefont{Takagi}},
  \bibinfo{author}{\bibfnamefont{D.-H.} \bibnamefont{Lee}}, \emph{et~al.},
  \bibinfo{year}{2007}, \bibinfo{journal}{Phys.\ Rev.\ Lett.}
  \textbf{\bibinfo{volume}{98}}(\bibinfo{number}{6}), \bibinfo{eid}{067004}.

\bibitem[{\citenamefont{Granath}(2004)}]{Granath04a}
\bibinfo{author}{\bibnamefont{Granath}, \bibfnamefont{M.}},
  \bibinfo{year}{2004}, \bibinfo{journal}{Phys.\ Rev.\ B}
  \textbf{\bibinfo{volume}{69}}(\bibinfo{number}{21}), \bibinfo{pages}{214433}.

\bibitem[{\citenamefont{Gupta} \emph{et~al.}(1989)\citenamefont{Gupta, Koren,
  Tsuei, Segm\"{u}ller, and McGuire}}]{Gupta89a}
\bibinfo{author}{\bibnamefont{Gupta}, \bibfnamefont{A.}},
  \bibinfo{author}{\bibfnamefont{G.}~\bibnamefont{Koren}},
  \bibinfo{author}{\bibfnamefont{C.~C.} \bibnamefont{Tsuei}},
  \bibinfo{author}{\bibfnamefont{A.}~\bibnamefont{Segm\"{u}ller}}, and
  \bibinfo{author}{\bibfnamefont{T.~R.} \bibnamefont{McGuire}},
  \bibinfo{year}{1989}, \bibinfo{journal}{Appl.\ Phys.\ Lett.}
  \textbf{\bibinfo{volume}{55}}(\bibinfo{number}{17}), \bibinfo{pages}{1795}.

\bibitem[{\citenamefont{Gurvitch and Fiory}(1987)}]{Gurvitch87a}
\bibinfo{author}{\bibnamefont{Gurvitch}, \bibfnamefont{M.}}, and
  \bibinfo{author}{\bibfnamefont{A.~T.} \bibnamefont{Fiory}},
  \bibinfo{year}{1987}, \bibinfo{journal}{Phys.\ Rev.\ Lett.}
  \textbf{\bibinfo{volume}{59}}(\bibinfo{number}{12}), \bibinfo{pages}{1337}.

\bibitem[{\citenamefont{Hackl and Sachdev}(2009)}]{Hackl09a}
\bibinfo{author}{\bibnamefont{Hackl}, \bibfnamefont{A.}}, and
  \bibinfo{author}{\bibfnamefont{S.}~\bibnamefont{Sachdev}},
  \bibinfo{year}{2009}, \bibinfo{journal}{Physical Review B (Condensed Matter
  and Materials Physics)} \textbf{\bibinfo{volume}{79}}(\bibinfo{number}{23}),
  \bibinfo{eid}{235124} (pages~\bibinfo{numpages}{9}).

\bibitem[{\citenamefont{Hagen} \emph{et~al.}(1991)\citenamefont{Hagen, Peng,
  Li, and Greene}}]{Hagen91a}
\bibinfo{author}{\bibnamefont{Hagen}, \bibfnamefont{S.~J.}},
  \bibinfo{author}{\bibfnamefont{J.~L.} \bibnamefont{Peng}},
  \bibinfo{author}{\bibfnamefont{Z.~Y.} \bibnamefont{Li}}, and
  \bibinfo{author}{\bibfnamefont{R.~L.} \bibnamefont{Greene}},
  \bibinfo{year}{1991}, \bibinfo{journal}{Phys.\ Rev.\ B}
  \textbf{\bibinfo{volume}{43}}(\bibinfo{number}{16}), \bibinfo{pages}{13606}.

\bibitem[{\citenamefont{Hanaguri} \emph{et~al.}(2004)\citenamefont{Hanaguri,
  Lupien, Kohsaka, Lee, Azuma, Takano, Takagi, and Davis}}]{Hanaguri04a}
\bibinfo{author}{\bibnamefont{Hanaguri}, \bibfnamefont{T.}},
  \bibinfo{author}{\bibfnamefont{C.}~\bibnamefont{Lupien}},
  \bibinfo{author}{\bibfnamefont{Y.}~\bibnamefont{Kohsaka}},
  \bibinfo{author}{\bibfnamefont{D.~H.} \bibnamefont{Lee}},
  \bibinfo{author}{\bibfnamefont{M.}~\bibnamefont{Azuma}},
  \bibinfo{author}{\bibfnamefont{M.}~\bibnamefont{Takano}},
  \bibinfo{author}{\bibfnamefont{H.}~\bibnamefont{Takagi}}, and
  \bibinfo{author}{\bibfnamefont{J.~C.} \bibnamefont{Davis}},
  \bibinfo{year}{2004}, \bibinfo{journal}{Nature}
  \textbf{\bibinfo{volume}{430}}, \bibinfo{pages}{1001}.

\bibitem[{\citenamefont{Hardy} \emph{et~al.}(1993)\citenamefont{Hardy, Bonn,
  Morgan, Liang, and Zhang}}]{Hardy93a}
\bibinfo{author}{\bibnamefont{Hardy}, \bibfnamefont{W.~N.}},
  \bibinfo{author}{\bibfnamefont{D.~A.} \bibnamefont{Bonn}},
  \bibinfo{author}{\bibfnamefont{D.~C.} \bibnamefont{Morgan}},
  \bibinfo{author}{\bibfnamefont{R.}~\bibnamefont{Liang}}, and
  \bibinfo{author}{\bibfnamefont{K.}~\bibnamefont{Zhang}},
  \bibinfo{year}{1993}, \bibinfo{journal}{Phys.\ Rev.\ Lett.}
  \textbf{\bibinfo{volume}{70}}(\bibinfo{number}{25}), \bibinfo{pages}{3999}.

\bibitem[{\citenamefont{Harima} \emph{et~al.}(2003)\citenamefont{Harima,
  Fujimori, Sugaya, and Terasaki}}]{Harima03a}
\bibinfo{author}{\bibnamefont{Harima}, \bibfnamefont{N.}},
  \bibinfo{author}{\bibfnamefont{A.}~\bibnamefont{Fujimori}},
  \bibinfo{author}{\bibfnamefont{T.}~\bibnamefont{Sugaya}}, and
  \bibinfo{author}{\bibfnamefont{I.}~\bibnamefont{Terasaki}},
  \bibinfo{year}{2003}, \bibinfo{journal}{Phys.\ Rev.\ B}
  \textbf{\bibinfo{volume}{67}}(\bibinfo{number}{17}), \bibinfo{pages}{172501}.

\bibitem[{\citenamefont{Harima} \emph{et~al.}(2001)\citenamefont{Harima,
  Matsuno, Fujimori, Onose, Taguchi, and Tokura}}]{Harima01a}
\bibinfo{author}{\bibnamefont{Harima}, \bibfnamefont{N.}},
  \bibinfo{author}{\bibfnamefont{J.}~\bibnamefont{Matsuno}},
  \bibinfo{author}{\bibfnamefont{A.}~\bibnamefont{Fujimori}},
  \bibinfo{author}{\bibfnamefont{Y.}~\bibnamefont{Onose}},
  \bibinfo{author}{\bibfnamefont{Y.}~\bibnamefont{Taguchi}}, and
  \bibinfo{author}{\bibfnamefont{Y.}~\bibnamefont{Tokura}},
  \bibinfo{year}{2001}, \bibinfo{journal}{Phys.\ Rev.\ B}
  \textbf{\bibinfo{volume}{64}}(\bibinfo{number}{22}), \bibinfo{pages}{220507}.

\bibitem[{\citenamefont{van Harlingen}(1995)}]{vanHarlingen95a}
\bibinfo{author}{\bibnamefont{van Harlingen}, \bibfnamefont{D.~J.}},
  \bibinfo{year}{1995}, \bibinfo{journal}{Rev.\ Mod.\ Phys.}
  \textbf{\bibinfo{volume}{67}}(\bibinfo{number}{2}), \bibinfo{pages}{515}.

\bibitem[{\citenamefont{Harshman} \emph{et~al.}(1988)\citenamefont{Harshman,
  Aeppli, Espinosa, Cooper, Remeika, Ansaldo, Riseman, Williams, Noakes,
  Ellman, and Rosenbaum}}]{Harshman88a}
\bibinfo{author}{\bibnamefont{Harshman}, \bibfnamefont{D.~R.}},
  \bibinfo{author}{\bibfnamefont{G.}~\bibnamefont{Aeppli}},
  \bibinfo{author}{\bibfnamefont{G.~P.} \bibnamefont{Espinosa}},
  \bibinfo{author}{\bibfnamefont{A.~S.} \bibnamefont{Cooper}},
  \bibinfo{author}{\bibfnamefont{J.~P.} \bibnamefont{Remeika}},
  \bibinfo{author}{\bibfnamefont{E.~J.} \bibnamefont{Ansaldo}},
  \bibinfo{author}{\bibfnamefont{T.~M.} \bibnamefont{Riseman}},
  \bibinfo{author}{\bibfnamefont{D.~L.} \bibnamefont{Williams}},
  \bibinfo{author}{\bibfnamefont{D.~R.} \bibnamefont{Noakes}},
  \bibinfo{author}{\bibfnamefont{B.}~\bibnamefont{Ellman}}, and
  \bibinfo{author}{\bibfnamefont{T.~F.} \bibnamefont{Rosenbaum}},
  \bibinfo{year}{1988}, \bibinfo{journal}{Phys.\ Rev.\ B}
  \textbf{\bibinfo{volume}{38}}(\bibinfo{number}{1}), \bibinfo{pages}{852}.

\bibitem[{\citenamefont{Helm} \emph{et~al.}(2009)\citenamefont{Helm,
  Kartsovnik, Bartkowiak, Bittner, Lambacher, Erb, Wosnitza, and
  Gross}}]{Helm09a}
\bibinfo{author}{\bibnamefont{Helm}, \bibfnamefont{T.}},
  \bibinfo{author}{\bibfnamefont{M.~V.} \bibnamefont{Kartsovnik}},
  \bibinfo{author}{\bibfnamefont{M.}~\bibnamefont{Bartkowiak}},
  \bibinfo{author}{\bibfnamefont{N.}~\bibnamefont{Bittner}},
  \bibinfo{author}{\bibfnamefont{M.}~\bibnamefont{Lambacher}},
  \bibinfo{author}{\bibfnamefont{A.}~\bibnamefont{Erb}},
  \bibinfo{author}{\bibfnamefont{J.}~\bibnamefont{Wosnitza}}, and
  \bibinfo{author}{\bibfnamefont{R.}~\bibnamefont{Gross}},
  \bibinfo{year}{2009}, \bibinfo{journal}{Phys.\ Rev.\ Lett.}
  \textbf{\bibinfo{volume}{103}}(\bibinfo{number}{15}), \bibinfo{eid}{157002}.

\bibitem[{\citenamefont{Heyen} \emph{et~al.}(1991)\citenamefont{Heyen, Liu,
  Cardona, Piol, Melville, Paul, Mor\'an, and Alario-Franco}}]{Heyen91a}
\bibinfo{author}{\bibnamefont{Heyen}, \bibfnamefont{E.~T.}},
  \bibinfo{author}{\bibfnamefont{R.}~\bibnamefont{Liu}},
  \bibinfo{author}{\bibfnamefont{M.}~\bibnamefont{Cardona}},
  \bibinfo{author}{\bibfnamefont{S.}~\bibnamefont{Piol}},
  \bibinfo{author}{\bibfnamefont{R.~J.} \bibnamefont{Melville}},
  \bibinfo{author}{\bibfnamefont{D.~M.} \bibnamefont{Paul}},
  \bibinfo{author}{\bibfnamefont{E.}~\bibnamefont{Mor\'an}}, and
  \bibinfo{author}{\bibfnamefont{M.~A.} \bibnamefont{Alario-Franco}},
  \bibinfo{year}{1991}, \bibinfo{journal}{Phys.\ Rev.\ B}
  \textbf{\bibinfo{volume}{43}}(\bibinfo{number}{4}), \bibinfo{pages}{2857}.

\bibitem[{\citenamefont{Hidaka and Suzuki}(1989)}]{Hidaka89a}
\bibinfo{author}{\bibnamefont{Hidaka}, \bibfnamefont{Y.}}, and
  \bibinfo{author}{\bibfnamefont{M.}~\bibnamefont{Suzuki}},
  \bibinfo{year}{1989}, \bibinfo{journal}{Nature}
  \textbf{\bibinfo{volume}{338}}, \bibinfo{pages}{635 }.

\bibitem[{\citenamefont{Higgins} \emph{et~al.}(2006)\citenamefont{Higgins,
  Dagan, Barr, Weaver, and Greene}}]{higgins06a}
\bibinfo{author}{\bibnamefont{Higgins}, \bibfnamefont{J.~S.}},
  \bibinfo{author}{\bibfnamefont{Y.}~\bibnamefont{Dagan}},
  \bibinfo{author}{\bibfnamefont{M.~C.} \bibnamefont{Barr}},
  \bibinfo{author}{\bibfnamefont{B.~D.} \bibnamefont{Weaver}}, and
  \bibinfo{author}{\bibfnamefont{R.~L.} \bibnamefont{Greene}},
  \bibinfo{year}{2006}, \bibinfo{journal}{Phys.\ Rev.\ B}
  \textbf{\bibinfo{volume}{73}}(\bibinfo{number}{10}), \bibinfo{eid}{104510}.

\bibitem[{\citenamefont{Hilgenkamp and Mannhart}(2002)}]{Hilgenkamp02a}
\bibinfo{author}{\bibnamefont{Hilgenkamp}, \bibfnamefont{H.}}, and
  \bibinfo{author}{\bibfnamefont{J.}~\bibnamefont{Mannhart}},
  \bibinfo{year}{2002}, \bibinfo{journal}{Rev.\ Mod.\ Phys.}
  \textbf{\bibinfo{volume}{74}}(\bibinfo{number}{2}), \bibinfo{eid}{485}.

\bibitem[{\citenamefont{Hill} \emph{et~al.}(2001)\citenamefont{Hill, Proust,
  Taillefer, Fournier, and Greene}}]{Hill01a}
\bibinfo{author}{\bibnamefont{Hill}, \bibfnamefont{R.~W.}},
  \bibinfo{author}{\bibfnamefont{C.}~\bibnamefont{Proust}},
  \bibinfo{author}{\bibfnamefont{L.}~\bibnamefont{Taillefer}},
  \bibinfo{author}{\bibfnamefont{P.}~\bibnamefont{Fournier}}, and
  \bibinfo{author}{\bibfnamefont{R.~L.} \bibnamefont{Greene}},
  \bibinfo{year}{2001}, \bibinfo{journal}{Nature}
  \textbf{\bibinfo{volume}{414}}, \bibinfo{pages}{711}.

\bibitem[{\citenamefont{Hirsch and Marsiglio}(1989)}]{Hirsch89a}
\bibinfo{author}{\bibnamefont{Hirsch}, \bibfnamefont{J.}}, and
  \bibinfo{author}{\bibfnamefont{F.}~\bibnamefont{Marsiglio}},
  \bibinfo{year}{1989}, \bibinfo{journal}{Physica C}
  \textbf{\bibinfo{volume}{162-164}}, \bibinfo{pages}{591}.

\bibitem[{\citenamefont{Hirsch}(1995)}]{Hirsch95a}
\bibinfo{author}{\bibnamefont{Hirsch}, \bibfnamefont{J.~E.}},
  \bibinfo{year}{1995}, \bibinfo{journal}{Physica C}
  \textbf{\bibinfo{volume}{243}}, \bibinfo{pages}{319}.

\bibitem[{\citenamefont{Hirschfeld and Goldenfeld}(1993)}]{Hirschfeld93a}
\bibinfo{author}{\bibnamefont{Hirschfeld}, \bibfnamefont{P.~J.}}, and
  \bibinfo{author}{\bibfnamefont{N.}~\bibnamefont{Goldenfeld}},
  \bibinfo{year}{1993}, \bibinfo{journal}{Phys.\ Rev.\ B}
  \textbf{\bibinfo{volume}{48}}(\bibinfo{number}{6}), \bibinfo{pages}{4219}.

\bibitem[{\citenamefont{Hlubina and Rice}(1995)}]{hlubina95}
\bibinfo{author}{\bibnamefont{Hlubina}, \bibfnamefont{R.}}, and
  \bibinfo{author}{\bibfnamefont{T.~M.} \bibnamefont{Rice}},
  \bibinfo{year}{1995}, \bibinfo{journal}{Phys.\ Rev.\ B}
  \textbf{\bibinfo{volume}{51}}(\bibinfo{number}{14}), \bibinfo{pages}{9253}.

\bibitem[{\citenamefont{Hodges} \emph{et~al.}(1971)\citenamefont{Hodges, Smith,
  and Wilkins}}]{Hodges71a}
\bibinfo{author}{\bibnamefont{Hodges}, \bibfnamefont{C.}},
  \bibinfo{author}{\bibfnamefont{H.}~\bibnamefont{Smith}}, and
  \bibinfo{author}{\bibfnamefont{J.~W.} \bibnamefont{Wilkins}},
  \bibinfo{year}{1971}, \bibinfo{journal}{Phys.\ Rev.\ B}
  \textbf{\bibinfo{volume}{4}}(\bibinfo{number}{2}), \bibinfo{pages}{302}.

\bibitem[{\citenamefont{Homes} \emph{et~al.}(1997)\citenamefont{Homes, Clayman,
  Peng, and Greene}}]{Homes97a}
\bibinfo{author}{\bibnamefont{Homes}, \bibfnamefont{C.~C.}},
  \bibinfo{author}{\bibfnamefont{B.~P.} \bibnamefont{Clayman}},
  \bibinfo{author}{\bibfnamefont{J.~L.} \bibnamefont{Peng}}, and
  \bibinfo{author}{\bibfnamefont{R.~L.} \bibnamefont{Greene}},
  \bibinfo{year}{1997}, \bibinfo{journal}{Phys.\ Rev.\ B}
  \textbf{\bibinfo{volume}{56}}(\bibinfo{number}{9}), \bibinfo{pages}{5525}.

\bibitem[{\citenamefont{Homes} \emph{et~al.}(2006)\citenamefont{Homes, Lobo,
  Fournier, Zimmers, and Greene}}]{Homes06a}
\bibinfo{author}{\bibnamefont{Homes}, \bibfnamefont{C.~C.}},
  \bibinfo{author}{\bibfnamefont{R.~P. S.~M.} \bibnamefont{Lobo}},
  \bibinfo{author}{\bibfnamefont{P.}~\bibnamefont{Fournier}},
  \bibinfo{author}{\bibfnamefont{A.}~\bibnamefont{Zimmers}}, and
  \bibinfo{author}{\bibfnamefont{R.~L.} \bibnamefont{Greene}},
  \bibinfo{year}{2006}, \bibinfo{journal}{Phys.\ Rev.\ B}
  \textbf{\bibinfo{volume}{74}}(\bibinfo{number}{21}), \bibinfo{eid}{214515}.

\bibitem[{\citenamefont{Howald} \emph{et~al.}(2001)\citenamefont{Howald,
  Fournier, and Kapitulnik}}]{Howald01a}
\bibinfo{author}{\bibnamefont{Howald}, \bibfnamefont{C.}},
  \bibinfo{author}{\bibfnamefont{P.}~\bibnamefont{Fournier}}, and
  \bibinfo{author}{\bibfnamefont{A.}~\bibnamefont{Kapitulnik}},
  \bibinfo{year}{2001}, \bibinfo{journal}{Phys.\ Rev.\ B}
  \textbf{\bibinfo{volume}{64}}(\bibinfo{number}{10}), \bibinfo{pages}{100504}.

\bibitem[{\citenamefont{Hozoi} \emph{et~al.}(2008)\citenamefont{Hozoi, Laad,
  and Fulde}}]{Hozoi08a}
\bibinfo{author}{\bibnamefont{Hozoi}, \bibfnamefont{L.}},
  \bibinfo{author}{\bibfnamefont{M.~S.} \bibnamefont{Laad}}, and
  \bibinfo{author}{\bibfnamefont{P.}~\bibnamefont{Fulde}},
  \bibinfo{year}{2008}, \bibinfo{journal}{Phys.\ Rev.\ B}
  \textbf{\bibinfo{volume}{78}}(\bibinfo{number}{16}), \bibinfo{eid}{165107}.

\bibitem[{\citenamefont{Hu}(1994)}]{Hu94a}
\bibinfo{author}{\bibnamefont{Hu}, \bibfnamefont{C.-R.}}, \bibinfo{year}{1994},
  \bibinfo{journal}{Phys.\ Rev.\ Lett.}
  \textbf{\bibinfo{volume}{72}}(\bibinfo{number}{10}), \bibinfo{pages}{1526}.

\bibitem[{\citenamefont{Hui and Berker}(1989)}]{Hui89a}
\bibinfo{author}{\bibnamefont{Hui}, \bibfnamefont{K.}}, and
  \bibinfo{author}{\bibfnamefont{A.~N.} \bibnamefont{Berker}},
  \bibinfo{year}{1989}, \bibinfo{journal}{Phys.\ Rev.\ Lett.}
  \textbf{\bibinfo{volume}{62}}(\bibinfo{number}{21}), \bibinfo{pages}{2507}.

\bibitem[{\citenamefont{Hundley} \emph{et~al.}(1989)\citenamefont{Hundley,
  Thompson, Cheong, Fisk, and Oseroff}}]{Hundley89a}
\bibinfo{author}{\bibnamefont{Hundley}, \bibfnamefont{M.~F.}},
  \bibinfo{author}{\bibfnamefont{J.~D.} \bibnamefont{Thompson}},
  \bibinfo{author}{\bibfnamefont{S.-W.} \bibnamefont{Cheong}},
  \bibinfo{author}{\bibfnamefont{Z.}~\bibnamefont{Fisk}}, and
  \bibinfo{author}{\bibfnamefont{B.}~\bibnamefont{Oseroff}},
  \bibinfo{year}{1989}, \bibinfo{journal}{Physica C}
  \textbf{\bibinfo{volume}{158}}, \bibinfo{pages}{102}.

\bibitem[{\citenamefont{Hybertsen} \emph{et~al.}(1990)\citenamefont{Hybertsen,
  Stechel, Schluter, and Jennison}}]{Hybertsen90a}
\bibinfo{author}{\bibnamefont{Hybertsen}, \bibfnamefont{M.~S.}},
  \bibinfo{author}{\bibfnamefont{E.~B.} \bibnamefont{Stechel}},
  \bibinfo{author}{\bibfnamefont{M.}~\bibnamefont{Schluter}}, and
  \bibinfo{author}{\bibfnamefont{D.~R.} \bibnamefont{Jennison}},
  \bibinfo{year}{1990}, \bibinfo{journal}{Phys.\ Rev.\ B}
  \textbf{\bibinfo{volume}{41}}(\bibinfo{number}{16}), \bibinfo{pages}{11068}.

\bibitem[{\citenamefont{Ignatov} \emph{et~al.}(1998)\citenamefont{Ignatov,
  Ivanov, Menushenkov, Iacobucci, and Lagarde}}]{Yu98a}
\bibinfo{author}{\bibnamefont{Ignatov}, \bibfnamefont{A.~Y.}},
  \bibinfo{author}{\bibfnamefont{A.~A.} \bibnamefont{Ivanov}},
  \bibinfo{author}{\bibfnamefont{A.~P.} \bibnamefont{Menushenkov}},
  \bibinfo{author}{\bibfnamefont{S.}~\bibnamefont{Iacobucci}}, and
  \bibinfo{author}{\bibfnamefont{P.}~\bibnamefont{Lagarde}},
  \bibinfo{year}{1998}, \bibinfo{journal}{Phys.\ Rev.\ B}
  \textbf{\bibinfo{volume}{57}}(\bibinfo{number}{14}), \bibinfo{pages}{8671}.

\bibitem[{\citenamefont{Ikeda}
  \emph{et~al.}(2009{\natexlab{a}})\citenamefont{Ikeda, Takizawa, Yoshida,
  Fujimori, Segawa, and Ando}}]{Ikeda09c}
\bibinfo{author}{\bibnamefont{Ikeda}, \bibfnamefont{M.}},
  \bibinfo{author}{\bibfnamefont{M.}~\bibnamefont{Takizawa}},
  \bibinfo{author}{\bibfnamefont{T.}~\bibnamefont{Yoshida}},
  \bibinfo{author}{\bibfnamefont{A.}~\bibnamefont{Fujimori}},
  \bibinfo{author}{\bibfnamefont{K.}~\bibnamefont{Segawa}}, and
  \bibinfo{author}{\bibfnamefont{Y.}~\bibnamefont{Ando}},
  \bibinfo{year}{2009}{\natexlab{a}}, \bibinfo{journal}{unpublished}
  \textbf{\bibinfo{volume}{arXiv:1001.0102v1}}.

\bibitem[{\citenamefont{Ikeda}
  \emph{et~al.}(2009{\natexlab{b}})\citenamefont{Ikeda, Yoshida, Fujimori,
  Kubota, Ono, Das, Saha-Dasgupta, Unozawa, Kaga, Sasagawa, and
  Takagi}}]{Ikeda09a}
\bibinfo{author}{\bibnamefont{Ikeda}, \bibfnamefont{M.}},
  \bibinfo{author}{\bibfnamefont{T.}~\bibnamefont{Yoshida}},
  \bibinfo{author}{\bibfnamefont{A.}~\bibnamefont{Fujimori}},
  \bibinfo{author}{\bibfnamefont{M.}~\bibnamefont{Kubota}},
  \bibinfo{author}{\bibfnamefont{K.}~\bibnamefont{Ono}},
  \bibinfo{author}{\bibfnamefont{H.}~\bibnamefont{Das}},
  \bibinfo{author}{\bibfnamefont{T.}~\bibnamefont{Saha-Dasgupta}},
  \bibinfo{author}{\bibfnamefont{K.}~\bibnamefont{Unozawa}},
  \bibinfo{author}{\bibfnamefont{Y.}~\bibnamefont{Kaga}},
  \bibinfo{author}{\bibfnamefont{T.}~\bibnamefont{Sasagawa}}, and
  \bibinfo{author}{\bibfnamefont{H.}~\bibnamefont{Takagi}},
  \bibinfo{year}{2009}{\natexlab{b}}, \bibinfo{journal}{Phys.\ Rev.\ B}
  \textbf{\bibinfo{volume}{80}}(\bibinfo{number}{1}), \bibinfo{eid}{014510}.

\bibitem[{\citenamefont{Ikeda} \emph{et~al.}(2007)\citenamefont{Ikeda, Yoshida,
  Fujimori, Kubota, Ono, Unozawa, Sasagawa, and Takagi}}]{Ikeda07a}
\bibinfo{author}{\bibnamefont{Ikeda}, \bibfnamefont{M.}},
  \bibinfo{author}{\bibfnamefont{T.}~\bibnamefont{Yoshida}},
  \bibinfo{author}{\bibfnamefont{A.}~\bibnamefont{Fujimori}},
  \bibinfo{author}{\bibfnamefont{M.}~\bibnamefont{Kubota}},
  \bibinfo{author}{\bibfnamefont{K.}~\bibnamefont{Ono}},
  \bibinfo{author}{\bibfnamefont{K.}~\bibnamefont{Unozawa}},
  \bibinfo{author}{\bibfnamefont{T.}~\bibnamefont{Sasagawa}}, and
  \bibinfo{author}{\bibfnamefont{H.}~\bibnamefont{Takagi}},
  \bibinfo{year}{2007}, \bibinfo{journal}{J. Supercond. Nov. Mag.}
  \textbf{\bibinfo{volume}{20}}, \bibinfo{pages}{563}.

\bibitem[{\citenamefont{Ikeda} \emph{et~al.}(1993)\citenamefont{Ikeda, Hiroi,
  Azuma, Takano, Bando, and Takeda}}]{Ikeda93a}
\bibinfo{author}{\bibnamefont{Ikeda}, \bibfnamefont{N.}},
  \bibinfo{author}{\bibfnamefont{Z.}~\bibnamefont{Hiroi}},
  \bibinfo{author}{\bibfnamefont{M.}~\bibnamefont{Azuma}},
  \bibinfo{author}{\bibfnamefont{M.}~\bibnamefont{Takano}},
  \bibinfo{author}{\bibfnamefont{Y.}~\bibnamefont{Bando}}, and
  \bibinfo{author}{\bibfnamefont{Y.}~\bibnamefont{Takeda}},
  \bibinfo{year}{1993}, \bibinfo{journal}{Physica C}
  \textbf{\bibinfo{volume}{210}}(\bibinfo{number}{3-4}), \bibinfo{pages}{367 }.

\bibitem[{\citenamefont{Imai} \emph{et~al.}(1995)\citenamefont{Imai, Slichter,
  Cobb, and Markert}}]{imai95a}
\bibinfo{author}{\bibnamefont{Imai}, \bibfnamefont{T.}},
  \bibinfo{author}{\bibfnamefont{C.~P.} \bibnamefont{Slichter}},
  \bibinfo{author}{\bibfnamefont{J.~L.} \bibnamefont{Cobb}}, and
  \bibinfo{author}{\bibfnamefont{J.~T.} \bibnamefont{Markert}},
  \bibinfo{year}{1995}, \bibinfo{journal}{J.\ Phys.\ Chem.\ Solids}
  \textbf{\bibinfo{volume}{56}}, \bibinfo{pages}{1921}.

\bibitem[{\citenamefont{Imry and Wortis}(1979)}]{Imry79a}
\bibinfo{author}{\bibnamefont{Imry}, \bibfnamefont{Y.}}, and
  \bibinfo{author}{\bibfnamefont{M.}~\bibnamefont{Wortis}},
  \bibinfo{year}{1979}, \bibinfo{journal}{Phys.\ Rev.\ B}
  \textbf{\bibinfo{volume}{19}}(\bibinfo{number}{7}), \bibinfo{pages}{3580}.

\bibitem[{\citenamefont{Ino} \emph{et~al.}(2000)\citenamefont{Ino, Kim,
  Nakamura, Yoshida, Mizokawa, Shen, Fujimori, Kakeshita, Eisaki, and
  Uchida}}]{Ino00a}
\bibinfo{author}{\bibnamefont{Ino}, \bibfnamefont{A.}},
  \bibinfo{author}{\bibfnamefont{C.}~\bibnamefont{Kim}},
  \bibinfo{author}{\bibfnamefont{M.}~\bibnamefont{Nakamura}},
  \bibinfo{author}{\bibfnamefont{T.}~\bibnamefont{Yoshida}},
  \bibinfo{author}{\bibfnamefont{T.}~\bibnamefont{Mizokawa}},
  \bibinfo{author}{\bibfnamefont{Z.-X.} \bibnamefont{Shen}},
  \bibinfo{author}{\bibfnamefont{A.}~\bibnamefont{Fujimori}},
  \bibinfo{author}{\bibfnamefont{T.}~\bibnamefont{Kakeshita}},
  \bibinfo{author}{\bibfnamefont{H.}~\bibnamefont{Eisaki}}, and
  \bibinfo{author}{\bibfnamefont{S.}~\bibnamefont{Uchida}},
  \bibinfo{year}{2000}, \bibinfo{journal}{Phys.\ Rev.\ B}
  \textbf{\bibinfo{volume}{62}}(\bibinfo{number}{6}), \bibinfo{pages}{4137}.

\bibitem[{\citenamefont{Ishii} \emph{et~al.}(1989)\citenamefont{Ishii,
  Koshizawa, Kataura, Hanyu, Takai, Mizoguchi, Kume, Shiozaki, and
  Yamaguchi}}]{Ishii89a}
\bibinfo{author}{\bibnamefont{Ishii}, \bibfnamefont{H.}},
  \bibinfo{author}{\bibfnamefont{T.}~\bibnamefont{Koshizawa}},
  \bibinfo{author}{\bibfnamefont{H.}~\bibnamefont{Kataura}},
  \bibinfo{author}{\bibfnamefont{T.}~\bibnamefont{Hanyu}},
  \bibinfo{author}{\bibfnamefont{H.}~\bibnamefont{Takai}},
  \bibinfo{author}{\bibfnamefont{K.}~\bibnamefont{Mizoguchi}},
  \bibinfo{author}{\bibfnamefont{K.}~\bibnamefont{Kume}},
  \bibinfo{author}{\bibfnamefont{I.}~\bibnamefont{Shiozaki}}, and
  \bibinfo{author}{\bibfnamefont{S.}~\bibnamefont{Yamaguchi}},
  \bibinfo{year}{1989}, \bibinfo{journal}{Jpn.\ J.\ Appl.\ Phys}
  \textbf{\bibinfo{volume}{28}}, \bibinfo{pages}{L1952}.

\bibitem[{\citenamefont{Ismer} \emph{et~al.}(2007)\citenamefont{Ismer, Eremin,
  Rossi, and Morr}}]{Ismer07a}
\bibinfo{author}{\bibnamefont{Ismer}, \bibfnamefont{J.-P.}},
  \bibinfo{author}{\bibfnamefont{I.}~\bibnamefont{Eremin}},
  \bibinfo{author}{\bibfnamefont{E.}~\bibnamefont{Rossi}}, and
  \bibinfo{author}{\bibfnamefont{D.~K.} \bibnamefont{Morr}},
  \bibinfo{year}{2007}, \bibinfo{journal}{Phys.\ Rev.\ Lett.}
  \textbf{\bibinfo{volume}{99}}(\bibinfo{number}{4}), \bibinfo{eid}{047005}.

\bibitem[{\citenamefont{James} \emph{et~al.}(1989)\citenamefont{James, Zahurak,
  and Murphy}}]{James89a}
\bibinfo{author}{\bibnamefont{James}, \bibfnamefont{A.~C. W.~P.}},
  \bibinfo{author}{\bibfnamefont{S.~M.} \bibnamefont{Zahurak}}, and
  \bibinfo{author}{\bibfnamefont{D.~W.} \bibnamefont{Murphy}},
  \bibinfo{year}{1989}, \bibinfo{journal}{Nature}
  \textbf{\bibinfo{volume}{338}}, \bibinfo{pages}{240}.

\bibitem[{\citenamefont{Jandl} \emph{et~al.}(1996)\citenamefont{Jandl, Dufour,
  Strach, Ruf, Cardona, Nekvasil, Chen, Wanklyn, and Pi\~nol}}]{Jandl96a}
\bibinfo{author}{\bibnamefont{Jandl}, \bibfnamefont{S.}},
  \bibinfo{author}{\bibfnamefont{P.}~\bibnamefont{Dufour}},
  \bibinfo{author}{\bibfnamefont{T.}~\bibnamefont{Strach}},
  \bibinfo{author}{\bibfnamefont{T.}~\bibnamefont{Ruf}},
  \bibinfo{author}{\bibfnamefont{M.}~\bibnamefont{Cardona}},
  \bibinfo{author}{\bibfnamefont{V.}~\bibnamefont{Nekvasil}},
  \bibinfo{author}{\bibfnamefont{C.}~\bibnamefont{Chen}},
  \bibinfo{author}{\bibfnamefont{B.~M.} \bibnamefont{Wanklyn}}, and
  \bibinfo{author}{\bibfnamefont{S.}~\bibnamefont{Pi\~nol}},
  \bibinfo{year}{1996}, \bibinfo{journal}{Phys.\ Rev.\ B}
  \textbf{\bibinfo{volume}{53}}(\bibinfo{number}{13}), \bibinfo{pages}{8632}.

\bibitem[{\citenamefont{Jandl} \emph{et~al.}(1993)\citenamefont{Jandl, Iliev,
  Thomsen, Ruf, Cardona, Wanklyn, and Chen}}]{Jandl93a}
\bibinfo{author}{\bibnamefont{Jandl}, \bibfnamefont{S.}},
  \bibinfo{author}{\bibfnamefont{M.}~\bibnamefont{Iliev}},
  \bibinfo{author}{\bibfnamefont{C.}~\bibnamefont{Thomsen}},
  \bibinfo{author}{\bibfnamefont{T.}~\bibnamefont{Ruf}},
  \bibinfo{author}{\bibfnamefont{M.}~\bibnamefont{Cardona}},
  \bibinfo{author}{\bibfnamefont{B.~M.} \bibnamefont{Wanklyn}}, and
  \bibinfo{author}{\bibfnamefont{C.}~\bibnamefont{Chen}}, \bibinfo{year}{1993},
  \bibinfo{journal}{Solid State Comm.}
  \textbf{\bibinfo{volume}{87}}(\bibinfo{number}{7}), \bibinfo{pages}{609}.

\bibitem[{\citenamefont{Jandl} \emph{et~al.}(1999)\citenamefont{Jandl, Richard,
  Nekvasil, Zhigunov, Barilo, and Shiryaev}}]{Jandl99a}
\bibinfo{author}{\bibnamefont{Jandl}, \bibfnamefont{S.}},
  \bibinfo{author}{\bibfnamefont{P.}~\bibnamefont{Richard}},
  \bibinfo{author}{\bibfnamefont{V.}~\bibnamefont{Nekvasil}},
  \bibinfo{author}{\bibfnamefont{D.~I.} \bibnamefont{Zhigunov}},
  \bibinfo{author}{\bibfnamefont{S.~N.} \bibnamefont{Barilo}}, and
  \bibinfo{author}{\bibfnamefont{S.~V.} \bibnamefont{Shiryaev}},
  \bibinfo{year}{1999}, \bibinfo{journal}{Physica C}
  \textbf{\bibinfo{volume}{314}}, \bibinfo{pages}{189}.

\bibitem[{\citenamefont{Jenkins}
  \emph{et~al.}(2009{\natexlab{a}})\citenamefont{Jenkins, Schmadel, Bach,
  Greene, B\'{e}champ-Lagani\`{e}re, Roberge, Fournier, and Drew}}]{Jenkins09b}
\bibinfo{author}{\bibnamefont{Jenkins}, \bibfnamefont{G.~S.}},
  \bibinfo{author}{\bibfnamefont{D.~C.} \bibnamefont{Schmadel}},
  \bibinfo{author}{\bibfnamefont{P.~L.} \bibnamefont{Bach}},
  \bibinfo{author}{\bibfnamefont{R.~L.} \bibnamefont{Greene}},
  \bibinfo{author}{\bibfnamefont{X.}~\bibnamefont{B\'{e}champ-Lagani\`{e}re}},
  \bibinfo{author}{\bibfnamefont{G.}~\bibnamefont{Roberge}},
  \bibinfo{author}{\bibfnamefont{P.}~\bibnamefont{Fournier}}, and
  \bibinfo{author}{\bibfnamefont{H.~D.} \bibnamefont{Drew}},
  \bibinfo{year}{2009}{\natexlab{a}}, \bibinfo{journal}{Phys.\ Rev.\ B}
  \textbf{\bibinfo{volume}{79}}(\bibinfo{number}{22}), \bibinfo{pages}{224525}.

\bibitem[{\citenamefont{Jenkins}
  \emph{et~al.}(2009{\natexlab{b}})\citenamefont{Jenkins, Schmadel, Bach,
  Greene, Roberge, Fournier, Kontani, and Drew}}]{Jenkins09a}
\bibinfo{author}{\bibnamefont{Jenkins}, \bibfnamefont{G.~S.}},
  \bibinfo{author}{\bibfnamefont{D.~C.} \bibnamefont{Schmadel}},
  \bibinfo{author}{\bibfnamefont{P.~L.} \bibnamefont{Bach}},
  \bibinfo{author}{\bibfnamefont{R.~L.} \bibnamefont{Greene}},
  \bibinfo{author}{\bibfnamefont{X.~B.-L.~G.} \bibnamefont{Roberge}},
  \bibinfo{author}{\bibfnamefont{P.}~\bibnamefont{Fournier}},
  \bibinfo{author}{\bibfnamefont{H.}~\bibnamefont{Kontani}}, and
  \bibinfo{author}{\bibfnamefont{H.~D.} \bibnamefont{Drew}},
  \bibinfo{year}{2009}{\natexlab{b}}, \bibinfo{journal}{arXiv:0901.1701v1} .

\bibitem[{\citenamefont{Jiang} \emph{et~al.}(1994)\citenamefont{Jiang, Mao, Xi,
  Jiang, Peng, Venkatesan, Lobb, and Greene}}]{Jiang94a}
\bibinfo{author}{\bibnamefont{Jiang}, \bibfnamefont{W.}},
  \bibinfo{author}{\bibfnamefont{S.~N.} \bibnamefont{Mao}},
  \bibinfo{author}{\bibfnamefont{X.~X.} \bibnamefont{Xi}},
  \bibinfo{author}{\bibfnamefont{X.}~\bibnamefont{Jiang}},
  \bibinfo{author}{\bibfnamefont{J.~L.} \bibnamefont{Peng}},
  \bibinfo{author}{\bibfnamefont{T.}~\bibnamefont{Venkatesan}},
  \bibinfo{author}{\bibfnamefont{C.~J.} \bibnamefont{Lobb}}, and
  \bibinfo{author}{\bibfnamefont{R.~L.} \bibnamefont{Greene}},
  \bibinfo{year}{1994}, \bibinfo{journal}{Phys.\ Rev.\ Lett.}
  \textbf{\bibinfo{volume}{73}}(\bibinfo{number}{9}), \bibinfo{pages}{1291}.

\bibitem[{\citenamefont{Jin} \emph{et~al.}(2009)\citenamefont{Jin, Zhang, Bach,
  and Greene}}]{Jin09a}
\bibinfo{author}{\bibnamefont{Jin}, \bibfnamefont{K.}},
  \bibinfo{author}{\bibfnamefont{X.~H.} \bibnamefont{Zhang}},
  \bibinfo{author}{\bibfnamefont{P.}~\bibnamefont{Bach}}, and
  \bibinfo{author}{\bibfnamefont{R.~L.} \bibnamefont{Greene}},
  \bibinfo{year}{2009}, \bibinfo{journal}{Phys.\ Rev.\ B}
  \textbf{\bibinfo{volume}{80}}(\bibinfo{number}{1}), \bibinfo{eid}{012501}.

\bibitem[{\citenamefont{Jin} \emph{et~al.}(2003)\citenamefont{Jin, Onose,
  Tokura, Mandrus, Dai, and Sales}}]{Jin03a}
\bibinfo{author}{\bibnamefont{Jin}, \bibfnamefont{R.}},
  \bibinfo{author}{\bibfnamefont{Y.}~\bibnamefont{Onose}},
  \bibinfo{author}{\bibfnamefont{Y.}~\bibnamefont{Tokura}},
  \bibinfo{author}{\bibfnamefont{D.}~\bibnamefont{Mandrus}},
  \bibinfo{author}{\bibfnamefont{P.}~\bibnamefont{Dai}}, and
  \bibinfo{author}{\bibfnamefont{B.~C.} \bibnamefont{Sales}},
  \bibinfo{year}{2003}, \bibinfo{journal}{Phys.\ Rev.\ Lett.}
  \textbf{\bibinfo{volume}{91}}(\bibinfo{number}{14}), \bibinfo{eid}{146601}.

\bibitem[{\citenamefont{Jorgensen} \emph{et~al.}(1993)\citenamefont{Jorgensen,
  Radaelli, Hinks, Wagner, Kikkawa, Er, and Kanamaru}}]{Jorgensen93a}
\bibinfo{author}{\bibnamefont{Jorgensen}, \bibfnamefont{J.~D.}},
  \bibinfo{author}{\bibfnamefont{P.~G.} \bibnamefont{Radaelli}},
  \bibinfo{author}{\bibfnamefont{D.~G.} \bibnamefont{Hinks}},
  \bibinfo{author}{\bibfnamefont{J.~L.} \bibnamefont{Wagner}},
  \bibinfo{author}{\bibfnamefont{S.}~\bibnamefont{Kikkawa}},
  \bibinfo{author}{\bibfnamefont{G.}~\bibnamefont{Er}}, and
  \bibinfo{author}{\bibfnamefont{F.}~\bibnamefont{Kanamaru}},
  \bibinfo{year}{1993}, \bibinfo{journal}{Phys.\ Rev.\ B}
  \textbf{\bibinfo{volume}{47}}(\bibinfo{number}{21}), \bibinfo{pages}{14654}.

\bibitem[{\citenamefont{Joynt and Taillefer}(2002)}]{Joynt02a}
\bibinfo{author}{\bibnamefont{Joynt}, \bibfnamefont{R.}}, and
  \bibinfo{author}{\bibfnamefont{L.}~\bibnamefont{Taillefer}},
  \bibinfo{year}{2002}, \bibinfo{journal}{Rev.\ Mod.\ Phys.}
  \textbf{\bibinfo{volume}{74}}(\bibinfo{number}{1}), \bibinfo{eid}{235}.

\bibitem[{\citenamefont{Jung} \emph{et~al.}(2002)\citenamefont{Jung, Kim, Park,
  Kim, Kim, Lee, and Lee}}]{Jung02a}
\bibinfo{author}{\bibnamefont{Jung}, \bibfnamefont{C.~U.}},
  \bibinfo{author}{\bibfnamefont{J.~Y.} \bibnamefont{Kim}},
  \bibinfo{author}{\bibfnamefont{M.-S.} \bibnamefont{Park}},
  \bibinfo{author}{\bibfnamefont{M.-S.} \bibnamefont{Kim}},
  \bibinfo{author}{\bibfnamefont{H.-J.} \bibnamefont{Kim}},
  \bibinfo{author}{\bibfnamefont{S.~Y.} \bibnamefont{Lee}}, and
  \bibinfo{author}{\bibfnamefont{S.-I.} \bibnamefont{Lee}},
  \bibinfo{year}{2002}, \bibinfo{journal}{Phys.\ Rev.\ B}
  \textbf{\bibinfo{volume}{65}}(\bibinfo{number}{17}), \bibinfo{pages}{172501}.

\bibitem[{\citenamefont{Kadono}
  \emph{et~al.}(2004{\natexlab{a}})\citenamefont{Kadono, Higemoto, Koda,
  Larkin, Luke, Savici, Uemura, Kojima, Okamoto, Kakeshita, Uchida, Ito}
  \emph{et~al.}}]{Kadono04b}
\bibinfo{author}{\bibnamefont{Kadono}, \bibfnamefont{R.}},
  \bibinfo{author}{\bibfnamefont{W.}~\bibnamefont{Higemoto}},
  \bibinfo{author}{\bibfnamefont{A.}~\bibnamefont{Koda}},
  \bibinfo{author}{\bibfnamefont{M.~I.} \bibnamefont{Larkin}},
  \bibinfo{author}{\bibfnamefont{G.~M.} \bibnamefont{Luke}},
  \bibinfo{author}{\bibfnamefont{A.~T.} \bibnamefont{Savici}},
  \bibinfo{author}{\bibfnamefont{Y.~J.} \bibnamefont{Uemura}},
  \bibinfo{author}{\bibfnamefont{K.~M.} \bibnamefont{Kojima}},
  \bibinfo{author}{\bibfnamefont{T.}~\bibnamefont{Okamoto}},
  \bibinfo{author}{\bibfnamefont{T.}~\bibnamefont{Kakeshita}},
  \bibinfo{author}{\bibfnamefont{S.}~\bibnamefont{Uchida}},
  \bibinfo{author}{\bibfnamefont{T.}~\bibnamefont{Ito}}, \emph{et~al.},
  \bibinfo{year}{2004}{\natexlab{a}}, \bibinfo{journal}{Phys.\ Rev.\ B}
  \textbf{\bibinfo{volume}{69}}(\bibinfo{number}{10}), \bibinfo{eid}{104523}.

\bibitem[{\citenamefont{Kadono}
  \emph{et~al.}(2004{\natexlab{b}})\citenamefont{Kadono, Ohishi, Koda, ,
  Higemoto, Kojima, ichi Kuroshima, Fujita, and Yamada}}]{Kadono04a}
\bibinfo{author}{\bibnamefont{Kadono}, \bibfnamefont{R.}},
  \bibinfo{author}{\bibfnamefont{K.}~\bibnamefont{Ohishi}},
  \bibinfo{author}{\bibfnamefont{A.}~\bibnamefont{Koda}}, ,
  \bibinfo{author}{\bibfnamefont{W.}~\bibnamefont{Higemoto}},
  \bibinfo{author}{\bibfnamefont{K.~M.} \bibnamefont{Kojima}},
  \bibinfo{author}{\bibfnamefont{S.}~\bibnamefont{ichi Kuroshima}},
  \bibinfo{author}{\bibfnamefont{M.}~\bibnamefont{Fujita}}, and
  \bibinfo{author}{\bibfnamefont{K.}~\bibnamefont{Yamada}},
  \bibinfo{year}{2004}{\natexlab{b}}, \bibinfo{journal}{J.\ Phys.\ Soc.\ Japan}
  \textbf{\bibinfo{volume}{73}}(\bibinfo{number}{11}), \bibinfo{pages}{2944}.

\bibitem[{\citenamefont{Kadono} \emph{et~al.}(2005)\citenamefont{Kadono,
  Ohishi, Koda, Saha, Higemoto, Fujita, and Yamada}}]{Kadono05a}
\bibinfo{author}{\bibnamefont{Kadono}, \bibfnamefont{R.}},
  \bibinfo{author}{\bibfnamefont{K.}~\bibnamefont{Ohishi}},
  \bibinfo{author}{\bibfnamefont{A.}~\bibnamefont{Koda}},
  \bibinfo{author}{\bibfnamefont{S.~R.} \bibnamefont{Saha}},
  \bibinfo{author}{\bibfnamefont{W.}~\bibnamefont{Higemoto}},
  \bibinfo{author}{\bibfnamefont{M.}~\bibnamefont{Fujita}}, and
  \bibinfo{author}{\bibfnamefont{K.}~\bibnamefont{Yamada}},
  \bibinfo{year}{2005}, \bibinfo{journal}{J.\ Phys.\ Soc.\ Japan}
  \textbf{\bibinfo{volume}{74}}(\bibinfo{number}{10}), \bibinfo{pages}{2806}.

\bibitem[{\citenamefont{Kakuyanagi}
  \emph{et~al.}(2002)\citenamefont{Kakuyanagi, Kumagai, and
  Matsuda}}]{Kakuyanagi02a}
\bibinfo{author}{\bibnamefont{Kakuyanagi}, \bibfnamefont{K.}},
  \bibinfo{author}{\bibfnamefont{K.-i.} \bibnamefont{Kumagai}}, and
  \bibinfo{author}{\bibfnamefont{Y.}~\bibnamefont{Matsuda}},
  \bibinfo{year}{2002}, \bibinfo{journal}{Phys.\ Rev.\ B}
  \textbf{\bibinfo{volume}{65}}(\bibinfo{number}{6}), \bibinfo{pages}{060503}.

\bibitem[{\citenamefont{Kancharla} \emph{et~al.}(2008)\citenamefont{Kancharla,
  Kyung, Senechal, Civelli, Capone, Kotliar, and Tremblay}}]{Kancharla08a}
\bibinfo{author}{\bibnamefont{Kancharla}, \bibfnamefont{S.~S.}},
  \bibinfo{author}{\bibfnamefont{B.}~\bibnamefont{Kyung}},
  \bibinfo{author}{\bibfnamefont{D.}~\bibnamefont{Senechal}},
  \bibinfo{author}{\bibfnamefont{M.}~\bibnamefont{Civelli}},
  \bibinfo{author}{\bibfnamefont{M.}~\bibnamefont{Capone}},
  \bibinfo{author}{\bibfnamefont{G.}~\bibnamefont{Kotliar}}, and
  \bibinfo{author}{\bibfnamefont{A.-M.~S.} \bibnamefont{Tremblay}},
  \bibinfo{year}{2008}, \bibinfo{journal}{Phys.\ Rev.\ B}
  \textbf{\bibinfo{volume}{77}}(\bibinfo{number}{18}), \bibinfo{eid}{184516}.

\bibitem[{\citenamefont{Kaneko} \emph{et~al.}(1999)\citenamefont{Kaneko,
  Hidaka, Hosoya, Yamada, Endoh, Takekawa, and Kitamura}}]{Kaneko99a}
\bibinfo{author}{\bibnamefont{Kaneko}, \bibfnamefont{N.}},
  \bibinfo{author}{\bibfnamefont{Y.}~\bibnamefont{Hidaka}},
  \bibinfo{author}{\bibfnamefont{S.}~\bibnamefont{Hosoya}},
  \bibinfo{author}{\bibfnamefont{K.}~\bibnamefont{Yamada}},
  \bibinfo{author}{\bibfnamefont{Y.}~\bibnamefont{Endoh}},
  \bibinfo{author}{\bibfnamefont{S.}~\bibnamefont{Takekawa}}, and
  \bibinfo{author}{\bibfnamefont{K.}~\bibnamefont{Kitamura}},
  \bibinfo{year}{1999}, \bibinfo{journal}{J.\ Crystal Growth}
  \textbf{\bibinfo{volume}{197}}(\bibinfo{number}{4}), \bibinfo{pages}{818 }.

\bibitem[{\citenamefont{Kang} \emph{et~al.}(2007)\citenamefont{Kang, Dai,
  Campbell, Chupas, Rosenkranz, Lee, Huang, Li, Komiya, and Ando}}]{Kang07a}
\bibinfo{author}{\bibnamefont{Kang}, \bibfnamefont{H.~J.}},
  \bibinfo{author}{\bibfnamefont{P.}~\bibnamefont{Dai}},
  \bibinfo{author}{\bibfnamefont{B.~J.} \bibnamefont{Campbell}},
  \bibinfo{author}{\bibfnamefont{P.~J.} \bibnamefont{Chupas}},
  \bibinfo{author}{\bibfnamefont{S.}~\bibnamefont{Rosenkranz}},
  \bibinfo{author}{\bibfnamefont{P.~L.} \bibnamefont{Lee}},
  \bibinfo{author}{\bibfnamefont{Q.}~\bibnamefont{Huang}},
  \bibinfo{author}{\bibfnamefont{S.}~\bibnamefont{Li}},
  \bibinfo{author}{\bibfnamefont{S.}~\bibnamefont{Komiya}}, and
  \bibinfo{author}{\bibfnamefont{Y.}~\bibnamefont{Ando}}, \bibinfo{year}{2007},
  \bibinfo{journal}{Nat Mater}
  \textbf{\bibinfo{volume}{6}}(\bibinfo{number}{3}), \bibinfo{pages}{224}.

\bibitem[{\citenamefont{Kang}
  \emph{et~al.}(2003{\natexlab{a}})\citenamefont{Kang, Dai, Lynn, Matsuura,
  Thompson, Zhang, Argyriou, Onose, and Tokura}}]{Kang03a}
\bibinfo{author}{\bibnamefont{Kang}, \bibfnamefont{H.~J.}},
  \bibinfo{author}{\bibfnamefont{P.}~\bibnamefont{Dai}},
  \bibinfo{author}{\bibfnamefont{J.~W.} \bibnamefont{Lynn}},
  \bibinfo{author}{\bibfnamefont{M.}~\bibnamefont{Matsuura}},
  \bibinfo{author}{\bibfnamefont{J.~R.} \bibnamefont{Thompson}},
  \bibinfo{author}{\bibfnamefont{S.-C.} \bibnamefont{Zhang}},
  \bibinfo{author}{\bibfnamefont{D.~N.} \bibnamefont{Argyriou}},
  \bibinfo{author}{\bibfnamefont{Y.}~\bibnamefont{Onose}}, and
  \bibinfo{author}{\bibfnamefont{Y.}~\bibnamefont{Tokura}},
  \bibinfo{year}{2003}{\natexlab{a}}, \bibinfo{journal}{Nature}
  \textbf{\bibinfo{volume}{423}}(\bibinfo{number}{6939}), \bibinfo{pages}{522}.

\bibitem[{\citenamefont{Kang}
  \emph{et~al.}(2003{\natexlab{b}})\citenamefont{Kang, Dai, Lynn, Matsuura,
  Thompson, Zhang, Argyriouk, Onose, and Tokura}}]{Kang03b}
\bibinfo{author}{\bibnamefont{Kang}, \bibfnamefont{H.~J.}},
  \bibinfo{author}{\bibfnamefont{P.}~\bibnamefont{Dai}},
  \bibinfo{author}{\bibfnamefont{J.~W.} \bibnamefont{Lynn}},
  \bibinfo{author}{\bibfnamefont{M.}~\bibnamefont{Matsuura}},
  \bibinfo{author}{\bibfnamefont{J.~R.} \bibnamefont{Thompson}},
  \bibinfo{author}{\bibfnamefont{S.-C.} \bibnamefont{Zhang}},
  \bibinfo{author}{\bibfnamefont{D.~N.} \bibnamefont{Argyriouk}},
  \bibinfo{author}{\bibfnamefont{Y.}~\bibnamefont{Onose}}, and
  \bibinfo{author}{\bibfnamefont{Y.}~\bibnamefont{Tokura}},
  \bibinfo{year}{2003}{\natexlab{b}}, \bibinfo{journal}{Nature}
  \textbf{\bibinfo{volume}{429}}, \bibinfo{eid}{140}.

\bibitem[{\citenamefont{Kang} \emph{et~al.}(2002)\citenamefont{Kang, Dai,
  Mandrus, Jin, Mook, Adroja, Bennington, Lee, and Lynn}}]{Kang02a}
\bibinfo{author}{\bibnamefont{Kang}, \bibfnamefont{H.~J.}},
  \bibinfo{author}{\bibfnamefont{P.}~\bibnamefont{Dai}},
  \bibinfo{author}{\bibfnamefont{D.}~\bibnamefont{Mandrus}},
  \bibinfo{author}{\bibfnamefont{R.}~\bibnamefont{Jin}},
  \bibinfo{author}{\bibfnamefont{H.~A.} \bibnamefont{Mook}},
  \bibinfo{author}{\bibfnamefont{D.~T.} \bibnamefont{Adroja}},
  \bibinfo{author}{\bibfnamefont{S.~M.} \bibnamefont{Bennington}},
  \bibinfo{author}{\bibfnamefont{S.-H.} \bibnamefont{Lee}}, and
  \bibinfo{author}{\bibfnamefont{J.~W.} \bibnamefont{Lynn}},
  \bibinfo{year}{2002}, \bibinfo{journal}{Phys.\ Rev.\ B}
  \textbf{\bibinfo{volume}{66}}(\bibinfo{number}{6}), \bibinfo{pages}{064506}.

\bibitem[{\citenamefont{Kang} \emph{et~al.}(2005)\citenamefont{Kang, Dai, Mook,
  Argyriou, Sikolenko, Lynn, Kurita, Komiya, and Ando}}]{Kang05a}
\bibinfo{author}{\bibnamefont{Kang}, \bibfnamefont{H.~J.}},
  \bibinfo{author}{\bibfnamefont{P.}~\bibnamefont{Dai}},
  \bibinfo{author}{\bibfnamefont{H.~A.} \bibnamefont{Mook}},
  \bibinfo{author}{\bibfnamefont{D.~N.} \bibnamefont{Argyriou}},
  \bibinfo{author}{\bibfnamefont{V.}~\bibnamefont{Sikolenko}},
  \bibinfo{author}{\bibfnamefont{J.~W.} \bibnamefont{Lynn}},
  \bibinfo{author}{\bibfnamefont{Y.}~\bibnamefont{Kurita}},
  \bibinfo{author}{\bibfnamefont{S.}~\bibnamefont{Komiya}}, and
  \bibinfo{author}{\bibfnamefont{Y.}~\bibnamefont{Ando}}, \bibinfo{year}{2005},
  \bibinfo{journal}{Phys.\ Rev.\ B}
  \textbf{\bibinfo{volume}{71}}(\bibinfo{number}{21}), \bibinfo{eid}{214512}.

\bibitem[{\citenamefont{ichi Karimoto and Naito}(2004)}]{Karimoto04a}
\bibinfo{author}{\bibnamefont{ichi Karimoto}, \bibfnamefont{S.}}, and
  \bibinfo{author}{\bibfnamefont{M.}~\bibnamefont{Naito}},
  \bibinfo{year}{2004}, \bibinfo{journal}{Appl.\ Phys.\ Lett.}
  \textbf{\bibinfo{volume}{84}}(\bibinfo{number}{12}), \bibinfo{pages}{2136}.

\bibitem[{\citenamefont{Kashiwaya} \emph{et~al.}(1998)\citenamefont{Kashiwaya,
  Ito, Oka, Ueno, Takashima, Koyanagi, Tanaka, and Kajimura}}]{Kashiwaya98a}
\bibinfo{author}{\bibnamefont{Kashiwaya}, \bibfnamefont{S.}},
  \bibinfo{author}{\bibfnamefont{T.}~\bibnamefont{Ito}},
  \bibinfo{author}{\bibfnamefont{K.}~\bibnamefont{Oka}},
  \bibinfo{author}{\bibfnamefont{S.}~\bibnamefont{Ueno}},
  \bibinfo{author}{\bibfnamefont{H.}~\bibnamefont{Takashima}},
  \bibinfo{author}{\bibfnamefont{M.}~\bibnamefont{Koyanagi}},
  \bibinfo{author}{\bibfnamefont{Y.}~\bibnamefont{Tanaka}}, and
  \bibinfo{author}{\bibfnamefont{K.}~\bibnamefont{Kajimura}},
  \bibinfo{year}{1998}, \bibinfo{journal}{Phys.\ Rev.\ B}
  \textbf{\bibinfo{volume}{57}}(\bibinfo{number}{14}), \bibinfo{pages}{8680}.

\bibitem[{\citenamefont{Kashiwaya} \emph{et~al.}(1995)\citenamefont{Kashiwaya,
  Tanaka, Koyanagi, Takashima, and Kajimura}}]{Kashiwaya95a}
\bibinfo{author}{\bibnamefont{Kashiwaya}, \bibfnamefont{S.}},
  \bibinfo{author}{\bibfnamefont{Y.}~\bibnamefont{Tanaka}},
  \bibinfo{author}{\bibfnamefont{M.}~\bibnamefont{Koyanagi}},
  \bibinfo{author}{\bibfnamefont{H.}~\bibnamefont{Takashima}}, and
  \bibinfo{author}{\bibfnamefont{K.}~\bibnamefont{Kajimura}},
  \bibinfo{year}{1995}, \bibinfo{journal}{Phys.\ Rev.\ B}
  \textbf{\bibinfo{volume}{51}}(\bibinfo{number}{2}), \bibinfo{pages}{1350}.

\bibitem[{\citenamefont{Kastner} \emph{et~al.}(1998)\citenamefont{Kastner,
  Birgeneau, Shirane, and Endoh}}]{Kastner98a}
\bibinfo{author}{\bibnamefont{Kastner}, \bibfnamefont{M.~A.}},
  \bibinfo{author}{\bibfnamefont{R.~J.} \bibnamefont{Birgeneau}},
  \bibinfo{author}{\bibfnamefont{G.}~\bibnamefont{Shirane}}, and
  \bibinfo{author}{\bibfnamefont{Y.}~\bibnamefont{Endoh}},
  \bibinfo{year}{1998}, \bibinfo{journal}{Rev.\ Mod.\ Phys.}
  \textbf{\bibinfo{volume}{70}}(\bibinfo{number}{3}), \bibinfo{pages}{897}.

\bibitem[{\citenamefont{Katano} \emph{et~al.}(2000)\citenamefont{Katano, Sato,
  Yamada, Suzuki, and Fukase}}]{Katano00a}
\bibinfo{author}{\bibnamefont{Katano}, \bibfnamefont{S.}},
  \bibinfo{author}{\bibfnamefont{M.}~\bibnamefont{Sato}},
  \bibinfo{author}{\bibfnamefont{K.}~\bibnamefont{Yamada}},
  \bibinfo{author}{\bibfnamefont{T.}~\bibnamefont{Suzuki}}, and
  \bibinfo{author}{\bibfnamefont{T.}~\bibnamefont{Fukase}},
  \bibinfo{year}{2000}, \bibinfo{journal}{Phys.\ Rev.\ B}
  \textbf{\bibinfo{volume}{62}}(\bibinfo{number}{22}), \bibinfo{pages}{R14677}.

\bibitem[{\citenamefont{Kawakami} \emph{et~al.}(2006)\citenamefont{Kawakami,
  Shibauchi, Terao, and Suzuki}}]{Kawakami06a}
\bibinfo{author}{\bibnamefont{Kawakami}, \bibfnamefont{T.}},
  \bibinfo{author}{\bibfnamefont{T.}~\bibnamefont{Shibauchi}},
  \bibinfo{author}{\bibfnamefont{Y.}~\bibnamefont{Terao}}, and
  \bibinfo{author}{\bibfnamefont{M.}~\bibnamefont{Suzuki}},
  \bibinfo{year}{2006}, \bibinfo{journal}{Phys.\ Rev.\ B}
  \textbf{\bibinfo{volume}{74}}(\bibinfo{number}{14}), \bibinfo{eid}{144520}.

\bibitem[{\citenamefont{Kawakami} \emph{et~al.}(2005)\citenamefont{Kawakami,
  Shibauchi, Terao, Suzuki, and Krusin-Elbaum}}]{Kawakami05a}
\bibinfo{author}{\bibnamefont{Kawakami}, \bibfnamefont{T.}},
  \bibinfo{author}{\bibfnamefont{T.}~\bibnamefont{Shibauchi}},
  \bibinfo{author}{\bibfnamefont{Y.}~\bibnamefont{Terao}},
  \bibinfo{author}{\bibfnamefont{M.}~\bibnamefont{Suzuki}}, and
  \bibinfo{author}{\bibfnamefont{L.}~\bibnamefont{Krusin-Elbaum}},
  \bibinfo{year}{2005}, \bibinfo{journal}{Phys.\ Rev.\ Lett.}
  \textbf{\bibinfo{volume}{95}}(\bibinfo{number}{1}), \bibinfo{eid}{017001}.

\bibitem[{\citenamefont{Keimer} \emph{et~al.}(1992)\citenamefont{Keimer,
  Aharony, Auerbach, Birgeneau, Cassanho, Endoh, Erwin, Kastner, and
  Shirane}}]{Keimer92a}
\bibinfo{author}{\bibnamefont{Keimer}, \bibfnamefont{B.}},
  \bibinfo{author}{\bibfnamefont{A.}~\bibnamefont{Aharony}},
  \bibinfo{author}{\bibfnamefont{A.}~\bibnamefont{Auerbach}},
  \bibinfo{author}{\bibfnamefont{R.~J.} \bibnamefont{Birgeneau}},
  \bibinfo{author}{\bibfnamefont{A.}~\bibnamefont{Cassanho}},
  \bibinfo{author}{\bibfnamefont{Y.}~\bibnamefont{Endoh}},
  \bibinfo{author}{\bibfnamefont{R.~W.} \bibnamefont{Erwin}},
  \bibinfo{author}{\bibfnamefont{M.~A.} \bibnamefont{Kastner}}, and
  \bibinfo{author}{\bibfnamefont{G.}~\bibnamefont{Shirane}},
  \bibinfo{year}{1992}, \bibinfo{journal}{Phys.\ Rev.\ B}
  \textbf{\bibinfo{volume}{45}}(\bibinfo{number}{13}), \bibinfo{pages}{7430}.

\bibitem[{\citenamefont{Kendziora} \emph{et~al.}(2001)\citenamefont{Kendziora,
  Nachumi, Fournier, Li, Greene, and Hinks}}]{Kendziora01a}
\bibinfo{author}{\bibnamefont{Kendziora}, \bibfnamefont{C.}},
  \bibinfo{author}{\bibfnamefont{B.}~\bibnamefont{Nachumi}},
  \bibinfo{author}{\bibfnamefont{P.}~\bibnamefont{Fournier}},
  \bibinfo{author}{\bibfnamefont{Z.~Y.} \bibnamefont{Li}},
  \bibinfo{author}{\bibfnamefont{R.~L.} \bibnamefont{Greene}}, and
  \bibinfo{author}{\bibfnamefont{D.~G.} \bibnamefont{Hinks}},
  \bibinfo{year}{2001}, \bibinfo{journal}{Physica C}
  \textbf{\bibinfo{volume}{364-365}}, \bibinfo{pages}{541}.

\bibitem[{\citenamefont{Khasanov} \emph{et~al.}(2008)\citenamefont{Khasanov,
  Shengelaya, Maisuradze, Castro, Savi\'{c}, Weyeneth, Park, Jang, Lee, and
  Keller}}]{Khasanov08a}
\bibinfo{author}{\bibnamefont{Khasanov}, \bibfnamefont{R.}},
  \bibinfo{author}{\bibfnamefont{A.}~\bibnamefont{Shengelaya}},
  \bibinfo{author}{\bibfnamefont{A.}~\bibnamefont{Maisuradze}},
  \bibinfo{author}{\bibfnamefont{D.~D.} \bibnamefont{Castro}},
  \bibinfo{author}{\bibfnamefont{I.~M.} \bibnamefont{Savi\'{c}}},
  \bibinfo{author}{\bibfnamefont{S.}~\bibnamefont{Weyeneth}},
  \bibinfo{author}{\bibfnamefont{M.~S.} \bibnamefont{Park}},
  \bibinfo{author}{\bibfnamefont{D.~J.} \bibnamefont{Jang}},
  \bibinfo{author}{\bibfnamefont{S.-I.} \bibnamefont{Lee}}, and
  \bibinfo{author}{\bibfnamefont{H.}~\bibnamefont{Keller}},
  \bibinfo{year}{2008}, \bibinfo{journal}{Phys.\ Rev.\ B}
  \textbf{\bibinfo{volume}{77}}(\bibinfo{number}{18}), \bibinfo{pages}{184512}.

\bibitem[{\citenamefont{Khaykovich}
  \emph{et~al.}(2002)\citenamefont{Khaykovich, Lee, Erwin, Lee, Wakimoto,
  Thomas, Kastner, and Birgeneau}}]{Khaykovich02a}
\bibinfo{author}{\bibnamefont{Khaykovich}, \bibfnamefont{B.}},
  \bibinfo{author}{\bibfnamefont{Y.~S.} \bibnamefont{Lee}},
  \bibinfo{author}{\bibfnamefont{R.~W.} \bibnamefont{Erwin}},
  \bibinfo{author}{\bibfnamefont{S.-H.} \bibnamefont{Lee}},
  \bibinfo{author}{\bibfnamefont{S.}~\bibnamefont{Wakimoto}},
  \bibinfo{author}{\bibfnamefont{K.~J.} \bibnamefont{Thomas}},
  \bibinfo{author}{\bibfnamefont{M.~A.} \bibnamefont{Kastner}}, and
  \bibinfo{author}{\bibfnamefont{R.~J.} \bibnamefont{Birgeneau}},
  \bibinfo{year}{2002}, \bibinfo{journal}{Phys.\ Rev.\ B}
  \textbf{\bibinfo{volume}{66}}(\bibinfo{number}{1}), \bibinfo{eid}{014528}.

\bibitem[{\citenamefont{Khurana}(1989)}]{Khurana89a}
\bibinfo{author}{\bibnamefont{Khurana}, \bibfnamefont{A.}},
  \bibinfo{year}{1989}, \bibinfo{journal}{Physics Today}
  \textbf{\bibinfo{volume}{42}}, \bibinfo{pages}{17}.

\bibitem[{\citenamefont{Kikkawa} \emph{et~al.}(1992)\citenamefont{Kikkawa,
  Kanamaru, Miyamoto, Tanaka, Sera, Sato, Hiroi, Takano, and
  Bando}}]{Kikkawa92a}
\bibinfo{author}{\bibnamefont{Kikkawa}, \bibfnamefont{G.}},
  \bibinfo{author}{\bibfnamefont{F.}~\bibnamefont{Kanamaru}},
  \bibinfo{author}{\bibfnamefont{Y.}~\bibnamefont{Miyamoto}},
  \bibinfo{author}{\bibfnamefont{S.}~\bibnamefont{Tanaka}},
  \bibinfo{author}{\bibfnamefont{M.}~\bibnamefont{Sera}},
  \bibinfo{author}{\bibfnamefont{M.}~\bibnamefont{Sato}},
  \bibinfo{author}{\bibfnamefont{Z.}~\bibnamefont{Hiroi}},
  \bibinfo{author}{\bibfnamefont{M.}~\bibnamefont{Takano}}, and
  \bibinfo{author}{\bibfnamefont{Y.}~\bibnamefont{Bando}},
  \bibinfo{year}{1992}, \bibinfo{journal}{Physica C}
  \textbf{\bibinfo{volume}{196}}(\bibinfo{number}{3-4}), \bibinfo{pages}{271 }.

\bibitem[{\citenamefont{Kim and Gaskell}(1993)}]{Kim93a}
\bibinfo{author}{\bibnamefont{Kim}, \bibfnamefont{J.~S.}}, and
  \bibinfo{author}{\bibfnamefont{D.~R.} \bibnamefont{Gaskell}},
  \bibinfo{year}{1993}, \bibinfo{journal}{Physica C}
  \textbf{\bibinfo{volume}{209}}(\bibinfo{number}{4}), \bibinfo{pages}{381}.

\bibitem[{\citenamefont{Kim} \emph{et~al.}(2002)\citenamefont{Kim, Lemberger,
  Jung, Choi, Kim, Kim, and Lee}}]{Kim02a}
\bibinfo{author}{\bibnamefont{Kim}, \bibfnamefont{M.-S.}},
  \bibinfo{author}{\bibfnamefont{T.~R.} \bibnamefont{Lemberger}},
  \bibinfo{author}{\bibfnamefont{C.~U.} \bibnamefont{Jung}},
  \bibinfo{author}{\bibfnamefont{J.-H.} \bibnamefont{Choi}},
  \bibinfo{author}{\bibfnamefont{J.~Y.} \bibnamefont{Kim}},
  \bibinfo{author}{\bibfnamefont{H.-J.} \bibnamefont{Kim}}, and
  \bibinfo{author}{\bibfnamefont{S.-I.} \bibnamefont{Lee}},
  \bibinfo{year}{2002}, \bibinfo{journal}{Phys.\ Rev.\ B}
  \textbf{\bibinfo{volume}{66}}(\bibinfo{number}{21}), \bibinfo{pages}{214509}.

\bibitem[{\citenamefont{Kim} \emph{et~al.}(2003)\citenamefont{Kim, Skinta,
  Lemberger, Tsukada, and Naito}}]{Kim03a}
\bibinfo{author}{\bibnamefont{Kim}, \bibfnamefont{M.-S.}},
  \bibinfo{author}{\bibfnamefont{J.~A.} \bibnamefont{Skinta}},
  \bibinfo{author}{\bibfnamefont{T.~R.} \bibnamefont{Lemberger}},
  \bibinfo{author}{\bibfnamefont{A.}~\bibnamefont{Tsukada}}, and
  \bibinfo{author}{\bibfnamefont{M.}~\bibnamefont{Naito}},
  \bibinfo{year}{2003}, \bibinfo{journal}{Phys.\ Rev.\ Lett.}
  \textbf{\bibinfo{volume}{91}}(\bibinfo{number}{8}), \bibinfo{eid}{087001}.

\bibitem[{\citenamefont{King} \emph{et~al.}(1993)\citenamefont{King, Shen,
  Dessau, Wells, Spicer, Arko, Marshall, DiCarlo, Loeser, Park, Ratner, Peng}
  \emph{et~al.}}]{King93a}
\bibinfo{author}{\bibnamefont{King}, \bibfnamefont{D.~M.}},
  \bibinfo{author}{\bibfnamefont{Z.-X.} \bibnamefont{Shen}},
  \bibinfo{author}{\bibfnamefont{D.~S.} \bibnamefont{Dessau}},
  \bibinfo{author}{\bibfnamefont{B.~O.} \bibnamefont{Wells}},
  \bibinfo{author}{\bibfnamefont{W.~E.} \bibnamefont{Spicer}},
  \bibinfo{author}{\bibfnamefont{A.~J.} \bibnamefont{Arko}},
  \bibinfo{author}{\bibfnamefont{D.~S.} \bibnamefont{Marshall}},
  \bibinfo{author}{\bibfnamefont{J.}~\bibnamefont{DiCarlo}},
  \bibinfo{author}{\bibfnamefont{A.~G.} \bibnamefont{Loeser}},
  \bibinfo{author}{\bibfnamefont{C.~H.} \bibnamefont{Park}},
  \bibinfo{author}{\bibfnamefont{E.~R.} \bibnamefont{Ratner}},
  \bibinfo{author}{\bibfnamefont{J.~L.} \bibnamefont{Peng}}, \emph{et~al.},
  \bibinfo{year}{1993}, \bibinfo{journal}{Phys.\ Rev.\ Lett.}
  \textbf{\bibinfo{volume}{70}}(\bibinfo{number}{20}), \bibinfo{pages}{3159}.

\bibitem[{\citenamefont{Kirtley} \emph{et~al.}(1996)\citenamefont{Kirtley,
  Tsuei, Rupp, Sun, Yu-Jahnes, Gupta, Ketchen, Moler, and
  Bhushan}}]{Kirtley96a}
\bibinfo{author}{\bibnamefont{Kirtley}, \bibfnamefont{J.~R.}},
  \bibinfo{author}{\bibfnamefont{C.~C.} \bibnamefont{Tsuei}},
  \bibinfo{author}{\bibfnamefont{M.}~\bibnamefont{Rupp}},
  \bibinfo{author}{\bibfnamefont{J.~Z.} \bibnamefont{Sun}},
  \bibinfo{author}{\bibfnamefont{L.~S.} \bibnamefont{Yu-Jahnes}},
  \bibinfo{author}{\bibfnamefont{A.}~\bibnamefont{Gupta}},
  \bibinfo{author}{\bibfnamefont{M.~B.} \bibnamefont{Ketchen}},
  \bibinfo{author}{\bibfnamefont{K.~A.} \bibnamefont{Moler}}, and
  \bibinfo{author}{\bibfnamefont{M.}~\bibnamefont{Bhushan}},
  \bibinfo{year}{1996}, \bibinfo{journal}{Phys.\ Rev.\ Lett.}
  \textbf{\bibinfo{volume}{76}}(\bibinfo{number}{8}), \bibinfo{pages}{1336}.

\bibitem[{\citenamefont{Kivelson} \emph{et~al.}(2003)\citenamefont{Kivelson,
  Bindloss, Fradkin, Oganesyan, Tranquada, Kapitulnik, and
  Howald}}]{Kivelson03a}
\bibinfo{author}{\bibnamefont{Kivelson}, \bibfnamefont{S.~A.}},
  \bibinfo{author}{\bibfnamefont{I.~P.} \bibnamefont{Bindloss}},
  \bibinfo{author}{\bibfnamefont{E.}~\bibnamefont{Fradkin}},
  \bibinfo{author}{\bibfnamefont{V.}~\bibnamefont{Oganesyan}},
  \bibinfo{author}{\bibfnamefont{J.~M.} \bibnamefont{Tranquada}},
  \bibinfo{author}{\bibfnamefont{A.}~\bibnamefont{Kapitulnik}}, and
  \bibinfo{author}{\bibfnamefont{C.}~\bibnamefont{Howald}},
  \bibinfo{year}{2003}, \bibinfo{journal}{Rev.\ Mod.\ Phys.}
  \textbf{\bibinfo{volume}{75}}(\bibinfo{number}{4}), \bibinfo{pages}{1201}.

\bibitem[{\citenamefont{Klamut} \emph{et~al.}(1997)\citenamefont{Klamut,
  Sikora, Bukowski, Dabrowski, and Klamut}}]{Klamut97a}
\bibinfo{author}{\bibnamefont{Klamut}, \bibfnamefont{P.}},
  \bibinfo{author}{\bibfnamefont{A.}~\bibnamefont{Sikora}},
  \bibinfo{author}{\bibfnamefont{Z.}~\bibnamefont{Bukowski}},
  \bibinfo{author}{\bibfnamefont{B.}~\bibnamefont{Dabrowski}}, and
  \bibinfo{author}{\bibfnamefont{J.}~\bibnamefont{Klamut}},
  \bibinfo{year}{1997}, \bibinfo{journal}{Physica C}
  \textbf{\bibinfo{volume}{282-287}}, \bibinfo{pages}{541}.

\bibitem[{\citenamefont{Kleefisch} \emph{et~al.}(2001)\citenamefont{Kleefisch,
  Welter, Marx, Alff, Gross, and Naito}}]{Kleefisch01a}
\bibinfo{author}{\bibnamefont{Kleefisch}, \bibfnamefont{S.}},
  \bibinfo{author}{\bibfnamefont{B.}~\bibnamefont{Welter}},
  \bibinfo{author}{\bibfnamefont{A.}~\bibnamefont{Marx}},
  \bibinfo{author}{\bibfnamefont{L.}~\bibnamefont{Alff}},
  \bibinfo{author}{\bibfnamefont{R.}~\bibnamefont{Gross}}, and
  \bibinfo{author}{\bibfnamefont{M.}~\bibnamefont{Naito}},
  \bibinfo{year}{2001}, \bibinfo{journal}{Phys.\ Rev.\ B}
  \textbf{\bibinfo{volume}{63}}(\bibinfo{number}{10}), \bibinfo{pages}{100507}.

\bibitem[{\citenamefont{Koike} \emph{et~al.}(1992)\citenamefont{Koike,
  Kakimoto, Mochida, Sato, Noji, Kato, and Saito}}]{Koike92a}
\bibinfo{author}{\bibnamefont{Koike}, \bibfnamefont{Y.}},
  \bibinfo{author}{\bibfnamefont{A.}~\bibnamefont{Kakimoto}},
  \bibinfo{author}{\bibfnamefont{M.}~\bibnamefont{Mochida}},
  \bibinfo{author}{\bibfnamefont{H.}~\bibnamefont{Sato}},
  \bibinfo{author}{\bibfnamefont{T.}~\bibnamefont{Noji}},
  \bibinfo{author}{\bibfnamefont{M.}~\bibnamefont{Kato}}, and
  \bibinfo{author}{\bibfnamefont{Y.}~\bibnamefont{Saito}},
  \bibinfo{year}{1992}, \bibinfo{journal}{Jpn.\ J.\ Appl.\ Phys}
  \textbf{\bibinfo{volume}{31}}, \bibinfo{pages}{2721}.

\bibitem[{\citenamefont{Koitzsch} \emph{et~al.}(2003)\citenamefont{Koitzsch,
  Blumberg, Gozar, Dennis, Fournier, and Greene}}]{Koitzsch03a}
\bibinfo{author}{\bibnamefont{Koitzsch}, \bibfnamefont{A.}},
  \bibinfo{author}{\bibfnamefont{G.}~\bibnamefont{Blumberg}},
  \bibinfo{author}{\bibfnamefont{A.}~\bibnamefont{Gozar}},
  \bibinfo{author}{\bibfnamefont{B.}~\bibnamefont{Dennis}},
  \bibinfo{author}{\bibfnamefont{P.}~\bibnamefont{Fournier}}, and
  \bibinfo{author}{\bibfnamefont{R.}~\bibnamefont{Greene}},
  \bibinfo{year}{2003}, \bibinfo{journal}{Phys.\ Rev.\ B}
  \textbf{\bibinfo{volume}{67}}, \bibinfo{pages}{184522}.

\bibitem[{\citenamefont{Kokales} \emph{et~al.}(2000)\citenamefont{Kokales,
  Fournier, Mercaldo, Talanov, Greene, and Anlage}}]{Kokales00a}
\bibinfo{author}{\bibnamefont{Kokales}, \bibfnamefont{J.~D.}},
  \bibinfo{author}{\bibfnamefont{P.}~\bibnamefont{Fournier}},
  \bibinfo{author}{\bibfnamefont{L.~V.} \bibnamefont{Mercaldo}},
  \bibinfo{author}{\bibfnamefont{V.~V.} \bibnamefont{Talanov}},
  \bibinfo{author}{\bibfnamefont{R.~L.} \bibnamefont{Greene}}, and
  \bibinfo{author}{\bibfnamefont{S.~M.} \bibnamefont{Anlage}},
  \bibinfo{year}{2000}, \bibinfo{journal}{Phys.\ Rev.\ Lett.}
  \textbf{\bibinfo{volume}{85}}, \bibinfo{pages}{3696}.

\bibitem[{\citenamefont{Kontani}(2008)}]{Kotani08a}
\bibinfo{author}{\bibnamefont{Kontani}, \bibfnamefont{H.}},
  \bibinfo{year}{2008}, \bibinfo{journal}{Rep.\ Prog.\ Phys.}
  \textbf{\bibinfo{volume}{71}}, \bibinfo{pages}{026501}.

\bibitem[{\citenamefont{Kordyuk} \emph{et~al.}(2002)\citenamefont{Kordyuk,
  Borisenko, Golden, Legner, Nenkov, Knupfer, Fink, Berger, Forr\'o, and
  Follath}}]{Kordyuk02a}
\bibinfo{author}{\bibnamefont{Kordyuk}, \bibfnamefont{A.~A.}},
  \bibinfo{author}{\bibfnamefont{S.~V.} \bibnamefont{Borisenko}},
  \bibinfo{author}{\bibfnamefont{M.~S.} \bibnamefont{Golden}},
  \bibinfo{author}{\bibfnamefont{S.}~\bibnamefont{Legner}},
  \bibinfo{author}{\bibfnamefont{K.~A.} \bibnamefont{Nenkov}},
  \bibinfo{author}{\bibfnamefont{M.}~\bibnamefont{Knupfer}},
  \bibinfo{author}{\bibfnamefont{J.}~\bibnamefont{Fink}},
  \bibinfo{author}{\bibfnamefont{H.}~\bibnamefont{Berger}},
  \bibinfo{author}{\bibfnamefont{L.}~\bibnamefont{Forr\'o}}, and
  \bibinfo{author}{\bibfnamefont{R.}~\bibnamefont{Follath}},
  \bibinfo{year}{2002}, \bibinfo{journal}{Phys.\ Rev.\ B}
  \textbf{\bibinfo{volume}{66}}(\bibinfo{number}{1}), \bibinfo{pages}{014502}.

\bibitem[{\citenamefont{Kramers}(1930)}]{Kramers30a}
\bibinfo{author}{\bibnamefont{Kramers}, \bibfnamefont{H.}},
  \bibinfo{year}{1930}, \bibinfo{journal}{Proc. Amsterdam Acad.}
  \textbf{\bibinfo{volume}{33}}, \bibinfo{pages}{959}.

\bibitem[{\citenamefont{Krockenberger}
  \emph{et~al.}(2008)\citenamefont{Krockenberger, Kurian, Winkler, Tsukada,
  Naito, and Alff}}]{Krockenberger08a}
\bibinfo{author}{\bibnamefont{Krockenberger}, \bibfnamefont{Y.}},
  \bibinfo{author}{\bibfnamefont{J.}~\bibnamefont{Kurian}},
  \bibinfo{author}{\bibfnamefont{A.}~\bibnamefont{Winkler}},
  \bibinfo{author}{\bibfnamefont{A.}~\bibnamefont{Tsukada}},
  \bibinfo{author}{\bibfnamefont{M.}~\bibnamefont{Naito}}, and
  \bibinfo{author}{\bibfnamefont{L.}~\bibnamefont{Alff}}, \bibinfo{year}{2008},
  \bibinfo{journal}{Phys.\ Rev.\ B}
  \textbf{\bibinfo{volume}{77}}(\bibinfo{number}{6}), \bibinfo{eid}{060505}.

\bibitem[{\citenamefont{Kr\"{u}ger}
  \emph{et~al.}(2007)\citenamefont{Kr\"{u}ger, Wilson, Shan, Li, Huang, Wen,
  Zhang, Dai, and Zaanen}}]{Kruger07a}
\bibinfo{author}{\bibnamefont{Kr\"{u}ger}, \bibfnamefont{F.}},
  \bibinfo{author}{\bibfnamefont{S.~D.} \bibnamefont{Wilson}},
  \bibinfo{author}{\bibfnamefont{L.}~\bibnamefont{Shan}},
  \bibinfo{author}{\bibfnamefont{S.}~\bibnamefont{Li}},
  \bibinfo{author}{\bibfnamefont{Y.}~\bibnamefont{Huang}},
  \bibinfo{author}{\bibfnamefont{H.-H.} \bibnamefont{Wen}},
  \bibinfo{author}{\bibfnamefont{S.-C.} \bibnamefont{Zhang}},
  \bibinfo{author}{\bibfnamefont{P.}~\bibnamefont{Dai}}, and
  \bibinfo{author}{\bibfnamefont{J.}~\bibnamefont{Zaanen}},
  \bibinfo{year}{2007}, \bibinfo{journal}{Phys.\ Rev.\ B}
  \textbf{\bibinfo{volume}{76}}(\bibinfo{number}{9}), \bibinfo{eid}{094506}.

\bibitem[{\citenamefont{Kuiper} \emph{et~al.}(1988)\citenamefont{Kuiper,
  Kruizinga, Ghijsen, Grioni, Weijs, de~Groot, Sawatzky, Verweij, Feiner, and
  Petersen}}]{Kuiper88a}
\bibinfo{author}{\bibnamefont{Kuiper}, \bibfnamefont{P.}},
  \bibinfo{author}{\bibfnamefont{G.}~\bibnamefont{Kruizinga}},
  \bibinfo{author}{\bibfnamefont{J.}~\bibnamefont{Ghijsen}},
  \bibinfo{author}{\bibfnamefont{M.}~\bibnamefont{Grioni}},
  \bibinfo{author}{\bibfnamefont{P.~J.~W.} \bibnamefont{Weijs}},
  \bibinfo{author}{\bibfnamefont{F.~M.~F.} \bibnamefont{de~Groot}},
  \bibinfo{author}{\bibfnamefont{G.~A.} \bibnamefont{Sawatzky}},
  \bibinfo{author}{\bibfnamefont{H.}~\bibnamefont{Verweij}},
  \bibinfo{author}{\bibfnamefont{L.~F.} \bibnamefont{Feiner}}, and
  \bibinfo{author}{\bibfnamefont{H.}~\bibnamefont{Petersen}},
  \bibinfo{year}{1988}, \bibinfo{journal}{Phys.\ Rev.\ B}
  \textbf{\bibinfo{volume}{38}}(\bibinfo{number}{10}), \bibinfo{pages}{6483}.

\bibitem[{\citenamefont{Kurahashi} \emph{et~al.}(2002)\citenamefont{Kurahashi,
  Matsushita, Fujita, and Yamada}}]{Kurahashi02a}
\bibinfo{author}{\bibnamefont{Kurahashi}, \bibfnamefont{K.}},
  \bibinfo{author}{\bibfnamefont{H.}~\bibnamefont{Matsushita}},
  \bibinfo{author}{\bibfnamefont{M.}~\bibnamefont{Fujita}}, and
  \bibinfo{author}{\bibfnamefont{K.}~\bibnamefont{Yamada}},
  \bibinfo{year}{2002}, \bibinfo{journal}{J.\ Phys.\ Soc.\ Japan}
  \textbf{\bibinfo{volume}{71}}(\bibinfo{number}{3}), \bibinfo{pages}{910}.

\bibitem[{\citenamefont{Kuroshima} \emph{et~al.}(2003)\citenamefont{Kuroshima,
  Fujita, Uefuji, Matsuda, and Yamada}}]{Kuroshima03a}
\bibinfo{author}{\bibnamefont{Kuroshima}, \bibfnamefont{S.}},
  \bibinfo{author}{\bibfnamefont{M.}~\bibnamefont{Fujita}},
  \bibinfo{author}{\bibfnamefont{T.}~\bibnamefont{Uefuji}},
  \bibinfo{author}{\bibfnamefont{M.}~\bibnamefont{Matsuda}}, and
  \bibinfo{author}{\bibfnamefont{K.}~\bibnamefont{Yamada}},
  \bibinfo{year}{2003}, \bibinfo{journal}{Physica C}
  \textbf{\bibinfo{volume}{392--396}}, \bibinfo{pages}{216}.

\bibitem[{\citenamefont{Kusko} \emph{et~al.}(2002)\citenamefont{Kusko,
  Markiewicz, Lindroos, and Bansil}}]{Kusko02a}
\bibinfo{author}{\bibnamefont{Kusko}, \bibfnamefont{C.}},
  \bibinfo{author}{\bibfnamefont{R.~S.} \bibnamefont{Markiewicz}},
  \bibinfo{author}{\bibfnamefont{M.}~\bibnamefont{Lindroos}}, and
  \bibinfo{author}{\bibfnamefont{A.}~\bibnamefont{Bansil}},
  \bibinfo{year}{2002}, \bibinfo{journal}{Phys.\ Rev.\ B}
  \textbf{\bibinfo{volume}{66}}(\bibinfo{number}{14}), \bibinfo{eid}{140513}.

\bibitem[{\citenamefont{Kwei} \emph{et~al.}(1989)\citenamefont{Kwei, Cheong,
  Fisk, Garzon, Goldstone, and Thompson}}]{Kwei89a}
\bibinfo{author}{\bibnamefont{Kwei}, \bibfnamefont{G.~H.}},
  \bibinfo{author}{\bibfnamefont{S.-W.} \bibnamefont{Cheong}},
  \bibinfo{author}{\bibfnamefont{Z.}~\bibnamefont{Fisk}},
  \bibinfo{author}{\bibfnamefont{F.~H.} \bibnamefont{Garzon}},
  \bibinfo{author}{\bibfnamefont{J.~A.} \bibnamefont{Goldstone}}, and
  \bibinfo{author}{\bibfnamefont{J.~D.} \bibnamefont{Thompson}},
  \bibinfo{year}{1989}, \bibinfo{journal}{Phys.\ Rev.\ B}
  \textbf{\bibinfo{volume}{40}}(\bibinfo{number}{13}), \bibinfo{pages}{9370}.

\bibitem[{\citenamefont{Kyung} \emph{et~al.}(2004)\citenamefont{Kyung,
  Hankevych, Dare, and Tremblay}}]{Kyung04a}
\bibinfo{author}{\bibnamefont{Kyung}, \bibfnamefont{B.}},
  \bibinfo{author}{\bibfnamefont{V.}~\bibnamefont{Hankevych}},
  \bibinfo{author}{\bibfnamefont{A.-M.} \bibnamefont{Dare}}, and
  \bibinfo{author}{\bibfnamefont{A.-M.~S.} \bibnamefont{Tremblay}},
  \bibinfo{year}{2004}, \bibinfo{journal}{Phys.\ Rev.\ Lett.}
  \textbf{\bibinfo{volume}{93}}(\bibinfo{number}{14}), \bibinfo{eid}{147004}.

\bibitem[{\citenamefont{Kyung} \emph{et~al.}(2003)\citenamefont{Kyung, Landry,
  and Tremblay}}]{Kyung03a}
\bibinfo{author}{\bibnamefont{Kyung}, \bibfnamefont{B.}},
  \bibinfo{author}{\bibfnamefont{J.-S.} \bibnamefont{Landry}}, and
  \bibinfo{author}{\bibfnamefont{A.-M.~S.} \bibnamefont{Tremblay}},
  \bibinfo{year}{2003}, \bibinfo{journal}{Phys.\ Rev.\ B}
  \textbf{\bibinfo{volume}{68}}(\bibinfo{number}{17}), \bibinfo{eid}{174502}.

\bibitem[{\citenamefont{Lake} \emph{et~al.}(2001)\citenamefont{Lake, Aeppli,
  Clausen, McMorrow, Lefmann, Hussey, Mangkorntong, Nohara, Takagi, Mason, and
  Schroeder}}]{Lake01a}
\bibinfo{author}{\bibnamefont{Lake}, \bibfnamefont{B.}},
  \bibinfo{author}{\bibfnamefont{G.}~\bibnamefont{Aeppli}},
  \bibinfo{author}{\bibfnamefont{K.~N.} \bibnamefont{Clausen}},
  \bibinfo{author}{\bibfnamefont{D.~F.} \bibnamefont{McMorrow}},
  \bibinfo{author}{\bibfnamefont{K.}~\bibnamefont{Lefmann}},
  \bibinfo{author}{\bibfnamefont{N.~E.} \bibnamefont{Hussey}},
  \bibinfo{author}{\bibfnamefont{N.}~\bibnamefont{Mangkorntong}},
  \bibinfo{author}{\bibfnamefont{M.}~\bibnamefont{Nohara}},
  \bibinfo{author}{\bibfnamefont{H.}~\bibnamefont{Takagi}},
  \bibinfo{author}{\bibfnamefont{T.~E.} \bibnamefont{Mason}}, and
  \bibinfo{author}{\bibfnamefont{A.}~\bibnamefont{Schroeder}},
  \bibinfo{year}{2001}, \bibinfo{journal}{Science}
  \textbf{\bibinfo{volume}{291}}(\bibinfo{number}{5509}), \bibinfo{pages}{1759
  }.

\bibitem[{\citenamefont{Lake} \emph{et~al.}(2002)\citenamefont{Lake, R$\o$nnow,
  Christensen, Aeppli, Lefmann, McMorrow, Vorderwisch, Smeibidl, Mangkorntong,
  Sasagawa, Nohara, Takagi} \emph{et~al.}}]{Lake02a}
\bibinfo{author}{\bibnamefont{Lake}, \bibfnamefont{B.}},
  \bibinfo{author}{\bibfnamefont{H.~M.} \bibnamefont{R$\o$nnow}},
  \bibinfo{author}{\bibfnamefont{N.~B.} \bibnamefont{Christensen}},
  \bibinfo{author}{\bibfnamefont{G.}~\bibnamefont{Aeppli}},
  \bibinfo{author}{\bibfnamefont{K.}~\bibnamefont{Lefmann}},
  \bibinfo{author}{\bibfnamefont{D.~F.} \bibnamefont{McMorrow}},
  \bibinfo{author}{\bibfnamefont{P.}~\bibnamefont{Vorderwisch}},
  \bibinfo{author}{\bibfnamefont{P.}~\bibnamefont{Smeibidl}},
  \bibinfo{author}{\bibfnamefont{N.}~\bibnamefont{Mangkorntong}},
  \bibinfo{author}{\bibfnamefont{T.}~\bibnamefont{Sasagawa}},
  \bibinfo{author}{\bibfnamefont{M.}~\bibnamefont{Nohara}},
  \bibinfo{author}{\bibfnamefont{H.}~\bibnamefont{Takagi}}, \emph{et~al.},
  \bibinfo{year}{2002}, \bibinfo{journal}{Nature}
  \textbf{\bibinfo{volume}{415}}, \bibinfo{pages}{299}.

\bibitem[{\citenamefont{Lanfredi} \emph{et~al.}(2006)\citenamefont{Lanfredi,
  Sergeenkov, and Araujo-Moreira}}]{Lanfredi06a}
\bibinfo{author}{\bibnamefont{Lanfredi}, \bibfnamefont{A.}},
  \bibinfo{author}{\bibfnamefont{S.}~\bibnamefont{Sergeenkov}}, and
  \bibinfo{author}{\bibfnamefont{F.}~\bibnamefont{Araujo-Moreira}},
  \bibinfo{year}{2006}, \bibinfo{journal}{Physica C}
  \textbf{\bibinfo{volume}{450}}(\bibinfo{number}{1-2}), \bibinfo{pages}{40}.

\bibitem[{\citenamefont{Lang} \emph{et~al.}(2002)\citenamefont{Lang, Madhavan,
  Hoffman, Hudson, Eisaki, Uchida, and Davis}}]{Lang02a}
\bibinfo{author}{\bibnamefont{Lang}, \bibfnamefont{K.}},
  \bibinfo{author}{\bibfnamefont{V.}~\bibnamefont{Madhavan}},
  \bibinfo{author}{\bibfnamefont{J.}~\bibnamefont{Hoffman}},
  \bibinfo{author}{\bibfnamefont{E.}~\bibnamefont{Hudson}},
  \bibinfo{author}{\bibfnamefont{H.}~\bibnamefont{Eisaki}},
  \bibinfo{author}{\bibfnamefont{S.}~\bibnamefont{Uchida}}, and
  \bibinfo{author}{\bibfnamefont{J.}~\bibnamefont{Davis}},
  \bibinfo{year}{2002}, \bibinfo{journal}{Nature}
  \textbf{\bibinfo{volume}{415}}, \bibinfo{pages}{412}.

\bibitem[{\citenamefont{Lanzara} \emph{et~al.}(2001)\citenamefont{Lanzara,
  Bogdanov, X.~J. Zhou~and, Feng, Lu, Yoshida, Eisaki, Fujimori, Kishio,
  Shimoyama, Noda, Uchida} \emph{et~al.}}]{Lanzara01a}
\bibinfo{author}{\bibnamefont{Lanzara}, \bibfnamefont{A.}},
  \bibinfo{author}{\bibfnamefont{P.~V.} \bibnamefont{Bogdanov}},
  \bibinfo{author}{\bibfnamefont{S.~A.~K.} \bibnamefont{X.~J. Zhou~and}},
  \bibinfo{author}{\bibfnamefont{D.~L.} \bibnamefont{Feng}},
  \bibinfo{author}{\bibfnamefont{E.~D.} \bibnamefont{Lu}},
  \bibinfo{author}{\bibfnamefont{T.}~\bibnamefont{Yoshida}},
  \bibinfo{author}{\bibfnamefont{H.}~\bibnamefont{Eisaki}},
  \bibinfo{author}{\bibfnamefont{A.}~\bibnamefont{Fujimori}},
  \bibinfo{author}{\bibfnamefont{K.}~\bibnamefont{Kishio}},
  \bibinfo{author}{\bibfnamefont{J.-I.} \bibnamefont{Shimoyama}},
  \bibinfo{author}{\bibfnamefont{T.}~\bibnamefont{Noda}},
  \bibinfo{author}{\bibfnamefont{S.}~\bibnamefont{Uchida}}, \emph{et~al.},
  \bibinfo{year}{2001}, \bibinfo{journal}{Nature}
  \textbf{\bibinfo{volume}{412}}, \bibinfo{pages}{510}.

\bibitem[{\citenamefont{Lavrov} \emph{et~al.}(2004)\citenamefont{Lavrov, Kang,
  Kurita, Suzuki, Komiya, Lynn, Lee, Dai, and Ando}}]{Lavrov04a}
\bibinfo{author}{\bibnamefont{Lavrov}, \bibfnamefont{A.~N.}},
  \bibinfo{author}{\bibfnamefont{H.~J.} \bibnamefont{Kang}},
  \bibinfo{author}{\bibfnamefont{Y.}~\bibnamefont{Kurita}},
  \bibinfo{author}{\bibfnamefont{T.}~\bibnamefont{Suzuki}},
  \bibinfo{author}{\bibfnamefont{S.}~\bibnamefont{Komiya}},
  \bibinfo{author}{\bibfnamefont{J.~W.} \bibnamefont{Lynn}},
  \bibinfo{author}{\bibfnamefont{S.-H.} \bibnamefont{Lee}},
  \bibinfo{author}{\bibfnamefont{P.}~\bibnamefont{Dai}}, and
  \bibinfo{author}{\bibfnamefont{Y.}~\bibnamefont{Ando}}, \bibinfo{year}{2004},
  \bibinfo{journal}{Phys.\ Rev.\ Lett.}
  \textbf{\bibinfo{volume}{92}}(\bibinfo{number}{22}), \bibinfo{eid}{227003}.

\bibitem[{\citenamefont{LeBoeuf} \emph{et~al.}(2007)\citenamefont{LeBoeuf,
  Doiron-Leyraud, Levallois, Daou, Bonnemaison, Hussey, andB.J. Ramshaw, Liang,
  Bonn, Hardy, Proust, and Taillefer}}]{LeBoeuf07a}
\bibinfo{author}{\bibnamefont{LeBoeuf}, \bibfnamefont{D.}},
  \bibinfo{author}{\bibfnamefont{N.}~\bibnamefont{Doiron-Leyraud}},
  \bibinfo{author}{\bibfnamefont{J.}~\bibnamefont{Levallois}},
  \bibinfo{author}{\bibfnamefont{R.}~\bibnamefont{Daou}},
  \bibinfo{author}{\bibfnamefont{J.-B.} \bibnamefont{Bonnemaison}},
  \bibinfo{author}{\bibfnamefont{N.}~\bibnamefont{Hussey}},
  \bibinfo{author}{\bibfnamefont{L.~B.} \bibnamefont{andB.J. Ramshaw}},
  \bibinfo{author}{\bibfnamefont{R.}~\bibnamefont{Liang}},
  \bibinfo{author}{\bibfnamefont{D.}~\bibnamefont{Bonn}},
  \bibinfo{author}{\bibfnamefont{W.}~\bibnamefont{Hardy}},
  \bibinfo{author}{\bibfnamefont{C.}~\bibnamefont{Proust}}, and
  \bibinfo{author}{\bibfnamefont{L.}~\bibnamefont{Taillefer}},
  \bibinfo{year}{2007}, \bibinfo{journal}{Nature}
  \textbf{\bibinfo{volume}{450}}, \bibinfo{pages}{533}.

\bibitem[{\citenamefont{Leca} \emph{et~al.}(2006)\citenamefont{Leca, Blank,
  Rijnders, Bals, and van Tendeloo}}]{Leca06a}
\bibinfo{author}{\bibnamefont{Leca}, \bibfnamefont{V.}},
  \bibinfo{author}{\bibfnamefont{D.~H.~A.} \bibnamefont{Blank}},
  \bibinfo{author}{\bibfnamefont{G.}~\bibnamefont{Rijnders}},
  \bibinfo{author}{\bibfnamefont{S.}~\bibnamefont{Bals}}, and
  \bibinfo{author}{\bibfnamefont{G.}~\bibnamefont{van Tendeloo}},
  \bibinfo{year}{2006}, \bibinfo{journal}{Appl.\ Phys.\ Lett.}
  \textbf{\bibinfo{volume}{89}}(\bibinfo{number}{9}), \bibinfo{pages}{092504}.

\bibitem[{\citenamefont{Lee and Kivelson}(2003)}]{Lee03a}
\bibinfo{author}{\bibnamefont{Lee}, \bibfnamefont{D.-H.}}, and
  \bibinfo{author}{\bibfnamefont{S.~A.} \bibnamefont{Kivelson}},
  \bibinfo{year}{2003}, \bibinfo{journal}{Phys.\ Rev.\ B}
  \textbf{\bibinfo{volume}{67}}(\bibinfo{number}{2}), \bibinfo{pages}{024506}.

\bibitem[{\citenamefont{Lee}
  \emph{et~al.}(2006{\natexlab{a}})\citenamefont{Lee, Fujita, McElroy, Slezak,
  Wang, Aiura, Bando, Ishikado, Masui, Zhu, Balatsky, Eisaki}
  \emph{et~al.}}]{JLee06a}
\bibinfo{author}{\bibnamefont{Lee}, \bibfnamefont{J.}},
  \bibinfo{author}{\bibfnamefont{K.}~\bibnamefont{Fujita}},
  \bibinfo{author}{\bibfnamefont{K.}~\bibnamefont{McElroy}},
  \bibinfo{author}{\bibfnamefont{J.~A.} \bibnamefont{Slezak}},
  \bibinfo{author}{\bibfnamefont{M.}~\bibnamefont{Wang}},
  \bibinfo{author}{\bibfnamefont{Y.}~\bibnamefont{Aiura}},
  \bibinfo{author}{\bibfnamefont{H.}~\bibnamefont{Bando}},
  \bibinfo{author}{\bibfnamefont{M.}~\bibnamefont{Ishikado}},
  \bibinfo{author}{\bibfnamefont{T.}~\bibnamefont{Masui}},
  \bibinfo{author}{\bibfnamefont{J.~X.} \bibnamefont{Zhu}},
  \bibinfo{author}{\bibfnamefont{A.~V.} \bibnamefont{Balatsky}},
  \bibinfo{author}{\bibfnamefont{H.}~\bibnamefont{Eisaki}}, \emph{et~al.},
  \bibinfo{year}{2006}{\natexlab{a}}, \bibinfo{journal}{Nature}
  \textbf{\bibinfo{volume}{442}}(\bibinfo{number}{7102}), \bibinfo{pages}{546}.

\bibitem[{\citenamefont{Lee}
  \emph{et~al.}(2006{\natexlab{b}})\citenamefont{Lee, Nagaosa, and
  Wen}}]{Lee06a}
\bibinfo{author}{\bibnamefont{Lee}, \bibfnamefont{P.~A.}},
  \bibinfo{author}{\bibfnamefont{N.}~\bibnamefont{Nagaosa}}, and
  \bibinfo{author}{\bibfnamefont{X.-G.} \bibnamefont{Wen}},
  \bibinfo{year}{2006}{\natexlab{b}}, \bibinfo{journal}{Rev.\ Mod.\ Phys.}
  \textbf{\bibinfo{volume}{78}}(\bibinfo{number}{1}), \bibinfo{eid}{17}.

\bibitem[{\citenamefont{Lefebvre} \emph{et~al.}(2000)\citenamefont{Lefebvre,
  Wzietek, Brown, Bourbonnais, J\'erome, M\'ezi\`ere, Fourmigu\'e, and
  Batail}}]{Lefebvre00a}
\bibinfo{author}{\bibnamefont{Lefebvre}, \bibfnamefont{S.}},
  \bibinfo{author}{\bibfnamefont{P.}~\bibnamefont{Wzietek}},
  \bibinfo{author}{\bibfnamefont{S.}~\bibnamefont{Brown}},
  \bibinfo{author}{\bibfnamefont{C.}~\bibnamefont{Bourbonnais}},
  \bibinfo{author}{\bibfnamefont{D.}~\bibnamefont{J\'erome}},
  \bibinfo{author}{\bibfnamefont{C.}~\bibnamefont{M\'ezi\`ere}},
  \bibinfo{author}{\bibfnamefont{M.}~\bibnamefont{Fourmigu\'e}}, and
  \bibinfo{author}{\bibfnamefont{P.}~\bibnamefont{Batail}},
  \bibinfo{year}{2000}, \bibinfo{journal}{Phys.\ Rev.\ Lett.}
  \textbf{\bibinfo{volume}{85}}(\bibinfo{number}{25}), \bibinfo{pages}{5420}.

\bibitem[{\citenamefont{Li} \emph{et~al.}(2003)\citenamefont{Li, Zhang, and
  Luo}}]{Li03a}
\bibinfo{author}{\bibnamefont{Li}, \bibfnamefont{J.-X.}},
  \bibinfo{author}{\bibfnamefont{J.}~\bibnamefont{Zhang}}, and
  \bibinfo{author}{\bibfnamefont{J.}~\bibnamefont{Luo}}, \bibinfo{year}{2003},
  \bibinfo{journal}{Phys.\ Rev.\ B}
  \textbf{\bibinfo{volume}{68}}(\bibinfo{number}{22}), \bibinfo{pages}{224503}.

\bibitem[{\citenamefont{Li} \emph{et~al.}(2007{\natexlab{a}})\citenamefont{Li,
  Balakirev, and Greene}}]{Li07b}
\bibinfo{author}{\bibnamefont{Li}, \bibfnamefont{P.}},
  \bibinfo{author}{\bibfnamefont{F.~F.} \bibnamefont{Balakirev}}, and
  \bibinfo{author}{\bibfnamefont{R.~L.} \bibnamefont{Greene}},
  \bibinfo{year}{2007}{\natexlab{a}}, \bibinfo{journal}{Phys.\ Rev.\ Lett.}
  \textbf{\bibinfo{volume}{99}}(\bibinfo{number}{4}), \bibinfo{eid}{047003}.

\bibitem[{\citenamefont{Li} \emph{et~al.}(2007{\natexlab{b}})\citenamefont{Li,
  Balakirev, and Greene}}]{PLi07c}
\bibinfo{author}{\bibnamefont{Li}, \bibfnamefont{P.}},
  \bibinfo{author}{\bibfnamefont{F.~F.} \bibnamefont{Balakirev}}, and
  \bibinfo{author}{\bibfnamefont{R.~L.} \bibnamefont{Greene}},
  \bibinfo{year}{2007}{\natexlab{b}}, \bibinfo{journal}{Phys.\ Rev.\ B}
  \textbf{\bibinfo{volume}{75}}(\bibinfo{number}{17}), \bibinfo{eid}{172508}.

\bibitem[{\citenamefont{Li} \emph{et~al.}(2007{\natexlab{c}})\citenamefont{Li,
  Behnia, and Greene}}]{PLi07a}
\bibinfo{author}{\bibnamefont{Li}, \bibfnamefont{P.}},
  \bibinfo{author}{\bibfnamefont{K.}~\bibnamefont{Behnia}}, and
  \bibinfo{author}{\bibfnamefont{R.~L.} \bibnamefont{Greene}},
  \bibinfo{year}{2007}{\natexlab{c}}, \bibinfo{journal}{Phys.\ Rev.\ B}
  \textbf{\bibinfo{volume}{75}}(\bibinfo{number}{2}), \bibinfo{eid}{020506}.

\bibitem[{\citenamefont{Li and Greene}(2007)}]{Li07a}
\bibinfo{author}{\bibnamefont{Li}, \bibfnamefont{P.}}, and
  \bibinfo{author}{\bibfnamefont{R.~L.} \bibnamefont{Greene}},
  \bibinfo{year}{2007}, \bibinfo{journal}{Phys.\ Rev.\ B}
  \textbf{\bibinfo{volume}{76}}(\bibinfo{number}{17}), \bibinfo{eid}{174512}.

\bibitem[{\citenamefont{Li} \emph{et~al.}(2008{\natexlab{a}})\citenamefont{Li,
  Chi, Zhao, Wen, Stone, Lynn, and Dai}}]{Li08a}
\bibinfo{author}{\bibnamefont{Li}, \bibfnamefont{S.}},
  \bibinfo{author}{\bibfnamefont{S.}~\bibnamefont{Chi}},
  \bibinfo{author}{\bibfnamefont{J.}~\bibnamefont{Zhao}},
  \bibinfo{author}{\bibfnamefont{H.-H.} \bibnamefont{Wen}},
  \bibinfo{author}{\bibfnamefont{M.~B.} \bibnamefont{Stone}},
  \bibinfo{author}{\bibfnamefont{J.~W.} \bibnamefont{Lynn}}, and
  \bibinfo{author}{\bibfnamefont{P.}~\bibnamefont{Dai}},
  \bibinfo{year}{2008}{\natexlab{a}}, \bibinfo{journal}{Phys.\ Rev.\ B}
  \textbf{\bibinfo{volume}{78}}(\bibinfo{number}{1}), \bibinfo{eid}{014520}.

\bibitem[{\citenamefont{Li} \emph{et~al.}(2005{\natexlab{a}})\citenamefont{Li,
  Wilson, Mandrus, Zhao, Onose, Tokura, and Dai}}]{Li05a}
\bibinfo{author}{\bibnamefont{Li}, \bibfnamefont{S.}},
  \bibinfo{author}{\bibfnamefont{S.~D.} \bibnamefont{Wilson}},
  \bibinfo{author}{\bibfnamefont{D.}~\bibnamefont{Mandrus}},
  \bibinfo{author}{\bibfnamefont{B.}~\bibnamefont{Zhao}},
  \bibinfo{author}{\bibfnamefont{Y.}~\bibnamefont{Onose}},
  \bibinfo{author}{\bibfnamefont{Y.}~\bibnamefont{Tokura}}, and
  \bibinfo{author}{\bibfnamefont{P.}~\bibnamefont{Dai}},
  \bibinfo{year}{2005}{\natexlab{a}}, \bibinfo{journal}{Phys.\ Rev.\ B}
  \textbf{\bibinfo{volume}{71}}(\bibinfo{number}{5}), \bibinfo{eid}{054505}.

\bibitem[{\citenamefont{Li} \emph{et~al.}(2005{\natexlab{b}})\citenamefont{Li,
  Taillefer, Wang, and Chen}}]{Li05b}
\bibinfo{author}{\bibnamefont{Li}, \bibfnamefont{S.~Y.}},
  \bibinfo{author}{\bibfnamefont{L.}~\bibnamefont{Taillefer}},
  \bibinfo{author}{\bibfnamefont{C.~H.} \bibnamefont{Wang}}, and
  \bibinfo{author}{\bibfnamefont{X.~H.} \bibnamefont{Chen}},
  \bibinfo{year}{2005}{\natexlab{b}}, \bibinfo{journal}{Phys.\ Rev.\ Lett.}
  \textbf{\bibinfo{volume}{95}}(\bibinfo{number}{15}), \bibinfo{eid}{156603}.

\bibitem[{\citenamefont{Li} \emph{et~al.}(2008{\natexlab{b}})\citenamefont{Li,
  Bal\'edent, Barisic, Cho, Fauqu\'e, Sidis, Yu, Zhao, Bourges, and
  Greven}}]{YLi08a}
\bibinfo{author}{\bibnamefont{Li}, \bibfnamefont{Y.}},
  \bibinfo{author}{\bibfnamefont{V.}~\bibnamefont{Bal\'edent}},
  \bibinfo{author}{\bibfnamefont{N.}~\bibnamefont{Barisic}},
  \bibinfo{author}{\bibfnamefont{Y.}~\bibnamefont{Cho}},
  \bibinfo{author}{\bibfnamefont{B.}~\bibnamefont{Fauqu\'e}},
  \bibinfo{author}{\bibfnamefont{Y.}~\bibnamefont{Sidis}},
  \bibinfo{author}{\bibfnamefont{G.}~\bibnamefont{Yu}},
  \bibinfo{author}{\bibfnamefont{X.}~\bibnamefont{Zhao}},
  \bibinfo{author}{\bibfnamefont{P.}~\bibnamefont{Bourges}}, and
  \bibinfo{author}{\bibfnamefont{M.}~\bibnamefont{Greven}},
  \bibinfo{year}{2008}{\natexlab{b}}, \bibinfo{journal}{Nature}
  \textbf{\bibinfo{volume}{455}}, \bibinfo{pages}{372}.

\bibitem[{\citenamefont{Li} \emph{et~al.}(2009)\citenamefont{Li, Jovanovic,
  Raffy, and Megtert}}]{ZZLi08a}
\bibinfo{author}{\bibnamefont{Li}, \bibfnamefont{Z.}},
  \bibinfo{author}{\bibfnamefont{V.}~\bibnamefont{Jovanovic}},
  \bibinfo{author}{\bibfnamefont{H.}~\bibnamefont{Raffy}}, and
  \bibinfo{author}{\bibfnamefont{S.}~\bibnamefont{Megtert}},
  \bibinfo{year}{2009}, \bibinfo{journal}{Physica C}
  \textbf{\bibinfo{volume}{469}}(\bibinfo{number}{2-3}), \bibinfo{pages}{73 }.

\bibitem[{\citenamefont{Liang} \emph{et~al.}(1995)\citenamefont{Liang, Guo,
  Badresingh, Xu, Tang, Croft, Chen, Sahiner, O, and Markert}}]{Liang94a}
\bibinfo{author}{\bibnamefont{Liang}, \bibfnamefont{G.}},
  \bibinfo{author}{\bibfnamefont{Y.}~\bibnamefont{Guo}},
  \bibinfo{author}{\bibfnamefont{D.}~\bibnamefont{Badresingh}},
  \bibinfo{author}{\bibfnamefont{W.}~\bibnamefont{Xu}},
  \bibinfo{author}{\bibfnamefont{Y.}~\bibnamefont{Tang}},
  \bibinfo{author}{\bibfnamefont{M.}~\bibnamefont{Croft}},
  \bibinfo{author}{\bibfnamefont{J.}~\bibnamefont{Chen}},
  \bibinfo{author}{\bibfnamefont{A.}~\bibnamefont{Sahiner}},
  \bibinfo{author}{\bibfnamefont{B.-h.} \bibnamefont{O}}, and
  \bibinfo{author}{\bibfnamefont{J.~T.} \bibnamefont{Markert}},
  \bibinfo{year}{1995}, \bibinfo{journal}{Phys.\ Rev.\ B}
  \textbf{\bibinfo{volume}{51}}(\bibinfo{number}{2}), \bibinfo{pages}{1258}.

\bibitem[{\citenamefont{Lin and Millis}(2005)}]{Lin05a}
\bibinfo{author}{\bibnamefont{Lin}, \bibfnamefont{J.}}, and
  \bibinfo{author}{\bibfnamefont{A.~J.} \bibnamefont{Millis}},
  \bibinfo{year}{2005}, \bibinfo{journal}{Phys.\ Rev.\ B}
  \textbf{\bibinfo{volume}{72}}(\bibinfo{number}{21}), \bibinfo{eid}{214506}.

\bibitem[{\citenamefont{Liu and Wu}(2007)}]{Liu07a}
\bibinfo{author}{\bibnamefont{Liu}, \bibfnamefont{C.~S.}}, and
  \bibinfo{author}{\bibfnamefont{W.~C.} \bibnamefont{Wu}},
  \bibinfo{year}{2007}, \bibinfo{journal}{Phys.\ Rev.\ B}
  \textbf{\bibinfo{volume}{76}}(\bibinfo{number}{22}), \bibinfo{eid}{220504}.

\bibitem[{\citenamefont{Liu} \emph{et~al.}(2008)\citenamefont{Liu, Liu, Zhang,
  Zhao, Meng, Xiaowen~Jia, Lu, Wang, Zhou, Zhu, Wang, Wu}
  \emph{et~al.}}]{Liu08b}
\bibinfo{author}{\bibnamefont{Liu}, \bibfnamefont{H.}},
  \bibinfo{author}{\bibfnamefont{G.}~\bibnamefont{Liu}},
  \bibinfo{author}{\bibfnamefont{W.}~\bibnamefont{Zhang}},
  \bibinfo{author}{\bibfnamefont{L.}~\bibnamefont{Zhao}},
  \bibinfo{author}{\bibfnamefont{J.}~\bibnamefont{Meng}},
  \bibinfo{author}{\bibfnamefont{X.~D.} \bibnamefont{Xiaowen~Jia}},
  \bibinfo{author}{\bibfnamefont{W.}~\bibnamefont{Lu}},
  \bibinfo{author}{\bibfnamefont{G.}~\bibnamefont{Wang}},
  \bibinfo{author}{\bibfnamefont{Y.}~\bibnamefont{Zhou}},
  \bibinfo{author}{\bibfnamefont{Y.}~\bibnamefont{Zhu}},
  \bibinfo{author}{\bibfnamefont{X.}~\bibnamefont{Wang}},
  \bibinfo{author}{\bibfnamefont{T.}~\bibnamefont{Wu}}, \emph{et~al.},
  \bibinfo{year}{2008}, \bibinfo{journal}{arXiv:0808.0802v1} .

\bibitem[{\citenamefont{Liu} \emph{et~al.}(2001)\citenamefont{Liu, Chen,
  Nachimuthu, Gundakaram, Jung, Kim, and Lee}}]{Liu01a}
\bibinfo{author}{\bibnamefont{Liu}, \bibfnamefont{R.}},
  \bibinfo{author}{\bibfnamefont{J.}~\bibnamefont{Chen}},
  \bibinfo{author}{\bibfnamefont{P.}~\bibnamefont{Nachimuthu}},
  \bibinfo{author}{\bibfnamefont{R.}~\bibnamefont{Gundakaram}},
  \bibinfo{author}{\bibfnamefont{C.}~\bibnamefont{Jung}},
  \bibinfo{author}{\bibfnamefont{J.}~\bibnamefont{Kim}}, and
  \bibinfo{author}{\bibfnamefont{S.}~\bibnamefont{Lee}}, \bibinfo{year}{2001},
  \bibinfo{journal}{Solid State Comm.} \textbf{\bibinfo{volume}{118}},
  \bibinfo{pages}{367}.

\bibitem[{\citenamefont{Liu} \emph{et~al.}(2005)\citenamefont{Liu, Wen, Shan,
  Yang, Lu, Gao, Park, Jung, and Lee}}]{Liu05a}
\bibinfo{author}{\bibnamefont{Liu}, \bibfnamefont{Z.~Y.}},
  \bibinfo{author}{\bibfnamefont{H.~H.} \bibnamefont{Wen}},
  \bibinfo{author}{\bibfnamefont{L.}~\bibnamefont{Shan}},
  \bibinfo{author}{\bibfnamefont{H.~P.} \bibnamefont{Yang}},
  \bibinfo{author}{\bibfnamefont{X.~F.} \bibnamefont{Lu}},
  \bibinfo{author}{\bibfnamefont{H.}~\bibnamefont{Gao}},
  \bibinfo{author}{\bibfnamefont{M.-S.} \bibnamefont{Park}},
  \bibinfo{author}{\bibfnamefont{C.~U.} \bibnamefont{Jung}}, and
  \bibinfo{author}{\bibfnamefont{S.-I.} \bibnamefont{Lee}},
  \bibinfo{year}{2005}, \bibinfo{journal}{Europhys.\ Lett.}
  \textbf{\bibinfo{volume}{69}}(\bibinfo{number}{2}), \bibinfo{eid}{263}.

\bibitem[{\citenamefont{Lofwander} \emph{et~al.}(2001)\citenamefont{Lofwander,
  Shumeiko, and Wendin}}]{Lofwander01a}
\bibinfo{author}{\bibnamefont{Lofwander}, \bibfnamefont{T.}},
  \bibinfo{author}{\bibfnamefont{V.~S.} \bibnamefont{Shumeiko}}, and
  \bibinfo{author}{\bibfnamefont{G.}~\bibnamefont{Wendin}},
  \bibinfo{year}{2001}, \bibinfo{journal}{Supercond.\ Sci.\ Technol.}
  \textbf{\bibinfo{volume}{14}}(\bibinfo{number}{5}), \bibinfo{pages}{R53}.

\bibitem[{\citenamefont{Luke} \emph{et~al.}(1990)\citenamefont{Luke, Le,
  Sternlieb, Uemura, Brewer, Kadono, Kiefl, Kreitzman, Riseman, Stronach,
  Davis, Uchida} \emph{et~al.}}]{Luke90a}
\bibinfo{author}{\bibnamefont{Luke}, \bibfnamefont{G.~M.}},
  \bibinfo{author}{\bibfnamefont{L.~P.} \bibnamefont{Le}},
  \bibinfo{author}{\bibfnamefont{B.~J.} \bibnamefont{Sternlieb}},
  \bibinfo{author}{\bibfnamefont{Y.~J.} \bibnamefont{Uemura}},
  \bibinfo{author}{\bibfnamefont{J.~H.} \bibnamefont{Brewer}},
  \bibinfo{author}{\bibfnamefont{R.}~\bibnamefont{Kadono}},
  \bibinfo{author}{\bibfnamefont{R.~F.} \bibnamefont{Kiefl}},
  \bibinfo{author}{\bibfnamefont{S.~R.} \bibnamefont{Kreitzman}},
  \bibinfo{author}{\bibfnamefont{T.~M.} \bibnamefont{Riseman}},
  \bibinfo{author}{\bibfnamefont{C.~E.} \bibnamefont{Stronach}},
  \bibinfo{author}{\bibfnamefont{M.~R.} \bibnamefont{Davis}},
  \bibinfo{author}{\bibfnamefont{S.}~\bibnamefont{Uchida}}, \emph{et~al.},
  \bibinfo{year}{1990}, \bibinfo{journal}{Phys.\ Rev.\ B}
  \textbf{\bibinfo{volume}{42}}(\bibinfo{number}{13}), \bibinfo{pages}{7981}.

\bibitem[{\citenamefont{Luo and Xiang}(2005)}]{Luo05a}
\bibinfo{author}{\bibnamefont{Luo}, \bibfnamefont{H.~G.}}, and
  \bibinfo{author}{\bibfnamefont{T.}~\bibnamefont{Xiang}},
  \bibinfo{year}{2005}, \bibinfo{journal}{Phys.\ Rev.\ Lett.}
  \textbf{\bibinfo{volume}{94}}(\bibinfo{number}{2}), \bibinfo{eid}{027001}.

\bibitem[{\citenamefont{Lupi} \emph{et~al.}(1998)\citenamefont{Lupi, Capizzi,
  Calvani, Ruzicka, Maselli, Dore, and Paolone}}]{Lupi98a}
\bibinfo{author}{\bibnamefont{Lupi}, \bibfnamefont{S.}},
  \bibinfo{author}{\bibfnamefont{M.}~\bibnamefont{Capizzi}},
  \bibinfo{author}{\bibfnamefont{P.}~\bibnamefont{Calvani}},
  \bibinfo{author}{\bibfnamefont{B.}~\bibnamefont{Ruzicka}},
  \bibinfo{author}{\bibfnamefont{P.}~\bibnamefont{Maselli}},
  \bibinfo{author}{\bibfnamefont{P.}~\bibnamefont{Dore}}, and
  \bibinfo{author}{\bibfnamefont{A.}~\bibnamefont{Paolone}},
  \bibinfo{year}{1998}, \bibinfo{journal}{Phys.\ Rev.\ B}
  \textbf{\bibinfo{volume}{57}}(\bibinfo{number}{2}), \bibinfo{pages}{1248}.

\bibitem[{\citenamefont{Lupi} \emph{et~al.}(1999)\citenamefont{Lupi, Maselli,
  Capizzi, Calvani, Giura, and Roy}}]{Lupi99a}
\bibinfo{author}{\bibnamefont{Lupi}, \bibfnamefont{S.}},
  \bibinfo{author}{\bibfnamefont{P.}~\bibnamefont{Maselli}},
  \bibinfo{author}{\bibfnamefont{M.}~\bibnamefont{Capizzi}},
  \bibinfo{author}{\bibfnamefont{P.}~\bibnamefont{Calvani}},
  \bibinfo{author}{\bibfnamefont{P.}~\bibnamefont{Giura}}, and
  \bibinfo{author}{\bibfnamefont{P.}~\bibnamefont{Roy}}, \bibinfo{year}{1999},
  \bibinfo{journal}{Phys.\ Rev.\ Lett.}
  \textbf{\bibinfo{volume}{83}}(\bibinfo{number}{23}), \bibinfo{pages}{4852}.

\bibitem[{\citenamefont{Luttinger}(1960)}]{Luttinger60a}
\bibinfo{author}{\bibnamefont{Luttinger}, \bibfnamefont{J.~M.}},
  \bibinfo{year}{1960}, \bibinfo{journal}{Phys.\ Rev.}
  \textbf{\bibinfo{volume}{119}}(\bibinfo{number}{4}), \bibinfo{pages}{1153}.

\bibitem[{\citenamefont{Lynn and Skanthakumar}(2001)}]{Lynn01a}
\bibinfo{author}{\bibnamefont{Lynn}, \bibfnamefont{J.}}, and
  \bibinfo{author}{\bibfnamefont{S.}~\bibnamefont{Skanthakumar}},
  \bibinfo{year}{2001}, in \emph{\bibinfo{booktitle}{Handbook on the Physics
  and Chemistry of Rare Earths}}, edited by
  \bibinfo{editor}{\bibfnamefont{L.~E.} \bibnamefont{K.A.~Gschneidner},
  \bibfnamefont{Jr.}} and
  \bibinfo{editor}{\bibfnamefont{M.}~\bibnamefont{Maple}}
  (\bibinfo{publisher}{Elsevier Science B.V}), volume~\bibinfo{volume}{31}, p.
  \bibinfo{pages}{313}.

\bibitem[{\citenamefont{Lynn} \emph{et~al.}(1990)\citenamefont{Lynn, Sumarlin,
  Skanthakumar, Li, Shelton, Peng, Fisk, and Cheong}}]{Lynn90a}
\bibinfo{author}{\bibnamefont{Lynn}, \bibfnamefont{J.~W.}},
  \bibinfo{author}{\bibfnamefont{I.~W.} \bibnamefont{Sumarlin}},
  \bibinfo{author}{\bibfnamefont{S.}~\bibnamefont{Skanthakumar}},
  \bibinfo{author}{\bibfnamefont{W.-H.} \bibnamefont{Li}},
  \bibinfo{author}{\bibfnamefont{R.~N.} \bibnamefont{Shelton}},
  \bibinfo{author}{\bibfnamefont{J.~L.} \bibnamefont{Peng}},
  \bibinfo{author}{\bibfnamefont{Z.}~\bibnamefont{Fisk}}, and
  \bibinfo{author}{\bibfnamefont{S.-W.} \bibnamefont{Cheong}},
  \bibinfo{year}{1990}, \bibinfo{journal}{Phys.\ Rev.\ B}
  \textbf{\bibinfo{volume}{41}}(\bibinfo{number}{4}), \bibinfo{pages}{2569}.

\bibitem[{\citenamefont{Lyons} \emph{et~al.}(1988)\citenamefont{Lyons, Fleury,
  Remeika, Cooper, and Negran}}]{Lyons88a}
\bibinfo{author}{\bibnamefont{Lyons}, \bibfnamefont{K.~B.}},
  \bibinfo{author}{\bibfnamefont{P.~A.} \bibnamefont{Fleury}},
  \bibinfo{author}{\bibfnamefont{J.~P.} \bibnamefont{Remeika}},
  \bibinfo{author}{\bibfnamefont{A.~S.} \bibnamefont{Cooper}}, and
  \bibinfo{author}{\bibfnamefont{T.~J.} \bibnamefont{Negran}},
  \bibinfo{year}{1988}, \bibinfo{journal}{Phys.\ Rev.\ B}
  \textbf{\bibinfo{volume}{37}}(\bibinfo{number}{4}), \bibinfo{pages}{2353}.

\bibitem[{\citenamefont{Ma} \emph{et~al.}(1993)\citenamefont{Ma, Taber,
  Lombardo, Kapitulnik, Beasley, Merchant, Eom, Hou, and Phillips}}]{Ma93a}
\bibinfo{author}{\bibnamefont{Ma}, \bibfnamefont{Z.}},
  \bibinfo{author}{\bibfnamefont{R.~C.} \bibnamefont{Taber}},
  \bibinfo{author}{\bibfnamefont{L.~W.} \bibnamefont{Lombardo}},
  \bibinfo{author}{\bibfnamefont{A.}~\bibnamefont{Kapitulnik}},
  \bibinfo{author}{\bibfnamefont{M.~R.} \bibnamefont{Beasley}},
  \bibinfo{author}{\bibfnamefont{P.}~\bibnamefont{Merchant}},
  \bibinfo{author}{\bibfnamefont{C.~B.} \bibnamefont{Eom}},
  \bibinfo{author}{\bibfnamefont{S.~Y.} \bibnamefont{Hou}}, and
  \bibinfo{author}{\bibfnamefont{J.~M.} \bibnamefont{Phillips}},
  \bibinfo{year}{1993}, \bibinfo{journal}{Phys.\ Rev.\ Lett.}
  \textbf{\bibinfo{volume}{71}}(\bibinfo{number}{5}), \bibinfo{pages}{781}.

\bibitem[{\citenamefont{Machida}(1989)}]{Machida89a}
\bibinfo{author}{\bibnamefont{Machida}, \bibfnamefont{K.}},
  \bibinfo{year}{1989}, \bibinfo{journal}{Physica C}
  \textbf{\bibinfo{volume}{158}}(\bibinfo{number}{1-2}), \bibinfo{pages}{192}.

\bibitem[{\citenamefont{Macridin} \emph{et~al.}(2006)\citenamefont{Macridin,
  Jarrell, Maier, Kent, and D'Azevedo}}]{Macridin06a}
\bibinfo{author}{\bibnamefont{Macridin}, \bibfnamefont{A.}},
  \bibinfo{author}{\bibfnamefont{M.}~\bibnamefont{Jarrell}},
  \bibinfo{author}{\bibfnamefont{T.}~\bibnamefont{Maier}},
  \bibinfo{author}{\bibfnamefont{P.~R.~C.} \bibnamefont{Kent}}, and
  \bibinfo{author}{\bibfnamefont{E.}~\bibnamefont{D'Azevedo}},
  \bibinfo{year}{2006}, \bibinfo{journal}{Phys.\ Rev.\ Lett.}
  \textbf{\bibinfo{volume}{97}}(\bibinfo{number}{3}), \bibinfo{eid}{036401}.

\bibitem[{\citenamefont{Maier} \emph{et~al.}(2008)\citenamefont{Maier,
  Poilblanc, and Scalapino}}]{Maier08a}
\bibinfo{author}{\bibnamefont{Maier}, \bibfnamefont{T.~A.}},
  \bibinfo{author}{\bibfnamefont{D.}~\bibnamefont{Poilblanc}}, and
  \bibinfo{author}{\bibfnamefont{D.~J.} \bibnamefont{Scalapino}},
  \bibinfo{year}{2008}, \bibinfo{journal}{Phys.\ Rev.\ Lett.}
  \textbf{\bibinfo{volume}{100}}(\bibinfo{number}{23}), \bibinfo{eid}{237001}.

\bibitem[{\citenamefont{Maiser} \emph{et~al.}(1998)\citenamefont{Maiser,
  Fournier, Peng, Araujo-Moreira, Venkatesan, Greene, and Czjzek}}]{Maiser98a}
\bibinfo{author}{\bibnamefont{Maiser}, \bibfnamefont{E.}},
  \bibinfo{author}{\bibfnamefont{P.}~\bibnamefont{Fournier}},
  \bibinfo{author}{\bibfnamefont{J.-L.} \bibnamefont{Peng}},
  \bibinfo{author}{\bibfnamefont{F.~M.} \bibnamefont{Araujo-Moreira}},
  \bibinfo{author}{\bibfnamefont{T.}~\bibnamefont{Venkatesan}},
  \bibinfo{author}{\bibfnamefont{R.}~\bibnamefont{Greene}}, and
  \bibinfo{author}{\bibfnamefont{G.}~\bibnamefont{Czjzek}},
  \bibinfo{year}{1998}, \bibinfo{journal}{Physica C}
  \textbf{\bibinfo{volume}{297}}, \bibinfo{pages}{15}.

\bibitem[{\citenamefont{Maljuk} \emph{et~al.}(1996)\citenamefont{Maljuk,
  Emel'chenko, and Kosenko}}]{Maljuk96a}
\bibinfo{author}{\bibnamefont{Maljuk}, \bibfnamefont{A.~N.}},
  \bibinfo{author}{\bibfnamefont{G.~A.} \bibnamefont{Emel'chenko}}, and
  \bibinfo{author}{\bibfnamefont{A.~V.} \bibnamefont{Kosenko}},
  \bibinfo{year}{1996}, \bibinfo{journal}{Journal of Alloys and Compounds}
  \textbf{\bibinfo{volume}{234}}(\bibinfo{number}{1}), \bibinfo{pages}{52}.

\bibitem[{\citenamefont{Maljuk} \emph{et~al.}(2000)\citenamefont{Maljuk,
  Jokhov, Naumenko, Bdikin, Zver'kov, and Emel'chenko}}]{Maljuk00a}
\bibinfo{author}{\bibnamefont{Maljuk}, \bibfnamefont{A.~N.}},
  \bibinfo{author}{\bibfnamefont{A.~A.} \bibnamefont{Jokhov}},
  \bibinfo{author}{\bibfnamefont{I.~G.} \bibnamefont{Naumenko}},
  \bibinfo{author}{\bibfnamefont{I.~K.} \bibnamefont{Bdikin}},
  \bibinfo{author}{\bibfnamefont{S.~A.} \bibnamefont{Zver'kov}}, and
  \bibinfo{author}{\bibfnamefont{G.~A.} \bibnamefont{Emel'chenko}},
  \bibinfo{year}{2000}, \bibinfo{journal}{Physica C}
  \textbf{\bibinfo{volume}{329}}(\bibinfo{number}{1}), \bibinfo{pages}{51}.

\bibitem[{\citenamefont{Mang}
  \emph{et~al.}(2004{\natexlab{a}})\citenamefont{Mang, Larochelle, Mehta, Vajk,
  Erickson, Lu, Buyers, Marshall, Prokes, and Greven}}]{Mang04b}
\bibinfo{author}{\bibnamefont{Mang}, \bibfnamefont{P.~K.}},
  \bibinfo{author}{\bibfnamefont{S.}~\bibnamefont{Larochelle}},
  \bibinfo{author}{\bibfnamefont{A.}~\bibnamefont{Mehta}},
  \bibinfo{author}{\bibfnamefont{O.~P.} \bibnamefont{Vajk}},
  \bibinfo{author}{\bibfnamefont{A.~S.} \bibnamefont{Erickson}},
  \bibinfo{author}{\bibfnamefont{L.}~\bibnamefont{Lu}},
  \bibinfo{author}{\bibfnamefont{W.~J.~L.} \bibnamefont{Buyers}},
  \bibinfo{author}{\bibfnamefont{A.~F.} \bibnamefont{Marshall}},
  \bibinfo{author}{\bibfnamefont{K.}~\bibnamefont{Prokes}}, and
  \bibinfo{author}{\bibfnamefont{M.}~\bibnamefont{Greven}},
  \bibinfo{year}{2004}{\natexlab{a}}, \bibinfo{journal}{Phys.\ Rev.\ B}
  \textbf{\bibinfo{volume}{70}}(\bibinfo{number}{9}), \bibinfo{eid}{094507}.

\bibitem[{\citenamefont{Mang} \emph{et~al.}(2003)\citenamefont{Mang,
  Larochelle, and M.Greven}}]{mang03a}
\bibinfo{author}{\bibnamefont{Mang}, \bibfnamefont{P.~K.}},
  \bibinfo{author}{\bibfnamefont{S.}~\bibnamefont{Larochelle}}, and
  \bibinfo{author}{\bibnamefont{M.Greven}}, \bibinfo{year}{2003},
  \bibinfo{journal}{Nature} \textbf{\bibinfo{volume}{429}}, \bibinfo{eid}{139}.

\bibitem[{\citenamefont{Mang}
  \emph{et~al.}(2004{\natexlab{b}})\citenamefont{Mang, Vajk, Arvanitaki, Lynn,
  and Greven}}]{Mang04a}
\bibinfo{author}{\bibnamefont{Mang}, \bibfnamefont{P.~K.}},
  \bibinfo{author}{\bibfnamefont{O.~P.} \bibnamefont{Vajk}},
  \bibinfo{author}{\bibfnamefont{A.}~\bibnamefont{Arvanitaki}},
  \bibinfo{author}{\bibfnamefont{J.~W.} \bibnamefont{Lynn}}, and
  \bibinfo{author}{\bibfnamefont{M.}~\bibnamefont{Greven}},
  \bibinfo{year}{2004}{\natexlab{b}}, \bibinfo{journal}{Phys.\ Rev.\ Lett.}
  \textbf{\bibinfo{volume}{93}}(\bibinfo{number}{2}), \bibinfo{eid}{027002}.

\bibitem[{\citenamefont{Manske}
  \emph{et~al.}(2001{\natexlab{a}})\citenamefont{Manske, Eremin, and
  Bennemann}}]{Manske01a}
\bibinfo{author}{\bibnamefont{Manske}, \bibfnamefont{D.}},
  \bibinfo{author}{\bibfnamefont{I.}~\bibnamefont{Eremin}}, and
  \bibinfo{author}{\bibfnamefont{K.~H.} \bibnamefont{Bennemann}},
  \bibinfo{year}{2001}{\natexlab{a}}, \bibinfo{journal}{Phys. Rev. B}
  \textbf{\bibinfo{volume}{63}}(\bibinfo{number}{5}), \bibinfo{pages}{054517}.

\bibitem[{\citenamefont{Manske}
  \emph{et~al.}(2001{\natexlab{b}})\citenamefont{Manske, Eremin, and
  Bennemann}}]{Manske01b}
\bibinfo{author}{\bibnamefont{Manske}, \bibfnamefont{D.}},
  \bibinfo{author}{\bibfnamefont{I.}~\bibnamefont{Eremin}}, and
  \bibinfo{author}{\bibfnamefont{K.~H.} \bibnamefont{Bennemann}},
  \bibinfo{year}{2001}{\natexlab{b}}, \bibinfo{journal}{Europhys.\ Lett.}
  \textbf{\bibinfo{volume}{53}}(\bibinfo{number}{3}), \bibinfo{pages}{371}.

\bibitem[{\citenamefont{Mao} \emph{et~al.}(1992)\citenamefont{Mao, Xi,
  Bhattacharya, Li, Venkatesan, Peng, Greene, Mao, Wu, and Anlage}}]{Mao92a}
\bibinfo{author}{\bibnamefont{Mao}, \bibfnamefont{S.~N.}},
  \bibinfo{author}{\bibfnamefont{X.~X.} \bibnamefont{Xi}},
  \bibinfo{author}{\bibfnamefont{S.}~\bibnamefont{Bhattacharya}},
  \bibinfo{author}{\bibfnamefont{Q.}~\bibnamefont{Li}},
  \bibinfo{author}{\bibfnamefont{T.}~\bibnamefont{Venkatesan}},
  \bibinfo{author}{\bibfnamefont{J.~L.} \bibnamefont{Peng}},
  \bibinfo{author}{\bibfnamefont{R.~L.} \bibnamefont{Greene}},
  \bibinfo{author}{\bibfnamefont{J.}~\bibnamefont{Mao}},
  \bibinfo{author}{\bibfnamefont{D.~H.} \bibnamefont{Wu}}, and
  \bibinfo{author}{\bibfnamefont{S.~M.} \bibnamefont{Anlage}},
  \bibinfo{year}{1992}, \bibinfo{journal}{Appl.\ Phys.\ Lett.}
  \textbf{\bibinfo{volume}{61}}(\bibinfo{number}{19}), \bibinfo{pages}{2356}.

\bibitem[{\citenamefont{Mao} \emph{et~al.}(1994)\citenamefont{Mao, Xi, Li,
  Venkatesan, Beesabathina, Salamanca-Riba, and Wu}}]{Mao94b}
\bibinfo{author}{\bibnamefont{Mao}, \bibfnamefont{S.~N.}},
  \bibinfo{author}{\bibfnamefont{X.~X.} \bibnamefont{Xi}},
  \bibinfo{author}{\bibfnamefont{Q.}~\bibnamefont{Li}},
  \bibinfo{author}{\bibfnamefont{T.}~\bibnamefont{Venkatesan}},
  \bibinfo{author}{\bibfnamefont{D.~P.} \bibnamefont{Beesabathina}},
  \bibinfo{author}{\bibfnamefont{L.}~\bibnamefont{Salamanca-Riba}}, and
  \bibinfo{author}{\bibfnamefont{X.~D.} \bibnamefont{Wu}},
  \bibinfo{year}{1994}, \bibinfo{journal}{J.\ of Appl.\ Phys.}
  \textbf{\bibinfo{volume}{75}}(\bibinfo{number}{4}), \bibinfo{pages}{2119}.

\bibitem[{\citenamefont{Maple}(1990)}]{Maple90a}
\bibinfo{author}{\bibnamefont{Maple}, \bibfnamefont{M.~B.}},
  \bibinfo{year}{1990}, \bibinfo{journal}{Mater.\ Res.\ Bull.}
  \textbf{\bibinfo{volume}{15}}, \bibinfo{pages}{60}.

\bibitem[{\citenamefont{Marcenat} \emph{et~al.}(1993)\citenamefont{Marcenat,
  Calemczuk, Khoder, Bonjour, Marin, and Henry}}]{Marcenat93a}
\bibinfo{author}{\bibnamefont{Marcenat}, \bibfnamefont{C.}},
  \bibinfo{author}{\bibfnamefont{R.}~\bibnamefont{Calemczuk}},
  \bibinfo{author}{\bibfnamefont{A.~F.} \bibnamefont{Khoder}},
  \bibinfo{author}{\bibfnamefont{E.}~\bibnamefont{Bonjour}},
  \bibinfo{author}{\bibfnamefont{C.}~\bibnamefont{Marin}}, and
  \bibinfo{author}{\bibfnamefont{J.~Y.} \bibnamefont{Henry}},
  \bibinfo{year}{1993}, \bibinfo{journal}{Physica C}
  \textbf{\bibinfo{volume}{216}}(\bibinfo{number}{3-4}), \bibinfo{pages}{443}.

\bibitem[{\citenamefont{Marcenat} \emph{et~al.}(1994)\citenamefont{Marcenat,
  Henry, and Calemczuk}}]{Marcenat94a}
\bibinfo{author}{\bibnamefont{Marcenat}, \bibfnamefont{C.}},
  \bibinfo{author}{\bibfnamefont{J.~Y.} \bibnamefont{Henry}}, and
  \bibinfo{author}{\bibfnamefont{R.}~\bibnamefont{Calemczuk}},
  \bibinfo{year}{1994}, \bibinfo{journal}{Physica C}
  \textbf{\bibinfo{volume}{235-240}}(\bibinfo{number}{Part 3}),
  \bibinfo{pages}{1747}.

\bibitem[{\citenamefont{Marin} \emph{et~al.}(1993)\citenamefont{Marin, Henry,
  and Boucherle}}]{Marin93a}
\bibinfo{author}{\bibnamefont{Marin}, \bibfnamefont{C.}},
  \bibinfo{author}{\bibfnamefont{J.~Y.} \bibnamefont{Henry}}, and
  \bibinfo{author}{\bibfnamefont{J.~X.} \bibnamefont{Boucherle}},
  \bibinfo{year}{1993}, \bibinfo{journal}{Solid State Comm.}
  \textbf{\bibinfo{volume}{86}}(\bibinfo{number}{7}), \bibinfo{pages}{425}.

\bibitem[{\citenamefont{Markert} \emph{et~al.}(1990)\citenamefont{Markert,
  Beille, Neumeier, Early, Seaman, Moran, and Maple}}]{Markert90a}
\bibinfo{author}{\bibnamefont{Markert}, \bibfnamefont{J.~T.}},
  \bibinfo{author}{\bibfnamefont{J.}~\bibnamefont{Beille}},
  \bibinfo{author}{\bibfnamefont{J.~J.} \bibnamefont{Neumeier}},
  \bibinfo{author}{\bibfnamefont{E.~A.} \bibnamefont{Early}},
  \bibinfo{author}{\bibfnamefont{C.~L.} \bibnamefont{Seaman}},
  \bibinfo{author}{\bibfnamefont{T.}~\bibnamefont{Moran}}, and
  \bibinfo{author}{\bibfnamefont{M.~B.} \bibnamefont{Maple}},
  \bibinfo{year}{1990}, \bibinfo{journal}{Phys.\ Rev.\ Lett.}
  \textbf{\bibinfo{volume}{64}}(\bibinfo{number}{1}), \bibinfo{pages}{80}.

\bibitem[{\citenamefont{Marshall} \emph{et~al.}(1996)\citenamefont{Marshall,
  Dessau, Loeser, Park, Matsuura, Eckstein, Bozovic, Fournier, Kapitulnik,
  Spicer, and Shen}}]{marshall96a}
\bibinfo{author}{\bibnamefont{Marshall}, \bibfnamefont{D.~S.}},
  \bibinfo{author}{\bibfnamefont{D.~S.} \bibnamefont{Dessau}},
  \bibinfo{author}{\bibfnamefont{A.~G.} \bibnamefont{Loeser}},
  \bibinfo{author}{\bibfnamefont{C.-H.} \bibnamefont{Park}},
  \bibinfo{author}{\bibfnamefont{A.~Y.} \bibnamefont{Matsuura}},
  \bibinfo{author}{\bibfnamefont{J.~N.} \bibnamefont{Eckstein}},
  \bibinfo{author}{\bibfnamefont{I.}~\bibnamefont{Bozovic}},
  \bibinfo{author}{\bibfnamefont{P.}~\bibnamefont{Fournier}},
  \bibinfo{author}{\bibfnamefont{A.}~\bibnamefont{Kapitulnik}},
  \bibinfo{author}{\bibfnamefont{W.~E.} \bibnamefont{Spicer}}, and
  \bibinfo{author}{\bibfnamefont{Z.-X.} \bibnamefont{Shen}},
  \bibinfo{year}{1996}, \bibinfo{journal}{Phys.\ Rev.\ Lett.}
  \textbf{\bibinfo{volume}{76}}(\bibinfo{number}{25}), \bibinfo{pages}{4841}.

\bibitem[{\citenamefont{{Massidda}}
  \emph{et~al.}(1989)\citenamefont{{Massidda}, {Hamada}, {Yu}, and
  {Freeman}}}]{Massidda89a}
\bibinfo{author}{\bibnamefont{{Massidda}}, \bibfnamefont{S.}},
  \bibinfo{author}{\bibfnamefont{N.}~\bibnamefont{{Hamada}}},
  \bibinfo{author}{\bibfnamefont{J.}~\bibnamefont{{Yu}}}, and
  \bibinfo{author}{\bibfnamefont{A.~J.} \bibnamefont{{Freeman}}},
  \bibinfo{year}{1989}, \bibinfo{journal}{Physica C}
  \textbf{\bibinfo{volume}{157}}, \bibinfo{pages}{571}.

\bibitem[{\citenamefont{Matsuda} \emph{et~al.}(1991)\citenamefont{Matsuda,
  Endoh, and Hidaka}}]{Matsuda91a}
\bibinfo{author}{\bibnamefont{Matsuda}, \bibfnamefont{M.}},
  \bibinfo{author}{\bibfnamefont{Y.}~\bibnamefont{Endoh}}, and
  \bibinfo{author}{\bibfnamefont{Y.}~\bibnamefont{Hidaka}},
  \bibinfo{year}{1991}, \bibinfo{journal}{Physica C}
  \textbf{\bibinfo{volume}{179}}(\bibinfo{number}{4-6}), \bibinfo{pages}{347}.

\bibitem[{\citenamefont{Matsuda} \emph{et~al.}(1992)\citenamefont{Matsuda,
  Endoh, Yamada, Kojima, Tanaka, Birgeneau, Kastner, and Shirane}}]{Matsuda92a}
\bibinfo{author}{\bibnamefont{Matsuda}, \bibfnamefont{M.}},
  \bibinfo{author}{\bibfnamefont{Y.}~\bibnamefont{Endoh}},
  \bibinfo{author}{\bibfnamefont{K.}~\bibnamefont{Yamada}},
  \bibinfo{author}{\bibfnamefont{H.}~\bibnamefont{Kojima}},
  \bibinfo{author}{\bibfnamefont{I.}~\bibnamefont{Tanaka}},
  \bibinfo{author}{\bibfnamefont{R.~J.} \bibnamefont{Birgeneau}},
  \bibinfo{author}{\bibfnamefont{M.~A.} \bibnamefont{Kastner}}, and
  \bibinfo{author}{\bibfnamefont{G.}~\bibnamefont{Shirane}},
  \bibinfo{year}{1992}, \bibinfo{journal}{Phys.\ Rev.\ B}
  \textbf{\bibinfo{volume}{45}}(\bibinfo{number}{21}), \bibinfo{pages}{12548}.

\bibitem[{\citenamefont{Matsuda} \emph{et~al.}(2002)\citenamefont{Matsuda,
  Fujita, Yamada, Birgeneau, Endoh, and Shirane}}]{Matsuda02a}
\bibinfo{author}{\bibnamefont{Matsuda}, \bibfnamefont{M.}},
  \bibinfo{author}{\bibfnamefont{M.}~\bibnamefont{Fujita}},
  \bibinfo{author}{\bibfnamefont{K.}~\bibnamefont{Yamada}},
  \bibinfo{author}{\bibfnamefont{R.~J.} \bibnamefont{Birgeneau}},
  \bibinfo{author}{\bibfnamefont{Y.}~\bibnamefont{Endoh}}, and
  \bibinfo{author}{\bibfnamefont{G.}~\bibnamefont{Shirane}},
  \bibinfo{year}{2002}, \bibinfo{journal}{Phys.\ Rev.\ B}
  \textbf{\bibinfo{volume}{65}}(\bibinfo{number}{13}), \bibinfo{pages}{134515}.

\bibitem[{\citenamefont{Matsuda} \emph{et~al.}(1990)\citenamefont{Matsuda,
  Yamada, Kakurai, Kadowaki, Thurston, Endoh, Hidaka, Birgeneau, Kastner,
  Gehring, Moudden, and Shirane}}]{Matsuda90a}
\bibinfo{author}{\bibnamefont{Matsuda}, \bibfnamefont{M.}},
  \bibinfo{author}{\bibfnamefont{K.}~\bibnamefont{Yamada}},
  \bibinfo{author}{\bibfnamefont{K.}~\bibnamefont{Kakurai}},
  \bibinfo{author}{\bibfnamefont{H.}~\bibnamefont{Kadowaki}},
  \bibinfo{author}{\bibfnamefont{T.~R.} \bibnamefont{Thurston}},
  \bibinfo{author}{\bibfnamefont{Y.}~\bibnamefont{Endoh}},
  \bibinfo{author}{\bibfnamefont{Y.}~\bibnamefont{Hidaka}},
  \bibinfo{author}{\bibfnamefont{R.~J.} \bibnamefont{Birgeneau}},
  \bibinfo{author}{\bibfnamefont{M.~A.} \bibnamefont{Kastner}},
  \bibinfo{author}{\bibfnamefont{P.~M.} \bibnamefont{Gehring}},
  \bibinfo{author}{\bibfnamefont{A.~H.} \bibnamefont{Moudden}}, and
  \bibinfo{author}{\bibfnamefont{G.}~\bibnamefont{Shirane}},
  \bibinfo{year}{1990}, \bibinfo{journal}{Phys.\ Rev.\ B}
  \textbf{\bibinfo{volume}{42}}(\bibinfo{number}{16}), \bibinfo{pages}{10098}.

\bibitem[{\citenamefont{Matsui} \emph{et~al.}(2007)\citenamefont{Matsui,
  Takahashi, Sato, Terashima, Ding, Uefuji, and Yamada}}]{Matsui07a}
\bibinfo{author}{\bibnamefont{Matsui}, \bibfnamefont{H.}},
  \bibinfo{author}{\bibfnamefont{T.}~\bibnamefont{Takahashi}},
  \bibinfo{author}{\bibfnamefont{T.}~\bibnamefont{Sato}},
  \bibinfo{author}{\bibfnamefont{K.}~\bibnamefont{Terashima}},
  \bibinfo{author}{\bibfnamefont{H.}~\bibnamefont{Ding}},
  \bibinfo{author}{\bibfnamefont{T.}~\bibnamefont{Uefuji}}, and
  \bibinfo{author}{\bibfnamefont{K.}~\bibnamefont{Yamada}},
  \bibinfo{year}{2007}, \bibinfo{journal}{Phys.\ Rev.\ B}
  \textbf{\bibinfo{volume}{75}}(\bibinfo{number}{22}), \bibinfo{eid}{224514}.

\bibitem[{\citenamefont{Matsui}
  \emph{et~al.}(2005{\natexlab{a}})\citenamefont{Matsui, Terashima, Sato,
  Takahashi, Fujita, and Yamada}}]{Matsui05a}
\bibinfo{author}{\bibnamefont{Matsui}, \bibfnamefont{H.}},
  \bibinfo{author}{\bibfnamefont{K.}~\bibnamefont{Terashima}},
  \bibinfo{author}{\bibfnamefont{T.}~\bibnamefont{Sato}},
  \bibinfo{author}{\bibfnamefont{T.}~\bibnamefont{Takahashi}},
  \bibinfo{author}{\bibfnamefont{M.}~\bibnamefont{Fujita}}, and
  \bibinfo{author}{\bibfnamefont{K.}~\bibnamefont{Yamada}},
  \bibinfo{year}{2005}{\natexlab{a}}, \bibinfo{journal}{Phys.\ Rev.\ Lett.}
  \textbf{\bibinfo{volume}{95}}(\bibinfo{number}{1}), \bibinfo{eid}{017003}.

\bibitem[{\citenamefont{Matsui}
  \emph{et~al.}(2005{\natexlab{b}})\citenamefont{Matsui, Terashima, Sato,
  Takahashi, Wang, Yang, Ding, Uefuji, and Yamada}}]{Matsui05b}
\bibinfo{author}{\bibnamefont{Matsui}, \bibfnamefont{H.}},
  \bibinfo{author}{\bibfnamefont{K.}~\bibnamefont{Terashima}},
  \bibinfo{author}{\bibfnamefont{T.}~\bibnamefont{Sato}},
  \bibinfo{author}{\bibfnamefont{T.}~\bibnamefont{Takahashi}},
  \bibinfo{author}{\bibfnamefont{S.-C.} \bibnamefont{Wang}},
  \bibinfo{author}{\bibfnamefont{H.-B.} \bibnamefont{Yang}},
  \bibinfo{author}{\bibfnamefont{H.}~\bibnamefont{Ding}},
  \bibinfo{author}{\bibfnamefont{T.}~\bibnamefont{Uefuji}}, and
  \bibinfo{author}{\bibfnamefont{K.}~\bibnamefont{Yamada}},
  \bibinfo{year}{2005}{\natexlab{b}}, \bibinfo{journal}{Phys.\ Rev.\ Lett.}
  \textbf{\bibinfo{volume}{94}}(\bibinfo{number}{4}), \bibinfo{eid}{047005}.

\bibitem[{\citenamefont{Matsumoto} \emph{et~al.}(2009)\citenamefont{Matsumoto,
  Utsuki, Tsukada, Yamamoto, Manabe, and Naito}}]{Matsumoto09a}
\bibinfo{author}{\bibnamefont{Matsumoto}, \bibfnamefont{O.}},
  \bibinfo{author}{\bibfnamefont{A.}~\bibnamefont{Utsuki}},
  \bibinfo{author}{\bibfnamefont{A.}~\bibnamefont{Tsukada}},
  \bibinfo{author}{\bibfnamefont{H.}~\bibnamefont{Yamamoto}},
  \bibinfo{author}{\bibfnamefont{T.}~\bibnamefont{Manabe}}, and
  \bibinfo{author}{\bibfnamefont{M.}~\bibnamefont{Naito}},
  \bibinfo{year}{2009}, \bibinfo{journal}{Phys.\ Rev.\ B}
  \textbf{\bibinfo{volume}{79}}(\bibinfo{number}{10}), \bibinfo{eid}{100508}.

\bibitem[{\citenamefont{Matsuura} \emph{et~al.}(2003)\citenamefont{Matsuura,
  Dai, Kang, Lynn, Argyriou, Prokes, Onose, and Tokura}}]{Matsuura03a}
\bibinfo{author}{\bibnamefont{Matsuura}, \bibfnamefont{M.}},
  \bibinfo{author}{\bibfnamefont{P.}~\bibnamefont{Dai}},
  \bibinfo{author}{\bibfnamefont{H.~J.} \bibnamefont{Kang}},
  \bibinfo{author}{\bibfnamefont{J.~W.} \bibnamefont{Lynn}},
  \bibinfo{author}{\bibfnamefont{D.~N.} \bibnamefont{Argyriou}},
  \bibinfo{author}{\bibfnamefont{K.}~\bibnamefont{Prokes}},
  \bibinfo{author}{\bibfnamefont{Y.}~\bibnamefont{Onose}}, and
  \bibinfo{author}{\bibfnamefont{Y.}~\bibnamefont{Tokura}},
  \bibinfo{year}{2003}, \bibinfo{journal}{Phys.\ Rev.\ B}
  \textbf{\bibinfo{volume}{68}}(\bibinfo{number}{14}), \bibinfo{eid}{144503}.

\bibitem[{\citenamefont{Matsuyama} \emph{et~al.}(1989)\citenamefont{Matsuyama,
  Takahashi, Katayama-Yoshida, Kashiwakura, Okabe, Sato, Kosugi, Yagishita,
  Tanaka, Fujimoto, and Inokuchi}}]{Matsuyama89a}
\bibinfo{author}{\bibnamefont{Matsuyama}, \bibfnamefont{H.}},
  \bibinfo{author}{\bibfnamefont{T.}~\bibnamefont{Takahashi}},
  \bibinfo{author}{\bibfnamefont{H.}~\bibnamefont{Katayama-Yoshida}},
  \bibinfo{author}{\bibfnamefont{T.}~\bibnamefont{Kashiwakura}},
  \bibinfo{author}{\bibfnamefont{Y.}~\bibnamefont{Okabe}},
  \bibinfo{author}{\bibfnamefont{S.}~\bibnamefont{Sato}},
  \bibinfo{author}{\bibfnamefont{N.}~\bibnamefont{Kosugi}},
  \bibinfo{author}{\bibfnamefont{A.}~\bibnamefont{Yagishita}},
  \bibinfo{author}{\bibfnamefont{K.}~\bibnamefont{Tanaka}},
  \bibinfo{author}{\bibfnamefont{H.}~\bibnamefont{Fujimoto}}, and
  \bibinfo{author}{\bibfnamefont{H.}~\bibnamefont{Inokuchi}},
  \bibinfo{year}{1989}, \bibinfo{journal}{Physica C}
  \textbf{\bibinfo{volume}{160}}(\bibinfo{number}{5-6}), \bibinfo{pages}{567 }.

\bibitem[{\citenamefont{McKenzie}(1997)}]{McKenzie97a}
\bibinfo{author}{\bibnamefont{McKenzie}, \bibfnamefont{R.~H.}},
  \bibinfo{year}{1997}, \bibinfo{journal}{Science}
  \textbf{\bibinfo{volume}{278}}(\bibinfo{number}{5339}), \bibinfo{pages}{820}.

\bibitem[{\citenamefont{McQueeney} \emph{et~al.}(1999)\citenamefont{McQueeney,
  Petrov, Egami, Yethiraj, Shirane, and Endoh}}]{McQueeney99a}
\bibinfo{author}{\bibnamefont{McQueeney}, \bibfnamefont{R.~J.}},
  \bibinfo{author}{\bibfnamefont{Y.}~\bibnamefont{Petrov}},
  \bibinfo{author}{\bibfnamefont{T.}~\bibnamefont{Egami}},
  \bibinfo{author}{\bibfnamefont{M.}~\bibnamefont{Yethiraj}},
  \bibinfo{author}{\bibfnamefont{G.}~\bibnamefont{Shirane}}, and
  \bibinfo{author}{\bibfnamefont{Y.}~\bibnamefont{Endoh}},
  \bibinfo{year}{1999}, \bibinfo{journal}{Phys.\ Rev.\ Lett.}
  \textbf{\bibinfo{volume}{82}}(\bibinfo{number}{3}), \bibinfo{pages}{628}.

\bibitem[{\citenamefont{McQueeney} \emph{et~al.}(2001)\citenamefont{McQueeney,
  Sarrao, Pagliuso, Stephens, and Osborn}}]{McQueeney01a}
\bibinfo{author}{\bibnamefont{McQueeney}, \bibfnamefont{R.~J.}},
  \bibinfo{author}{\bibfnamefont{J.~L.} \bibnamefont{Sarrao}},
  \bibinfo{author}{\bibfnamefont{P.~G.} \bibnamefont{Pagliuso}},
  \bibinfo{author}{\bibfnamefont{P.~W.} \bibnamefont{Stephens}}, and
  \bibinfo{author}{\bibfnamefont{R.}~\bibnamefont{Osborn}},
  \bibinfo{year}{2001}, \bibinfo{journal}{Phys.\ Rev.\ Lett.}
  \textbf{\bibinfo{volume}{87}}(\bibinfo{number}{7}), \bibinfo{pages}{077001}.

\bibitem[{\citenamefont{Meevasana} \emph{et~al.}(2007)\citenamefont{Meevasana,
  Zhou, Sahrakorpi, Lee, Yang, Tanaka, Mannella, Yoshida, Lu, Chen, He, Lin}
  \emph{et~al.}}]{Meevasana07a}
\bibinfo{author}{\bibnamefont{Meevasana}, \bibfnamefont{W.}},
  \bibinfo{author}{\bibfnamefont{X.~J.} \bibnamefont{Zhou}},
  \bibinfo{author}{\bibfnamefont{S.}~\bibnamefont{Sahrakorpi}},
  \bibinfo{author}{\bibfnamefont{W.~S.} \bibnamefont{Lee}},
  \bibinfo{author}{\bibfnamefont{W.~L.} \bibnamefont{Yang}},
  \bibinfo{author}{\bibfnamefont{K.}~\bibnamefont{Tanaka}},
  \bibinfo{author}{\bibfnamefont{N.}~\bibnamefont{Mannella}},
  \bibinfo{author}{\bibfnamefont{T.}~\bibnamefont{Yoshida}},
  \bibinfo{author}{\bibfnamefont{D.~H.} \bibnamefont{Lu}},
  \bibinfo{author}{\bibfnamefont{Y.~L.} \bibnamefont{Chen}},
  \bibinfo{author}{\bibfnamefont{R.~H.} \bibnamefont{He}},
  \bibinfo{author}{\bibfnamefont{H.}~\bibnamefont{Lin}}, \emph{et~al.},
  \bibinfo{year}{2007}, \bibinfo{journal}{Phys.\ Rev.\ B}
  \textbf{\bibinfo{volume}{75}}(\bibinfo{number}{17}), \bibinfo{eid}{174506}.

\bibitem[{\citenamefont{Meinders} \emph{et~al.}(1993)\citenamefont{Meinders,
  Eskes, and Sawatzky}}]{Meinders93a}
\bibinfo{author}{\bibnamefont{Meinders}, \bibfnamefont{M.~B.~J.}},
  \bibinfo{author}{\bibfnamefont{H.}~\bibnamefont{Eskes}}, and
  \bibinfo{author}{\bibfnamefont{G.~A.} \bibnamefont{Sawatzky}},
  \bibinfo{year}{1993}, \bibinfo{journal}{Phys.\ Rev.\ B}
  \textbf{\bibinfo{volume}{48}}(\bibinfo{number}{6}), \bibinfo{pages}{3916}.

\bibitem[{\citenamefont{Miller} \emph{et~al.}(2002)\citenamefont{Miller, Kiefl,
  Brewer, Sonier, Chakhalian, Dunsiger, Morris, Price, Bonn, Hardy, and
  Liang}}]{Miller02a}
\bibinfo{author}{\bibnamefont{Miller}, \bibfnamefont{R.~I.}},
  \bibinfo{author}{\bibfnamefont{R.~F.} \bibnamefont{Kiefl}},
  \bibinfo{author}{\bibfnamefont{J.~H.} \bibnamefont{Brewer}},
  \bibinfo{author}{\bibfnamefont{J.~E.} \bibnamefont{Sonier}},
  \bibinfo{author}{\bibfnamefont{J.}~\bibnamefont{Chakhalian}},
  \bibinfo{author}{\bibfnamefont{S.}~\bibnamefont{Dunsiger}},
  \bibinfo{author}{\bibfnamefont{G.~D.} \bibnamefont{Morris}},
  \bibinfo{author}{\bibfnamefont{A.~N.} \bibnamefont{Price}},
  \bibinfo{author}{\bibfnamefont{D.~A.} \bibnamefont{Bonn}},
  \bibinfo{author}{\bibfnamefont{W.~H.} \bibnamefont{Hardy}}, and
  \bibinfo{author}{\bibfnamefont{R.}~\bibnamefont{Liang}},
  \bibinfo{year}{2002}, \bibinfo{journal}{Phys.\ Rev.\ Lett.}
  \textbf{\bibinfo{volume}{88}}(\bibinfo{number}{13}), \bibinfo{pages}{137002}.

\bibitem[{\citenamefont{Mira} \emph{et~al.}(1995)\citenamefont{Mira, Rivas,
  Fiorani, Caciuffo, Rinaldi, V\'azquez-V\'azquez, Mah\'ia, L\'opez-Quintela,
  and Oseroff}}]{Mira95a}
\bibinfo{author}{\bibnamefont{Mira}, \bibfnamefont{J.}},
  \bibinfo{author}{\bibfnamefont{J.}~\bibnamefont{Rivas}},
  \bibinfo{author}{\bibfnamefont{D.}~\bibnamefont{Fiorani}},
  \bibinfo{author}{\bibfnamefont{R.}~\bibnamefont{Caciuffo}},
  \bibinfo{author}{\bibfnamefont{D.}~\bibnamefont{Rinaldi}},
  \bibinfo{author}{\bibfnamefont{C.}~\bibnamefont{V\'azquez-V\'azquez}},
  \bibinfo{author}{\bibfnamefont{J.}~\bibnamefont{Mah\'ia}},
  \bibinfo{author}{\bibfnamefont{M.~A.} \bibnamefont{L\'opez-Quintela}}, and
  \bibinfo{author}{\bibfnamefont{S.~B.} \bibnamefont{Oseroff}},
  \bibinfo{year}{1995}, \bibinfo{journal}{Phys.\ Rev.\ B}
  \textbf{\bibinfo{volume}{52}}(\bibinfo{number}{22}), \bibinfo{pages}{16020}.

\bibitem[{\citenamefont{Mitrovic} \emph{et~al.}(2001)\citenamefont{Mitrovic,
  Sigmund, Eschrig, Bachman, Halperin, Reyes, Kuhns, and
  Moulton}}]{Mitrovic01a}
\bibinfo{author}{\bibnamefont{Mitrovic}, \bibfnamefont{V.~F.}},
  \bibinfo{author}{\bibfnamefont{E.~E.} \bibnamefont{Sigmund}},
  \bibinfo{author}{\bibfnamefont{M.}~\bibnamefont{Eschrig}},
  \bibinfo{author}{\bibfnamefont{H.~N.} \bibnamefont{Bachman}},
  \bibinfo{author}{\bibfnamefont{W.~P.} \bibnamefont{Halperin}},
  \bibinfo{author}{\bibfnamefont{A.~P.} \bibnamefont{Reyes}},
  \bibinfo{author}{\bibfnamefont{P.}~\bibnamefont{Kuhns}}, and
  \bibinfo{author}{\bibfnamefont{W.~G.} \bibnamefont{Moulton}},
  \bibinfo{year}{2001}, \bibinfo{journal}{Nature}
  \textbf{\bibinfo{volume}{413}}, \bibinfo{pages}{501}.

\bibitem[{\citenamefont{Moler} \emph{et~al.}(1994)\citenamefont{Moler, Baar,
  Urbach, Liang, Hardy, and Kapitulnik}}]{Moler94a}
\bibinfo{author}{\bibnamefont{Moler}, \bibfnamefont{K.~A.}},
  \bibinfo{author}{\bibfnamefont{D.~J.} \bibnamefont{Baar}},
  \bibinfo{author}{\bibfnamefont{J.~S.} \bibnamefont{Urbach}},
  \bibinfo{author}{\bibfnamefont{R.}~\bibnamefont{Liang}},
  \bibinfo{author}{\bibfnamefont{W.~N.} \bibnamefont{Hardy}}, and
  \bibinfo{author}{\bibfnamefont{A.}~\bibnamefont{Kapitulnik}},
  \bibinfo{year}{1994}, \bibinfo{journal}{Phys.\ Rev.\ Lett.}
  \textbf{\bibinfo{volume}{73}}(\bibinfo{number}{20}), \bibinfo{pages}{2744}.

\bibitem[{\citenamefont{Moler} \emph{et~al.}(1997)\citenamefont{Moler, Sisson,
  Urbach, Beasley, Kapitulnik, Baar, Liang, and Hardy}}]{Moler97a}
\bibinfo{author}{\bibnamefont{Moler}, \bibfnamefont{K.~A.}},
  \bibinfo{author}{\bibfnamefont{D.~L.} \bibnamefont{Sisson}},
  \bibinfo{author}{\bibfnamefont{J.~S.} \bibnamefont{Urbach}},
  \bibinfo{author}{\bibfnamefont{M.~R.} \bibnamefont{Beasley}},
  \bibinfo{author}{\bibfnamefont{A.}~\bibnamefont{Kapitulnik}},
  \bibinfo{author}{\bibfnamefont{D.~J.} \bibnamefont{Baar}},
  \bibinfo{author}{\bibfnamefont{R.}~\bibnamefont{Liang}}, and
  \bibinfo{author}{\bibfnamefont{W.~N.} \bibnamefont{Hardy}},
  \bibinfo{year}{1997}, \bibinfo{journal}{Phys.\ Rev.\ B}
  \textbf{\bibinfo{volume}{55}}(\bibinfo{number}{6}), \bibinfo{pages}{3954}.

\bibitem[{\citenamefont{Mook} \emph{et~al.}(1998)\citenamefont{Mook, Dai,
  Hayden, Aeppli, Perring, and Dogan}}]{Mook98a}
\bibinfo{author}{\bibnamefont{Mook}, \bibfnamefont{H.~A.}},
  \bibinfo{author}{\bibfnamefont{P.}~\bibnamefont{Dai}},
  \bibinfo{author}{\bibfnamefont{S.~M.} \bibnamefont{Hayden}},
  \bibinfo{author}{\bibfnamefont{G.}~\bibnamefont{Aeppli}},
  \bibinfo{author}{\bibfnamefont{T.~G.} \bibnamefont{Perring}}, and
  \bibinfo{author}{\bibfnamefont{F.}~\bibnamefont{Dogan}},
  \bibinfo{year}{1998}, \bibinfo{journal}{Nature}
  \textbf{\bibinfo{volume}{395}}, \bibinfo{pages}{580}.

\bibitem[{\citenamefont{Mook} \emph{et~al.}(2002)\citenamefont{Mook, Dai, and
  Do\ifmmode~\breve{g}\else \u{g}\fi{}an}}]{Mook02a}
\bibinfo{author}{\bibnamefont{Mook}, \bibfnamefont{H.~A.}},
  \bibinfo{author}{\bibfnamefont{P.}~\bibnamefont{Dai}}, and
  \bibinfo{author}{\bibfnamefont{F.}~\bibnamefont{Do\ifmmode~\breve{g}\else
  \u{g}\fi{}an}}, \bibinfo{year}{2002}, \bibinfo{journal}{Phys.\ Rev.\ Lett.}
  \textbf{\bibinfo{volume}{88}}(\bibinfo{number}{9}), \bibinfo{pages}{097004}.

\bibitem[{\citenamefont{Moon and Sachdev}(2009)}]{Moon09a}
\bibinfo{author}{\bibnamefont{Moon}, \bibfnamefont{E.~G.}}, and
  \bibinfo{author}{\bibfnamefont{S.}~\bibnamefont{Sachdev}},
  \bibinfo{year}{2009}, \bibinfo{journal}{arXiv:0905.2608} .

\bibitem[{\citenamefont{Moran} \emph{et~al.}(1989)\citenamefont{Moran, Nazzal,
  Huang, and Torrance}}]{Moran89a}
\bibinfo{author}{\bibnamefont{Moran}, \bibfnamefont{E.}},
  \bibinfo{author}{\bibfnamefont{A.~I.} \bibnamefont{Nazzal}},
  \bibinfo{author}{\bibfnamefont{T.~C.} \bibnamefont{Huang}}, and
  \bibinfo{author}{\bibfnamefont{J.~B.} \bibnamefont{Torrance}},
  \bibinfo{year}{1989}, \bibinfo{journal}{Physica C}
  \textbf{\bibinfo{volume}{160}}, \bibinfo{pages}{30}.

\bibitem[{\citenamefont{Moritz} \emph{et~al.}(2009)\citenamefont{Moritz,
  Schmitt, Meevasana, Johnston, Motoyama, Greven, Lu, Kim, Scalettar, Shen, and
  Devereaux}}]{Moritz08a}
\bibinfo{author}{\bibnamefont{Moritz}, \bibfnamefont{B.}},
  \bibinfo{author}{\bibfnamefont{F.}~\bibnamefont{Schmitt}},
  \bibinfo{author}{\bibfnamefont{W.}~\bibnamefont{Meevasana}},
  \bibinfo{author}{\bibfnamefont{S.}~\bibnamefont{Johnston}},
  \bibinfo{author}{\bibfnamefont{E.~M.} \bibnamefont{Motoyama}},
  \bibinfo{author}{\bibfnamefont{M.}~\bibnamefont{Greven}},
  \bibinfo{author}{\bibfnamefont{D.~H.} \bibnamefont{Lu}},
  \bibinfo{author}{\bibfnamefont{C.}~\bibnamefont{Kim}},
  \bibinfo{author}{\bibfnamefont{R.~T.} \bibnamefont{Scalettar}},
  \bibinfo{author}{\bibfnamefont{Z.-X.} \bibnamefont{Shen}}, and
  \bibinfo{author}{\bibfnamefont{T.~P.} \bibnamefont{Devereaux}},
  \bibinfo{year}{2009}, \bibinfo{journal}{New Journal of Physics}
  \textbf{\bibinfo{volume}{11}}(\bibinfo{number}{9}), \bibinfo{pages}{093020}.

\bibitem[{\citenamefont{Motoyama} \emph{et~al.}(2006)\citenamefont{Motoyama,
  Mang, Petitgrand, Yu, Vajk, Vishik, and Greven}}]{Motoyama06a}
\bibinfo{author}{\bibnamefont{Motoyama}, \bibfnamefont{E.~M.}},
  \bibinfo{author}{\bibfnamefont{P.~K.} \bibnamefont{Mang}},
  \bibinfo{author}{\bibfnamefont{D.}~\bibnamefont{Petitgrand}},
  \bibinfo{author}{\bibfnamefont{G.}~\bibnamefont{Yu}},
  \bibinfo{author}{\bibfnamefont{O.~P.} \bibnamefont{Vajk}},
  \bibinfo{author}{\bibfnamefont{I.~M.} \bibnamefont{Vishik}}, and
  \bibinfo{author}{\bibfnamefont{M.}~\bibnamefont{Greven}},
  \bibinfo{year}{2006}, \bibinfo{journal}{Phys.\ Rev.\ Lett.}
  \textbf{\bibinfo{volume}{96}}(\bibinfo{number}{13}), \bibinfo{eid}{137002}.

\bibitem[{\citenamefont{Motoyama} \emph{et~al.}(2007)\citenamefont{Motoyama,
  Yu, Vishik, Vajk, Mang, and Greven}}]{Motoyama07a}
\bibinfo{author}{\bibnamefont{Motoyama}, \bibfnamefont{E.~M.}},
  \bibinfo{author}{\bibfnamefont{G.}~\bibnamefont{Yu}},
  \bibinfo{author}{\bibfnamefont{I.~M.} \bibnamefont{Vishik}},
  \bibinfo{author}{\bibfnamefont{O.~P.} \bibnamefont{Vajk}},
  \bibinfo{author}{\bibfnamefont{P.~K.} \bibnamefont{Mang}}, and
  \bibinfo{author}{\bibfnamefont{M.}~\bibnamefont{Greven}},
  \bibinfo{year}{2007}, \bibinfo{journal}{Nature}
  \textbf{\bibinfo{volume}{445}}(\bibinfo{number}{7124}), \bibinfo{pages}{186}.

\bibitem[{\citenamefont{Muller-Buschbaum and Wollschlager}(1975)}]{Muller75a}
\bibinfo{author}{\bibnamefont{Muller-Buschbaum}, \bibfnamefont{X.}}, and
  \bibinfo{author}{\bibfnamefont{X.}~\bibnamefont{Wollschlager}},
  \bibinfo{year}{1975}, \bibinfo{journal}{Z. Anorg. Allg. Chem.}
  \textbf{\bibinfo{volume}{414}}, \bibinfo{pages}{76}.

\bibitem[{\citenamefont{Naito and Hepp}(2000)}]{Naito00a}
\bibinfo{author}{\bibnamefont{Naito}, \bibfnamefont{M.}}, and
  \bibinfo{author}{\bibfnamefont{M.}~\bibnamefont{Hepp}}, \bibinfo{year}{2000},
  \bibinfo{journal}{Jap.\ J.\ Appl.\ Phys.}
  \textbf{\bibinfo{volume}{39}}(\bibinfo{number}{Part 2, No. 6A}),
  \bibinfo{pages}{L485}.

\bibitem[{\citenamefont{Naito} \emph{et~al.}(2002)\citenamefont{Naito,
  Karimoto, and Tsukada2}}]{Naito02a}
\bibinfo{author}{\bibnamefont{Naito}, \bibfnamefont{M.}},
  \bibinfo{author}{\bibfnamefont{S.}~\bibnamefont{Karimoto}}, and
  \bibinfo{author}{\bibfnamefont{A.}~\bibnamefont{Tsukada2}},
  \bibinfo{year}{2002}, \bibinfo{journal}{Supercond.\ Sci.\ Technol.}
  \textbf{\bibinfo{volume}{15}}, \bibinfo{pages}{1663–}.

\bibitem[{\citenamefont{Naito} \emph{et~al.}(1997)\citenamefont{Naito, Sato,
  and Yamamoto}}]{Naito97a}
\bibinfo{author}{\bibnamefont{Naito}, \bibfnamefont{M.}},
  \bibinfo{author}{\bibfnamefont{H.}~\bibnamefont{Sato}}, and
  \bibinfo{author}{\bibfnamefont{H.}~\bibnamefont{Yamamoto}},
  \bibinfo{year}{1997}, \bibinfo{journal}{Physica C}
  \textbf{\bibinfo{volume}{293}}(\bibinfo{number}{1-4}), \bibinfo{pages}{36 },
  \bibinfo{note}{intrinsic Josephson Effects and THz Plasma Oscillations in
  High-Tc Superconductors}.

\bibitem[{\citenamefont{Nakamae} \emph{et~al.}(2003)\citenamefont{Nakamae,
  Behnia, Mangkorntong, Nohara, Takagi, Yates, and Hussey}}]{Nakamae03a}
\bibinfo{author}{\bibnamefont{Nakamae}, \bibfnamefont{S.}},
  \bibinfo{author}{\bibfnamefont{K.}~\bibnamefont{Behnia}},
  \bibinfo{author}{\bibfnamefont{N.}~\bibnamefont{Mangkorntong}},
  \bibinfo{author}{\bibfnamefont{M.}~\bibnamefont{Nohara}},
  \bibinfo{author}{\bibfnamefont{H.}~\bibnamefont{Takagi}},
  \bibinfo{author}{\bibfnamefont{S.~J.~C.} \bibnamefont{Yates}}, and
  \bibinfo{author}{\bibfnamefont{N.~E.} \bibnamefont{Hussey}},
  \bibinfo{year}{2003}, \bibinfo{journal}{Phys.\ Rev.\ B}
  \textbf{\bibinfo{volume}{68}}(\bibinfo{number}{10}), \bibinfo{pages}{100502}.

\bibitem[{\citenamefont{Namatame} \emph{et~al.}(1990)\citenamefont{Namatame,
  Fujimori, Tokura, Nakamura, Yamaguchi, Misu, Matsubara, Suga, Eisaki, Ito,
  Takagi, and Uchida}}]{Namatame90a}
\bibinfo{author}{\bibnamefont{Namatame}, \bibfnamefont{H.}},
  \bibinfo{author}{\bibfnamefont{A.}~\bibnamefont{Fujimori}},
  \bibinfo{author}{\bibfnamefont{Y.}~\bibnamefont{Tokura}},
  \bibinfo{author}{\bibfnamefont{M.}~\bibnamefont{Nakamura}},
  \bibinfo{author}{\bibfnamefont{K.}~\bibnamefont{Yamaguchi}},
  \bibinfo{author}{\bibfnamefont{A.}~\bibnamefont{Misu}},
  \bibinfo{author}{\bibfnamefont{H.}~\bibnamefont{Matsubara}},
  \bibinfo{author}{\bibfnamefont{S.}~\bibnamefont{Suga}},
  \bibinfo{author}{\bibfnamefont{H.}~\bibnamefont{Eisaki}},
  \bibinfo{author}{\bibfnamefont{T.}~\bibnamefont{Ito}},
  \bibinfo{author}{\bibfnamefont{H.}~\bibnamefont{Takagi}}, and
  \bibinfo{author}{\bibfnamefont{S.}~\bibnamefont{Uchida}},
  \bibinfo{year}{1990}, \bibinfo{journal}{Phys.\ Rev.\ B}
  \textbf{\bibinfo{volume}{41}}(\bibinfo{number}{10}), \bibinfo{pages}{7205}.

\bibitem[{\citenamefont{Navarro} \emph{et~al.}(2001)\citenamefont{Navarro,
  Jaque, Villegas, Martyn, Serquis, Prado, Caneiro, , and Vicent}}]{Navarro01a}
\bibinfo{author}{\bibnamefont{Navarro}, \bibfnamefont{E.}},
  \bibinfo{author}{\bibfnamefont{D.}~\bibnamefont{Jaque}},
  \bibinfo{author}{\bibfnamefont{J.}~\bibnamefont{Villegas}},
  \bibinfo{author}{\bibfnamefont{J.}~\bibnamefont{Martyn}},
  \bibinfo{author}{\bibfnamefont{A.}~\bibnamefont{Serquis}},
  \bibinfo{author}{\bibfnamefont{F.}~\bibnamefont{Prado}},
  \bibinfo{author}{\bibfnamefont{A.}~\bibnamefont{Caneiro}}, , and
  \bibinfo{author}{\bibfnamefont{J.}~\bibnamefont{Vicent}},
  \bibinfo{year}{2001}, \bibinfo{journal}{J.\ of Alloys and Compounds}
  \textbf{\bibinfo{volume}{323–-324}}, \bibinfo{pages}{580}.

\bibitem[{\citenamefont{Nedil'ko}(1982)}]{Nedilko82a}
\bibinfo{author}{\bibnamefont{Nedil'ko}, \bibfnamefont{A.}},
  \bibinfo{year}{1982}, \bibinfo{journal}{Russian J.\ Inorg.\ Chem.}
  \textbf{\bibinfo{volume}{27}}, \bibinfo{pages}{634}.

\bibitem[{\citenamefont{Nekvasil and Divis}(2001)}]{Nekvasil01a}
\bibinfo{author}{\bibnamefont{Nekvasil}, \bibfnamefont{V.}}, and
  \bibinfo{author}{\bibfnamefont{M.}~\bibnamefont{Divis}},
  \bibinfo{year}{2001}, in \emph{\bibinfo{booktitle}{Encyclopedia of Materials:
  Science and Technology}}, edited by \bibinfo{editor}{\bibfnamefont{K.~H.~J.}
  \bibnamefont{Buschow}}, \bibinfo{editor}{\bibfnamefont{R.~W.}
  \bibnamefont{Cahn}}, \bibinfo{editor}{\bibfnamefont{M.~C.}
  \bibnamefont{Flemings}}, \bibinfo{editor}{\bibfnamefont{B.~I.}
  \bibnamefont{(print)}}, \bibinfo{editor}{\bibfnamefont{E.~J.}
  \bibnamefont{Kramer}},
  \bibinfo{editor}{\bibfnamefont{S.}~\bibnamefont{Mahajan}}, , and
  \bibinfo{editor}{\bibfnamefont{P.~V.} \bibnamefont{(updates)}}
  (\bibinfo{publisher}{Elsevier}, \bibinfo{address}{Oxford}), pp.
  \bibinfo{pages}{4613 -- 4627}.

\bibitem[{\citenamefont{Nie} \emph{et~al.}(2003)\citenamefont{Nie, Badica,
  Hirai, Sundaresan, Crisan, Kito, Terada, Kodama, Iyo, Tanaka, and
  Ihara}}]{Nie03a}
\bibinfo{author}{\bibnamefont{Nie}, \bibfnamefont{J.~C.}},
  \bibinfo{author}{\bibfnamefont{P.}~\bibnamefont{Badica}},
  \bibinfo{author}{\bibfnamefont{M.}~\bibnamefont{Hirai}},
  \bibinfo{author}{\bibfnamefont{A.}~\bibnamefont{Sundaresan}},
  \bibinfo{author}{\bibfnamefont{A.}~\bibnamefont{Crisan}},
  \bibinfo{author}{\bibfnamefont{H.}~\bibnamefont{Kito}},
  \bibinfo{author}{\bibfnamefont{N.}~\bibnamefont{Terada}},
  \bibinfo{author}{\bibfnamefont{Y.}~\bibnamefont{Kodama}},
  \bibinfo{author}{\bibfnamefont{A.}~\bibnamefont{Iyo}},
  \bibinfo{author}{\bibfnamefont{Y.}~\bibnamefont{Tanaka}}, and
  \bibinfo{author}{\bibfnamefont{H.}~\bibnamefont{Ihara}},
  \bibinfo{year}{2003}, \bibinfo{journal}{Supercond.\ Sci.\ Technol.}
  \textbf{\bibinfo{volume}{16}}(\bibinfo{number}{1}), \bibinfo{pages}{L1}.

\bibitem[{\citenamefont{Niestemski}
  \emph{et~al.}(2007)\citenamefont{Niestemski, Kunwar, Zhou, Li, Ding, Wang,
  Dai, and Madhavan}}]{Niestemski07a}
\bibinfo{author}{\bibnamefont{Niestemski}, \bibfnamefont{F.~C.}},
  \bibinfo{author}{\bibfnamefont{S.}~\bibnamefont{Kunwar}},
  \bibinfo{author}{\bibfnamefont{S.}~\bibnamefont{Zhou}},
  \bibinfo{author}{\bibfnamefont{S.}~\bibnamefont{Li}},
  \bibinfo{author}{\bibfnamefont{H.}~\bibnamefont{Ding}},
  \bibinfo{author}{\bibfnamefont{Z.}~\bibnamefont{Wang}},
  \bibinfo{author}{\bibfnamefont{P.}~\bibnamefont{Dai}}, and
  \bibinfo{author}{\bibfnamefont{V.}~\bibnamefont{Madhavan}},
  \bibinfo{year}{2007}, \bibinfo{journal}{Nature}
  \textbf{\bibinfo{volume}{450}}(\bibinfo{number}{7172}),
  \bibinfo{pages}{1058}.

\bibitem[{\citenamefont{Norman} \emph{et~al.}(2005)\citenamefont{Norman, Pines,
  and Kallin}}]{Norman05a}
\bibinfo{author}{\bibnamefont{Norman}, \bibfnamefont{M.~R.}},
  \bibinfo{author}{\bibfnamefont{D.}~\bibnamefont{Pines}}, and
  \bibinfo{author}{\bibfnamefont{C.}~\bibnamefont{Kallin}},
  \bibinfo{year}{2005}, \bibinfo{journal}{Advances In Physics}
  \textbf{\bibinfo{volume}{54}}(\bibinfo{number}{8}), \bibinfo{pages}{715},
  ISSN \bibinfo{issn}{0001-8732}.

\bibitem[{\citenamefont{N\"ucker} \emph{et~al.}(1989)\citenamefont{N\"ucker,
  Adelmann, Alexander, Romberg, Nakai, Fink, Rietschel, Roth, Schmidt, and
  Spille}}]{Nucker89a}
\bibinfo{author}{\bibnamefont{N\"ucker}, \bibfnamefont{N.}},
  \bibinfo{author}{\bibfnamefont{P.}~\bibnamefont{Adelmann}},
  \bibinfo{author}{\bibfnamefont{M.}~\bibnamefont{Alexander}},
  \bibinfo{author}{\bibfnamefont{H.}~\bibnamefont{Romberg}},
  \bibinfo{author}{\bibfnamefont{S.}~\bibnamefont{Nakai}},
  \bibinfo{author}{\bibfnamefont{J.}~\bibnamefont{Fink}},
  \bibinfo{author}{\bibfnamefont{H.}~\bibnamefont{Rietschel}},
  \bibinfo{author}{\bibfnamefont{G.}~\bibnamefont{Roth}},
  \bibinfo{author}{\bibfnamefont{H.}~\bibnamefont{Schmidt}}, and
  \bibinfo{author}{\bibfnamefont{H.}~\bibnamefont{Spille}},
  \bibinfo{year}{1989}, \bibinfo{journal}{Z. Phys. B}
  \textbf{\bibinfo{volume}{75}}, \bibinfo{pages}{421}.

\bibitem[{\citenamefont{Ohta} \emph{et~al.}(1991)\citenamefont{Ohta, Tohyama,
  and Maekawa}}]{Ohta91a}
\bibinfo{author}{\bibnamefont{Ohta}, \bibfnamefont{Y.}},
  \bibinfo{author}{\bibfnamefont{T.}~\bibnamefont{Tohyama}}, and
  \bibinfo{author}{\bibfnamefont{S.}~\bibnamefont{Maekawa}},
  \bibinfo{year}{1991}, \bibinfo{journal}{Phys. Rev. B}
  \textbf{\bibinfo{volume}{43}}(\bibinfo{number}{4}), \bibinfo{pages}{2968}.

\bibitem[{\citenamefont{Oka} \emph{et~al.}(2003)\citenamefont{Oka, Shibata,
  Kashiwaya, and Eisaki}}]{Oka03a}
\bibinfo{author}{\bibnamefont{Oka}, \bibfnamefont{K.}},
  \bibinfo{author}{\bibfnamefont{H.}~\bibnamefont{Shibata}},
  \bibinfo{author}{\bibfnamefont{S.}~\bibnamefont{Kashiwaya}}, and
  \bibinfo{author}{\bibfnamefont{H.}~\bibnamefont{Eisaki}},
  \bibinfo{year}{2003}, \bibinfo{journal}{Physica C}
  \textbf{\bibinfo{volume}{388}}, \bibinfo{pages}{389}.

\bibitem[{\citenamefont{Oka and Unoki}(1990)}]{Oka90a}
\bibinfo{author}{\bibnamefont{Oka}, \bibfnamefont{K.}}, and
  \bibinfo{author}{\bibfnamefont{H.}~\bibnamefont{Unoki}},
  \bibinfo{year}{1990}, \bibinfo{journal}{Jap.\ J.\ Appl.\ Phys.}
  \textbf{\bibinfo{volume}{29}}, \bibinfo{pages}{L909}.

\bibitem[{\citenamefont{Okada} \emph{et~al.}(1990)\citenamefont{Okada, Seino,
  and Kotani}}]{Okada90a}
\bibinfo{author}{\bibnamefont{Okada}, \bibfnamefont{K.}},
  \bibinfo{author}{\bibfnamefont{Y.}~\bibnamefont{Seino}}, and
  \bibinfo{author}{\bibfnamefont{A.}~\bibnamefont{Kotani}},
  \bibinfo{year}{1990}, \bibinfo{journal}{J.\ Phys.\ Soc.\ Japan}
  \textbf{\bibinfo{volume}{59}}(\bibinfo{number}{8}), \bibinfo{pages}{2639}.

\bibitem[{\citenamefont{Onose} \emph{et~al.}(1999)\citenamefont{Onose, Taguchi,
  Ishikawa, Shinomori, Ishizaka, and Tokura}}]{Onose99a}
\bibinfo{author}{\bibnamefont{Onose}, \bibfnamefont{Y.}},
  \bibinfo{author}{\bibfnamefont{Y.}~\bibnamefont{Taguchi}},
  \bibinfo{author}{\bibfnamefont{T.}~\bibnamefont{Ishikawa}},
  \bibinfo{author}{\bibfnamefont{S.}~\bibnamefont{Shinomori}},
  \bibinfo{author}{\bibfnamefont{K.}~\bibnamefont{Ishizaka}}, and
  \bibinfo{author}{\bibfnamefont{Y.}~\bibnamefont{Tokura}},
  \bibinfo{year}{1999}, \bibinfo{journal}{Phys.\ Rev.\ Lett.}
  \textbf{\bibinfo{volume}{82}}(\bibinfo{number}{25}), \bibinfo{pages}{5120}.

\bibitem[{\citenamefont{Onose} \emph{et~al.}(2001)\citenamefont{Onose, Taguchi,
  Ishizaka, and Tokura}}]{Onose01a}
\bibinfo{author}{\bibnamefont{Onose}, \bibfnamefont{Y.}},
  \bibinfo{author}{\bibfnamefont{Y.}~\bibnamefont{Taguchi}},
  \bibinfo{author}{\bibfnamefont{K.}~\bibnamefont{Ishizaka}}, and
  \bibinfo{author}{\bibfnamefont{Y.}~\bibnamefont{Tokura}},
  \bibinfo{year}{2001}, \bibinfo{journal}{Phys.\ Rev.\ Lett.}
  \textbf{\bibinfo{volume}{87}}(\bibinfo{number}{21}), \bibinfo{eid}{217001}.

\bibitem[{\citenamefont{Onose} \emph{et~al.}(2004)\citenamefont{Onose, Taguchi,
  Ishizaka, and Tokura}}]{Onose04a}
\bibinfo{author}{\bibnamefont{Onose}, \bibfnamefont{Y.}},
  \bibinfo{author}{\bibfnamefont{Y.}~\bibnamefont{Taguchi}},
  \bibinfo{author}{\bibfnamefont{K.}~\bibnamefont{Ishizaka}}, and
  \bibinfo{author}{\bibfnamefont{Y.}~\bibnamefont{Tokura}},
  \bibinfo{year}{2004}, \bibinfo{journal}{Phys.\ Rev.\ B}
  \textbf{\bibinfo{volume}{69}}(\bibinfo{number}{2}), \bibinfo{eid}{024504}.

\bibitem[{\citenamefont{Orenstein and Millis}(2000)}]{Orenstein00a}
\bibinfo{author}{\bibnamefont{Orenstein}, \bibfnamefont{J.}}, and
  \bibinfo{author}{\bibfnamefont{A.~J.} \bibnamefont{Millis}},
  \bibinfo{year}{2000}, \bibinfo{journal}{Science}
  \textbf{\bibinfo{volume}{288}}(\bibinfo{number}{5465}), \bibinfo{pages}{468}.

\bibitem[{\citenamefont{Oseroff} \emph{et~al.}(1990)\citenamefont{Oseroff, Rao,
  Wright, Vier, Schultz, Thompson, Fisk, Cheong, Hundley, and
  Tovar}}]{Oseroff90a}
\bibinfo{author}{\bibnamefont{Oseroff}, \bibfnamefont{S.~B.}},
  \bibinfo{author}{\bibfnamefont{D.}~\bibnamefont{Rao}},
  \bibinfo{author}{\bibfnamefont{F.}~\bibnamefont{Wright}},
  \bibinfo{author}{\bibfnamefont{D.~C.} \bibnamefont{Vier}},
  \bibinfo{author}{\bibfnamefont{S.}~\bibnamefont{Schultz}},
  \bibinfo{author}{\bibfnamefont{J.~D.} \bibnamefont{Thompson}},
  \bibinfo{author}{\bibfnamefont{Z.}~\bibnamefont{Fisk}},
  \bibinfo{author}{\bibfnamefont{S.-W.} \bibnamefont{Cheong}},
  \bibinfo{author}{\bibfnamefont{M.~F.} \bibnamefont{Hundley}}, and
  \bibinfo{author}{\bibfnamefont{M.}~\bibnamefont{Tovar}},
  \bibinfo{year}{1990}, \bibinfo{journal}{Phys.\ Rev.\ B}
  \textbf{\bibinfo{volume}{41}}(\bibinfo{number}{4}), \bibinfo{pages}{1934}.

\bibitem[{\citenamefont{Pan} \emph{et~al.}(2006)\citenamefont{Pan, Richard,
  Fedrov, Kondo, Takeuchi, Li, Dai, Gu, Ku, Wang, and Ding}}]{Pan06a}
\bibinfo{author}{\bibnamefont{Pan}, \bibfnamefont{Z.-H.}},
  \bibinfo{author}{\bibfnamefont{P.}~\bibnamefont{Richard}},
  \bibinfo{author}{\bibfnamefont{A.~V.} \bibnamefont{Fedrov}},
  \bibinfo{author}{\bibfnamefont{T.}~\bibnamefont{Kondo}},
  \bibinfo{author}{\bibfnamefont{T.}~\bibnamefont{Takeuchi}},
  \bibinfo{author}{\bibfnamefont{S.~L.} \bibnamefont{Li}},
  \bibinfo{author}{\bibfnamefont{P.}~\bibnamefont{Dai}},
  \bibinfo{author}{\bibfnamefont{G.}~\bibnamefont{Gu}},
  \bibinfo{author}{\bibfnamefont{W.}~\bibnamefont{Ku}},
  \bibinfo{author}{\bibfnamefont{Z.}~\bibnamefont{Wang}}, and
  \bibinfo{author}{\bibfnamefont{H.}~\bibnamefont{Ding}}, \bibinfo{year}{2006},
  \bibinfo{journal}{arXiv:cond-mat/0610442} .

\bibitem[{\citenamefont{Park} \emph{et~al.}(2007)\citenamefont{Park, Roh, Yoon,
  Leem, Kim, Kim, Koh, Eisaki, Armitage, and Kim}}]{Park07a}
\bibinfo{author}{\bibnamefont{Park}, \bibfnamefont{S.~R.}},
  \bibinfo{author}{\bibfnamefont{Y.~S.} \bibnamefont{Roh}},
  \bibinfo{author}{\bibfnamefont{Y.~K.} \bibnamefont{Yoon}},
  \bibinfo{author}{\bibfnamefont{C.~S.} \bibnamefont{Leem}},
  \bibinfo{author}{\bibfnamefont{J.~H.} \bibnamefont{Kim}},
  \bibinfo{author}{\bibfnamefont{B.~J.} \bibnamefont{Kim}},
  \bibinfo{author}{\bibfnamefont{H.}~\bibnamefont{Koh}},
  \bibinfo{author}{\bibfnamefont{H.}~\bibnamefont{Eisaki}},
  \bibinfo{author}{\bibfnamefont{N.~P.} \bibnamefont{Armitage}}, and
  \bibinfo{author}{\bibfnamefont{C.}~\bibnamefont{Kim}}, \bibinfo{year}{2007},
  \bibinfo{journal}{Phys.\ Rev.\ B}
  \textbf{\bibinfo{volume}{75}}(\bibinfo{number}{6}), \bibinfo{eid}{060501}.

\bibitem[{\citenamefont{Park} \emph{et~al.}(2008)\citenamefont{Park, Song,
  Leem, Kim, Kim, Kim, and Eisaki}}]{Park08a}
\bibinfo{author}{\bibnamefont{Park}, \bibfnamefont{S.~R.}},
  \bibinfo{author}{\bibfnamefont{D.~J.} \bibnamefont{Song}},
  \bibinfo{author}{\bibfnamefont{C.~S.} \bibnamefont{Leem}},
  \bibinfo{author}{\bibfnamefont{C.}~\bibnamefont{Kim}},
  \bibinfo{author}{\bibfnamefont{C.}~\bibnamefont{Kim}},
  \bibinfo{author}{\bibfnamefont{B.~J.} \bibnamefont{Kim}}, and
  \bibinfo{author}{\bibfnamefont{H.}~\bibnamefont{Eisaki}},
  \bibinfo{year}{2008}, \bibinfo{journal}{Phys.\ Rev.\ Lett.}
  \textbf{\bibinfo{volume}{101}}(\bibinfo{number}{11}), \bibinfo{eid}{117006}.

\bibitem[{\citenamefont{Pathak} \emph{et~al.}(2009)\citenamefont{Pathak,
  Shenoy, Randeria, and Trivedi}}]{Pathak09a}
\bibinfo{author}{\bibnamefont{Pathak}, \bibfnamefont{S.}},
  \bibinfo{author}{\bibfnamefont{V.~B.} \bibnamefont{Shenoy}},
  \bibinfo{author}{\bibfnamefont{M.}~\bibnamefont{Randeria}}, and
  \bibinfo{author}{\bibfnamefont{N.}~\bibnamefont{Trivedi}},
  \bibinfo{year}{2009}, \bibinfo{journal}{Phys.\ Rev.\ Lett.}
  \textbf{\bibinfo{volume}{102}}(\bibinfo{number}{2}), \bibinfo{eid}{027002}.

\bibitem[{\citenamefont{Pellegrin} \emph{et~al.}(1993)\citenamefont{Pellegrin,
  N\"ucker, Fink, Molodtsov, Guti\'errez, Navas, Strebel, Hu, Domke, Kaindl,
  Uchida, Nakamura} \emph{et~al.}}]{Pellegrin93a}
\bibinfo{author}{\bibnamefont{Pellegrin}, \bibfnamefont{E.}},
  \bibinfo{author}{\bibfnamefont{N.}~\bibnamefont{N\"ucker}},
  \bibinfo{author}{\bibfnamefont{J.}~\bibnamefont{Fink}},
  \bibinfo{author}{\bibfnamefont{S.~L.} \bibnamefont{Molodtsov}},
  \bibinfo{author}{\bibfnamefont{A.}~\bibnamefont{Guti\'errez}},
  \bibinfo{author}{\bibfnamefont{E.}~\bibnamefont{Navas}},
  \bibinfo{author}{\bibfnamefont{O.}~\bibnamefont{Strebel}},
  \bibinfo{author}{\bibfnamefont{Z.}~\bibnamefont{Hu}},
  \bibinfo{author}{\bibfnamefont{M.}~\bibnamefont{Domke}},
  \bibinfo{author}{\bibfnamefont{G.}~\bibnamefont{Kaindl}},
  \bibinfo{author}{\bibfnamefont{S.}~\bibnamefont{Uchida}},
  \bibinfo{author}{\bibfnamefont{Y.}~\bibnamefont{Nakamura}}, \emph{et~al.},
  \bibinfo{year}{1993}, \bibinfo{journal}{Phys.\ Rev.\ B}
  \textbf{\bibinfo{volume}{47}}(\bibinfo{number}{6}), \bibinfo{pages}{3354}.

\bibitem[{\citenamefont{Peng} \emph{et~al.}(1991)\citenamefont{Peng, Li, and
  Greene}}]{Peng91a}
\bibinfo{author}{\bibnamefont{Peng}, \bibfnamefont{J.~L.}},
  \bibinfo{author}{\bibfnamefont{Z.~Y.} \bibnamefont{Li}}, and
  \bibinfo{author}{\bibfnamefont{R.~L.} \bibnamefont{Greene}},
  \bibinfo{year}{1991}, \bibinfo{journal}{Physica C}
  \textbf{\bibinfo{volume}{177}}(\bibinfo{number}{1-3}), \bibinfo{pages}{79}.

\bibitem[{\citenamefont{Peters} \emph{et~al.}(1988)\citenamefont{Peters,
  Birgeneau, Kastner, Yoshizawa, Endoh, Tranquada, Shirane, Hidaka, Oda,
  Suzuki, and Murakami}}]{Peters88a}
\bibinfo{author}{\bibnamefont{Peters}, \bibfnamefont{C.~J.}},
  \bibinfo{author}{\bibfnamefont{R.~J.} \bibnamefont{Birgeneau}},
  \bibinfo{author}{\bibfnamefont{M.~A.} \bibnamefont{Kastner}},
  \bibinfo{author}{\bibfnamefont{H.}~\bibnamefont{Yoshizawa}},
  \bibinfo{author}{\bibfnamefont{Y.}~\bibnamefont{Endoh}},
  \bibinfo{author}{\bibfnamefont{J.}~\bibnamefont{Tranquada}},
  \bibinfo{author}{\bibfnamefont{G.}~\bibnamefont{Shirane}},
  \bibinfo{author}{\bibfnamefont{Y.}~\bibnamefont{Hidaka}},
  \bibinfo{author}{\bibfnamefont{M.}~\bibnamefont{Oda}},
  \bibinfo{author}{\bibfnamefont{M.}~\bibnamefont{Suzuki}}, and
  \bibinfo{author}{\bibfnamefont{T.}~\bibnamefont{Murakami}},
  \bibinfo{year}{1988}, \bibinfo{journal}{Phys.\ Rev.\ B}
  \textbf{\bibinfo{volume}{37}}(\bibinfo{number}{16}), \bibinfo{pages}{9761}.

\bibitem[{\citenamefont{Petitgrand}
  \emph{et~al.}(1999)\citenamefont{Petitgrand, Maleyev, Bourges, and
  Ivanov}}]{Petitgrand99a}
\bibinfo{author}{\bibnamefont{Petitgrand}, \bibfnamefont{D.}},
  \bibinfo{author}{\bibfnamefont{S.~V.} \bibnamefont{Maleyev}},
  \bibinfo{author}{\bibfnamefont{P.}~\bibnamefont{Bourges}}, and
  \bibinfo{author}{\bibfnamefont{A.~S.} \bibnamefont{Ivanov}},
  \bibinfo{year}{1999}, \bibinfo{journal}{Phys.\ Rev.\ B}
  \textbf{\bibinfo{volume}{59}}(\bibinfo{number}{2}), \bibinfo{pages}{1079}.

\bibitem[{\citenamefont{Pfleiderer}(2009)}]{Pfleiderer09a}
\bibinfo{author}{\bibnamefont{Pfleiderer}, \bibfnamefont{C.}},
  \bibinfo{year}{2009}, \bibinfo{journal}{Rev.\ Mod.\ Phys.} .

\bibitem[{\citenamefont{Pimenov} \emph{et~al.}(2000)\citenamefont{Pimenov,
  Pronin, Loidl, Michelucci, Kampf, Krasnosvobodtsev, Nozdrin, and
  Rainer}}]{Pimenov00}
\bibinfo{author}{\bibnamefont{Pimenov}, \bibfnamefont{A.}},
  \bibinfo{author}{\bibfnamefont{A.~V.} \bibnamefont{Pronin}},
  \bibinfo{author}{\bibfnamefont{A.}~\bibnamefont{Loidl}},
  \bibinfo{author}{\bibfnamefont{U.}~\bibnamefont{Michelucci}},
  \bibinfo{author}{\bibfnamefont{A.~P.} \bibnamefont{Kampf}},
  \bibinfo{author}{\bibfnamefont{S.~I.} \bibnamefont{Krasnosvobodtsev}},
  \bibinfo{author}{\bibfnamefont{V.~S.} \bibnamefont{Nozdrin}}, and
  \bibinfo{author}{\bibfnamefont{D.}~\bibnamefont{Rainer}},
  \bibinfo{year}{2000}, \bibinfo{journal}{Phys.\ Rev.\ B}
  \textbf{\bibinfo{volume}{62}}(\bibinfo{number}{14}), \bibinfo{pages}{9822}.

\bibitem[{\citenamefont{Pinol} \emph{et~al.}(1990)\citenamefont{Pinol,
  Fontcuberta, Miravitlles, and Paul}}]{Pinol90b}
\bibinfo{author}{\bibnamefont{Pinol}, \bibfnamefont{S.}},
  \bibinfo{author}{\bibfnamefont{J.}~\bibnamefont{Fontcuberta}},
  \bibinfo{author}{\bibfnamefont{C.}~\bibnamefont{Miravitlles}}, and
  \bibinfo{author}{\bibfnamefont{D.~M.} \bibnamefont{Paul}},
  \bibinfo{year}{1990}, \bibinfo{journal}{Physica C}
  \textbf{\bibinfo{volume}{165}}(\bibinfo{number}{3-4}), \bibinfo{pages}{265}.

\bibitem[{\citenamefont{Pintschovius and Braden}(1999)}]{Pintschovius99a}
\bibinfo{author}{\bibnamefont{Pintschovius}, \bibfnamefont{L.}}, and
  \bibinfo{author}{\bibfnamefont{M.}~\bibnamefont{Braden}},
  \bibinfo{year}{1999}, \bibinfo{journal}{Phys.\ Rev.\ B}
  \textbf{\bibinfo{volume}{60}}(\bibinfo{number}{22}), \bibinfo{pages}{R15039}.

\bibitem[{\citenamefont{Pintschovius}
  \emph{et~al.}(2006)\citenamefont{Pintschovius, Reznik, and
  Yamada}}]{Pintschovius06a}
\bibinfo{author}{\bibnamefont{Pintschovius}, \bibfnamefont{L.}},
  \bibinfo{author}{\bibfnamefont{D.}~\bibnamefont{Reznik}}, and
  \bibinfo{author}{\bibfnamefont{K.}~\bibnamefont{Yamada}},
  \bibinfo{year}{2006}, \bibinfo{journal}{Phys.\ Rev.\ B}
  \textbf{\bibinfo{volume}{74}}(\bibinfo{number}{17}), \bibinfo{eid}{174514}.

\bibitem[{\citenamefont{Plakhty} \emph{et~al.}(2003)\citenamefont{Plakhty,
  Maleyev, Gavrilov, Bourdarot, Pouget, and Barilo}}]{Plakhty03a}
\bibinfo{author}{\bibnamefont{Plakhty}, \bibfnamefont{V.~P.}},
  \bibinfo{author}{\bibfnamefont{S.~V.} \bibnamefont{Maleyev}},
  \bibinfo{author}{\bibfnamefont{S.~V.} \bibnamefont{Gavrilov}},
  \bibinfo{author}{\bibfnamefont{E.}~\bibnamefont{Bourdarot}},
  \bibinfo{author}{\bibfnamefont{S.}~\bibnamefont{Pouget}}, and
  \bibinfo{author}{\bibfnamefont{S.~N.} \bibnamefont{Barilo}},
  \bibinfo{year}{2003}, \bibinfo{journal}{Europhys.\ Lett.}
  \textbf{\bibinfo{volume}{61}}, \bibinfo{pages}{534}.

\bibitem[{\citenamefont{Podolsky} \emph{et~al.}(2007)\citenamefont{Podolsky,
  Raghu, and Vishwanath}}]{Podolsky07a}
\bibinfo{author}{\bibnamefont{Podolsky}, \bibfnamefont{D.}},
  \bibinfo{author}{\bibfnamefont{S.}~\bibnamefont{Raghu}}, and
  \bibinfo{author}{\bibfnamefont{A.}~\bibnamefont{Vishwanath}},
  \bibinfo{year}{2007}, \bibinfo{journal}{Phys.\ Rev.\ Lett.}
  \textbf{\bibinfo{volume}{99}}(\bibinfo{number}{11}), \bibinfo{eid}{117004}.

\bibitem[{\citenamefont{Ponomarev} \emph{et~al.}(2004)\citenamefont{Ponomarev,
  Charikova, Ignatenkov, Tashlykov, and Ivanov}}]{Ponomarev04a}
\bibinfo{author}{\bibnamefont{Ponomarev}, \bibfnamefont{A.~I.}},
  \bibinfo{author}{\bibfnamefont{T.~B.} \bibnamefont{Charikova}},
  \bibinfo{author}{\bibfnamefont{A.~N.} \bibnamefont{Ignatenkov}},
  \bibinfo{author}{\bibfnamefont{A.~O.} \bibnamefont{Tashlykov}}, and
  \bibinfo{author}{\bibfnamefont{A.~A.} \bibnamefont{Ivanov}},
  \bibinfo{year}{2004}, \bibinfo{journal}{Low Temp.\ Phys.}
  \textbf{\bibinfo{volume}{30}}(\bibinfo{number}{11}), \bibinfo{pages}{885}.

\bibitem[{\citenamefont{Prijamboedi and Kashiwaya}(2006)}]{Prijamboedi06a}
\bibinfo{author}{\bibnamefont{Prijamboedi}, \bibfnamefont{B.}}, and
  \bibinfo{author}{\bibfnamefont{S.}~\bibnamefont{Kashiwaya}},
  \bibinfo{year}{2006}, \bibinfo{journal}{Journal of Materials Science:
  Materials in Electronics} \textbf{\bibinfo{volume}{17}}(\bibinfo{number}{7}),
  \bibinfo{pages}{483}.

\bibitem[{\citenamefont{Pronin} \emph{et~al.}(2003)\citenamefont{Pronin,
  Pimenov, Loidl, Tsukada, and Naito}}]{Pronin03a}
\bibinfo{author}{\bibnamefont{Pronin}, \bibfnamefont{A.~V.}},
  \bibinfo{author}{\bibfnamefont{A.}~\bibnamefont{Pimenov}},
  \bibinfo{author}{\bibfnamefont{A.}~\bibnamefont{Loidl}},
  \bibinfo{author}{\bibfnamefont{A.}~\bibnamefont{Tsukada}}, and
  \bibinfo{author}{\bibfnamefont{M.}~\bibnamefont{Naito}},
  \bibinfo{year}{2003}, \bibinfo{journal}{Phys.\ Rev.\ B}
  \textbf{\bibinfo{volume}{68}}(\bibinfo{number}{5}), \bibinfo{pages}{054511}.

\bibitem[{\citenamefont{Proust} \emph{et~al.}(2005)\citenamefont{Proust,
  Behnia, Bel, Maude, and Vedeneev}}]{Proust05a}
\bibinfo{author}{\bibnamefont{Proust}, \bibfnamefont{C.}},
  \bibinfo{author}{\bibfnamefont{K.}~\bibnamefont{Behnia}},
  \bibinfo{author}{\bibfnamefont{R.}~\bibnamefont{Bel}},
  \bibinfo{author}{\bibfnamefont{D.}~\bibnamefont{Maude}}, and
  \bibinfo{author}{\bibfnamefont{S.~I.} \bibnamefont{Vedeneev}},
  \bibinfo{year}{2005}, \bibinfo{journal}{Phys.\ Rev.\ B}
  \textbf{\bibinfo{volume}{72}}(\bibinfo{number}{21}), \bibinfo{eid}{214511}.

\bibitem[{\citenamefont{Proust} \emph{et~al.}(2002)\citenamefont{Proust,
  Boaknin, Hill, Taillefer, and Mackenzie}}]{Proust02a}
\bibinfo{author}{\bibnamefont{Proust}, \bibfnamefont{C.}},
  \bibinfo{author}{\bibfnamefont{E.}~\bibnamefont{Boaknin}},
  \bibinfo{author}{\bibfnamefont{R.~W.} \bibnamefont{Hill}},
  \bibinfo{author}{\bibfnamefont{L.}~\bibnamefont{Taillefer}}, and
  \bibinfo{author}{\bibfnamefont{A.~P.} \bibnamefont{Mackenzie}},
  \bibinfo{year}{2002}, \bibinfo{journal}{Phys.\ Rev.\ Lett.}
  \textbf{\bibinfo{volume}{89}}(\bibinfo{number}{14}), \bibinfo{pages}{147003}.

\bibitem[{\citenamefont{Prozorov} \emph{et~al.}(2000)\citenamefont{Prozorov,
  Giannetta, Fournier, and Greene}}]{Prozorov00b}
\bibinfo{author}{\bibnamefont{Prozorov}, \bibfnamefont{R.}},
  \bibinfo{author}{\bibfnamefont{R.}~\bibnamefont{Giannetta}},
  \bibinfo{author}{\bibfnamefont{P.}~\bibnamefont{Fournier}}, and
  \bibinfo{author}{\bibfnamefont{R.~L.} \bibnamefont{Greene}},
  \bibinfo{year}{2000}, \bibinfo{journal}{Phys.\ Rev.\ Lett.}
  \textbf{\bibinfo{volume}{85}}, \bibinfo{pages}{3700}.

\bibitem[{\citenamefont{Puchkov} \emph{et~al.}(1996)\citenamefont{Puchkov, DN,
  and Timusk}}]{Puchkov96a}
\bibinfo{author}{\bibnamefont{Puchkov}, \bibfnamefont{A.}},
  \bibinfo{author}{\bibfnamefont{D.~B.} \bibnamefont{DN}}, and
  \bibinfo{author}{\bibfnamefont{T.}~\bibnamefont{Timusk}},
  \bibinfo{year}{1996}, \bibinfo{journal}{J.\ Phys.\ Condens.\ Matter}
  \textbf{\bibinfo{volume}{8}}(\bibinfo{number}{10}), \bibinfo{pages}{10049}.

\bibitem[{\citenamefont{Qazilbash} \emph{et~al.}(2003)\citenamefont{Qazilbash,
  Biswas, Dagan, Ott, and Greene}}]{Qazilbash03a}
\bibinfo{author}{\bibnamefont{Qazilbash}, \bibfnamefont{M.~M.}},
  \bibinfo{author}{\bibfnamefont{A.}~\bibnamefont{Biswas}},
  \bibinfo{author}{\bibfnamefont{Y.}~\bibnamefont{Dagan}},
  \bibinfo{author}{\bibfnamefont{R.~A.} \bibnamefont{Ott}}, and
  \bibinfo{author}{\bibfnamefont{R.~L.} \bibnamefont{Greene}},
  \bibinfo{year}{2003}, \bibinfo{journal}{Phys.\ Rev.\ B}
  \textbf{\bibinfo{volume}{68}}(\bibinfo{number}{2}), \bibinfo{eid}{024502}.

\bibitem[{\citenamefont{Qazilbash} \emph{et~al.}(2005)\citenamefont{Qazilbash,
  Koitzsch, Dennis, Gozar, Balci, Kendziora, Greene, and
  Blumberg}}]{Qazilbash05a}
\bibinfo{author}{\bibnamefont{Qazilbash}, \bibfnamefont{M.~M.}},
  \bibinfo{author}{\bibfnamefont{A.}~\bibnamefont{Koitzsch}},
  \bibinfo{author}{\bibfnamefont{B.~S.} \bibnamefont{Dennis}},
  \bibinfo{author}{\bibfnamefont{A.}~\bibnamefont{Gozar}},
  \bibinfo{author}{\bibfnamefont{H.}~\bibnamefont{Balci}},
  \bibinfo{author}{\bibfnamefont{C.~A.} \bibnamefont{Kendziora}},
  \bibinfo{author}{\bibfnamefont{R.~L.} \bibnamefont{Greene}}, and
  \bibinfo{author}{\bibfnamefont{G.}~\bibnamefont{Blumberg}},
  \bibinfo{year}{2005}, \bibinfo{journal}{Phys.\ Rev.\ B}
  \textbf{\bibinfo{volume}{72}}(\bibinfo{number}{21}), \bibinfo{eid}{214510}.

\bibitem[{\citenamefont{Radaelli} \emph{et~al.}(1994)\citenamefont{Radaelli,
  Jorgensen, Schultz, Peng, and Greene}}]{Radaelli94a}
\bibinfo{author}{\bibnamefont{Radaelli}, \bibfnamefont{P.~G.}},
  \bibinfo{author}{\bibfnamefont{J.~D.} \bibnamefont{Jorgensen}},
  \bibinfo{author}{\bibfnamefont{A.~J.} \bibnamefont{Schultz}},
  \bibinfo{author}{\bibfnamefont{J.~L.} \bibnamefont{Peng}}, and
  \bibinfo{author}{\bibfnamefont{R.~L.} \bibnamefont{Greene}},
  \bibinfo{year}{1994}, \bibinfo{journal}{Phys.\ Rev.\ B}
  \textbf{\bibinfo{volume}{49}}(\bibinfo{number}{21}), \bibinfo{pages}{15322}.

\bibitem[{\citenamefont{Randeria}(2007)}]{Randeria97a}
\bibinfo{author}{\bibnamefont{Randeria}, \bibfnamefont{M.}},
  \bibinfo{year}{2007}, in \emph{\bibinfo{booktitle}{Models and Phenomenology
  for Conventional and High-temperature Superconductivity}}, edited by
  \bibinfo{editor}{\bibfnamefont{G.}~\bibnamefont{Iadonisi}},
  \bibinfo{editor}{\bibfnamefont{J.~R.} \bibnamefont{Schrieffer}}, and
  \bibinfo{editor}{\bibfnamefont{M.}~\bibnamefont{Chiofalo}}
  (\bibinfo{publisher}{Ios Pr Inc}), pp. \bibinfo{pages}{53--75}.

\bibitem[{\citenamefont{Reznik} \emph{et~al.}(2006)\citenamefont{Reznik,
  Pintschovius, Ito, Iikubo, Sato, Goka, Fujita, Yamada, Gu, and
  Tranquada}}]{Reznik06a}
\bibinfo{author}{\bibnamefont{Reznik}, \bibfnamefont{D.}},
  \bibinfo{author}{\bibfnamefont{L.}~\bibnamefont{Pintschovius}},
  \bibinfo{author}{\bibfnamefont{M.}~\bibnamefont{Ito}},
  \bibinfo{author}{\bibfnamefont{S.}~\bibnamefont{Iikubo}},
  \bibinfo{author}{\bibfnamefont{M.}~\bibnamefont{Sato}},
  \bibinfo{author}{\bibfnamefont{H.}~\bibnamefont{Goka}},
  \bibinfo{author}{\bibfnamefont{M.}~\bibnamefont{Fujita}},
  \bibinfo{author}{\bibfnamefont{K.}~\bibnamefont{Yamada}},
  \bibinfo{author}{\bibfnamefont{G.}~\bibnamefont{Gu}}, and
  \bibinfo{author}{\bibfnamefont{J.}~\bibnamefont{Tranquada}},
  \bibinfo{year}{2006}, \bibinfo{journal}{Nature}
  \textbf{\bibinfo{volume}{440}}(\bibinfo{number}{1170}),
  \bibinfo{pages}{1170}.

\bibitem[{\citenamefont{Richard}
  \emph{et~al.}(2005{\natexlab{a}})\citenamefont{Richard, Jandl, Poirier,
  Fournier, Nekvasil, and Sadowski}}]{Richard05b}
\bibinfo{author}{\bibnamefont{Richard}, \bibfnamefont{P.}},
  \bibinfo{author}{\bibfnamefont{S.}~\bibnamefont{Jandl}},
  \bibinfo{author}{\bibfnamefont{M.}~\bibnamefont{Poirier}},
  \bibinfo{author}{\bibfnamefont{P.}~\bibnamefont{Fournier}},
  \bibinfo{author}{\bibfnamefont{V.}~\bibnamefont{Nekvasil}}, and
  \bibinfo{author}{\bibfnamefont{M.~L.} \bibnamefont{Sadowski}},
  \bibinfo{year}{2005}{\natexlab{a}}, \bibinfo{journal}{Phys.\ Rev.\ B}
  \textbf{\bibinfo{volume}{72}}(\bibinfo{number}{1}), \bibinfo{eid}{014506}.

\bibitem[{\citenamefont{Richard} \emph{et~al.}(2007)\citenamefont{Richard,
  Neupane, Xu, Fournier, Li, Dai, Wang, and Ding}}]{Richard07a}
\bibinfo{author}{\bibnamefont{Richard}, \bibfnamefont{P.}},
  \bibinfo{author}{\bibfnamefont{M.}~\bibnamefont{Neupane}},
  \bibinfo{author}{\bibfnamefont{Y.-M.} \bibnamefont{Xu}},
  \bibinfo{author}{\bibfnamefont{P.}~\bibnamefont{Fournier}},
  \bibinfo{author}{\bibfnamefont{S.}~\bibnamefont{Li}},
  \bibinfo{author}{\bibfnamefont{P.}~\bibnamefont{Dai}},
  \bibinfo{author}{\bibfnamefont{Z.}~\bibnamefont{Wang}}, and
  \bibinfo{author}{\bibfnamefont{H.}~\bibnamefont{Ding}}, \bibinfo{year}{2007},
  \bibinfo{journal}{Phys.\ Rev.\ Lett.}
  \textbf{\bibinfo{volume}{99}}(\bibinfo{number}{15}), \bibinfo{eid}{157002}.

\bibitem[{\citenamefont{Richard}
  \emph{et~al.}(2005{\natexlab{b}})\citenamefont{Richard, Poirier, and
  Jandl}}]{Richard05a}
\bibinfo{author}{\bibnamefont{Richard}, \bibfnamefont{P.}},
  \bibinfo{author}{\bibfnamefont{M.}~\bibnamefont{Poirier}}, and
  \bibinfo{author}{\bibfnamefont{S.}~\bibnamefont{Jandl}},
  \bibinfo{year}{2005}{\natexlab{b}}, \bibinfo{journal}{Phys.\ Rev.\ B}
  \textbf{\bibinfo{volume}{71}}(\bibinfo{number}{14}), \bibinfo{eid}{144425}.

\bibitem[{\citenamefont{Richard} \emph{et~al.}(2004)\citenamefont{Richard,
  Riou, Hetel, Jandl, Poirier, and Fournier}}]{Richard04a}
\bibinfo{author}{\bibnamefont{Richard}, \bibfnamefont{P.}},
  \bibinfo{author}{\bibfnamefont{G.}~\bibnamefont{Riou}},
  \bibinfo{author}{\bibfnamefont{I.}~\bibnamefont{Hetel}},
  \bibinfo{author}{\bibfnamefont{S.}~\bibnamefont{Jandl}},
  \bibinfo{author}{\bibfnamefont{M.}~\bibnamefont{Poirier}}, and
  \bibinfo{author}{\bibfnamefont{P.}~\bibnamefont{Fournier}},
  \bibinfo{year}{2004}, \bibinfo{journal}{Phys.\ Rev.\ B}
  \textbf{\bibinfo{volume}{70}}(\bibinfo{number}{6}), \bibinfo{pages}{064513}.

\bibitem[{\citenamefont{Riou} \emph{et~al.}(2001)\citenamefont{Riou, Jandl,
  Poirier, Nekvasil, Divis, Fournier, Greene, Zhigunov, and Barilo}}]{Riou01a}
\bibinfo{author}{\bibnamefont{Riou}, \bibfnamefont{G.}},
  \bibinfo{author}{\bibfnamefont{S.}~\bibnamefont{Jandl}},
  \bibinfo{author}{\bibfnamefont{M.}~\bibnamefont{Poirier}},
  \bibinfo{author}{\bibfnamefont{V.}~\bibnamefont{Nekvasil}},
  \bibinfo{author}{\bibfnamefont{M.}~\bibnamefont{Divis}},
  \bibinfo{author}{\bibfnamefont{P.}~\bibnamefont{Fournier}},
  \bibinfo{author}{\bibfnamefont{R.}~\bibnamefont{Greene}},
  \bibinfo{author}{\bibfnamefont{D.}~\bibnamefont{Zhigunov}}, and
  \bibinfo{author}{\bibfnamefont{S.}~\bibnamefont{Barilo}},
  \bibinfo{year}{2001}, \bibinfo{journal}{Eur.\ Phys.\ J.\ B}
  \textbf{\bibinfo{volume}{23}}, \bibinfo{pages}{179}.

\bibitem[{\citenamefont{Riou} \emph{et~al.}(2004)\citenamefont{Riou, Richard,
  Jandl, Poirier, Fournier, Nekvasil, Barilo, and Kurnevich}}]{Riou04a}
\bibinfo{author}{\bibnamefont{Riou}, \bibfnamefont{G.}},
  \bibinfo{author}{\bibfnamefont{P.}~\bibnamefont{Richard}},
  \bibinfo{author}{\bibfnamefont{S.}~\bibnamefont{Jandl}},
  \bibinfo{author}{\bibfnamefont{M.}~\bibnamefont{Poirier}},
  \bibinfo{author}{\bibfnamefont{P.}~\bibnamefont{Fournier}},
  \bibinfo{author}{\bibfnamefont{V.}~\bibnamefont{Nekvasil}},
  \bibinfo{author}{\bibfnamefont{S.~N.} \bibnamefont{Barilo}}, and
  \bibinfo{author}{\bibfnamefont{L.~A.} \bibnamefont{Kurnevich}},
  \bibinfo{year}{2004}, \bibinfo{journal}{Phys.\ Rev.\ B}
  \textbf{\bibinfo{volume}{69}}(\bibinfo{number}{2}), \bibinfo{pages}{024511}.

\bibitem[{\citenamefont{Roberge} \emph{et~al.}(2009)\citenamefont{Roberge,
  Charpentier, Godin-Proulx, Rauwel, Truong, and Fournier}}]{Roberge09a}
\bibinfo{author}{\bibnamefont{Roberge}, \bibfnamefont{G.}},
  \bibinfo{author}{\bibfnamefont{S.}~\bibnamefont{Charpentier}},
  \bibinfo{author}{\bibfnamefont{S.}~\bibnamefont{Godin-Proulx}},
  \bibinfo{author}{\bibfnamefont{P.}~\bibnamefont{Rauwel}},
  \bibinfo{author}{\bibfnamefont{K.}~\bibnamefont{Truong}}, and
  \bibinfo{author}{\bibfnamefont{P.}~\bibnamefont{Fournier}},
  \bibinfo{year}{2009}, \bibinfo{journal}{J.\ Crystal Growth}
  \textbf{\bibinfo{volume}{311}}(\bibinfo{number}{5}), \bibinfo{pages}{1340 }.

\bibitem[{\citenamefont{Ronning} \emph{et~al.}(1998)\citenamefont{Ronning, Kim,
  Feng, Marshall, Loeser, Miller, Eckstein, Bozovic, and Shen}}]{Ronning98a}
\bibinfo{author}{\bibnamefont{Ronning}, \bibfnamefont{F.}},
  \bibinfo{author}{\bibfnamefont{C.}~\bibnamefont{Kim}},
  \bibinfo{author}{\bibfnamefont{D.~L.} \bibnamefont{Feng}},
  \bibinfo{author}{\bibfnamefont{D.~S.} \bibnamefont{Marshall}},
  \bibinfo{author}{\bibfnamefont{A.~G.} \bibnamefont{Loeser}},
  \bibinfo{author}{\bibfnamefont{L.~L.} \bibnamefont{Miller}},
  \bibinfo{author}{\bibfnamefont{J.~N.} \bibnamefont{Eckstein}},
  \bibinfo{author}{\bibfnamefont{I.}~\bibnamefont{Bozovic}}, and
  \bibinfo{author}{\bibfnamefont{Z.-X.} \bibnamefont{Shen}},
  \bibinfo{year}{1998}, \bibinfo{journal}{Science}
  \textbf{\bibinfo{volume}{282}}(\bibinfo{number}{5396}),
  \bibinfo{pages}{2067}.

\bibitem[{\citenamefont{Ronning} \emph{et~al.}(2003)\citenamefont{Ronning,
  Sasagawa, Kohsaka, Shen, Damascelli, Kim, Yoshida, Armitage, Lu, Feng,
  Miller, Takagi} \emph{et~al.}}]{Ronning03a}
\bibinfo{author}{\bibnamefont{Ronning}, \bibfnamefont{F.}},
  \bibinfo{author}{\bibfnamefont{T.}~\bibnamefont{Sasagawa}},
  \bibinfo{author}{\bibfnamefont{Y.}~\bibnamefont{Kohsaka}},
  \bibinfo{author}{\bibfnamefont{K.~M.} \bibnamefont{Shen}},
  \bibinfo{author}{\bibfnamefont{A.}~\bibnamefont{Damascelli}},
  \bibinfo{author}{\bibfnamefont{C.}~\bibnamefont{Kim}},
  \bibinfo{author}{\bibfnamefont{T.}~\bibnamefont{Yoshida}},
  \bibinfo{author}{\bibfnamefont{N.~P.} \bibnamefont{Armitage}},
  \bibinfo{author}{\bibfnamefont{D.~H.} \bibnamefont{Lu}},
  \bibinfo{author}{\bibfnamefont{D.~L.} \bibnamefont{Feng}},
  \bibinfo{author}{\bibfnamefont{L.~L.} \bibnamefont{Miller}},
  \bibinfo{author}{\bibfnamefont{H.}~\bibnamefont{Takagi}}, \emph{et~al.},
  \bibinfo{year}{2003}, \bibinfo{journal}{Phys.\ Rev.\ B}
  \textbf{\bibinfo{volume}{67}}(\bibinfo{number}{16}), \bibinfo{pages}{165101}.

\bibitem[{\citenamefont{Ronning} \emph{et~al.}(2005)\citenamefont{Ronning,
  Shen, Armitage, Damascelli, Lu, Shen, Miller, and Kim}}]{Ronning05a}
\bibinfo{author}{\bibnamefont{Ronning}, \bibfnamefont{F.}},
  \bibinfo{author}{\bibfnamefont{K.~M.} \bibnamefont{Shen}},
  \bibinfo{author}{\bibfnamefont{N.~P.} \bibnamefont{Armitage}},
  \bibinfo{author}{\bibfnamefont{A.}~\bibnamefont{Damascelli}},
  \bibinfo{author}{\bibfnamefont{D.~H.} \bibnamefont{Lu}},
  \bibinfo{author}{\bibfnamefont{Z.-X.} \bibnamefont{Shen}},
  \bibinfo{author}{\bibfnamefont{L.~L.} \bibnamefont{Miller}}, and
  \bibinfo{author}{\bibfnamefont{C.}~\bibnamefont{Kim}}, \bibinfo{year}{2005},
  \bibinfo{journal}{Phys.\ Rev.\ B}
  \textbf{\bibinfo{volume}{71}}(\bibinfo{number}{9}), \bibinfo{eid}{094518}.

\bibitem[{\citenamefont{Rossat-Mignod}
  \emph{et~al.}(1991)\citenamefont{Rossat-Mignod, Regnault, Vettier, Bourges,
  Burlet, Bossy, Henry, and Lapertot}}]{rossat91a}
\bibinfo{author}{\bibnamefont{Rossat-Mignod}, \bibfnamefont{J.~M.}},
  \bibinfo{author}{\bibfnamefont{L.~P.} \bibnamefont{Regnault}},
  \bibinfo{author}{\bibfnamefont{C.}~\bibnamefont{Vettier}},
  \bibinfo{author}{\bibfnamefont{P.}~\bibnamefont{Bourges}},
  \bibinfo{author}{\bibfnamefont{P.}~\bibnamefont{Burlet}},
  \bibinfo{author}{\bibfnamefont{J.}~\bibnamefont{Bossy}},
  \bibinfo{author}{\bibfnamefont{J.~Y.} \bibnamefont{Henry}}, and
  \bibinfo{author}{\bibfnamefont{G.}~\bibnamefont{Lapertot}},
  \bibinfo{year}{1991}, \bibinfo{journal}{Physica C}
  \textbf{\bibinfo{volume}{185}}, \bibinfo{eid}{86}.

\bibitem[{\citenamefont{Rullier-Albenque}
  \emph{et~al.}(2008)\citenamefont{Rullier-Albenque, Alloul, Balakirev, and
  Proust}}]{Rullier08a}
\bibinfo{author}{\bibnamefont{Rullier-Albenque}, \bibfnamefont{F.}},
  \bibinfo{author}{\bibfnamefont{H.}~\bibnamefont{Alloul}},
  \bibinfo{author}{\bibfnamefont{F.}~\bibnamefont{Balakirev}}, and
  \bibinfo{author}{\bibfnamefont{C.}~\bibnamefont{Proust}},
  \bibinfo{year}{2008}, \bibinfo{journal}{Europhys.\ Lett.}
  \textbf{\bibinfo{volume}{81}}(\bibinfo{number}{3}), \bibinfo{pages}{37008}.

\bibitem[{\citenamefont{Rullier-Albenque}
  \emph{et~al.}(2003)\citenamefont{Rullier-Albenque, Alloul, and
  Tourbot}}]{Rullier03a}
\bibinfo{author}{\bibnamefont{Rullier-Albenque}, \bibfnamefont{F.}},
  \bibinfo{author}{\bibfnamefont{H.}~\bibnamefont{Alloul}}, and
  \bibinfo{author}{\bibfnamefont{R.}~\bibnamefont{Tourbot}},
  \bibinfo{year}{2003}, \bibinfo{journal}{Phys.\ Rev.\ Lett.}
  \textbf{\bibinfo{volume}{91}}(\bibinfo{number}{4}), \bibinfo{eid}{047001}.

\bibitem[{\citenamefont{Sachdev}(2003)}]{Sachdev03a}
\bibinfo{author}{\bibnamefont{Sachdev}, \bibfnamefont{S.}},
  \bibinfo{year}{2003}, \bibinfo{journal}{Rev.\ Mod.\ Phys.}
  \textbf{\bibinfo{volume}{75}}(\bibinfo{number}{3}), \bibinfo{pages}{913}.

\bibitem[{\citenamefont{Sachdev} \emph{et~al.}(2009)\citenamefont{Sachdev,
  Metlitski, Qi, and Xu}}]{Sachdev09a}
\bibinfo{author}{\bibnamefont{Sachdev}, \bibfnamefont{S.}},
  \bibinfo{author}{\bibfnamefont{M.~A.} \bibnamefont{Metlitski}},
  \bibinfo{author}{\bibfnamefont{Y.}~\bibnamefont{Qi}}, and
  \bibinfo{author}{\bibfnamefont{C.}~\bibnamefont{Xu}}, \bibinfo{year}{2009},
  \bibinfo{journal}{Phys.\ Rev.\ B}
  \textbf{\bibinfo{volume}{80}}(\bibinfo{number}{15}), \bibinfo{eid}{155129}.

\bibitem[{\citenamefont{Sachidanandam}
  \emph{et~al.}(1997)\citenamefont{Sachidanandam, Yildirim, Harris, Aharony,
  and Entin-Wohlman}}]{Sachidanandam97a}
\bibinfo{author}{\bibnamefont{Sachidanandam}, \bibfnamefont{R.}},
  \bibinfo{author}{\bibfnamefont{T.}~\bibnamefont{Yildirim}},
  \bibinfo{author}{\bibfnamefont{A.~B.} \bibnamefont{Harris}},
  \bibinfo{author}{\bibfnamefont{A.}~\bibnamefont{Aharony}}, and
  \bibinfo{author}{\bibfnamefont{O.}~\bibnamefont{Entin-Wohlman}},
  \bibinfo{year}{1997}, \bibinfo{journal}{Phys.\ Rev.\ B}
  \textbf{\bibinfo{volume}{56}}(\bibinfo{number}{1}), \bibinfo{pages}{260}.

\bibitem[{\citenamefont{Sadori and Grilli}(2000)}]{Sadori00a}
\bibinfo{author}{\bibnamefont{Sadori}, \bibfnamefont{A.}}, and
  \bibinfo{author}{\bibfnamefont{M.}~\bibnamefont{Grilli}},
  \bibinfo{year}{2000}, \bibinfo{journal}{Phys. Rev. Lett.}
  \textbf{\bibinfo{volume}{84}}(\bibinfo{number}{23}), \bibinfo{pages}{5375}.

\bibitem[{\citenamefont{Sadowski} \emph{et~al.}(1990)\citenamefont{Sadowski,
  Hagemann, François, Bill, Peter, Walker, and Yvon}}]{Sadowski90a}
\bibinfo{author}{\bibnamefont{Sadowski}, \bibfnamefont{W.}},
  \bibinfo{author}{\bibfnamefont{H.}~\bibnamefont{Hagemann}},
  \bibinfo{author}{\bibfnamefont{M.}~\bibnamefont{François}},
  \bibinfo{author}{\bibfnamefont{H.}~\bibnamefont{Bill}},
  \bibinfo{author}{\bibfnamefont{M.}~\bibnamefont{Peter}},
  \bibinfo{author}{\bibfnamefont{E.}~\bibnamefont{Walker}}, and
  \bibinfo{author}{\bibfnamefont{K.}~\bibnamefont{Yvon}}, \bibinfo{year}{1990},
  \bibinfo{journal}{Physica C}
  \textbf{\bibinfo{volume}{170}}(\bibinfo{number}{1-2}), \bibinfo{pages}{103 }.

\bibitem[{\citenamefont{Sakisaka} \emph{et~al.}(1990)\citenamefont{Sakisaka,
  Maruyama, Morikawa, Kato, Edamoto, Okusawa, Aiura, Yanashima, Terashima,
  Bando, Iijima, Yamamoto} \emph{et~al.}}]{Sakisaka90a}
\bibinfo{author}{\bibnamefont{Sakisaka}, \bibfnamefont{Y.}},
  \bibinfo{author}{\bibfnamefont{T.}~\bibnamefont{Maruyama}},
  \bibinfo{author}{\bibfnamefont{Y.}~\bibnamefont{Morikawa}},
  \bibinfo{author}{\bibfnamefont{H.}~\bibnamefont{Kato}},
  \bibinfo{author}{\bibfnamefont{K.}~\bibnamefont{Edamoto}},
  \bibinfo{author}{\bibfnamefont{M.}~\bibnamefont{Okusawa}},
  \bibinfo{author}{\bibfnamefont{Y.}~\bibnamefont{Aiura}},
  \bibinfo{author}{\bibfnamefont{H.}~\bibnamefont{Yanashima}},
  \bibinfo{author}{\bibfnamefont{T.}~\bibnamefont{Terashima}},
  \bibinfo{author}{\bibfnamefont{Y.}~\bibnamefont{Bando}},
  \bibinfo{author}{\bibfnamefont{K.}~\bibnamefont{Iijima}},
  \bibinfo{author}{\bibfnamefont{K.}~\bibnamefont{Yamamoto}}, \emph{et~al.},
  \bibinfo{year}{1990}, \bibinfo{journal}{Phys.\ Rev.\ B}
  \textbf{\bibinfo{volume}{42}}(\bibinfo{number}{7}), \bibinfo{pages}{4189}.

\bibitem[{\citenamefont{Santander-Syro}
  \emph{et~al.}(2009)\citenamefont{Santander-Syro, Kondo, Chang, Kaminski,
  Pailhes, Shi, Patthey, Zimmers, Liang, Li, and Greene}}]{Santander09a}
\bibinfo{author}{\bibnamefont{Santander-Syro}, \bibfnamefont{A.~F.}},
  \bibinfo{author}{\bibfnamefont{T.}~\bibnamefont{Kondo}},
  \bibinfo{author}{\bibfnamefont{J.}~\bibnamefont{Chang}},
  \bibinfo{author}{\bibfnamefont{A.}~\bibnamefont{Kaminski}},
  \bibinfo{author}{\bibfnamefont{S.}~\bibnamefont{Pailhes}},
  \bibinfo{author}{\bibfnamefont{M.}~\bibnamefont{Shi}},
  \bibinfo{author}{\bibfnamefont{L.}~\bibnamefont{Patthey}},
  \bibinfo{author}{\bibfnamefont{A.}~\bibnamefont{Zimmers}},
  \bibinfo{author}{\bibfnamefont{B.}~\bibnamefont{Liang}},
  \bibinfo{author}{\bibfnamefont{P.}~\bibnamefont{Li}}, and
  \bibinfo{author}{\bibfnamefont{R.~L.} \bibnamefont{Greene}},
  \bibinfo{year}{2009}, \bibinfo{journal}{arXiv:0903.3413v1} .

\bibitem[{\citenamefont{Sato} \emph{et~al.}(2001)\citenamefont{Sato, Kamiyama,
  Takahashi, Kurahashi, and Yamada}}]{Sato01a}
\bibinfo{author}{\bibnamefont{Sato}, \bibfnamefont{T.}},
  \bibinfo{author}{\bibfnamefont{T.}~\bibnamefont{Kamiyama}},
  \bibinfo{author}{\bibfnamefont{T.}~\bibnamefont{Takahashi}},
  \bibinfo{author}{\bibfnamefont{K.}~\bibnamefont{Kurahashi}}, and
  \bibinfo{author}{\bibfnamefont{K.}~\bibnamefont{Yamada}},
  \bibinfo{year}{2001}, \bibinfo{journal}{Science}
  \textbf{\bibinfo{volume}{291}}(\bibinfo{number}{5508}),
  \bibinfo{pages}{1517}.

\bibitem[{\citenamefont{Savici} \emph{et~al.}(2005)\citenamefont{Savici,
  Fukaya, Gat-Malureanu, Ito, Russo, Uemura, Wiebe, Kyriakou, MacDougall,
  Rovers, Luke, Kojima} \emph{et~al.}}]{Savici05a}
\bibinfo{author}{\bibnamefont{Savici}, \bibfnamefont{A.~T.}},
  \bibinfo{author}{\bibfnamefont{A.}~\bibnamefont{Fukaya}},
  \bibinfo{author}{\bibfnamefont{I.~M.} \bibnamefont{Gat-Malureanu}},
  \bibinfo{author}{\bibfnamefont{T.}~\bibnamefont{Ito}},
  \bibinfo{author}{\bibfnamefont{P.~L.} \bibnamefont{Russo}},
  \bibinfo{author}{\bibfnamefont{Y.~J.} \bibnamefont{Uemura}},
  \bibinfo{author}{\bibfnamefont{C.~R.} \bibnamefont{Wiebe}},
  \bibinfo{author}{\bibfnamefont{P.~P.} \bibnamefont{Kyriakou}},
  \bibinfo{author}{\bibfnamefont{G.~J.} \bibnamefont{MacDougall}},
  \bibinfo{author}{\bibfnamefont{M.~T.} \bibnamefont{Rovers}},
  \bibinfo{author}{\bibfnamefont{G.~M.} \bibnamefont{Luke}},
  \bibinfo{author}{\bibfnamefont{K.~M.} \bibnamefont{Kojima}}, \emph{et~al.},
  \bibinfo{year}{2005}, \bibinfo{journal}{Phys.\ Rev.\ Lett.}
  \textbf{\bibinfo{volume}{95}}(\bibinfo{number}{15}), \bibinfo{eid}{157001}.

\bibitem[{\citenamefont{Sawa} \emph{et~al.}(2002)\citenamefont{Sawa, Kawasaki,
  Takagi, and Tokura}}]{Sawa02a}
\bibinfo{author}{\bibnamefont{Sawa}, \bibfnamefont{A.}},
  \bibinfo{author}{\bibfnamefont{M.}~\bibnamefont{Kawasaki}},
  \bibinfo{author}{\bibfnamefont{H.}~\bibnamefont{Takagi}}, and
  \bibinfo{author}{\bibfnamefont{Y.}~\bibnamefont{Tokura}},
  \bibinfo{year}{2002}, \bibinfo{journal}{Phys.\ Rev.\ B}
  \textbf{\bibinfo{volume}{66}}(\bibinfo{number}{1}), \bibinfo{pages}{014531}.

\bibitem[{\citenamefont{Scalapino}(1995)}]{Scalapino95a}
\bibinfo{author}{\bibnamefont{Scalapino}, \bibfnamefont{D.~J.}},
  \bibinfo{year}{1995}, \bibinfo{journal}{Physics Reports}
  \textbf{\bibinfo{volume}{250}}, \bibinfo{pages}{329}.

\bibitem[{\citenamefont{Schachinger}
  \emph{et~al.}(2008)\citenamefont{Schachinger, Homes, Lobo, and
  Carbotte}}]{Schachinger08a}
\bibinfo{author}{\bibnamefont{Schachinger}, \bibfnamefont{E.}},
  \bibinfo{author}{\bibfnamefont{C.~C.} \bibnamefont{Homes}},
  \bibinfo{author}{\bibfnamefont{R.~P. S.~M.} \bibnamefont{Lobo}}, and
  \bibinfo{author}{\bibfnamefont{J.~P.} \bibnamefont{Carbotte}},
  \bibinfo{year}{2008}, \bibinfo{journal}{Phys.\ Rev.\ B}
  \textbf{\bibinfo{volume}{78}}(\bibinfo{number}{13}), \bibinfo{eid}{134522}.

\bibitem[{\citenamefont{Schachinger}
  \emph{et~al.}(2003)\citenamefont{Schachinger, Tu, and
  Carbotte}}]{Schachinger03a}
\bibinfo{author}{\bibnamefont{Schachinger}, \bibfnamefont{E.}},
  \bibinfo{author}{\bibfnamefont{J.~J.} \bibnamefont{Tu}}, and
  \bibinfo{author}{\bibfnamefont{J.~P.} \bibnamefont{Carbotte}},
  \bibinfo{year}{2003}, \bibinfo{journal}{Phys.\ Rev.\ B}
  \textbf{\bibinfo{volume}{67}}(\bibinfo{number}{21}), \bibinfo{pages}{214508}.

\bibitem[{\citenamefont{Schmitt} \emph{et~al.}(2008)\citenamefont{Schmitt, Lee,
  Lu, Meevasana, Motoyama, Greven, and Shen}}]{Schmitt08a}
\bibinfo{author}{\bibnamefont{Schmitt}, \bibfnamefont{F.}},
  \bibinfo{author}{\bibfnamefont{W.~S.} \bibnamefont{Lee}},
  \bibinfo{author}{\bibfnamefont{D.-H.} \bibnamefont{Lu}},
  \bibinfo{author}{\bibfnamefont{W.}~\bibnamefont{Meevasana}},
  \bibinfo{author}{\bibfnamefont{E.}~\bibnamefont{Motoyama}},
  \bibinfo{author}{\bibfnamefont{M.}~\bibnamefont{Greven}}, and
  \bibinfo{author}{\bibfnamefont{Z.-X.} \bibnamefont{Shen}},
  \bibinfo{year}{2008}, \bibinfo{journal}{Phys.\ Rev.\ B}
  \textbf{\bibinfo{volume}{78}}(\bibinfo{number}{10}), \bibinfo{eid}{100505}.

\bibitem[{\citenamefont{Schneider} \emph{et~al.}(1994)\citenamefont{Schneider,
  Barber, Evetts, Mao, Xi, and Venkatesan}}]{Schneider94a}
\bibinfo{author}{\bibnamefont{Schneider}, \bibfnamefont{C.~W.}},
  \bibinfo{author}{\bibfnamefont{Z.~H.} \bibnamefont{Barber}},
  \bibinfo{author}{\bibfnamefont{J.~E.} \bibnamefont{Evetts}},
  \bibinfo{author}{\bibfnamefont{S.~N.} \bibnamefont{Mao}},
  \bibinfo{author}{\bibfnamefont{X.~X.} \bibnamefont{Xi}}, and
  \bibinfo{author}{\bibfnamefont{T.}~\bibnamefont{Venkatesan}},
  \bibinfo{year}{1994}, \bibinfo{journal}{Physica C}
  \textbf{\bibinfo{volume}{233}}, \bibinfo{pages}{77}.

\bibitem[{\citenamefont{Schultz} \emph{et~al.}(1996)\citenamefont{Schultz,
  Jorgensen, Peng, and Greene}}]{Schultz96a}
\bibinfo{author}{\bibnamefont{Schultz}, \bibfnamefont{A.~J.}},
  \bibinfo{author}{\bibfnamefont{J.~D.} \bibnamefont{Jorgensen}},
  \bibinfo{author}{\bibfnamefont{J.~L.} \bibnamefont{Peng}}, and
  \bibinfo{author}{\bibfnamefont{R.~L.} \bibnamefont{Greene}},
  \bibinfo{year}{1996}, \bibinfo{journal}{Phys.\ Rev.\ B}
  \textbf{\bibinfo{volume}{53}}(\bibinfo{number}{9}), \bibinfo{pages}{5157}.

\bibitem[{\citenamefont{Segawa and Ando}(2006)}]{Segawa06a}
\bibinfo{author}{\bibnamefont{Segawa}, \bibfnamefont{K.}}, and
  \bibinfo{author}{\bibfnamefont{Y.}~\bibnamefont{Ando}}, \bibinfo{year}{2006},
  \bibinfo{journal}{Phys.\ Rev.\ B}
  \textbf{\bibinfo{volume}{74}}(\bibinfo{number}{10}), \bibinfo{eid}{100508}.

\bibitem[{\citenamefont{Sekitani} \emph{et~al.}(2003)\citenamefont{Sekitani,
  Naito, and Miura}}]{Sekitani03a}
\bibinfo{author}{\bibnamefont{Sekitani}, \bibfnamefont{T.}},
  \bibinfo{author}{\bibfnamefont{M.}~\bibnamefont{Naito}}, and
  \bibinfo{author}{\bibfnamefont{N.}~\bibnamefont{Miura}},
  \bibinfo{year}{2003}, \bibinfo{journal}{Phys.\ Rev.\ B}
  \textbf{\bibinfo{volume}{67}}(\bibinfo{number}{17}), \bibinfo{eid}{174503}.

\bibitem[{\citenamefont{Senechal and Tremblay}(2004)}]{Senechal04a}
\bibinfo{author}{\bibnamefont{Senechal}, \bibfnamefont{D.}}, and
  \bibinfo{author}{\bibfnamefont{A.-M.~S.} \bibnamefont{Tremblay}},
  \bibinfo{year}{2004}, \bibinfo{journal}{Phys.\ Rev.\ Lett.}
  \textbf{\bibinfo{volume}{92}}(\bibinfo{number}{12}), \bibinfo{eid}{126401}.

\bibitem[{\citenamefont{Seo} \emph{et~al.}(2007)\citenamefont{Seo, Chen, and
  Hu}}]{Seo07a}
\bibinfo{author}{\bibnamefont{Seo}, \bibfnamefont{K.}},
  \bibinfo{author}{\bibfnamefont{H.-D.} \bibnamefont{Chen}}, and
  \bibinfo{author}{\bibfnamefont{J.}~\bibnamefont{Hu}}, \bibinfo{year}{2007},
  \bibinfo{journal}{Phys.\ Rev.\ B}
  \textbf{\bibinfo{volume}{76}}(\bibinfo{number}{2}), \bibinfo{eid}{020511}.

\bibitem[{\citenamefont{Shan} \emph{et~al.}(2005)\citenamefont{Shan, Huang,
  Gao, Wang, Li, Dai, Zhou, Xiong, Ti, and Wen}}]{Shan05a}
\bibinfo{author}{\bibnamefont{Shan}, \bibfnamefont{L.}},
  \bibinfo{author}{\bibfnamefont{Y.}~\bibnamefont{Huang}},
  \bibinfo{author}{\bibfnamefont{H.}~\bibnamefont{Gao}},
  \bibinfo{author}{\bibfnamefont{Y.}~\bibnamefont{Wang}},
  \bibinfo{author}{\bibfnamefont{S.~L.} \bibnamefont{Li}},
  \bibinfo{author}{\bibfnamefont{P.~C.} \bibnamefont{Dai}},
  \bibinfo{author}{\bibfnamefont{F.}~\bibnamefont{Zhou}},
  \bibinfo{author}{\bibfnamefont{J.~W.} \bibnamefont{Xiong}},
  \bibinfo{author}{\bibfnamefont{W.~X.} \bibnamefont{Ti}}, and
  \bibinfo{author}{\bibfnamefont{H.~H.} \bibnamefont{Wen}},
  \bibinfo{year}{2005}, \bibinfo{journal}{Phys.\ Rev.\ B}
  \textbf{\bibinfo{volume}{72}}(\bibinfo{number}{14}), \bibinfo{eid}{144506}.

\bibitem[{\citenamefont{Shan}
  \emph{et~al.}(2008{\natexlab{a}})\citenamefont{Shan, Huang, Wang, Li, Zhao,
  Dai, Zhang, Ren, and Wen}}]{Shan08a}
\bibinfo{author}{\bibnamefont{Shan}, \bibfnamefont{L.}},
  \bibinfo{author}{\bibfnamefont{Y.}~\bibnamefont{Huang}},
  \bibinfo{author}{\bibfnamefont{Y.~L.} \bibnamefont{Wang}},
  \bibinfo{author}{\bibfnamefont{S.}~\bibnamefont{Li}},
  \bibinfo{author}{\bibfnamefont{J.}~\bibnamefont{Zhao}},
  \bibinfo{author}{\bibfnamefont{P.}~\bibnamefont{Dai}},
  \bibinfo{author}{\bibfnamefont{Y.}~\bibnamefont{Zhang}},
  \bibinfo{author}{\bibfnamefont{C.}~\bibnamefont{Ren}}, and
  \bibinfo{author}{\bibfnamefont{H.~H.} \bibnamefont{Wen}},
  \bibinfo{year}{2008}{\natexlab{a}},
  \bibinfo{journal}{http://arxiv.org/abs/cond-mat/0703256} .

\bibitem[{\citenamefont{Shan}
  \emph{et~al.}(2008{\natexlab{b}})\citenamefont{Shan, Wang, Huang, Li, Zhao,
  Dai, and Wen}}]{Shan08b}
\bibinfo{author}{\bibnamefont{Shan}, \bibfnamefont{L.}},
  \bibinfo{author}{\bibfnamefont{Y.~L.} \bibnamefont{Wang}},
  \bibinfo{author}{\bibfnamefont{Y.}~\bibnamefont{Huang}},
  \bibinfo{author}{\bibfnamefont{S.~L.} \bibnamefont{Li}},
  \bibinfo{author}{\bibfnamefont{J.}~\bibnamefont{Zhao}},
  \bibinfo{author}{\bibfnamefont{P.}~\bibnamefont{Dai}}, and
  \bibinfo{author}{\bibfnamefont{H.~H.} \bibnamefont{Wen}},
  \bibinfo{year}{2008}{\natexlab{b}}, \bibinfo{journal}{Phys.\ Rev.\ B}
  \textbf{\bibinfo{volume}{78}}(\bibinfo{number}{1}), \bibinfo{eid}{014505}.

\bibitem[{\citenamefont{Shannon}(1976)}]{Shannon76a}
\bibinfo{author}{\bibnamefont{Shannon}, \bibfnamefont{R.~D.}},
  \bibinfo{year}{1976}, \bibinfo{journal}{Acta Crystallographica Section A}
  \textbf{\bibinfo{volume}{32}}, \bibinfo{pages}{751}.

\bibitem[{\citenamefont{Shen} \emph{et~al.}(2004)\citenamefont{Shen, Ronning,
  Lu, Lee, Ingle, Meevasana, Baumberger, Damascelli, Armitage, Miller, Kohsaka,
  Azuma} \emph{et~al.}}]{Shen04a}
\bibinfo{author}{\bibnamefont{Shen}, \bibfnamefont{K.~M.}},
  \bibinfo{author}{\bibfnamefont{F.}~\bibnamefont{Ronning}},
  \bibinfo{author}{\bibfnamefont{D.~H.} \bibnamefont{Lu}},
  \bibinfo{author}{\bibfnamefont{W.~S.} \bibnamefont{Lee}},
  \bibinfo{author}{\bibfnamefont{N.~J.~C.} \bibnamefont{Ingle}},
  \bibinfo{author}{\bibfnamefont{W.}~\bibnamefont{Meevasana}},
  \bibinfo{author}{\bibfnamefont{F.}~\bibnamefont{Baumberger}},
  \bibinfo{author}{\bibfnamefont{A.}~\bibnamefont{Damascelli}},
  \bibinfo{author}{\bibfnamefont{N.~P.} \bibnamefont{Armitage}},
  \bibinfo{author}{\bibfnamefont{L.~L.} \bibnamefont{Miller}},
  \bibinfo{author}{\bibfnamefont{Y.}~\bibnamefont{Kohsaka}},
  \bibinfo{author}{\bibfnamefont{M.}~\bibnamefont{Azuma}}, \emph{et~al.},
  \bibinfo{year}{2004}, \bibinfo{journal}{Phys.\ Rev.\ Lett.}
  \textbf{\bibinfo{volume}{93}}(\bibinfo{number}{26}), \bibinfo{eid}{267002}.

\bibitem[{\citenamefont{Shen} \emph{et~al.}(1993)\citenamefont{Shen, Dessau,
  Wells, King, Spicer, Arko, Marshall, Lombardo, Kapitulnik, Dickinson,
  Doniach, DiCarlo} \emph{et~al.}}]{Shen93a}
\bibinfo{author}{\bibnamefont{Shen}, \bibfnamefont{Z.-X.}},
  \bibinfo{author}{\bibfnamefont{D.~S.} \bibnamefont{Dessau}},
  \bibinfo{author}{\bibfnamefont{B.~O.} \bibnamefont{Wells}},
  \bibinfo{author}{\bibfnamefont{D.~M.} \bibnamefont{King}},
  \bibinfo{author}{\bibfnamefont{W.~E.} \bibnamefont{Spicer}},
  \bibinfo{author}{\bibfnamefont{A.~J.} \bibnamefont{Arko}},
  \bibinfo{author}{\bibfnamefont{D.}~\bibnamefont{Marshall}},
  \bibinfo{author}{\bibfnamefont{L.~W.} \bibnamefont{Lombardo}},
  \bibinfo{author}{\bibfnamefont{A.}~\bibnamefont{Kapitulnik}},
  \bibinfo{author}{\bibfnamefont{P.}~\bibnamefont{Dickinson}},
  \bibinfo{author}{\bibfnamefont{S.}~\bibnamefont{Doniach}},
  \bibinfo{author}{\bibfnamefont{J.}~\bibnamefont{DiCarlo}}, \emph{et~al.},
  \bibinfo{year}{1993}, \bibinfo{journal}{Phys.\ Rev.\ Lett.}
  \textbf{\bibinfo{volume}{70}}(\bibinfo{number}{10}), \bibinfo{pages}{1553}.

\bibitem[{\citenamefont{Shengelaya}
  \emph{et~al.}(2005)\citenamefont{Shengelaya, Khasanov, Eshchenko, Castro,
  Savic, Park, Kim, Lee, Muller, and Keller}}]{shengelaya05a}
\bibinfo{author}{\bibnamefont{Shengelaya}, \bibfnamefont{A.}},
  \bibinfo{author}{\bibfnamefont{R.}~\bibnamefont{Khasanov}},
  \bibinfo{author}{\bibfnamefont{D.~G.} \bibnamefont{Eshchenko}},
  \bibinfo{author}{\bibfnamefont{D.~D.} \bibnamefont{Castro}},
  \bibinfo{author}{\bibfnamefont{I.~M.} \bibnamefont{Savic}},
  \bibinfo{author}{\bibfnamefont{M.~S.} \bibnamefont{Park}},
  \bibinfo{author}{\bibfnamefont{K.~H.} \bibnamefont{Kim}},
  \bibinfo{author}{\bibfnamefont{S.-I.} \bibnamefont{Lee}},
  \bibinfo{author}{\bibfnamefont{K.~A.} \bibnamefont{Muller}}, and
  \bibinfo{author}{\bibfnamefont{H.}~\bibnamefont{Keller}},
  \bibinfo{year}{2005}, \bibinfo{journal}{Phys.\ Rev.\ Lett.}
  \textbf{\bibinfo{volume}{94}}(\bibinfo{number}{12}), \bibinfo{eid}{127001}.

\bibitem[{\citenamefont{Shibauchi} \emph{et~al.}(2001)\citenamefont{Shibauchi,
  Krusin-Elbaum, Li, Maley, and Kes}}]{Shibauchi01a}
\bibinfo{author}{\bibnamefont{Shibauchi}, \bibfnamefont{T.}},
  \bibinfo{author}{\bibfnamefont{L.}~\bibnamefont{Krusin-Elbaum}},
  \bibinfo{author}{\bibfnamefont{M.}~\bibnamefont{Li}},
  \bibinfo{author}{\bibfnamefont{M.~P.} \bibnamefont{Maley}}, and
  \bibinfo{author}{\bibfnamefont{P.~H.} \bibnamefont{Kes}},
  \bibinfo{year}{2001}, \bibinfo{journal}{Phys.\ Rev.\ Lett.}
  \textbf{\bibinfo{volume}{86}}(\bibinfo{number}{25}), \bibinfo{pages}{5763}.

\bibitem[{\citenamefont{Siegrist} \emph{et~al.}(1988)\citenamefont{Siegrist,
  Zahurak, Murphy, , and Roth}}]{Siegrist88a}
\bibinfo{author}{\bibnamefont{Siegrist}, \bibfnamefont{T.}},
  \bibinfo{author}{\bibfnamefont{S.~M.} \bibnamefont{Zahurak}},
  \bibinfo{author}{\bibfnamefont{D.}~\bibnamefont{Murphy}}, , and
  \bibinfo{author}{\bibfnamefont{R.~S.} \bibnamefont{Roth}},
  \bibinfo{year}{1988}, \bibinfo{journal}{Nature}
  \textbf{\bibinfo{volume}{334}}, \bibinfo{eid}{231}.

\bibitem[{\citenamefont{Singer} \emph{et~al.}(1999)\citenamefont{Singer, Hunt,
  Cederstr\"om, and Imai}}]{Singer99a}
\bibinfo{author}{\bibnamefont{Singer}, \bibfnamefont{P.~M.}},
  \bibinfo{author}{\bibfnamefont{A.~W.} \bibnamefont{Hunt}},
  \bibinfo{author}{\bibfnamefont{A.~F.} \bibnamefont{Cederstr\"om}}, and
  \bibinfo{author}{\bibfnamefont{T.}~\bibnamefont{Imai}}, \bibinfo{year}{1999},
  \bibinfo{journal}{Phys.\ Rev.\ B}
  \textbf{\bibinfo{volume}{60}}(\bibinfo{number}{22}), \bibinfo{pages}{15345}.

\bibitem[{\citenamefont{Singh and Ghosh}(2002)}]{Singh02a}
\bibinfo{author}{\bibnamefont{Singh}, \bibfnamefont{A.}}, and
  \bibinfo{author}{\bibfnamefont{H.}~\bibnamefont{Ghosh}},
  \bibinfo{year}{2002}, \bibinfo{journal}{Phys.\ Rev.\ B}
  \textbf{\bibinfo{volume}{65}}(\bibinfo{number}{13}), \bibinfo{pages}{134414}.

\bibitem[{\citenamefont{Singh} \emph{et~al.}(1989)\citenamefont{Singh, Fleury,
  Lyons, and Sulewski}}]{Singh89a}
\bibinfo{author}{\bibnamefont{Singh}, \bibfnamefont{R.~R.~P.}},
  \bibinfo{author}{\bibfnamefont{P.~A.} \bibnamefont{Fleury}},
  \bibinfo{author}{\bibfnamefont{K.~B.} \bibnamefont{Lyons}}, and
  \bibinfo{author}{\bibfnamefont{P.~E.} \bibnamefont{Sulewski}},
  \bibinfo{year}{1989}, \bibinfo{journal}{Phys.\ Rev.\ Lett.}
  \textbf{\bibinfo{volume}{62}}(\bibinfo{number}{23}), \bibinfo{pages}{2736}.

\bibitem[{\citenamefont{Singley} \emph{et~al.}(2001)\citenamefont{Singley,
  Basov, Kurahashi, Uefuji, and Yamada}}]{Singley01a}
\bibinfo{author}{\bibnamefont{Singley}, \bibfnamefont{E.~J.}},
  \bibinfo{author}{\bibfnamefont{D.~N.} \bibnamefont{Basov}},
  \bibinfo{author}{\bibfnamefont{K.}~\bibnamefont{Kurahashi}},
  \bibinfo{author}{\bibfnamefont{T.}~\bibnamefont{Uefuji}}, and
  \bibinfo{author}{\bibfnamefont{K.}~\bibnamefont{Yamada}},
  \bibinfo{year}{2001}, \bibinfo{journal}{Phys.\ Rev.\ B}
  \textbf{\bibinfo{volume}{64}}(\bibinfo{number}{22}), \bibinfo{eid}{224503}.

\bibitem[{\citenamefont{Skanthakumar}
  \emph{et~al.}(1991)\citenamefont{Skanthakumar, Lynn, Peng, and
  Li}}]{Skanthakumar93a}
\bibinfo{author}{\bibnamefont{Skanthakumar}, \bibfnamefont{S.}},
  \bibinfo{author}{\bibfnamefont{J.~W.} \bibnamefont{Lynn}},
  \bibinfo{author}{\bibfnamefont{J.~L.} \bibnamefont{Peng}}, and
  \bibinfo{author}{\bibfnamefont{Z.~Y.} \bibnamefont{Li}},
  \bibinfo{year}{1991}, \bibinfo{journal}{J.\ Appl.\ Phys.}
  \textbf{\bibinfo{volume}{69}}(\bibinfo{number}{8}), \bibinfo{pages}{4866}.

\bibitem[{\citenamefont{Skanthakumar}
  \emph{et~al.}(1993)\citenamefont{Skanthakumar, Lynn, Peng, and
  Li}}]{Skanthakumar93b}
\bibinfo{author}{\bibnamefont{Skanthakumar}, \bibfnamefont{S.}},
  \bibinfo{author}{\bibfnamefont{J.~W.} \bibnamefont{Lynn}},
  \bibinfo{author}{\bibfnamefont{J.~L.} \bibnamefont{Peng}}, and
  \bibinfo{author}{\bibfnamefont{Z.~Y.} \bibnamefont{Li}},
  \bibinfo{year}{1993}, \bibinfo{journal}{Phys.\ Rev.\ B}
  \textbf{\bibinfo{volume}{47}}(\bibinfo{number}{10}), \bibinfo{pages}{6173}.

\bibitem[{\citenamefont{Skanthakumar}
  \emph{et~al.}(1995)\citenamefont{Skanthakumar, Lynn, and
  Sumarlin}}]{Skanthakumar95a}
\bibinfo{author}{\bibnamefont{Skanthakumar}, \bibfnamefont{S.}},
  \bibinfo{author}{\bibfnamefont{J.~W.} \bibnamefont{Lynn}}, and
  \bibinfo{author}{\bibfnamefont{I.~W.} \bibnamefont{Sumarlin}},
  \bibinfo{year}{1995}, \bibinfo{journal}{Phys. Rev. Lett.}
  \textbf{\bibinfo{volume}{74}}(\bibinfo{number}{14}), \bibinfo{pages}{2842}.

\bibitem[{\citenamefont{Skelton} \emph{et~al.}(1994)\citenamefont{Skelton,
  Drews, Osofsky, Qadi, Hu, Vanderah, Peng, and Greene}}]{Skelton94a}
\bibinfo{author}{\bibnamefont{Skelton}, \bibfnamefont{E.~F.}},
  \bibinfo{author}{\bibfnamefont{A.~R.} \bibnamefont{Drews}},
  \bibinfo{author}{\bibfnamefont{M.~S.} \bibnamefont{Osofsky}},
  \bibinfo{author}{\bibfnamefont{S.~B.} \bibnamefont{Qadi}},
  \bibinfo{author}{\bibfnamefont{J.~Z.} \bibnamefont{Hu}},
  \bibinfo{author}{\bibfnamefont{T.~A.} \bibnamefont{Vanderah}},
  \bibinfo{author}{\bibfnamefont{J.~L.} \bibnamefont{Peng}}, and
  \bibinfo{author}{\bibfnamefont{R.~L.} \bibnamefont{Greene}},
  \bibinfo{year}{1994}, \bibinfo{journal}{Science}
  \textbf{\bibinfo{volume}{263}}, \bibinfo{pages}{1416}.

\bibitem[{\citenamefont{Skinta} \emph{et~al.}(2002)\citenamefont{Skinta, Kim,
  Lemberger, Greibe, and Naito}}]{Skinta02a}
\bibinfo{author}{\bibnamefont{Skinta}, \bibfnamefont{J.~A.}},
  \bibinfo{author}{\bibfnamefont{M.-S.} \bibnamefont{Kim}},
  \bibinfo{author}{\bibfnamefont{T.~R.} \bibnamefont{Lemberger}},
  \bibinfo{author}{\bibfnamefont{T.}~\bibnamefont{Greibe}}, and
  \bibinfo{author}{\bibfnamefont{M.}~\bibnamefont{Naito}},
  \bibinfo{year}{2002}, \bibinfo{journal}{Phys.\ Rev.\ Lett.}
  \textbf{\bibinfo{volume}{88}}(\bibinfo{number}{20}), \bibinfo{eid}{207005}.

\bibitem[{\citenamefont{Smith} \emph{et~al.}(2005)\citenamefont{Smith,
  Paglione, Walker, and Taillefer}}]{Smith05a}
\bibinfo{author}{\bibnamefont{Smith}, \bibfnamefont{M.~F.}},
  \bibinfo{author}{\bibfnamefont{J.}~\bibnamefont{Paglione}},
  \bibinfo{author}{\bibfnamefont{M.~B.} \bibnamefont{Walker}}, and
  \bibinfo{author}{\bibfnamefont{L.}~\bibnamefont{Taillefer}},
  \bibinfo{year}{2005}, \bibinfo{journal}{Phys.\ Rev.\ B}
  \textbf{\bibinfo{volume}{71}}(\bibinfo{number}{1}), \bibinfo{eid}{014506}.

\bibitem[{\citenamefont{Smith} \emph{et~al.}(1991)\citenamefont{Smith,
  Manthiram, Zhou, Goodenough, and Markert}}]{Smith91a}
\bibinfo{author}{\bibnamefont{Smith}, \bibfnamefont{M.~G.}},
  \bibinfo{author}{\bibfnamefont{A.}~\bibnamefont{Manthiram}},
  \bibinfo{author}{\bibfnamefont{J.}~\bibnamefont{Zhou}},
  \bibinfo{author}{\bibfnamefont{J.~B.} \bibnamefont{Goodenough}}, and
  \bibinfo{author}{\bibfnamefont{J.~T.} \bibnamefont{Markert}},
  \bibinfo{year}{1991}, \bibinfo{journal}{Nature}
  \textbf{\bibinfo{volume}{351}}(\bibinfo{number}{6327}), \bibinfo{pages}{549}.

\bibitem[{\citenamefont{Snezhko} \emph{et~al.}(2004)\citenamefont{Snezhko,
  Prozorov, Lawrie, Giannetta, Gauthier, Renaud, and Fournier}}]{Snezhko04a}
\bibinfo{author}{\bibnamefont{Snezhko}, \bibfnamefont{A.}},
  \bibinfo{author}{\bibfnamefont{R.}~\bibnamefont{Prozorov}},
  \bibinfo{author}{\bibfnamefont{D.~D.} \bibnamefont{Lawrie}},
  \bibinfo{author}{\bibfnamefont{R.}~\bibnamefont{Giannetta}},
  \bibinfo{author}{\bibfnamefont{J.}~\bibnamefont{Gauthier}},
  \bibinfo{author}{\bibfnamefont{J.}~\bibnamefont{Renaud}}, and
  \bibinfo{author}{\bibfnamefont{P.}~\bibnamefont{Fournier}},
  \bibinfo{year}{2004}, \bibinfo{journal}{Phys.\ Rev.\ Lett.}
  \textbf{\bibinfo{volume}{92}}, \bibinfo{pages}{157005}.

\bibitem[{\citenamefont{{Sondheimer}}(1948)}]{Sondheimer48a}
\bibinfo{author}{\bibnamefont{{Sondheimer}}, \bibfnamefont{E.~H.}},
  \bibinfo{year}{1948}, \bibinfo{journal}{Royal Society of London Proceedings
  Series A} \textbf{\bibinfo{volume}{193}}, \bibinfo{pages}{484}.

\bibitem[{\citenamefont{Sonier} \emph{et~al.}(2000)\citenamefont{Sonier,
  Brewer, and Kiefl}}]{Sonier00a}
\bibinfo{author}{\bibnamefont{Sonier}, \bibfnamefont{J.~E.}},
  \bibinfo{author}{\bibfnamefont{J.~H.} \bibnamefont{Brewer}}, and
  \bibinfo{author}{\bibfnamefont{R.~F.} \bibnamefont{Kiefl}},
  \bibinfo{year}{2000}, \bibinfo{journal}{Rev.\ Mod.\ Phys.}
  \textbf{\bibinfo{volume}{72}}(\bibinfo{number}{3}), \bibinfo{pages}{769}.

\bibitem[{\citenamefont{Sonier} \emph{et~al.}(2003)\citenamefont{Sonier, Poon,
  Luke, Kyriakou, Miller, Liang, Wiebe, Fournier, and Greene}}]{Sonier03a}
\bibinfo{author}{\bibnamefont{Sonier}, \bibfnamefont{J.~E.}},
  \bibinfo{author}{\bibfnamefont{K.~F.} \bibnamefont{Poon}},
  \bibinfo{author}{\bibfnamefont{G.~M.} \bibnamefont{Luke}},
  \bibinfo{author}{\bibfnamefont{P.}~\bibnamefont{Kyriakou}},
  \bibinfo{author}{\bibfnamefont{R.~I.} \bibnamefont{Miller}},
  \bibinfo{author}{\bibfnamefont{R.}~\bibnamefont{Liang}},
  \bibinfo{author}{\bibfnamefont{C.~R.} \bibnamefont{Wiebe}},
  \bibinfo{author}{\bibfnamefont{P.}~\bibnamefont{Fournier}}, and
  \bibinfo{author}{\bibfnamefont{R.~L.} \bibnamefont{Greene}},
  \bibinfo{year}{2003}, \bibinfo{journal}{Phys.\ Rev.\ Lett.}
  \textbf{\bibinfo{volume}{91}}(\bibinfo{number}{14}), \bibinfo{pages}{147002}.

\bibitem[{\citenamefont{Sooryakumar and Klein}(1980)}]{Sooryakumar80a}
\bibinfo{author}{\bibnamefont{Sooryakumar}, \bibfnamefont{R.}}, and
  \bibinfo{author}{\bibfnamefont{M.~V.} \bibnamefont{Klein}},
  \bibinfo{year}{1980}, \bibinfo{journal}{Phys.\ Rev.\ Lett.}
  \textbf{\bibinfo{volume}{45}}(\bibinfo{number}{8}), \bibinfo{pages}{660}.

\bibitem[{\citenamefont{Sooryakumar and Klein}(1981)}]{Sooryakumar81a}
\bibinfo{author}{\bibnamefont{Sooryakumar}, \bibfnamefont{R.}}, and
  \bibinfo{author}{\bibfnamefont{M.~V.} \bibnamefont{Klein}},
  \bibinfo{year}{1981}, \bibinfo{journal}{Phys.\ Rev.\ B}
  \textbf{\bibinfo{volume}{23}}(\bibinfo{number}{7}), \bibinfo{pages}{3213}.

\bibitem[{\citenamefont{Stadlober} \emph{et~al.}(1995)\citenamefont{Stadlober,
  Krug, Nemetschek, Hackl, Cobb, and Markert}}]{Stadlober95a}
\bibinfo{author}{\bibnamefont{Stadlober}, \bibfnamefont{B.}},
  \bibinfo{author}{\bibfnamefont{G.}~\bibnamefont{Krug}},
  \bibinfo{author}{\bibfnamefont{R.}~\bibnamefont{Nemetschek}},
  \bibinfo{author}{\bibfnamefont{R.}~\bibnamefont{Hackl}},
  \bibinfo{author}{\bibfnamefont{J.~L.} \bibnamefont{Cobb}}, and
  \bibinfo{author}{\bibfnamefont{J.~T.} \bibnamefont{Markert}},
  \bibinfo{year}{1995}, \bibinfo{journal}{Phys.\ Rev.\ Lett.}
  \textbf{\bibinfo{volume}{74}}(\bibinfo{number}{24}), \bibinfo{pages}{4911}.

\bibitem[{\citenamefont{Steeneken} \emph{et~al.}(2003)\citenamefont{Steeneken,
  Tjeng, Sawatzky, Tanaka, Tjernberg, Ghiringhelli, Brookes, Nugroho, and
  Menovsky}}]{Steeneken03a}
\bibinfo{author}{\bibnamefont{Steeneken}, \bibfnamefont{P.~G.}},
  \bibinfo{author}{\bibfnamefont{L.~H.} \bibnamefont{Tjeng}},
  \bibinfo{author}{\bibfnamefont{G.~A.} \bibnamefont{Sawatzky}},
  \bibinfo{author}{\bibfnamefont{A.}~\bibnamefont{Tanaka}},
  \bibinfo{author}{\bibfnamefont{O.}~\bibnamefont{Tjernberg}},
  \bibinfo{author}{\bibfnamefont{G.}~\bibnamefont{Ghiringhelli}},
  \bibinfo{author}{\bibfnamefont{N.~B.} \bibnamefont{Brookes}},
  \bibinfo{author}{\bibfnamefont{A.~A.} \bibnamefont{Nugroho}}, and
  \bibinfo{author}{\bibfnamefont{A.~A.} \bibnamefont{Menovsky}},
  \bibinfo{year}{2003}, \bibinfo{journal}{Phys.\ Rev.\ Lett.}
  \textbf{\bibinfo{volume}{90}}(\bibinfo{number}{24}), \bibinfo{pages}{247005}.

\bibitem[{\citenamefont{Stepanov} \emph{et~al.}(1993)\citenamefont{Stepanov,
  Wyder, Chattopadhyay, Brown, Fillion, Vitebsky, Deville, Gaillard, Barilo,
  and Zhigunov}}]{Stepanov93a}
\bibinfo{author}{\bibnamefont{Stepanov}, \bibfnamefont{A.~A.}},
  \bibinfo{author}{\bibfnamefont{P.}~\bibnamefont{Wyder}},
  \bibinfo{author}{\bibfnamefont{T.}~\bibnamefont{Chattopadhyay}},
  \bibinfo{author}{\bibfnamefont{P.~J.} \bibnamefont{Brown}},
  \bibinfo{author}{\bibfnamefont{G.}~\bibnamefont{Fillion}},
  \bibinfo{author}{\bibfnamefont{I.~M.} \bibnamefont{Vitebsky}},
  \bibinfo{author}{\bibfnamefont{A.}~\bibnamefont{Deville}},
  \bibinfo{author}{\bibfnamefont{B.}~\bibnamefont{Gaillard}},
  \bibinfo{author}{\bibfnamefont{S.~N.} \bibnamefont{Barilo}}, and
  \bibinfo{author}{\bibfnamefont{D.~I.} \bibnamefont{Zhigunov}},
  \bibinfo{year}{1993}, \bibinfo{journal}{Phys.\ Rev.\ B}
  \textbf{\bibinfo{volume}{48}}(\bibinfo{number}{17}), \bibinfo{pages}{12979}.

\bibitem[{\citenamefont{Sugai} \emph{et~al.}(1989)\citenamefont{Sugai,
  Kobayashi, and Akimitsu}}]{Sugai89a}
\bibinfo{author}{\bibnamefont{Sugai}, \bibfnamefont{S.}},
  \bibinfo{author}{\bibfnamefont{T.}~\bibnamefont{Kobayashi}}, and
  \bibinfo{author}{\bibfnamefont{J.}~\bibnamefont{Akimitsu}},
  \bibinfo{year}{1989}, \bibinfo{journal}{Phys.\ Rev.\ B}
  \textbf{\bibinfo{volume}{40}}(\bibinfo{number}{4}), \bibinfo{pages}{2686}.

\bibitem[{\citenamefont{Sulewski} \emph{et~al.}(1990)\citenamefont{Sulewski,
  Fleury, Lyons, Cheong, and Fisk}}]{Sulewski90a}
\bibinfo{author}{\bibnamefont{Sulewski}, \bibfnamefont{P.~E.}},
  \bibinfo{author}{\bibfnamefont{P.~A.} \bibnamefont{Fleury}},
  \bibinfo{author}{\bibfnamefont{K.~B.} \bibnamefont{Lyons}},
  \bibinfo{author}{\bibfnamefont{S.-W.} \bibnamefont{Cheong}}, and
  \bibinfo{author}{\bibfnamefont{Z.}~\bibnamefont{Fisk}}, \bibinfo{year}{1990},
  \bibinfo{journal}{Phys.\ Rev.\ B}
  \textbf{\bibinfo{volume}{41}}(\bibinfo{number}{1}), \bibinfo{pages}{225}.

\bibitem[{\citenamefont{Sumarlin} \emph{et~al.}(1995)\citenamefont{Sumarlin,
  Lynn, Chattopadhyay, Barilo, Zhigunov, and Peng}}]{Sumarlin95a}
\bibinfo{author}{\bibnamefont{Sumarlin}, \bibfnamefont{I.~W.}},
  \bibinfo{author}{\bibfnamefont{J.~W.} \bibnamefont{Lynn}},
  \bibinfo{author}{\bibfnamefont{T.}~\bibnamefont{Chattopadhyay}},
  \bibinfo{author}{\bibfnamefont{S.~N.} \bibnamefont{Barilo}},
  \bibinfo{author}{\bibfnamefont{D.~I.} \bibnamefont{Zhigunov}}, and
  \bibinfo{author}{\bibfnamefont{J.~L.} \bibnamefont{Peng}},
  \bibinfo{year}{1995}, \bibinfo{journal}{Phys.\ Rev.\ B}
  \textbf{\bibinfo{volume}{51}}(\bibinfo{number}{9}), \bibinfo{pages}{5824}.

\bibitem[{\citenamefont{Sumarlin} \emph{et~al.}(1992)\citenamefont{Sumarlin,
  Skanthakumar, Lynn, Peng, Li, Jiang, and Greene}}]{Sumarlin92a}
\bibinfo{author}{\bibnamefont{Sumarlin}, \bibfnamefont{I.~W.}},
  \bibinfo{author}{\bibfnamefont{S.}~\bibnamefont{Skanthakumar}},
  \bibinfo{author}{\bibfnamefont{J.~W.} \bibnamefont{Lynn}},
  \bibinfo{author}{\bibfnamefont{J.~L.} \bibnamefont{Peng}},
  \bibinfo{author}{\bibfnamefont{Z.~Y.} \bibnamefont{Li}},
  \bibinfo{author}{\bibfnamefont{W.}~\bibnamefont{Jiang}}, and
  \bibinfo{author}{\bibfnamefont{R.~L.} \bibnamefont{Greene}},
  \bibinfo{year}{1992}, \bibinfo{journal}{Phys.\ Rev.\ Lett.}
  \textbf{\bibinfo{volume}{68}}(\bibinfo{number}{14}), \bibinfo{pages}{2228}.

\bibitem[{\citenamefont{Sun} \emph{et~al.}(2008)\citenamefont{Sun, Yang, Cheng,
  Wang, Xu, Ke, and Cao}}]{Sun08a}
\bibinfo{author}{\bibnamefont{Sun}, \bibfnamefont{C.}},
  \bibinfo{author}{\bibfnamefont{H.}~\bibnamefont{Yang}},
  \bibinfo{author}{\bibfnamefont{L.}~\bibnamefont{Cheng}},
  \bibinfo{author}{\bibfnamefont{J.}~\bibnamefont{Wang}},
  \bibinfo{author}{\bibfnamefont{X.}~\bibnamefont{Xu}},
  \bibinfo{author}{\bibfnamefont{S.}~\bibnamefont{Ke}}, and
  \bibinfo{author}{\bibfnamefont{L.}~\bibnamefont{Cao}}, \bibinfo{year}{2008},
  \bibinfo{journal}{Phys.\ Rev.\ B}
  \textbf{\bibinfo{volume}{78}}(\bibinfo{number}{10}), \bibinfo{pages}{104518}.

\bibitem[{\citenamefont{Sun} \emph{et~al.}(2004)\citenamefont{Sun, Kurita,
  Suzuki, Komiya, and Ando}}]{Sun04a}
\bibinfo{author}{\bibnamefont{Sun}, \bibfnamefont{X.~F.}},
  \bibinfo{author}{\bibfnamefont{Y.}~\bibnamefont{Kurita}},
  \bibinfo{author}{\bibfnamefont{T.}~\bibnamefont{Suzuki}},
  \bibinfo{author}{\bibfnamefont{S.}~\bibnamefont{Komiya}}, and
  \bibinfo{author}{\bibfnamefont{Y.}~\bibnamefont{Ando}}, \bibinfo{year}{2004},
  \bibinfo{journal}{Phys.\ Rev.\ Lett.}
  \textbf{\bibinfo{volume}{92}}(\bibinfo{number}{4}), \bibinfo{eid}{047001}.

\bibitem[{\citenamefont{Suzuki} \emph{et~al.}(1990)\citenamefont{Suzuki,
  Kishio, Hasegawa, and Kitazawa}}]{Suzuki90a}
\bibinfo{author}{\bibnamefont{Suzuki}, \bibfnamefont{K.}},
  \bibinfo{author}{\bibfnamefont{K.}~\bibnamefont{Kishio}},
  \bibinfo{author}{\bibfnamefont{T.}~\bibnamefont{Hasegawa}}, and
  \bibinfo{author}{\bibfnamefont{K.}~\bibnamefont{Kitazawa}},
  \bibinfo{year}{1990}, \bibinfo{journal}{Physica C}
  \textbf{\bibinfo{volume}{166}}, \bibinfo{pages}{357}.

\bibitem[{\citenamefont{Taguchi} \emph{et~al.}(2005)\citenamefont{Taguchi,
  Chainani, Horiba, Takata, Yabashi, Tamasaku, Nishino, Miwa, Ishikawa,
  Takeuchi, Yamamoto, Matsunami} \emph{et~al.}}]{Taguchi05a}
\bibinfo{author}{\bibnamefont{Taguchi}, \bibfnamefont{M.}},
  \bibinfo{author}{\bibfnamefont{A.}~\bibnamefont{Chainani}},
  \bibinfo{author}{\bibfnamefont{K.}~\bibnamefont{Horiba}},
  \bibinfo{author}{\bibfnamefont{Y.}~\bibnamefont{Takata}},
  \bibinfo{author}{\bibfnamefont{M.}~\bibnamefont{Yabashi}},
  \bibinfo{author}{\bibfnamefont{K.}~\bibnamefont{Tamasaku}},
  \bibinfo{author}{\bibfnamefont{Y.}~\bibnamefont{Nishino}},
  \bibinfo{author}{\bibfnamefont{D.}~\bibnamefont{Miwa}},
  \bibinfo{author}{\bibfnamefont{T.}~\bibnamefont{Ishikawa}},
  \bibinfo{author}{\bibfnamefont{T.}~\bibnamefont{Takeuchi}},
  \bibinfo{author}{\bibfnamefont{K.}~\bibnamefont{Yamamoto}},
  \bibinfo{author}{\bibfnamefont{M.}~\bibnamefont{Matsunami}}, \emph{et~al.},
  \bibinfo{year}{2005}, \bibinfo{journal}{Phys.\ Rev.\ Lett.}
  \textbf{\bibinfo{volume}{95}}(\bibinfo{number}{17}), \bibinfo{eid}{177002}.

\bibitem[{\citenamefont{Taillefer} \emph{et~al.}(1997)\citenamefont{Taillefer,
  Lussier, Gagnon, Behnia, and Aubin}}]{Taillefer97a}
\bibinfo{author}{\bibnamefont{Taillefer}, \bibfnamefont{L.}},
  \bibinfo{author}{\bibfnamefont{B.}~\bibnamefont{Lussier}},
  \bibinfo{author}{\bibfnamefont{R.}~\bibnamefont{Gagnon}},
  \bibinfo{author}{\bibfnamefont{K.}~\bibnamefont{Behnia}}, and
  \bibinfo{author}{\bibfnamefont{H.}~\bibnamefont{Aubin}},
  \bibinfo{year}{1997}, \bibinfo{journal}{Phys.\ Rev.\ Lett.}
  \textbf{\bibinfo{volume}{79}}(\bibinfo{number}{3}), \bibinfo{pages}{483}.

\bibitem[{\citenamefont{Takagi} \emph{et~al.}(1989)\citenamefont{Takagi,
  Uchida, and Tokura}}]{Takagi89a}
\bibinfo{author}{\bibnamefont{Takagi}, \bibfnamefont{H.}},
  \bibinfo{author}{\bibfnamefont{S.}~\bibnamefont{Uchida}}, and
  \bibinfo{author}{\bibfnamefont{Y.}~\bibnamefont{Tokura}},
  \bibinfo{year}{1989}, \bibinfo{journal}{Phys.\ Rev.\ Lett.}
  \textbf{\bibinfo{volume}{62}}, \bibinfo{pages}{1197}.

\bibitem[{\citenamefont{Takayama-Muromachi}
  \emph{et~al.}(1989)\citenamefont{Takayama-Muromachi, Izumi, Uchida, Kato, and
  Asano}}]{Takayama89a}
\bibinfo{author}{\bibnamefont{Takayama-Muromachi}, \bibfnamefont{E.}},
  \bibinfo{author}{\bibfnamefont{F.}~\bibnamefont{Izumi}},
  \bibinfo{author}{\bibfnamefont{Y.}~\bibnamefont{Uchida}},
  \bibinfo{author}{\bibfnamefont{K.}~\bibnamefont{Kato}}, and
  \bibinfo{author}{\bibfnamefont{H.}~\bibnamefont{Asano}},
  \bibinfo{year}{1989}, \bibinfo{journal}{Physica C}
  \textbf{\bibinfo{volume}{159}}, \bibinfo{pages}{634}.

\bibitem[{\citenamefont{Tan} \emph{et~al.}(1990)\citenamefont{Tan, Budnick,
  Bouldin, Woicik, Cheong, Cooper, Espinosa, and Fisk}}]{Tan90a}
\bibinfo{author}{\bibnamefont{Tan}, \bibfnamefont{Z.}},
  \bibinfo{author}{\bibfnamefont{J.~I.} \bibnamefont{Budnick}},
  \bibinfo{author}{\bibfnamefont{C.~E.} \bibnamefont{Bouldin}},
  \bibinfo{author}{\bibfnamefont{J.~C.} \bibnamefont{Woicik}},
  \bibinfo{author}{\bibfnamefont{S.-W.} \bibnamefont{Cheong}},
  \bibinfo{author}{\bibfnamefont{A.~S.} \bibnamefont{Cooper}},
  \bibinfo{author}{\bibfnamefont{G.~P.} \bibnamefont{Espinosa}}, and
  \bibinfo{author}{\bibfnamefont{Z.}~\bibnamefont{Fisk}}, \bibinfo{year}{1990},
  \bibinfo{journal}{Phys.\ Rev.\ B}
  \textbf{\bibinfo{volume}{42}}(\bibinfo{number}{1}), \bibinfo{pages}{1037}.

\bibitem[{\citenamefont{Tanaka} \emph{et~al.}(1991)\citenamefont{Tanaka,
  Watanabe, Komai, and Kojima}}]{Tanaka91a}
\bibinfo{author}{\bibnamefont{Tanaka}, \bibfnamefont{I.}},
  \bibinfo{author}{\bibfnamefont{T.}~\bibnamefont{Watanabe}},
  \bibinfo{author}{\bibfnamefont{N.}~\bibnamefont{Komai}}, and
  \bibinfo{author}{\bibfnamefont{H.}~\bibnamefont{Kojima}},
  \bibinfo{year}{1991}, \bibinfo{journal}{Physica C}
  \textbf{\bibinfo{volume}{185-189}}(\bibinfo{number}{Part 1}),
  \bibinfo{pages}{437}.

\bibitem[{\citenamefont{Tanaka and Kashiwaya}(1995)}]{Tanaka95a}
\bibinfo{author}{\bibnamefont{Tanaka}, \bibfnamefont{Y.}}, and
  \bibinfo{author}{\bibfnamefont{S.}~\bibnamefont{Kashiwaya}},
  \bibinfo{year}{1995}, \bibinfo{journal}{Phys.\ Rev.\ Lett.}
  \textbf{\bibinfo{volume}{74}}(\bibinfo{number}{17}), \bibinfo{pages}{3451}.

\bibitem[{\citenamefont{Tanatar} \emph{et~al.}(2007)\citenamefont{Tanatar,
  Paglione, Petrovic, and Taillefer}}]{Tantar07a}
\bibinfo{author}{\bibnamefont{Tanatar}, \bibfnamefont{M.~A.}},
  \bibinfo{author}{\bibfnamefont{J.}~\bibnamefont{Paglione}},
  \bibinfo{author}{\bibfnamefont{C.}~\bibnamefont{Petrovic}}, and
  \bibinfo{author}{\bibfnamefont{L.}~\bibnamefont{Taillefer}},
  \bibinfo{year}{2007}, \bibinfo{journal}{Science}
  \textbf{\bibinfo{volume}{316}}(\bibinfo{number}{5829}),
  \bibinfo{pages}{1320}.

\bibitem[{\citenamefont{Tanda} \emph{et~al.}(1992)\citenamefont{Tanda, Ohzeki,
  and Nakayama}}]{Tanda92a}
\bibinfo{author}{\bibnamefont{Tanda}, \bibfnamefont{S.}},
  \bibinfo{author}{\bibfnamefont{S.}~\bibnamefont{Ohzeki}}, and
  \bibinfo{author}{\bibfnamefont{T.}~\bibnamefont{Nakayama}},
  \bibinfo{year}{1992}, \bibinfo{journal}{Phys.\ Rev.\ Lett.}
  \textbf{\bibinfo{volume}{69}}(\bibinfo{number}{3}), \bibinfo{pages}{530}.

\bibitem[{\citenamefont{Tarascon} \emph{et~al.}(1989)\citenamefont{Tarascon,
  Wang, Greene, Bagley, Hull, D'Egidio, Miceli, Wang, Jing, Clayhold, Brawner,
  and Ong}}]{Tarascon89a}
\bibinfo{author}{\bibnamefont{Tarascon}, \bibfnamefont{J.-M.}},
  \bibinfo{author}{\bibfnamefont{E.}~\bibnamefont{Wang}},
  \bibinfo{author}{\bibfnamefont{L.~H.} \bibnamefont{Greene}},
  \bibinfo{author}{\bibfnamefont{B.~G.} \bibnamefont{Bagley}},
  \bibinfo{author}{\bibfnamefont{G.~W.} \bibnamefont{Hull}},
  \bibinfo{author}{\bibfnamefont{S.~M.} \bibnamefont{D'Egidio}},
  \bibinfo{author}{\bibfnamefont{P.~F.} \bibnamefont{Miceli}},
  \bibinfo{author}{\bibfnamefont{Z.~Z.} \bibnamefont{Wang}},
  \bibinfo{author}{\bibfnamefont{T.~W.} \bibnamefont{Jing}},
  \bibinfo{author}{\bibfnamefont{J.}~\bibnamefont{Clayhold}},
  \bibinfo{author}{\bibfnamefont{D.}~\bibnamefont{Brawner}}, and
  \bibinfo{author}{\bibfnamefont{N.~P.} \bibnamefont{Ong}},
  \bibinfo{year}{1989}, \bibinfo{journal}{Phys.\ Rev.\ B}
  \textbf{\bibinfo{volume}{40}}(\bibinfo{number}{7}), \bibinfo{pages}{4494}.

\bibitem[{\citenamefont{Thompson} \emph{et~al.}(1989)\citenamefont{Thompson,
  Cheong, Brown, Fisk, Oseroff, Tovar, Vier, and Schultz}}]{Thompson89a}
\bibinfo{author}{\bibnamefont{Thompson}, \bibfnamefont{J.~D.}},
  \bibinfo{author}{\bibfnamefont{S.-W.} \bibnamefont{Cheong}},
  \bibinfo{author}{\bibfnamefont{S.~E.} \bibnamefont{Brown}},
  \bibinfo{author}{\bibfnamefont{Z.}~\bibnamefont{Fisk}},
  \bibinfo{author}{\bibfnamefont{S.~B.} \bibnamefont{Oseroff}},
  \bibinfo{author}{\bibfnamefont{M.}~\bibnamefont{Tovar}},
  \bibinfo{author}{\bibfnamefont{D.~C.} \bibnamefont{Vier}}, and
  \bibinfo{author}{\bibfnamefont{S.}~\bibnamefont{Schultz}},
  \bibinfo{year}{1989}, \bibinfo{journal}{Phys.\ Rev.\ B}
  \textbf{\bibinfo{volume}{39}}(\bibinfo{number}{10}), \bibinfo{pages}{6660}.

\bibitem[{\citenamefont{Thurston} \emph{et~al.}(1990)\citenamefont{Thurston,
  Matsuda, Kakurai, Yamada, Endoh, Birgeneau, Gehring, Hidaka, Kastner,
  Murakami, and Shirane}}]{Thurston90a}
\bibinfo{author}{\bibnamefont{Thurston}, \bibfnamefont{T.~R.}},
  \bibinfo{author}{\bibfnamefont{M.}~\bibnamefont{Matsuda}},
  \bibinfo{author}{\bibfnamefont{K.}~\bibnamefont{Kakurai}},
  \bibinfo{author}{\bibfnamefont{K.}~\bibnamefont{Yamada}},
  \bibinfo{author}{\bibfnamefont{Y.}~\bibnamefont{Endoh}},
  \bibinfo{author}{\bibfnamefont{R.~J.} \bibnamefont{Birgeneau}},
  \bibinfo{author}{\bibfnamefont{P.~M.} \bibnamefont{Gehring}},
  \bibinfo{author}{\bibfnamefont{Y.}~\bibnamefont{Hidaka}},
  \bibinfo{author}{\bibfnamefont{M.~A.} \bibnamefont{Kastner}},
  \bibinfo{author}{\bibfnamefont{T.}~\bibnamefont{Murakami}}, and
  \bibinfo{author}{\bibfnamefont{G.}~\bibnamefont{Shirane}},
  \bibinfo{year}{1990}, \bibinfo{journal}{Phys.\ Rev.\ Lett.}
  \textbf{\bibinfo{volume}{65}}(\bibinfo{number}{2}), \bibinfo{pages}{263}.

\bibitem[{\citenamefont{Timusk and Statt}(1999)}]{Timusk99a}
\bibinfo{author}{\bibnamefont{Timusk}, \bibfnamefont{T.}}, and
  \bibinfo{author}{\bibfnamefont{B.}~\bibnamefont{Statt}},
  \bibinfo{year}{1999}, \bibinfo{journal}{Rep.\ Prog.\ Phys.}
  \textbf{\bibinfo{volume}{62}}(\bibinfo{number}{1}), \bibinfo{pages}{61}.

\bibitem[{\citenamefont{Tinkham}(1996)}]{Tinkham96a}
\bibinfo{author}{\bibnamefont{Tinkham}, \bibfnamefont{M.}},
  \bibinfo{year}{1996}, \emph{\bibinfo{title}{Introduction to
  superconductivity}} (\bibinfo{publisher}{McGraw-Hill}),
  \bibinfo{edition}{second edition} edition.

\bibitem[{\citenamefont{Tohyama}(2004)}]{Tohyama04a}
\bibinfo{author}{\bibnamefont{Tohyama}, \bibfnamefont{T.}},
  \bibinfo{year}{2004}, \bibinfo{journal}{Phys.\ Rev.\ B}
  \textbf{\bibinfo{volume}{70}}(\bibinfo{number}{17}), \bibinfo{pages}{174517}.

\bibitem[{\citenamefont{Tohyama and Maekawa}(1990)}]{Tohyama90a}
\bibinfo{author}{\bibnamefont{Tohyama}, \bibfnamefont{T.}}, and
  \bibinfo{author}{\bibfnamefont{S.}~\bibnamefont{Maekawa}},
  \bibinfo{year}{1990}, \bibinfo{journal}{J.\ Phys.\ Soc.\ Jap.}
  \textbf{\bibinfo{volume}{59}}, \bibinfo{pages}{1760}.

\bibitem[{\citenamefont{Tohyama and Maekawa}(1994)}]{Tohyama94a}
\bibinfo{author}{\bibnamefont{Tohyama}, \bibfnamefont{T.}}, and
  \bibinfo{author}{\bibfnamefont{S.}~\bibnamefont{Maekawa}},
  \bibinfo{year}{1994}, \bibinfo{journal}{Phys.\ Rev.\ B}
  \textbf{\bibinfo{volume}{49}}(\bibinfo{number}{5}), \bibinfo{pages}{3596}.

\bibitem[{\citenamefont{Tohyama and Maekawa}(2001)}]{Tohyama01a}
\bibinfo{author}{\bibnamefont{Tohyama}, \bibfnamefont{T.}}, and
  \bibinfo{author}{\bibfnamefont{S.}~\bibnamefont{Maekawa}},
  \bibinfo{year}{2001}, \bibinfo{journal}{Phys.\ Rev.\ B}
  \textbf{\bibinfo{volume}{64}}(\bibinfo{number}{21}), \bibinfo{pages}{212505}.

\bibitem[{\citenamefont{Tokura}
  \emph{et~al.}(1989{\natexlab{a}})\citenamefont{Tokura, Fujimori, Matsubara,
  Watabe, Takagi, Uchida, Sakai, Ikeda, Okuda, and Tanaka}}]{Tokura89b}
\bibinfo{author}{\bibnamefont{Tokura}, \bibfnamefont{Y.}},
  \bibinfo{author}{\bibfnamefont{A.}~\bibnamefont{Fujimori}},
  \bibinfo{author}{\bibfnamefont{H.}~\bibnamefont{Matsubara}},
  \bibinfo{author}{\bibfnamefont{H.}~\bibnamefont{Watabe}},
  \bibinfo{author}{\bibfnamefont{H.}~\bibnamefont{Takagi}},
  \bibinfo{author}{\bibfnamefont{S.}~\bibnamefont{Uchida}},
  \bibinfo{author}{\bibfnamefont{M.}~\bibnamefont{Sakai}},
  \bibinfo{author}{\bibfnamefont{H.}~\bibnamefont{Ikeda}},
  \bibinfo{author}{\bibfnamefont{S.}~\bibnamefont{Okuda}}, and
  \bibinfo{author}{\bibfnamefont{S.}~\bibnamefont{Tanaka}},
  \bibinfo{year}{1989}{\natexlab{a}}, \bibinfo{journal}{Phys.\ Rev.\ B}
  \textbf{\bibinfo{volume}{39}}(\bibinfo{number}{13}), \bibinfo{pages}{9704}.

\bibitem[{\citenamefont{Tokura} \emph{et~al.}(1990)\citenamefont{Tokura,
  Koshihara, Arima, Takagi, Ishibashi, Ido, and Uchida}}]{Tokura90a}
\bibinfo{author}{\bibnamefont{Tokura}, \bibfnamefont{Y.}},
  \bibinfo{author}{\bibfnamefont{S.}~\bibnamefont{Koshihara}},
  \bibinfo{author}{\bibfnamefont{T.}~\bibnamefont{Arima}},
  \bibinfo{author}{\bibfnamefont{H.}~\bibnamefont{Takagi}},
  \bibinfo{author}{\bibfnamefont{S.}~\bibnamefont{Ishibashi}},
  \bibinfo{author}{\bibfnamefont{T.}~\bibnamefont{Ido}}, and
  \bibinfo{author}{\bibfnamefont{S.}~\bibnamefont{Uchida}},
  \bibinfo{year}{1990}, \bibinfo{journal}{Phys.\ Rev.\ B}
  \textbf{\bibinfo{volume}{41}}(\bibinfo{number}{16}), \bibinfo{pages}{11657}.

\bibitem[{\citenamefont{Tokura}
  \emph{et~al.}(1989{\natexlab{b}})\citenamefont{Tokura, Takagi, and
  Uchida}}]{Tokura89a}
\bibinfo{author}{\bibnamefont{Tokura}, \bibfnamefont{Y.}},
  \bibinfo{author}{\bibfnamefont{H.}~\bibnamefont{Takagi}}, and
  \bibinfo{author}{\bibfnamefont{S.}~\bibnamefont{Uchida}},
  \bibinfo{year}{1989}{\natexlab{b}}, \bibinfo{journal}{Nature}
  \textbf{\bibinfo{volume}{337}}, \bibinfo{pages}{345}.

\bibitem[{\citenamefont{Torrance and Metzger}(1989)}]{Torrance89a}
\bibinfo{author}{\bibnamefont{Torrance}, \bibfnamefont{J.~B.}}, and
  \bibinfo{author}{\bibfnamefont{R.~M.} \bibnamefont{Metzger}},
  \bibinfo{year}{1989}, \bibinfo{journal}{Phys.\ Rev.\ Lett.}
  \textbf{\bibinfo{volume}{63}}(\bibinfo{number}{14}), \bibinfo{pages}{1515}.

\bibitem[{\citenamefont{Tranquada} \emph{et~al.}(1995)\citenamefont{Tranquada,
  B.J. Sternlieb J.D.~Axe, and Uchida}}]{Tranquada95a}
\bibinfo{author}{\bibnamefont{Tranquada}, \bibfnamefont{J.}},
  \bibinfo{author}{\bibfnamefont{Y.~N.} \bibnamefont{B.J. Sternlieb J.D.~Axe}},
  and \bibinfo{author}{\bibfnamefont{S.}~\bibnamefont{Uchida}},
  \bibinfo{year}{1995}, \bibinfo{journal}{Nature}
  \textbf{\bibinfo{volume}{375}}, \bibinfo{pages}{561}.

\bibitem[{\citenamefont{Tranquada} \emph{et~al.}(1989)\citenamefont{Tranquada,
  Heald, Moodenbaugh, Liang, and Croft}}]{Tranquada89a}
\bibinfo{author}{\bibnamefont{Tranquada}, \bibfnamefont{J.}},
  \bibinfo{author}{\bibfnamefont{S.}~\bibnamefont{Heald}},
  \bibinfo{author}{\bibfnamefont{A.}~\bibnamefont{Moodenbaugh}},
  \bibinfo{author}{\bibfnamefont{G.}~\bibnamefont{Liang}}, and
  \bibinfo{author}{\bibfnamefont{M.}~\bibnamefont{Croft}},
  \bibinfo{year}{1989}, \bibinfo{journal}{Nature}
  \textbf{\bibinfo{volume}{337}}, \bibinfo{pages}{720}.

\bibitem[{\citenamefont{Tranquada}
  \emph{et~al.}(2004{\natexlab{a}})\citenamefont{Tranquada, Lee, Yamada, Lee,
  Regnault, and Ronnow}}]{Tranquada04b}
\bibinfo{author}{\bibnamefont{Tranquada}, \bibfnamefont{J.~M.}},
  \bibinfo{author}{\bibfnamefont{C.~H.} \bibnamefont{Lee}},
  \bibinfo{author}{\bibfnamefont{K.}~\bibnamefont{Yamada}},
  \bibinfo{author}{\bibfnamefont{Y.~S.} \bibnamefont{Lee}},
  \bibinfo{author}{\bibfnamefont{L.~P.} \bibnamefont{Regnault}}, and
  \bibinfo{author}{\bibfnamefont{H.~M.} \bibnamefont{Ronnow}},
  \bibinfo{year}{2004}{\natexlab{a}}, \bibinfo{journal}{Phys.\ Rev.\ B}
  \textbf{\bibinfo{volume}{69}}(\bibinfo{number}{17}), \bibinfo{eid}{174507}.

\bibitem[{\citenamefont{Tranquada}
  \emph{et~al.}(2004{\natexlab{b}})\citenamefont{Tranquada, Woo, Perring, Goka,
  Gu, Xu, Fujita, and Yamada}}]{tranquada04a}
\bibinfo{author}{\bibnamefont{Tranquada}, \bibfnamefont{J.~M.}},
  \bibinfo{author}{\bibfnamefont{H.}~\bibnamefont{Woo}},
  \bibinfo{author}{\bibfnamefont{T.~G.} \bibnamefont{Perring}},
  \bibinfo{author}{\bibfnamefont{H.}~\bibnamefont{Goka}},
  \bibinfo{author}{\bibfnamefont{G.~D.} \bibnamefont{Gu}},
  \bibinfo{author}{\bibfnamefont{G.}~\bibnamefont{Xu}},
  \bibinfo{author}{\bibfnamefont{M.}~\bibnamefont{Fujita}}, and
  \bibinfo{author}{\bibfnamefont{K.}~\bibnamefont{Yamada}},
  \bibinfo{year}{2004}{\natexlab{b}}, \bibinfo{journal}{Nature}
  \textbf{\bibinfo{volume}{429}}(\bibinfo{number}{6991}), \bibinfo{pages}{534}.

\bibitem[{\citenamefont{Tremblay} \emph{et~al.}(2006)\citenamefont{Tremblay,
  Kyung, and Senechal}}]{Tremblay06a}
\bibinfo{author}{\bibnamefont{Tremblay}, \bibfnamefont{A.~M.~S.}},
  \bibinfo{author}{\bibfnamefont{B.}~\bibnamefont{Kyung}}, and
  \bibinfo{author}{\bibfnamefont{D.}~\bibnamefont{Senechal}},
  \bibinfo{year}{2006}, \bibinfo{journal}{Low Temp.\ Phys.}
  \textbf{\bibinfo{volume}{32}}(\bibinfo{number}{4-5}), \bibinfo{pages}{424}.

\bibitem[{\citenamefont{Tsuei} \emph{et~al.}(1989)\citenamefont{Tsuei, Gupta,
  and Koren}}]{Tsuei89a}
\bibinfo{author}{\bibnamefont{Tsuei}, \bibfnamefont{C.~C.}},
  \bibinfo{author}{\bibfnamefont{A.}~\bibnamefont{Gupta}}, and
  \bibinfo{author}{\bibfnamefont{G.}~\bibnamefont{Koren}},
  \bibinfo{year}{1989}, \bibinfo{journal}{Physica C}
  \textbf{\bibinfo{volume}{161}}, \bibinfo{pages}{415}.

\bibitem[{\citenamefont{Tsuei and Kirtley}(2000{\natexlab{a}})}]{Tsuei00b}
\bibinfo{author}{\bibnamefont{Tsuei}, \bibfnamefont{C.~C.}}, and
  \bibinfo{author}{\bibfnamefont{J.~R.} \bibnamefont{Kirtley}},
  \bibinfo{year}{2000}{\natexlab{a}}, \bibinfo{journal}{Rev.\ Mod.\ Phys.}
  \textbf{\bibinfo{volume}{72}}(\bibinfo{number}{4}), \bibinfo{pages}{969}.

\bibitem[{\citenamefont{Tsuei and Kirtley}(2000{\natexlab{b}})}]{Tsuei00a}
\bibinfo{author}{\bibnamefont{Tsuei}, \bibfnamefont{C.~C.}}, and
  \bibinfo{author}{\bibfnamefont{J.~R.} \bibnamefont{Kirtley}},
  \bibinfo{year}{2000}{\natexlab{b}}, \bibinfo{journal}{Phys.\ Rev.\ Lett.}
  \textbf{\bibinfo{volume}{85}}(\bibinfo{number}{1}), \bibinfo{pages}{182}.

\bibitem[{\citenamefont{Tsukada} \emph{et~al.}(2005)\citenamefont{Tsukada,
  Krockenberger, Noda, Yamamoto, Manske, Alff, and Naito}}]{Tsukada05a}
\bibinfo{author}{\bibnamefont{Tsukada}, \bibfnamefont{A.}},
  \bibinfo{author}{\bibfnamefont{Y.}~\bibnamefont{Krockenberger}},
  \bibinfo{author}{\bibfnamefont{M.}~\bibnamefont{Noda}},
  \bibinfo{author}{\bibfnamefont{H.}~\bibnamefont{Yamamoto}},
  \bibinfo{author}{\bibfnamefont{D.}~\bibnamefont{Manske}},
  \bibinfo{author}{\bibfnamefont{L.}~\bibnamefont{Alff}}, and
  \bibinfo{author}{\bibfnamefont{M.}~\bibnamefont{Naito}},
  \bibinfo{year}{2005}, \bibinfo{journal}{Solid State Comm.}
  \textbf{\bibinfo{volume}{133}}, \bibinfo{pages}{427}.

\bibitem[{\citenamefont{Tsukada} \emph{et~al.}(2007)\citenamefont{Tsukada,
  Naito, and Yamamoto}}]{Tsukada07a}
\bibinfo{author}{\bibnamefont{Tsukada}, \bibfnamefont{A.}},
  \bibinfo{author}{\bibfnamefont{M.}~\bibnamefont{Naito}}, and
  \bibinfo{author}{\bibfnamefont{H.}~\bibnamefont{Yamamoto}},
  \bibinfo{year}{2007}, \bibinfo{journal}{Physica C}
  \textbf{\bibinfo{volume}{463-465}}, \bibinfo{pages}{64 },
  \bibinfo{note}{proceedings of the 19th International Symposium on
  Superconductivity (ISS 2006)}.

\bibitem[{\citenamefont{Tsunekawa} \emph{et~al.}(2008)\citenamefont{Tsunekawa,
  Sekiyama, Kasai, Imada, Fujiwara, Muro, Onose, Tokura, , and
  Suga}}]{Tsunekawa08a}
\bibinfo{author}{\bibnamefont{Tsunekawa}, \bibfnamefont{M.}},
  \bibinfo{author}{\bibfnamefont{A.}~\bibnamefont{Sekiyama}},
  \bibinfo{author}{\bibfnamefont{S.}~\bibnamefont{Kasai}},
  \bibinfo{author}{\bibfnamefont{S.}~\bibnamefont{Imada}},
  \bibinfo{author}{\bibfnamefont{H.}~\bibnamefont{Fujiwara}},
  \bibinfo{author}{\bibfnamefont{T.}~\bibnamefont{Muro}},
  \bibinfo{author}{\bibfnamefont{Y.}~\bibnamefont{Onose}},
  \bibinfo{author}{\bibfnamefont{Y.}~\bibnamefont{Tokura}}, , and
  \bibinfo{author}{\bibfnamefont{S.}~\bibnamefont{Suga}}, \bibinfo{year}{2008},
  \bibinfo{journal}{New Journal of Physics} \textbf{\bibinfo{volume}{10}},
  \bibinfo{pages}{073005}.

\bibitem[{\citenamefont{Uchiyama} \emph{et~al.}(2004)\citenamefont{Uchiyama,
  Baron, Tsutsui, Tanaka, Hu, Yamamoto, Tajima, and Endoh}}]{Uchiyama04a}
\bibinfo{author}{\bibnamefont{Uchiyama}, \bibfnamefont{H.}},
  \bibinfo{author}{\bibfnamefont{A.~Q.~R.} \bibnamefont{Baron}},
  \bibinfo{author}{\bibfnamefont{S.}~\bibnamefont{Tsutsui}},
  \bibinfo{author}{\bibfnamefont{Y.}~\bibnamefont{Tanaka}},
  \bibinfo{author}{\bibfnamefont{W.-Z.} \bibnamefont{Hu}},
  \bibinfo{author}{\bibfnamefont{A.}~\bibnamefont{Yamamoto}},
  \bibinfo{author}{\bibfnamefont{S.}~\bibnamefont{Tajima}}, and
  \bibinfo{author}{\bibfnamefont{Y.}~\bibnamefont{Endoh}},
  \bibinfo{year}{2004}, \bibinfo{journal}{Phys.\ Rev.\ Lett.}
  \textbf{\bibinfo{volume}{92}}(\bibinfo{number}{19}), \bibinfo{eid}{197005}.

\bibitem[{\citenamefont{Uefuji} \emph{et~al.}(2001)\citenamefont{Uefuji, Kubo,
  Yamada, Fujita, Kurahashi, Watanabe, and Nagamine}}]{Uefuji01a}
\bibinfo{author}{\bibnamefont{Uefuji}, \bibfnamefont{T.}},
  \bibinfo{author}{\bibfnamefont{T.}~\bibnamefont{Kubo}},
  \bibinfo{author}{\bibfnamefont{K.}~\bibnamefont{Yamada}},
  \bibinfo{author}{\bibfnamefont{M.}~\bibnamefont{Fujita}},
  \bibinfo{author}{\bibfnamefont{K.}~\bibnamefont{Kurahashi}},
  \bibinfo{author}{\bibfnamefont{I.}~\bibnamefont{Watanabe}}, and
  \bibinfo{author}{\bibfnamefont{K.}~\bibnamefont{Nagamine}},
  \bibinfo{year}{2001}, \bibinfo{journal}{Physica C}
  \textbf{\bibinfo{volume}{357-360}}(\bibinfo{number}{Part 1}),
  \bibinfo{pages}{208}.

\bibitem[{\citenamefont{Uemura} \emph{et~al.}(1991)\citenamefont{Uemura, Le,
  Luke, Sternlieb, Wu, Brewer, Riseman, Seaman, Maple, Ishikawa, Hinks,
  Jorgensen} \emph{et~al.}}]{Uemura91a}
\bibinfo{author}{\bibnamefont{Uemura}, \bibfnamefont{Y.~J.}},
  \bibinfo{author}{\bibfnamefont{L.~P.} \bibnamefont{Le}},
  \bibinfo{author}{\bibfnamefont{G.~M.} \bibnamefont{Luke}},
  \bibinfo{author}{\bibfnamefont{B.~J.} \bibnamefont{Sternlieb}},
  \bibinfo{author}{\bibfnamefont{W.~D.} \bibnamefont{Wu}},
  \bibinfo{author}{\bibfnamefont{J.~H.} \bibnamefont{Brewer}},
  \bibinfo{author}{\bibfnamefont{T.~M.} \bibnamefont{Riseman}},
  \bibinfo{author}{\bibfnamefont{C.~L.} \bibnamefont{Seaman}},
  \bibinfo{author}{\bibfnamefont{M.~B.} \bibnamefont{Maple}},
  \bibinfo{author}{\bibfnamefont{M.}~\bibnamefont{Ishikawa}},
  \bibinfo{author}{\bibfnamefont{D.~G.} \bibnamefont{Hinks}},
  \bibinfo{author}{\bibfnamefont{J.~D.} \bibnamefont{Jorgensen}},
  \emph{et~al.}, \bibinfo{year}{1991}, \bibinfo{journal}{Phys.\ Rev.\ Lett.}
  \textbf{\bibinfo{volume}{66}}(\bibinfo{number}{20}), \bibinfo{pages}{2665}.

\bibitem[{\citenamefont{Uemura} \emph{et~al.}(1989)\citenamefont{Uemura, Luke,
  Sternlieb, Brewer, Carolan, Hardy, Kadono, Kempton, Kiefl, Kreitzman,
  Mulhern, Riseman} \emph{et~al.}}]{Uemura89a}
\bibinfo{author}{\bibnamefont{Uemura}, \bibfnamefont{Y.~J.}},
  \bibinfo{author}{\bibfnamefont{G.~M.} \bibnamefont{Luke}},
  \bibinfo{author}{\bibfnamefont{B.~J.} \bibnamefont{Sternlieb}},
  \bibinfo{author}{\bibfnamefont{J.~H.} \bibnamefont{Brewer}},
  \bibinfo{author}{\bibfnamefont{J.~F.} \bibnamefont{Carolan}},
  \bibinfo{author}{\bibfnamefont{W.~N.} \bibnamefont{Hardy}},
  \bibinfo{author}{\bibfnamefont{R.}~\bibnamefont{Kadono}},
  \bibinfo{author}{\bibfnamefont{J.~R.} \bibnamefont{Kempton}},
  \bibinfo{author}{\bibfnamefont{R.~F.} \bibnamefont{Kiefl}},
  \bibinfo{author}{\bibfnamefont{S.~R.} \bibnamefont{Kreitzman}},
  \bibinfo{author}{\bibfnamefont{P.}~\bibnamefont{Mulhern}},
  \bibinfo{author}{\bibfnamefont{T.~M.} \bibnamefont{Riseman}}, \emph{et~al.},
  \bibinfo{year}{1989}, \bibinfo{journal}{Phys.\ Rev.\ Lett.}
  \textbf{\bibinfo{volume}{62}}(\bibinfo{number}{19}), \bibinfo{pages}{2317}.

\bibitem[{\citenamefont{Uzumaki} \emph{et~al.}(1991)\citenamefont{Uzumaki,
  Kamehura, and Niwa}}]{Uzumaki91a}
\bibinfo{author}{\bibnamefont{Uzumaki}, \bibfnamefont{X.}},
  \bibinfo{author}{\bibfnamefont{X.}~\bibnamefont{Kamehura}}, and
  \bibinfo{author}{\bibfnamefont{X.}~\bibnamefont{Niwa}}, \bibinfo{year}{1991},
  \bibinfo{journal}{Jap.\ J.\ Appl.\ Phys.} \textbf{\bibinfo{volume}{30}},
  \bibinfo{pages}{981}.

\bibitem[{\citenamefont{Varma}(1997)}]{Varma97a}
\bibinfo{author}{\bibnamefont{Varma}, \bibfnamefont{C.~M.}},
  \bibinfo{year}{1997}, \bibinfo{journal}{Phys.\ Rev.\ B}
  \textbf{\bibinfo{volume}{55}}(\bibinfo{number}{21}), \bibinfo{pages}{14554}.

\bibitem[{\citenamefont{Varma}(1999)}]{Varma99a}
\bibinfo{author}{\bibnamefont{Varma}, \bibfnamefont{C.~M.}},
  \bibinfo{year}{1999}, \bibinfo{journal}{Phys.\ Rev.\ Lett.}
  \textbf{\bibinfo{volume}{83}}(\bibinfo{number}{17}), \bibinfo{pages}{3538}.

\bibitem[{\citenamefont{Varma} \emph{et~al.}(1987)\citenamefont{Varma,
  Schmitt-Rink, and Abrahams}}]{Varma87a}
\bibinfo{author}{\bibnamefont{Varma}, \bibfnamefont{C.~M.}},
  \bibinfo{author}{\bibfnamefont{S.}~\bibnamefont{Schmitt-Rink}}, and
  \bibinfo{author}{\bibfnamefont{E.}~\bibnamefont{Abrahams}},
  \bibinfo{year}{1987}, \bibinfo{journal}{Solid State Comm.}
  \textbf{\bibinfo{volume}{62}}, \bibinfo{pages}{681}.

\bibitem[{\citenamefont{Venturini} \emph{et~al.}(2003)\citenamefont{Venturini,
  Hackl, and Michelucci}}]{Venturini03a}
\bibinfo{author}{\bibnamefont{Venturini}, \bibfnamefont{F.}},
  \bibinfo{author}{\bibfnamefont{R.}~\bibnamefont{Hackl}}, and
  \bibinfo{author}{\bibfnamefont{U.}~\bibnamefont{Michelucci}},
  \bibinfo{year}{2003}, \bibinfo{journal}{Phys.\ Rev.\ Lett.}
  \textbf{\bibinfo{volume}{90}}(\bibinfo{number}{14}), \bibinfo{eid}{149701}.

\bibitem[{\citenamefont{Vigoureux}(1995)}]{Vigoureux95a}
\bibinfo{author}{\bibnamefont{Vigoureux}, \bibfnamefont{P.}},
  \bibinfo{year}{1995}, \emph{\bibinfo{title}{\'Etudes structurales des
  compos\'es supraconducteurs de type n : R$_{2-x}$Ce$_x$CuO$_{4±d}$ (R = Gd,
  Eu, Sm, Nd, Pr); influences des traitements chimiques sur les propri\'et\'es
  physiques}}, Ph.D. thesis, \bibinfo{school}{Universit\'e Paris XI, Orsay,
  France}.

\bibitem[{\citenamefont{Vigoureux} \emph{et~al.}(1997)\citenamefont{Vigoureux,
  Braden, Gukasov, Paulus, Bourges, Cousson, Petitgrand, Lauriat, Meven,
  Barilo, Zhigunov, Adelmann} \emph{et~al.}}]{Vigoureux97a}
\bibinfo{author}{\bibnamefont{Vigoureux}, \bibfnamefont{P.}},
  \bibinfo{author}{\bibfnamefont{M.}~\bibnamefont{Braden}},
  \bibinfo{author}{\bibfnamefont{A.}~\bibnamefont{Gukasov}},
  \bibinfo{author}{\bibfnamefont{W.}~\bibnamefont{Paulus}},
  \bibinfo{author}{\bibfnamefont{P.}~\bibnamefont{Bourges}},
  \bibinfo{author}{\bibfnamefont{A.}~\bibnamefont{Cousson}},
  \bibinfo{author}{\bibfnamefont{D.}~\bibnamefont{Petitgrand}},
  \bibinfo{author}{\bibfnamefont{J.-P.} \bibnamefont{Lauriat}},
  \bibinfo{author}{\bibfnamefont{M.}~\bibnamefont{Meven}},
  \bibinfo{author}{\bibfnamefont{S.~N.} \bibnamefont{Barilo}},
  \bibinfo{author}{\bibfnamefont{D.}~\bibnamefont{Zhigunov}},
  \bibinfo{author}{\bibfnamefont{P.}~\bibnamefont{Adelmann}}, \emph{et~al.},
  \bibinfo{year}{1997}, \bibinfo{journal}{Physica C}
  \textbf{\bibinfo{volume}{273}}, \bibinfo{pages}{239 }.

\bibitem[{\citenamefont{Volovik}(1993)}]{Volovik93a}
\bibinfo{author}{\bibnamefont{Volovik}, \bibfnamefont{G.~E.}},
  \bibinfo{year}{1993}, \bibinfo{journal}{JETP Lett.}
  \textbf{\bibinfo{volume}{58}}, \bibinfo{pages}{469}.

\bibitem[{\citenamefont{Wagenknecht}
  \emph{et~al.}(2008)\citenamefont{Wagenknecht, Koelle, Kleiner, Graser,
  Schopohl, Chesca, Tsukada, Goennenwein, and Gross}}]{Wagenknecht08a}
\bibinfo{author}{\bibnamefont{Wagenknecht}, \bibfnamefont{M.}},
  \bibinfo{author}{\bibfnamefont{D.}~\bibnamefont{Koelle}},
  \bibinfo{author}{\bibfnamefont{R.}~\bibnamefont{Kleiner}},
  \bibinfo{author}{\bibfnamefont{S.}~\bibnamefont{Graser}},
  \bibinfo{author}{\bibfnamefont{N.}~\bibnamefont{Schopohl}},
  \bibinfo{author}{\bibfnamefont{B.}~\bibnamefont{Chesca}},
  \bibinfo{author}{\bibfnamefont{A.}~\bibnamefont{Tsukada}},
  \bibinfo{author}{\bibfnamefont{S.~T.~B.} \bibnamefont{Goennenwein}}, and
  \bibinfo{author}{\bibfnamefont{R.}~\bibnamefont{Gross}},
  \bibinfo{year}{2008}, \bibinfo{journal}{Phys.\ Rev.\ Lett.}
  \textbf{\bibinfo{volume}{100}}(\bibinfo{number}{22}), \bibinfo{eid}{227001}.

\bibitem[{\citenamefont{Wang} \emph{et~al.}(2005)\citenamefont{Wang, Wang, Wu,
  Feng, Luo, and Chen}}]{CHWang05a}
\bibinfo{author}{\bibnamefont{Wang}, \bibfnamefont{C.~H.}},
  \bibinfo{author}{\bibfnamefont{G.~Y.} \bibnamefont{Wang}},
  \bibinfo{author}{\bibfnamefont{T.}~\bibnamefont{Wu}},
  \bibinfo{author}{\bibfnamefont{Z.}~\bibnamefont{Feng}},
  \bibinfo{author}{\bibfnamefont{X.~G.} \bibnamefont{Luo}}, and
  \bibinfo{author}{\bibfnamefont{X.~H.} \bibnamefont{Chen}},
  \bibinfo{year}{2005}, \bibinfo{journal}{Phys.\ Rev.\ B}
  \textbf{\bibinfo{volume}{72}}(\bibinfo{number}{13}), \bibinfo{eid}{132506}.

\bibitem[{\citenamefont{Wang}
  \emph{et~al.}(2006{\natexlab{a}})\citenamefont{Wang, Li, Wu, Chen, Wang, and
  Ding}}]{Wang06a}
\bibinfo{author}{\bibnamefont{Wang}, \bibfnamefont{N.~L.}},
  \bibinfo{author}{\bibfnamefont{G.}~\bibnamefont{Li}},
  \bibinfo{author}{\bibfnamefont{D.}~\bibnamefont{Wu}},
  \bibinfo{author}{\bibfnamefont{X.~H.} \bibnamefont{Chen}},
  \bibinfo{author}{\bibfnamefont{C.~H.} \bibnamefont{Wang}}, and
  \bibinfo{author}{\bibfnamefont{H.}~\bibnamefont{Ding}},
  \bibinfo{year}{2006}{\natexlab{a}}, \bibinfo{journal}{Phys.\ Rev.\ B}
  \textbf{\bibinfo{volume}{73}}(\bibinfo{number}{18}), \bibinfo{eid}{184502}.

\bibitem[{\citenamefont{Wang}
  \emph{et~al.}(2006{\natexlab{b}})\citenamefont{Wang, Li, and
  Ong}}]{YayuWang06a}
\bibinfo{author}{\bibnamefont{Wang}, \bibfnamefont{Y.}},
  \bibinfo{author}{\bibfnamefont{L.}~\bibnamefont{Li}}, and
  \bibinfo{author}{\bibfnamefont{N.~P.} \bibnamefont{Ong}},
  \bibinfo{year}{2006}{\natexlab{b}}, \bibinfo{journal}{Phys.\ Rev.\ B}
  \textbf{\bibinfo{volume}{73}}(\bibinfo{number}{2}), \bibinfo{eid}{024510}.

\bibitem[{\citenamefont{Wang} \emph{et~al.}(2001)\citenamefont{Wang, Revaz,
  Erb, and Junod}}]{Wang01a}
\bibinfo{author}{\bibnamefont{Wang}, \bibfnamefont{Y.}},
  \bibinfo{author}{\bibfnamefont{B.}~\bibnamefont{Revaz}},
  \bibinfo{author}{\bibfnamefont{A.}~\bibnamefont{Erb}}, and
  \bibinfo{author}{\bibfnamefont{A.}~\bibnamefont{Junod}},
  \bibinfo{year}{2001}, \bibinfo{journal}{Phys.\ Rev.\ B}
  \textbf{\bibinfo{volume}{63}}(\bibinfo{number}{9}), \bibinfo{pages}{094508}.

\bibitem[{\citenamefont{Wang} \emph{et~al.}(1991)\citenamefont{Wang, Chien,
  Ong, Tarascon, and Wang}}]{Wang91a}
\bibinfo{author}{\bibnamefont{Wang}, \bibfnamefont{Z.~Z.}},
  \bibinfo{author}{\bibfnamefont{T.~R.} \bibnamefont{Chien}},
  \bibinfo{author}{\bibfnamefont{N.~P.} \bibnamefont{Ong}},
  \bibinfo{author}{\bibfnamefont{J.~M.} \bibnamefont{Tarascon}}, and
  \bibinfo{author}{\bibfnamefont{E.}~\bibnamefont{Wang}}, \bibinfo{year}{1991},
  \bibinfo{journal}{Phys.\ Rev.\ B}
  \textbf{\bibinfo{volume}{43}}(\bibinfo{number}{4}), \bibinfo{pages}{3020}.

\bibitem[{\citenamefont{Weber} \emph{et~al.}(2009)\citenamefont{Weber, Haule,
  and Kotliar}}]{Weber09a}
\bibinfo{author}{\bibnamefont{Weber}, \bibfnamefont{C.}},
  \bibinfo{author}{\bibfnamefont{K.}~\bibnamefont{Haule}}, and
  \bibinfo{author}{\bibfnamefont{G.}~\bibnamefont{Kotliar}},
  \bibinfo{year}{2009}, \bibinfo{journal}{preprint} .

\bibitem[{\citenamefont{Wells} \emph{et~al.}(1995)\citenamefont{Wells, Shen,
  Matsuura, King, Kastner, Greven, and Birgeneau}}]{Wells95a}
\bibinfo{author}{\bibnamefont{Wells}, \bibfnamefont{B.~O.}},
  \bibinfo{author}{\bibfnamefont{Z.-X.} \bibnamefont{Shen}},
  \bibinfo{author}{\bibfnamefont{A.}~\bibnamefont{Matsuura}},
  \bibinfo{author}{\bibfnamefont{D.~M.} \bibnamefont{King}},
  \bibinfo{author}{\bibfnamefont{M.~A.} \bibnamefont{Kastner}},
  \bibinfo{author}{\bibfnamefont{M.}~\bibnamefont{Greven}}, and
  \bibinfo{author}{\bibfnamefont{R.~J.} \bibnamefont{Birgeneau}},
  \bibinfo{year}{1995}, \bibinfo{journal}{Phys.\ Rev.\ Lett.}
  \textbf{\bibinfo{volume}{74}}(\bibinfo{number}{6}), \bibinfo{pages}{964}.

\bibitem[{\citenamefont{Williams} \emph{et~al.}(2005)\citenamefont{Williams,
  Haase, Park, Kim, and Lee}}]{Williams05a}
\bibinfo{author}{\bibnamefont{Williams}, \bibfnamefont{G.~V.~M.}},
  \bibinfo{author}{\bibfnamefont{J.}~\bibnamefont{Haase}},
  \bibinfo{author}{\bibfnamefont{M.-S.} \bibnamefont{Park}},
  \bibinfo{author}{\bibfnamefont{K.~H.} \bibnamefont{Kim}}, and
  \bibinfo{author}{\bibfnamefont{S.-I.} \bibnamefont{Lee}},
  \bibinfo{year}{2005}, \bibinfo{journal}{Phys.\ Rev.\ B}
  \textbf{\bibinfo{volume}{72}}(\bibinfo{number}{21}), \bibinfo{pages}{212511}.

\bibitem[{\citenamefont{Wilson}
  \emph{et~al.}(2006{\natexlab{a}})\citenamefont{Wilson, Dai, Li, Chi, Kang,
  and Lynn}}]{Wilson06b}
\bibinfo{author}{\bibnamefont{Wilson}, \bibfnamefont{S.~D.}},
  \bibinfo{author}{\bibfnamefont{P.}~\bibnamefont{Dai}},
  \bibinfo{author}{\bibfnamefont{S.}~\bibnamefont{Li}},
  \bibinfo{author}{\bibfnamefont{S.}~\bibnamefont{Chi}},
  \bibinfo{author}{\bibfnamefont{H.~J.} \bibnamefont{Kang}}, and
  \bibinfo{author}{\bibfnamefont{J.~W.} \bibnamefont{Lynn}},
  \bibinfo{year}{2006}{\natexlab{a}}, \bibinfo{journal}{Nature}
  \textbf{\bibinfo{volume}{442}}, \bibinfo{pages}{59 }.

\bibitem[{\citenamefont{Wilson}
  \emph{et~al.}(2006{\natexlab{b}})\citenamefont{Wilson, Li, Dai, Bao, Chung,
  Kang, Lee, Komiya, Ando, and Si}}]{Wilson06c}
\bibinfo{author}{\bibnamefont{Wilson}, \bibfnamefont{S.~D.}},
  \bibinfo{author}{\bibfnamefont{S.}~\bibnamefont{Li}},
  \bibinfo{author}{\bibfnamefont{P.}~\bibnamefont{Dai}},
  \bibinfo{author}{\bibfnamefont{W.}~\bibnamefont{Bao}},
  \bibinfo{author}{\bibfnamefont{J.-H.} \bibnamefont{Chung}},
  \bibinfo{author}{\bibfnamefont{H.~J.} \bibnamefont{Kang}},
  \bibinfo{author}{\bibfnamefont{S.-H.} \bibnamefont{Lee}},
  \bibinfo{author}{\bibfnamefont{S.}~\bibnamefont{Komiya}},
  \bibinfo{author}{\bibfnamefont{Y.}~\bibnamefont{Ando}}, and
  \bibinfo{author}{\bibfnamefont{Q.}~\bibnamefont{Si}},
  \bibinfo{year}{2006}{\natexlab{b}}, \bibinfo{journal}{Phys.\ Rev.\ B}
  \textbf{\bibinfo{volume}{74}}(\bibinfo{number}{14}), \bibinfo{eid}{144514}.

\bibitem[{\citenamefont{Wilson}
  \emph{et~al.}(2006{\natexlab{c}})\citenamefont{Wilson, Li, Woo, Dai, Mook,
  Frost, Komiya, and Ando}}]{Wilson06a}
\bibinfo{author}{\bibnamefont{Wilson}, \bibfnamefont{S.~D.}},
  \bibinfo{author}{\bibfnamefont{S.}~\bibnamefont{Li}},
  \bibinfo{author}{\bibfnamefont{H.}~\bibnamefont{Woo}},
  \bibinfo{author}{\bibfnamefont{P.}~\bibnamefont{Dai}},
  \bibinfo{author}{\bibfnamefont{H.~A.} \bibnamefont{Mook}},
  \bibinfo{author}{\bibfnamefont{C.~D.} \bibnamefont{Frost}},
  \bibinfo{author}{\bibfnamefont{S.}~\bibnamefont{Komiya}}, and
  \bibinfo{author}{\bibfnamefont{Y.}~\bibnamefont{Ando}},
  \bibinfo{year}{2006}{\natexlab{c}}, \bibinfo{journal}{Phys.\ Rev.\ Lett.}
  \textbf{\bibinfo{volume}{96}}(\bibinfo{number}{15}), \bibinfo{eid}{157001}.

\bibitem[{\citenamefont{Wilson} \emph{et~al.}(2007)\citenamefont{Wilson, Li,
  Zhao, Mu, Wen, Lynn, Freeman, Regnault, Habicht, and Dai}}]{Wilson07a}
\bibinfo{author}{\bibnamefont{Wilson}, \bibfnamefont{S.~D.}},
  \bibinfo{author}{\bibfnamefont{S.}~\bibnamefont{Li}},
  \bibinfo{author}{\bibfnamefont{J.}~\bibnamefont{Zhao}},
  \bibinfo{author}{\bibfnamefont{G.}~\bibnamefont{Mu}},
  \bibinfo{author}{\bibfnamefont{H.-H.} \bibnamefont{Wen}},
  \bibinfo{author}{\bibfnamefont{J.~W.} \bibnamefont{Lynn}},
  \bibinfo{author}{\bibfnamefont{P.~G.} \bibnamefont{Freeman}},
  \bibinfo{author}{\bibfnamefont{L.-P.} \bibnamefont{Regnault}},
  \bibinfo{author}{\bibfnamefont{K.}~\bibnamefont{Habicht}}, and
  \bibinfo{author}{\bibfnamefont{P.}~\bibnamefont{Dai}}, \bibinfo{year}{2007},
  \bibinfo{journal}{Proceedings of the National Academy of Sciences}
  \textbf{\bibinfo{volume}{104}}(\bibinfo{number}{39}), \bibinfo{pages}{15259}.

\bibitem[{\citenamefont{Woods} \emph{et~al.}(1998)\citenamefont{Woods, Katz,
  de~Andrade, Herrmann, Maple, and Dynes}}]{Woods98a}
\bibinfo{author}{\bibnamefont{Woods}, \bibfnamefont{S.~I.}},
  \bibinfo{author}{\bibfnamefont{A.~S.} \bibnamefont{Katz}},
  \bibinfo{author}{\bibfnamefont{M.~C.} \bibnamefont{de~Andrade}},
  \bibinfo{author}{\bibfnamefont{J.}~\bibnamefont{Herrmann}},
  \bibinfo{author}{\bibfnamefont{M.~B.} \bibnamefont{Maple}}, and
  \bibinfo{author}{\bibfnamefont{R.~C.} \bibnamefont{Dynes}},
  \bibinfo{year}{1998}, \bibinfo{journal}{Phys.\ Rev.\ B}
  \textbf{\bibinfo{volume}{58}}(\bibinfo{number}{13}), \bibinfo{pages}{8800}.

\bibitem[{\citenamefont{Woods} \emph{et~al.}(2002)\citenamefont{Woods, Katz,
  Applebaum, de~Andrade, Maple, and Dynes}}]{Woods02a}
\bibinfo{author}{\bibnamefont{Woods}, \bibfnamefont{S.~I.}},
  \bibinfo{author}{\bibfnamefont{A.~S.} \bibnamefont{Katz}},
  \bibinfo{author}{\bibfnamefont{S.~I.} \bibnamefont{Applebaum}},
  \bibinfo{author}{\bibfnamefont{M.~C.} \bibnamefont{de~Andrade}},
  \bibinfo{author}{\bibfnamefont{M.~B.} \bibnamefont{Maple}}, and
  \bibinfo{author}{\bibfnamefont{R.~C.} \bibnamefont{Dynes}},
  \bibinfo{year}{2002}, \bibinfo{journal}{Phys.\ Rev.\ B}
  \textbf{\bibinfo{volume}{66}}(\bibinfo{number}{1}), \bibinfo{pages}{014538}.

\bibitem[{\citenamefont{Woynarovich}(1982)}]{Woynarovich82a}
\bibinfo{author}{\bibnamefont{Woynarovich}, \bibfnamefont{F.}},
  \bibinfo{year}{1982}, \bibinfo{journal}{J.\ Phys.\ C: Solid State Physics}
  \textbf{\bibinfo{volume}{15}}(\bibinfo{number}{1}), \bibinfo{pages}{97}.

\bibitem[{\citenamefont{Wu} \emph{et~al.}(2009)\citenamefont{Wu, Jin, Yuan,
  Wang, Hatano, Zhao, and Zhu}}]{Wu09a}
\bibinfo{author}{\bibnamefont{Wu}, \bibfnamefont{B.}},
  \bibinfo{author}{\bibfnamefont{K.}~\bibnamefont{Jin}},
  \bibinfo{author}{\bibfnamefont{J.}~\bibnamefont{Yuan}},
  \bibinfo{author}{\bibfnamefont{H.}~\bibnamefont{Wang}},
  \bibinfo{author}{\bibfnamefont{T.}~\bibnamefont{Hatano}},
  \bibinfo{author}{\bibfnamefont{B.}~\bibnamefont{Zhao}}, and
  \bibinfo{author}{\bibfnamefont{B.}~\bibnamefont{Zhu}}, \bibinfo{year}{2009},
  \bibinfo{journal}{Physica C} \textbf{\bibinfo{volume}{469}},
  \bibinfo{pages}{1945}.

\bibitem[{\citenamefont{Wu} \emph{et~al.}(1993)\citenamefont{Wu, Mao, Mao,
  Peng, Xi, Venkatesan, Greene, and Anlage}}]{Wu93a}
\bibinfo{author}{\bibnamefont{Wu}, \bibfnamefont{D.~H.}},
  \bibinfo{author}{\bibfnamefont{J.}~\bibnamefont{Mao}},
  \bibinfo{author}{\bibfnamefont{S.~N.} \bibnamefont{Mao}},
  \bibinfo{author}{\bibfnamefont{J.~L.} \bibnamefont{Peng}},
  \bibinfo{author}{\bibfnamefont{X.~X.} \bibnamefont{Xi}},
  \bibinfo{author}{\bibfnamefont{T.}~\bibnamefont{Venkatesan}},
  \bibinfo{author}{\bibfnamefont{R.~L.} \bibnamefont{Greene}}, and
  \bibinfo{author}{\bibfnamefont{S.~M.} \bibnamefont{Anlage}},
  \bibinfo{year}{1993}, \bibinfo{journal}{Phys.\ Rev.\ Lett.}
  \textbf{\bibinfo{volume}{70}}(\bibinfo{number}{1}), \bibinfo{pages}{85}.

\bibitem[{\citenamefont{Wu} \emph{et~al.}(2006)\citenamefont{Wu, Zhao, Yuan,
  Cao, Zhong, Gao, Xu, Dai, Zhu, Qiu, and Zhao}}]{HWu06a}
\bibinfo{author}{\bibnamefont{Wu}, \bibfnamefont{H.}},
  \bibinfo{author}{\bibfnamefont{L.}~\bibnamefont{Zhao}},
  \bibinfo{author}{\bibfnamefont{J.}~\bibnamefont{Yuan}},
  \bibinfo{author}{\bibfnamefont{L.~X.} \bibnamefont{Cao}},
  \bibinfo{author}{\bibfnamefont{J.~P.} \bibnamefont{Zhong}},
  \bibinfo{author}{\bibfnamefont{L.~J.} \bibnamefont{Gao}},
  \bibinfo{author}{\bibfnamefont{B.}~\bibnamefont{Xu}},
  \bibinfo{author}{\bibfnamefont{P.~C.} \bibnamefont{Dai}},
  \bibinfo{author}{\bibfnamefont{B.~Y.} \bibnamefont{Zhu}},
  \bibinfo{author}{\bibfnamefont{X.~G.} \bibnamefont{Qiu}}, and
  \bibinfo{author}{\bibfnamefont{B.~R.} \bibnamefont{Zhao}},
  \bibinfo{year}{2006}, \bibinfo{journal}{Phys.\ Rev.\ B}
  \textbf{\bibinfo{volume}{73}}(\bibinfo{number}{10}), \bibinfo{eid}{104512}.

\bibitem[{\citenamefont{Wu} \emph{et~al.}(2008)\citenamefont{Wu, Wang, Wu,
  Fang, Luo, Liu, and Chen}}]{TWu08a}
\bibinfo{author}{\bibnamefont{Wu}, \bibfnamefont{T.}},
  \bibinfo{author}{\bibfnamefont{C.~H.} \bibnamefont{Wang}},
  \bibinfo{author}{\bibfnamefont{G.}~\bibnamefont{Wu}},
  \bibinfo{author}{\bibfnamefont{D.~F.} \bibnamefont{Fang}},
  \bibinfo{author}{\bibfnamefont{J.~L.} \bibnamefont{Luo}},
  \bibinfo{author}{\bibfnamefont{G.~T.} \bibnamefont{Liu}}, and
  \bibinfo{author}{\bibfnamefont{X.~H.} \bibnamefont{Chen}},
  \bibinfo{year}{2008}, \bibinfo{journal}{J.\ Phys.\ Condens.\ Matter}
  \textbf{\bibinfo{volume}{20}}(\bibinfo{number}{27}), \bibinfo{pages}{275226}.

\bibitem[{\citenamefont{Xia} \emph{et~al.}(2008)\citenamefont{Xia, Schemm,
  Deutscher, Kivelson, Bonn, Hardy, Liang, Siemons, Koster, Fejer, and
  Kapitulnik}}]{Xia08a}
\bibinfo{author}{\bibnamefont{Xia}, \bibfnamefont{J.}},
  \bibinfo{author}{\bibfnamefont{E.}~\bibnamefont{Schemm}},
  \bibinfo{author}{\bibfnamefont{G.}~\bibnamefont{Deutscher}},
  \bibinfo{author}{\bibfnamefont{S.~A.} \bibnamefont{Kivelson}},
  \bibinfo{author}{\bibfnamefont{D.~A.} \bibnamefont{Bonn}},
  \bibinfo{author}{\bibfnamefont{W.~N.} \bibnamefont{Hardy}},
  \bibinfo{author}{\bibfnamefont{R.}~\bibnamefont{Liang}},
  \bibinfo{author}{\bibfnamefont{W.}~\bibnamefont{Siemons}},
  \bibinfo{author}{\bibfnamefont{G.}~\bibnamefont{Koster}},
  \bibinfo{author}{\bibfnamefont{M.~M.} \bibnamefont{Fejer}}, and
  \bibinfo{author}{\bibfnamefont{A.}~\bibnamefont{Kapitulnik}},
  \bibinfo{year}{2008}, \bibinfo{journal}{Phys.\ Rev.\ Lett.}
  \textbf{\bibinfo{volume}{100}}(\bibinfo{number}{12}), \bibinfo{eid}{127002}.

\bibitem[{\citenamefont{Xiang} \emph{et~al.}(2008)\citenamefont{Xiang, Luo, Lu,
  Shen, and Shen}}]{Xiang08a}
\bibinfo{author}{\bibnamefont{Xiang}, \bibfnamefont{T.}},
  \bibinfo{author}{\bibfnamefont{H.~G.} \bibnamefont{Luo}},
  \bibinfo{author}{\bibfnamefont{D.~H.} \bibnamefont{Lu}},
  \bibinfo{author}{\bibfnamefont{K.~M.} \bibnamefont{Shen}}, and
  \bibinfo{author}{\bibfnamefont{Z.~X.} \bibnamefont{Shen}},
  \bibinfo{year}{2008}, \bibinfo{journal}{arXiv:0807.2498} .

\bibitem[{\citenamefont{Xu} \emph{et~al.}(1996)\citenamefont{Xu, Mao, Jiang,
  Peng, and Greene}}]{Xu96a}
\bibinfo{author}{\bibnamefont{Xu}, \bibfnamefont{X.~Q.}},
  \bibinfo{author}{\bibfnamefont{S.~N.} \bibnamefont{Mao}},
  \bibinfo{author}{\bibfnamefont{W.}~\bibnamefont{Jiang}},
  \bibinfo{author}{\bibfnamefont{J.~L.} \bibnamefont{Peng}}, and
  \bibinfo{author}{\bibfnamefont{R.~L.} \bibnamefont{Greene}},
  \bibinfo{year}{1996}, \bibinfo{journal}{Phys.\ Rev.\ B}
  \textbf{\bibinfo{volume}{53}}(\bibinfo{number}{2}), \bibinfo{pages}{871}.

\bibitem[{\citenamefont{Yagi} \emph{et~al.}(2006)\citenamefont{Yagi, Yoshida,
  Fujimori, Kohsaka, Misawa, Sasagawa, Takagi, Azuma, and Takano}}]{Yagi06a}
\bibinfo{author}{\bibnamefont{Yagi}, \bibfnamefont{H.}},
  \bibinfo{author}{\bibfnamefont{T.}~\bibnamefont{Yoshida}},
  \bibinfo{author}{\bibfnamefont{A.}~\bibnamefont{Fujimori}},
  \bibinfo{author}{\bibfnamefont{Y.}~\bibnamefont{Kohsaka}},
  \bibinfo{author}{\bibfnamefont{M.}~\bibnamefont{Misawa}},
  \bibinfo{author}{\bibfnamefont{T.}~\bibnamefont{Sasagawa}},
  \bibinfo{author}{\bibfnamefont{H.}~\bibnamefont{Takagi}},
  \bibinfo{author}{\bibfnamefont{M.}~\bibnamefont{Azuma}}, and
  \bibinfo{author}{\bibfnamefont{M.}~\bibnamefont{Takano}},
  \bibinfo{year}{2006}, \bibinfo{journal}{Phys.\ Rev.\ B}
  \textbf{\bibinfo{volume}{73}}(\bibinfo{number}{17}), \bibinfo{eid}{172503}.

\bibitem[{\citenamefont{Yamada} \emph{et~al.}(1999)\citenamefont{Yamada,
  Kurahashi, Endoh, Birgeneau, and Shirane}}]{Yamada99a}
\bibinfo{author}{\bibnamefont{Yamada}, \bibfnamefont{K.}},
  \bibinfo{author}{\bibfnamefont{K.}~\bibnamefont{Kurahashi}},
  \bibinfo{author}{\bibfnamefont{Y.}~\bibnamefont{Endoh}},
  \bibinfo{author}{\bibfnamefont{R.~J.} \bibnamefont{Birgeneau}}, and
  \bibinfo{author}{\bibfnamefont{G.}~\bibnamefont{Shirane}},
  \bibinfo{year}{1999}, \bibinfo{journal}{J.\ Phys.\ Chem.\ Sol.}
  \textbf{\bibinfo{volume}{8-9}}(\bibinfo{number}{10}), \bibinfo{pages}{1025}.

\bibitem[{\citenamefont{Yamada} \emph{et~al.}(2003)\citenamefont{Yamada,
  Kurahashi, Uefuji, Fujita, Park, Lee, and Endoh}}]{Yamada03a}
\bibinfo{author}{\bibnamefont{Yamada}, \bibfnamefont{K.}},
  \bibinfo{author}{\bibfnamefont{K.}~\bibnamefont{Kurahashi}},
  \bibinfo{author}{\bibfnamefont{T.}~\bibnamefont{Uefuji}},
  \bibinfo{author}{\bibfnamefont{M.}~\bibnamefont{Fujita}},
  \bibinfo{author}{\bibfnamefont{S.}~\bibnamefont{Park}},
  \bibinfo{author}{\bibfnamefont{S.-H.} \bibnamefont{Lee}}, and
  \bibinfo{author}{\bibfnamefont{Y.}~\bibnamefont{Endoh}},
  \bibinfo{year}{2003}, \bibinfo{journal}{Phys.\ Rev.\ Lett.}
  \textbf{\bibinfo{volume}{90}}(\bibinfo{number}{13}), \bibinfo{eid}{137004}.

\bibitem[{\citenamefont{Yamada} \emph{et~al.}(1998)\citenamefont{Yamada, Lee,
  Kurahashi, Wada, Wakimoto, Ueki, Kimura, Endoh, Hosoya, Shirane, Birgeneau,
  Greven} \emph{et~al.}}]{Yamada98a}
\bibinfo{author}{\bibnamefont{Yamada}, \bibfnamefont{K.}},
  \bibinfo{author}{\bibfnamefont{C.~H.} \bibnamefont{Lee}},
  \bibinfo{author}{\bibfnamefont{K.}~\bibnamefont{Kurahashi}},
  \bibinfo{author}{\bibfnamefont{J.}~\bibnamefont{Wada}},
  \bibinfo{author}{\bibfnamefont{S.}~\bibnamefont{Wakimoto}},
  \bibinfo{author}{\bibfnamefont{S.}~\bibnamefont{Ueki}},
  \bibinfo{author}{\bibfnamefont{H.}~\bibnamefont{Kimura}},
  \bibinfo{author}{\bibfnamefont{Y.}~\bibnamefont{Endoh}},
  \bibinfo{author}{\bibfnamefont{S.}~\bibnamefont{Hosoya}},
  \bibinfo{author}{\bibfnamefont{G.}~\bibnamefont{Shirane}},
  \bibinfo{author}{\bibfnamefont{R.~J.} \bibnamefont{Birgeneau}},
  \bibinfo{author}{\bibfnamefont{M.}~\bibnamefont{Greven}}, \emph{et~al.},
  \bibinfo{year}{1998}, \bibinfo{journal}{Phys.\ Rev.\ B}
  \textbf{\bibinfo{volume}{57}}(\bibinfo{number}{10}), \bibinfo{pages}{6165}.

\bibitem[{\citenamefont{Yamada} \emph{et~al.}(1995)\citenamefont{Yamada,
  Wakimoto, Shirane, Lee, Kastner, Hosoya, Greven, Endoh, and
  Birgeneau}}]{Yamada95a}
\bibinfo{author}{\bibnamefont{Yamada}, \bibfnamefont{K.}},
  \bibinfo{author}{\bibfnamefont{S.}~\bibnamefont{Wakimoto}},
  \bibinfo{author}{\bibfnamefont{G.}~\bibnamefont{Shirane}},
  \bibinfo{author}{\bibfnamefont{C.~H.} \bibnamefont{Lee}},
  \bibinfo{author}{\bibfnamefont{M.~A.} \bibnamefont{Kastner}},
  \bibinfo{author}{\bibfnamefont{S.}~\bibnamefont{Hosoya}},
  \bibinfo{author}{\bibfnamefont{M.}~\bibnamefont{Greven}},
  \bibinfo{author}{\bibfnamefont{Y.}~\bibnamefont{Endoh}}, and
  \bibinfo{author}{\bibfnamefont{R.~J.} \bibnamefont{Birgeneau}},
  \bibinfo{year}{1995}, \bibinfo{journal}{Phys.\ Rev.\ Lett.}
  \textbf{\bibinfo{volume}{75}}(\bibinfo{number}{8}), \bibinfo{pages}{1626}.

\bibitem[{\citenamefont{Yamada} \emph{et~al.}(1994)\citenamefont{Yamada,
  Kinoshita, and Shibata}}]{Yamada94a}
\bibinfo{author}{\bibnamefont{Yamada}, \bibfnamefont{T.}},
  \bibinfo{author}{\bibfnamefont{K.}~\bibnamefont{Kinoshita}}, and
  \bibinfo{author}{\bibfnamefont{H.}~\bibnamefont{Shibata}},
  \bibinfo{year}{1994}, \bibinfo{journal}{Jap.\ J.\ Appl.\ Phys.}
  \textbf{\bibinfo{volume}{33}}(\bibinfo{number}{Part 2, No. 2A}),
  \bibinfo{pages}{L168}.

\bibitem[{\citenamefont{Yamamoto} \emph{et~al.}(1997)\citenamefont{Yamamoto,
  Naito, and Sato}}]{Yamamoto97a}
\bibinfo{author}{\bibnamefont{Yamamoto}, \bibfnamefont{H.}},
  \bibinfo{author}{\bibfnamefont{M.}~\bibnamefont{Naito}}, and
  \bibinfo{author}{\bibfnamefont{H.}~\bibnamefont{Sato}}, \bibinfo{year}{1997},
  \bibinfo{journal}{Phys.\ Rev.\ B}
  \textbf{\bibinfo{volume}{56}}(\bibinfo{number}{5}), \bibinfo{pages}{2852}.

\bibitem[{\citenamefont{Yildirim} \emph{et~al.}(1994)\citenamefont{Yildirim,
  Harris, Entin-Wohlman, and Aharony}}]{Yildirim94a}
\bibinfo{author}{\bibnamefont{Yildirim}, \bibfnamefont{T.}},
  \bibinfo{author}{\bibfnamefont{A.~B.} \bibnamefont{Harris}},
  \bibinfo{author}{\bibfnamefont{O.}~\bibnamefont{Entin-Wohlman}}, and
  \bibinfo{author}{\bibfnamefont{A.}~\bibnamefont{Aharony}},
  \bibinfo{year}{1994}, \bibinfo{journal}{Phys. Rev. Lett.}
  \textbf{\bibinfo{volume}{72}}(\bibinfo{number}{23}), \bibinfo{pages}{3710}.

\bibitem[{\citenamefont{Yildirim} \emph{et~al.}(1996)\citenamefont{Yildirim,
  Harris, and Shender}}]{Yildirim96a}
\bibinfo{author}{\bibnamefont{Yildirim}, \bibfnamefont{T.}},
  \bibinfo{author}{\bibfnamefont{A.~B.} \bibnamefont{Harris}}, and
  \bibinfo{author}{\bibfnamefont{E.~F.} \bibnamefont{Shender}},
  \bibinfo{year}{1996}, \bibinfo{journal}{Phys. Rev. B}
  \textbf{\bibinfo{volume}{53}}(\bibinfo{number}{10}), \bibinfo{pages}{6455}.

\bibitem[{\citenamefont{Yu} \emph{et~al.}(2008)\citenamefont{Yu, Greven,
  Yamamoto} \emph{et~al.}}]{Yu08a}
\bibinfo{author}{\bibnamefont{Yu}, \bibfnamefont{G.}},
  \bibinfo{author}{\bibfnamefont{M.}~\bibnamefont{Greven}},
  \bibinfo{author}{\bibfnamefont{E.}~\bibnamefont{Yamamoto}}, \emph{et~al.},
  \bibinfo{year}{2008}, \bibinfo{journal}{arXiv:0803.3250} .

\bibitem[{\citenamefont{Yu} \emph{et~al.}(1992)\citenamefont{Yu, Salamon, Lu,
  and Lee}}]{Yu92a}
\bibinfo{author}{\bibnamefont{Yu}, \bibfnamefont{R.~C.}},
  \bibinfo{author}{\bibfnamefont{M.~B.} \bibnamefont{Salamon}},
  \bibinfo{author}{\bibfnamefont{J.~P.} \bibnamefont{Lu}}, and
  \bibinfo{author}{\bibfnamefont{W.~C.} \bibnamefont{Lee}},
  \bibinfo{year}{1992}, \bibinfo{journal}{Phys.\ Rev.\ Lett.}
  \textbf{\bibinfo{volume}{69}}(\bibinfo{number}{9}), \bibinfo{pages}{1431}.

\bibitem[{\citenamefont{Yu} \emph{et~al.}(2007{\natexlab{a}})\citenamefont{Yu,
  Higgins, Bach, and Greene}}]{WYu07b}
\bibinfo{author}{\bibnamefont{Yu}, \bibfnamefont{W.}},
  \bibinfo{author}{\bibfnamefont{J.~S.} \bibnamefont{Higgins}},
  \bibinfo{author}{\bibfnamefont{P.}~\bibnamefont{Bach}}, and
  \bibinfo{author}{\bibfnamefont{R.~L.} \bibnamefont{Greene}},
  \bibinfo{year}{2007}{\natexlab{a}}, \bibinfo{journal}{Phys.\ Rev.\ B}
  \textbf{\bibinfo{volume}{76}}(\bibinfo{number}{2}), \bibinfo{eid}{020503}.

\bibitem[{\citenamefont{Yu} \emph{et~al.}(2007{\natexlab{b}})\citenamefont{Yu,
  Higgins, Bach, and Greene}}]{Yu07a}
\bibinfo{author}{\bibnamefont{Yu}, \bibfnamefont{W.}},
  \bibinfo{author}{\bibfnamefont{J.~S.} \bibnamefont{Higgins}},
  \bibinfo{author}{\bibfnamefont{P.}~\bibnamefont{Bach}}, and
  \bibinfo{author}{\bibfnamefont{R.~L.} \bibnamefont{Greene}},
  \bibinfo{year}{2007}{\natexlab{b}}, \bibinfo{journal}{Phys.\ Rev.\ B}
  \textbf{\bibinfo{volume}{76}}(\bibinfo{number}{2}), \bibinfo{eid}{020503}.

\bibitem[{\citenamefont{Yu} \emph{et~al.}(2005)\citenamefont{Yu, Liang, and
  Greene}}]{Yu05a}
\bibinfo{author}{\bibnamefont{Yu}, \bibfnamefont{W.}},
  \bibinfo{author}{\bibfnamefont{B.}~\bibnamefont{Liang}}, and
  \bibinfo{author}{\bibfnamefont{R.~L.} \bibnamefont{Greene}},
  \bibinfo{year}{2005}, \bibinfo{journal}{Phys.\ Rev.\ B}
  \textbf{\bibinfo{volume}{72}}(\bibinfo{number}{21}), \bibinfo{eid}{212512}.

\bibitem[{\citenamefont{Yu} \emph{et~al.}(2006)\citenamefont{Yu, Liang, and
  Greene}}]{Yu06a}
\bibinfo{author}{\bibnamefont{Yu}, \bibfnamefont{W.}},
  \bibinfo{author}{\bibfnamefont{B.}~\bibnamefont{Liang}}, and
  \bibinfo{author}{\bibfnamefont{R.~L.} \bibnamefont{Greene}},
  \bibinfo{year}{2006}, \bibinfo{journal}{Phys.\ Rev.\ B}
  \textbf{\bibinfo{volume}{74}}(\bibinfo{number}{21}), \bibinfo{eid}{212504}.

\bibitem[{\citenamefont{Yuan}
  \emph{et~al.}(2006{\natexlab{a}})\citenamefont{Yuan, Yan, and
  Ting}}]{Yuan06a}
\bibinfo{author}{\bibnamefont{Yuan}, \bibfnamefont{Q.}},
  \bibinfo{author}{\bibfnamefont{X.-Z.} \bibnamefont{Yan}}, and
  \bibinfo{author}{\bibfnamefont{C.~S.} \bibnamefont{Ting}},
  \bibinfo{year}{2006}{\natexlab{a}}, \bibinfo{journal}{Phys.\ Rev.\ B}
  \textbf{\bibinfo{volume}{74}}(\bibinfo{number}{21}), \bibinfo{eid}{214503}.

\bibitem[{\citenamefont{Yuan} \emph{et~al.}(2005)\citenamefont{Yuan, Yuan, and
  Ting}}]{Yuan05a}
\bibinfo{author}{\bibnamefont{Yuan}, \bibfnamefont{Q.}},
  \bibinfo{author}{\bibfnamefont{F.}~\bibnamefont{Yuan}}, and
  \bibinfo{author}{\bibfnamefont{C.~S.} \bibnamefont{Ting}},
  \bibinfo{year}{2005}, \bibinfo{journal}{Phys.\ Rev.\ B}
  \textbf{\bibinfo{volume}{72}}(\bibinfo{number}{5}), \bibinfo{eid}{054504}.

\bibitem[{\citenamefont{Yuan}
  \emph{et~al.}(2006{\natexlab{b}})\citenamefont{Yuan, Yuan, and
  Ting}}]{Yuan06b}
\bibinfo{author}{\bibnamefont{Yuan}, \bibfnamefont{Q.}},
  \bibinfo{author}{\bibfnamefont{F.}~\bibnamefont{Yuan}}, and
  \bibinfo{author}{\bibfnamefont{C.~S.} \bibnamefont{Ting}},
  \bibinfo{year}{2006}{\natexlab{b}}, \bibinfo{journal}{Phys.\ Rev.\ B}
  \textbf{\bibinfo{volume}{73}}(\bibinfo{number}{5}), \bibinfo{eid}{054501}.

\bibitem[{\citenamefont{Zaanen and Gunnarsson}(1989)}]{Zaanen89a}
\bibinfo{author}{\bibnamefont{Zaanen}, \bibfnamefont{J.}}, and
  \bibinfo{author}{\bibfnamefont{O.}~\bibnamefont{Gunnarsson}},
  \bibinfo{year}{1989}, \bibinfo{journal}{Phys.\ Rev.\ B}
  \textbf{\bibinfo{volume}{40}}(\bibinfo{number}{10}), \bibinfo{pages}{7391}.

\bibitem[{\citenamefont{Zaanen} \emph{et~al.}(1985)\citenamefont{Zaanen,
  Sawatzky, and Allen}}]{Zaanen85a}
\bibinfo{author}{\bibnamefont{Zaanen}, \bibfnamefont{J.}},
  \bibinfo{author}{\bibfnamefont{G.~A.} \bibnamefont{Sawatzky}}, and
  \bibinfo{author}{\bibfnamefont{J.~W.} \bibnamefont{Allen}},
  \bibinfo{year}{1985}, \bibinfo{journal}{Phys.\ Rev.\ Lett.}
  \textbf{\bibinfo{volume}{55}}(\bibinfo{number}{4}), \bibinfo{pages}{418}.

\bibitem[{\citenamefont{Zamborszky}
  \emph{et~al.}(2004)\citenamefont{Zamborszky, Wu, Shinagawa, Yu, Balci,
  Greene, Clark, and Brown}}]{Zamborszky04a}
\bibinfo{author}{\bibnamefont{Zamborszky}, \bibfnamefont{F.}},
  \bibinfo{author}{\bibfnamefont{G.}~\bibnamefont{Wu}},
  \bibinfo{author}{\bibfnamefont{J.}~\bibnamefont{Shinagawa}},
  \bibinfo{author}{\bibfnamefont{W.}~\bibnamefont{Yu}},
  \bibinfo{author}{\bibfnamefont{H.}~\bibnamefont{Balci}},
  \bibinfo{author}{\bibfnamefont{R.~L.} \bibnamefont{Greene}},
  \bibinfo{author}{\bibfnamefont{W.~G.} \bibnamefont{Clark}}, and
  \bibinfo{author}{\bibfnamefont{S.~E.} \bibnamefont{Brown}},
  \bibinfo{year}{2004}, \bibinfo{journal}{Phys.\ Rev.\ Lett.}
  \textbf{\bibinfo{volume}{92}}(\bibinfo{number}{4}), \bibinfo{eid}{047003}.

\bibitem[{\citenamefont{Zhang and Rice}(1988)}]{Zhang88a}
\bibinfo{author}{\bibnamefont{Zhang}, \bibfnamefont{F.~C.}}, and
  \bibinfo{author}{\bibfnamefont{T.~M.} \bibnamefont{Rice}},
  \bibinfo{year}{1988}, \bibinfo{journal}{Phys.\ Rev.\ B}
  \textbf{\bibinfo{volume}{37}}(\bibinfo{number}{7}), \bibinfo{pages}{3759}.

\bibitem[{\citenamefont{Zhang}(1997)}]{Zhang97a}
\bibinfo{author}{\bibnamefont{Zhang}, \bibfnamefont{S.-C.}},
  \bibinfo{year}{1997}, \bibinfo{journal}{Science}
  \textbf{\bibinfo{volume}{275}}(\bibinfo{number}{5303}),
  \bibinfo{pages}{1089}.

\bibitem[{\citenamefont{Zhao} \emph{et~al.}(2007)\citenamefont{Zhao, Dai, Li,
  Freeman, Onose, and Tokura}}]{Zhao07a}
\bibinfo{author}{\bibnamefont{Zhao}, \bibfnamefont{J.}},
  \bibinfo{author}{\bibfnamefont{P.}~\bibnamefont{Dai}},
  \bibinfo{author}{\bibfnamefont{S.}~\bibnamefont{Li}},
  \bibinfo{author}{\bibfnamefont{P.~G.} \bibnamefont{Freeman}},
  \bibinfo{author}{\bibfnamefont{Y.}~\bibnamefont{Onose}}, and
  \bibinfo{author}{\bibfnamefont{Y.}~\bibnamefont{Tokura}},
  \bibinfo{year}{2007}, \bibinfo{journal}{Phys.\ Rev.\ Lett.}
  \textbf{\bibinfo{volume}{99}}(\bibinfo{number}{1}), \bibinfo{pages}{017001}.

\bibitem[{\citenamefont{Zhao} \emph{et~al.}(2004)\citenamefont{Zhao, Wu, Miao,
  Yang, Zhang, Qiu, and Zhao}}]{Zhao04a}
\bibinfo{author}{\bibnamefont{Zhao}, \bibfnamefont{L.}},
  \bibinfo{author}{\bibfnamefont{H.}~\bibnamefont{Wu}},
  \bibinfo{author}{\bibfnamefont{J.}~\bibnamefont{Miao}},
  \bibinfo{author}{\bibfnamefont{H.}~\bibnamefont{Yang}},
  \bibinfo{author}{\bibfnamefont{F.~C.} \bibnamefont{Zhang}},
  \bibinfo{author}{\bibfnamefont{X.~G.} \bibnamefont{Qiu}}, and
  \bibinfo{author}{\bibfnamefont{B.~R.} \bibnamefont{Zhao}},
  \bibinfo{year}{2004}, \bibinfo{journal}{Supercond.\ Sci.\ Technol.}
  \textbf{\bibinfo{volume}{17}}(\bibinfo{number}{11}), \bibinfo{pages}{1361}.

\bibitem[{\citenamefont{Zheng} \emph{et~al.}(2003)\citenamefont{Zheng, Sato,
  Kitaoka, Fujita, and Yamada}}]{Zheng03a}
\bibinfo{author}{\bibnamefont{Zheng}, \bibfnamefont{G.-q.}},
  \bibinfo{author}{\bibfnamefont{T.}~\bibnamefont{Sato}},
  \bibinfo{author}{\bibfnamefont{Y.}~\bibnamefont{Kitaoka}},
  \bibinfo{author}{\bibfnamefont{M.}~\bibnamefont{Fujita}}, and
  \bibinfo{author}{\bibfnamefont{K.}~\bibnamefont{Yamada}},
  \bibinfo{year}{2003}, \bibinfo{journal}{Phys.\ Rev.\ Lett.}
  \textbf{\bibinfo{volume}{90}}(\bibinfo{number}{19}), \bibinfo{pages}{197005}.

\bibitem[{\citenamefont{Zimmers} \emph{et~al.}(2004)\citenamefont{Zimmers,
  Lobo, Bontemps, Homes, Barr, Dagan, and Greene}}]{Zimmers04a}
\bibinfo{author}{\bibnamefont{Zimmers}, \bibfnamefont{A.}},
  \bibinfo{author}{\bibfnamefont{R.~P. S.~M.} \bibnamefont{Lobo}},
  \bibinfo{author}{\bibfnamefont{N.}~\bibnamefont{Bontemps}},
  \bibinfo{author}{\bibfnamefont{C.~C.} \bibnamefont{Homes}},
  \bibinfo{author}{\bibfnamefont{M.~C.} \bibnamefont{Barr}},
  \bibinfo{author}{\bibfnamefont{Y.}~\bibnamefont{Dagan}}, and
  \bibinfo{author}{\bibfnamefont{R.~L.} \bibnamefont{Greene}},
  \bibinfo{year}{2004}, \bibinfo{journal}{Phys.\ Rev.\ B}
  \textbf{\bibinfo{volume}{70}}(\bibinfo{number}{13}), \bibinfo{eid}{132502}.

\bibitem[{\citenamefont{Zimmers}
  \emph{et~al.}(2007{\natexlab{a}})\citenamefont{Zimmers, Noat, Cren, Sacks,
  Roditchev, Liang, and Greene}}]{Zimmers07a}
\bibinfo{author}{\bibnamefont{Zimmers}, \bibfnamefont{A.}},
  \bibinfo{author}{\bibfnamefont{Y.}~\bibnamefont{Noat}},
  \bibinfo{author}{\bibfnamefont{T.}~\bibnamefont{Cren}},
  \bibinfo{author}{\bibfnamefont{W.}~\bibnamefont{Sacks}},
  \bibinfo{author}{\bibfnamefont{D.}~\bibnamefont{Roditchev}},
  \bibinfo{author}{\bibfnamefont{B.}~\bibnamefont{Liang}}, and
  \bibinfo{author}{\bibfnamefont{R.~L.} \bibnamefont{Greene}},
  \bibinfo{year}{2007}{\natexlab{a}}, \bibinfo{journal}{Phys.\ Rev.\ B}
  \textbf{\bibinfo{volume}{76}}(\bibinfo{number}{13}), \bibinfo{eid}{132505}.

\bibitem[{\citenamefont{Zimmers}
  \emph{et~al.}(2007{\natexlab{b}})\citenamefont{Zimmers, Shi, Schmadel,
  Fisher, Greene, Drew, Houseknecht, Acbas, Kim, Yang, Cerne, Lin}
  \emph{et~al.}}]{Zimmers07b}
\bibinfo{author}{\bibnamefont{Zimmers}, \bibfnamefont{A.}},
  \bibinfo{author}{\bibfnamefont{L.}~\bibnamefont{Shi}},
  \bibinfo{author}{\bibfnamefont{D.~C.} \bibnamefont{Schmadel}},
  \bibinfo{author}{\bibfnamefont{W.~M.} \bibnamefont{Fisher}},
  \bibinfo{author}{\bibfnamefont{R.~L.} \bibnamefont{Greene}},
  \bibinfo{author}{\bibfnamefont{H.~D.} \bibnamefont{Drew}},
  \bibinfo{author}{\bibfnamefont{M.}~\bibnamefont{Houseknecht}},
  \bibinfo{author}{\bibfnamefont{G.}~\bibnamefont{Acbas}},
  \bibinfo{author}{\bibfnamefont{M.-H.} \bibnamefont{Kim}},
  \bibinfo{author}{\bibfnamefont{M.-H.} \bibnamefont{Yang}},
  \bibinfo{author}{\bibfnamefont{J.}~\bibnamefont{Cerne}},
  \bibinfo{author}{\bibfnamefont{J.}~\bibnamefont{Lin}}, \emph{et~al.},
  \bibinfo{year}{2007}{\natexlab{b}}, \bibinfo{journal}{Phys.\ Rev.\ B}
  \textbf{\bibinfo{volume}{76}}(\bibinfo{number}{6}), \bibinfo{eid}{064515}.

\bibitem[{\citenamefont{Zimmers} \emph{et~al.}(2005)\citenamefont{Zimmers,
  Tomczak, Lobo, Bontemps, Hill, Barr, Dagan, Greene, Millis, and
  Homes}}]{Zimmers05a}
\bibinfo{author}{\bibnamefont{Zimmers}, \bibfnamefont{A.}},
  \bibinfo{author}{\bibfnamefont{J.~M.} \bibnamefont{Tomczak}},
  \bibinfo{author}{\bibfnamefont{R.~P. S.~M.} \bibnamefont{Lobo}},
  \bibinfo{author}{\bibfnamefont{N.}~\bibnamefont{Bontemps}},
  \bibinfo{author}{\bibfnamefont{C.~P.} \bibnamefont{Hill}},
  \bibinfo{author}{\bibfnamefont{M.~C.} \bibnamefont{Barr}},
  \bibinfo{author}{\bibfnamefont{Y.}~\bibnamefont{Dagan}},
  \bibinfo{author}{\bibfnamefont{R.~L.} \bibnamefont{Greene}},
  \bibinfo{author}{\bibfnamefont{A.~J.} \bibnamefont{Millis}}, and
  \bibinfo{author}{\bibfnamefont{C.~C.} \bibnamefont{Homes}},
  \bibinfo{year}{2005}, \bibinfo{journal}{Europhys.\ Lett.}
  \textbf{\bibinfo{volume}{70}}(\bibinfo{number}{2}), \bibinfo{pages}{225}.

\end{thebibliography}

\end{document}